\documentclass[11pt,a4paper]{book}
\usepackage[utf8x]{inputenc}
\usepackage{mathtools}
\usepackage{phdthesis}
\usepackage{hyperref}
\usepackage{caption}
\usepackage{subcaption}
\usepackage{slashed}
\usepackage{bm,graphicx}
\usepackage{dsfont}
\usepackage{amsmath}
\usepackage{amssymb}
\usepackage{color}
\usepackage[utf8x]{inputenc}
\usepackage{layout}

\usepackage{psboxit}
\PScommands
\usepackage{fancyhdr}
\usepackage[avantgarde]{quotchap}
\usepackage{epsfig,epsf}
\usepackage{amsthm}
\usepackage{amsfonts}
\usepackage{lscape}

\usepackage[all]{xy}
\usepackage{enumitem}
\usepackage{cite}
\usepackage{doi}
\usepackage{url}

\usepackage{array}

\usepackage{float}
\newfloat{figtab}{thp}{lop}

\newcommand{\F}{\mathcal{F}}
\newcommand{\Fpm}{\mathcal{F}_\updownarrow}
\newcommand{\Fm}{\mathcal{F}_\uparrow}
\newcommand{\Fp}{\mathcal{F}_\downarrow}

\renewcommand{\S}{\mathcal{S}}
\newcommand{\Sq}{\mathcal{S}_q}
\newcommand{\Sl}{\mathcal{S}_l}

\newcommand{\tcf}{\mathcal{F}}
\newcommand{\tcs}{\mathcal{S}}
\newcommand{\Fup}{\tcf_\uparrow}
\newcommand{\Fdn}{\tcf_\downarrow}
\newcommand{\Fupdn}{\tcf_\updownarrow}
\newcommand{\Fupbar}{\bar{\tcf}_\uparrow}
\newcommand{\Fdnbar}{\bar{\tcf}_\downarrow}

\newcommand{\TC}{\mathrm{TC}}
\newcommand{\hc}{\; + \; \mathrm{h.c.} \;}
\newcommand{\SM}{\mathrm{SM}}

\newcommand{\ubar}{{\bar{u}}}
\newcommand{\dbar}{{\bar{d}}}
\newcommand{\ebar}{{\bar{e}}}
\newcommand{\nubar}{{\bar{\nu}}}

\newcommand{\spur}[2]{\psi^{#1}\phantom{}_{#2} }
\newcommand{\spurbar}[2]{{\psi^\dagger}^{#1 #2}}

\newcommand{\transpose}{^{\mathrm{T}}}

\usepackage{multirow}

\usepackage{tabularx,colortbl}
\usepackage[all]{nowidow}

\newcolumntype{L}[1]{>{\raggedright\let\newline\\\arraybackslash\hspace{0pt}}m{#1}}
\newcolumntype{C}[1]{>{\centering\let\newline\\\arraybackslash\hspace{0pt}}m{#1}}
\newcolumntype{R}[1]{>{\raggedleft\let\newline\\\arraybackslash\hspace{0pt}}m{#1}}

\newcommand{\SU}[1]{{\rm SU}(#1)}
\newcommand{\SUs}[2]{{\rm SU}(#1)_{\rm #2}}
\newcommand{\U}[1]{{\rm U}(#1)}
\newcommand{\Us}[2]{{\rm U}(#1)_{\rm #2}}
\newcommand{\SO}[1]{{\rm SO}(#1)}

\newcommand{\Sp}[1]{{\rm Sp}(#1)}

\DeclareFontFamily{OT1}{pzc}{}
\DeclareFontShape{OT1}{pzc}{m}{it}{<-> s * [1.10] pzcmi7t}{}
\DeclareMathAlphabet{\mathpzc}{OT1}{pzc}{m}{it}

\DeclareFontFamily{U}{mathc}{}
\DeclareFontShape{U}{mathc}{m}{it}{<->s*[1.03] mathc10}{}
\DeclareMathAlphabet{\mathscr}{U}{mathc}{m}{it}

\newcommand{\T}{\mathsf{T}}
\usepackage{rotating}

\newcommand{\axial}{a}
\newcommand{\gb}[1]{{\hat #1}}
\newcommand{\sh}{s_h}
\newcommand{\ch}{c_h}

\newcommand{\setatilde}{\widetilde{s}_\eta}
\newcommand{\cetatilde}{\widetilde{c}_\eta}

\usepackage{feynmp}
\makeatletter
\renewcommand*\env@matrix[1][\arraystretch]{%
  \edef\arraystretch{#1}%
  \hskip -\arraycolsep
  \let\@ifnextchar\new@ifnextchar
  \array{*\c@MaxMatrixCols c}}
\makeatother

\makeatletter
\def\endfmffile{%
  \fmfcmd{\p@rcent\space the end.^^J%
          end.^^J%
          endinput;}%
  \if@fmfio
    \immediate\closeout\@outfmf
  \fi
  \ifnum\pdfshellescape=\@ne
    \immediate\write18{mpost \thefmffile}%
  \fi}
\makeatother

\newcommand{\doublerightxyarrow}{\ar@{-}[rr] |-{\SelectTips{eu}{}\object@{>}}}
\newcommand{\doublerightdownxyarrow}{\ar@{-}[rdr] |-{\SelectTips{eu}{}\object@{>}}}
\newcommand{\doublerightupxyarrow}{\ar@{-}[rur] |-{\SelectTips{eu}{}\object@{>}}}

\newcommand{\doubleleftxyarrow}{\ar@{-}[ll] |-{\SelectTips{eu}{}\object@{>}}}
\newcommand{\tripleleftxyarrow}{\ar@{-}[lll] |-{\SelectTips{eu}{}\object@{>}}}
\newcommand{\triplerightxyarrow}{\ar@{-}[rrr] |-{\SelectTips{eu}{}\object@{>}}}

\newcommand{\rightxyarrow}{\ar@{-}[r] |-{\SelectTips{eu}{}\object@{>}}}
\newcommand{\rightdownxyarrow}{\ar@{-}[rd] |-{\SelectTips{eu}{}\object@{>}}}
\newcommand{\downxyarrow}{\ar@{-}[d] |-{\SelectTips{eu}{}\object@{>}}}
\newcommand{\rightupxyarrow}{\ar@{-}[ru] |-{\SelectTips{eu}{}\object@{>}}}
\newcommand{\upxyarrow}{\ar@{-}[u] |-{\SelectTips{eu}{}\object@{>}}}
\newcommand{\leftupxyarrow}{\ar@{-}[lu] |-{\SelectTips{eu}{}\object@{>}}}
\newcommand{\leftdownxyarrow}{\ar@{-}[ld] |-{\SelectTips{eu}{}\object@{>}}}
\newcommand{\leftxyarrow}{\ar@{-}[l] |-{\SelectTips{eu}{}\object@{>}}}

\newcommand{\longdashedrightxyarrow}{\ar@{--}[rr] |-{\SelectTips{eu}{}\object@{>}}}
\newcommand{\longdashedrightdownxyarrow}{\ar@{--}[rdrd] |-{\SelectTips{eu}{}\object@{>}}}
\newcommand{\longdasheddownxyarrow}{\ar@{--}[dd] |-{\SelectTips{eu}{}\object@{>}}}
\newcommand{\longdashedrightupxyarrow}{\ar@{--}[ruru] |-{\SelectTips{eu}{}\object@{>}}}
\newcommand{\longdashedupxyarrow}{\ar@{--}[uu] |-{\SelectTips{eu}{}\object@{>}}}
\newcommand{\longdashedleftupxyarrow}{\ar@{--}[lulu] |-{\SelectTips{eu}{}\object@{>}}}
\newcommand{\longdashedleftdownxyarrow}{\ar@{--}[ldld] |-{\SelectTips{eu}{}\object@{>}}}
\newcommand{\longdashedleftxyarrow}{\ar@{--}[ll] |-{\SelectTips{eu}{}\object@{>}}}

\newcommand{\altleftxyarrow}{\ar@{-}[r] |-{\SelectTips{eu}{}\object@{<}}}

\def \labeltest #1 {\label{#1}}

\def \tr {\mbox{tr}}

\def \e {\mbox{e}}

\allowdisplaybreaks

\fancypagestyle{plain}{
\fancyhf{}
\fancyfoot[RO]{\psboxit{box 1 setgray fill}
{\hspace{\textwidth}\psboxit{box 0.8 setgray fill}
{\framebox[10mm][c]{\rule[-1mm]{0.0cm}{5mm}\color{black}{\bfseries \thepage}}}}}

}
\makeatletter
\def\cleardoublepage{\clearpage\if@twoside \ifodd\c@page\else
        \hbox{}
        \vspace*{\fill}
        \thispagestyle{empty}
        \newpage
        \if@twocolumn\hbox{}\newpage\fi\fi\fi}
\makeatother

\allowdisplaybreaks

\begin{document}
\pagestyle{empty}
\newpage
\pagestyle{empty}
\begin{titlepage}
\begin{center}

\vspace*{-60pt}
\begin{picture}(440,130)
\put(0,0){\includegraphics[height=130pt]{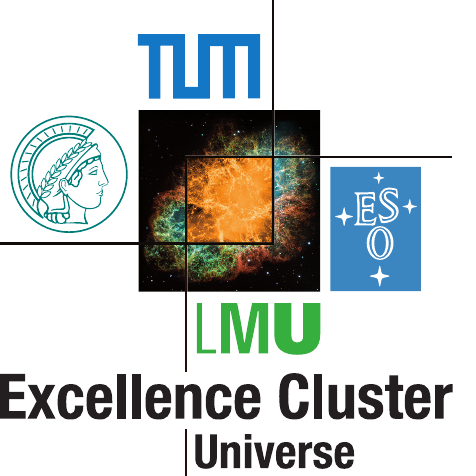}}
\put(360,85){\includegraphics[width=80pt]{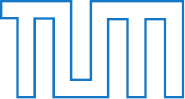}}
\put(370,0){\includegraphics[width=60pt]{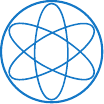}}
\end{picture}
\end{center}
\vspace*{50pt}
\begin{center}
\begin{tabular}{C{400pt}}
\hline\hline\\
{\huge\bf Direct Constraints, Flavor Physics,}\vspace{0.4cm}\\
{\huge\bf and Flavor Anomalies}\vspace{0.4cm}\\
{\huge\bf in Composite Higgs Models}\vspace{0.4cm}\\
\hline\hline
\end{tabular}
\end{center}
\vspace*{30pt}
\begin{center}

{\LARGE \bf Peter Paul Stangl}\\
\end{center}
\vspace*{30pt}
\noindent
Vollst\"andiger Abdruck der von der Fakult\"at f\"ur Physik der Technischen Universit\"at
M\"unchen zur Erlangung des akademischen Grades eines
\vspace*{10pt}
\\
{\bf Doktors der Naturwissenschaften}
\vspace*{10pt}
\\
genehmigten Dissertation.
\vspace*{40pt}
\\
\begin{tabular}{lll}
{\bf Vorsitzender:} && Prof. Dr. Bastian M\"arkisch \\
{\bf Pr\"ufer der Dissertation:} & 1. & Prof. Dr. Andreas Weiler \\
			   & 2. & TUM Junior Fellow Dr. Danny van Dyk
\end{tabular}

\vspace*{40pt}
\noindent
Die Dissertation wurde am 20.02.2018 bei der Technischen Universit\"at M\"unchen
\mbox{eingereicht} und durch die Fakult\"at f\"ur Physik am 08.03.2018 angenommen.

\end{titlepage}

\cleardoublepage

\pagenumbering{roman}

\hyphenation{Aus-schluss}
\hyphenation{zusammen-ge-setzte}

\section*{Zusammenfassung}
Modelle mit zusammengesetztem Higgs-Boson, sogenannte Composite-Higgs-Modelle~(CHM),
bieten eine elegante Lösung für das Natürlichkeitsproblem des Standardmodells~(SM). Ihre direkten Effekte an Teilchenbeschleunigern wie dem Large Hadron Collider (LHC) sind daher von zentralem Interesse.
Während bisher noch keine direkten Effekte beobachtet wurden, gibt es neuere indirekte Hinweise auf neue Physik (NP), die aus Messungen seltener $B$-Meson-Zerfälle resultieren.
Diese Arbeit untersucht sowohl direkte Einschränkungen von CHM durch Experimente an Teilchenbeschleunigern als auch die Frage, ob diese Modelle die Hinweise auf NP in seltenen $B$-Zerfällen erklären können.
Der erste Teil dieser Arbeit gibt eine in sich abgeschlossene Einführung in alle Hauptkonzepte von CHM, die in der weiteren Arbeit verwendet werden.
Diese Konzepte werden dann in mehreren phänomenologischen Analysen angewendet.
Im Rahmen globaler numerischer Analysen von zwei expliziten CHM werden detailliert die direkten Einschränkungen untersucht, die aus experimentellen Suchen nach Vektor-, Fermion- und Skalar-Resonanzen resultieren.
Die Aussichten auf eine Beobachtung oder den Ausschluss momentan noch realisierbarer Parameterpunkte der betrachteten Modelle werden für verschiedene Zerfallskanäle diskutiert.
Modellunabhängige Analysen der Hinweise auf NP in seltenen $B$-Zerfällen werden durchgeführt, die sowohl die ${B\to K^* \mu^ + \mu^-}$-Anomalie untersuchen, als auch die Hinweise auf eine Verletzung der Lepton-Flavor-Univer-salität (LFU), die in Messungen der Observablen $R_{K^{(*)}}$ gefunden wurden.
Ein einfaches CHM wird vorgestellt, das die Anomalien in seltenen $B$-Zerfällen durch teilweise zusammengesetzte linkshändige Myonen erklären kann.
Die Flavor-Physik eines viel ehrgeizigeren Modells, das auf einer UV-Vervollständigung effektiver CHM basiert, die als Fundamental-Partial-Compositness~(FPC) bezeichnet wird, wird im Detail untersucht.
Unter Berücksichtigung aller relevanten Einschränkungen durch die Physik an der elektroschwachen Skala und die Niedrigenergie-Flavorobservablen wird gezeigt, dass dieses Modell die Anomalien in seltenen $B$-Zerfällen erklären kann.

\cleardoublepage
\section*{Abstract}
Composite Higgs models (CHMs) offer an elegant solution to the naturalness problem of the Standard Model (SM). Their direct effects at particle colliders like the Large Hadron Collider~(LHC) are thus of central interest.
While no direct effects have been observed so far, there are recent indirect hints for new physics (NP) coming from measurements of rare $B$~meson decays.
This thesis studies direct collider constraints on CHMs as well as the question if these models can explain the hints for NP in rare $B$ decays.
The first part of this thesis gives a self-contained introduction to all main concepts of CHMs used in the remainder of the work.
These concepts are then applied in several phenomenological analyses.
In the context of global numerical analyses of two explicit CHMs, the direct constraints on vector, fermion and scalar resonances are studied in detail.
The prospects of various decay channels for observing or excluding still viable parameter points of the considered models are discussed.
Model independent analyses of the hints for NP in rare $B$ decays are performed in the context of the $B\to K^*\mu^+\mu^-$ anomaly as well as the hints for violation of lepton flavor universality~(LFU) found in measurements of the observables $R_{K^{(*)}}$.
A simple CHM is presented that can explain the anomalies in rare $B$ decays by partially composite left-handed muons.
The flavor physics of a much more ambitious model, which is based on a UV completion of effective CHMs called fundamental partial compositeness (FPC), is investigated in detail.
Taking into account all relevant constraints from electroweak scale physics and low-energy flavor observables, it is shown that this model can explain the anomalies found in rare $B$~decays.

\cleardoublepage

\thispagestyle{empty}
\vspace*{\fill}
\section*{Publications within the context of this dissertation}
\begin{itemize}

  \item [\cite{Niehoff:2015bfa}]
C.~Niehoff, P.~Stangl and D.~M. Straub, \emph{{Violation of lepton flavour
  universality in composite Higgs models}},
  \href{https://doi.org/10.1016/j.physletb.2015.05.063}{\emph{Phys. Lett.}
  {\bfseries B747} (2015) 182--186},
  [\href{https://arxiv.org/abs/1503.03865}{{\ttfamily 1503.03865}}].

  \item [\cite{Niehoff:2015iaa}]
C.~Niehoff, P.~Stangl and D.~M. Straub, \emph{{Direct and indirect signals of
  natural composite Higgs models}},
  \href{https://doi.org/10.1007/JHEP01(2016)119}{\emph{JHEP} {\bfseries 01}
  (2016) 119}, [\href{https://arxiv.org/abs/1508.00569}{{\ttfamily
  1508.00569}}].

  \item [\cite{Niehoff:2016zso}]
C.~Niehoff, P.~Stangl and D.~M. Straub, \emph{{Electroweak symmetry breaking
  and collider signatures in the next-to-minimal composite Higgs model}},
  \href{https://doi.org/10.1007/JHEP04(2017)117}{\emph{JHEP} {\bfseries 04}
  (2017) 117}, [\href{https://arxiv.org/abs/1611.09356}{{\ttfamily
  1611.09356}}].

  \item [\cite{Altmannshofer:2017fio}]
W.~Altmannshofer, C.~Niehoff, P.~Stangl and D.~M. Straub, \emph{{Status of the
  $B\rightarrow K^*\mu ^+\mu ^-$ anomaly after Moriond 2017}},
  \href{https://doi.org/10.1140/epjc/s10052-017-4952-0}{\emph{Eur. Phys. J.}
  {\bfseries C77} (2017) 377},
  [\href{https://arxiv.org/abs/1703.09189}{{\ttfamily 1703.09189}}].

 \item [\cite{Altmannshofer:2017yso}]
W.~Altmannshofer, P.~Stangl and D.~M. Straub, \emph{{Interpreting Hints for
  Lepton Flavor Universality Violation}},
  \href{https://doi.org/10.1103/PhysRevD.96.055008}{\emph{Phys. Rev.}
  {\bfseries D96} (2017) 055008},
  [\href{https://arxiv.org/abs/1704.05435}{{\ttfamily 1704.05435}}].

 \item [\cite{Sannino:2017utc}]
F.~Sannino, P.~Stangl, D.~M. Straub and A.~E. Thomsen, \emph{{Flavor Physics
  and Flavor Anomalies in Minimal Fundamental Partial Compositeness}},
  \href{https://arxiv.org/abs/1712.07646}{{\ttfamily 1712.07646}}.

\end{itemize}

\cleardoublepage

\vspace*{\fill}
\begin{quote}
\sffamily
{There is a sort of general distinction between the way to look for weakly interacting theories of spontaneous symmetry-breaking and strongly interacting
theories of spontaneous symmetry-breaking. When things are strongly interacting it shows up in deviations in the properties of the particles, in the set of
particles that you have already produced, from what you expect.
If things are weakly interacting, new physics first shows up in the form of new particles.
}
\qauthor{Howard Georgi, 1994}
\end{quote}
\vspace*{\fill}

\cleardoublepage

\pagestyle{fancy}
\tableofcontents
\cleardoublepage
\pagestyle{fancy}
\pagenumbering{arabic}

\chapter{Introduction}
%
Learning about the physics of the most fundamental particles in nature has always been an interplay between experimental and theoretical efforts.
There has been progress that was mainly driven by experimental data, showing completely new, often unexpected phenomena, and necessitating theoretical explanations.
A historic example is the large amount of hadrons found in 1950s and 1960s accelerator experiments, which led to their classification in terms of representations of an SU$(3)$ flavor symmetry~\cite{Neeman:1961jhl,GellMann:1962xb} that paved the way for the quark model~\cite{GellMann:1964nj,Zweig:352337,Zweig:1964jf}, and led to the theory of the strong interaction now known as quantum chromodynamics~(QCD).
On the other hand, theoretical considerations led to predictions that served as guidance for experimental explorations.
The probably most famous example from recent history is the
Higgs mechanism\footnote{%
This mechanism for giving mass to vector bosons was described independently by Anderson~\cite{Anderson:1963pc}, Englert and Brout~\cite{Englert:1964et}, Higgs~\cite{Higgs:1964ia,Higgs:1964pj}, and Guralnik, Hagen and Kibble~\cite{Guralnik:1964eu}, but is for simplicity usually only called the ``Higgs mechanism''.
}~\cite{Anderson:1963pc,Englert:1964et,Higgs:1964ia,Higgs:1964pj,Guralnik:1964eu}
that is responsible for electroweak symmetry breaking (EWSB) in the Standard Model (SM) and has been a main motivation for building the Large Hadron Collider (LHC)~\cite{Asner:1984jv}.
This in turn led to the experimental discovery of a particle compatible with the SM Higgs boson~\cite{Aad:2012tfa,Chatrchyan:2012xdj}.
The Higgs mechanism in the SM was more or less a guarantee for finding something new at the LHC, either the Higgs boson or something else.
Although there is no such strong case for further new discoveries at the LHC, there is at least a theoretical consideration that indicates new physics (NP) beyond the SM at a scale of the order of some TeV: the so called {\it naturalness problem}.
And while there is no such clear sign of new particles at accessible energies like in the 1950s and 1960s, there are at least some recent experimental results in flavor physics that hint at NP below a scale of roughly 100~TeV.

Among the prime candidates for solving the naturalness problem are models in which the Higgs boson is a composite object.
These composite Higgs models (CHMs) will be analysed in detail in this thesis.
While they are mainly motivated by the naturalness problem, it is interesting to investigate whether also the experimental hints for NP can be explained in the CHM context.
%
%
%
%
At least, if these hints should get established, any viable NP model has to accommodate them.
%
%
%
%
%

\section{The SM Higgs sector and the naturalness problem}
The origin of the naturalness problem is closely related to the structure of the SM and its Higgs sector.
Among the first steps towards the SM was the unification of the electromagnetic and the weak interaction.
The mediators of the weak interaction, $W_\mu^+$, $W_\mu^-$, and $Z_\mu$, and the mediator of the  electromagnetic interaction, the photon $A_\mu$, were described in terms of the four generators of the unified electroweak (EW) symmetry group ${\rm SU}(2)_{\rm L}\times {\rm U}(1)_{\rm Y}$ already in the early 1960s~\cite{Glashow:1961tr}.
However, an integral part in constructing the SM was to incorporate the Higgs mechanism as a means to give mass to $W_\mu^{\pm}$ and $Z_\mu$ in a renormalizable way~\cite{Weinberg:1967tq,Salam:1968rm}.
In doing this, the Higgs mechanism builds upon the notion of spontaneous symmetry breaking.
When the Lagrangian of a given theory possesses a global symmetry\footnote{The symmetry groups considered here are always assumed to be connected and compact Lie groups.} $G$, but the vacuum of the theory is only invariant under a subgroup $H\subset G$, the global symmetry $G$ is said to be broken spontaneously to $H$.
In this case, the theory contains a massless scalar for each broken generator of $G$, called a {\it Nambu-Goldstone boson} (NGB)~\cite{Nambu:1961tp,Goldstone:1961eq,Goldstone:1962es}.
There are, however, ways to prevent massless NGBs from appearing.
One way is having terms in the Lagrangian that explicitly break the $G$ symmetry weakly.
This can lead to an effective potential that yields a small mass for a NGB, which is then called {\it pseudo Nambu-Goldstone boson} (pNGB)~\cite{Weinberg:1972fn}.
Another way is to gauge broken generators.
In this case, a NGB corresponding to a gauged broken generator can be removed from the theory by choosing an appropriate gauge, called the {\it unitary gauge}, and such a NGB is called {\it would-be Nambu-Goldstone boson} or {\it fictitious Nambu-Goldstone boson}.
At the same time, the gauge boson associated with the broken generator becomes massive.
This mechanism that gives mass to a gauge boson and turns a NGB into an unphysical would-be NGB is nothing but the Higgs mechanism~\cite{Anderson:1963pc,Englert:1964et,Higgs:1964ia,Higgs:1964pj,Guralnik:1964eu}.

For the three vector bosons $W^+$, $W^-$, and $Z$ to become massive, three of the four generators of ${\rm SU}(2)_{\rm L}\times {\rm U}(1)_{\rm Y}$ have to be spontaneously broken, while the ${\rm U}(1)_{\rm Q}$ generator $Q$ that corresponds to the massless photon must be unbroken.
The latter is found to be
\begin{equation}\label{eq:intro:hypercharge}
 Q = t_3 + Y,
\end{equation}
where $Y$ is the generator of ${\rm U}(1)_{\rm Y}$, $t_a=\sigma_a/2$, $a\in\{1,2,3\}$ are the generators of ${\rm SU}(2)_{\rm L}$, and $\sigma_a$ are the Pauli matrices.
An economical way to achieve
EWSB
is to employ a complex scalar field $\Phi(x)$, the so called {\it Higgs field}.
The Higgs field transforms as a doublet under ${\rm SU}(2)_{\rm L}$ and has a ${\rm U}(1)_{\rm Y}$ charge $q_{\rm Y}=1/2$.
This implies that one component of the doublet has vanishing electric charge and is therefore invariant under ${\rm U}(1)_{\rm Q}$ (cf.~eq.~(\ref{eq:intro:hypercharge})).
If this component has a finite value at the minimum of the Higgs potential, which corresponds to the vacuum of the theory, then the EW symmetry is broken spontaneously to its electromagnetic subgroup.
This can be realized by the Higgs Lagrangian
\begin{equation}\label{eq:intro:L_Higgs}
 \mathcal{L}_{\rm Higgs} = \big(D_\mu\,\Phi(x)\big)^\dagger\, \big(D^\mu\,\Phi(x)\big)
 -\lambda\,\bigg(\Phi^\dagger(x)\,\Phi(x)-\frac{v^2}{2}\bigg)^2,
\end{equation}
with
the covariant derivative\footnote{%
Repeated indices only appearing on one side of an equation are assumed to be summed over.
}
\begin{equation}
 D_\mu\,\Phi(x) = \left(\partial_\mu-i\,g\,t_a\,W_\mu^a-i\,g'\,\frac{1}{2}\,B_\mu\right)\,\Phi(x),
\end{equation}
where $g$ and $g'$ are the gauge couplings of the gauge fields $W_\mu^a$ and $B_\mu$, which are associated with ${\rm SU}(2)_{\rm L}$ and ${\rm U}(1)_{\rm Y}$, respectively.
The minimum of the potential term in $\mathcal{L}_{\rm Higgs}$ is found for $|\Phi(x)|=\frac{v}{\sqrt{2}}$ and the choice of the generator basis in terms of the Pauli matrices implies that the vacuum configuration of the Higgs field that breaks the EW symmetry to ${\rm U}(1)_{\rm Q}$ is\footnote{%
Any other choice of $\Phi_0$ that satisfies $|\Phi_0|=\frac{v}{\sqrt{2}}$ is physically equivalent but requires a different choice of the generator basis of ${\rm SU}(2)_{\rm L}$.
The generator $t_3$ is fixed via eq.~(\ref{eq:intro:hypercharge}) by the requirement that $\Phi_0$ is invariant under ${\rm U}(1)_{\rm Q}$, i.e.\ $Q\,\Phi_0=0$.
}
\begin{equation}
 \Phi_0=\frac{1}{\sqrt{2}}\begin{pmatrix}
         0\\
         v
        \end{pmatrix}.
\end{equation}
The quantity $v$ is called the {\it vacuum expectation value} (VEV)\footnote{%
While $\Phi(x)$ is treated here merely as a classical field, the term ``expectation value'' is used for convenience and corresponds to an actual quantum mechanical expectation value when $\Phi(x)$ is promoted to a field operator.
} of the Higgs field.
The four real degrees of freedom of the complex Higgs doublet are reduced to only one by employing the unitary gauge, where the three would-be NGBs, which are associated with the three gauged broken generators, are removed.
The remaining real scalar field~$h(x)$ describes a massive particle called the {\it Higgs boson}, which corresponds to fluctuations about the vacuum in the ${\rm U}(1)_{\rm Q}$ invariant component of $\Phi(x)$, i.e.\ in unitary gauge, one finds
\begin{equation}\label{eq:intro:Higgs_field_unitary_gauge}
 \Phi(x)=\frac{1}{\sqrt{2}}\begin{pmatrix}
         0\\
         v+h(x)
        \end{pmatrix}.
\end{equation}
Plugging this into the Higgs Lagrangian, eq.~(\ref{eq:intro:L_Higgs}), yields mass terms for the EW gauge bosons.
The mass eigenstates in terms of the $W_\mu^a$ and $B_\mu$ fields are
\begin{equation}
 W^{\pm}_\mu=\frac{1}{\sqrt{2}}(W_\mu^1\mp i\, W_\mu^2),
 \quad
 Z_\mu = \cos\theta_W\,W_\mu^3 - \sin\theta_W\,B_\mu,
 \quad
 A_\mu = \cos\theta_W\,B_\mu + \sin\theta_W\,W_\mu^3,
\end{equation}
where $\cos\theta_W = \frac{g}{\sqrt{g^2+g'^2}}$ and $\sin\theta_W = \frac{g'}{\sqrt{g^2+g'^2}}$,
and the masses are
\begin{equation}\label{eq:intro:gauge_boson_masses}
 m_W = \frac{1}{2}\,g\,v,
 \quad\quad
 m_Z
 = \frac{m_W}{\cos\theta_W},
 \quad\quad
 m_A = 0.
\end{equation}
As expected, the gauge bosons associated with the broken generators become massive,
while the photon $A_\mu$ that corresponds to the unbroken ${\rm U}(1)_{\rm Q}$ stays massless.

However, not only the gauge bosons become massive, but also the Higgs boson receives a mass proportional to the Higgs VEV:
\begin{equation}
 m_h = \sqrt{2\,\lambda}\,v.
\end{equation}
In the SM, also the fermions are coupled to the Higgs field via Yukawa interactions.
Hence, they also gain masses proportional to the Higgs VEV.
Actually every single massive elementary particle in the SM has a mass proportional to the Higgs VEV.
This has an important consequence that can readily be inferred from eq.~(\ref{eq:intro:Higgs_field_unitary_gauge}) and may be regarded as the origin of the naturalness problem: all these particles couple to the elementary scalar Higgs boson~$h(x)$ with coupling strengths proportional to their masses, and this also means that they all contribute to the Higgs mass via quantum corrections.
Because there must be some kind of NP beyond the SM, at least around the Planck scale $M_{\rm Pl}\approx10^{19}$~GeV where gravity cannot be neglected, the SM is usually considered to be an effective theory, only viable up to a cutoff scale $\Lambda$.
The contributions due to quantum corrections that the SM Higgs mass~$m_h$ receives to its bare mass%
~$m_0$ depend on this cutoff, such that\footnote{%
%
%
%
The discussion of the naturalness problem presented here is based on \cite{Susskind:1978ms,Giudice:2008bi}.
}
\begin{equation}\label{eq:intro:Higgs_mass_correction}
 m_h^2 = m_0^2 + \kappa\,\Lambda^2,
\end{equation}
where $\kappa$ is some dimensionless constant typically of order $10^{-2}$~\cite{Giudice:2008bi}.
The ratio of the bare Higgs mass to the cutoff scale is a dimensionless parameter of the UV theory that replaces the SM above the scale~$\Lambda$.
%
From eq.~(\ref{eq:intro:Higgs_mass_correction}), this parameter is found to be
\begin{equation}\label{eq:intro:mass_lambda_ratio}
 \frac{m_0^2}{\Lambda^2} = -\kappa\,\left(1-\frac{m_h^2}{\kappa\,\Lambda^2}\right).
\end{equation}
The Higgs mass $m_h$ in the SM has to be around the EW scale, i.e. of the order of 100~GeV, and a particle compatible with this requirement has been found at 125~GeV~\cite{Patrignani:2016xqp}.
Plugging this into eq.~(\ref{eq:intro:mass_lambda_ratio}) and assuming that there is no NP up to the Planck scale, i.e.\ $\Lambda = M_{\rm Pl}$, one finds
\begin{equation}
 \frac{m_0^2}{\Lambda^2} = -\kappa\,\left(1-10^{-32}\right).
\end{equation}
This means that $\frac{m_0^2}{\Lambda^2}$
has to be tuned to the 32nd decimal place to yield the correct Higgs mass.
Such an extreme tuning is considered to be unnatural for a parameter in the UV theory.
%
A natural theory in this sense requires that a small change to a parameter like $\frac{m_0^2}{\Lambda^2}$ does not have a dramatic consequence on an observable like the Higgs mass.
The naturalness problem of the SM therefore is the problem that it seemingly leads to an unnatural theory in the UV.
%

An obvious solution to the naturalness problem of the SM is a cutoff much lower than the Planck scale, say around some TeV.
But then the UV theory above $\Lambda$ has to contain some mechanism that protects the Higgs mass from further quantum corrections stemming from higher scales.
An example for such a theory is one that is supersymmetric%
\footnote{%
For an introduction to supersymmetric theories, see e.g.~\cite{Martin:1997ns}.
}.
In a supersymmetric theory, quantum corrections to the Higgs mass that are induced by fermions are exactly canceled by those stemming from their bosonic superpartners, and vice
versa, at least if the superpartners are not much heavier than the EW scale.
There are many other possible solutions to the naturalness problem.
Among them are e.g.\ ideas like that the Planck scale is actually close to the EW scale and just seems much larger due to extra dimensions~\cite{ArkaniHamed:1998rs,Randall:1999ee}, or that dynamics in the early universe are responsible for a Higgs boson mass much smaller than the cutoff~\cite{Graham:2015cka}, or that the scalar Higgs sector itself dynamically generates a new cutoff scale of the order of 100~TeV~\cite{Khoze:2017tjt}.

There is another class of models in which the naturalness problem is avoided by not introducing the elementary scalar Higgs field as a means to break the EW symmetry.
%
%
%
%
%
Notable examples of this kind are so called {\it Technicolor} (TC) models~\cite{Weinberg:1975gm,Susskind:1978ms} that
do not require a Higgs boson at all
and
break the EW symmetry by a condensate that is due to a new strong interaction.
While their early versions were disfavored because of problems with experimentally excluded large flavor-changing neutral currents (FCNCs)~\cite{Eichten:1979ah,Dimopoulos:1980fj}, the discovery of a scalar particle with properties similar to the SM Higgs boson completely rules out any Higgsless TC model.
The Higgs boson, however, might not be an elementary particle but actually a bound state of some TC like strong interaction that is also responsible for EWSB.
The virtue of such a composite Higgs boson is that it does not suffer at all from the naturalness problem of an elementary scalar
.
Any possible corrections to its mass are cut off at the scale where a more fundamental theory in terms of its constituents takes over.
A strongly interacting theory that yields a composite Higgs bound state generically predicts also other bound states that have not been observed so far.
However, an explanation for the  lightness of the Higgs boson compared to other bound states can be provided by assuming that the Higgs is a composite pNGB~\cite{Kaplan:1983fs,Kaplan:1983sm,Banks:1984gj,Georgi:1984ef,Georgi:1984af,Dugan:1984hq}.
In the following, when referring to composite Higgs models (CHMs), this assumption will always be made.
%
In addition to solving the naturalness problem, these CHMs have also other interesting features:
While the Higgs sector in the SM is introduced more or less ad hoc for breaking the EW symmetry, its origin is explained in CHMs in terms of a more fundamental, strongly coupled theory.
This also means that SM parameters like the Higgs mass are in principle calculable in CHMs.
Because masses and mixings of SM fermions are intimately interrelated with the Higgs sector, CHMs offer the possibility to get some insights into the generation of fermion masses and mixings.

Since CHMs generically predict composite vector and fermion resonances that might be accessible at the LHC, direct constraints from experimental analyses are an important test of the viability of CHMs.
Studying the impact of these constraints on realistic CHMs is one of the main topics of this thesis.
In light of the afore mentioned experimental hints for NP, another main topic is to investigate whether these hints can be explained by CHMs.
%
%

\section{Flavor anomalies and hints for new physics}
Already in 2013, the LHCb collaboration reported a tension between their measurement of the $B\to K^*\mu^+\mu^-$ angular observable $P_5'$ and the SM prediction~\cite{Aaij:2013qta}.
Several groups subsequently showed that this could be explained by a NP contribution to a single Wilson coefficient~(WC) of the weak effective Hamiltonian (WEH)\footnote{%
The weak effective Hamiltonian describes an effective theory derived from the SM by integrating out all degrees of freedom heavier than the bottom quark, i.e.\ the weak gauge bosons, the Higgs boson, and the top quark.
}, modifying the quark-level transition ${b\to s\,\ell^+\ell^-}$~\cite{Descotes-Genon:2013wba,Altmannshofer:2013foa,Beaujean:2013soa,Hurth:2013ssa}.
Various branching ratio measurements of processes that involve the same quark-level transition have since showed deviations from the SM~\cite{Aaij:2014pli,Aaij:2015esa} and the LHCb measurement of $P_5'$ with the full Run 1 dataset, presented in 2015, confirmed the previously found tensions~\cite{Aaij:2015oid}.
While updated global analyses showed the consistency of all these tensions with a NP explanation, hadronic uncertainties in SM calculations cannot be excluded as the origin of the tensions~\cite{Altmannshofer:2014rta,Descotes-Genon:2015uva,Hurth:2016fbr}.

However, ratios of $b\to s\,\ell^+\ell^-$ branching ratios with different leptons in the final states are practically free of hadronic uncertainties (cf.~\cite{Hiller:2003js,Bobeth:2007dw,Hiller:2014ula}).
In addition, these observables are sensitive to a violation of lepton flavor universality (LFU), and observation of LFU violation (LFUV) would be a clear sign of NP.
Interestingly, first measurements of two of such ratios, the LFU observables $R_K$ and $R_{K^*}$, show a deviation from the SM, each with a significance of around 2$\sigma$~\cite{Aaij:2014ora,Aaij:2017vbb}.
This result is tantalizing because it is fully compatible with the NP explanation of the other tensions found in angular observables and branching ratios and can be further tested with new measurements at LHCb and the upcoming Belle~II experiment.

Apart from the neutral current transition $b\to s\,\ell^+\ell^-$, there are also hints for NP in charged current $b\to c\,\ell^-\,\nu$ transitions, namely in the LFU observables $R_D$ and $R_{D^*}$.
With first evidence reported by the BaBar collaboration in 2012 and 2013~\cite{Lees:2012xj,Lees:2013uzd}, several measurements by the Belle and LHCb collaborations have since confirmed a tension with the SM prediction~\cite{Aaij:2015yra,Huschle:2015rga,Sato:2016svk,Hirose:2016wfn,Aaij:2017uff}, and a recent average is found to deviate from the SM by about $4\sigma$~\cite{Amhis:2016xyh}.

Although NP explanations of the tensions found in both $b\to c\,\ell^-\,\nu$ and $b\to s\,\ell^+\ell^-$ transitions require contributions to different WCs of the WEH, it is intriguing that all the tensions seem to indicate LFUV, and thus might be connected.
However, while CHMs in principle offer the possibility for LFUV, the $b\to c\,\ell^-\,\nu$ anomalies seem to be harder to accommodate.
This thesis therefore focuses on the $b\to s\,\ell^+\ell^-$ anomalies and their possible explanation in terms of CHMs.

\section{Outline and scope}
Modern CHMs are based on a plethora of different concepts, ideas, and formalisms.
Among them are the description of NGBs, hidden local symmetries, partial compositeness of fermions and vector bosons, flavor symmetries, vacuum alignment, collective breaking of symmetries, extra dimensions, etc.
One aim of this thesis is therefore to provide, in chapter~\ref{chap:CHMs}, a self-contained introduction to the main concepts\footnote{%
For lecture notes that also introduce many of these concepts, see e.g.~\cite{Panico:2015jxa,Contino:2010rs}.
} that are needed to construct realistic CHMs.
The focus primarily lies on the concepts and not on specific models, which are discussed later.
The idea is to start with the arguably most central aspect of CHMs, namely NGBs, and then to proceed step by step in building upon the description of NGBs.
In particular, some effort is made to relate new concepts to those previously introduced.
To this end, section~\ref{sec:NGBs} discusses the formalisms needed to construct Lagrangians that describe NGBs.
%
In section~\ref{sec:Vectorres}, it is shown how these formalisms can be extended in a very natural way to include vector resonances, and how the inclusion of more and more resonances ultimately leads to a theory with an extra dimension.
After introducing the notion of fermion partial compositeness, section~\ref{sec:Fermions} shows how it automatically arises from a theory with an extra dimension.
This is then used to include fermions into the models describing NGBs and vector resonances.
In addition, flavor symmetries that allow for building phenomenologically viable fermion sectors are discussed.
Having introduced all particle species, section~\ref{sec:EWSB} turns to the mechanism of EWSB by vacuum misalignment and discusses collective breaking as a means to arrive at a finite one-loop scalar potential.
In section~\ref{sec:FPC}, finally, a possible UV completion of the effective low-energy description is presented.

The concepts introduced in chapter~\ref{chap:CHMs} are applied in chapter~\ref{chap:direct_constraints}.
In the context of our global analyses~\cite{Niehoff:2015iaa} and~\cite{Niehoff:2016zso} of two specific CHMs, the direct collider constraints on these models are discussed.
After describing the numerical method and the constraints used in these analyses in sections~\ref{sec:direct_constraints:numerical_strategy} and~\ref{sec:direct_constraints:constraints}, the two concrete models are specified in section~\ref{sec:direct_constraints:models}.
Section~\ref{sec:direct_constraints:direct_constraints} details in a model independent way how direct collider constraints are included into our global analyses.
In particular, the calculations of cross sections and branching ratios for generic scalar, vector and fermion resonances are described in sections~\ref{sec:direct_constraints:direct_constraints:widths_and_BRs} and~\ref{sec:direct_constraints:direct_constraints:boson_xsec}, while their comparison with experimental data is discussed in section~\ref{sec:direct_constraints:direct_constraints:exp_bounds}.
Results of our global analyses are finally presented in section~\ref{sec:direct_constraints:results}.
The production and decay of the particles present in the two models are analyzed, the most promising decay channels are identified, and the prospects for probing parameter points of the models are discussed.
This is done in section~\ref{sec:direct_constraints:results:quarks} for quark resonances and in section~\ref{sec:direct_constraints:results:vectors} for vector resonances.
Section~\ref{sec:direct_constraints:results:eta} describes the collider constraints for a particle only present in one of the two models: a scalar singlet pNGB that can mix with the Higgs and generically is considerably lighter than the vector and fermion resonances.

Apart from direct constraints, a central topic of this thesis are the anomalies in rare $B$ decays mentioned above.
The recent experimental data on these decays is interpreted in a model independent way in chapter~\ref{chap:anomalies}.
By performing global fits in~\cite{Altmannshofer:2017fio} and~\cite{Altmannshofer:2017yso}, we have analyzed possible explanations of the measurements in terms of NP contributions to WCs of the WEH.
The relevant part of the WEH and the numerical method employed in these fits are described in section~\ref{sec:anomalies:WEH}.
These methods are then first applied to measurements of processes only involving the $b\to s\,\mu^+\mu^+$ transition in section~\ref{sec:anomalies:bsmumu}.
Subsequently, the hints for violation of LFU are interpreted in section~\ref{sec:anomalies:LFUV}.

Motivated by the results from the global fits shown in chapter~\ref{chap:anomalies}, a possible explanation of the anomalies in rare $B$ decays in the context of CHMs is presented in chapter~\ref{chap:LUFV_in_CHMs}.
Based on our proposal in~\cite{Niehoff:2015bfa}, a simple model is described in section~\ref{sec:LFUV:model}, and constraints from electroweak precision tests and quark flavor physics are discussed in sections~\ref{sec:LFUV:EWPT} and~\ref{sec:LFUV:quark_flavor_contraints}, respectively.
The ranges of the parameters of the model that can explain the $b\to s\,\ell^+\ell^-$ anomalies while satisfying the constraints are presented in section~\ref{sec:LFUV:results}.

After presenting the simplified model in chapter~\ref{chap:LUFV_in_CHMs},
the flavor phenomenology of a much more ambitious model is discussed in chapter~\ref{chap:Flavor_MFPC}.
This model, which we have analyzed in~\cite{Sannino:2017utc}, is a UV completion of CHMs described already in section~\ref{sec:FPC} and known as minimal fundamental partial compositeness.
Working in an effective low-energy description, the model's contributions to observables at the electroweak scale and to low-energy flavor observables are discussed in detail in section~\ref{sec:FlavMFPC:observables}.
The numerical method we have used in our analysis as well as the relevant parameters are described in section~\ref{sec:FlavMFPC:numerical_analysis}.
The results are presented in~\ref{sec:FlavMFPC:results}, discussing meson-antimeson mixing, charged current semi-leptonic decays and neutral current semi-leptonic decays.
It is shown that while an explanation of the anomalies in charged current decays is in conflict with precise measurements of $Z$ decays,  the model is actually capable of explaining the anomalies in rare $B$ decays.

\chapter{Composite Higgs models}\label{chap:CHMs}

The central idea of CHMs and TC models is that the EW symmetry is not broken by the VEV of an elementary scalar field, but by strong interactions.
In both CHMs and TC models,
there is a strongly interacting sector that has a global symmetry $G$ into which the EW symmetry is embedded.
The symmetry group $G$ is then spontaneously broken to a subgroup~$H$ by strong dynamics.
In traditional TC models, the remaining unbroken group $H$ can only accommodate ${\rm U}(1)_{\rm Q}$, and thus the EW symmetry is broken as~$G$ is broken to~$H$.
In CHMs on the other hand, $H$ can in principle accommodate the whole EW symmetry group, but it could also be aligned inside~$G$ in such a way that only ${\rm U}(1)_{\rm Q}$ lies in~$H$.
The actual alignment of~$H$ in~$G$, or in other words the misalignment between~$H$ and the EW symmetry group, is not determined by the strong dynamics that break $G$ to $H$ but by weak interactions external to the strong sector~\cite{Weinberg:1975gm,Peskin:1980gc,Preskill:1980mz}.
These weak interactions generate an effective scalar potential for the NGBs that arise in the $G$ to $H$ breaking, and minimizing this potential yields the actual alignment of~$H$ in~$G$.
The amount of misalignment controls the scale of EWSB, such that this mechanism of {\it symmetry breaking by vacuum misalignment} allows to raise the TC scale while keeping the EWSB scale at its observed value.
This was used in the early CHMs of the mid 1980s~\cite{Kaplan:1983fs,Kaplan:1983sm,Banks:1984gj,Georgi:1984ef,Georgi:1984af,Dugan:1984hq} to ameliorate problems with large FCNCs that traditional TC models had~\cite{Eichten:1979ah,Dimopoulos:1980fj}.
As a byproduct of EWSB by vacuum misalignment, the theory contains a pNGB that resembles the properties of the SM Higgs boson.
In CHMs, where the $G$ to $H$ breaking is assumed to be due to strong dynamics, this pNGB is a composite object.
Hence, these models are called {\it composite Higgs models}.

In the first years after CHMs were proposed,
not much attention was paid to them.
This was probably due to the ``first superstring revolution'' taking place also in the mid 1980s~\cite{Green:1984sg,Witten:1984dg,Gross:1984dd,Candelas:1985en}, the popularity that grand unified supersymmetric explanations of the naturalness problem had gained~\cite{Dimopoulos:1981yj,Dimopoulos:1981zb,Sakai:1981gr}, and the intrinsic difficulties of performing calculations in strongly coupled theories (cf.~\cite{Newman:1996fza}, in particular \cite{Lane:1994vu}).
This changed about 15 years later.
In the late 1990s and early 2000s, the AdS/CFT correspondence~\cite{Maldacena:1997re,Gubser:1998bc,Witten:1998qj} was discovered and models with extra dimensions became increasingly popular (see e.g.~\cite{ArkaniHamed:1998rs,Antoniadis:1998ig,Randall:1999ee,Randall:1999vf}).
By discretizing the extra dimension of certain extra dimensional models, which is also known as dimensional deconstruction~\cite{ArkaniHamed:2001ca}, 4D~models were obtained in which a pNGB plays the role of the Higgs boson and the EW symmetry is broken by vacuum misalignment~\cite{ArkaniHamed:2001nc}.
In contrast to the early CHMs, these deconstructed models contain an additional mechanism for protecting the NGBs' scalar potential from UV dependent quantum corrections, which is known as {\it collective breaking}.
This led to a plethora of models with a pNGB Higgs, the so called {\it little Higgs models}~\cite{ArkaniHamed:2002pa,ArkaniHamed:2002qx,ArkaniHamed:2002qy,Gregoire:2002ra,Low:2002ws,Kaplan:2003uc,Chang:2003un,Skiba:2003yf,Chang:2003zn,Cheng:2003ju,Cheng:2004yc,Schmaltz:2004de,Low:2004xc}.
By applying holographic methods based on the AdS/CFT correspondence to extra dimensional models, they can be interpreted as duals of strongly interacting theories in four dimensions~\cite{ArkaniHamed:2000ds,Rattazzi:2000hs}.
This was used to construct {\it holographic composite Higgs models} that have a 4D description similar to conventional CHMs but were formulated in terms of 5D theories~\cite{Contino:2003ve,Agashe:2004rs,Contino:2006qr,Panico:2007qd}.
In the 5D formulation, the composite pNGB Higgs becomes the extra dimensional component of a 5D gauge field, which is known as {\it gauge-Higgs unification} (cf.\ section~\ref{sec:vectorres_contunuum_limit}).
While extra dimensional models of this kind were actually already discussed in the late 1970s and 1980s~\cite{Manton:1979kb,Fairlie:1979at,Hosotani:1983xw,Hosotani:1988bm}, their equivalence to CHMs was only noticed about 20 years later.
In the 4D deconstructed description of models with gauge-Higgs unification, the 4D components of the extra dimensional gauge fields yield the SM gauge bosons as well as massive spin one states that can be interpreted as composite resonances.
When putting fermions in the extra dimensional bulk, the deconstructed version contains
massless chiral fermions as well as massive Dirac fermions with the same quantum numbers~\cite{Grossman:1999ra,Gherghetta:2000qt}.
The latter can again be interpreted as composite resonances and are allowed to mix with the chiral fermions.
Interestingly, this implements
the so called {\it partial compositeness} mechanism for giving mass to the chiral fermions~\cite{Kaplan:1991dc}.
This mechanism was actually already proposed in the early 90s in the context of TC and CHMs as a means to solve the FCNC problem, but it essentially got no attention until
its rediscovery in models with extra dimensions.
%
%
%
%
%

For phenomenological applications, several kinds of 4D CHMs based on extra dimensional models have been constructed.
Notable examples\footnote{%
For a comparison of various CHM constructions see~\cite{Bellazzini:2014yua}.
} are simplified models that mainly focus on the pNGB nature of the Higgs~\cite{Giudice:2007fh,Barbieri:2007bh} or on heavy resonances and partial compositeness~\cite{Contino:2006nn}
and more complete {\it multi-site models}\footnote{%
When an extra dimension is discretized, or latticized, each point in the resulting lattice is called a {\it site} and the number of heavy resonances depends on the number of sites (cf.\ section~\ref{sec:Vectorres}).
Multi-site models therefore always contain heavy resonances.
} featuring a pNGB Higgs, vector and fermion resonances, and partial compositeness~\cite{Panico:2011pw,DeCurtis:2011yx,Marzocca:2012zn}.
The multi-site models are especially suited for studying the dynamically generated scalar potential of the NGBs and direct collider constraints on the heavy resonances.
They are also more general in the sense that the simplified models can be derived from them by integrating out heavy resonances or taking specific limits.

When constructing CHMs, a general source of model dependence is the choice of the global symmetry groups $G$ and $H$.
The minimal requirements for EWSB by vacuum misalignment are that the EW symmetry group can be embedded into $H$ and that the $G$ to $H$ breaking yields NGBs in a complex ${\rm SU}(2)_{\rm L}$ doublet with ${\rm U}(1)_{\rm Y}$ charge $q_{\rm Y}=1/2$.
The latter is necessary to provide the degrees of freedom for the massive weak gauge bosons and a pNGB Higgs.

For phenomenological reasons, the former requirement is usually extended, i.e.\ one requires that not only the EW group ${\rm SU}(2)_{\rm L}\times {\rm U}(1)_{\rm Y}$ can be embedded into $H$, but the larger group ${\rm SU}(2)_{\rm L}\times {\rm SU}(2)_{\rm R}$.
This enforces the ratio of the tree-level masses of $W^\pm_\mu$ and $Z_\mu$ to be exactly the same like in the SM~\cite{Weinberg:1975gm_addendum,Susskind:1978ms}.
This can be understood as follows.
The vacuum alignment that breaks the EW symmetry to ${\rm U}(1)_{\rm Q}$ breaks the global ${{\rm SU}(2)_{\rm L}\times {\rm SU}(2)_{\rm R}}$ to its diagonal subgroup ${\rm SU}(2)_{{\rm L}+{\rm R}}$.
A completely unbroken ${\rm SU}(2)_{{\rm L}+{\rm R}}$ would imply identical masses for $W^\pm_\mu$ and $Z_\mu$.
While ${\rm SU}(2)_{{\rm L}+{\rm R}}$ is explicitly broken by the gauging of the ${\rm U}(1)_{\rm Y}$ subgroup of ${\rm SU}(2)_{\rm R}$, the above requirement guarantees that the non-zero ${\rm U}(1)_{\rm Y}$ gauge coupling $g'$ is the only source of ${\rm SU}(2)_{{\rm L}+{\rm R}}$ breaking in the gauge sector\footnote{%
The couplings in the fermion sector also break the ${\rm SU}(2)_{{\rm L}+{\rm R}}$.
However, this does not affect the gauge boson masses at tree level.
}.
In particular, the strong sector is exactly invariant under ${\rm SU}(2)_{{\rm L}+{\rm R}}$ and one finds
\begin{equation}
 m_Z^2
 =
 m_W^2 \cdot \left(1+\frac{g'^2}{g^2}\right)
 =
 \frac{m_W^2}{\cos^2\theta_W},
\end{equation}
i.e.\ the difference between the masses of $W^\pm_\mu$ and $Z_\mu$ is only due to the non-zero $g'$ and vanishes for $g'\to 0$.
This reproduces the SM result from eq.~(\ref{eq:intro:gauge_boson_masses}).
The ${\rm SU}(2)_{{\rm L}+{\rm R}}$ symmetry protects the ratio of the weak gauge boson masses from tree-level corrections and is therefore also called a {\it custodial symmetry}~\cite{Sikivie:1980hm}.
In models containing strong couplings, a custodial symmetry is also necessary to prohibit loop-level contributions that are enhanced by the strong couplings and cannot be neglected.
A composite sector that is invariant under a custodial symmetry is thus essential for most CHMs.
The breaking of custodial symmetry is usually parameterized by the Peskin–Takeuchi parameters $S$ and $T$~\cite{Peskin:1990zt,Marciano:1990dp,Kennedy:1990ib,Altarelli:1990zd}, and their experimentally measured values can put strong constraints on NP models.
For an analysis of the contributions to the $S$ and $T$ parameters in models discussed in this thesis, see~\cite{Niehoff:2017thesis,Niehoff:2015iaa}.

Including the custodial symmetry, the requirements on the $G$ and $H$ groups used in the following are:
\begin{enumerate}
 \item\label{enum:CHMs:custodial_sym} ${\rm SU}(2)_{\rm L}\times {\rm SU}(2)_{\rm R}\cong {\rm SO}(4)$ is a subgroup of $H$.
 \item\label{enum:CHMS:Higgs_NGBS} The $G$ to $H$ breaking yields NGBs in a complex ${\rm SU}(2)_{\rm L}$ doublet with ${\rm U}(1)_{\rm Y}$ charge $q_{\rm Y}=1/2$.
\end{enumerate}
The minimal choice for satisfying requirement~\ref{enum:CHMs:custodial_sym} is $H={\rm SO}(4)$.
In this case, the smallest group $G$ that satisfies requirement~\ref{enum:CHMS:Higgs_NGBS} is $G={\rm SO}(5)$:
the spontaneous symmetry breaking ${\rm SO}(5)\to {\rm SO}(4)$ exactly yields the desired ${\rm SU}(2)_{\rm L}$ doublet of NGBs and no other additional NGBs.
In a possible UV completion, a dynamical mechanism for the $G$ to $H$ breaking has to be present.
The most common one is chiral symmetry breaking where a bilinear of TC fermions forms a condensate due to strong dynamics.
The breaking pattern ${\rm SO}(5)\to {\rm SO}(4)$ of the minimal CHM unfortunately is not simply realized via chiral symmetry breaking and requires a quite involved construction (cf.~\cite{Caracciolo:2012je}).
However, the next-to-minimal breaking pattern ${\rm SO}(6)\to {\rm SO}(5)$ can be realized in a ${\rm Sp}(N)$ gauge theory with fermions in the fundamental representation (cf.\ section~\ref{sec:FPC}).
It only yields one scalar singlet pNGB in addition to the required complex doublet and actually contains a limit where it resembles the minimal CHM (cf.\ section~\ref{sec:direct_constraints:models}).

In chapter~\ref{chap:direct_constraints} and \ref{chap:Flavor_MFPC},
explicit models featuring both the minimal and the next-to-minimal breaking pattern are analyzed.
However, for the discussion of the structure of CHMs, it is instructive to consider the general $G$ to $H$ breaking case without specifying the groups explicitly.
The construction of 4D CHMs, namely the multi-site models is discussed in detail in the following sections.

\section{Nambu-Goldstone Bosons}\label{sec:NGBs}

For introducing CHMs, it is instructive to start with their central aspect: NGBs.
They arise from a $G$ invariant Lagrangian whose vacuum is only invariant under a subgroup $H$ of $G$, one NGB for each generator of $G$ that is not a generator of $H$.
%
%
%
The most general $G$-invariant Lagrangian that describes the NGB degrees of freedom can be constructed by a formalism due to Callan, Coleman, Wess and Zumino (CCWZ)~\cite{Coleman:1969sm,Callan:1969sn}.
In the following, after an introductory example, the CCWZ formalism is described and some important special cases are discussed.
In addition, the language of the hidden local symmetry (HLS)~\cite{Bando:1984ej,Bando:1987ym,Bando:1987br} is introduced.

\subsection{A first example: The linear sigma model}\label{sec:NGB_LSM}
Before describing to the CCWZ formalism, it is instructive to first discuss the essential properties of NGBs using a concrete example.
To this end, following \cite{Manohar:1996cq}, the linear sigma model\footnote{%
A model of this kind was introduced in \cite{GellMann:1960np} and contains a field named $\sigma$.
} is considered.
Its Lagrangian is given by
\begin{equation}\label{eq:LSM_lagrangian}
 \mathcal{L} =
 \frac{1}{2}\, \partial_\mu \vec{\phi}(x)\cdot \partial^\mu \vec{\phi}(x)
 - \lambda \left({\vec{\phi}}^{\,2}(x)-f^2\right)^2
 \quad,\quad
 \vec{\phi}(x)=\begin{pmatrix}
             \phi_1(x)\\
             \vdots\\
             \phi_N(x)
            \end{pmatrix},
\end{equation}
with $N\geq2$.
The scalar fields parametrized by the vector $\vec{\phi}(x)$ transform under the fundamental representation of the group $G = {\rm O}(N)$, whose elements correspond to the length-preserving transformations in an $N$-dimensional real vector space.
Since $\vec{\phi}(x)$ enters the Lagrangian only inside scalar products, the theory is obviously invariant under $G$.

The vacuum of the theory is defined by the minimum of the potential $V = \lambda \left({\vec{\phi}}^{\,2}(x)-f^2\right)^2$.
This minimum is found to be $|\vec{\phi}(x)|=f$, i.e.\ the vacuum corresponds to a fixed length of $\vec{\phi}(x)$.
So $f$ plays the role of the VEV of $\vec{\phi}(x)$.
All the different field configurations that satisfy the vacuum condition $|\vec{\phi}(x)|=f$ form a manifold called the {\it vacuum manifold}.
In the given case, one finds
\begin{equation}\label{eq:LSM_vacuum}
 |\vec{\phi}(x)|=f
 \quad
 \Rightarrow
 \quad
 \phi_1^2(x)+\dots+\phi_N^2(x)=f^2,
\end{equation}
which parametrizes the N-1 dimensional sphere $S^{N-1}$.
Since the points inside the vacuum manifold all yield the same potential energy, the possible vacua are degenerate and physically equivalent.
Choosing any of them, one finds that a given element $\vec{\phi}_0$ of the vacuum manifold, i.e.\ a specific choice of vacuum, is not invariant under a general $G$ transformation.
There are, however, elements of $G$ that actually leave the vacuum $\vec{\phi}_0$ invariant.
They form a subgroup\footnote{%
If a group $G$ is acting (transitively) on a manifold $M$, its subgroup $H$ that leaves a specific point $\vec{\phi}_0$ in this manifold fixed is called the {\it isotropy subgroup} of $G$ at $\vec{\phi}_0$~\cite{helgason1979differential}, or simply the {\it isotropy group} of $\vec{\phi}_0$ (other synonyms are {\it little group} or {\it stabilizer} of $\vec{\phi}_0$~\cite{nakahara2003geometry}).
} $H\subset G$ that in the present case can be identified with $H={\rm O}(N-1)$.
Since the vacuum $\vec{\phi}_0$ is not invariant under $G$, but invariant under $H$,
the symmetry $G$ is said to be spontaneously broken to $H$.

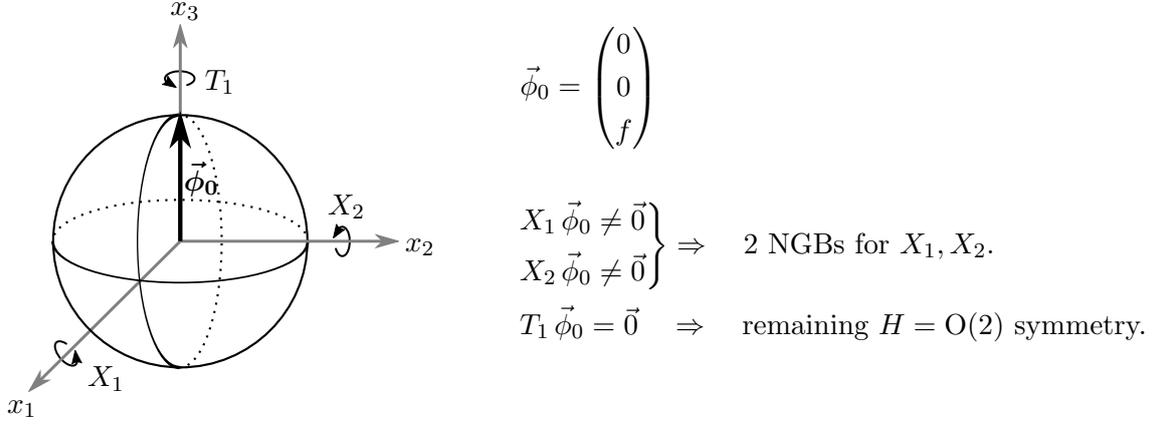
\begin{figure}[t]
\centering
\begin{picture}(430,160)
\put(8,8){\includegraphics{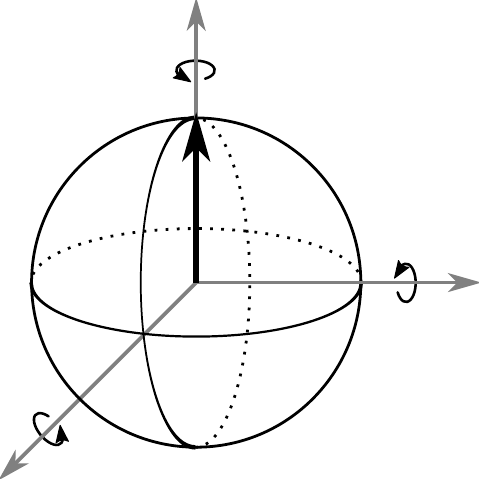}}
\put(0,0){$x_1$}
\put(149,61){$x_2$}
\put(61,150){$x_3$}
\put(30,10){$X_1$}
\put(120,75){$X_2$}
\put(74,122){$T_1$}
\put(66,84){$\bm{\vec{\phi}_0}$}
\put(192,120){$\vec{\phi}_0=\begin{pmatrix}0\\0\\f\end{pmatrix}$}
\put(190,50){$%
\begin{aligned}
& \negthickspace
 \begin{rcases}
  X_1\,\vec{\phi}_0 \neq \vec{0} \\
  X_2\,\vec{\phi}_0 \neq \vec{0} \\
 \end{rcases}\Rightarrow\quad \parbox{100pt}{2 NGBs for $X_1, X_2$.}\\
& T_1\,\vec{\phi}_0 = \vec{0}\quad\Rightarrow \quad \text{remaining $H={\rm O}(2)$ symmetry.}
\end{aligned}
$}
\end{picture}
\caption{${\rm O}(3)\to {\rm O}(2)$ spontaneous symmetry breaking.
The two-sphere around the origin is the vacuum manifold where $|\vec{\phi}(x)|=f$.
When the vacuum $\vec{\phi}_0$ is chosen, the $G={\rm O}(3)$ symmetry is broken to $H={\rm O}(2)$.
Only the rotations induced by $T_1$ (around the $x_3$ axis) leave $\vec{\phi}_0$ (chosen in the direction of the $x_3$ axis) invariant and thus $T_1$ is the generator of the unbroken ${\rm O}(2)$.
$X_1$ and $X_2$ (inducing rotations around the $x_1$ and $x_2$ axes, respectively) are the broken generators and yield two NGBs.}
\label{fig:SSB}
\end{figure}

The spontaneously broken symmetry group $G$ is however not really broken.
While the vacuum $\vec{\phi}_0$ is only invariant under $H$, there is, however, still another invariance left.
This is the invariance of the vacuum potential energy under the choice of a specific vacuum inside the vacuum manifold.
These two different kinds of invariance divide the generators of $G$ into two sets:
\begin{itemize}
 \item The generators $T^a$ of $H$ can be defined for some specific reference vacuum $\vec{\phi}_0$ by ${T^a\,\vec{\phi}_0 = 0}$.
Their associated group elements leave $\vec{\phi}_0$ invariant and they correspond to the former invariance.
 \item The remaining generators $X^a$ that satisfy $X^a\,\vec{\phi}_0 \neq 0$ and are thus also called broken generators are associated with elements of $G$ that transform $\vec{\phi}_0$ to another, but physically equivalent vacuum $\vec{\phi}_0'$.
 The action of these group elements has the same effect as choosing a different reference vacuum and therefore the generators $X^a$ correspond to the latter invariance.
\end{itemize}
For illustration, the vacuum manifold, a specific vacuum $\vec{\phi}_0$, and the generators of $G$ are visualized in figure~\ref{fig:SSB} for the $N=3$ case.
%

The $G$ symmetry is thus actually preserved, but for a given vacuum $\vec{\phi_0}$ only the $H$ subgroup is linearly realized\footnote{%
As will be discussed in section~\ref{sec:NGBs_CCWZ}, the whole $G$ symmetry is in this case non-linearly realized.
}.
It is interesting to note that the vacuum manifold happens to be isomorphic to the coset space $G/H$ (i.e.\ the manifold formed by the set of left cosets of $H$ in $G$) and that the  tangent space of $G/H$ at the point $H$ (i.e.\ at the left coset $\mathscr{e}H$ where $\mathscr{e}$ is the identity in $G$), or equivalently, the tangent space of the vacuum manifold at $\vec{\phi_0}$, is isomorphic to the vector space spanned by the broken generators $X^a$ (see e.g.~\cite{MR2011126}).
In this sense, the group elements associated to the $X^a$ generators correspond to the elements of the coset space $G/H$ (at least for group elements close to the identity in $G$).
Moreover, the above observations have the immediate consequence that the vacuum manifold as well as the $G/H$ coset space both have a dimensionality equal to the number of broken generators.
That this number also coincides with the number of NGBs gets obvious by considering fluctuations around the minimum $\vec{\phi}_0$.
There are two distinct cases:
\begin{itemize}
 \item Fluctuations inside the vacuum manifold do not change the length of $\vec{\phi}(x)$ and therefore also do not change the potential energy.
 They thus correspond to massless degrees of freedom.
 The number of these degrees of freedom is given by the dimensionality of the vacuum manifold and each of them can be associated with a rotation induced by one of the broken generators $X^a$.
 The fluctuations inside the vacuum manifold can therefore be identified with the massless NGBs $\pi^a(x)$.
 \item A radial excitation changes the length of $\vec{\phi}(x)$ and hence the potential energy.
 Consequently, it corresponds to a massive scalar $\sigma(x)$.
\end{itemize}
It is convenient to parametrize $\vec{\phi}(x)$ by ``polar coordinates'' using the $\pi^a(x)$ and $\sigma(x)$ fluctuations around the minimum $\vec{\phi}_0$. This yields
\begin{equation}\label{eq:LSM_sigma_pi_parametrization}
 \vec{\phi}(x) = \left(1+\sigma(x)/f\right)\,e^{i\frac{\sqrt{2}}{f}\,\pi^a(x)\,X^a}\,\vec{\phi}_0.
\end{equation}
Plugging this parametrization into the initial Lagrangian from eq.~(\ref{eq:LSM_lagrangian}), and using a specific vacuum, e.g.\ ${\vec{\phi}_0=(0,\dots,0,f)^T}$, one finds
\begin{equation}
 \mathcal{L} =
 \frac{1}{2}\,\partial_\mu\sigma(x)\,\partial^\mu\sigma(x)
 -\lambda\,(\sigma(x)^2+2\,\sigma(x)\,f)^2
 +\frac{1}{2}\,(f+\sigma(x))^2\,
 \left[\partial_\mu U^T(x)\,\partial^\mu U(x)\right]_{NN},
\end{equation}
where $U(x)$ is the NGB matrix defined by
\begin{equation}\label{eq:LSM_NGB_matrix}
  U(x)=e^{i\frac{\sqrt{2}}{f}\,\pi^a(x)\,X^a}.
\end{equation}
Given the above discussed relation between the elements of the coset space $G/H$ and the generators $X^a$, with a slight abuse of terminology, the coset space $G/H$ will be used in the following to refer to the subset of $G$ that consists of the group elements associated to the $X^a$ generators.
In this sense, the NGB matrix $U(x)$ is an element of the coset space $G/H$.

Because the fields $\sigma(x)$ and $\pi^a(x)$ are perturbations around the vacuum, they have a vanishing VEV.
While this is obvious for $\sigma(x)$ from its definition in eq.~(\ref{eq:LSM_sigma_pi_parametrization}) and the vacuum condition $|\vec{\phi}(x)|=f$, this condition does not forbid a VEV for the $\pi^a(x)$ fields.
Such a VEV, however, just corresponds to a transformation of the vacuum $\vec{\phi}_0$ to another equivalent vacuum $\vec{\phi}_0'$ and is therefore not physical.
In the parameterization of the scalar fields in terms of $\sigma(x)$ and $\pi^a(x)$, the quantity $f$ thus loses its interpretation as a VEV.
In this context, it is usually called the {\it decay constant}\footnote{%
The term ``decay constant'' stems from the description of pions as (pseudo) NGBs in a low-energy effective theory of QCD, where the strength of leptonic pion decay depends on $f$.
} and enters several interaction terms as well as the mass of the radial excitation $\sigma(x)$.
This mass is found to be
\begin{equation}
 m_\sigma = \sqrt{8\,\lambda}\,f.
\end{equation}
At energies well below $m_\sigma$, the massive scalar $\sigma(x)$ decouples and the low-energy degrees of freedom can be described by the pure NGB Lagrangian
\begin{equation}\label{eq:LSM_NGB_lagrangian}
 \mathcal{L}_{NGB} =\frac{f^2}{2}\,\left[\partial_\mu U^T(x)\,\partial^\mu U(x)\right]_{NN}.
\end{equation}
Some important properties of the NGBs can be read off from this Lagrangian:
\begin{itemize}
 \item It is non-linear in the NGBs $\pi^a(x)$ which enter via the NGB matrix $U(x)$, eq.~(\ref{eq:LSM_NGB_matrix}).
 The model containing only NGB degrees of freedom is thus also called the {\it non-linear sigma model}\footnote{%
 The NGBs not only enter the Lagrangian non-linearly, they also transform non-linearly under $G$, cf.\ section~\ref{sec:NGBs_CCWZ}.
 }.
 \item The NGBs are derivatively coupled. This implies that the Lagrangian is invariant under constant shifts of $\pi^a(x)$.
 This invariance under constant shifts is nothing but the invariance under the choice of a specific vacuum that is discussed above: constant shifts correspond to choosing a different vacuum inside the vacuum manifold.
 The shift symmetry can also be seen as a reason for the masslessness of the NGBs since it does not allow a mass term.
 \item Via the Fourier transformation, the derivative coupling corresponds to a momentum dependent coupling.
 The theory is thus weakly coupled for low momentum and so in this regime an expansion in the momentum can be performed.
\end{itemize}
It is worth mentioning that the pure NGB Lagrangian is non-renormalizable and only valid for a description of the theory at energies below $\Lambda\approx 4\pi f$.
At higher energies it has to be UV-completed, e.g.\ by a linear sigma model as described above (by ``integrating in'' the $\sigma(x)$ field), or by a strongly coupled theory that yields NGBs as bound states of more fundamental degrees of freedom.
However, the great benefit of the NGB Lagrangian is that it does not depend on the actual UV completion, but only on the symmetry breaking pattern.

\subsection{The CCWZ formalism}\label{sec:NGBs_CCWZ}
The Lagrangian in eq.~(\ref{eq:LSM_NGB_lagrangian}) describes the NGBs for a ${\rm O}(N)\to {\rm O}(N-1)$ spontaneous symmetry breaking.
For a generic group $G$ that is spontaneously broken to a subgroup $H$, the question arises how the most general NGB Lagrangian that is invariant under $G$ can be constructed.
The answer to this question is the CCWZ formalism~\cite{Coleman:1969sm,Callan:1969sn} presented in the following%
\footnote{This discussion of the CCWZ formalism is loosely based on~\cite{Bando:1987br,Manohar:1996cq}.}.
To this end, it is useful to first fix the notation.
The generators of $G$ that are elements of the Lie-algebra $\mathfrak{g}$ will be denoted by $T^a$ and $X^a$, where $T^a\in\mathfrak{h}$ are the generators of the remaining $H$ symmetry and elements of the Lie-algebra $\mathfrak{h}$, while $X^a\in\mathfrak{g}-\mathfrak{h}$ are the broken generators\footnote{%
$\mathfrak{g}-\mathfrak{h}$ is the orthogonal complement of the $\mathfrak{h}$ subalgebra of $\mathfrak{g}$, i.e.\ $\mathfrak{g} = \mathfrak{h}\oplus(\mathfrak{g}-\mathfrak{h})$.
}.
The generators are normalized such that $\tr[T^a\,T^b]=\tr[X^a\,X^b]=\delta^{ab}$.

As a starting point, the NGBs are parametrized by fluctuations $\pi^a(x)$ around a $G$-breaking minimum $\phi_0$ inside the vacuum manifold\footnote{
In contrast to the field $\vec{\phi}(x)$ from the previous section, the field $\phi(x)$ is defined to be dimensionless for convenience.
Both are related by $f\,\phi(x) \sim \vec{\phi}(x)$.
},
\begin{equation}\label{eq:CCWZ_U}
 \phi(x) = U[\pi(x)]\,\phi_0,
 \quad\quad
  U[\pi(x)]=e^{i\frac{\sqrt{2}}{f}\,\pi^a(x)\,X^a}\in G/H,
\end{equation}
where  $\pi(x)=\pi^a(x)\,X^a$ and $U[\pi(x)]$ is the NGB matrix (cf.\ section~\ref{sec:NGB_LSM}).
Performing a $G$ transformation on $\phi(x)$ yields
\begin{equation}\label{eq:CCWZ_G_transformation}
 G:\,\phi(x)\to \phi'(x) = \mathscr{g}\,\phi(x) = \mathscr{g}\,U[\pi(x)]\,\phi_0,
\end{equation}
where $\mathscr{g}$ is an element of $G$.
In general, the object $\mathscr{g}\,U[\pi(x)]$ is not an element of the coset space $G/H$ and thus cannot be expressed simply in terms of a matrix $U[\pi'(x)]$ that depends on a transformed $\pi'(x)$.
Using the fact that any $\mathscr{g}\in G$ can be decomposed as $\mathscr{g}=\mathscr{g}_X\,\mathscr{g}_h$ where $\mathscr{g}_X\in G/H$ and $\mathscr{g}_h\in H$, one can decompose $\mathscr{g}\,U[\pi(x)]$ as follows:
\begin{equation}\label{eq:CCWZ_decomposition}
 \underbrace{\mathscr{g}\,U[\pi(x)]}_{\in G} = \underbrace{U[\pi'(x)]}_{\in G/H}\,\underbrace{\mathscr{h}[\pi(x),\mathscr{g}]}_{\in H},
\end{equation}
where $U[\pi'(x)]\in G/H$ depends on a transformed $\pi'(x)$ and $\mathscr{h}[\pi(x),\mathscr{g}]\in H$ depends on both $\pi(x)$ and $\mathscr{g}$.
Plugging this decomposition into eq.~(\ref{eq:CCWZ_G_transformation}) yields
\begin{equation}
 G:\,\phi(x)\to \phi'(x) = \mathscr{g}\,\phi(x) = U[\pi'(x)]\,\mathscr{h}[\pi(x),\mathscr{g}]\,\phi_0 = U[\pi'(x)]\,\phi_0,
\end{equation}
where in the last step the invariance of the vacuum $\phi_0$ under $H$ is used, i.e.\ $\mathscr{h}[\pi(x),\mathscr{g}]\,\phi_0 = \phi_0$ for any $\pi(x)$ and $\mathscr{g}$.
So it is actually possible to express the transformed $\phi'(x)$ only in terms of a NGB matrix $U[\pi'(x)]$ and the vacuum $\phi_0$.
The transformation $G:\,\phi(x)\to \phi'(x)$ thus corresponds to transforming the NGB matrix $U[\pi(x)]$ and the NGB fields $\pi(x)$ in a non-linear way.
Using eq.~(\ref{eq:CCWZ_decomposition}), one finds
\begin{equation}\label{eq:CCWZ_non_linear_U_transformation}
 G:\,U[\pi(x)]\to U[\pi'(x)]=\mathscr{g}\,U[\pi(x)]\,\mathscr{h}^{-1}[\pi(x),\mathscr{g}]
\end{equation}
and
\begin{equation}
 G:\,\pi(x)\to \pi'(x)
 \quad\text{where $\pi'(x)$ is defined by eq.~(\ref{eq:CCWZ_non_linear_U_transformation})}.
\end{equation}
Due to the non-linear transformation of $U[\pi(x)]$ and $\pi(x)$ under $G$, they are also called {\it non-linear realizations of G}.

Having laid down the transformation properties of the NGB matrix $U[\pi(x)]$, it can now be used to construct a $G$-invariant Lagrangian.
It is instructive to first try a naive approach and see why this fails.
The arguably simplest Lorentz invariant Lagrangian for $U(x)=U[\pi(x)]$ is
\begin{equation}\label{eq:CCWZ_L_try}
 \mathcal{L}_{\rm try}=\frac{f^2}{4}\,\tr\left[\partial_\mu U^{-1}(x)\,\partial^\mu U(x)\right],
\end{equation}
where the prefactor $\frac{f^2}{4}$ is included to yield a canonically normalized kinetic term for the $\pi^a(x)$ fields.
Under a $G$ transformation, according to eq.~(\ref{eq:CCWZ_non_linear_U_transformation}), $\mathcal{L}_{\rm try}$ transforms as
\begin{equation}\label{eq:CCWZ_L_try_transformation}
 G:\,\mathcal{L}_{\rm try}\to\mathcal{L}'_{\rm try} = \frac{f^2}{4}\,\tr\left[\partial_\mu\left(\mathscr{h}(x)\,U^{-1}(x)\,\mathscr{g}^{-1}\right)\,\partial^\mu\left(\mathscr{g}\,U(x)\,\mathscr{h}^{-1}(x)\right)\right],
\end{equation}
where $\mathscr{h}(x)=\mathscr{h}[\pi(x),\mathscr{g}]$.
Due to the $x$-dependence of $\mathscr{h}(x)$, it does not commute with the partial derivatives and thus $\mathcal{L}'_{\rm try}\neq\mathcal{L}_{\rm try}$, i.e.\ $\mathcal{L}_{\rm try}$ is not invariant under a general $G$-transformation.
To investigate this in more detail, it is convenient to rewrite $\mathcal{L}_{\rm try}$ in the following way.
First, one can introduce a $\mathds{1}=U(x)\,U^{-1}(x)$ inside the trace, such that
\begin{equation}
 \mathcal{L}_{\rm try}=\frac{f^2}{4}\,\tr\left[\left(\partial_\mu U^{-1}(x)\right)\,U(x)\,U^{-1}(x)\,\left(\partial^\mu U(x)\right)\right].
\end{equation}
From the relation%
\footnote{The relation $\left(\partial_\mu\,U^{-1}(x)\right)U(x) =- U^{-1}(x)\,\partial_\mu U(x)$ can readily be derived from $\partial_\mu\left(U^{-1}(x)\,U(x)\right)=0$.}
\begin{equation}\label{eq:CCWZ_partial_UU}
\left(\partial_\mu\,U^{-1}(x)\right)U(x) =- U^{-1}(x)\,\partial_\mu U(x),
\end{equation}
one gets
\begin{equation}
 \mathcal{L}_{\rm try}=\frac{f^2}{4}\,\tr\left[-\left(U^{-1}(x)\,\partial_\mu U(x)\right)\,\left(U^{-1}(x)\,\partial^\mu U(x)\right)\right],
\end{equation}
where in comparison to eq.~(\ref{eq:CCWZ_L_try}) the partial derivatives now both act on $U(x)$.
The objects inside the parentheses are called {\it Maurer-Cartan-forms} and may be written as
\begin{equation}\label{eq:CCWZ_Maurer_Cartan_form}
 a_\mu[U]=i\,U^{-1}(x)\,\partial_\mu U(x).
\end{equation}
The Maurer-Cartan-form has the useful property that it is Lie-algebra valued, i.e.\ $a_\mu[U]\in\mathfrak{g}$.
Using $a_\mu[U]$, the Lagrangian can be expressed as
\begin{equation}
 \mathcal{L}_{\rm try}=\frac{f^2}{4}\,\tr\left[a_\mu[U]\,a^\mu[U]\right].
\end{equation}
It should therefore be possible to trace back the fact that $\mathcal{L}_{\rm try}$ is not invariant under $G$ to the transformation properties of $a_\mu[U]$ under a $G$ transformation.
Using the definition of $a_\mu[U]$, eq.~(\ref{eq:CCWZ_Maurer_Cartan_form}), one finds
\begin{equation}\label{eq:CCWZ_a_mu_transformation}
\begin{aligned}
 G:\,a_\mu[U]
 \to\ &i\,\mathscr{h}(x)\,U^{-1}(x)\,\mathscr{g}^{-1}\,\partial_\mu\left(\mathscr{g}\,U(x)\,\mathscr{h}^{-1}(x)\right)\\
   =\ &i\,\mathscr{h}(x)\,U^{-1}(x)\,\left(\partial_\mu U(x)\right)\,\mathscr{h}^{-1}(x) + i\,\mathscr{h}(x)\,\partial_\mu \mathscr{h}^{-1}(x)\\
   =\ & \mathscr{h}(x)\,a_\mu[U]\,\mathscr{h}^{-1}(x) + a_\mu[\mathscr{h}^{-1}]
\end{aligned}
\end{equation}
and obviously the appearance of $a_\mu[\mathscr{h}^{-1}]$ spoils the invariance of $\mathcal{L}_{\rm try}$.
The reason for both, the problem with the $x$-dependence of $\mathscr{h}(x)$ in eq.~(\ref{eq:CCWZ_L_try_transformation}) as well as the appearance of $a_\mu[\mathscr{h}^{-1}]$ in eq.~(\ref{eq:CCWZ_a_mu_transformation}), is that the partial derivative $\partial_\mu U(x)$ does not transform in the same way as $U(x)$ and therefore in general $a_\mu[U]\notin \mathfrak{g}-\mathfrak{h}$.
Being an element of $\mathfrak{g}$, it is however possible to expand $a_\mu[U]$ in terms of the generators $T^a$ and $X^a$ such that
\begin{equation}\label{eq:CCWZ_a_mu_expansion}
 a_\mu[U] = d_\mu[U]^a\,X^a + e_\mu[U]^a\,T^a,
\end{equation}
and one can define
\begin{equation}\label{eq:CCWZ_d_mu_definition}
 \begin{aligned}
  d_\mu[U] &= d_\mu[U]^a\,X^a \in \mathfrak{g}-\mathfrak{h},\\
  e_\mu[U] &= e_\mu[U]^a\,T^a \in \mathfrak{h}.
 \end{aligned}
\end{equation}
For the transformation properties of $d_\mu[U]$ and $e_\mu[U]$ under a $G$ transformation one then finds
\begin{equation}
\begin{aligned}\label{eq:CCWZ_d_mu_transformation}
 G:\,d_\mu[U]&\to \mathscr{h}(x)\, d_\mu[U]\,\mathscr{h}^{-1}(x),\\
 G:\,e_\mu[U]&\to \mathscr{h}(x)\, e_\mu[U]\,\mathscr{h}^{-1}(x) + a_\mu[\mathscr{h}^{-1}]
 = \mathscr{h}(x)\left( e_\mu[U]+i\partial_\mu\right) \mathscr{h}^{-1}(x),
\end{aligned}
\end{equation}
i.e.\ $a_\mu[\mathscr{h}]$ only appears in the transformation of $e_\mu[U]$ while $d_\mu[U]$ transforms homogeneously.
Constructing a Lagrangian using $d_\mu[U]$ instead of $a_\mu[U]$ yields
\begin{equation}\label{eq:CCWZ_L_2}
 \mathcal{L}_2 = \frac{f^2}{4}\,\tr\left[d_\mu[U]\,d^{\mu}[U]\right],
\end{equation}
and using the transformation property of $d_\mu[U]$ under $G$, eq.~(\ref{eq:CCWZ_d_mu_transformation}), one immediately sees that $\mathcal{L}_2$ is invariant under $G$-transformations.
That this Lagrangian is indeed the most general one that is $G$-invariant and leading order in an expansion in the number of derivatives was shown in~\cite{Coleman:1969sm,Callan:1969sn}.
The subscript ``2'' on $\mathcal{L}_2$ indicates that this Lagrangian only contains the leading order terms with two derivatives.
For the remainder of this thesis, higher order terms will not be necessary.
However, it should be noted that not only $d_\mu[U]$, but also $e_\mu[U]$ enters the higher order terms.
The transformation property of $e_\mu[U]$ under $G$, eq.~(\ref{eq:CCWZ_d_mu_transformation}), suggests that it transforms like a gauge connection.
Consequently, it enters the $G$-invariant Lagrangian in terms of a covariant derivative replacing the partial derivative according to $i\partial_\mu\to i\partial_\mu+e_\mu[U]$.

In the above discussion, $G$ is assumed to be a global symmetry.
It is however straightforward to include gauge fields for a subgroup $E\subseteq G$ (where $E$ could also be the whole group $G$).
Denoting the generators of $E$ by $P^a$, one just has to add the kinetic terms for the gauge fields $A_\mu = A_\mu^a\,P^a$ to the Lagrangian and promote the partial derivatives to gauge-covariant derivatives, i.e.\ $i\partial_\mu\to i\partial_\mu+g_E\,A_\mu$.
The Maurer-Cartan-form then also gets covariantized as
\begin{equation}
\begin{aligned}
 a_\mu[U]
 \to
 \tilde{a}_\mu[U] &= U^{-1}(x)\,(i\,\partial_\mu+g_E\,A_\mu) U(x)
 \\
 &= a_\mu[U] + g_E\,U^{-1}(x)\,A_\mu\,U(x),
\end{aligned}
\end{equation}
which can also be thought of as adding a term corresponding to the gauge field $A_\mu$ dressed by the NGB matrix $U(x)$ to yield the correct transformation properties.
The covariantized Maurer-Cartan-form $\tilde{a}_\mu[U]$ can again be decomposed in terms of unbroken and broken generators as
\begin{equation}
 \tilde{a}_\mu[U] = \tilde{d}_\mu[U]^a\,X^a + \tilde{e}_\mu[U]^a\,T^a,
\end{equation}
such that the leading order Lagrangian including gauge fields is given by
\begin{equation}
 \mathcal{L}_2
 = \frac{f^2}{4}\,\tr\left[\tilde{d}_\mu[U]\,\tilde{d}^{\mu}[U]\right]
 - \frac{1}{4}\,\tr[F^{\mu\nu}\,F_{\mu\nu}],
\end{equation}
where $F_{\mu\nu}=\partial_\mu\,A_\nu-\partial_\nu\,A_\mu-i\,g_E\,[A_\mu,A_\nu]$.
If $E\subseteq H$, all the gauge bosons are massless.
On the other hand, for each of the gauged generators $P^a$ that is also an element of the coset generators, i.e.\ $P^a\in\mathfrak{g}-\mathfrak{h}$, the associated gauge boson acquires mass via the Higgs mechanism.
The NGBs corresponding to these gauged broken generators are unphysical would-be NGBs.
Since the gauging of a subgroup $E\subseteq G$ explicitly breaks the global $G$ symmetry, some of the physical NGBs may pick up mass terms and become pNGBs (cf.\ section~\ref{sec:EWSB}).

In addition to employing the transformation properties of the NGB matrix $U[\pi(x)]$ for dressing gauge fields, this can also be done with other fields that do not transform linearly under the full $G$ group.
In particular, any field $\psi(x)$ transforming linearly under a representation $D(\mathscr{h})$ of the $H$-transformation $\mathscr{h}$ can be included in the Lagrangian with the help of $U[\pi(x)]$.
Assuming for simplicity that $D(\mathscr{h})$ is the fundamental representation, i.e. $\psi(x)$ transforms as
\begin{equation}
 H: \psi(x)\to \psi'(x) = \mathscr{h}\,\psi(x),
\end{equation}
one can consider the dressed field $U[\pi(x)]\,\psi(x)$.
Performing a $G$ transformation $\mathscr{g}$, one finds
\begin{equation}
 G: U(x)\,\psi(x)\to \mathscr{g}\,U[\pi(x)]\,\mathscr{h}^{-1}[\pi(x),\mathscr{g}]\,\psi'(x),
\end{equation}
such that the requirement that $U[\pi(x)]\,\psi(x)$ transforms linearly leads to $\psi(x)$ transforming non-linearly as
\begin{equation}
 G: \psi(x)\to \psi'(x) = \mathscr{h}[\pi(x),\mathscr{g}]\,\psi(x).
\end{equation}
This can be generalized to arbitrary representations of $H$, such that a $\psi(x)$ transforming under $D(\mathscr{h})$ transforms under $G$ non-linearly as
\begin{equation}
 G: \psi(x)\to \psi'(x) = D(\mathscr{h}[\pi(x),\mathscr{g}])\,\psi(x),
\end{equation}
and by dressing it with the NGB matrix $U[\pi(x)]$, one can include it in a $G$-invariant Lagrangian.

\subsection{Symmetric spaces}\label{sec:symmetric_spaces}
The object $d_\mu[U]$ defined in eqs.~(\ref{eq:CCWZ_a_mu_expansion}) and (\ref{eq:CCWZ_d_mu_definition}) that enters the result for the leading order NGB Lagrangian $\mathcal{L}_2$ can in general not easily be expressed in terms of the NGB matrix $U(x)$.
In the special case where $G/H$ is a {\it symmetric space}, $d_\mu[U]$ and thus $\mathcal{L}_2$ however have a simple form in terms of $U(x)$.
This is shown in the following discussion that is loosely based on~\cite{Manohar:1996cq,Thaler:2005kr}.
$G/H$ is called a symmetric space if in addition to the commutator relation $[T^a,T^b]\propto T^c$, which is just a consequence of $\mathfrak{h}$ being a Lie subalgebra of $\mathfrak{g}$, the generators $T^a$ and $X^a$ also satisfy $[T^a,X^b]\propto X^c$ and $[X^a,X^b]\propto T^c$.
In this case, the transformation
\begin{equation}
\tau:\quad
 T^a\to T^a,
 \quad
 X^a\to -X^a
\end{equation}
leaves the above commutator relations invariant and thus constitutes an automorphism.
Applying this automorphism to the NGB matrix, one finds
\begin{equation}\label{eq:CCWZ_tau_U}
 \tau:\quad
 U(x)\to \mathring{U}(x) = U^{-1}(x)\,,
\end{equation}
where $\mathring{U}(x)$ is the image of $U(x)$ under $\tau$.
From the expansion of $a_\mu[U]$ in terms of unbroken and broken generators, eq.~(\ref{eq:CCWZ_a_mu_expansion}), it then follows
\begin{equation}
 \tau:\quad
 a_\mu[U] \to
 a_\mu[U^{-1}]=
 -d_\mu[U]^a\,X^a +e_\mu[U]^a\,T^a,
\end{equation}
and hence
\begin{equation}
 d_\mu[U] = \frac{1}{2}\left(a_\mu[U]-a_\mu[U^{-1}]\right)
 = \frac{i}{2}\left(U^{-1}(x)\,\partial_\mu U(x)-U(x)\,\partial_\mu U^{-1}(x)\right).
\end{equation}
Plugging this into the Lagrangian $\mathcal{L}_2$ then yields
\begin{equation}
 \mathcal{L}_2
 = \frac{f^2}{4}\,\tr\left[d_\mu[U]\,d^{\mu}[U]\right]
 = \frac{f^2}{16}\,\tr\left[\partial_\mu (U^{-1}(x)U^{-1}(x))\,\partial^\mu (U(x)U(x))\right].
\end{equation}
Using the definition
\begin{equation}
 \Sigma(x) = U(x)U(x) = e^{i\frac{2\sqrt{2}}{f}\pi^a(x)\, X^a},
\end{equation}
the Lagrangian can be further simplified to
\begin{equation}\label{eq:CCWZ_L_2_Usquared}
 \mathcal{L}_2
 =
 \frac{f^2}{16}\,\tr\left[\partial_\mu \Sigma^{-1}(x)\,\partial^\mu \Sigma(x)\right].
\end{equation}
So one finds that for $G/H$ being a symmetric space, the NGB Lagrangian has a simple form in terms of the squared NGB matrix $\Sigma(x)$.
The transformation properties of $\Sigma(x)$ can be derived in the following way.
Applying the automorphism $\tau$ to the $G$-transformation of $U(x)$, eq.~(\ref{eq:CCWZ_non_linear_U_transformation}), one gets
\begin{equation}
 G:\,\mathring{U}(x)\to \mathring{\mathscr{g}}\,\mathring{U}(x)\,\mathring{\mathscr{h}}^{-1}(x),
\end{equation}
where $\mathring{\mathscr{g}}$ and $\mathring{\mathscr{h}}^{-1}(x)$ are the images of $\mathscr{g}$ and $\mathscr{h}^{-1}(x)$ under $\tau$.
Since the generators of $H$ transform trivially under $\tau$, one gets $\mathring{\mathscr{h}}^{-1}(x)=\mathscr{h}^{-1}(x)$.
Using in addition that $\mathring{U}(x) = U^{-1}(x)$, one arrives at
\begin{equation}
 G:\,U^{-1}(x)\to \mathring{\mathscr{g}}\,U^{-1}(x)\,\mathscr{h}^{-1}(x).
\end{equation}
and thus
\begin{equation}
 G:\,U(x)\to \mathscr{h}(x)\,U(x)\,\mathring{\mathscr{g}}^{-1}.
\end{equation}
From this, it follows that the squared NGB matrix $\Sigma(x)$ transforms linearly, i.e.\
\begin{equation}\label{eq:CCWZ_linear_Usquared_transformation}
 G:\,\Sigma(x) = U(x)U(x)\to \mathscr{g}\,U(x)\,\mathscr{h}^{-1}(x)\,\mathscr{h}(x)\,U(x)\,\mathring{\mathscr{g}}^{-1}
 = \mathscr{g}\,\Sigma(x)\,\mathring{\mathscr{g}}^{-1}
\end{equation}
and hence, using $\Sigma(x)$, not only the Lagrangian but also the $G$-transformation is tremendously simplified.

A common example where one encounters a symmetric space is the case where a global symmetry that is a direct product of two groups isomorphic to each other is broken to its diagonal subgroup.
To be specific, the global symmetry%
\footnote{In the following, objects that correspond to a direct product of two isomorphic groups are written with a hat. This distinguishes them from objects that correspond to one of the two isomorphic groups themselves.}
$\hat{G}=G_L\times G_R$ is broken to $\hat{H}=G_D$, where $G_L\cong G_R\cong G_D$ are all isomorphic to each other.
Then $\hat{G}/\hat{H}=(G_L\times G_R)/G_D$ is a symmetric space.
It is useful for the later discussion in sections~\ref{sec:CCWZ_HLS} and \ref{sec:Vectorres} to investigate this case in more detail.
Denoting the generators of the $G_L$ and $G_R$ parts of $\hat{G}$ by $\hat{S}_L^a$ and $\hat{S}_R^a$ respectively, the generators of $\hat{H}$ are given by $\hat{T}^a = \frac{1}{\sqrt{2}}(\hat{S}_L^a+\hat{S}_R^a)$ and the broken generators are $\hat{X}^a=\frac{1}{\sqrt{2}}(\hat{S}_L^a-\hat{S}_R^a)$.
Since $\hat{G}$ is a direct product and $G_L$ and $G_R$ are isomorphic to each other, one can employ a matrix notation such that
\begin{equation}\label{eq:CCWZ_symspace_matrix_notation}
 \hat{S}_L^a = \begin{pmatrix}
          S^a & 0\\
          0 & 0\\
         \end{pmatrix},
 \quad
 \hat{S}_R^a = \begin{pmatrix}
          0 & 0\\
          0 & S^a\\
         \end{pmatrix},
 \quad
 \hat{T}^a = \frac{1}{\sqrt{2}}\begin{pmatrix}
          S^a & 0\\
          0 & S^a\\
         \end{pmatrix},
 \quad
 \hat{X}^a = \frac{1}{\sqrt{2}}\begin{pmatrix}
          S^a & 0\\
          0 & -S^a\\
         \end{pmatrix},
\end{equation}
where $S^a$ are the generators of the group $G_L \cong G_R$.
Using this notation, the NGB matrix $\hat{U}(x)$ is given by
\begin{equation}\label{eq:CCWZ_GLxGR_U}
 \hat{U}(x) =
e^{i\frac{\sqrt{2}}{\hat{f}}\,\pi^a(x)\,\hat{X}^a}
=
\begin{pmatrix}
        e^{\frac{i}{\hat{f}}\,\pi^a(x)\,S^a} & 0\\
        0 & e^{-\frac{i}{\hat{f}}\,\pi^a(x)\,S^a}\\
     \end{pmatrix}
=
\begin{pmatrix}
        u(x) & 0\\
        0 & u^{-1}(x)\\
     \end{pmatrix},
\end{equation}
where $u(x) = e^{\frac{i}{\hat{f}}\,\pi^a(x)\,S^a}$.
To infer the transformation properties of $u(x)$, one can start with the $\hat{G}$-transformation of the NGB matrix,
\begin{equation}\label{eq:CCWZ_symspace_Uhat_transformation}
 \hat{G}:\,\hat{U}(x)\to \hat{\mathscr{g}}\,\hat{U}(x)\,\hat{\mathscr{h}}^{-1}(x),
\end{equation}
and use the facts that any $\hat{\mathscr{g}}\in G_L\times G_R$ can be decomposed as $\hat{\mathscr{g}}=\hat{\mathscr{g}}_L\,\hat{\mathscr{g}}_R$ with
\begin{equation}\label{eq:CCWZ_g_gL_gR}
 \hat{\mathscr{g}} = \begin{pmatrix}
        L & 0\\
        0 & R\\
       \end{pmatrix},
\quad
 \hat{\mathscr{g}}_L = \begin{pmatrix}
        L & 0\\
        0 & 1\\
       \end{pmatrix},
\quad
 \hat{\mathscr{g}}_R = \begin{pmatrix}
        1 & 0\\
        0 & R\\
       \end{pmatrix},
\end{equation}
where $L\in G_L$ and $R\in G_R$ and that $\hat{\mathscr{h}}(x)$ can be written as
\begin{equation}
 \hat{\mathscr{h}}(x) = \begin{pmatrix}
        \mathscr{k}(x) & 0\\
        0 & \mathscr{k}(x)\\
       \end{pmatrix}.
\end{equation}
This then yields
\begin{equation}\label{eq:CCWZ_xi_transformation}
\hat{G}:\,u(x) \to L\,u(x)\,\mathscr{k}^{-1}(x) = \mathscr{k}(x)\,u(x)\,R^{-1}.
\end{equation}
Since $\hat{G}/\hat{H}$ is a symmetric space, there is an automorphism $\tau$ of $\hat{G}$ as discussed above.
In the case where $\hat{G}=G_L\times G_R$, this automorphism just corresponds to exchanging $G_L$ and $G_R$.
Using the above matrix notation, the automorphism $\tau$ can be represented by a multiplication from the right by the matrix $\kappa$ given by
\begin{equation}
 \kappa= \begin{pmatrix}
         0 & 1\\
         1 & 0\\
        \end{pmatrix}.
\end{equation}
For an element of $\hat{G}$ one then finds
\begin{equation}\label{eq:CCWZ_tau_g}
 \tau: \hat{\mathscr{g}}\to \hspace{0.2em}\mathring{\hat{\hspace{-0.2em}\mathscr{g}}} = \hat{\mathscr{g}}\,\kappa =
 \begin{pmatrix}
        R & 0\\
        0 & L\\
       \end{pmatrix},
\end{equation}
while for the NGB matrix one gets
\begin{equation}
 \tau: \hat{U}(x)\to \mathring{\hat{U}}(x) = \hat{U}(x)\,\kappa =
\begin{pmatrix}
        u^{-1}(x) & 0\\
        0 & u(x)\\
     \end{pmatrix}
     = \hat{U}^{-1}(x),
\end{equation}
which reproduces the general result from eq.~(\ref{eq:CCWZ_tau_U}).
As the next step, the squared NGB matrix $\hat{\Sigma}(x)$ that transforms linearly via eq.~(\ref{eq:CCWZ_linear_Usquared_transformation}) can be constructed.
In the matrix notation one finds
\begin{equation}\label{eq:CCWZ_symspace_UU_Omega}
 \hat{\Sigma}(x) = \hat{U}(x)\,\hat{U}(x)= \begin{pmatrix}
        u(x)\,u(x) & 0\\
        0 & u^{-1}(x)\,u^{-1}(x)\\
     \end{pmatrix}
     = \begin{pmatrix}
        \Omega(x) & 0\\
        0 & \Omega^{-1}(x)\\
     \end{pmatrix},
\end{equation}
where $\Omega(x)$ is defined as
\begin{equation}
\Omega(x) =u(x)\,u(x) = e^{i\frac{2}{\hat{f}}\,\pi^a(x)\,S^a}\,.
\end{equation}
The transformation properties of $\Omega(x)$ are readily derived by either using those of $\hat{\Sigma}(x)$, eq.~(\ref{eq:CCWZ_linear_Usquared_transformation}), together with the explicit matrix form of $\hat{\mathscr{g}}$, eq.~(\ref{eq:CCWZ_g_gL_gR}), and $\hspace{0.2em}\mathring{\hat{\hspace{-0.2em}\mathscr{g}}}$, eq.~(\ref{eq:CCWZ_tau_g}), or alternatively by employing the transformation properties of $u(x)$, eq.~(\ref{eq:CCWZ_xi_transformation}).
In any case, one gets
\begin{equation}\label{eq:CCWZ_Omega_transformation}
  G_L\times G_R: \Omega(x) \to L\,\Omega(x)\,R^{-1},
\end{equation}
i.e.\ under a $G_L\times G_R$ transformation $\hat{\mathscr{g}}$, the matrix $\Omega(x)$ transforms linearly with $L\in G_L$ and $R\in G_R$ that both constitute $\hat{\mathscr{g}}$ via eq.~(\ref{eq:CCWZ_g_gL_gR}).

Given these transformation properties, it is useful to introduce the so called {\it moose diagrams}~\cite{Georgi:1985hf} to describe the theory.
They are a diagrammatic tool to visualize the symmetry structure and particle content of a theory.
This thesis adopts the notation of \cite{Thaler:2005kr} and depicts global symmetry groups (that may contain gauged subgroups) as circles and fields transforming under these groups as lines connected to the circles.
An arrow on the lines is used to indicate whether the corresponding field transforms under the fundamental (arrow pointing away from the circle) or anti-fundamental (arrow pointing towards the circle) representation of the group associated to the circle.
A circle describing a symmetry group is also called a {\it site}\footnote{%
The term {\it site} is derived from {\it lattice site}, a point in a lattice.
A latticized extra dimension can actually be described by a moose diagram such that the lattice sites correspond to the moose diagram's sites (cf.\ section~\ref{sec:vectorres_contunuum_limit})
}.
The theory of the $(G_L\times G_R)/G_D$ NGBs parametrized by $\Omega(x)$ can thus be described by the following moose diagram with two sites,
\begin{equation}\label{eq:CCWZ_symspace_GL_GR_GD_moose}
\begin{tabular}{c}
\xy
\xymatrix@R=.4pc@C=1.4pc{
\mathrm{Global:} & G_L && G_R  \\
& *=<20pt>[o][F]{} \doublerightxyarrow^{\mbox{\raisebox{1.5ex}{$\Omega$}}} && *=<20pt>[o][F]{}\\
\mathrm{Gauged:} & \emptyset && \emptyset
}
\endxy
\end{tabular},
\end{equation}
i.e.\ $\Omega(x)$ transforms as in eq.~(\ref{eq:CCWZ_Omega_transformation}) and no subgroups of $G_L$ and $G_R$ are gauged.

Using $\Omega(x)$, the Lagrangian $\mathcal{L}_2$ that is written in terms of $\hat{\Sigma}(x)$ in eq.~(\ref{eq:CCWZ_L_2_Usquared}) can be expressed as
\begin{equation}
  \mathcal{L}_2
 =
 \frac{{\hat{f}}^2}{8}\,\tr\left[\partial_\mu \Omega^{-1}(x)\,\partial^\mu \Omega(x)\right].
\end{equation}
There are important cases (as will become clearer in section \ref{sec:CCWZ_HLS}), where one encounters two NGB matrices, one from a $G_L\times G_R\to G_D$ breaking conveniently described by $\Omega(x)$ and one from a $G\to H$ breaking described by $U(x)$, where the groups $G_L\cong G_R \cong G_D \cong G$ are all isomorphic to each other.
The broken generators $X^a$ appearing in the $G/H$ NGB matrix $U(x)$ as well as the generators $T^a$ of $H$ are then each a subset of the generators $S^a$ that appear in the $(G_L\times G_R)/G_D$ NGB matrix $\Omega(x)$.
It is then useful to define the decay constant entering $\Omega(x)$ such that the normalization in the exponent of $\Omega(x)$ matches the one of $U(x)$, eq.~(\ref{eq:CCWZ_U}).
This is achieved by employing
\begin{equation}
 \hat{f} = \sqrt{2}\,f,
\end{equation}
which then yields
\begin{equation}\label{eq:CCWZ_Omega}
  \mathcal{L}_2
 =
 \frac{f^2}{4}\,\tr\left[\partial_\mu \Omega^{-1}(x)\,\partial^\mu \Omega(x)\right],
 \quad
 \Omega(x) = e^{i\frac{\sqrt{2}}{f}\,\pi^a(x)\,S^a},
\end{equation}
where $\Omega(x)$ of course still transforms as in eq.~(\ref{eq:CCWZ_Omega_transformation}).

While the above discussion leads to a simplified Lagrangian in the case where $G/H$ is a symmetric space, in CHMs one often encounters cases where $G/H$ is not a symmetric space.
For practical purposes, it is then often convenient to work with an explicit vacuum state instead of using the $d_\mu[U]$ objects appearing in the CCWZ formalism.
This approach is further discussed in appendix~\ref{sec:vacuum_states}.

\subsection{Hidden local symmetry}\label{sec:CCWZ_HLS}
It is instructive and also useful for the later discussion of vector resonances to introduce the language of the hidden local symmetry (HLS).
The following discussion is loosely based on~\cite{Bando:1984ej,Bando:1987br,Thaler:2005kr} and shows how the $G/H$ non-linear sigma model is equivalent to a linear model with a global $G$ and a local $H$ symmetry.
First, consider the $G/H$ NGB matrix $U(x)$, given by
\begin{equation}
U(x) = e^{i\frac{\sqrt{2}}{f}\,\pi^a_U(x)\,X^a},
\end{equation}
transforming under $G$ as
\begin{equation}
  G: U(x) \to \mathscr{g}\,U(x)\,\mathscr{h}^{-1}(x,\mathscr{g}),
\end{equation}
where $\mathscr{g}\in G$ and $\mathscr{h}(x,\mathscr{g})\in H$ depends on $\mathscr{g}$.
The generators of $G$ and $H$ are the elements of the Lie algebras $\mathfrak{g}$ and $\mathfrak{h}$ and are denoted by $S^a\in\mathfrak{g}$ and $T^a\in\mathfrak{h}$, respectively. $X^a\in\mathfrak{g}-\mathfrak{h}$ are the generators broken by the $G\to H$ spontaneous symmetry breaking.

In addition to the above NGBs, consider a model describing NGBs from a $G_L\times G_R\to G_D$ spontaneous symmetry breaking where $G_L\cong G_R\cong G_D\cong G$ are all isomorphic to each other. This model is parametrized as in eq.~(\ref{eq:CCWZ_Omega}) by the NGB matrix
\begin{equation}
\Omega(x) = e^{i\frac{\sqrt{2}}{f}\,\pi^a_\Omega(x)\,S^a},
\end{equation}
transforming linearly under $G_L\times G_R$ as
\begin{equation}
  G_L\times G_R: \Omega(x) \to L\,\Omega(x)\,R^{-1},
\end{equation}
where $L\in G_L$ and $R\in G_R$.
It is interesting to observe what happens if one gauges a subgroup $H_R\subset G_R$ that is isomorphic to $H$.
This model is then conveniently described by the following moose diagram:
\begin{equation}
\begin{tabular}{c}
\xy
\xymatrix@R=.4pc@C=1.4pc{
\mathrm{Global:} & G_L && G_R  \\
& *=<20pt>[o][F]{} \doublerightxyarrow^{\mbox{\raisebox{1.5ex}{$\Omega$}}} && *=<20pt>[o][F]{}\\
\mathrm{Gauged:} & \emptyset && H_R
}
\endxy
\end{tabular}.
\end{equation}
Since no subgroup of $G_L$ is gauged, the $G_L\times G_R\to G_D$ breaking also spontaneously breaks the $H_R$ gauge symmetry and thus all its gauge bosons acquire mass via the Higgs mechanism.
In addition, the gauging explicitly breaks the $G_L\times G_R$ symmetry to $G_L\times H_R$~\footnote{It is assumed here that there is no other subgroup $E_R\subset G_R$ that commutes with $H_R$.} under which the NGB matrix $\Omega(x)$ transforms as
\begin{equation}\label{eq:CCWZ_HLS_Omega_transformation}
  G_L\times H_R: \Omega(x) \to L\,\Omega(x)\,\mathscr{h}_R^{-1}(x),
\end{equation}
where $L\in G_L$ as before and $\mathscr{h}_R(x)\in H_R$ is a gauge transformation.
This resembles the transformation properties of $U(x)$.
But while $G_L\cong G$ and $H_R\cong H$, of course $U(x)$ transforms non-linearly under the group $G$ and $\Omega(x)$ still transforms linearly under the group $G_L\times H_R$.
Another obvious difference between $\Omega(x)$ and $U(x)$ is that the former contains all the generators $S^a\in\mathfrak{g}$ while the latter only contains the subset $X^a\in\mathfrak{g}-\mathfrak{h}$.
This difference can be made more explicit by decomposing $\Omega(x)$ as
\begin{equation}\label{eq:CCWZ_HLS_Omega_decomposition}
 \Omega(x) = \widetilde{U}(x)\,\widetilde{\Xi}(x)
 \quad
 \text{where}
 \quad
 \widetilde{U}(x) = e^{i\frac{\sqrt{2}}{f}\,\pi^a_{\widetilde{U}}(x)\,X^a},
 \quad
 \widetilde{\Xi}(x)= e^{i\frac{\sqrt{2}}{f}\,\pi^a_{\widetilde{\Xi}}(x)\,T^a}.
\end{equation}
Now if the element of the $G_L/H_R$ coset $\widetilde{U}(x)$ is identified with the element of the $G/H$ coset $U(x)$, the difference between $\Omega(x)$ and $U(x)$ is due to the presence of $\widetilde{\Xi}(x)$.
Since $\widetilde{\Xi}(x)$ is an element of $H_R$, it is however possible to remove it by performing a gauge transformation.
Using a specific gauge transformation $\widetilde{\mathscr{h}}_R(x)=\widetilde{\Xi}(x)$ to fix the gauge, one finds
\begin{equation}\label{eq:CCWZ_HLS_unitary_gauge}
 \Omega(x) \to \Omega(x)\,\widetilde{\mathscr{h}}_R^{-1}(x) = \widetilde{U}(x),
\end{equation}
which is nothing but the unitary gauge where the unphysical would-be NGBs $\pi^a_{\widetilde{\Xi}}(x)$ are removed from the spectrum.
This gauge fixing is however not respected by the global $G_L$ transformations.
Applying a $G_L$ transformation to $\widetilde{U}(x)=\Omega(x)\,\widetilde{\mathscr{h}}_R^{-1}(x)$ yields
\begin{equation}
  G_L: \widetilde{U}(x)=\Omega(x)\,\widetilde{\mathscr{h}}_R^{-1}(x) \to L\,\Omega(x)\,\widetilde{\mathscr{h}}_R^{-1}(x) = \Omega'(x)\,\widetilde{\mathscr{h}}_R^{-1}(x),
\end{equation}
where $\Omega'(x)$ can again be decomposed as
\begin{equation}
 \Omega'(x) = \widetilde{U}'(x)\,\widetilde{\Xi}'(x)
 \quad
 \text{where}
 \quad
 \widetilde{U}'(x) = e^{i\frac{\sqrt{2}}{f}\,\pi'^a_{\widetilde{U}}(x)\,X^a},
 \quad
 \widetilde{\Xi}'(x)= e^{i\frac{\sqrt{2}}{f}\,\pi'^a_{\widetilde{\Xi}}(x)\,T^a}.
\end{equation}
Using this decomposition, the transformation of $\widetilde{U}(x)$ under $G_L$ is found to be
\begin{equation}
 G_L: \widetilde{U}(x) \to \widetilde{U}'(x)\,\widetilde{\Xi}'(x)\,\widetilde{\mathscr{h}}_R^{-1}(x).
\end{equation}
Since $\widetilde{\mathscr{h}}_R^{-1}(x)$ has been chosen such that it removes $\widetilde{\Xi}(x)$, it can not be used to also remove a generic $\widetilde{\Xi}'(x)$ and thus $\widetilde{U}'(x)\,\widetilde{\Xi}'(x)\,\widetilde{\mathscr{h}}_R^{-1}(x)$ is in general not an element of the $G_L/H_R$ coset and would-be NGBs are reintroduced.
To fix this, one can apply just another gauge transformation that removes the $H_R$ element $\widetilde{\Xi}'(x)\,\widetilde{\mathscr{h}}_R^{-1}(x)$.
Different $G_L$ transformations $L$ yield different $\widetilde{\Xi}'(x)$.
Depending on $L$, different gauge transformations are thus needed for removing $\widetilde{\Xi}'(x)\,\widetilde{\mathscr{h}}_R^{-1}(x)$.
These $L$-dependent gauge transformations $\mathscr{h}_R(x,L)$ are then defined by
\begin{equation}
 \widetilde{\Xi}'(x)\,\widetilde{\mathscr{h}}_R^{-1}(x)\,\mathscr{h}_R^{-1}(x,L) = \mathds{1}.
\end{equation}
So to stay in the unitary gauge, after each global $G_L$ transformation of $\widetilde{U}(x)$, the gauge transformation $\mathscr{h}_R(x,L)$ has to be applied.
The $G_L$ transformation of $\widetilde{U}(x)$ that keeps the unitary gauge fixed can thus be defined as
\begin{equation}\label{eq:HLS_Uhat_transformation}
 G_L: \widetilde{U}(x)\to L\,\widetilde{U}(x)\,\mathscr{h}_R^{-1}(x,L) = \widetilde{U}'(x).
\end{equation}
Under this $G_L$ transformation that includes the prescription for the gauge fixing, $\widetilde{U}(x)$ evidently transforms non-linearly.
The $H_R$ gauge symmetry is then only used for keeping the unitary gauge fixed when performing a $G_L$ transformation and is otherwise explicitly broken by the gauge fixing.
This is the reason why $H_R$ is said to be a hidden local symmetry.
The equivalence of the transformation properties of $U(x)$ under $G$ and those of $\widetilde{U}(x)$ under $G_L$ is now manifest and can be made a one-to-one correspondence by identifying
\begin{equation}
 G\leftrightarrow G_L
 ,\quad
 \mathscr{g}\leftrightarrow L
 ,\quad
 H\leftrightarrow H_R
 ,\quad
 \mathscr{h}(x,\mathscr{g})\leftrightarrow \mathscr{h}_R(x,L)
 ,\quad
 U(x)\leftrightarrow \widetilde{U}(x).
\end{equation}
An obvious difference between the two models discussed above is of course that in contrast to the non-linear sigma model, the HLS model contains massive gauge bosons.
By taking the $H_R$ gauge coupling $g_{H_R}\to \infty$, the gauge bosons can however be made infinitely heavy and thus decouple.
In this limit, both models are completely equivalent, i.e.\ the HLS model becomes a non-linear sigma model describing NGBs in the $G_L/H_R$ coset with no additional gauge bosons present.
Using the language of moose diagrams, this can be expressed as
\begin{equation}
\begin{tabular}{c}
\xy
\xymatrix@R=.4pc@C=1.4pc{
\mathrm{Global:} & G_L && G_R  \\
& *=<20pt>[o][F]{} \doublerightxyarrow^{\mbox{\raisebox{1.5ex}{$\Omega$}}} && *=<20pt>[o][F]{}\\
\mathrm{Gauged:} & \emptyset && H_R
}
\endxy
\end{tabular}
\quad\quad
\mathop{\Longrightarrow}^{\mbox{\raisebox{1.5ex}{$g_{H_R}\to \infty$}}}
\quad\quad
\begin{tabular}{c}
\xy
\xymatrix@R=.4pc@C=1.4pc{
G_L  &&&  \\
 *=<20pt>[o][F]{} \doublerightxyarrow^{\mbox{\raisebox{1.5ex}{$\widetilde{U}$}}} && *=<0pt,20pt>[l][F]{} & *-<0pt,20pt>[l]{H_R}\\
\emptyset &&&
}
\endxy
\end{tabular}.
\end{equation}
The moose diagram on the right-hand side describes a non-linear sigma model where a global symmetry $G_L$ is spontaneously broken to $H_R$, no subgroup of $G_L$ is gauged and the NGBs in the $G_L/H_R$ coset are parametrized by the NGB matrix $\widetilde{U}(x)$, transforming as in eq.~(\ref{eq:HLS_Uhat_transformation}).
$\widetilde{U}(x)$ is nothing but $\Omega(x)$ in the unitary gauge, cf.\ eq.~\eqref{eq:CCWZ_HLS_unitary_gauge}.
The two-site diagram on the left-hand side is thus reduced to a one-site diagram by taking $g_{H_R}\to\infty.$

If in addition to $H_R$, also a subgroup $E_L \subseteq G_L$ is gauged, in the limit $g_{H_R}\to \infty$ this exactly corresponds to the gauging of $E \subseteq G$ in the $G/H$ non-linear sigma model discussed in section~\ref{sec:NGBs_CCWZ}.
All generators of $E_L$ that are also generators of $H_R$ yield linear combinations of the $E_L$ and $H_R$ generators that are unbroken by the spontaneous symmetry breaking $G_L\times G_R\to G_D$ and correspond to massless gauge bosons.
The orthogonal linear combinations of these generators are however spontaneously broken and yield massive gauge bosons that become infinitely heavy as $g_{H_R}\to \infty$.
The generators of $E_L$ that are not generators of $H_R$ are associated to gauge bosons that acquire mass via the Higgs mechanism and the corresponding NGBs in the $G_L/H_R$ coset become unphysical would-be NGBs that can be gauged away.
This is thus completely analogous to the discussion in section~\ref{sec:NGBs_CCWZ}.

\section{Vector resonances}\label{sec:Vectorres}
In models where the spontaneous symmetry breaking that yields NGBs is due to a confining strong interaction, like it is the case for QCD and CHMs, in addition to the NGBs one also expects heavy spin one resonances.
These resonances have to respect the same global symmetries as the NGBs.
An example of a model where this is realized is the above discussed HLS model if the gauge coupling $g_{H}$ is not taken to infinity but kept finite.
In this case, it contains both NGBs in a $G/H$ coset as well as heavy resonances in the adjoint representation of $H$.
In the present section, this construction is first investigated in more detail and subsequently generalized to include several multiplets of resonances.
Finally, a connection to models with extra dimensions is made.

\subsection{Hidden local symmetry and a first level of heavy resonances}\label{sec:Vectorres_NLSM_HLS}
In section~\ref{sec:CCWZ_HLS}, it is shown that the HLS model with a global $G$ and a local $H$ symmetry can be reduced to the non-linear sigma model describing NGBs in a $G/H$ coset.
This is done by taking the gauge coupling $g_H\to\infty$ and thus effectively removing the heavy $H$ gauge bosons.
This procedure can be reversed to add spin one resonances in an adjoint representation of $H$ to a non-linear sigma model describing NGBs in a $G/H$ coset.
Starting from the $G/H$ non-linear sigma model, one just has to promote it to a HLS model with a global $G$ and a local $H$ symmetry and keep the gauge coupling $g_H$ finite.
In terms of moose diagrams, this corresponds to
\begin{equation}\label{eq:Vectorres_HLS_extension}
\begin{tabular}{c}
\xy
\xymatrix@R=.4pc@C=1.4pc{
\mathrm{Global:} & G  &&&  \\
& *=<20pt>[o][F]{} \doublerightxyarrow^{\mbox{\raisebox{1.5ex}{$U$}}} && *=<0pt,20pt>[l][F]{} & *-<0pt,20pt>[l]{H}\\
\mathrm{Gauged:} & E &&&
}
\endxy
\end{tabular}
\quad
\mathop{\longrightarrow}^{\mbox{\raisebox{1.5ex}{extend to HLS}}}
\quad
\quad
\begin{tabular}{c}
\xy
\xymatrix@R=.4pc@C=1.4pc{
G && G'  \\
 *=<20pt>[o][F]{} \doublerightxyarrow^{\mbox{\raisebox{1.5ex}{$\Omega$}}} && *=<20pt>[o][F]{}\\
E && H
}
\endxy
\end{tabular},
\end{equation}
where for generality also a subgroup $E\subseteq G$ is gauged.
The global symmetry $G'$ on the right site of the HLS moose diagram could also be chosen to be just $H$.
It is however convenient to describe it as a global $G'\cong G$ symmetry that is explicitly broken to its subgroup $H$ by the gauging%
\footnote{Using $G'\cong G$, one can think about $\Omega(x)$ also as the NGB matrix in a theory with a global $G\times G'$ symmetry broken to its diagonal subgroup (cf.\ sections \ref{sec:symmetric_spaces} and \ref{sec:CCWZ_HLS}).}.
The field $\Omega(x)$ transforms under the global $G$ and the local $H$ symmetry as
\begin{equation}
 G\times H : \Omega(x)\to \mathscr{g}\,\Omega(x)\,\mathscr{h}^{-1}(x),
\end{equation}
where $\mathscr{g}\in G$ and $\mathscr{h}(x)\in H$ is a gauge transformation.
It can be decomposed as
\begin{equation}\label{eq:Vectorres_Omega_decomposition}
 \Omega(x) = U(x)\,\Xi(x),
 \quad\quad
 U(x) = e^{i\frac{\sqrt{2}}{f}\,\pi^a(x)\,X^a},
 \quad\quad
 \Xi(x)= e^{i\frac{\sqrt{2}}{f_{\Xi}}\,\pi^a_{\Xi}(x)\,T^a},
\end{equation}
where $U(x)$ is the NGB matrix in the non-linear sigma model on the left-hand side of eq.~(\ref{eq:Vectorres_HLS_extension}) and $\Xi(x)\in H$ contains the would-be NGBs that can be gauged away by a $H$ gauge transformation (cf.\ section~\ref{sec:CCWZ_HLS}).
The decay constant $f_\Xi$ of the would-be NGBs and the decay constant $f$ of the non-linear sigma model are in general different%
\footnote{The constant $f_\Xi$ is relevant because even if the would-be NGBs are gauged away, $f_\Xi$ enters the mass terms of the heavy gauge bosons (cf.\ eq.~(\ref{eq:Vectorres_mass_terms})).
In section~\ref{sec:CCWZ_HLS}, it is assumed that $\Omega(x)$ parametrizes the NGBs of a spontaneous symmetry breaking where the coset is a symmetric space.
This construction forces $f_{\Xi}=f$.
In the limit $g_H\to\infty$ considered in section~\ref{sec:CCWZ_HLS}, there is however no difference between the general construction and the one stemming from the symmetric space: if $\Xi(x)$ is gauged away and the heavy gauge bosons are decoupled, no dependence on $f_{\Xi}$ is left.}.

The HLS model in eq.~(\ref{eq:Vectorres_HLS_extension}) is now investigated in more detail.
To this end, it is useful to fix the notation.
Like in the previous sections, the generators of $H$ are called $T^a$, while the generators of the $G/H$ coset are called $X^a$.
In addition, it is convenient for the following discussion to denote the generators of the intersection $E\cap H$, i.e.\ those that are both generators of $H$ and of $E$ by
$K^a$, the generators of $E$ that are not generators of $H$ by $P^{\hat{a}}$ and the generators of $H$ that are not generators of $E$ by $T^{\hat{a}}$.
In terms of the Lie algebras $\mathfrak{h}$ and $\mathfrak{e}$ of $H$ and $E$, this reads
\begin{equation}
 P^{\hat{a}},K^a\in\mathfrak{e},
 \quad
 T^{\hat{a}},K^a\in\mathfrak{h}.
\end{equation}
Using the above defined generators, the covariant derivative of the matrix $\Omega(x)$ is given by
\begin{equation}
\begin{aligned}
 i\,D_\mu\,\Omega(x) = i\,\partial_\mu\,\Omega(x)
 &+g_{E}\left((\hat{A}_E)_\mu^{\hat{a}}\,P^{\hat{a}}+(A_E)_\mu^a\,K^a\right)\Omega(x)
 \\
 &-g_{H}\,\Omega(x)\left((\hat{A}_H)_\mu^{\hat{a}}\,T^{\hat{a}}+(A_H)_\mu^a\,K^a\right),
\end{aligned}
\end{equation}
where $g_E$ and $g_H$ are the coupling constants of the $E$ and $H$ gauge groups.
The leading order Lagrangian then contains the following mass terms for the gauge bosons (cf.~\cite{Harada:2003jx}):
\begin{equation}\label{eq:Vectorres_mass_terms}
\begin{aligned}
 \mathcal{L}_2\supset
 &\phantom{{}+{}}\frac{f_\Xi^2\,g_{E}^2}{4}\left(\frac{f^2}{f_\Xi^2}\,(\hat{A}_E)_\mu^{\hat{a}} (\hat{A}_E)^{\hat{a}\,\mu} + (A_E)_\mu^a (A_E)^{a\,\mu}\right)
 \\
 &+\frac{f_\Xi^2\,g_{H}^2}{4}\left((\hat{A}_H)_\mu^{\hat{a}} (\hat{A}_H)^{\hat{a}\,\mu} + (A_H)_\mu^a (A_H)^{a\,\mu}\right)
 \\
 &-\frac{f_\Xi^2\,g_{H}\,g_{E}}{2}\,(A_H)_\mu^a (A_E)^{a\,\mu}.
\end{aligned}
\end{equation}
The last line corresponds to a mixing term for the gauge bosons $(A_H)_\mu^a$ and $(A_E)_\mu^a$ which are therefore not mass eigenstates.
It is however straightforward to rewrite the mass terms as
\begin{equation}\label{eq:Vectorres_mass_terms_diag}
\begin{aligned}
 \mathcal{L}_2\supset
 &\phantom{{}+{}}\frac{m_{E}^2}{2}\,(\hat{A}_E)_\mu^{\hat{a}} (\hat{A}_E)^{\hat{a}\,\mu}
  +\frac{m_{H}^2}{2}\,(\hat{A}_H)_\mu^{\hat{a}} (\hat{A}_H)^{\hat{a}\,\mu}
  \\
 &+\frac{m_{K}^2}{2}\,\left(
 \cos\theta_{\rm mix}\,(A_H)_\mu^a - \sin\theta_{\rm mix}\,(A_E)_\mu^a
 \right)
 \left(
 \cos\theta_{\rm mix}\,(A_H)^{a\,\mu} - \sin\theta_{\rm mix}\,(A_E)^{a\,\mu}
 \right),
\end{aligned}
\end{equation}
where the masses are given by
\begin{equation}\label{eq:Vectorres_gauge_boson_masses}
m_{E}^2 = \frac{f^2\,g_{E}^2}{2}
,\quad
m_{H}^2 = \frac{f_\Xi^2\,g_{H}^2}{2}
,\quad
m_{K}^2 = \frac{f_\Xi^2\,(g_{H}^2+g_{E}^2)}{2},
\end{equation}
and the mixing angle $\theta_{\rm mix}$ is defined by
\begin{equation}
\cos\theta_{\rm mix} = \frac{g_{H}}{\sqrt{g_{H}^2+g_{E}^2}}
,\quad
\sin\theta_{\rm mix} = \frac{g_{E}}{\sqrt{g_{H}^2+g_{E}^2}}.
\end{equation}
From the Lagrangian in eq.~(\ref{eq:Vectorres_mass_terms_diag}), one can read off that the linear combinations of gauge fields
\begin{equation}
 (A_K)_\mu^a = \cos\theta_{\rm mix}\,(A_H)_\mu^a - \sin\theta_{\rm mix}\,(A_E)_\mu^a
\end{equation}
have masses $m_K$, while there are no mass terms for the orthogonal linear combinations
\begin{equation}
 (A_0)_\mu^a = \sin\theta_{\rm mix}\,(A_H)_\mu^a + \cos\theta_{\rm mix}\,(A_E)_\mu^a,
\end{equation}
which are consequently massless.
To summarize, for each generator $P^{\hat{a}}$ there is a massive gauge boson $(\hat{A}_E)_\mu^{\hat{a}}$ with mass $m_{E}$, for each $T^{\hat{a}}$ there is a massive gauge boson $(\hat{A}_H)_\mu^{\hat{a}}$ with mass $m_{H}$ and for each $K^a$ there is one massive gauge boson $(A_K)_\mu^a$ with mass $m_K$ and one massless gauge boson $(A_0)_\mu^a$.
$(A_K)_\mu^a K^a$ and $(A_0)_\mu^a K^a$ are in the adjoint representation of $E\cap H$.
It is instructive to investigate several different limits and special cases of this model.
\begin{enumerate}
 \item Taking $g_{H}\to \infty$, one finds
 \begin{equation}
  m_{H}\to\infty
  ,\quad
  m_{K}\to\infty
  ,\quad
  (A_K)_\mu^a \to (A_H)_\mu^a
  ,\quad
  (A_0)_\mu^a \to (A_E)_\mu^a,
 \end{equation}
 i.e.\ in this case the gauge bosons $(\hat{A}_H)_\mu^{\hat{a}}$ and $(A_K)_\mu^a$ become infinitely heavy and decouple.
 There is no mixing between $(A_H)_\mu^a$ and $(A_E)_\mu^a$.
 The former become exactly the infinitely heavy $(A_K)_\mu^a$ gauge bosons while the latter become exactly the massless $(A_0)_\mu^a$.
 There are NGBs in the $G/H$ coset, of which those corresponding to the $P^{\hat{a}}$ generators associated with the massive $(\hat{A}_E)_\mu^{\hat{a}}$ gauge bosons are unphysical and can be gauged away.
 This is of course just again the non-linear sigma model for a $G/H$ coset with $E \subseteq G$ gauged, i.e.\ the model discussed in sections \ref{sec:NGBs_CCWZ} and \ref{sec:CCWZ_HLS}.
 \item Assuming $g_{E}\ll g_{H}$, the mixing angle $\theta_{\rm mix}$ is small and one arrives at a model where the massless $(A_0)_\mu^a$ are approximately the $(A_E)_\mu^a$ gauge bosons, while the massive $(A_K)_\mu^a$ are approximately the $(A_H)_\mu^a$.
 $(A_K)_\mu^a\,K^a$ and $(\hat{A}_H)_\mu^{\hat{a}}\,T^{\hat{a}}$ together thus approximately constitute a full adjoint representation of $H$.
 While their squared masses differ by $m_K^2-m_H^2=\tfrac{f_\Xi^2}{f^2} m_E^2$ (cf.\ eq.~(\ref{eq:Vectorres_gauge_boson_masses})), assuming $g_{E}\ll g_{H}$, this difference is however small compared to the values of $m_K^2$ and $m_H^2$ and thus using $m_K/m_H\approx 1$ is a good approximation.
 Although $(A_0)_\mu^a$ and $(\hat{A}_E)_\mu^{\hat{a}}$ would also constitute a full adjoint representation of $E$ if their masses were equal, this is not a good approximation since the former is massless and the latter has a finite mass $m_E$ and hence the relative difference of their masses is large.
 \label{item:Vectorres_small_mixing}
 \item  Taking $g_{E}\to 0$, $(A_K)_\mu^a\,K^a$ is exactly $(A_H)_\mu^a\,K^a$ and together with $(\hat{A}_H)_\mu^{\hat{a}}\,T^{\hat{a}}$ it constitutes an adjoint representation of $H$.
 The approximation from the previous case is now exact since the mass difference $m_K^2-m_H^2\to 0$.
 $(\hat{A}_E)_\mu^{\hat{a}}$ and $(A_E)_\mu^a$ (which is now exactly $(A_0)_\mu^a$) do not couple if $g_{E}\to 0$ and hence this limit is equivalent to removing them altogether.
 While taking $g_{E}\to 0$ has exactly the same consequences as $E\to \emptyset$, the interpretation in the latter case is however slightly different: there are no $P^{\hat{a}}$ generators, all the $K^a$ become $T^{\hat{a}}$ generators and the spin one spectrum consists of only the massive $(\hat{A}_H)_\mu^{\hat{a}}\,T^{\hat{a}}$ in the adjoint representation of $H$.
 But in any case one is left with a spectrum consisting of massive gauge bosons in the adjoint representation of $H$ with mass $m_{H}$ and NGBs in the $G/H$ coset that are all physical.
 \item\label{item:Vectorres_E_subgroup_H} In the case where $E$ is isomorphic to a subgroup of $H$, there is no $P^{\hat{a}}$ generator because $E\cap H = E$ and thus all generators of $E$ are also generators of $H$.
 Consequently, there are no $(\hat{A}_E)_\mu^{\hat{a}}$ gauge bosons and the spectrum consists of the massive $(\hat{A}_H)_\mu^{\hat{a}}$ and $(A_K)_\mu^a$ as well as the massless $(A_0)_\mu^a$ gauge bosons.
 The $(A_K)_\mu^a\,K^a$ and $(A_0)_\mu^a\,K^a$ are in the adjoint representation of $E$ while the $(\hat{A}_H)_\mu^{\hat{a}}\,T^{\hat{a}}$ are elements of $\mathfrak{h}-\mathfrak{e}$.
 All NGBs in the $G/H$ coset are physical.
 \item In the case where $H\cong G$, all NGBs correspond to either a $K^a$ or a $T^{\hat{a}}$ generator, are unphysical and can be gauged away.
 Apart from the fact that there are no NGBs left, this case is similar to the previous one since in this case one always has $E\cap H = E$.
 So again, the spin one spectrum consists of the $(A_K)_\mu^a\,K^a$ and $(A_0)_\mu^a\,K^a$ in the adjoint representation of $E$ and the $(\hat{A}_H)_\mu^{\hat{a}}\,T^{\hat{a}}$ which are now elements of $\mathfrak{g}-\mathfrak{e}$.
\end{enumerate}
Given all the different limits and cases discussed above, the HLS model in eq.~(\ref{eq:Vectorres_HLS_extension}) can serve as a starting point for a wide variety of models, from one containing only NGBs where $E=\emptyset$ and $g_H\to\infty$ to a model featuring NGBs, massive spin one states and also massless gauge bosons.
It is thus the prototype model for describing the lightest resonances in a low-energy effective description of a strongly coupled confining theory.

In the context of QCD, it has been applied to describe pions as NGBs and the $\rho$ mesons as the lightest spin one states \cite{Bando:1984ej}.
In this case one has $G={\rm SU}(2)_{\rm L}\times {\rm SU}(2)_{\rm R}$ and $H={\rm SU}(2)_{\rm D}$.
The photon can be included in this construction by gauging an appropriate $E={\rm U}(1)_{\rm Q}$ subgroup of $G$.
The corresponding moose diagram is thus written as
\begin{equation}\label{eq:Vectorres_rho_moose}
\begin{tabular}{c}
\xy
\xymatrix@R=.4pc@C=1.4pc{
\mathrm{Global:} & {\rm SU}(2)_{\rm L}\times {\rm SU}(2)_{\rm R} && {\rm SU}(2)_{\rm D}  \\
& *=<20pt>[o][F]{} \doublerightxyarrow^{\mbox{\raisebox{1.5ex}{$\Omega$}}} && *=<20pt>[o][F]{}\\
\mathrm{Gauged:} & {\rm U}(1)_{\rm Q} && {\rm SU}(2)_{\rm D}
}
\endxy
\end{tabular}.
\end{equation}
While the massless $(A_0)_\mu$ linear combination of $(A_E)_\mu$ and $(A_H)_\mu$ can be identified with the physical photon, the $\rho_\mu^0$ corresponds to the orthogonal linear combination $(A_K)_\mu$ and its mass is enhanced compared to the charged $\rho_\mu^\pm$ that correspond to $(A_H)_\mu^{\hat{a}}$ (cf.\ eq.~(\ref{eq:Vectorres_gauge_boson_masses})).
Modern versions of this construction use $G={\rm U}(3)_{\rm L}\times {\rm U}(3)_{\rm R}$ and $G'=H={\rm U}(3)_{\rm V}$ to model $\pi$, K, $\eta$ and $\eta'$ as NGBs as well as $\rho$, $\omega$, $K^*$ and $\phi$ as spin one resonances and also include terms that model the breaking of the ${\rm U}(3)$ flavor symmetries due to the different quark masses~\cite{Harada:2003jx,Benayoun:2011mm}.

In the context of CHMs, the HLS model in eq.~(\ref{eq:Vectorres_HLS_extension}) is known from the two-site Discrete Composite Higgs Model (DCHM)~\cite{Panico:2011pw} where in the minimal construction $G={\rm SO}(5)_{\rm L}$, $G'={\rm SO}(5)_{\rm R}$ and $H={\rm SO}(4)$, yielding a Higgs doublet in the ${\rm SO}(5)_{\rm L}/{\rm SO}(4)$ coset and spin one resonances approximately%
\footnote{The massive spin one resonances are only approximately in an adjoint representation of ${\rm SO}(4)$ due to the mixing with ${\rm SU}(2)_{\rm L}^0\times {\rm U}(1)_{\rm Y}^0$ gauge bosons, cf.\ \#\ref{item:Vectorres_small_mixing} of the above limits and special cases.}
in an adjoint representation of ${\rm SO}(4)$.
In addition, an $E={\rm SU}(2)_{\rm L}^0\times {\rm U}(1)_{\rm Y}^0$ subgroup of $G$ is gauged to include the electroweak gauge bosons.
The corresponding moose diagram thus reads
\begin{equation}\label{eq:Vectorres_two_site_DCHM_moose}
\begin{tabular}{c}
\xy
\xymatrix@R=.4pc@C=1.4pc{
\mathrm{Global:} & {\rm SO}(5)_{\rm L} && {\rm SO}(5)_{\rm R}  \\
& *=<20pt>[o][F]{} \doublerightxyarrow^{\mbox{\raisebox{1.5ex}{$\Omega$}}} && *=<20pt>[o][F]{}\\
\mathrm{Gauged:} & {\rm SU}(2)_{\rm L}^0\times {\rm U}(1)_{\rm Y}^0 && {\rm SO}(4)
}
\endxy
\end{tabular}.
\end{equation}
Here, the massless electroweak gauge bosons (before electroweak symmetry breaking) are the $(A_0)^a_\mu$ linear combinations of  of $(A_E)^a_\mu$ and $(A_H)^a_\mu$ while the $(A_K)^a_\mu$ and $(A_H)_\mu^{\hat{a}}$ are the massive spin one resonances.

In both examples above, an $H$ subgroup of $G'$ as well as an $E$ subgroup of $G$ are gauged.
The gauging of $H$ is a central ingredient of the HLS construction.
It allows to make the connection to the $G/H$ non-linear sigma model by removing the would-be NGBs (cf.\ section~\ref{sec:CCWZ_HLS}).
Due to this construction, its gauge bosons are always massive.
Assuming a strongly coupled confining theory as UV completion of the HLS model, these massive gauge bosons as well as the NGBs in the $G/H$ coset are then naturally interpreted as composite objects.
The gauging of $E$ on the other hand is different.
In the limit where the heavy $H$ gauge bosons are decoupled and the HLS model becomes the non-linear sigma model, the gauging of $E$ corresponds to a gauging of some of the global symmetries of the non-linear sigma model.
Assuming again a strongly coupled confining UV completion, these global symmetries are those of the UV theory before the spontaneous symmetry breaking.
If some of these symmetries are gauged, the corresponding gauge bosons are thus also present in the UV theory and can therefore not be interpreted as composite objects but have to be considered as being elementary.
As expected for gauge bosons present in the UV theory, they are found to be massless in the effective non-linear sigma model description if they correspond to symmetries that are not spontaneously broken by the strong interactions.
If the heavy $H$ gauge bosons in the HLS model are not decoupled, these composite resonances mix with the elementary $E$ gauge bosons.
If in addition the model corresponds to case \#\ref{item:Vectorres_small_mixing} of the above discussed special cases, i.e.\ $g_E\ll g_H$, then all massless gauge bosons in this model mainly consist of elementary gauge bosons with a small admixture of composite resonances.
They are thus said to be partially composite.
This {\it partial compositeness} is encountered in both examples above.
It applies to the physical photon in the QCD case as well as to the electroweak gauge bosons in the two-site DCHM.

\subsection{Adding higher levels of heavy resonances}\label{sec:vectorres_higher_levels}
Starting from NGBs in a $G/H$ coset, the previous section shows how to include heavy resonances in an adjoint representation of $H$ by using the HLS construction.
In the present section, this construction is extended to both higher levels of resonances as well as resonances in an adjoint representation of the full $G$ group.
Both can actually be done at once.
As a starting point, consider the moose diagram
\begin{equation}\label{eq:Vectorres_GL_GR_GD_U_NLSM_moose}
\begin{tabular}{c}
\xy
\xymatrix@R=.4pc@C=1.4pc{
\mathrm{Global:} & G_L\times G_R  &&&  \\
& *=<20pt>[o][F]{} \doublerightxyarrow^{\mbox{\raisebox{1.5ex}{$\hat{U}$}}} && *=<0pt,20pt>[l][F]{} & *-<0pt,20pt>[l]{G_D}\\
\mathrm{Gauged:} & \emptyset &&&
}
\endxy
\end{tabular},
\end{equation}
describing NGBs in a $(G_L\times G_R)/G_D$ coset.
This is just the model already discussed in section~\ref{sec:symmetric_spaces} where the coset is a symmetric space and $\hat{U}(x)$ transforms under a $G_L\times G_R$ transformation $\hat{\mathscr{g}}$ non-linearly as in eq.~(\ref{eq:CCWZ_symspace_Uhat_transformation}), which is repeated here for convenience:
\begin{equation}
 G_L\times G_R:\,\hat{U}(x)\to \hat{\mathscr{g}}\,\hat{U}(x)\,\hat{\mathscr{h}}^{-1}(x).
\end{equation}
After the discussion in section~\ref{sec:Vectorres_NLSM_HLS}, it is now easy to add heavy resonances in the adjoint representation of $G_D$ to this model.
One just has to extend it to an HLS model.
The corresponding moose diagram is given by
\begin{equation}\label{eq:Vectorres_GL_GR_GD_U_HLS_moose}
\begin{tabular}{c}
\xy
\xymatrix@R=.4pc@C=1.4pc{
\mathrm{Global:} & G_L\times G_R && G_L'\times G_R'  \\
& *=<20pt>[o][F]{} \doublerightxyarrow^{\mbox{\raisebox{1.5ex}{$\hat{\Omega}$}}} && *=<20pt>[o][F]{}\\
\mathrm{Gauged:} & \emptyset && G_D
}
\endxy
\end{tabular},
\end{equation}
where $\hat{\Omega}(x)=\hat{U}(x)\,\hat{\Xi}(x)$.
$\hat{\Omega}(x)$ transforms under the full global symmetry $G_L\times G_R\times G'_L\times G'_R$ linearly as
\begin{equation}\label{eq:Vectorres_Omegahat_transformation_properties}
 G_L\times G_R\times G'_L\times G'_R:\,\hat{\Omega}(x)\to\hat{\mathscr{g}}\,\hat{\Omega}\,\hat{\mathscr{g}}'^{-1},
\end{equation}
where $\hat{\mathscr{g}}\in G_L\times G_R$ and $\hat{\mathscr{g}}'\in G'_L\times G'_R$.
The $G_D$ gauge symmetry allows going to unitary gauge by gauging away $\hat{\Xi}(x)$ such that $\hat{\Omega}(x)\to\hat{U}(x)$.
In the unitary gauge, $\hat{U}(x)$ transforms non-linearly under a $G_L\times G_R$ transformation $\hat{\mathscr{g}}$ as
\begin{equation}
 G_L\times G_R:\,\hat{U}(x)\to \hat{\mathscr{g}}\,\hat{U}(x)\,\hat{\mathscr{g}}_{G_D}^{-1}(x,\hat{\mathscr{g}}),
\end{equation}
where $\hat{\mathscr{g}}_{G_D}(x,\hat{\mathscr{g}})$ is the element of the $G_D$ gauge symmetry that removes the would-be NGBs from $\hat{U}(x)$ to stay in unitary gauge (cf.\ section~\ref{sec:CCWZ_HLS}).
For the further discussion, it is however useful to keep $\hat{\Xi}(x)$ such that the linear transformation properties of $\hat{\Omega}(x)$ can be used.

In section~\ref{sec:symmetric_spaces}, a matrix notation is presented that makes it possible to treat the direct factors of the direct product group in a way separately.
Doing this is the crucial step in the present discussion.
Employing the matrix notation, the objects $\hat{U}(x)$ and $\hat{\Xi}(x)$ are given by
\begin{equation}
 \hat{U}(x) =
\begin{pmatrix}
        u(x) & 0\\
        0 & u^{-1}(x)\\
     \end{pmatrix},
 \quad\quad
 \hat{\Xi}(x) =
\begin{pmatrix}
        \xi(x) & 0\\
        0 & \xi(x)\\
     \end{pmatrix},
\end{equation}
where $u(x) = e^{\frac{i}{\hat{f}}\,\pi^a(x)\,S^a}$ and $\xi(x) = e^{\frac{i}{\hat{f}}\,\pi_\Xi^a(x)\,S^a}$ (cf.\ eqs.~(\ref{eq:CCWZ_symspace_matrix_notation}) and~(\ref{eq:CCWZ_GLxGR_U})).
Defining
\begin{equation}\label{eq:Vectorres_omega1_omega2_definition}
 \omega_1(x) = u(x)\,\xi(x)
\quad{\rm and}\quad
\omega_2(x) = \xi^{-1}(x)\,u(x),
\end{equation}
the matrix notation can also be applied to $\hat{\Omega}(x)$.
This yields
\begin{equation}
 \hat{\Omega}(x)=
\begin{pmatrix}
        \omega_1(x) & 0\\
        0 & \omega_2^{-1}(x)\\
     \end{pmatrix}.
\end{equation}
Using the matrix notation for $\hat{\Omega}(x)$ together with the matrix notation for $\hat{\mathscr{g}}$ and $\hat{\mathscr{g}}'$ (cf.\ eq.~(\ref{eq:CCWZ_g_gL_gR})), the transformations properties of $\omega_1(x)$ and $\omega_2(x)$ under the full global symmetry are readily derived from those of $\hat{\Omega}(x)$, eq.~(\ref{eq:Vectorres_Omegahat_transformation_properties}).
One finds
\begin{equation}
\begin{aligned}
 G_L\times G_R\times G'_L\times G'_R: \omega_1(x) &\to L\, \omega_1(x)\,L'^{-1},
 \\
 G_L\times G_R\times G'_L\times G'_R: \omega_2(x) &\to R'\, \omega_1(x)\,R^{-1},
\end{aligned}
\end{equation}
i.e.\ $\omega_1(x)$ and $\omega_2(x)$ only transform non-trivially under $G_L\times G'_L$ and $G_R\times G'_R$, respectively.
Like $\hat{\Omega}(x)$, they also transform linearly.
Comparing this to the discussion in section~\ref{sec:symmetric_spaces},
$\omega_1(x)$ and $\omega_2(x)$ can thus be thought of as describing the NGBs of global $G_L\times G'_L$ and $G_R\times G'_R$ symmetries, that are each spontaneously broken to their diagonal subgroups.
Would it not be for the gauging of $G_D\subset (G'_L\times G'_R)$, this model would therefore just correspond to two copies of the model described in eq.~(\ref{eq:CCWZ_symspace_GL_GR_GD_moose})%
\footnote{
Starting with two copies of the model described in eq.~(\ref{eq:CCWZ_symspace_GL_GR_GD_moose}), i.e.\ a $(G_L\times G'_L)/{G_L}_D$ model with NGB matrix $\omega_1(x)$ and a $(G_R\times G'_R)/{G_R}_D$ model with NGB matrix $\omega_2(x)$ and connecting both copies to each other by gauging the diagonal subgroup of $G'_L\times G'_R$, one obviously arrives at essentially the same model as the one discussed in this section.
This can be used for building models containing heavy spin one resonances (see e.g.~\cite{Panico:2011pw,DeCurtis:2011yx}).}.
The transformation properties of $\omega_1(x)$ and $\omega_2(x)$ can be expressed by the moose diagram
\begin{equation}\label{eq:Vector_two_omega_split_moose}
\begin{tabular}{c}
\xy
\xymatrix@R=.4pc@C=1.4pc{
\mathrm{Global:} & G_L && G'_L & G'_R && G_R \\
& *=<20pt>[o][F]{} \doublerightxyarrow^{\mbox{\raisebox{1.5ex}{$\omega_1$}}} && *=<20pt>[o][F]{} & *=<20pt>[o][F]{}      \doublerightxyarrow^{\mbox{\raisebox{1.5ex}{$\omega_2$}}} && *=<20pt>[o][F]{} \\
\mathrm{Gauged:} &\emptyset &&\ar@{}[r]|{\mbox{\raisebox{0.0ex}{$G_D$}}} &&&  \emptyset
\save "2,4"-(5,5);"2,5"+(5,5) **\frm{--}  \restore
}
\endxy
\end{tabular},
\end{equation}
where the gauging of the $G_D$ diagonal subgroup of $G'_L\times G'_R$ is also shown.
This gauging explicitly breaks the global $G'_L\times G'_R$ to the local $G_D$ symmetry such that the transformations $\hat{\mathscr{g}}'\in G'_L\times G'_R$ are restricted to the $\hat{\mathscr{g}}_{G_D}(x)\in G_D$.
In the matrix notation, the effect of the gauging can be expressed as
\begin{equation}
 \hat{\mathscr{g}}'=
\begin{pmatrix}
        L' & 0\\
        0 & R'\\
     \end{pmatrix}
\quad
\mathop{\longrightarrow}^{\mbox{\raisebox{1.5ex}{gauging of $G_D$}}}
\quad
\hat{\mathscr{g}}_{G_D}(x)=
\begin{pmatrix}
        \mathscr{g}_{G_D}(x) & 0\\
        0 & \mathscr{g}_{G_D}(x)\\
     \end{pmatrix},
\end{equation}
i.e.\ in the diagonal subgroup $L'=R'$ and the gauging replaces them by the $x$-dependent $\mathscr{g}_{G_D}(x)$.
After gauging $G_D$, the remaining symmetry is $G_L\times G_R\times G_D$.
Under this symmetry, $\omega_1(x)$ and $\omega_2(x)$ transform as
\begin{equation}\label{eq:Vectorres_omega1_omega2_transformation_properties}
\begin{aligned}
 G_L\times G_R\times G_D:\, \omega_1(x) &\to L\, \omega_1(x)\,\mathscr{g}_{G_D}^{-1}(x),
 \\
 G_L\times G_R\times G_D:\, \omega_2(x) &\to \mathscr{g}_{G_D}(x)\, \omega_2(x)\,R^{-1},
\end{aligned}
\end{equation}
which can again be written in terms of a moose diagram:
\begin{equation}\label{eq:Vector_two_omega_combined_moose}
\begin{tabular}{c}
\xy
\xymatrix@R=.4pc@C=1.4pc{
\mathrm{Global:} & G_L && G_D && G_R  \\
& *=<20pt>[o][F]{} \doublerightxyarrow^{\mbox{\raisebox{1.5ex}{$\omega_1$}}} && *=<20pt>[o][F]{}      \doublerightxyarrow^{\mbox{\raisebox{1.5ex}{$\omega_2$}}} && *=<20pt>[o][F]{} \\
\mathrm{Gauged:} &\emptyset && G_D &&  \emptyset
}
\endxy
\end{tabular}.
\end{equation}
The transformation properties of $\omega_1(x)$ and $\omega_2(x)$, eq.~(\ref{eq:Vectorres_omega1_omega2_transformation_properties}), reveal that the product of both only transforms non-trivially under $G_L\times G_R$, i.e.\
\begin{equation}
 G_L\times G_R\times G_D:\,\omega_1(x)\,\omega_2(x)\to L\,\omega_1(x)\,\omega_2(x)\,R^{-1}.
\end{equation}
It is no coincidence that this resembles the transformation properties of the NGB matrix $\Omega(x)$, eq.~(\ref{eq:CCWZ_Omega_transformation}), that may be used as an alternative to $\hat{U}(x)$ for describing the $(G_L\times G_R)/G_D$ non-linear sigma model\footnote{$\Omega(x)$ and $\hat{U}(x)$ are related by $\hat{U}(x)\,\hat{U}(x) = \begin{pmatrix}
        \Omega(x) & 0\\
        0 & \Omega^{-1}(x)\\
     \end{pmatrix}$ (cf.\ eq.~(\ref{eq:CCWZ_symspace_UU_Omega})).}.
Employing the definition of $\omega_1(x)$ and $\omega_2(x)$, eq.~(\ref{eq:Vectorres_omega1_omega2_definition}), one finds
\begin{equation}
 \omega_1(x)\,\omega_2(x)=u(x)\,\xi(x)\,\xi^{-1}(x)\,u(x)=u(x)\,u(x)=\Omega(x),
\end{equation}
i.e.\ the product of $\omega_1(x)$ and $\omega_2(x)$ is actually nothing but $\Omega(x)$.

The description of the $(G_L\times G_R)/G_D$ non-linear sigma model in terms of $\hat{U}$, eq.~(\ref{eq:Vectorres_GL_GR_GD_U_NLSM_moose}), makes it possible to readily extend it to a HLS model parametrized by $\hat{\Omega}=\hat{U}\,\hat{\Xi}$, eq.~(\ref{eq:Vectorres_GL_GR_GD_U_HLS_moose}), by applying the prescription from section \ref{sec:Vectorres_NLSM_HLS}, namely eq.~(\ref{eq:Vectorres_HLS_extension}).
This model then contains heavy spin one resonances in the adjoint representation of $G_D$ in addition to the NGBs in the $(G_L\times G_R)/G_D$ coset.
Employing the matrix notation to separate the direct factors of the direct product groups,
the non-linear sigma model can be expressed in terms of $\Omega(x)$ and the HLS model in terms of $\omega_1(x)$ and $\omega_2(x)$.
With this parametrization one thus finds
\begin{equation}\label{eq:Vectorres_symspace_HLS_extension}
\begin{tabular}{c}
\xy
\xymatrix@R=.4pc@C=1.4pc{
\mathrm{Global:} & G_L && G_R  \\
&*=<20pt>[o][F]{} \doublerightxyarrow^{\mbox{\raisebox{1.5ex}{$\Omega$}}} && *=<20pt>[o][F]{}\\
\mathrm{Gauged:} &\emptyset && \emptyset
}
\endxy
\end{tabular}
\quad
\mathop{\longrightarrow}^{\mbox{\raisebox{1.5ex}{extend to HLS}}}
\quad
\begin{tabular}{c}
\xy
\xymatrix@R=.4pc@C=1.4pc{
G_L && G_D && G_R  \\
 *=<20pt>[o][F]{} \doublerightxyarrow^{\mbox{\raisebox{1.5ex}{$\omega_1$}}} && *=<20pt>[o][F]{}      \doublerightxyarrow^{\mbox{\raisebox{1.5ex}{$\omega_2$}}} && *=<20pt>[o][F]{} \\
\emptyset && G_D &&  \emptyset
}
\endxy
\end{tabular},
\end{equation}
where $\omega_1(x)\,\omega_2(x) = \Omega(x)$.
This can be interpreted in the following way: By introducing the $G_D$ gauge symmetry, the NGB matrix $\Omega(x)$ is split into two NGB matrices $\omega_1(x)$ and $\omega_2(x)$ which are both connected to each other by $G_D$.
The $G_D$ gauge bosons acquire mass via the Higgs mechanism and half of the NGBs in $\omega_1(x)$ and $\omega_2(x)$ are would-be NGBs that can be gauged away.
How the actual NGBs and the would-be NGBs are distributed among $\omega_1(x)$ and $\omega_2(x)$ depends on the gauge.
It is e.g.\ possible to apply a specific gauge transformation $\tilde{\mathscr{g}}_{G_D}(x)=\omega_2^{-1}(x)$ such that
\begin{equation}
\begin{aligned}
 \omega_1(x) &\to \omega_1(x)\,\tilde{\mathscr{g}}_{G_D}^{-1}(x) = \omega_1(x)\,\omega_2(x) = \Omega(x),
 \\
 \omega_2(x) &\to \tilde{\mathscr{g}}_{G_D}(x)\, \omega_2(x) = \omega_2^{-1}(x)\,\omega_2(x) = \mathds{1},
\end{aligned}
\end{equation}
i.e.\ the NGBs in $\omega_1(x)$ all become actual NGBs and $\omega_1(x)$ becomes equal to $\Omega(x)$, whereas the NGBs in $\omega_2(x)$ all become would-be NGBs and are gauged away.
But independent of the chosen gauge, the number of actual NGBs of course always stays the same and they can always be parametrized by the gauge-independent product $\Omega(x) = \omega_1(x)\,\omega_2(x)$.

The two models in eq.~(\ref{eq:Vectorres_symspace_HLS_extension}) describe NGBs in a $(G_L\times G_R)/G_D$ coset.
It is however known from the discussion in section~\ref{sec:CCWZ_HLS}, that by gauging an $H_R$ subgroup of $G_R$, the non-linear sigma model on the left-hand side can be turned into a HLS model that describes NGBs in a $G_L/H_R$ coset as well as heavy spin one resonances in an adjoint representation of $H_R$.
On the right hand-side of eq.~(\ref{eq:Vectorres_symspace_HLS_extension}), the gauging of $H_R$ thus corresponds to a model describing NGBs in a $G_L/H_R$ coset and heavy spin one resonances in adjoint representations of both $G_D$ and $H_R$.
It is therefore a model that contains two levels of resonances where one of them comes in an adjoint representation of the full $G_D\cong G_L$ group.
While $\omega_1(x)$ and $\omega_2(x)$ are related to each other due to the construction stemming from the symmetric space and they share the same decay constant, the model can be generalized as described in the following.

Starting from a $G/H$ non-linear sigma model parametrized by $U(x)$, where for generality a subgroup $E\subset G$ is gauged, one can extend this to a HLS model containing spin one resonances in an adjoint representation of $H$ as described in section~\ref{sec:Vectorres_NLSM_HLS}.
The NGBs in the HLS model are parametrized by $\Omega(x)=U(x)\,\Xi(x)$, where in unitary gauge $\Xi(x)$ can be removed.
This model can then further be extended by the construction discussed above to include heavy spin one resonances in the adjoint representation of $G_1\cong G$.
In terms of moose diagrams, this procedure is described by
\begin{equation}
\begin{tabular}{c}
\xy
\xymatrix@R=.4pc@C=1.4pc{
\mathrm{Global:}\hspace{-15pt}& G  &  \\
& *=<20pt>[o][F]{} \rightxyarrow^{\mbox{\raisebox{1.5ex}{$U$}}} & *=<0pt,20pt>[l][F]{} & *-<0pt,20pt>[l]{H}\\
\mathrm{Gauged:}\hspace{-15pt} & E &
}
\endxy
\end{tabular}
\mathop{\longrightarrow}^{\mbox{\raisebox{1.5ex}{add $H$}}}_{\mbox{\raisebox{-1.5ex}{resonances}}}
\begin{tabular}{c}
\xy
\xymatrix@R=.4pc@C=1.4pc{
 G & G'  \\
*=<20pt>[o][F]{} \rightxyarrow^{\mbox{\raisebox{1.5ex}{$\Omega$}}} & *=<20pt>[o][F]{}\\
E & H
}
\endxy
\end{tabular}
\mathop{\longrightarrow}^{\mbox{\raisebox{1.5ex}{add $G_1$}}}_{\mbox{\raisebox{-1.5ex}{resonances}}}
\begin{tabular}{c}
\xy
\xymatrix@R=.4pc@C=1.4pc{
G & G_1 & G'  \\
 *=<20pt>[o][F]{} \rightxyarrow^{\mbox{\raisebox{1.5ex}{$\Omega_1$}}} & *=<20pt>[o][F]{}      \rightxyarrow^{\mbox{\raisebox{1.5ex}{$\Omega_2$}}} & *=<20pt>[o][F]{} \\
E & G_1 &  H
}
\endxy
\end{tabular}.
\end{equation}
When adding the $G_1$ resonances, $\Omega(x)$ is split into two NGB matrices that for clarity are now called $\Omega_1(x)$ and $\Omega_2(x)$.
In contrast to $\omega_1(x)$ and $\omega_2(x)$ used above, they are not related by a construction stemming from a symmetric space and in general do not share the same decay constant.
But as long as the product of $\Omega_1(x)$ and $\Omega_2(x)$ gives the NGB matrix $\Omega(x)$, the model still contains the NGBs in the $G/H$ coset.
It is possible to add further resonances in the adjoint representation of $G_3\cong G$ by simply splitting $\Omega_1(x)$ or $\Omega_2(x)$ exactly like it was done for $\Omega(x)$.
This can then be repeated successively to add more and more resonances.
Which one of the $\Omega_k(x)$ is split for adding a new level of resonances does actually not matter at all and all different possibilities are equivalent.
$\Omega(x)$ is always given by the product of all the introduced NGB matrices $\Omega_k(x)$.
So one can equivalently also just build a model with $N-1$ levels of resonances in adjoint representations of $G_k\cong G$ by splitting $\Omega(x)$ into a product of $N$ matrices $\Omega_k(x)$ such that
\begin{equation}
 \Omega(x)=\prod_{k=1}^N \Omega_k(x).
\end{equation}
The $\Omega_k(x)$ then transform under the global $G$, the $N-1$ gauge groups $G_k$, and the gauged $H$ subgroup of $G'$ as
\begin{equation}\label{eq:Vectorres_Omegak_gauge_transformations}
\begin{aligned}
 G\times G_1 :\, \Omega_1(x) &\to \mathscr{g}\,\Omega_1(x)\,\mathscr{g}_1^{-1}(x),
 \\
 G_{k-1}\times G_k :\, \Omega_k(x) &\to \mathscr{g}_{k-1}(x)\,\Omega_k(x)\,\mathscr{g}_k^{-1}(x)\quad {\rm for}\ 2 \leq k \leq N-1,
 \\
 G_{N-1}\times H :\, \Omega_N(x) &\to \mathscr{g}_{N-1}(x)\,\Omega_N(x)\,\mathscr{h}^{-1}(x),
\end{aligned}
\end{equation}
while $\Omega(x)$ transforms as ever only non-trivially under $G$ and $H$, i.e.\
\begin{equation}
 G\times H :\, \Omega(x) \to \mathscr{g}\,\Omega(x)\,\mathscr{h}^{-1}(x).
\end{equation}
This model with NGBs in a $G/H$ coset, one level of resonances in the adjoint representation of $H$ and $N-1$ levels of resonances in the adjoint representations of $G_k\cong G$ corresponds to the moose diagram
\begin{equation}\label{eq:Vectorres_general_moose}
\begin{tabular}{c}
\xy
\xymatrix@R=.4pc@C=1.4pc{
\mathrm{Global:}
& G
&& G_1
&& G_2
&
& G_{N-1}
&& G'
\\
& *=<20pt>[o][F]{} \doublerightxyarrow^{\mbox{\raisebox{1.5ex}{$\Omega_1$}}}
&& *=<20pt>[o][F]{} \doublerightxyarrow^{\mbox{\raisebox{1.5ex}{$\Omega_2$}}}
&& *=<20pt>[o][F]{} \rightxyarrow
& *=<20pt>[o]{\cdots} \rightxyarrow
& *=<20pt>[o][F]{} \doublerightxyarrow^{\mbox{\raisebox{1.5ex}{$\Omega_N$}}}
&& *=<20pt>[o][F]{}
\\
\mathrm{Gauged:}
&E
&& G_1
&& G_2
&
& G_{N-1}
&&  H
}
\endxy
\end{tabular}.
\end{equation}
In the following, it is assumed that in the decomposition $\Omega(x)=U(x)\,\Xi(x)$ both $U(x)$ and $\Xi(x)$ depend on the same decay constant, i.e.\ $f=f_\Xi$.
The same is also assumed for $U_k(x)$ and $\Xi_k(x)$ in the decompositions%
\footnote{As before, $U(x)$ and $U_k(x)$ are elements of the $G/H$ coset, while $\Xi(x)$ and $\Xi_k(x)$ are elements of $H$.}
$\Omega_k(x)=U_k(x)\,\Xi_k(x)$, i.e $f_k={f_\Xi}_k$.
The decay constants $f_k$ for different indices $k$ are however in general independent of each other.
This has the effect that the couplings and mass terms of the heavy gauge bosons are allowed to be different at each site.
The relation $f_k={f_\Xi}_k$ on the other hand implies that among the resonances of a given site, those transforming in the adjoint representation of $H$ do not have different mass terms and couplings to those corresponding to the coset $G/H$.
Since the actual mass eigenstates in general do not correspond to the gauge bosons at separate sites but are a mixture of them, especially the gauging of $H$ can of course lead to different masses of spin one resonances in adjoint representations of $H$ and those corresponding to the $G/H$ coset.
Given the above assumptions, $\Omega(x)$ can be expressed in the canonical form
\begin{equation}\label{eq:Vectorres_Omega_canonical_form}
 \Omega(x) = e^{i\frac{\sqrt{2}}{f}\,\pi_{\Omega}^a(x)\,S^a}
\end{equation}
and the Lagrangian reads%
\footnote{The assumptions above allow to interpret $\Omega(x)$ and the $\Omega_k(x)$ as describing NGBs of global $G\times G'$ and $G_{k-1}\times G_k$ symmetries (where $G_0=G$ and $G_N=G'$) that are each spontaneously broken to their diagonal subgroups.
The NGB Lagrangian can thus be written in the simplified form discussed in section~\ref{sec:symmetric_spaces}.
This is the construction employed e.g.\ in \cite{Panico:2011pw,DeCurtis:2011yx,Son:2003et}.
A more general construction can be found in \cite{DeCurtis:2014oza}.}%
\begin{equation}\label{eq:Vectorres_N_moose_lagrangian}
 \mathcal{L}_2 =
 \sum_{k=1}^N
 \frac{f_k^2}{4}\,\tr\left[D_\mu \Omega_k^{-1}(x)\,D^\mu \Omega_k(x)\right]
 - \sum_{k=0}^N\frac{1}{4}\,\tr\left[{F_k}_{\mu\nu}(x)\,F_k^{\mu\nu}(x)\right]
 ,
\end{equation}
where the gauge covariant derivatives are defined as
\begin{equation}
 i\,D_\mu\,\Omega_k(x) = i\,\partial_\mu\,\Omega_k(x)
 +g_{k-1}\,A^{k-1}_\mu(x)\,\Omega_k(x)
 -g_{k}\,\Omega_k(x)\,A^k_\mu(x).
\end{equation}
In the above expression, the short-hand notation
\begin{equation}\label{eq:Vectorres_shorthand_notation}
\begin{aligned}
 g_0&=g_E,\quad &A^0_\mu(x)&=(A_E)_\mu^a(x)\,P^a,
 \\
 g_k&=g_{G_k},\quad &A^k_\mu(x)&=(A_k)_\mu^a(x)\,S^a,
 \quad {\rm for}\
 k\in[1,N-1]
 ,\\
 g_N&=g_H,\quad &A^N_\mu(x)&=(A_H)_\mu^a(x)\,T^a
\end{aligned}
\end{equation}
is used.
As before, $S^a$ denotes the generators of $G\cong G'\cong G_k$, $P^a$ the generators of $E$ and $T^a$ the generators of $H$.
Employing the short-hand notation, the field strength tensors are given by
\begin{equation}
 {F_k}_{\mu\nu}(x)=\partial_\mu\,A^k_\nu(x)-\partial_\nu\,A^k_\mu(x)-i\,g_k\,[A^k_\mu(x),A^k_\nu(x)].
\end{equation}
While the expression for $\Omega(x)$ can in general be given by eq.~(\ref{eq:Vectorres_Omega_canonical_form}), the explicit form of the $\Omega_k(x)$ depends on the gauge.
One possibility is a gauge where the $G/H$ NGBs in $\Omega(x)$ do not mix with the gauge fields
(see e.g.~\cite{Son:2003et}):
\begin{equation}\label{eq:Vectorres_gauge_no_mixing}
 \Omega_k(x)= e^{i\,\sqrt{2}\,\frac{f}{f_k^2}\,\pi_\Omega^a(x)\,S^a}.
\end{equation}
This gauge will be denoted in the following as the {\it non-mixing gauge}%
\footnote{It is actually possible to reach the non-mixing gauge from one where $\Omega(x)=\prod_{k=1}^N \widetilde{\Omega}_k(x)$ and the $\widetilde{\Omega}_k(x)$ are in the canonical form also used for $\Omega(x)$.
In this case, each $\widetilde{\Omega}_k(x)$ depends on a separate ${\pi_\Omega}_k^a(x)$ and is given by
$\widetilde{\Omega}_k(x)=e^{i\frac{\sqrt{2}}{f_k}\,{\pi_\Omega}_k^a(x)\,S^a}$.
Using the gauge transformations
\begin{equation}
 g_k(x)=\left(\prod_{i=1}^k\Omega_i(x)\right)^{-1}\prod_{i=1}^k\widetilde{\Omega}_i(x),
 \quad\quad
 \mathscr{h}(x)=\mathds{1},
 \quad\quad
 {\rm with}\ k\in[1,N-1],
\end{equation}
and employing eq.~(\ref{eq:Vectorres_f_fk_relation}), the non-mixing gauge is reached from the canonical form via eq.~(\ref{eq:Vectorres_Omegak_gauge_transformations}).
}.
The requirement of a canonically normalized kinetic term for the $\pi^a(x)$ fields then yields the relation
\begin{equation}\label{eq:Vectorres_f_fk_relation}
 \frac{1}{f^2}=\sum_{k=1}^N\frac{1}{f_k^2}.
\end{equation}
In the non-mixing gauge, $\Omega(x)$ as well as the $\Omega_k(x)$ transform linearly under global $G$ transformations.
In the case of a non-trivial gauge group $H$, one can however go to a kind of unitary gauge where each $\Omega_k(x)$ only contains NGBs in the $G/H$ coset (see e.g.~\cite{DeCurtis:2011yx}), i.e.\
\begin{equation}
 \Omega_k(x)=U_k(x)=e^{i\,\sqrt{2}\,\frac{f}{f_k^2}\,\pi^a(x)\,X^a}.
\end{equation}
For distinguishing this gauge from the previous one, it will be denoted as the {\it unitary non-mixing gauge} in the following.
To reach it from the non-mixing gauge, one has to apply the gauge transformations
\begin{equation}
 g_k(x)=\left(\prod_{i=1}^k U_i(x)\right)^{-1}\prod_{i=1}^{k}\Omega_i(x),
 \quad\quad
 \mathscr{h}(x)=\Xi(x),
 \quad\quad
 {\rm with}\ k\in[1,N-1].
\end{equation}
In the unitary non-mixing gauge, one automatically gets $\Omega(x)=U(x)$.
Since this gauge is not respected by the global $G$ transformations, $U(x)$ as well as all the $U_k(x)$ transform non-linearly under a $G$ transformation $\mathscr{g}$, i.e.\ $\mathscr{g}$-dependent gauge transformations have to be applied to stay in the unitary non-mixing gauge.
Another gauge frequently employed in the literature is the so-called {\it holographic gauge} (see e.g~\cite{Panico:2007qd,Panico:2011pw,Marzocca:2012zn}) where only $\Omega_1(x)$ contains the actual NGBs and all would-be NGBs are gauged away, i.e.\
\begin{equation}\label{eq:Vectorres_gauge_holographic_standard}
\begin{aligned}
 \Omega_1(x)&=U(x)=e^{i\frac{\sqrt{2}}{f}\,\pi^a(x)\,X^a},
 \\
 \Omega_k(x)&=\mathds{1}\quad {\rm for}\ 2 \leq k \leq N.
\end{aligned}
\end{equation}
The holographic gauge is reached from the unitary non-mixing gauge by employing the gauge transformations
\begin{equation}
 g_k(x)=U^{-1}(x)\prod_{i=1}^{k}U_i(x),
 \quad\quad
 \mathscr{h}(x)=\mathds{1},
 \quad\quad
 {\rm with}\ k\in[1,N-1].
\end{equation}
An equivalent prescription for defining the holographic gauge is to employ
\begin{equation}\label{eq:Vectorres_gauge_holographic_f1}
\begin{aligned}
 \Omega_1(x)&=U_1(x)=e^{i\frac{\sqrt{2}}{f_1}\,\pi_1^a(x)\,X^a},
 \\
 \Omega_k(x)&=\mathds{1}\quad {\rm for}\ 2 \leq k \leq N,
\end{aligned}
\end{equation}
i.e.\ $\Omega_1(x)$ is simply set to its unitary gauge canonical form and all the remaining $\Omega_k(x)$ are set to the identity matrix.
To coincide with eq.~\eqref{eq:Vectorres_gauge_holographic_standard} and to get a canonically normalized kinetic term, a field redefinition $\pi_1^a(x)=\frac{f_1}{f}\pi^a(x)$ has to be performed in this case (cf.~\cite{Marzocca:2012zn,Niehoff:2015iaa,Niehoff:2016zso}).
Independent of how the holographic gauge is defined (i.e.\ either by eq~(\ref{eq:Vectorres_gauge_holographic_standard} or~(\ref{eq:Vectorres_gauge_holographic_f1})), this gauge leads to a mixing between the NGBs and the heavy gauge bosons corresponding to the $G/H$ coset.
This mixing can however be removed by field shifts of the $G/H$ gauge bosons%
\footnote{An example of such a field shift can be found in eq.~(\ref{eq:decays:global_analyses:MCHM_field_shift}).
}.

The model from eq.~(\ref{eq:Vectorres_general_moose}) contains in addition to spin one resonances in adjoint representations of $G_k\cong G$ also resonances in the adjoint representation of $H$.
For constructing a model that only contains resonances in adjoint representations of the full $G$ group, one can take the limit $g_H\to\infty$ and decouple the heavy $H$ gauge bosons.
The corresponding moose diagram then reads
\begin{equation}\label{eq:Vectorres_general_moose_noH}
\begin{tabular}{c}
\xy
\xymatrix@R=.4pc@C=1.4pc{
\mathrm{Global:}
& G
&& G_1
&& G_2
&
& G_{N-1}
&&
\\
& *=<20pt>[o][F]{} \doublerightxyarrow^{\mbox{\raisebox{1.5ex}{$\Omega_1$}}}
&& *=<20pt>[o][F]{} \doublerightxyarrow^{\mbox{\raisebox{1.5ex}{$\Omega_2$}}}
&& *=<20pt>[o][F]{} \rightxyarrow
& *=<20pt>[o]{\cdots} \rightxyarrow
& *=<20pt>[o][F]{} \doublerightxyarrow^{\mbox{\raisebox{1.5ex}{$\Omega_N$}}}
&& *=<0pt,20pt>[l][F]{} & *-<0pt,20pt>[l]{H}
\\
\mathrm{Gauged:}
&E
&& G_1
&& G_2
&
& G_{N-1}
&&
}
\endxy
\end{tabular}.
\end{equation}
While the $H$ gauge symmetry is hidden in this case, it still is used in the non-linear transformations of the $U_k(x)$ and of $U(x)$ in the unitary non-mixing gauge or the holographic gauge.
So all the results given above for the model in eq.~(\ref{eq:Vectorres_general_moose}) still apply.

In the context of CHMs, the construction in eq.~(\ref{eq:Vectorres_general_moose}) is employed in the DCHM~\cite{Panico:2011pw} where $G\cong G'\cong G_k\cong {\rm SO}(5)$, and $H\cong {\rm SO}(4)$.
While the moose diagram of the two-site DCHM is already shown in eq.~(\ref{eq:Vectorres_two_site_DCHM_moose}), the three-site DCHM is given by
\begin{equation}\label{eq:Vectorres_three_site_DCHM_moose}
\begin{tabular}{c}
\xy
\xymatrix@R=.4pc@C=1.4pc{
\mathrm{Global:}
& {\rm SO}(5)_{\rm L}^1
&& {\rm SO}(5)_{\rm D}
&& {\rm SO}(5)_{\rm R}^2
\\
& *=<20pt>[o][F]{} \doublerightxyarrow^{\mbox{\raisebox{1.5ex}{$\Omega_1$}}}
&& *=<20pt>[o][F]{} \doublerightxyarrow^{\mbox{\raisebox{1.5ex}{$\Omega_2$}}}
&& *=<20pt>[o][F]{}
\\
\mathrm{Gauged:}
& {\rm SU}(2)_{\rm L}^0\times {\rm U}(1)_{\rm Y}^0
&& {\rm SO}(5)_{\rm D}
&& {\rm SO}(4)
}
\endxy
\end{tabular}.
\end{equation}
A similar construction but based on the model in eq.~(\ref{eq:Vectorres_general_moose_noH}) is the so called 4D composite Higgs model (4DCHM)~\cite{DeCurtis:2011yx}, where the minimal two-site 4DCHM is described by
\begin{equation}\label{eq:Vectorres_two_site_4DCHM_moose}
\begin{tabular}{c}
\xy
\xymatrix@R=.4pc@C=1.4pc{
\mathrm{Global:}
& {\rm SO}(5)
&& {\rm SO}(5)_1
&&
\\
& *=<20pt>[o][F]{} \doublerightxyarrow^{\mbox{\raisebox{1.5ex}{$\Omega_1$}}}
&& *=<20pt>[o][F]{} \doublerightxyarrow^{\mbox{\raisebox{1.5ex}{$\Omega_2$}}}
&& *=<0pt,20pt>[l][F]{} & *-<0pt,5pt>[l]{{\rm SO}(4)}
\\
\mathrm{Gauged:}
& {\rm SU}(2)_{\rm L}\times {\rm U}(1)_{\rm Y}
&& {\rm SO}(5)_1
&&
}
\endxy
\end{tabular}.
\end{equation}
While both the three-site DCHM as well as the two-site 4DCHM contain NGBs in an ${\rm SO}(5)/{\rm SO}(4)$ coset and heavy spin one resonances in an adjoint representation of ${\rm SO}(5)$, the three-site DCHM in addition contains also spin one resonances in an adjoint representation of the unbroken ${\rm SO}(4)$ and therefore has a slightly larger particle content while being otherwise similar.

\subsection{The continuum limit and the fifth dimension}\label{sec:vectorres_contunuum_limit}
The model described in the last section that contains N levels of heavy resonances connected by sigma model fields intriguingly resembles a model of a discretized extra dimension.
That an extra dimension actually arises from a moose diagram similar to the one in eq.~(\ref{eq:Vectorres_general_moose}) when taking the continuum limit $N\to\infty$ was shown
in~\cite{ArkaniHamed:2001ca}.
A latticized extra dimension inspired by HLS and its continuum limit was also discussed independently and
contemporaneously
in~\cite{Hill:2000mu}.
Similar constructions have subsequently been presented in e.g.~\cite{Sfetsos:2001qb,Abe:2002rj,Falkowski:2002cm,Randall:2002qr,Son:2003et,Becciolini:2009fu}.

To see how an extra dimension can be constructed from the model described by the moose diagram in eq.~(\ref{eq:Vectorres_general_moose}), the continuum limit $N\to\infty$ is performed in detail in the following.
To this end, the discrete indices $k$ have to be replaced by a continuous variable that will be called $u$ in the following.
The indices $k=0$ and $k=N$ then can be chosen to correspond to the coordinates $u=u_0$ and $u=u_N$, respectively.
The distance between the coordinates $u_0$ and $u_N$ may then be defined as $L=u_N-u_0$ and each index $k$ can be associated with a value $u=u_k$.
To summarize, the continuous variable $u$ is related to the discrete indices $k$ by
\begin{equation}
L = u_N - u_0
\quad\quad
\Delta u=\frac{L}{N},
\quad\quad
u_k=u_0+\Delta u\,k.
\end{equation}
Keeping the distance $L$ fixed when performing the limit $N\to\infty$, the distance $\Delta u$ between adjacent coordinates goes to zero, i.e.\
\begin{equation}
 N\to\infty
 \
 \Leftrightarrow
 \
 \Delta u \to 0.
\end{equation}
Before performing the continuum limit, it is convenient to redefine the gauge boson fields as
\begin{equation}
 g_k\,(A_k)_\mu\to (A_k)_\mu
\end{equation}
such that the action for the Lagrangian from eq.~(\ref{eq:Vectorres_N_moose_lagrangian}) reads
\begin{equation}
 S_2 = \int dx^4\,
 \left\{
 \sum_{k=1}^N
 \frac{f_k^2}{4}\,\tr\left[D_\mu \Omega_k^{-1}(x)\,D^\mu \Omega_k(x)\right]
 -\sum_{k=0}^N
 \frac{1}{4\,g_k^2}\,\tr\left[{F_k}_{\mu\nu}(x)\,F_k^{\mu\nu}(x)\right]
 \right\},
\end{equation}
where the gauge covariant derivatives are now given by
\begin{equation}\label{eq:Vectorres_Omega_cov_deriv_redefined_A}
 i\,D_\mu\,\Omega_k(x) = i\,\partial_\mu\,\Omega_k(x)
 +A^{k-1}_\mu(x)\,\Omega_k(x)
 -\Omega_k(x)\,A^k_\mu(x).
\end{equation}
In addition, it is useful to define the fields $A_5^k(x)$ to parametrize the NGBs inside the $\Omega_k(x)$ (in the non-mixing gauge) as
\begin{equation}\label{eq:Vectorres_A5_pi_definition}
 A^k_5(x) = -\frac{\sqrt{2}}{\Delta u}\,\frac{f}{f_k^2}\,\pi_\Omega^a(x)\,S^a.
\end{equation}
In terms of the $A_5^k(x)$, the NGB matrices $\Omega_k(x)$ are then given by
\begin{equation}\label{eq:Vectorres_Omega_k_A5_k}
 \Omega_k(x) = e^{-i\,\Delta u\,A^k_5(x)}
 =\mathds{1}-i\,\Delta u\,A^k_5(x)+\mathcal{O}(\Delta u^2).
\end{equation}
Plugging this new parametrization of the $\Omega_k(x)$ into the definition of their covariant derivatives, eq.~(\ref{eq:Vectorres_Omega_cov_deriv_redefined_A}), one finds
\begin{equation}
 i\,D_\mu\,\Omega_k(x) =
 \Delta u\,{F_k}_{\mu 5}(x)
 +\mathcal{O}(\Delta u^2),
\end{equation}
where ${F_k}_{\mu 5}(x)$ is defined as%
\footnote{Note that $A_\mu^{k-1}(x)=A_\mu^{k}(x)-\Delta u\frac{A_\mu^{k}(x)-A_\mu^{k-1}(x)}{\Delta u}=A_\mu^{k}(x)+\mathcal{O}(\Delta u)$,
where the difference quotient is treated as $\mathcal{O}(1)$ since $\lim_{\Delta u\to 0}\frac{A_\mu^{k}(x)-A_\mu^{k-1}(x)}{\Delta u} = \lim_{\Delta u\to 0}\frac{A_\mu(x,u_k)-A_\mu(x,u_k-\Delta u)}{\Delta u} = \partial_5 A_\mu(x,u)$.}
\begin{equation}
 {F_k}_{\mu 5}(x)=
 \partial_\mu A_5^k(x)
 -\frac{A_\mu^k(x)-A_\mu^{k-1}(x)}{\Delta u}
 -i[A_\mu^k(x),A_5^k(x)].
\end{equation}
Expressing the covariant derivatives of the $\Omega_k(x)$ in terms of ${F_k}_{\mu 5}(x)$ and factoring out $\Delta u$, the action reads
\begin{equation}
\begin{aligned}
 S_2 &= \int dx^4
 \left\{
 \sum_{k=1}^N \Delta u
 \frac{\Delta u\,f_k^2}{4}\,\tr\left[{F_k}_{\mu5}(x)\,F_k^{\mu5}(x)\right]
 - \sum_{k=0}^N \Delta u
 \frac{1}{4\,\Delta u\,g_k^2}\,\tr\left[{F_k}_{\mu\nu}(x)\,F_k^{\mu\nu}(x)\right]
 \right\}
 \\&+\mathcal{O}(\Delta u^2).
\end{aligned}
\end{equation}
To finally perform the continuum limit, all objects that depend on an index $k$ have to be replaced by objects that depend on the continuous variable $u$ instead.
For the 4D fields $A^k_5(x)$ and $A^k_\mu(x)$, it is straightforward to define the 5D fields $A_5(x,u)$ and $A_\mu(x,u)$ by
\begin{equation}\label{eq:Vectorres_5D_4D_field_dict}
 A_5(x,u_k) = A^k_5(x),
 \quad\quad
 A_\mu(x,u_k) = A^k_\mu(x),
\end{equation}
such that one finds $F_{\mu\nu}(x,u_k)={F_k}_{\mu \nu}(x)$ and
\begin{equation}
F_{\mu 5}(x,u)
  =
 \lim_{\Delta u\to 0}{F_k}_{\mu 5}(x)
  =
 \partial_\mu A_5(x,u)
 -\partial_5 A_\mu(x,u)
 -i[A_\mu(x,u),A_5(x,u)].
\end{equation}
In addition, continuous versions of the decay constants $f_k$ and the gauge couplings $g_k$ are conveniently defined as
\begin{equation}
 f(u_k) = \sqrt{\frac{\Delta u}{2}}\,f_k,
 \quad\quad
 g(u_k) = \sqrt{\Delta u}\,g_k.
\end{equation}
Plugging all these definitions into the action and taking the limit $\Delta u\to 0$, one arrives at
\begin{equation}\label{eq:Vectorres_action_continuum_limit}
 \lim_{\Delta u\to 0} S_2 = \int dx^4
 \int_{u_0}^{u_N}\hspace{-.5em} du\,
 \left\{
 \frac{f^2(u)}{2}\,\tr\left[F_{\mu5}(x,u)\,F^{\mu5}(x,u)\right]
 - \frac{1}{4\,g^2(u)}\,\tr\left[F_{\mu\nu}(x,u)\,F^{\mu\nu}(x,u)\right]
 \right\},
\end{equation}
This action obviously describes the 5D fields $A_5(x,u)$ and $A_\mu(x,u)$ in a space-time that is 4D Minkowski space times a fifth dimension interval of length $L$ with boundaries at $u_0$ and $u_N$.
It is instructive to compare this action to one of a 5D gauge field in a generic space-time with the fifth dimension being an interval of length $L$.
This actions reads
\begin{equation}
 S_{5D}
 =
 -\frac{1}{4\,g_5^2}\int dx^4\,\int_{u_0}^{u_N}\hspace{-.3em} du\,
 \sqrt{|G|}\,G^{MR}\,G^{NP}\,\tr\left[F_{MN}(x,u)\,F_{RP}(x,u)\right],
\end{equation}
where $M,N,R,P\in\{0,1,2,3,5\}$ are the 5D space-time indices, $G^{MN}$ is the inverse metric tensor, $\sqrt{|G|}$ is the square root of the determinant of the metric, $g_5$ is a 5D gauge coupling of mass dimension $-\frac{1}{2}$ and the 5D field strength tensor is given by
\begin{equation}
 F_{MN}(x,u) = \partial_M A_N(x,u)
 -\partial_N A_M(x,u)
 -i[A_M(x,u),A_N(x,u)].
\end{equation}
The line element of a 5D space-time with one time and four space dimensions that has 4D Poincaré invariance can be expressed as
\begin{equation}\label{eq:Vectorres_metric_u}
 ds^2=a^2(u)\,\eta_{\mu\nu}\,dx^\mu dx^\nu-b^2(u)\,du^2,
\end{equation}
where the infinitesimal displacements in 4D Minkowski space are denoted by $dx^\mu$ and the one in the fifth dimension by $du$. The functions $a(u)$ and $b(u)$ determine how distances in 4D Minkowski space and the extra dimension, respectively, change with the position in the extra dimension.
While $b(u)$ can in principle be absorbed into the extra dimensional coordinate by a coordinate transformation, the $u$-dependence of $a(u)$ implies a warped extra dimension. $a(u)$ is thus also called the warp factor.
The metric in eq.~(\ref{eq:Vectorres_metric_u}) with a constant $a(u)$ on the other hand describes a flat extra dimension.
With the above definition of $ds^2$, the inverse metric $G^{MN}$ and $\sqrt{|G|}$ read
\begin{equation}
 G^{MN}=\frac{1}{a^2(u)}\eta^{\mu\nu}\delta_\mu^M\delta_\nu^N-\frac{1}{b^2(u)}\delta_5^M\delta_5^N,
 \quad\quad
 \sqrt{|G|}=a^4(u)\,b(u)
\end{equation}
and thus the action of the 5D gauge field can be expressed as
\begin{equation}\label{eq:Vectorres_action_5D_u_coords}
 S_{5D}
 =
 \int dx^4\,\int_{u_0}^{u_N}\hspace{-.3em} du
 \left\{
 \frac{a^2(u)}{2\,g_5^2\,b(u)}\,\tr\left[F_{\mu5}(x,u)\,F^{\mu5}(x,u)\right]
 - \frac{b(u)}{4\,g_5^2}\,\tr\left[F_{\mu\nu}(x,u)\,F^{\mu\nu}(x,u)\right]
 \right\}.
\end{equation}
Comparing this 5D gauge field action to the continuum limit action in eq.~(\ref{eq:Vectorres_action_continuum_limit}), one finds that both are actually identical if one identifies
\begin{equation}
 f^2(u)=\frac{a^2(u)}{g_5^2\,b(u)},
 \quad\quad
 g^2(u)=\frac{g_5^2}{b(u)}.
\end{equation}
The metric in terms of the $u$-coordinate is thus fixed by
\begin{equation}
 a(u)=g_5^2\,\frac{f(u)}{g(u)},
 \quad\quad
 b(u)=\frac{g_5^2}{g^2(u)}.
\end{equation}
So by taking the continuum limit $N\to \infty$ for the model described by the moose diagram in eq.~(\ref{eq:Vectorres_N_moose_lagrangian}), one actually constructs an extra dimension and arrives at a model of a 5D gauge field in a warped background.
The dependence on the index $k$ of the decay constants $f_k$ and the gauge couplings $g_k$ in the 4D model is then reflected by the warp factor of the extra dimension%
\footnote{Actually only the ratio $\frac{f_k}{g_k}$ determines the warp factor.
Two models based on the moose diagram from eq.~(\ref{eq:Vectorres_N_moose_lagrangian}) that have different dependences of $f_k$ and $g_k$ on $k$ thus lead to the same 5D theory if the ratio $\frac{f_k}{g_k}$ has the same dependence on $k$ in both models.
}.
The Lagrangian and the metric however do not determine the 5D model completely.
Because it is defined on an interval with boundaries, it is also necessary to specify boundary conditions (BCs) for the 5D fields.
For the 4D components of the 5D gauge fields $A_\mu(x,u)$, the values at the boundaries are given by $A_\mu(x,u_0)=A^0_\mu(x)=(A_E)_\mu^a(x)\,P^a$ and $A_\mu(x,u_N)=A^N_\mu(x)=(A_H)_\mu^a(x)\,T^a$ (cf.\ eqs.~(\ref{eq:Vectorres_5D_4D_field_dict}),(\ref{eq:Vectorres_shorthand_notation})), i.e.\ they are determined by the gauge fields of the 4D moose model on the $k=0$ and $k=N$ sites.
For $k=0$, there are only gauge fields in the adjoint representation of $E\subseteq G$ and those in the $G/E$ coset are absent.
For $k=N$ only gauge fields in the adjoint representation of $H\subseteq G'$ are present.
From the 5D perspective, the 4D components of the 5D gauge fields in the $G/E$ coset vanish on the boundary at $u=u_0$ and those associated to the broken generators $X^a$ are set to zero on the boundary at $u=u_N$.
The bulk gauge symmetry $G$ is thus reduced to $E$ and $H$ at the $u=u_0$ and $u=u_N$ boundaries, respectively.
Considering for simplicity the case with no elementary gauge fields, i.e.\ $E=\emptyset$, the BCs for the 4D components of the 5D gauge fields thus read%
\footnote{
In a theory on an interval with coordinate $u$ that describes a massless field $\phi$ and contains no explicit boundary terms, for the variation of the action to vanish on the boundary, possible BCs for $\phi$ are (see e.g.~\cite{Csaki:2003dt,Csaki:2005vy}
):
\begin{itemize}
 \item Neumann BC $\partial_u \phi|_{\rm boundary}=0$, denoted by $(+)$.
 \item Dirichlet BC $\phi|_{\rm boundary}=0$, denoted by $(-)$.
\end{itemize}
}
\begin{equation}\label{eq:vectorres_5D_gauge_4D_BC}
\begin{aligned}
 &A_\mu^a(x,u_0)\,T^a = 0,
 \quad&
 \partial_5 &A_\mu^a(x,u)\,T^a\big|_{u=u_N} = 0
 \quad&
 &\Rightarrow\quad A_\mu^a(x,u)\,T^a\ (-+),
 \\
 &A_\mu^a(x,u_0)\,X^a = 0,
 \quad&
 &A_\mu^a(x,u_N)\,X^a = 0
 \quad&
 &\Rightarrow\quad A_\mu^a(x,u)\,X^a\ (--).
\end{aligned}
\end{equation}
The fifth components of the 5D gauge fields must have opposite boundary conditions compared to the 4D components (see e.g.~\cite{Csaki:2003dt,Contino:2010rs}).
This then yields
\begin{equation}
\begin{aligned}
 \partial_5 &A_5^a(x,u)\,T^a\big|_{u=u_0} = 0,
 \quad&
 &A_5^a(x,u_N)\,T^a = 0
 \quad&
 &\Rightarrow\quad &A_5^a(x,u)\,T^a\ (+-),
 \\
 \partial_5 &A_5^a(x,u)\,X^a\big|_{u=u_0} = 0,
 \quad&
 \partial_5 &A_5^a(x,u)\,X^a\big|_{u=u_N} = 0
 \quad&
 &\Rightarrow\quad &A_5^a(x,u)\,X^a\ (++).
\end{aligned}
\end{equation}
When performing a Kaluza-Klein (KK) decomposition of the 5D fields, one finds that only the fields having $(++)$ boundary conditions contain massless zero modes.
As expected from the discussion in section~\ref{sec:Vectorres_NLSM_HLS}, in the case $E=\emptyset$ there are thus no massless zero modes for the 4D components of the 5D gauge bosons.
The fifth components associated with the broken generators on the other hand contain massless zero modes which can be identified with the $G/H$ NGBs.
In this respect, it is worth noting how the NGB matrix $\Omega(x)$ arises in the 5D theory.
In the 4D theory, $\Omega(x)$ is given by the product of all $N$ NGB matrices $\Omega_k(x)$.
Using eq.~(\ref{eq:Vectorres_Omega_k_A5_k}) to express the $\Omega_k(x)$ by $A_5^k(x)$ and taking the continuum limit, $\Omega(x)$ can be written in terms of the fifth component of the 5D gauge field $A_5(x,u)$:
\begin{equation}
 \lim_{N\to\infty}\Omega(x)=\lim_{N\to\infty} \prod_{k=1}^N e^{-i\,A_5(x,u_k)\,\Delta u}
 =\mathcal{P} \exp\left(-i\,\int_{u_0}^{u_N}\hspace{-.3em} du\,A_5(x,u)\right),
\end{equation}
i.e.\ it can be identified with the Wilson line between the two boundaries of the extra dimension.
The path ordered exponential in the expression for the Wilson line can be evaluated by e.g.\ decomposing $A_5(x,u)$ into KK modes or by using an explicit parametrization in a specific gauge.
Doing the latter in the non-mixing gauge where $A_5(x,u)$ is given by (cf.\ eq.~(\ref{eq:Vectorres_A5_pi_definition}))
\begin{equation}
 A_5(x,u) = -\frac{f}{\sqrt{2}\,f^2(u)}\,\pi_\Omega^a(x)\,S^a,
\end{equation}
the path ordering is trivial and one finds%
\footnote{
In the last step of this derivation, the continuum version of eq.~(\ref{eq:Vectorres_f_fk_relation}), namely
\begin{equation}
  \frac{2}{f^2} = \int_{u_0}^{u_N}\hspace{-.3em} du\,\frac{1}{f^2(u)}
\end{equation}
is used.
The factor of $2$ in this relation is due to the normalization of $f(u)$ that was chosen to simplify expressions containing the warp factor.}
\begin{equation}
 \mathcal{P} \exp\left(-i\,\int_{u_0}^{u_N}\hspace{-.3em} du\,A_5(x,u)\right)
 =
 \exp\left(i\,\frac{f}{\sqrt{2}}\,\pi_\Omega^a(x)\,S^a\,\int_{u_0}^{u_N}\hspace{-.3em} du\,\frac{1}{f^2(u)}\right)
 =
 \e^{i\,\frac{\sqrt{2}}{f}\,\pi_\Omega^a(x)\,S^a},
\end{equation}
which is of course again nothing but $\Omega(x)$ as defined in eq.~(\ref{eq:Vectorres_Omega_canonical_form}).

To summarize this section, one observes that a non-linear sigma model describing NGBs in a $G/H$ coset that is supplemented by an infinite tower of heavy gauge bosons in the adjoint representation of $G$ is actually equivalent to a 5D gauge field in the adjoint representation of $G$ with boundary conditions that encode the $G\to H$ spontaneous symmetry breaking.
From the 5D point of view, both the NGBs as well as the heavy spin one resonances are part of a single 5D gauge field.
This is the notion of gauge-Higgs unification mentioned in the beginning of chapter~\ref{chap:CHMs}.

The most important phenomenological effects of theories with an extra dimension or an infinite tower of resonances are at low energies already captured by the lightest resonances.
In an HLS inspired extra-dimensional model for QCD, it was actually shown that the dependence on the specific geometry of the extra dimension only plays a minor role and that using a 4D model with only $\mathcal{O}(3)$ levels of resonances leads to an equivalently good fit to low-energy QCD data as a full 5D model~\cite{Becciolini:2009fu}.
For a generic effective description of a strongly coupled confining theory, it is not even guaranteed that the naive 5D model discussed above is a good approximation.
Like the non-linear sigma model, also the 5D theories are not renormalizable and only valid below some cutoff.
If the spectrum of heavy resonances only includes a few levels below the cutoff, taking the limit $N\to\infty$ might not be reasonable (cf.\ related discussions in~\cite{Sfetsos:2001qb,Falkowski:2002cm,Becciolini:2009fu,DeCurtis:2011yx}).
It is thus well motivated for a phenomenological analysis of the effects at low energies to consider a 4D moose-like model with only the lightest levels of resonances included.
Nevertheless, especially in the context of CHMs, the possibility to relate the 4D and the 5D models has proved to be a fruitful tool for model building (cf.\ e.g.~\cite{Contino:2003ve,Panico:2011pw,DeCurtis:2011yx}).

\section{Fermions}\label{sec:Fermions}
%
%
%
%
Including massless gauge bosons as well as heavy spin one resonances into a non-linear sigma model describing NGBs is in a way straightforward.
To add the former, one just has to gauge a subgroup of the global $G$ symmetry and the latter are introduced via the HLS prescription.
Apart from the number of resonance levels and the coupling and decays constants, there is not much model dependence in the HLS construction.
How well the NGBs and the vector resonances fit together in this framework is seen from their unification into a single 5D gauge field in the continuum limit discussed in the previous section.

The story for fermions is quite different.
It is of course possible to employ the CCWZ formalism to include fermions into a model with a global $G$ symmetry spontaneously broken to an $H$ subgroup.
There is however a peculiarity in CHMs where the coupling of SM fermions to the composite sector is responsible for generating their masses as well as their interaction with the physical Higgs boson.
The original mechanism for fermion mass generation from a composite sector was described in the context of (extended) technicolor%
\footnote{There are various names for a strongly coupled gauge group external to the SM and the particles charged under it.
In the preprint of \cite{Susskind:1978ms}, the name {\it technicolor} was used for the group and {\it techniquarks} for the particles, while in the published version the names were changed to {\it heavy-color} and {\it heavy-color quarks}.
In \cite{Dimopoulos:1979es}, again technicolor and techniquarks was used.
\cite{Eichten:1979ah} uses {\it hypercolor} and {\it hyperfermions} and \cite{Kaplan:1983sm} uses {\it ultracolor} and {\it ultrafermions}.
A larger gauge group in which the strongly coupled gauge group is embedded and under which also the SM fermions are charged is called {\it extended technicolor} in \cite{Dimopoulos:1979es}, {\it sideways} interaction in \cite{Eichten:1979ah} and {\it extended ultracolor} in \cite{Kaplan:1983sm}.
While historically technicolor was used for models where a condensate directly breaks EW symmetry and ultracolor was used in the context of composite Higgs models, in this thesis a strong interaction external to the SM is in general called technicolor and the particles charged under it {\it technifermions} (and {\it techniscalars}, cf. section~\ref{sec:FPC}).}
\cite{Weinberg:1975gm,Susskind:1978ms,Dimopoulos:1979es,Eichten:1979ah}.
It is due to four-fermion operators connecting two chiral SM fermions with two fermions charged under technicolor, the technifermions.
The condensation of the latter then yields a mass term for the SM fermions.
The same mechanism is employed in early CHMs \cite{Kaplan:1983fs,Kaplan:1983sm,Banks:1984gj,Georgi:1984ef,Georgi:1984af,Dugan:1984hq}.
The four-fermion operators themselves have to be due to some form of extended technicolor (ETC) model that produces them at a scale $\Lambda_{\rm ETC}$ quite above the confinement scale of the technifermions $\Lambda_{\rm TC}$.
These constructions have however a critical drawback:
extended technicolor interactions that are responsible for four-fermion operators of this kind also yield experimentally unacceptable large contributions to flavor-changing neutral currents (FCNCs)~\cite{Eichten:1979ah,Dimopoulos:1980fj}.
One could think about circumventing this problem by raising the $\Lambda_{\rm ETC}$ scale.
But this of course also affects the mass terms of the SM fermions.
Following~\cite{Panico:2015jxa}, such a mass term in general reads
\begin{equation}
 \mathcal{L}\supset \frac{\lambda(\Lambda_{\rm ETC})}{\Lambda_{\rm ETC}^{d-1}}
 \bar{f}_L\,\mathcal{O}_S\,f_R,
\end{equation}
where $f_L$ and $f_R$ are left-handed and right-handed versions of a SM fermion, $\lambda(\Lambda_{\rm ETC})$ is its Yukawa coupling at the scale $\Lambda_{\rm ETC}$, $\mathcal{O}_S$ is a scalar operator composed of technicolor charged fields and $d=\dim[\mathcal{O}_S]$ is the scaling dimension of the operator $\mathcal{O}_S$.
If $\mathcal{O}_S$ is given by a technifermion $\Psi$ bilinear, i.e. $\mathcal{O}_S=\bar{\Psi}\Psi$, the mass term corresponds just to the four-fermion operator discussed above and its canonical dimension is $d=3$.
If the TC theory above $\Lambda_{\rm TC}$ is asymptotically free and the running of the coupling between $\Lambda_{\rm TC}$ and $\Lambda_{\rm ETC}$ can be neglected, the mass term at $\Lambda_{\rm TC}$ is given by
\begin{equation}
 \mathcal{L}\supset \frac{\lambda(\Lambda_{\rm TC})}{\Lambda_{\rm TC}^{d-1}}
 \bar{f}_L\,\mathcal{O}_S\,f_R,
\end{equation}
such that the Yukawa coupling at the ETC scale can be related to the one at the scale where the SM fermion mass is generated by
\begin{equation}
 \lambda(\Lambda_{\rm ETC}) \approx \lambda(\Lambda_{\rm TC})\left(\frac{\Lambda_{\rm ETC}}{\Lambda_{\rm TC}}\right)^{d-1}.
\end{equation}
From this, it follows that raising the scale $\Lambda_{\rm ETC}$ by some factor, say $10$, to be save from FCNCs leads to an increase of the Yukawa coupling at $\Lambda_{\rm ETC}$ by a factor $10^{d-1}$.
Using $d=3$, this factor $100$ would be problematic especially for the large top quark Yukawa coupling.
It could easily make $\lambda(\Lambda_{\rm ETC})$ non-perturbative and the whole construction would be inconsistent.
One might think about giving the operator $\mathcal{O}_S$ a large anomalous dimension~$\gamma$ such that $d=3-\gamma$ is close to $1$.
A scaling dimension of $\mathcal{O}_S$ close to $1$ would however imply a scaling dimension of $\mathcal{O}_S^2$ close to $2$~\cite{Rattazzi:2008pe} and values below $4$ for the latter reintroduce the naturalness problem.
While there might still be a way so solve these problems (cf.\ e.g.~\cite{Luty:2004ye} and references therein), there is also another issue: the whole flavor structure of the SM and the mass hierarchies of SM fermions have to be generated in the ETC theory by producing different coefficients $\lambda$ for each of the SM fermions.
Again, there might be some way to solve this, e.g.\ by introducing several Higgs doublets that couple differently to different quarks or leptons \cite{Kaplan:1983sm}.
However, it seems that a solution to only some of the problems already requires an arbitrarily complicated construction.

\subsection{Fermion partial compositeness}\label{sec:Fermions:partial_compositeness}
In light of all the difficulties arising from the above discussed mechanism for SM fermion mass generation, a different mechanism was proposed 
in~\cite{Kaplan:1991dc}.
Instead of coupling a bilinear of left- and right-handed versions of a SM fermion to a single scalar composite operator $\mathcal{O}_S$, the idea is to couple each chirality independently to fermionic composite operators $\mathcal{O}_{F_L}$ and $\mathcal{O}_{F_R}$.
Again following~\cite{Panico:2015jxa}, such a coupling reads
\begin{equation}\label{eq:fermions_comp_elem_mixing_general}
 \mathcal{L}\supset
 \frac{\lambda_L(\Lambda_{\rm ETC})}{\Lambda_{\rm ETC}^{d_L-5/2}}
 \bar{f}_L\,\mathcal{O}_{F_L}
 +
 \frac{\lambda_R(\Lambda_{\rm ETC})}{\Lambda_{\rm ETC}^{d_R-5/2}}
 \bar{f}_R\,\mathcal{O}_{F_R},
\end{equation}
where now $f_L$ and $f_R$ each have separate coupling constants $\lambda_L(\Lambda_{\rm ETC})$ and $\lambda_R(\Lambda_{\rm ETC})$ at the scale $\Lambda_{\rm ETC}$.
Using the same arguments as before, the couplings at the ETC scale can be related to those at $\Lambda_{\rm TC}$ by
\begin{equation}
 \lambda_{L,R}(\Lambda_{\rm ETC}) \approx \lambda_{L,R}(\Lambda_{\rm TC})\left(\frac{\Lambda_{\rm ETC}}{\Lambda_{\rm TC}}\right)^{d_{L,R}-5/2}.
\end{equation}
If now the scaling dimensions $d_{L}=\dim[\mathcal{O}_{F_L}]$ and $d_{R}=\dim[\mathcal{O}_{F_R}]$ are close to $5/2$, $\Lambda_{\rm ETC}$ could be raised without suppressing the couplings and one could be save from the dangerous FCNCs.
It is especially encouraging that contrary to a scaling dimension of $1$ for a scalar operator, a scaling dimension of $5/2$ for a fermionic operator does not pose any general problems%
\footnote{The simplest version of fermionic composite operator might be due to a bound state of three technifermions, similar to a baryon in QCD.
The corresponding operator has a canonical dimension of $9/2$ and thus a large anomalous dimension would be needed to get a scaling dimension of $5/2$.
Such a large anomalous dimension seems to be unlikely in the cases explored so far~\cite{DeGrand:2015yna,Pica:2016rmv}.
An alternative is presented in section~\ref{sec:FPC}.}.

Assuming that the fermionic operators correspond to heavy composite fermions $F(x)$ and $\widetilde{F}(x)$,
one can use $\mathcal{O}_{F_L}=F(x)\,\Lambda_{\rm TC}^{d_L-3/2}$ and $\mathcal{O}_{F_R}=\widetilde{F}(x)\,\Lambda_{\rm TC}^{d_R-3/2}$.
Defining the abbreviations
\begin{equation}
 \Delta_{L}=\lambda_{L}(\Lambda_{\rm TC})\cdot\Lambda_{\rm TC},
 \quad\quad
 \Delta_{R}=\lambda_{R}(\Lambda_{\rm TC})\cdot\Lambda_{\rm TC},
\end{equation}
the linear operators from eq.~(\ref{eq:fermions_comp_elem_mixing_general})
and
the mass terms of the composite fermions
can be written as
\begin{equation}\label{eq:fermions_comp_elem_lagrangian}
\begin{aligned}
 \mathcal{L}\supset&
 -
 m_L\,\bar{F}_L(x)\,F_R(x)
 -
 m_R\,\bar{\widetilde{F}}_L(x)\,\widetilde{F}_R(x)
 \\
 &+
 \Delta_L\,\bar{f}_L(x)\,F_R(x)
 +
 \Delta_R\,\bar{f}_R(x)\,\widetilde{F}_L(x)
 \\
 &+{\rm h.c.},
\end{aligned}
\end{equation}
where the mass $m_L$ of the $F(x)$ that couples to $f_L(x)$ and the mass $m_R$ of the $\widetilde{F}(x)$ that couples to $f_R(x)$ are in general different from each other.
$F(x)$ and $\widetilde{F}(x)$ are massive Dirac fermions, i.e. they each contain both left-handed and right-handed fields.
Since this implies that their left- and right-handed fields couple to gauge bosons in the same way, they always couple via a vector-current and are therefore also called {\it vector-like fermions}.
However, only $F_R(x)$ and $\widetilde{F}_L(x)$ couple to the elementary chiral fields $f_L(x)$ and $f_R(x)$, respectively.
The linear couplings are clearly mass mixing terms between the composite and elementary fields.
Due to these mixing terms, the fields above are not the mass eigenstates.
Rotating them to the mass basis by performing a biunitary%
\footnote{In the one flavor case considered here, $\Delta_L$ and $\Delta_R$ can always be chosen to be real.
The transformation to the mass basis can thus be done using orthogonal matrices.}
transformation yields the mass eigenstate fields $f'_L(x)$, $f'_R(x)$, $F'(x)$ and $\widetilde{F}'(x)$ that are given in terms of the elementary and composite fields as
\begin{equation}\label{eq:fermions_mass_basis_fields}
\begin{aligned}
 f'_L(x) &= \cos \theta_L\,f_L(x) + \sin \theta_L F_L(x),
 \quad&\quad
 f'_R(x) &= \cos \theta_R\,f_R(x) + \sin \theta_R \widetilde{F}_R(x),
 \\
 F'_L(x) &= \cos \theta_L\,F_L(x) - \sin \theta_L f_L(x),
 \quad&\quad
 \widetilde{F}'_R(x) &= \cos \theta_R\,\widetilde{F}_R(x) - \sin \theta_R f_R(x),
 \\
 F'_R(x) &= F_R(x),
 \quad&\quad
 \widetilde{F}'_L(x) &= \widetilde{F}_L(x),
\end{aligned}
\end{equation}
where the mixing angles $\theta_L$ and $\theta_R$ are defined by
\begin{equation}\label{eq:fermions_degrees_of_compositeness}
 \cos \theta_{L,R} = \frac{m_{L,R}}{m_{L,R}'},
 \quad\quad
 \sin \theta_{L,R} = \frac{\Delta_{L,R}}{m_{L,R}'},
 \quad\quad
 m_{L,R}' = \sqrt{m_{L,R}^2+\Delta_{L,R}^2},
\end{equation}
and $m_L'$ and $m_R'$ are the masses of $F'(x)$ and $\widetilde{F}'(x)$, respectively.
The fields $f'_L(x)$ and $f'_R(x)$, which should now be identified with the actual SM fields, are massless.
If the mixing parameters $\Delta_{L,R}$ are small compared to the masses $m_{L,R}$, these fields are mainly composed of the elementary fields and get a small admixture of the composite fields. They are thus partially composite (cf.\ section~\ref{sec:Vectorres_NLSM_HLS} where the same concept is discussed for spin one particles%
\footnote{
Historically, the term ``partially composite'' is attributed to~\cite{Kaplan:1991dc}, where it appears in the context of fermion masses in technicolor theories.
It is, however, the same concept that is already observed in the SM, where the photon is in principle also partially composite due to its mixing with the $\rho$ meson.
}).
Since $\sin \theta_{L,R}$ controls the amount of compositeness of $f'_{L,R}(x)$, it is also called the {\it degree of compositeness} of the SM field and will be abbreviated in the following by $s_{L,R} = \sin \theta_{L,R}$.
While it is not a common name, $\cos \theta_{L,R}$ is a measure of the degree of elementarity of the SM field and will be abbreviated by $c_{L,R} = \cos \theta_{L,R}$.

But how do the partially composite fermions $f'_L(x)$ and $f'_R(x)$ get their mass and their coupling to a composite Higgs?
Since they now contain parts of the composite fermions, it is actually enough to couple $F(x)$ and $\widetilde{F}(x)$ to the composite Higgs.
This corresponds to adding to the Lagrangian in eq.~(\ref{eq:fermions_comp_elem_lagrangian}) a term
\begin{equation}\label{eq:fermions_yukawa_term}
\mathcal{L}\supset -Y_{\mathcal{H}}\,\bar{F}_L(x)\,\mathcal{H}(x)\,\widetilde{F}_R(x),
\end{equation}
where $Y_{\mathcal{H}}$ is a Yukawa coupling in the composite sector and $\mathcal{H}(x)$ is the composite Higgs%
\footnote{For simplicity, the Higgs is treated here as a singlet.
The generalization to a NGB Higgs is presented in section~\ref{sec:fermion_moose}.}.
After going to the mass basis%
\footnote{In this section, if not stated otherwise, ``mass basis'' refers to the mass basis before EWSB.
When the Higgs assumes its VEV, the actual mass basis has to be determined by taking the Yukawa coupling into account.
However, the structure of the mass matrix including the Yukawa coupling suggests to perform the necessary biunitary transformation only numerically.
Analytical formulas are therefore usually restricted to the mass basis before EWSB.},
one finds a term coupling the Higgs to $f'_L(x)$ and $f'_R(x)$:
\begin{equation}
\begin{aligned}
 \mathcal{L}\supset -Y_{\mathcal{H}}\,s_L\,s_R\,\bar{f}'_L(x)\,\mathcal{H}(x)\,f'_R(x)
 =
 -Y_f^{\rm SM}\,\bar{f}'_L(x)\,\mathcal{H}(x)\,f'_R(x)
 ,
\end{aligned}
\end{equation}
where the SM Yukawa coupling $Y_f^{\rm SM}$ of $f'(x)$ is identified as
\begin{equation}
 Y_f^{\rm SM} = Y_{\mathcal{H}}\,s_L\,s_R.
\end{equation}
When the Higgs assumes its VEV, this then yields a mass term for the SM fermion.
Some important features of the partial compositeness construction can be read off directly from this term:
\begin{itemize}
 \item The Yukawa couplings of different SM fermions depend on their degrees of compositeness and therefore also on the masses of the composite fermions they mix with as well as the strength of the mixing.
 \item Since chiral SM fermions transforming under different representations of the SM gauge group have to mix with different composite fermions, partial compositeness might quite naturally account for mass hierarchies and a non-trivial flavor structure of SM fermions.
 \item While for most of the SM fermions the degrees of compositeness can be relatively small, the large top quark Yukawa coupling requires a sizable degree of compositeness.
\end{itemize}
Given the advantages of partial compositeness compared to the traditional mechanism for fermion mass generation in technicolor theories and especially its interesting properties concerning the possibility to yield a non-trivial flavor structure, it has become a key ingredient of modern CHMs.

\subsection{The fermion moose diagram}\label{sec:fermion_moose}
It is clear that for employing fermion partial compositeness in a CHM, composite fermions are unavoidable.
But how do they fit into the picture of the N-site moose diagram that is used in section~\ref{sec:vectorres_higher_levels} to describe both NGBs as well as massless and heavy vector bosons?
Interestingly, the connection can be easily made by considering the continuum limit, i.e. the 5D model.
Fermions in 5D are however necessarily Dirac fermions~\cite{10.2307/2371218}, so one might ask how it is possible to get the chiral SM fermions from a fermionic 5D bulk field.
Considering a 5D theory on an interval, this can actually be done by choosing appropriate boundary conditions such that the 4D spectrum from the KK decomposition of a 5D fermion bulk field contains only a left-handed or a right-handed massless zero mode (see \cite{Grossman:1999ra,Gherghetta:2000qt,Csaki:2003sh}).
In addition, such a KK decomposition yields a tower of heavy vector-like fermions.
This is already very similar to the case discussed above: for each chirality of SM fermions, there are heavy composite fermions with the same quantum numbers.
Actually, the 5D model even automatically contains the partial compositeness mechanism for fermions.
This is easily seen by deriving a 4D model from a 5D action using dimensional deconstruction, i.e. by discretizing the extra dimension.
This just corresponds to inverting the continuum limit discussed in section~\ref{sec:vectorres_contunuum_limit}, but now with additional fermion fields in the extra-dimensional bulk.
Since a 5D fermion field only yields a single chiral zero mode, each SM field requires two 5D fermions $\Psi(x,u)$ and $\widetilde{\Psi}(x,u)$, one for each chirality.
The boundary conditions for $\Psi(x,u)$ and $\widetilde{\Psi}(x,u)$ are then chosen such that $\Psi(x,u)$ contains a left-handed and $\widetilde{\Psi}(x,u)$ contains a right-handed zero mode.
To satisfy the bulk gauge symmetry, both $\Psi(x,u)$ and $\widetilde{\Psi}(x,u)$ transform under a representation of the full $G$ group.
Unless stated otherwise, the fundamental representation is assumed in the following.

The actual discretization of the 5D action is similar to the spin one case.
In this thesis, the prescription from \cite{DeCurtis:2011yx} is used%
\footnote{
Other constructions for including fermions in a model with NGBs and spin one resonances are described e.g.\ in \cite{Panico:2011pw,Cheng:2006ht,Marzocca:2012zn}.
},
which corresponds to the spin one moose diagram in eq.~(\ref{eq:Vectorres_general_moose_noH}).
The fields $\Psi(x,u)$ and $\widetilde{\Psi}(x,u)$ are split into 4D fields at $N$ sites, parametrized by an index $k$.
The different sites are connected by the NGB matrices $\Omega_k(x)$ that act as link-fields and connect the fermions at the site $k-1$ with those at the site $k$.
The boundary conditions are chosen such that $\Psi(x,u_0) = \Psi_L^{0}(x) = f_L(x)$ and $\widetilde{\Psi}(x,u_0) = \widetilde{\Psi}_R^{0}(x) = f_R(x)$.
So the fields at site $k=0$ are identified with the chiral elementary fields.
This corresponds to the spin one case where the elementary gauge bosons are introduced by gauging the $E$ subgroup of the global $G$ symmetry on site $k=0$, which also corresponds to a boundary condition in the 5D case (cf.\ eq.~(\ref{eq:vectorres_5D_gauge_4D_BC})).
The fields $f_L(x)$ and $f_R(x)$ only transform under a representation of $E$.
So again, like in the spin one case, the elementary fields transform under a smaller group than the composite ones.
Nevertheless, one can keep the Lagrangian formally $G$ invariant  by embedding $f_L(x)$ and $f_R(x)$ into incomplete multiplets $\Psi_L^{0}(x)$ and $\widetilde{\Psi}_R^{0}(x)$ that formally transform under the full $G$ group.
On the second boundary, at site $k=N$, a Yukawa coupling term is added that connects $\Psi(x,u)$ with $\widetilde{\Psi}(x,u)$.
In terms of a moose diagram, this can be written as
\begin{equation}\label{eq:fermions_general_moose}
\begin{tabular}{c}
\xy
\xymatrix@R=.4pc@C=1.4pc{
\mathrm{Global:}
& G
& G_1
& G_2
&
& G_{N-1}
&
\\
& *=<20pt>[o][F]{} \rightxyarrow^{\mbox{\raisebox{1.5ex}{$\Omega_1$}}}
& *=<20pt>[o][F]{} \rightxyarrow^{\mbox{\raisebox{1.5ex}{$\Omega_2$}}}
& *=<20pt>[o][F]{} \rightxyarrow
& *=<20pt>[o]{\cdots} \rightxyarrow
& *=<20pt>[o][F]{} \rightxyarrow^{\mbox{\raisebox{1.5ex}{$\Omega_N$}}}
& *=<0pt,20pt>[l][F]{} & *-<0pt,20pt>[l]{H}
\\
\mathrm{Gauged:}
&E
& G_1
& G_2
&
& G_{N-1}
&
\\
\mathrm{Fermions:}
&
{\begin{matrix}[1.5]
 \Psi_L^{0}
 \\
 \widetilde{\Psi}_R^{0}
\end{matrix}}
&
{\begin{matrix}[1.5]
\Psi_R^{1} & \Psi_L^{1}
 \\
 \widetilde{\Psi}_L^{1} & \widetilde{\Psi}_R^{1}
\end{matrix}}
&
{\begin{matrix}[1.5]
\Psi_R^{2} & \Psi_L^{2}
 \\
 \widetilde{\Psi}_L^{2} & \widetilde{\Psi}_R^{2}
\end{matrix}}
&&
{\left.\begin{matrix}[1.5]
\Psi_R^{N-1} & \Psi_L^{N-1}
 \\
 \widetilde{\Psi}_L^{N-1} & \widetilde{\Psi}_R^{N-1}
\end{matrix}\ \ \right>\,Y_{\rm comp}}
\hspace{-50pt}
}
\endxy
\end{tabular},
\end{equation}
where it is indicated that each of the composite fermions is given in terms of a Dirac fermion containing left- and right-handed fields, while on site $k=0$ only one chirality for each of the $\Psi^0(x)$ and $\widetilde{\Psi}^0(x)$ is present.
The Yukawa coupling that connects $\Psi_L^{N-1}(x)$ and $\widetilde{\Psi}_R^{N-1}(x)$ is shown on the last site.
The corresponding fermion Lagrangian can then be written as
\begin{equation}\label{eq:fermions_general_lagrangian}
\begin{aligned}
 \mathcal{L}_{\Psi}
 &=
 \bar{\Psi}_L^{0}(x)\,i\slashed{D}\,\Psi_L^{0}(x)
 +
 \bar{\widetilde{\Psi}^{0}_R\hspace{-.6em}}\hspace{.6em}(x)\,i\slashed{D}\,\widetilde{\Psi}_R^{0}(x)
 \\
 &+\sum_{k=1}^{N-1}
 \begin{aligned}[t]
 \Bigg\{&
 \bar{\Psi}^{k}(x)\left(i\slashed{D}-m_L^{k}\right)\Psi^{k}(x)
 +
 \bar{\widetilde{\Psi}^{k}\hspace{-.5em}}\hspace{.5em}(x)\left(i\slashed{D}-m_R^{k}\right)\widetilde{\Psi}^{k}(x)
 \\
 +\,&
 \Delta_L^{k}\,\bar{\Psi}_L^{k-1}(x)\,\Omega_k(x)\,\Psi_R^{k}(x)
 +
 \Delta_R^{k}\,\bar{\widetilde{\Psi}_R^{k-1}\hspace{-1.5em}}\hspace{1.5em}(x)\,\Omega_k(x)\,\widetilde{\Psi}_L^{k}(x)
 +{\rm h.c.}
 \Bigg\}
 \end{aligned}
 \\
 &-
 Y_{\rm comp}\,\bar{\Psi}_L^{N-1}(x)\,\Omega_N(x)
 \,\phi_0\,\phi_0^\dagger\,
 \Omega_N^\dagger(x)\,\widetilde{\Psi}_R^{N-1}(x)
 -m_Y\,\bar{\Psi}_L^{N-1}(x)\,\widetilde{\Psi}_R^{N-1}(x)
 +{\rm h.c.}\ ,
\end{aligned}
\end{equation}
where the first line contains the kinetic terms of the massless chiral fields at site $k=0$, the second line contains the kinetic terms and masses of the $N-1$ composite fermions,
the fourth line contains mixing terms with mixing constants $\Delta_L^k$ and $\Delta_R^k$
and the last line contains the composite sector Yukawa coupling and a possible mass mixing between $\Psi_L^{N-1}(x)$ and $\widetilde{\Psi}_R^{N-1}(x)$.
The Yukawa coupling is written in terms of an explicit vacuum state $\phi_0$ (see \cite{DeCurtis:2011yx} and cf.\ appendix~\ref{sec:vacuum_states}).
The link-fields show up in the mixing terms as well as in the Yukawa coupling.
The covariant derivatives contained in the kinetic terms are given by
\begin{equation}
 i D_\mu\,\Psi^{k}(x) = \left(i\,\partial_\mu+g_k\,A_\mu^k(x)\right)\Psi^{k}(x),
\end{equation}
i.e.\ each fermion is only coupled to the gauge bosons at the same site.

That the above Lagrangian, eq.~(\ref{eq:fermions_general_lagrangian}), is indeed a generalization of the mass, mixing and Yukawa terms in eqs.~(\ref{eq:fermions_comp_elem_lagrangian}) and (\ref{eq:fermions_yukawa_term}) that are used in the previous section to introduce the concept of fermion partial compositeness is best shown by considering the case with one level of composite fermions, i.e.\ $N=2$.
The moose diagram is then given by
\begin{equation}\label{eq:fermion_moose:two-site_moose}
\begin{tabular}{c}
\xy
\xymatrix@R=.4pc@C=1.4pc{
\mathrm{Global:}
& G
&& G_1
&&
\\
& *=<20pt>[o][F]{} \doublerightxyarrow^{\mbox{\raisebox{1.5ex}{$\Omega_1$}}}
&& *=<20pt>[o][F]{} \doublerightxyarrow^{\mbox{\raisebox{1.5ex}{$\Omega_2$}}}
&& *=<0pt,20pt>[l][F]{} & *-<0pt,20pt>[l]{H}
\\
\mathrm{Gauged:}
&E
&& G_1
&&
\\
\mathrm{Fermions:}
&
{\begin{matrix}[1.5]
 \xi_L
 \\
 \xi_R
\end{matrix}}
&&
{\left.\begin{matrix}[1.5]
\Psi_R & \Psi_L
 \\
 \widetilde{\Psi}_L & \widetilde{\Psi}_R
\end{matrix}\ \ \hspace{20pt}\right>\,Y_{\rm comp}}
\hspace{-20pt}
\hspace{-50pt}
}
\endxy
\end{tabular},
\end{equation}
where for clarity $\Psi^0_L$ and $\widetilde{\Psi}^0_R$ are replaced by $\xi_L$ and $\xi_R$ to emphasize that these are incomplete multiplets into which the fields $f_L$ and $f_R$ are embedded.
Furthermore, in the $N=2$ case the indices on the composite fermions (and in the following also those on their masses and on the mixing coefficients) are dropped.
For a comparison with the previous section, it is useful to employ a version of the holographic gauge where $\Omega_1(x)=\mathds{1}$ and $\Omega_2(x)=\Omega(x)$, i.e.\ the complete dependence on the $G/H$ NGBs is contained in $\Omega_2(x)$ (cf.\ section~\ref{sec:vectorres_higher_levels}).
Omitting the kinetic terms and setting the mass mixing on the last site to zero, i.e.\ $m_Y\to0$, the corresponding Lagrangian then reads
\begin{equation}
\begin{aligned}
 \mathcal{L}\supset
 &-
 m_L\,\bar{\Psi}_L(x)\,\Psi_R(x)
 -
 m_R\,\bar{\widetilde{\Psi}}_L(x)\,\widetilde{\Psi}_R(x)
 \\
 &+
 \Delta_L\,\bar{\xi}_L(x)\,\Psi_R(x)
 +
 \Delta_R\,\bar{\xi}_R(x)\,\widetilde{\Psi}_L(x)
 \\
 &-
 Y_{\rm comp}\,\bar{\Psi}_L(x)\,\Omega(x)\,
 \phi_0\,\phi_0^\dagger\,
 \Omega^\dagger(x)\,\widetilde{\Psi}_R(x)
 \\
 &+{\rm h.c.}
\end{aligned}
\end{equation}
Due to the incomplete multiplets $\xi_L$ and $\xi_R$, the mixing terms actually only couple the embedded $f_L$ and $f_R$ to the corresponding components $F_R$ and $\widetilde{F}_L$ of $\Psi_R$ and $\widetilde{\Psi}_L$.
The mixing terms and the masses of the components $F$ and $\widetilde{F}$ are therefore equivalent to the terms in the Lagrangian in eq.~(\ref{eq:fermions_comp_elem_lagrangian}).
The last line on the other hand is a Yukawa coupling term containing the NGBs and can be regarded as a generalization of eq.~(\ref{eq:fermions_yukawa_term}).
Employing the definition of the mass eigenstates $f_L'$ and $f_R'$, eq.~(\ref{eq:fermions_mass_basis_fields}),
and embedding them in incomplete multiplets $\xi_L'$ and $\xi_R'$,
one finds a Yukawa term that couples them to the NGBs:
\begin{equation}
 \mathcal{L}\supset -Y_{comp}\,s_L\,s_R\,\bar{\xi}'_L(x)\,\Omega(x)\,\phi_0\,\phi_0^\dagger\,\Omega^\dagger(x)\,\xi'_R(x),
\end{equation}
where the degrees of compositeness $s_{L,R} = \sin \theta_{L,R}$ are defined in eq.~(\ref{eq:fermions_degrees_of_compositeness}).
The partial composite mechanism is thus found to arise directly from a 5D fermion field with appropriate boundary conditions.
Using the moose description derived from discretizing the extra dimension, fermions can readily be included in the models discussed in section~\ref{sec:Vectorres}.
Therefore, the effective theory of elementary (i.e.\ massless gauge fields and massless chiral fermions) and composite states (i.e.\ $G/H$ NGBs, heavy spin one resonances and vector-like fermions) can be described by a framework that features partial compositeness for both, the spin one and the fermionic fields.

\subsection{Flavor}\label{sec:Fermions:flavor}
One of the most interesting features of fermion partial compositeness is the possibility to endow the SM fermions with a non-trivial flavor structure.
To make use of this property, the single flavor case used for simplicity in the previous sections has to be generalized.
In the SM, the whole flavor structure is encoded in the $3\times 3$ Yukawa matrices $(Y_u^{\rm SM})^{ij}$, $(Y_d^{\rm SM})^{ij}$ and $(Y_e^{\rm SM})^{ij}$.
Since the SM does not contain flavor mixing in the lepton sector and $(Y_e^{\rm SM})^{ij}$ can be made diagonal%
\footnote{It is of course known that there is huge flavor mixing in the neutrino sector that is however not described by the SM (with only left-handed neutrinos).},
the following discussion will only consider the quark sector, i.e.\ a non-trivial structure of $(Y_u^{\rm SM})^{ij}$ and $(Y_d^{\rm SM})^{ij}$.
To generalize the construction in the previous section from the one flavor case to six quark flavors, one has to introduce two sets of fields with different quantum numbers: the up-type and the down-type quarks.
Furthermore, each of them consists of three generations, which is accounted for by promoting the fields to $3$-vectors and the masses, mixings and Yukawa couplings to $3\times 3$ matrices in generation, or flavor space.
Considering for simplicity again the case with one level of composite fermions, the moose diagram from the previous section is extended to
\begin{equation}\label{eq:fermions_2site_moose_ud}
\begin{tabular}{c}
\xy
\xymatrix@R=.4pc@C=1.4pc{
\mathrm{Global:}
& G
&& G_1
&&
\\
& *=<20pt>[o][F]{} \doublerightxyarrow^{\mbox{\raisebox{1.5ex}{$\Omega_1$}}}
&& *=<20pt>[o][F]{} \doublerightxyarrow^{\mbox{\raisebox{1.5ex}{$\Omega_2$}}}
&& *=<0pt,20pt>[l][F]{} & *-<0pt,20pt>[l]{H}
\\
\mathrm{Gauged:}
&E
&& G_1
&&
\\
\mathrm{Fermions:}
&
{
\begin{matrix}
\begin{matrix}[1.5]
 \xi_{uR}\, [u_R]
\end{matrix}
\\
\begin{matrix}[1.5]
 \xi_{uL}\, [q_L]
 \\
 \xi_{dL}\, [q_L]
\end{matrix}
\\
\begin{matrix}[1.5]
 \xi_{dR}\, [d_R]
\end{matrix}
\end{matrix}
}
&&
{
\begin{matrix}
\left.\begin{matrix}[1.5]
 \widetilde{\Psi}_{uL}\, [U_L] & \widetilde{\Psi}_{uR}\, [U_R]
 \\
\Psi_{uR}\, [Q^u_R] & \Psi_{uL}\, [Q^u_L]
\end{matrix}\ \ \hspace{20pt}\right>\,Y_u
\\
\left.\begin{matrix}[1.5]
\Psi_{dR}\, [Q^d_R] & \Psi_{dL}\, [Q^d_L]
 \\
 \widetilde{\Psi}_{dL}\, [D_L] & \widetilde{\Psi}_{dR}\, [D_R]
\end{matrix}\ \ \hspace{20pt}\right>\,Y_d
\end{matrix}
}
\hspace{-20pt}
\hspace{-40pt}
}
\endxy
\end{tabular}\quad,
\end{equation}
where brackets behind the $\xi$ fields show the elementary fields that are embedded into them, while brackets behind the $\Psi$ and $\widetilde{\Psi}$ fields show their components that mix with the elementary fields.
Following \cite{DeCurtis:2011yx,Contino:2006qr}, the elementary left-handed quark doublet $q_L$ is embedded into both $\xi_{uL}$ and $\xi_{dL}$ such that it can couple to the two composite fermions $\Psi_u$ and $\Psi_d$ of which the former has a Yukawa coupling to the composite partner of the elementary right-handed up-type quarks and the latter to the composite partner of the elementary right-handed down-type quarks.
While it is possible to reduce the fermion field content by employing only one composite partner for the quark doublet (cf.\ e.g.~\cite{Vignaroli:2012si})\footnote{%
The field content in the composite sector can also be reduced by coupling elementary left-handed and right-handed quarks to different components of a single composite quark representation~\cite{Panico:2011pw}.},
this would not allow to implement all different kinds of flavor symmetries discussed in the following (see \cite{Redi:2011zi}).

To investigate the quark flavor structure, it is convenient to make all flavor indices explicit.
The fermion Lagrangian corresponding to the moose diagram above then reads
\begin{equation}\label{eq:fermions_L_flavor}
\begin{aligned}
 \mathcal{L}\supset
 &-
 m_{uL}^{ij}\,\bar{\Psi}_{uL}^i(x)\,\Psi_{uR}^j(x)
 -
 m_{uR}^{ij}\,\bar{\widetilde{\Psi}_{uL}^{i}\hspace{-1.0em}}\hspace{1.0em}(x)\,\widetilde{\Psi}_{uR}^j(x)
 \\
 &+
 \Delta_{uL}^{ij}\,\bar{\xi}_{uL}^i(x)\,\Psi_{uR}^j(x)
 +
 \Delta_{uR}^{ij}\,\bar{\xi}_{uR}^i(x)\,\widetilde{\Psi}_{uL}^j(x)
 \\
 &-
 Y_{u}^{ij}\,\bar{\Psi}_{uL}^i(x)\,\Omega(x)\,
 \phi_0\,\phi_0^\dagger\,
 \Omega^\dagger(x)\,\widetilde{\Psi}_{uR}^j(x)
 \\
 &+ (u\rightarrow d)
 \\
 &+{\rm h.c.}\ ,
\end{aligned}
\end{equation}
where $i,j$ are the flavor indices.
The composite-elementary mixings, the composite fermion masses, and the composite Yukawa couplings are in general complex matrices, but for the following discussion, at least the composite fermion masses are assumed to be flavor universal, i.e.\ $m_\alpha^{ij}=m_\alpha\,\delta^{ij}$, $\alpha\in\{uL,uR,dL,dR\}$.
Nevertheless, there are still six complex matrices $\Delta_{uL}^{ij}$, $\Delta_{uR}^{ij}$, $Y_u^{ij}$, $\Delta_{dL}^{ij}$, $\Delta_{dR}^{ij}$ and $Y_d^{ij}$ in the model.
Any complex matrix $M$ can be decomposed by a singular value decomposition (SVD) into two unitary matrices $U$, $V$ and a diagonal matrix $D$:
\begin{equation}
 M = U\,D\,V^\dagger.
\end{equation}
Field redefinitions may be used to absorb some of the unitary matrices in the decomposition of the six complex mixing and Yukawa matrices, such that their full complex structure is not physical.
To investigate this further, it is useful to consider the flavor symmetries of the above Lagrangian in the absence of the mixings and Yukawa couplings.
In this case, each of the elementary and composite fields transforms under its own ${\rm U}(3)$ flavor symmetry.
Note that the two ${\rm U}(3)$ symmetries of the chiral components of each composite field are broken to their diagonal subgroup by the flavor universal mass terms, and that the two ${\rm U}(3)$ symmetries of $\xi_{uL}$ and $\xi_{dL}$ are broken to their diagonal subgroup by embedding $q_L$ in both $\xi_{uL}$ and $\xi_{dL}$.
The whole Lagrangian then has a global ${\rm U}(3)^7$ flavor symmetry\footnote{%
${\rm U}(3)^7={\rm U}(3)_{u_R} \times {\rm U}(3)_{U} \times {\rm U}(3)_{Q^{u}} \times {\rm U}(3)_{q_L} \times {\rm U}(3)_{Q^{d}} \times {\rm U}(3)_{D} \times {\rm U}(3)_{d_R}$.}.
The Lagrangian can be kept formally invariant under this global symmetry in the presence of the composite-elementary mixings and Yukawa couplings if they are treated as spurions, i.e.\ objects that formally transform under the global symmetries and break them only when assuming their background values.
To discuss the transformation properties of the six spurions under the seven ${\rm U}(3)$ factors of the global flavor symmetry group, it is convenient to once more employ the language of moose diagrams\footnote{
Recall that each group is represented by a circle, objects that transform under the fundamental representation of this group by an arrow pointing away from the circle and objects transforming under the anti-fundamental representation by an arrow pointing into the direction of the circle.}.
The flavor symmetry structure in the presence of the spurions can then be written as\footnote{
The flavor groups are defined such that the fields $u_R$, $U$, $Q^u$, $q_L$, $Q^d$, $D$, $d_R$ transform under the fundamental representation of their associated ${\rm U}(3)$ factor. The transformation properties of the spurions can then be read off from the Lagrangian in eq.~(\ref{eq:fermions_L_flavor}).}
\begin{equation}\label{eq:fermions_PC_flavor_moose}
\begin{tabular}{c}
\xy
\xymatrix@R=.4pc@C=1.4pc{
 *=<20pt>[o][F]{} \rightxyarrow^{\mbox{\raisebox{1.5ex}{$\Delta_{uR}$}}} &
 *=<20pt>[o][F]{} &
 *=<20pt>[o][F]{} \leftxyarrow_{\mbox{\raisebox{1.5ex}{$Y_u$}}} &
 *=<20pt>[o][F]{} \leftxyarrow_{\mbox{\raisebox{1.5ex}{$\Delta_{uL}$}}}
		  \rightxyarrow^{\mbox{\raisebox{1.5ex}{$\Delta_{dL}$}}} &
 *=<20pt>[o][F]{} \rightxyarrow^{\mbox{\raisebox{1.5ex}{$Y_d$}}} &
 *=<20pt>[o][F]{} &
 *=<20pt>[o][F]{} \leftxyarrow_{\mbox{\raisebox{1.5ex}{$\Delta_{dR}$}}} \\
{\rm U}(3)_{u_R} &
{\rm U}(3)_{U} &
{\rm U}(3)_{Q^{u}} &
{\rm U}(3)_{q_L} &
{\rm U}(3)_{Q^{d}} &
{\rm U}(3)_{D} &
{\rm U}(3)_{d_R}
}
\endxy
\end{tabular},
\end{equation}
so e.g.\ $Y_u$ transforms as a $(\mathbf{1}, \mathbf{\bar{3}}, \mathbf{3}, \mathbf{1}, \mathbf{1}, \mathbf{1}, \mathbf{1})$ under the ${\rm U}(3)^7$.
While the flavor symmetries are broken by the spurion background values, the symmetries can still be used to transform the fields in such a way that they absorb some of the unitary matrices that arise from the SVD of the spurion background values.
Each of the ${\rm U}(3)$ symmetries can be used to rotate away one of the unitary matrices, or equivalently, each of the fields associated to the ${\rm U}(3)$ symmetries can absorb one of these unitary matrices.
The six spurion background values are decomposed into six diagonal matrices and 12 unitary matrices of which seven can be rotated away.
This leaves six diagonal and five unitary physical matrices.
This is much more than what is present in the SM model, where the full quark flavor structure is due to the two Yukawa matrices $Y_u^{SM}$ and $Y_d^{SM}$.
When the SM Yukawa matrices are treated as spurions, the flavor symmetry in the quark sector is ${\rm U}(3)^3$ and the SM quark flavor moose diagram is therefore much simpler than the one of the above partial compositeness model.
It is given by
\begin{equation}\label{eq:fermions_SM_flavor_moose}
\begin{tabular}{c}
\xy
\xymatrix@R=.4pc@C=1.4pc{
 *=<20pt>[o][F]{} &
 *=<20pt>[o][F]{} \leftxyarrow_{\mbox{\raisebox{1.5ex}{$Y_u^{SM}$}}}
		  \rightxyarrow^{\mbox{\raisebox{1.5ex}{$Y_d^{SM}$}}} &
 *=<20pt>[o][F]{} \\
{\rm U}(3)_{u_R} &
{\rm U}(3)_{q_L} &
{\rm U}(3)_{d_R}
}
\endxy
\end{tabular}.
\end{equation}
Analogous to the discussion above, the SM Yukawa matrices can be decomposed into in total two diagonal and four unitary matrices of which three can be rotated away by the flavor symmetries.
This leaves two diagonal matrices that yield the hierarchical quark masses and one unitary matrix that is nothing but the Cabibbo–Kobayashi–Maskawa (CKM) matrix.
Experimental data puts stringent constraints on flavor violation that goes beyond what is predicted by the SM.
On the other hand, by assuming that the SM Yukawas are the only source of flavor violation, it can be shown that $\Delta F= 1$ and $\Delta F= 2$ flavor observables are sufficiently protected even in the presence of NP~\cite{Chivukula:1987py,DAmbrosio:2002vsn}.
This assumption is also known as Minimal Flavor Violation (MFV)~\cite{DAmbrosio:2002vsn}.
Applied to a generic NP model, MFV requires the flavor structure to be similar to the one in the SM in the sense that two spurions transforming under a ${\rm U}(3)^3$ flavor symmetry like in eq.~(\ref{eq:fermions_SM_flavor_moose}) are the only source of flavor violation.
The above partial composite model is therefore far from being MFV and experimentally unacceptable large flavor violation is expected.
Several paradigms that reduce the amount of flavor violating sources are discussed in the following.

\subsubsection{Anarchy}
It is well known from models with a warped extra dimension that the correct quark masses and CKM elements can arise from a 5D Yukawa coupling without any structure or hierarchy (see e.g.~\cite{Huber:2000ie,Gherghetta:2000qt}), which is therefore called {\it anarchic}.
The hierarchies in the SM Yukawa matrices are then due to the overlap of the profile functions of left- and right-handed fermion zero modes and the Higgs.
Translated to the 4D picture with one level of composite fermions, the overlap of the profile functions corresponds to the product of the degrees of compositeness of left- and right-handed SM fermions, and the 5D Yukawa coupling can be identified with a composite Yukawa coupling.
The central assumptions in 4D anarchic models are thus that all composite-elementary mixings $\Delta_{uL,uR,dL,dR}$ are diagonal, hierarchical matrices, while all entries of the composite Yukawa couplings $Y_{u,d}$ are of $\mathcal{O}(1)$.
It has however been shown that this structure leads to a ``flavor problem'', mainly due to tensions with experimental bounds from CP violation in kaon mixing that generically require implausibly high masses of composite resonances, or an undesired amount of tuning \cite{Csaki:2008zd,Blanke:2008zb,Bauer:2009cf}.
The reason for this may be traced back to the fact that while containing fewer sources of flavor violation than the most general model, the anarchic model still contains more than what would be required to fulfill the MFV assumption.
In contrast to two spurions that yield one unitary and two diagonal matrices, it contains six spurions that yield at least four diagonal and two unitary matrices.

\subsubsection{MFV implementations}
To ameliorate the flavor problem found in anarchic models, implementations of MFV that endow the composite sector with appropriate flavor symmetries have been proposed in~\cite{Cacciapaglia:2007fw,Barbieri:2008zt,Redi:2011zi}.
Investigating the moose diagram of the most general model, eq.~(\ref{eq:fermions_PC_flavor_moose}), one readily finds what is sufficient to construct an MFV model:
\begin{itemize}
 \item At least one of the spurions $\Delta_{uL}$, $\Delta_{uR}$, or $Y_u$ has to be non-trivial to yield the up-type quark masses.
 \item At least one of the spurions $\Delta_{dL}$, $\Delta_{dR}$, or $Y_d$ has to be non-trivial to yield the down-type quark masses.
 \item All but two spurions in total have to be flavor-universal to fulfill the MFV assumption.
\end{itemize}
The above criteria allow for in general nine different cases.
An additional requirement that is not necessary for MFV but seems reasonable is that the complete composite sector should be flavor universal.
This then forces $Y_u$ and $Y_d$ to be proportional to the unit matrix and reduces the possible cases to four.
With this assumption, the complete flavor structure has to stem from the composite-elementary mixings and is in this sense external to the composite sector.
Assuming flavor universal $Y_u$ and $Y_d$ and not treating them as spurions anymore breaks the ${\rm U}(3)_U\times {\rm U}(3)_{Q^u}$ and the ${\rm U}(3)_{Q^d}\times {\rm U}(3)_{D}$ to their diagonal subgroups ${\rm U}(3)_{U+Q^u}$ and ${\rm U}(3)_{Q^d+D}$.
The global flavor symmetry is thus reduced to ${\rm U}(3)^5$.
Among the four MFV cases that fulfill this assumption, there are two that are symmetric in the treatment of up-type and down-type quarks:
\begin{itemize}
 \item {\bf Left-compositeness} (LC):
 In this case, also the left-handed composite-elementary mixings are assumed to be flavor universal.
 This breaks the ${\rm U}(3)_{U+Q^u}\times {\rm U}(3)_q \times {\rm U}(3)_{Q^d+D}$ to its diagonal subgroup ${\rm U}(3)_{U+Q^u+q_L+Q^d+D}$ and the elementary left-handed quark doublet transforms under the same ${\rm U}(3)$ symmetry as the whole composite sector.
 Employing the moose diagram notation, the reduction of the generic model to the LC model by requiring the composite sector and the left-handed composite-elementary mixings to be flavor universal can be depicted as
\begin{equation}
\begin{tabular}{c}
\begin{tabular}{c}
\xy
\xymatrix@R=.4pc@C=1.4pc{
 *=<20pt>[o][F]{} \rightxyarrow^{\mbox{\raisebox{1.5ex}{$\Delta_{uR}$}}} &
 *=<20pt>[o][F]{} &
 *=<20pt>[o][F]{} \leftxyarrow_{\mbox{\raisebox{1.5ex}{$Y_u$}}} &
 *=<20pt>[o][F]{} \leftxyarrow_{\mbox{\raisebox{1.5ex}{$\Delta_{uL}$}}}
		  \rightxyarrow^{\mbox{\raisebox{1.5ex}{$\Delta_{dL}$}}} &
 *=<20pt>[o][F]{} \rightxyarrow^{\mbox{\raisebox{1.5ex}{$Y_d$}}} &
 *=<20pt>[o][F]{} &
 *=<20pt>[o][F]{} \leftxyarrow_{\mbox{\raisebox{1.5ex}{$\Delta_{dR}$}}} \\
{\rm U}(3)_{u_R} &
{\rm U}(3)_{U} &
{\rm U}(3)_{Q^{u}} &
{\rm U}(3)_{q_L} &
{\rm U}(3)_{Q^{d}} &
{\rm U}(3)_{D} &
{\rm U}(3)_{d_R} \\
&
\ar@{}[r]|{\mbox{\raisebox{0.0ex}{$Y_u\propto\mathds{1}$}}} &
\ar@{}[r]|{\mbox{\raisebox{0.0ex}{$\Delta_{u_L}\propto\mathds{1}$}}} &
\Bigg\Downarrow
\ar@{}[r]|{\mbox{\raisebox{0.0ex}{$\Delta_{d_L}\propto\mathds{1}$}}} &
\ar@{}[r]|{\mbox{\raisebox{0.0ex}{$Y_d\propto\mathds{1}$}}} &
}
\endxy
\end{tabular}
\\
\begin{tabular}{c}
\xy
\xymatrix@R=.4pc@C=1.4pc{
 *=<20pt>[o][F]{} \triplerightxyarrow^{\mbox{\raisebox{1.5ex}{$\Delta_{uR}$}}} &&&
 *=<20pt>[o][F]{} &&&
 *=<20pt>[o][F]{} \tripleleftxyarrow_{\mbox{\raisebox{1.5ex}{$\Delta_{dR}$}}}
 \\
{\rm U}(3)_{u_R} &&&
{\rm U}(3)_{U+Q^u+q_L+Q^d+D} &&&
{\rm U}(3)_{d_R} \\
}
\endxy
\end{tabular}
\end{tabular}.
\end{equation}
It is of course no coincidence that the flavor moose diagram of the LC model resembles the one of the SM, eq.~(\ref{eq:fermions_SM_flavor_moose}).
Actually, comparing both, $\Delta_{uR}^\dagger$ can be identified with the SM up-type Yukawa and $\Delta_{dR}^\dagger$ with the SM down-type Yukawa, up to flavor universal factors stemming from the composite Yukawas and left-handed composite-elementary mixings.
While being MFV, the LC model requires the first two generations of left-handed up-type and down-type quarks to have the same degree of compositeness as the left-handed top quark.
It therefore suffers from very strong constraints due to electroweak precision tests and CKM unitarity~\cite{Redi:2011zi,Barbieri:2012tu}.

 \item {\bf Right-compositeness} (RC):
 In this case, in addition to the composite sector, the right-handed composite-elementary mixings are assumed to be flavor universal.
 This breaks the ${\rm U}(3)_{u_R}\times {\rm U}(3)_{U+Q^u}$ and the ${\rm U}(3)_{Q^d+D}\times {\rm U}(3)_{d_R}$ to their diagonal subgroups ${\rm U}(3)_{u_R+U+Q^u}$ and ${\rm U}(3)_{Q^d+D+d_R}$.
 The elementary right-handed up-type quarks transform under the same ${\rm U}(3)$ symmetry as the composite up-type quarks, while the elementary right-handed down-type quarks transform under the same ${\rm U}(3)$ symmetry as the composite down-type quarks.
 In a diagrammatic way, this  case is represented by
\begin{equation}
\begin{tabular}{c}
\begin{tabular}{c}
\xy
\xymatrix@R=.4pc@C=1.4pc{
 *=<20pt>[o][F]{} \rightxyarrow^{\mbox{\raisebox{1.5ex}{$\Delta_{uR}$}}} &
 *=<20pt>[o][F]{} &
 *=<20pt>[o][F]{} \leftxyarrow_{\mbox{\raisebox{1.5ex}{$Y_u$}}} &
 *=<20pt>[o][F]{} \leftxyarrow_{\mbox{\raisebox{1.5ex}{$\Delta_{uL}$}}}
		  \rightxyarrow^{\mbox{\raisebox{1.5ex}{$\Delta_{dL}$}}} &
 *=<20pt>[o][F]{} \rightxyarrow^{\mbox{\raisebox{1.5ex}{$Y_d$}}} &
 *=<20pt>[o][F]{} &
 *=<20pt>[o][F]{} \leftxyarrow_{\mbox{\raisebox{1.5ex}{$\Delta_{dR}$}}} \\
{\rm U}(3)_{u_R} &
{\rm U}(3)_{U} &
{\rm U}(3)_{Q^{u}} &
{\rm U}(3)_{q_L} &
{\rm U}(3)_{Q^{d}} &
{\rm U}(3)_{D} &
{\rm U}(3)_{d_R} \\
&
\ar@{}[r]|{\mbox{\raisebox{0.0ex}{$\Delta_{u_R}\propto\mathds{1}$}}} &
\ar@{}[r]|{\mbox{\raisebox{0.0ex}{$Y_u\propto\mathds{1}$}}} &
\Bigg\Downarrow
\ar@{}[r]|{\mbox{\raisebox{0.0ex}{$Y_d\propto\mathds{1}$}}} &
\ar@{}[r]|{\mbox{\raisebox{0.0ex}{$\Delta_{d_R}\propto\mathds{1}$}}} &
}
\endxy
\end{tabular}
\\
\begin{tabular}{c}
\xy
\xymatrix@R=.4pc@C=1.4pc{
 *=<20pt>[o][F]{} &&&
 *=<20pt>[o][F]{} \tripleleftxyarrow_{\mbox{\raisebox{1.5ex}{$\Delta_{u_L}$}}}
		  \triplerightxyarrow^{\mbox{\raisebox{1.5ex}{$\Delta_{d_L}$}}} &&&
 *=<20pt>[o][F]{}
 \\
{\rm U}(3)_{u_R+U+Q^u} &&&
\hspace{1.5em}{\rm U}(3)_{q_L}\hspace{1.5em} &&&
{\rm U}(3)_{Q^d+D+d_R} \\
}
\endxy
\end{tabular}
\end{tabular}.
\end{equation}
Comparing the flavor moose diagram of the RC model to the SM, one can identify $\Delta_{u_L}$ with the SM up-type Yukawa and $\Delta_{d_L}$ with the SM down-type Yukawa, again up to flavor-universal factors.
RC is thus obviously only possible if the elementary quark doublet mixes with the two different composite fields $Q^u$ and $Q^d$.
An important difference to the LC model is that not all composite fields transform under the same ${\rm U}(3)$ symmetry.
In RC models, the composite up-type and down-type quarks transform under two separate ${\rm U}(3)$ symmetries.
The bounds are weaker than in the LC model, but RC is still considerably constrained, e.g. by dijet angular distributions~\cite{Redi:2011zi,Barbieri:2012tu}, because again the compositeness of light quarks is linked to the one of the top quark.
\end{itemize}
Some of the other possibilities to realize MFV apart from LC and RC are discussed in~\cite{Redi:2011zi,Redi:2012uj}.
The conclusion there is, however, that among the MFV models, RC is the one with the weakest experimental bounds.

\subsubsection{U($\mathbf{2}$)$\mathbf{^3}$ flavor symmetry}
Implementing the MFV assumption into models of partial compositeness mainly solves the flavor problems of the anarchic models.
There are however two interrelated caveats:
\begin{itemize}
 \item The flavor universal composite-elementary mixings connect the degrees of compositeness of light quarks with those of the third generation, especially the one of the top quark.
 The latter is required to be large to reproduce the top mass, while large values for the former lead to tensions with CKM unitarity and measurements of dijet angular distributions.
 \item Due to the third generation and especially the large top quark Yukawa coupling, the breaking of the global ${\rm U}(3)^3$ symmetry by the non-trivial composite-elementary mixings is not weak, i.e.\ ${\rm U}(3)^3$ is not a good approximate symmetry.
\end{itemize}
This suggests to treat the first two generations differently than the third one.
It has therefore been proposed in~\cite{Barbieri:2011ci,Barbieri:2012uh} to depart from the MFV assumption and to consider instead of a ${\rm U}(3)^3$ symmetry a ${\rm U}(2)^3={\rm U}(2)_{u_R}\times {\rm U}(2)_{q_L}\times {\rm U}(2)_{d_R}$ under which the first two generation quark fields transform as doublets, whereas the third generation quark fields transform as singlets.
While the full decomposition of the ${\rm U}(3)^3$ spurions $Y_u^{SM}=(\mathbf{\bar{3}}, \mathbf{3}, \mathbf{1})$ and $Y_d^{SM}=(\mathbf{1}, \mathbf{3}, \mathbf{\bar{3}})$ in terms of ${\rm U}(2)^3$ representations would yield many different spurions\footnote{\label{fn:fermions_spurion_decomposition}%
$(\mathbf{\bar{3}}, \mathbf{3}, \mathbf{1})\to
 (\mathbf{\bar{2}}, \mathbf{2}, \mathbf{1})
+(\mathbf{1}, \mathbf{2}, \mathbf{1})
+(\mathbf{\bar{2}}, \mathbf{1}, \mathbf{1})
+(\mathbf{1}, \mathbf{1}, \mathbf{1})$,
$(\mathbf{1}, \mathbf{3}, \mathbf{\bar{3}})\to
 (\mathbf{1}, \mathbf{2}, \mathbf{\bar{2}})
 +(\mathbf{1}, \mathbf{2}, \mathbf{1})
 +(\mathbf{1}, \mathbf{1}, \mathbf{\bar{2}})
 +(\mathbf{1}, \mathbf{1}, \mathbf{1})$
} that can break ${\rm U}(2)^3$, a minimal set that is sufficient to reproduce the quark masses and CKM elements, and at the same time allows for a weak breaking of ${\rm U}(2)^3$ consists of~\cite{Barbieri:2011ci,Barbieri:2012uh}
\begin{equation}
 \mathcal{Y}_u^{SM}=(\mathbf{\bar{2}}, \mathbf{2}, \mathbf{1}),
 \quad
 \mathcal{Y}_d^{SM}=(\mathbf{1}, \mathbf{2}, \mathbf{\bar{2}}),
 \quad
 \mathcal{V}^{SM}=(\mathbf{1}, \mathbf{2}, \mathbf{1}).
\end{equation}
By embedding the ${\rm U}(2)^3$ spurions into the ${\rm U}(3)^3$ spurions $Y_u^{SM}$ and $Y_d^{SM}$, the above discussed MFV LC and RC models can readily be turned into ${\rm U}(2)^3$ LC and RC models.
Writing $Y_u^{SM}$ and $Y_d^{SM}$ as matrices, the embedding reads~\cite{Barbieri:2012tu}
\begin{equation}
 Y_u^{SM}\to
 \begin{pmatrix}
  a_u\,\mathcal{Y}_u^{SM} & b_u\,e^{i\,\phi_u}\,\mathcal{V}^{SM}\\
  0 & c_u
 \end{pmatrix},
 \quad
 Y_d^{SM}\to
 \begin{pmatrix}
  a_d\,\mathcal{Y}_d^{SM} & b_d\,e^{i\,\phi_d}\,\mathcal{V}^{SM}\\
  0 & c_d
 \end{pmatrix},
\end{equation}
where $a_{u,d}$, $b_{u,d}$, $c_{u,d}$ and $\phi_{u,d}$ are real parameters.
So $\mathcal{V}^{SM}$ is embedded into the $(\mathbf{1}, \mathbf{2}, \mathbf{1})$ component of both $Y_u^{SM}$ and $Y_d^{SM}$, and their $(\mathbf{\bar{2}}, \mathbf{1}, \mathbf{1})$ and $(\mathbf{1}, \mathbf{1}, \mathbf{\bar{2}})$ components are set to zero (cf.\ footnote~\ref{fn:fermions_spurion_decomposition}).
The ${\rm U}(2)^3$ models are phenomenologically very interesting in the context of fermion partial compositeness.
On the one hand they suppress large flavor violating effects and can ameliorate the flavor problem of anarchic models, and on the other hand they allow for independent third generation composite-elementary mixings and thus reduce the tensions found in MFV models.

\section{Electroweak symmetry breaking}\label{sec:EWSB}
The main reason for considering CHMs is of course to replace the SM Higgs sector as the source of EWSB.
So after the discussion of the particle content of CHMs in the previous sections, this section is dedicated to the mechanism for breaking the electroweak symmetry.
A virtue of models featuring a strong interaction is that this breaking can be triggered dynamically.
This actually happens in QCD, where the quark condensate breaks the electroweak symmetry and gives mass to the electroweak gauge bosons.
Since this contribution is tiny compared to their actual masses, there has to be another source of EWSB, and in the SM it is provided by the Higgs VEV.
The example from QCD has however inspired the TC models, which are essentially up-scaled versions of QCD and also break the electroweak symmetry via a fermion condensate.
While CHMs are also based on a new strong interaction, the mechanism by which this strong interaction ultimately leads to EWSB is slightly different from that of QCD and traditional TC models.
As already sketched in the beginning of this chapter, CHMs employ the mechanism of symmetry breaking by vacuum misalignment.
If a a global symmetry $G$ is spontaneously broken to a subgroup $H\subset G$ and another subgroup $E\subset G$ is gauged, the gauge group $E$ can be spontaneously broken, depending on the vacuum alignment of $H$.
The multi-site moose models described so far actually already contain this symmetry structure.
The vacuum alignment is determined by a dynamically generated effective potential due to quantum corrections from all particles that couple to the scalar sector.
In particular, this involves gauge bosons and fermions, both elementary as well as composite.
This section first explains the mechanism of symmetry breaking by vacuum misalignment.
It then turns to the problem of calculating the effective one-loop potential and discusses the collective breaking mechanism that can render it finite.


%

\subsection{Vacuum (mis)alignment}\label{sec:EWSB:vac_allign}
A central property of NGBs, as discussed in section~\ref{sec:NGBs}, is their masslessness and the related shift symmetry.
Both stem from the degeneracy of the vacua in the vacuum manifold.
This degeneracy in turn is a consequence of the $G$-invariance of the Lagrangian, which implies that all the points in the vacuum manifold correspond to the same vacuum energy.
Consequently, there is no preference for choosing a specific vacuum.
Since the generators $T^a$ of the unbroken group $H$ are defined such that they leave the specific vacuum invariant,
there is also an ambiguity in choosing the generators $T^a$ among all the generators $S^a$ of $G$.
A different but physically equivalent specific vacuum corresponds to a different set of unbroken and broken generators $T^{\prime a}$ and $X^{\prime a}$.
The shift symmetry reflects this ambiguity: by a constant shift of the NGB fields $\pi^a(x)$, the NGBs parametrized by the broken generators $X^a$ can be turned into NGBs parametrized by different broken generators $X^{\prime a}$.

The situation changes dramatically if there is a term in the Lagrangian that does not treat all of the $S^a$ generators equivalently, which of course in turn implies an explicit breaking of the global $G$ invariance.
This is e.g.\ the case if a subset of the generators $S^a$ is gauged.
Recalling from section~\ref{sec:Vectorres_NLSM_HLS} that each gauged unbroken generator  $T^a$ yields a massless vector boson, while each gauged broken generator $X^a$ yields a massive vector boson, it is obvious that in the presence of gauging, constant shifts of the NGB fields $\pi^a$ could turn a massless vector boson into a massive one and therefore relate physically inequivalent vacua.
But among physically inequivalent vacua, only those with the minimal potential energy are true vacua.
The vacua of the ungauged case are thus divided into true vacua with a minimal potential energy and false vacua with a higher potential energy.
In general, not only gauging of some generators, but any term in the Lagrangian that explicitly breaks the $G$ invariance could divide the vacua of the $G$-invariant case into true and false vacua.
To do this, the explicitly $G$-breaking terms do not have to enter the tree-level potential.
It is in general up to loop corrections to the potential to determine which of the vacua are true and which are false.
The explicit $G$-breaking then has several important consequences:
\begin{itemize}
 \item The shift symmetry into directions that relate true vacua and false vacua is broken.
 Fluctuations around a true vacuum into any of these directions change the potential energy and hence correspond to massive degrees of freedom.
 They are called pseudo NGBs (pNGBs)~\cite{Weinberg:1972fn}.
 \item There might still be an infinite number of true vacua.
 The shift symmetry into directions that relate true vacua among themselves is unbroken.
 Fluctuations around a true vacuum into these directions still correspond to (true) NGBs.
 \item Each pNGB reduces the dimensionality of the vacuum manifold by one.
 Consequently, if there are only pNGBs left, but no true NGBs, the dimensionality of the vacuum manifold is zero, the degeneracy of the different vacua is completely lifted, and there is only a single true vacuum.
 \item The orientation of the true vacua in the directions associated to the pNGBs is fixed.
 This orientation is called {\it orientation of the vacuum} or {\it alignment of the vacuum}~\cite{Preskill:1980mz}.
 If there are no true NGBs left, the orientation of the true vacuum is completely fixed.
 \item The vacuum alignment is not only determined by the tree-level potential, but also by loop-corrections to the potential.
 The tree-level potential might actually be $G$-invariant, such that the vacuum alignment is determined solely by loop contributions.
\end{itemize}

The alignment of the vacuum has interesting effects on the properties of vector bosons that correspond to gauged generators of $G$.
It is instructive to investigate these effects using a concrete example that can be easily visualized: a global symmetry $G={\rm O}(3)$ that is spontaneously broken to its subgroup $H={\rm O}(2)$.
This example is discussed in section~\ref{sec:NGB_LSM}, but without considering $G$-breaking terms and gauging.
In this case, one finds a two-dimensional vacuum manifold parameterizing the degenerate vacua.
For a specific vacuum $\vec{\phi}_0$, the unbroken generator $T_1$ of ${\rm O}(2)$ is defined by $T_1\,\vec{\phi}_0=0$ and the remaining two broken generators $X_{1,2}$ correspond to two NGBs.
As discussed above, the presence of $G$-breaking terms in the Lagrangian divides the degenerate vacua into true and false vacua.
The details of the $G$-breaking terms and the loop induced corrections to the potential are not important for the following discussion, and it will just be assumed that $\vec{\phi}_0$ is a true vacuum.
In addition, it will be assumed that one of the generators of $G$, denoted by $R$, is gauged.
%
%
In this case, there are several possibilities concerning the relative alignment of the vacuum $\vec{\phi}_0$ with the gauged generator $R$.
It is important to note that this relative alignment is always fixed, because different relative alignments correspond to physically different cases.
For the ${\rm O}(3)\to {\rm O}(2)$ spontaneous symmetry breaking with one generator $R$ gauged, the following different cases can be distinguished:
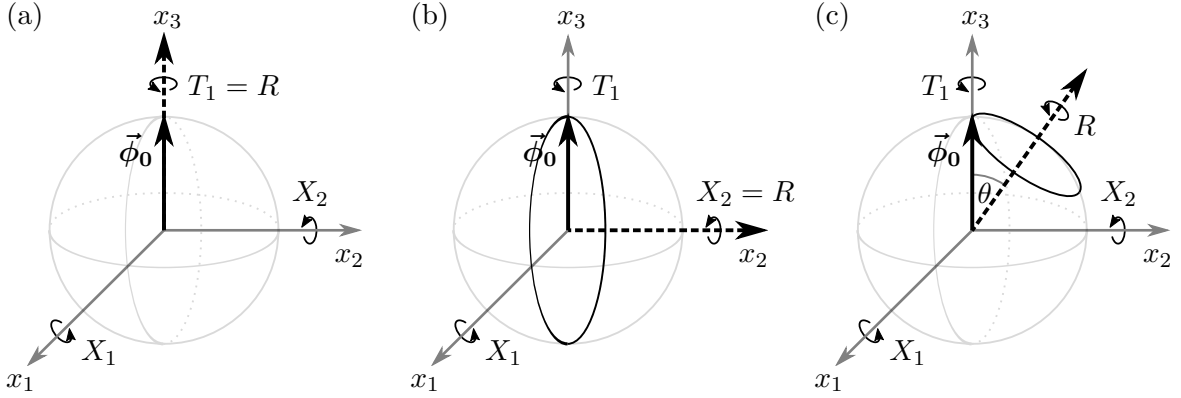
\begin{figure}[t]
\centering
\begin{picture}(132,160)
\put(8,8){\includegraphics[scale=0.9]{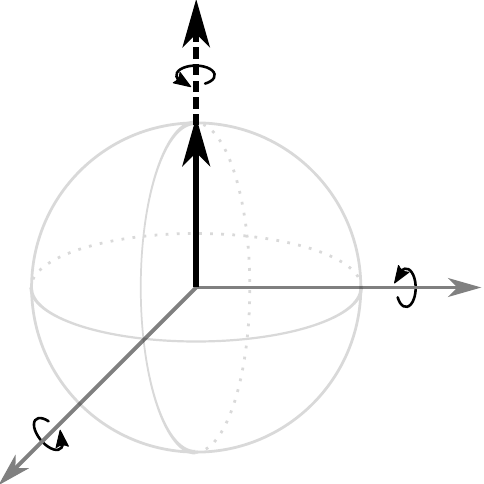}}
\put(0,137){(a)}
\put(0,0){$x_1$}
\put(123,47){$x_2$}
\put(55,137){$x_3$}
\put(28,10){$X_1$}
\put(107,70){$X_2$}
\put(68,110){$T_1=R$}
\put(42,85){$\bm{\vec{\phi}_0}$}
\end{picture}
\hspace{12pt}
\begin{picture}(132,160)
\put(8,8){\includegraphics[scale=0.9]{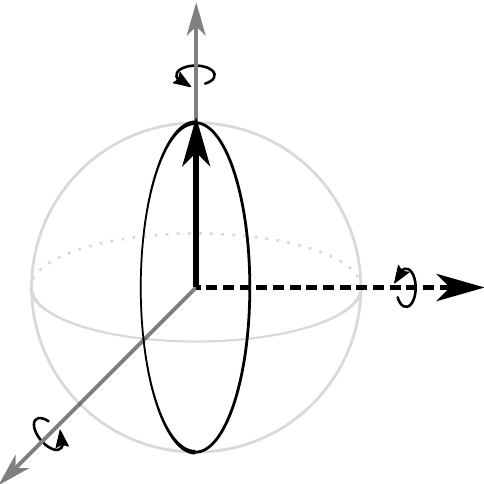}}
\put(0,137){(b)}
\put(0,0){$x_1$}
\put(123,47){$x_2$}
\put(55,137){$x_3$}
\put(28,10){$X_1$}
\put(107,70){$X_2=R$}
\put(68,110){$T_1$}
\put(42,85){$\bm{\vec{\phi}_0}$}
\end{picture}
\hspace{12pt}
\begin{picture}(132,160)
\put(8,8){\includegraphics[scale=0.9]{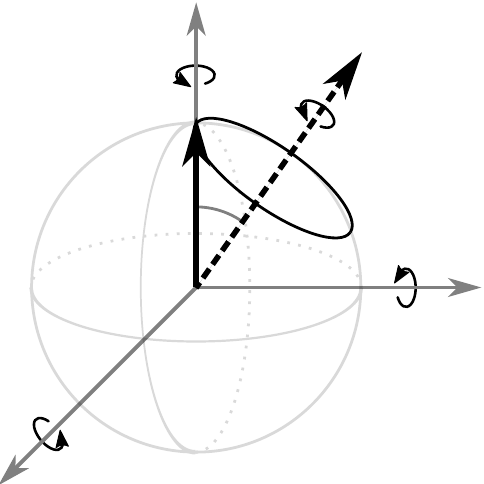}}
\put(0,137){(c)}
\put(0,0){$x_1$}
\put(123,47){$x_2$}
\put(55,137){$x_3$}
\put(28,10){$X_1$}
\put(107,70){$X_2$}
\put(40,110){$T_1$}
\put(97,95){$R$}
\put(42,85){$\bm{\vec{\phi}_0}$}
\put(61,69.2){$\theta$}
\end{picture}
\caption{Different vacuum alignments in ${\rm O}(3)\to {\rm O}(2)$ spontaneous symmetry breaking with one generator $R$ gauged.
(a) Gauging of unbroken generator: $R=T_1$.
(b) Gauging of broken generator: $R=X_2$.
(c) Gauging of linear combination of unbroken and broken generators: $R=\cos(\theta)\,T_1+\sin(\theta)\,X_2$.}
\label{fig:vac_alignment}
\end{figure}
\begin{enumerate}[label=(\alph*)]
\item One possibility is that the gauged generator $R$ is equal to the unbroken generator $T_1$.
This case is visualized in figure~\ref{fig:vac_alignment}a, where the direction of the gauged generator is depicted as a black dashed arrow.
The fact that the gauged generator is aligned with the true vacuum implies that the gauged subgroup of $G$ is not spontaneously broken.
The associated gauge boson $A_\mu\,T_1$ is therefore massless (cf.\ section~\ref{sec:Vectorres_NLSM_HLS}):
\begin{equation}
 m_A^2 = 0.
\end{equation}
This underpins that the relative alignment of the vacuum $\vec{\phi}_0$ with the gauged generator $R$ is indeed fixed, because any misalignment would not yield a massless gauge boson and hence correspond to a physically different case.
Consequently, there is only a single true vacuum and both NGBs associated with the generators $X_{1,2}$ pick up a mass and become pNGBs.
The particle spectrum thus consists of a massless gauge boson and two massive pNGBs.
\item Another possibility is that one of the broken generators is gauged, e.g.\ $R=X_2$.
This case is shown in in figure~\ref{fig:vac_alignment}b.
Because the gauge symmetry must not be explicitly broken, it seems like there is actually an infinite amount of true vacua, depicted in~\ref{fig:vac_alignment}b by a black solid circle perpendicular to the direction of the gauged generator.
While any point in this one-dimensional vacuum manifold corresponds to a true vacuum, there is, however, not an actual degeneracy of different vacua that would yield a NGB.
The reason for this is that the symmetry transformation connecting the different true vacua is a gauge symmetry.
The degeneracy is therefore lifted by any gauge fixing, which then yields only a single true vacuum.
This is equivalent to the statement that the NGB associated with the $X_2$ generator can be gauged away and is actually an unphysical would-be NGB.
This is of course nothing but the Higgs mechanism and the corresponding gauge boson $A_\mu\,X_2$ receives a mass
\begin{equation}
 m_A^2 = \frac{f^2\,g^2}{2},
\end{equation}
where $f=|\vec{\phi}_0|$ and $g$ is the gauge coupling constant (cf.\ section~\ref{sec:Vectorres_NLSM_HLS}).
Transformations induced by the $X_1$ generator on the other hand always relate true and false vacua, and consequently its associated NGB becomes a pNGB.
The particle spectrum in this case consists of a massive gauge boson and one massive pNGB.
\item A last possibility is that not either an unbroken or a broken generator is gauged, but a linear combination of both.
This case is visualized in figure~\ref{fig:vac_alignment}c and can be described with a gauged generator $R$ given by
\begin{equation}
 R = \cos(\theta)\,T_1 + \sin(\theta)\,X_2,
\end{equation}
where $\theta$ is a free parameter specifying the angle between the directions of the generators $T_1$ and $R$.
Like in case (b), the gauge symmetry implies a one-dimensional vacuum manifold prior to gauge fixing, which is again shown as a black solid circle perpendicular to the direction of the gauged generator $R$.
Also like in case (b), fluctuations around the vacuum $\vec{\phi}_0$ that lie inside this vacuum manifold can be parametrized by the generator $X_2$.
Its associated NGB can be gauged away and becomes a would-be NGB.
Again like in the other cases, the NGB associated with $X_1$ becomes a pNGB.
The important novelty, however, is that the gauge boson $A_\mu(\cos(\theta)\,T_1 + \sin(\theta)\,X_2)$ is a linear combination of a massive and a massless one.
Its mass is found to be
\begin{equation}
 m_A^2 = \frac{\sin(\theta)^2\,f^2\,g^2}{2} = \frac{v^2\,g^2}{2},
\end{equation}
such that the mass scale for this vector boson is not given by $f$, but by the effective scale $v = \sin(\theta)\,f$.
This effective scale is just the length of the component of $\vec{\phi}_0$ that is orthogonal to $R$ and coincides with the radius of the true vacuum manifold.
The parameter $\theta$ actually interpolates between the cases~(a) and~(b), which can be recovered for $\theta=0$ and $\theta=\frac{\pi}{2}$, respectively.
In case~(a), the radius of the true vacuum manifold goes to zero and the gauge boson becomes massless.
In case~(b), $\vec{\phi}_0$ is actually orthogonal to $R$ and the vector boson mass scale is simply given by the length of $\vec{\phi}_0$.
The parameter $\theta$ thus measures the misalignment between the true vacuum and the direction of the gauged generator, and the vector boson mass depends on this misalignment.
Apart from the $\theta=0$ case, the particle spectrum is the same as in case~(b), namely a massive gauge boson and a massive pNGB.
\end{enumerate}
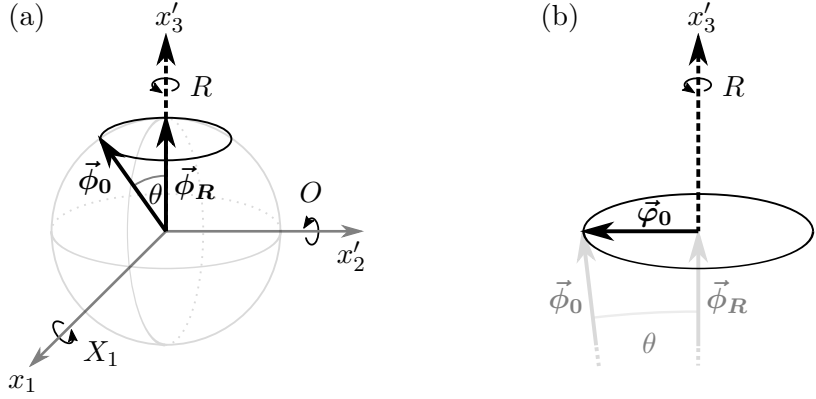
\begin{figure}[t]
\centering
\begin{picture}(132,160)
\put(8,8){\includegraphics[scale=0.9]{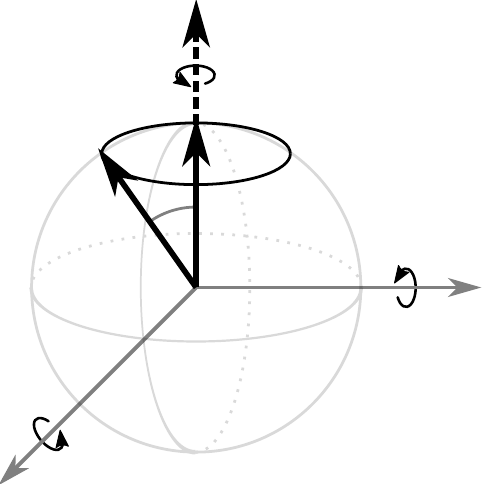}}
\put(0,137){(a)}
\put(0,0){$x_1$}
\put(123,47){$x_2'$}
\put(55,137){$x_3'$}
\put(28,10){$X_1$}
\put(109,70){$O$}
\put(68,110){$R$}
\put(26,75){$\bm{\vec{\phi}_0}$}
\put(62,73){$\bm{\vec{\phi}_R}$}
\put(52.5,69.2){$\theta$}
\end{picture}
\hspace{60pt}
\begin{picture}(132,160)
\put(8,8){\includegraphics[scale=0.9]{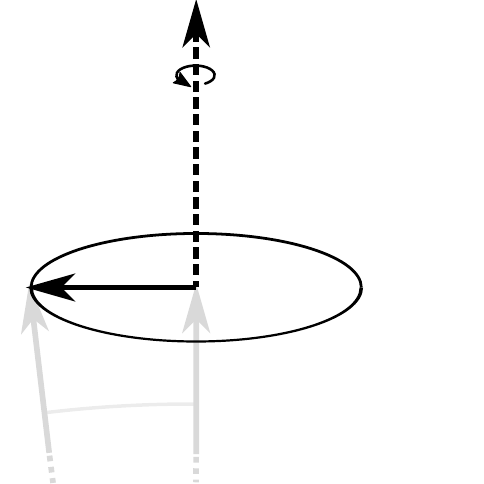}}
\put(0,137){(b)}
\put(55,137){$x_3'$}
\put(68,110){$R$}
\put(36,63){$\bm{\vec{\varphi}_0}$}
\definecolor{light-gray}{rgb}{0.5,0.5,0.5}
\put(3,30){$\color{light-gray} \bm{\vec{\phi}_0}$}
\put(62,30){$\color{light-gray} \bm{\vec{\phi}_R}$}
\put(38,13){$\color{light-gray} \theta$}
\end{picture}
\caption{(a) Vacuum alignment in ${\rm O}(3)\to {\rm O}(2)$ spontaneous symmetry breaking with one generator R gauged. The generator basis contains the gauged generator but not the unbroken generator.
(b) ${\rm O}(2)\to\emptyset$ spontaneous symmetry breaking with a gauged generator.}
\label{fig:vac_alignment2}
\end{figure}
While in the cases (a) and (b), it is natural to choose a basis for the generators of $G$ that consists of the unbroken $T_1$ and the broken $X_{1,2}$, there is another natural choice in case (c):
it might be convenient to replace $T_1$ and $X_2$ by the gauged generator $R$ and an additional generator $O = \cos(\theta)\,X_2 - \sin(\theta)\,T_1$ that is orthogonal to $R$ and $X_1$ (cf.\ figure~\ref{fig:vac_alignment2}a).
This choice of basis has the advantage that it is independent of the actual orientation of the vacuum, because it is already determined by defining which of the generators is gauged.
However, the field $\vec{\phi}(x)$ still parameterizes fluctuations around the true vacuum $\vec{\phi}_0$, and therefore it is obviously not independent of the vacuum alignment.
In a gauge where the $X_2$ would-be NGB is set to zero, $\vec{\phi}(x)$ is given by\footnote{%
It is assumed here that the $\sigma(x)$ field that corresponds to radial excitations is heavy and decouples, cf.\ section~\ref{sec:NGB_LSM}.
}
\begin{equation}\label{eq:vac_align:phi_phi0}
 \vec{\phi}(x)
 =
 e^{i\frac{\sqrt{2}}{f}\,\pi_1(x)\,X_1}\,\vec{\phi}_0.
\end{equation}
It can be made at least formally independent of $\vec{\phi}_0$ by introducing a vector $\vec{\phi}_R$ that is invariant under the gauge transformation and corresponds to the true vacuum in the $\theta=0$ case.
The vectors $\vec{\phi}_0$ and $\vec{\phi}_R$ are related by a rotation around the $x_1$ axis by an angle $\theta$ (cf.\ figure~\ref{fig:vac_alignment2}a), i.e.\
\begin{equation}
 \vec{\phi}_0
 =
 e^{i\,\sqrt{2}\,\theta\,X_1}\,\vec{\phi}_R.
\end{equation}
Plugging this into eq.~(\ref{eq:vac_align:phi_phi0}) and defining the field
\begin{equation}\label{eq:vac_align:varphi}
 \varphi(x) = \theta\,f+\pi_1(x),
\end{equation}
one finds
\begin{equation}\label{eq:vac_align:phi_phiR}
 \vec{\phi}(x)
 =
 e^{i\frac{\sqrt{2}}{f}\,\varphi(x)\,X_1}\,\vec{\phi}_R.
\end{equation}
Intriguingly, the actual dependence on the vacuum alignment is now completely parameterized by a VEV of the field $\varphi(x)$.
The physical pNGB $\pi_1(x)$ corresponds to fluctuations about that VEV.
If the potential of $\vec{\phi}(x)$ is expressed in terms of the parametrization in eq.~(\ref{eq:vac_align:phi_phiR}), the minimization that determines the alignment of the true vacuum can be performed with respect to the VEV of $\varphi(x)$.
This resembles the situation in the linear sigma model discussed in section~\ref{sec:NGB_LSM}, where the VEV of $\vec{\phi}(x)$ determines the vacuum manifold of the ${\rm O}(N)\to {\rm O}(N-1)$ spontaneous symmetry breaking.

The analogy with the linear sigma model can be made even more obvious by considering the case of a small misalignment angle, i.e.\ $\theta\ll1$.
This allows for approximating
\begin{equation}
 \vec{\phi}(x)
 \approx
 \vec{\phi}_R
 +
 \left(\theta\,f+\pi_1(x)\right)\,i\,\frac{\sqrt{2}}{f}\,X_1\,\vec{\phi}_R.
\end{equation}
Noting that for small $\theta$, the effective scale $v$ is linear in $\theta$, i.e.\
\begin{equation}
  v=\sin(\theta)\,f\approx\theta\,f,
\end{equation}
and defining the vector
\begin{equation}
 \vec{\varphi}_0=i\,\sqrt{2}\,\theta\,X_1\,\vec{\phi}_R
 \quad
 {\rm with}
 \quad
 |\vec{\varphi}_0|=\theta\,f\approx v,
\end{equation}
one can express $\vec{\phi}(x)$ as $\vec{\phi}(x)\approx\vec{\phi}_R+\vec{\varphi}(x)$ with (cf.\ figure~\ref{fig:vac_alignment2}b)
\begin{equation}
\vec{\varphi}(x)
=
 \left(1+\pi_1(x)/v\right)\,\vec{\varphi}_0.
\end{equation}
This is now completely analogous to the linear sigma model, eq.~\eqref{eq:LSM_sigma_pi_parametrization}.
The effective scale $v$ plays the role of the VEV of $\vec{\varphi}(x)$, and interestingly, the pNGB $\pi_1(x)$ corresponds to the radial excitation, which is called $\sigma(x)$ in section~\ref{sec:NGB_LSM}.
Because of the chosen gauge, there is no NGB present, but there is a massive vector boson with mass $m_A=\frac{g}{\sqrt{2}}\,v$.
This is just what one would expect in a linear sigma model with a gauged ${\rm O}(2)$ symmetry that is spontaneously broken by a VEV $v$.
So by using the parametrization in terms of $\varphi(x)$, the original ${\rm O}(3)/{\rm O}(2)$ non-linear sigma model with ${\rm O}(3)$-breaking terms and gauging has turned into a model that looks quite different.
It now resembles a linear sigma model describing an ${\rm O}(2)\to\emptyset$ spontaneous symmetry breaking that gives mass to the ${\rm O}(2)$ gauge boson via the Higgs mechanism.
This linear sigma model is of course only an approximation and its viability depends on the smallness of the misalignment angle $\theta$.
However, the parametrization of $\vec{\phi}(x)$ in terms of the field $\varphi(x)$, whose VEV $\theta f$ fixes the vacuum alignment, is also applicable in the case of a large angle $\theta$.

The toy model described above is actually already quite close to the minimal CHM.
To arrive at the latter, one just has to extend the global symmetries to $G={\rm SO}(5)$ and $H={\rm SO}(4)$ and to gauge an ${\rm SU}(2)_{\rm L}\times {\rm U}(1)_{\rm Y}$ subgroup of ${\rm SO}(4)$.
In the case where all generators of ${\rm SU}(2)_{\rm L}\times {\rm U}(1)_{\rm Y}$ are aligned with those of the ${\rm SO}(4)$ group that leaves the true vacuum $\vec{\phi}_0$ invariant, the EW group is unbroken and its four gauge bosons are massless, while the four NGBs of the ${\rm SO}(5)/{\rm SO}(4)$ coset become pNGBs.
This corresponds to the case shown in figure~\ref{fig:vac_alignment}a.
If, on the other hand, the true vacuum is only aligned with the generator of the ${\rm U}(1)_{\rm Q}$ subgroup of ${\rm SU}(2)_{\rm L}\times {\rm U}(1)_{\rm Y}$, three of the four gauge bosons (corresponding to $W_\mu^\pm$ and $Z_\mu$) become massive, three of the four NGBs are unphysical and can be removed by going to unitary gauge, and one NGB becomes a pNGB.
The broken generator associated with this pNGB induces exactly those rotations that relate the true vacuum $\vec{\phi}_0$ with the ${\rm SU}(2)_{\rm L}\times {\rm U}(1)_{\rm Y}$ invariant direction $\vec{\phi}_R$,
analogous to the case shown in figure~\ref{fig:vac_alignment2}a.
Denoting this broken generator by $X_1$, the field $\vec{\phi}(x)$ can be parameterized exactly as in eq.~(\ref{eq:vac_align:phi_phiR}).
The misalignment angle $\theta$ between $\vec{\phi}_0$ and $\vec{\phi}_R$ sets the scale $v=\sin(\theta)\,f$ for the masses of the weak gauge bosons and the VEV $\theta f$ of the scalar field $\varphi(x)$ (cf.\ eq.~\eqref{eq:vac_align:varphi}).
For $\theta\ll 1$ this resembles a linear sigma model in which the ${\rm SU}(2)_{\rm L}\times {\rm U}(1)_{\rm Y}$ symmetry is broken to ${\rm U}(1)_{\rm Q}$ by the VEV $\theta f\approx v$ and in which the weak gauge bosons acquire masses proportional to this VEV.
By keeping $v$ fixed and taking $\theta\to0$, one arrives at a linear sigma model that is nothing but the Higgs sector of the SM, with the pNGB $\pi_1(x)$ becoming the Higgs boson.

Interestingly, this means that the CHMs with EWSB due to vacuum misalignment contain a limiting case, in which for arbitrary small $\theta$ the SM is an arbitrary good approximation.
However, such a small $\theta$ with fixed $\theta f$ also implies an arbitrary large tuning among the different contributions to the potential.
%
In a CHM that should serve as a solution to the naturalness problem, this tuning should not be too large.
Otherwise, a new naturalness problem arises from $\theta$ being an unnatural small number, at least in the absence of some explicit mechanism for making a small angle $\theta$ natural.
So in realistic CHMs, deviations from the linear sigma model due to a finite angle $\theta$ are expected.

In the ${\rm O}(3)/{\rm O}(2)$ toy model as well as the minimal CHM with ${\rm SO}(5)/{\rm SO}(4)$ coset, only a single pNGB is left after the gauge group is spontaneously broken by the vacuum misalignment.
The true vacuum is thus completely determined by the VEV of the field $\varphi(x)$ corresponding to this pNGB, or equivalently by a single misalignment angle $\theta$.
In general, there could be additional pNGBs such that the true vacuum is determined by the VEVs of several fields $\varphi^a(x)$, each corresponding to a misalignment angle $\theta^a$.
This is e.g.\ the case in the next-to-minimal CHM with ${\rm SO}(6)/{\rm SO}(5)$ coset (cf.\ section~\ref{sec:direct_constraints:models}).
For clarity, it is useful to distinguish the scalar fields $\varphi^a(x)$ that develop a VEV from the physical pNGBs $\pi^a(x)$, both being related by
\begin{equation}
 \varphi^a(x) = \theta^a\,f+\pi^a(x).
\end{equation}
This is especially the case in the above discussion.
However, for simplicity, the field that develops a VEV will also be called just $\pi^a(x)$ in the following.
It is understood that in the case of a non-zero VEV, this $\pi^a(x)$ is shifted by
\begin{equation}
  \pi^a(x) \to \left<\pi^a\right> + \pi^a(x),
\end{equation}
where $\left<\pi^a\right> =\theta^a f$ is the VEV of $\pi^a(x)$.

\subsection{The effective potential and collective breaking}\label{sec:EWSB:eff_pot_coll_breaking}

For finding the actual alignment of the vacuum in the presence of $G$-breaking terms, one has to determine the effect of these terms on the scalar potential.
In CHMs, there are contributions from several sources.
The gauging of the electroweak ${\rm SU}(2)_{\rm L}\times {\rm U}(1)_{\rm Y}$ subgroup of $G$ explicitly breaks the $G$ invariance of the Lagrangian, and thus the EW gauge bosons contribute to the scalar potential via quantum corrections.
Via their mixing with the elementary gauge bosons, also the composite spin one resonances contribute.
Another source are the elementary fermions, which do not transform under the full $G$ group.
While they can be formally embedded into multiplets of $G$, this embedding is incomplete and explicitly breaks the $G$ invariance.
Consequently, they contribute to the scalar potential at the loop-level.
Like in the spin one case, the composite fermions also contribute via their mixing with the elementary fields.
Computing all one-loop contributions to the scalar potential is a problem that has been solved in general in~\cite{Coleman:1973jx}.
The result takes a simple form in terms of the generalized mass matrices of the contributing fields $M_i(\pi^a)$ that includes all the couplings to the scalar fields $\pi^a(x)$ and is given by
\begin{equation}
V_{\rm eff}^{\rm 1-loop} =
\sum_{i}
\frac{c_i}{64\,\pi^2}
\left\{
 \tr\left[M_i^4(\pi^a)\,\log\left(M_i^2(\pi^a)\right)\right]
 -\tr\left[M_i^4(\pi^a)\right]\log(\Lambda^2)
 +2\,\tr\left[M_i^2(\pi^a)\right]\Lambda^2
 \right\},
\end{equation}
where the index $i$ runs over all particle species that share a common mass matrix $M_i(\pi^a)$ and $c_i$ is a numerical prefactor accounting for spin and charge multiplicities.
In multi-site moose models, the scalar fields $\pi^a(x)$ correspond to the pNGBs.
The true vacuum is then found by minimizing the above potential with respect to their VEVs.
Plugging the VEVs back into the mass matrices yields, after a diagonalization, all the mass eigenstates and masses of vector bosons and fermions.
In addition, the mass terms for the pNGBs are given by the second derivatives at the minimum of the potential.
In the case of more than one pNGB, they are given in terms of the Hessian matrix at the minimum.
This matrix is in general not diagonal, such that the scalar mass eigenstates can be a mixture of the $\pi^a(x)$ and have to be determined by a diagonalization.

In general, the one-loop potential has a quadratical divergence proportional to $\tr\left[M_i^2(\pi^a)\right]$ and a logarithmic divergence proportional to $\tr\left[M_i^4(\pi^a)\right]$.
These divergences introduce a dependence on the cut-off $\Lambda$ that spoils the predictivity of the model.
However, the actual degrees of divergence of the one-loop contributions crucially depend on the structure of the Lagrangian.
One can use naive dimensional analysis (NDA) to determine the degree of divergence from the loop contributions to a generic operator $\mathcal{O}$.
Following~\cite{Panico:2011pw}, in the multi-site moose models one finds for the leading order Lagrangian\footnote{%
The notation used here is slightly different to the one used in~\cite{Panico:2011pw}. In particular, what is called ``$\eta$'' here is called ``$2\eta$'' there.
}
\begin{equation}
 \mathcal{O}\sim
 \Lambda^2\,f^2\,
 \left( \frac{\Lambda}{4\,\pi\,f} \right)^{2\,L}
 \left( \frac{\pi^a}{f} \right)^{E_\pi}
 \left( \frac{g\,A_\mu}{\Lambda} \right)^{E_A}
 \left( \frac{\psi}{\sqrt{\Lambda}\,f} \right)^{E_\psi}
 \left( \frac{g\,f}{\Lambda} \right)^{\eta}
 \left( \frac{\mu}{\Lambda} \right)^{\chi},
\end{equation}
where $L$ is the number of loops, $E_\pi$ the number of external NGB states, $E_A$ the number of external gauge fields, $d$ the number of derivatives, $\eta$ the number of gauge coupling insertions, and $\chi$ the number of fermion mass term insertions, where {\it mass term} stands for both masses and mass mixings.
The fields $\pi^a$, $A_\mu$, and $\psi$ denote arbitrary NGBs, vector bosons, and fermions, respectively.
Accordingly, $g$ and $\mu$ denote arbitrary gauge couplings and mass terms, respectively.
Considering the contributions to the scalar potential, only terms with no external gauge bosons and fermions are of interest.
Consequently, by setting $L=1$ and $E_A=E_\psi=0$, one finds
\begin{equation}\label{eq:coll_break:NDA_simplified}
 \mathcal{O}\sim
 \frac{\Lambda^4}{16\,\pi^2}
 \left( \frac{\pi^a}{f} \right)^{E_\pi}
 \left( \frac{g\,f}{\Lambda} \right)^{\eta}
 \left( \frac{\mu}{\Lambda} \right)^{\chi}
 \propto
 \Lambda^{4-\eta-\chi},
\end{equation}
such that, obviously, the degree of divergence crucially depends on the number of gauge coupling and mass term insertions $\eta$ and $\chi$.

These mass terms and gauge couplings are the only source of $G$-breaking.
The whole point of the one-loop potential is to communicate this breaking to the scalar NGB sector.
Without $G$-breaking, the NGBs would keep their shift symmetry and a potential would be forbidden at arbitrary loop order.
Therefore, any contribution to the one-loop potential must of course contain mass terms or gauge couplings, the former in the case of fermion loops, the latter in the case of gauge boson loops.
To characterize the actual breaking of the $G$ symmetry by the mass terms and gauge couplings, it is useful to promote them to spurions that formally preserve all global symmetries~\cite{ArkaniHamed:2001nc,Panico:2011pw}.
It is instructive to start with the simplest moose model, which is just the linear sigma model.
In addition, it is convenient to first only consider the gauge bosons and turn to the fermions later.
The spurion corresponding to the gauge coupling $g_E$ is denoted by $\mathcal{G}_{g_E}^a$.
It is an element of the Lie algebra of $G$ and transforms under the adjoint representation, i.e.\ a global $G$ transformation $\mathscr{g}$ yields
\begin{equation}\label{eq:coll_break:spurion_E_transform}
 G: \mathcal{G}_{g_E}^a \to \mathscr{g}\,\mathcal{G}_{g_E}^a\,\mathscr{g}^{-1}.
\end{equation}
When $\mathcal{G}_{g_E}^a$ assumes its background value, i.e.\
\begin{equation}
 \mathcal{G}_{g_E}^a \to g_E\,P^a,
\end{equation}
it only transforms under the adjoint representation of $E$ and thus breaks $G$ to $E$.
The symmetry structure of the model containing the spurion $\mathcal{G}_{g_E}^a$ can be conveniently visualized by the moose diagram
\begin{equation}
\begin{tabular}{c}
\xy
\xymatrix@R=.4pc@C=1.4pc{
\mathrm{Global:}
& G
&&
\\
& *=<20pt>[o][F]{} \doublerightxyarrow^{\mbox{\raisebox{1.5ex}{$U$}}}
&& *=<0pt,20pt>[l][F]{} & *-<0pt,20pt>[l]{H}
\\
{\begin{matrix}[.6]
\mathrm{gauge\ coupling}
 \\
\mathrm{spurion}
\end{matrix}}\mathrm{\ :}
& \mathcal{G}_{g_E}
&&
}
\endxy
\end{tabular}.
\end{equation}
Recalling that $U(x)$ transforms under a global $G$ transformation non-linearly as
\begin{equation}
G: U(x)\to \mathscr{g}\,U(x)\,\mathscr{h}^{-1}(\mathscr{g},x),
\end{equation}
the only type of operator generated at one loop that contains two NGB external states, is invariant under a global $G$ transformation, and depends on the spurion $\mathcal{G}_{g_E}$ is
\begin{equation}
 \mathcal{O}_{\mathcal{G}} \propto \tr\left[U^\dagger\,\mathcal{G}_{g_E}\,\mathcal{G}_{g_E}\,U\,\phi_0\,\phi_0^\dagger\right],
\end{equation}
where $\phi_0$ is an explicit vacuum state invariant under $H$.
From this, one finds $\eta=2$ such that the contribution is quadratically divergent

It is interesting to observe what happens if one adds a first level of resonances.
The resulting two-site moose model can be depicted as
\begin{equation}
\begin{tabular}{c}
\xy
\xymatrix@R=.4pc@C=1.4pc{
\mathrm{Global:}
& G
&& G'
\\
& *=<20pt>[o][F]{} \doublerightxyarrow^{\mbox{\raisebox{1.5ex}{$\Omega$}}}
&& *=<20pt>[o][F]{}
\\
{\begin{matrix}[.6]
\mathrm{gauge\ coupling}
 \\
\mathrm{spurions}
\end{matrix}}\mathrm{\ :}
& \mathcal{G}_{g_E}
&& \mathcal{G}_{g_H}
}
\endxy
\end{tabular}.
\end{equation}
The symmetry of this model is enhanced compared to the non-linear sigma model.
The NGB matrix $\Omega(x)$ transforms under the global symmetry $G\times G'$ as
\begin{equation}
G\times G': \Omega(x)\to \mathscr{g}\,\Omega(x)\,\mathscr{g}'^{-1},
\end{equation}
while the spurion $\mathcal{G}_{g_E}^a$ transforms as in eq.~(\ref{eq:coll_break:spurion_E_transform}), and $\mathcal{G}_{g_H}^a$ transform as
\begin{equation}\label{eq:coll_break:spurion_H_transform}
 G': \mathcal{G}_{g_H}^a \to \mathscr{g}'\,\mathcal{G}_{g_H}^a\,\mathscr{g}'^{-1}.
\end{equation}
Consequently, a one-loop contribution to the scalar potential that is compatible with the global symmetries is given by
\begin{equation}
 \mathcal{O}_{\mathcal{G}} \propto \tr\left[\Omega^\dagger\,\mathcal{G}_{g_E}\,\mathcal{G}_{g_E}\,\Omega\,\mathcal{G}_{g_H}\,\mathcal{G}_{g_H}\right].
\end{equation}
The larger global symmetry obviously has the effect that also the number of necessary gauge coupling insertions is larger, and one finds $\eta=4$.
While this still corresponds to a logarithmic divergence, the quadratical divergence is gone.

The reason for the reduction of the degree of divergence is that the NGB potential is now doubly protected by both the $G$ and the $G'$ symmetry.
Setting any of the two gauge couplings to zero would restore the complete shift symmetry of the NGBs, independently of the other gauge coupling.
Specifically, if $g_E\to 0$ and $g_H$ is finite, the NGBs corresponding to the $H$ generators are unphysical, while the $H$ gauge bosons are massive. But the NGBs in the $G/H$ coset are true massless NGBs.
If, on the other hand, $g_E$ is finite and $g_H\to 0$, the NGBs corresponding to the $E$ generators are unphysical and the $E$ gauge bosons are massive, while the NGBs in the $G/E$ coset remain true massless NGBs.
Consequently, a potential for the NGBs that yields massive pNGBs can only arise if both $g_E$ and $g_H$ are finite.
Thus, the shift symmetry can only be broken collectively by both gauge couplings.
This mechanism, which was first described in~\cite{ArkaniHamed:2001nc}, is therefore called {\it collective breaking}.

Motivated from the result that the quadratical divergence can be avoided by adding one level of resonances, one might be tempted to add another one.
As detailed in section~\ref{sec:vectorres_higher_levels}, this is done by splitting the NGB matrix $\Omega(x)$ into two matrices $\Omega_1(x)$ and $\Omega_2(x)$ that satisfy $\Omega_1(x)\,\Omega_2(x)=\Omega(x)$.
In addition, a new symmetry $G_1\cong G$ is added under which both $\Omega_1(x)$ and $\Omega_2(x)$ transform non-trivially.
The corresponding moose diagram is given by
\begin{equation}
\begin{tabular}{c}
\xy
\xymatrix@R=.4pc@C=1.4pc{
\mathrm{Global:}
& G
&& G_1
&& G'
\\
& *=<20pt>[o][F]{} \doublerightxyarrow^{\mbox{\raisebox{1.5ex}{$\Omega_1$}}}
&& *=<20pt>[o][F]{} \doublerightxyarrow^{\mbox{\raisebox{1.5ex}{$\Omega_2$}}}
&& *=<20pt>[o][F]{}
\\
{\begin{matrix}[.6]
\mathrm{gauge\ coupling}
 \\
\mathrm{spurions}
\end{matrix}}\mathrm{\ :}
& \mathcal{G}_{g_E}
&& \mathcal{G}_{g_1}
&& \mathcal{G}_{g_H}
}
\endxy
\end{tabular}.
\end{equation}
There is, however, a peculiarity that arises from splitting $\Omega(x)$ into two NGB matrices.
Setting $g_1\to 0$ actually makes $\Omega_1(x)$ and $\Omega_2(x)$ transform independently of each other.
The global symmetry $G_1$ therefore becomes a larger accidental ${G_1}_L\times {G_1}_R$ symmetry, which is broken to its diagonal subgroup $G_1$ by the gauging.
To take this into account, it is useful to make the larger symmetry manifest by actually introducing two different spurions $\mathcal{G}_{{g_1}_L}$ and $\mathcal{G}_{{g_1}_R}$ that both assume the same background value (cf.~\cite{Panico:2011pw}).
The moose diagram thus reads\footnote{%
This is analogous to the moose diagram in eq.~(\ref{eq:Vector_two_omega_split_moose}).
}
\begin{equation}
\begin{tabular}{c}
\xy
\xymatrix@R=.4pc@C=1.4pc{
\mathrm{Global:}
& G
&& {G_1}_L
& {G_1}_R
&& G'
\\
& *=<20pt>[o][F]{} \doublerightxyarrow^{\mbox{\raisebox{1.5ex}{$\Omega_1$}}}
&& *=<20pt>[o][F]{}
& *=<20pt>[o][F]{} \doublerightxyarrow^{\mbox{\raisebox{1.5ex}{$\Omega_2$}}}
&& *=<20pt>[o][F]{}
\\
{\begin{matrix}[.6]
\mathrm{gauge\ coupling}
 \\
\mathrm{spurions}
\end{matrix}}\mathrm{\ :}
& \mathcal{G}_{g_E}
&& \mathcal{G}_{{g_1}_L}
& \mathcal{G}_{{g_1}_R}
&& \mathcal{G}_{g_H}
\save "2,4"-(5,5);"2,5"+(5,5) **\frm{--}  \restore
}
\endxy
\end{tabular},
\end{equation}
where the accidental symmetry is explicitly shown.
The transformation properties can readily be read off from this diagram; it is now obvious that $\Omega_1(x)$ and $\Omega_2(x)$ transform independently.
Because it depends on the chosen gauge which of the two NGB matrices contains the actual NGBs, the one-loop contribution to the scalar potential has to depend on both $\Omega_1(x)$ and $\Omega_2(x)$.
Thus, one finds operators like
\begin{equation}
 \mathcal{O}_{\mathcal{G}} \propto \tr\left[\Omega_1^\dagger\,\mathcal{G}_{g_E}\,\mathcal{G}_{g_E}\,\Omega_1\,\mathcal{G}_{{g_1}_L}\,\mathcal{G}_{{g_1}_L}\right]
 \cdot
 \tr\left[\Omega_2^\dagger\,\mathcal{G}_{{g_1}_R}\,\mathcal{G}_{{g_1}_R}\,\Omega_2\,\mathcal{G}_{g_H}\,\mathcal{G}_{g_H}
 \right].
\end{equation}
The number of gauge coupling insertions is $\eta=8$.
Consequently, the gauge boson contributions to the potential are not only finite at one loop, but even finite at two loops.
Apparently, the splitting of $\Omega(x)$ leads to an additional increase of $\eta$, even larger compared to the introduction of the first level of resonances.
This can be understood by the emergence of the accidental symmetry that serves as an additional protection of the scalar potential.

Since $\eta=6$ would be sufficient for a model to feature finite gauge boson contributions to the one-loop potential, it is possible to employ a two-site model in which the $H$ gauge bosons are removed by formally taking $g_H\to\infty$.
The corresponding moose diagram is given by
\begin{equation}
\begin{tabular}{c}
\xy
\xymatrix@R=.4pc@C=1.4pc{
\mathrm{Global:}
& G
&& {G_1}_L
& {G_1}_R
&&
\\
& *=<20pt>[o][F]{} \doublerightxyarrow^{\mbox{\raisebox{1.5ex}{$\Omega_1$}}}
&& *=<20pt>[o][F]{}
& *=<20pt>[o][F]{} \doublerightxyarrow^{\mbox{\raisebox{1.5ex}{$U_2$}}}
&& *=<0pt,20pt>[l][F]{} & *-<0pt,20pt>[l]{H}
\\
{\begin{matrix}[.6]
\mathrm{gauge\ coupling}
 \\
\mathrm{spurions}
\end{matrix}}\mathrm{\ :}
& \mathcal{G}_{g_E}
&& \mathcal{G}_{{g_1}_L}
& \mathcal{G}_{{g_1}_R}
&&
\save "2,4"-(5,5);"2,5"+(5,5) **\frm{--}  \restore
}
\endxy
\end{tabular},
\end{equation}
where the NGB matrix $\Omega_2(x)=U_2(x)\,\Xi_2(x)$ has been turned into $U_2(x)$ by employing a hidden local symmetry transformation.
Similar to the first example featuring the non-linear sigma model, $U_2(x)$ transforms non-linearly under the ${G_1}_R$ symmetry.
By using an explicit vacuum state invariant under $H$, one finds one-loop contributions to the potential like
\begin{equation}
 \mathcal{O}_{\mathcal{G}} \propto \tr\left[\Omega_1^\dagger\,\mathcal{G}_{g_E}\,\mathcal{G}_{g_E}\,\Omega_1\,\mathcal{G}_{{g_1}_L}\,\mathcal{G}_{{g_1}_L}\right]
 \cdot
 \tr\left[U_2^\dagger\,\mathcal{G}_{{g_1}_R}\,\mathcal{G}_{{g_1}_R}\,U_2\,\phi_0\,\phi_0^\dagger
 \right].
\end{equation}
As expected, this corresponds to $\eta=6$, and thus the one-loop gauge boson contributions to the scalar potential are finite.
The virtue of this model is that the finiteness is already achieved with only two sites, i.e.\ a single level of resonances.
Furthermore, this is the kind of model for which the inclusion of fermions is discussed in section~\ref{sec:Fermions}.

To employ the spurion method used above also for analyzing the fermion contributions to the scalar potential, the fermion masses and mixings have to be promoted to spurions too.
Considering the two-site moose from section~\ref{sec:fermion_moose}, eq.~(\ref{eq:fermion_moose:two-site_moose}), one has to deal with two composite-elementary mixings $\Delta_L$ and $\Delta_R$, two composite fermion masses $m_L$ and $m_R$, the composite sector Yukawa coupling $Y_{\rm comp}$, as well as the mass term $m_Y$ that couples the same composite fermions as the Yukawa coupling but without involving the NGBs.
Thus, one introduces the spurions $\mathcal{M}_{\Delta_L}$,  $\mathcal{M}_{\Delta_R}$,  $\mathcal{M}_{m_L}$,  $\mathcal{M}_{m_R}$, and $\mathcal{M}_{Y}$.
Their transformation properties are chosen such that the Lagrangian has the same symmetries it would have when setting the fermion masses and mixings to zero.
In this case, the accidental symmetries also contain the SM group $G_{\rm SM}$.
This can be depicted by the moose diagram
\begin{equation}
\begin{tabular}{c}
\xy
\xymatrix@R=.4pc@C=1.4pc{
\mathrm{Global:}
& G_{\rm SM}
& G
&& {G_1}_L
& {G_1}_R
&&
\\
& *=<20pt>[o][F]{}
& *=<20pt>[o][F]{} \doublerightxyarrow^{\mbox{\raisebox{1.5ex}{$\Omega_1$}}}
&& *=<20pt>[o][F]{}
& *=<20pt>[o][F]{} \doublerightxyarrow^{\mbox{\raisebox{1.5ex}{$U_2$}}}
&& *=<0pt,20pt>[l][F]{} & *-<0pt,20pt>[l]{H}
\\
{\vphantom{
{\begin{matrix}[1.5]
. \\ .
\end{matrix}}
}}
{\begin{matrix}[.6]
\mathrm{mass\ term}
 \\
\mathrm{spurions}
\end{matrix}}\mathrm{\ :}
&\ar@{}[r]|{\mbox{\raisebox{0.0ex}{$
{\begin{matrix}[1.5]
 \mathcal{M}_{\Delta_L}
 \\
 \mathcal{M}_{\Delta_R}
\end{matrix}}
$}}}
&&&\ar@{}[r]|{\mbox{\raisebox{0.0ex}{$
{\left.\begin{matrix}[1.5]
\mathcal{M}_{m_L}
 \\
\mathcal{M}_{m_R}
\end{matrix}\ \ \hspace{40pt}\right>\,\mathcal{M}_{Y}}
\hspace{-40pt}
\hspace{-50pt}
$}}}
&&&
\save "2,5"-(5,5);"2,6"+(5,5) **\frm{--}  \restore
\save "2,2"-(5,5);"2,3"+(5,5) **\frm{--}  \restore
}
\endxy
\end{tabular}.
\end{equation}
While the elementary fields $\xi_{L}$ and $\xi_{R}$ correspond to incomplete $G$-multiplets, they transform properly under the SM group $G_{\rm SM}$.
Consequently, the spurions $\mathcal{M}_{\Delta_L}$ and $\mathcal{M}_{\Delta_R}$, which correspond to the composite-elementary mixings, have to transform under the accidental symmetry $G_{\rm SM}\times G$ as
\begin{equation}
 G_{\rm SM}\times G : \mathcal{M}_{\Delta_{L,R}} \to \mathscr{g}_{{\rm SM}_{L,R}}\,\mathcal{M}_{\Delta_{L,R}}\,\mathscr{g}^{-1},
\end{equation}
where $\mathscr{g}_{{\rm SM}_L}$ and $\mathscr{g}_{{\rm SM}_R}$ are the transformations under which $\xi_{L}$ and $\xi_{R}$ transform, respectively\footnote{%
$\xi_{L}$ contains ${\rm SU}(2)_{\rm L}$ doublets, while $\xi_{R}$ contains singlets; so they transform differently.
}.
This allows for making the composite-elementary mixing terms in the Lagrangian formally invariant under $G_{\rm SM}\times G$. They read
\begin{equation}
 \mathcal{L}\supset
 \bar{\xi}_L(x)\,\mathcal{M}_{\Delta_{L}}\,\Omega_1(x)\,\Psi_R(x)
 +
 \bar{\xi}_R(x)\,\mathcal{M}_{\Delta_{R}}\,\Omega_1(x)\,\widetilde{\Psi}_L(x),
\end{equation}
and they take their usual form when the spurions assume their background values
\begin{equation}
 \mathcal{M}_{\Delta_{L,R}}\to \Delta_{L,R}.
\end{equation}
This then breaks $G_{\rm SM}\times G$ to its diagonal subgroup, which is isomorphic to $G_{\rm SM}$, and thus explicitly breaks $G$.
Similarly, the spurions $\mathcal{M}_{m_{L,R}}$ transform under the accidental symmetry ${G_1}_L \times {G_1}_R$ and break it to $G_1$ when they assume their background values
\begin{equation}
 \mathcal{M}_{m_{L,R}}\to m_{L,R}.
\end{equation}
The last spurion $\mathcal{M}_Y$ is special in the sense that it actually transforms non-linearly under $G$, or equivalently, it transforms under the hidden local symmetry, i.e.\
\begin{equation}
 G : \mathcal{M}_Y \to \mathscr{h}(\mathscr{g},x)\,\mathcal{M}_Y\,\mathscr{h}^{-1}(\mathscr{g},x).
\end{equation}
It enters the fermion Lagrangian as
\begin{equation}
 \mathcal{L}\supset
 -
 \bar{\Psi}_L(x)\,U_2(x)\,
 \mathcal{M}_Y\,
 U_2^\dagger(x)\,\widetilde{\Psi}_R(x),
\end{equation}
which is manifestly invariant under the global symmetries even though $U_2(x)$ transforms non-linearly.
The spurion $\mathcal{M}_Y$ has the background value
\begin{equation}
 \mathcal{M}_Y\to Y_{\rm comp}\,\phi_0\,\phi_0^\dagger+m_Y.
\end{equation}
It is actually invariant under a hidden local symmetry transformation.
Therefore, no global symmetry is broken when $\mathcal{M}_Y$ assumes its background value.
However, it is still convenient to treat all mass and mixing terms on an equal footing by introducing spurions for all of them.

Having detailed the spurions for the fermionic Lagrangian and their transformation properties, one can now determine the degree of divergence of the fermion contributions to the scalar potential.
The operators that are generated at one loop take the form
\begin{equation}
\begin{aligned}
 \mathcal{O}_{\mathcal{M}_L} &\propto \tr\left[
 \mathcal{M}_{\Delta_{L}}\,\Omega_1\,\mathcal{M}_{m_{L}}\,U_2\,\mathcal{M}_Y\,
 \mathcal{M}_Y^\dagger\,U_2^\dagger\,\mathcal{M}_{m_{L}}^\dagger\,\Omega_1^\dagger\,\mathcal{M}_{\Delta_{L}}^\dagger
 \right],
 \\
 \mathcal{O}_{\mathcal{M}_R} &\propto \tr\left[
 \mathcal{M}_{\Delta_{R}}\,\Omega_1\,\mathcal{M}_{m_{R}}\,U_2\,\mathcal{M}_Y\,
 \mathcal{M}_Y^\dagger\,U_2^\dagger\,\mathcal{M}_{m_{R}}^\dagger\,\Omega_1^\dagger\,\mathcal{M}_{\Delta_{R}}^\dagger
 \right].
\end{aligned}
\end{equation}
Counting the number of mass term insertions, one finds $\chi=6$.
The one-loop fermion contributions to the scalar potential are therefore finite (cf.\ eq.~(\ref{eq:coll_break:NDA_simplified})).
With finite gauge boson and fermion contributions, the one-loop potential of the above two-site model takes the simple form
\begin{equation}\label{eq:EWSB:potential:finite}
V_{\rm eff}^{\rm 1-loop} =
\sum_{i}
\frac{c_i}{64\,\pi^2}\,
 \tr\left[M_i^4(\pi^a)\,\log\left(M_i^2(\pi^a)\right)\right].
\end{equation}
For the finiteness of the fermion contributions, it is crucial that the mass terms and mixings have exactly the form as described in section~\ref{sec:fermion_moose}, i.e.\ that only nearest-neighbor interactions are present.
While the symmetries would in general also allow other terms, this would spoil the finiteness of the one-loop potential of the two-site model.
However, by employing dimensional deconstruction, the nearest-neighbor interactions automatically arise; they are guaranteed by locality in the extra dimension.
This property of a Lagrangian, having only nearest-neighbor interactions, is thus also called {\it locality in theory space} (cf.\ e.g.~\cite{Kahn:2012as}).
It can be seen as the origin of collective breaking in the above discussed models.

\section{A UV completion: fundamental partial compositeness}\label{sec:FPC}

The central ingredients of the phenomenological multi-site CHMs are a pNGB Higgs and partial compositeness.
These models allow to solve some of the problems of traditional TC models and are found to be in good accordance with experimental data (cf.\ chapter~\ref{chap:direct_constraints}).
While their structure is mainly inspired by models with extra dimensions, the idea of a pNGB Higgs as well as partial compositeness are deeply rooted in TC-like 4D strongly coupled gauge theories.
In view of this, it is an interesting question if it is actually possible to construct a UV completion in terms of a strongly coupled 4D quantum field theory that incorporates both a pNGB Higgs and partial compositeness.

A first hurdle is the symmetry structure.
Any UV completion is required to break a global symmetry $G$ spontaneously to a subgroup $H$.
While this can be realized by dynamical chiral symmetry breaking, there are some non-trivial requirements on $G$ and $H$ as explained in the beginning of this chapter.
For convenience, these requirements are listed here again:
\begin{enumerate}
 \item ${\rm SU}(2)_{\rm L}\times {\rm SU}(2)_{\rm R}\cong {\rm SO}(4)$ is a subgroup of $H$.
 \item The $G$ to $H$ breaking yields NGBs in a complex ${\rm SU}(2)_{\rm L}$ doublet with ${\rm U}(1)_{\rm Y}$ charge $q_{\rm Y}=1/2$.
\end{enumerate}
To find a possible candidate, it is useful to consider general breaking patterns that can be realized by chiral symmetry breaking from a strongly coupled TC like theory.
One finds~\cite{Peskin:1980gc,Preskill:1980mz,Kosower:1984aw} (see also \cite{Cacciapaglia:2014uja} and references therein):
\begin{enumerate}
 \item ${\rm SU}(M)_{\rm L}\times {\rm SU}(M)_{\rm R}\to {\rm SU}(M)_{{\rm L}+{\rm R}}$, which requires TC fermions in a complex representation of the TC gauge group, e.g.\ the fundamental of an ${\rm SU}(N)$ gauge group.
 This is analogous to chiral symmetry breaking in QCD.
 \item ${\rm SU}(M)\to {\rm Sp}(M)$, which requires TC fermions in a pseudoreal representation of the TC gauge group, e.g.\ the fundamental of an ${\rm Sp}(N)$ gauge group.
 \item ${\rm SU}(M)\to {\rm SO}(M)$, which requires TC fermions in a real representation of the TC gauge group, e.g.\ the fundamental of an ${\rm SO}(N)$ gauge group.
\end{enumerate}
The next-to-minimal CHM is actually found to be among these cases.
Noting that ${\rm SU}(4)\cong {\rm SO}(6)$ and ${\rm Sp}(4)\cong {\rm SO}(5)$, the ${\rm SO}(6)\to {\rm SO}(5)$ breaking pattern is equivalent to ${\rm SU}(4)\to {\rm Sp}(4)$ and can be realized with an ${\rm Sp}(N)$ gauge group.
This breaking pattern was actually used in one of the very first CHMs described in the 1980s~\cite{Kaplan:1983sm},
but has also been discussed more recently in~\cite{Katz:2005au,Galloway:2010bp,Barnard:2013zea,Ferretti:2013kya,Cacciapaglia:2014uja,Sannino:2016sfx,Galloway:2016fuo,Agugliaro:2016clv}.
While any other symmetry breaking pattern necessarily yields more pNGBs, this is not a problem per se.
At least, as long as the pNGBs only couple to SM particles very weakly or are considerably heavier than the Higgs this does not impose strong experimental constraints on a given model (cf.\ also section~\ref{sec:direct_constraints:results:eta}).

Apparently, it is possible to overcome the first hurdle, so the next step would be to find a way for including partial compositeness.
In purely fermionic constructions, this necessarily requires bound states with the same quantum numbers as the SM fields that are composed of only fundamental fermions.
Models which could yield these bound states from fermion trilinears are discussed in~\cite{Barnard:2013zea,Ferretti:2013kya,Vecchi:2015fma}.
As motivated in section~\ref{sec:Fermions:partial_compositeness}, the fermionic operators $\mathcal{O}_F$ that mix with the SM fermions are required to have a scaling dimension ${\dim[\mathcal{O}_F]\approx 5/2}$.
Fermion trilinears have a canonical dimension of $9/2$ and therefore must have a large anomalous dimension.
This seems to be not possible in the cases explored so far~\cite{DeGrand:2015yna,Pica:2016rmv}.

However, the desired scaling dimension of $5/2$ is exactly the canonical dimension of a bound state formed by a TC fermion and a TC scalar.
A framework of fundamental partial compositeness (FPC) in which the composite fermionic operators $\mathcal{O}_F$ consist of these so called techniscalars and technifermions has been proposed in~\cite{Sannino:2016sfx}\footnote{%
The idea of fermionic bound states composed of strongly coupled scalars and fermions is much older, but it has been mainly considered in models of composite SM fermions, cf.\ e.g.~\cite{Greenberg:1980ri,Barbieri:1981cy,Fritzsch:1981zh,Casalbuoni:1981nd,Gerard:1981ep,Schrempp:1983qd}.
}.
It is argued that any model that contains fermionic operators of scaling dimension $5/2$ should behave as if these operators are made of a fermion and a scalar.
At higher scales, the scalars might themselves be composite bound states (see also~\cite{Cacciapaglia:2017cdi}).
This would also be a solution to the apparent new naturalness problem that arises from fundamental techniscalars.
Another one would be to supersymmetrize the TC theory (cf.~\cite{Antola:2010nt}).

\subsection{Minimal fundamental partial compositeness}
There are many possibilities to construct models out of technifermions and techniscalars that, confined by a new strong TC force, yield a composite pNGB Higgs and fermionic bound states that mix with SM fermions.
This is even the case if one considers techniscalars and technifermions only transforming under the fundamental representation of the TC gauge group (see~\cite{Sannino:2016sfx} for a classification  of economical models).
However, among these models is a minimal one that actually implements the symmetry structure of the next-to-minimal CHM: it contains a pNGB Higgs that arises from a global $\SU{4}$ symmetry being spontaneously broken to $\Sp{4}$ by a technifermion condensate.
This minimal fundamental partial compositeness (MFPC) model was proposed in~\cite{Sannino:2016sfx}.
It was further analyzed in~\cite{Cacciapaglia:2017cdi} from an effective field theory (EFT) perspective.
Its full flavor structure and its consequences for flavor physics have been worked out in~\cite{Sannino:2017utc} and are discussed in detail in chapter~\ref{chap:Flavor_MFPC}.

In addition to the SM fields, the model contains techni\-fermions and techni\-scalars, both transforming under the fundamental, pseudoreal representation of the new $G_{\rm TC}={\rm Sp}(N_{\rm TC})$ gauge group.
For the techniscalars and technifermions to form bound states that have the quantum numbers of the SM fermions, they themselves have to be charged under the SM gauge group.
To get a pNGB Higgs as a bound state of technifermions, they have to transform under a global $\SU{4}$ symmetry into which the EW gauge group is embedded.
This can be realized by considering four technifermions\footnote{%
All fermion fields used in this section are left-handed two-component Weyl spinors.
} that form the four-plet
\begin{equation}\label{eq:MFPC:F_fourplet}
\F = \begin{pmatrix}
      \Fup & \Fdn & \bar{\Fup} & \bar{\Fdn}
     \end{pmatrix}\transpose,
\end{equation}
where the first two components transform under $\SUs{2}{L}$ as a doublet (cf.\ table~\ref{tab:FPC:quantum_numbers}):
\begin{equation}
 \Fupdn=\begin{pmatrix}
	\Fup\\
	\Fdn
      \end{pmatrix}.
\end{equation}
To get composite partners for all three generations of SM fermions, one also needs three generations of either technifermions or techniscalars.
To restrict the global symmetry under which the technifermions transform to $\SU{4}$, there can only be one generation of technifermions, such that the techniscalars have to come in three generations.
In addition, also the QCD charge has to be carried by the techniscalars.
The most economical choice is to introduce, for each generation, two techniscalars $\S_q$ and  $\S_l$, where the former is a QCD anti-triplet and the latter is a QCD singlet (cf.\ table~\ref{tab:FPC:quantum_numbers}).
They can be embedded into the 12-plet
\begin{equation}\label{eq:MFPC:S_12plet}
 \S = \begin{pmatrix}
       \S_q\\
       \S_l
      \end{pmatrix},
\end{equation}
where the generation indices are implicit.
\begin{table}[t]
\centering
\renewcommand{\arraystretch}{1}
\setlength{\tabcolsep}{5pt}
\begin{tabular}{c|R{20pt}R{20pt}R{20pt}R{20pt}R{20pt}R{20pt}R{5pt}|R{20pt}R{20pt}R{20pt}R{5pt}|R{20pt}R{20pt}R{5pt}}
		  & $Q$	& $\bar{u}$	& $\bar{d}$	& $L$	& $\bar{\nu}$	& $\bar{e}$ && $\Fpm$ & $\bar{\Fm}$ & $\bar{\Fp}$ && $\Sq$ & $\Sl$ &\\
\hline
\hline
Sp$(N)_\text{TC}$
& $\bf{1}$	  & $\bf{1}$ & $\bf{1}$ & $\bf{1}$ & $\bf{1}$ & $\bf{1}$ && $\bf{N}$ & $\bf{N}$ & $\bf{N}$ && $\bf{N}$ & $\bf{N}$ &\\
SU$(3)_\text{C}$  & $\bf{3}$ 	  & $\bf{\bar{3}}$ & $\bf{\bar{3}}$ & $\bf{1}$ & $\bf{1}$ & $\bf{1}$ && $\bf{1}$ & $\bf{1}$ & $\bf{1}$ && $\bf{\bar{3}}$ & $\bf{1}$ &\\
SU$(2)_\text{L}$  & $\bf{2}$ 	  & $\bf{1}$	   & $\bf{1}$	    & $\bf{2}$ & $\bf{1}$ & $\bf{1}$ && $\bf{2}$ & $\bf{1}$ &
$\bf{1}$ &&  $\bf{1}$ & $\bf{1}$ &\\
U$(1)_\text{Y}$   & $\frac{1}{6}$ & $-\frac{2}{3}$ & $\frac{1}{3}$  & $-\frac{1}{2}$ & $0$ & $1$ && $0$ & $-\frac{1}{2}$ & $\frac{1}{2}$ && $-\frac{1}{6}$ & $\frac{1}{2}$ & \\
$N_{\rm g}$ 		  & 3  & 3 & 3 & 3 & 3 & 3 && 1 & 1 & 1 && 3 & 3 &\\
\end{tabular}
\caption{Quantum numbers of SM fields, TC fermions, and TC scalars in MFPC.
The last line gives the number of generations $N_{\rm g}$. All fermion fields are left-handed Weyl spinors.}
\label{tab:FPC:quantum_numbers}
\end{table}
In terms of the fields $\F$  and $\S$, the kinetic terms of the TC sector can be written as
\begin{equation}
\mathcal{L}_{\mathrm{kin}}^\TC =
-\tfrac{1}{4}\, \tr\left[\mathcal{G}_{\mu\nu} \mathcal{G}^{\mu\nu}\right]
+ i \tcf^\dagger \bar{\sigma}^\mu D_\mu \tcf
- \left(\tfrac{1}{2} \tcf\transpose m_\tcf\, \epsilon_\TC\, \tcf \hc\right)
+ \left(D_\mu \tcs\right)^{\dagger} \left(D^{\mu} \tcs\right)
-  \tcs^{\dagger} m^2_\tcs\, \tcs,
\label{eq:L_kin}
\end{equation}
where $\mathcal{G}_{\mu\nu}$ denotes the TC gauge bosons' field strength tensor, $m_\tcf$ and $m_\tcs$ are the technifermion and techniscalar mass matrices, and $\epsilon_\TC$ is the antisymmetric invariant tensor of $G_\TC$.

In the absence of the mass term $m_\tcf$, the TC sector has a global $\SU{4}$ symmetry, under which $\F$ transforms in the fundamental representation.
Because only the techni\-fermions transform non-trivially under this symmetry group, it will also be denoted by $\SU{4}_\F$ in the following.
In the case of a trivial mass matrix $m_\S$, one would naively expect that the 12 complex scalars have a global $\SU{12}_\S$ symmetry.
However, because the techniscalars transform under a pseudoreal representation of the TC gauge group, the TC sector actually has an accidental $\Sp{24}_\S$ symmetry (cf.~\cite{Sannino:2016sfx}).
This symmetry can be made manifest by arranging the techniscalars in terms of the field
\begin{equation}\label{eq:MFPC:Phi_24plet}
 \Phi = \begin{pmatrix}
 \tcs \\
 -\epsilon_\TC\, \tcs^{\ast}
 \end{pmatrix},
\end{equation}
which transforms under the fundamental representation of $\Sp{24}_\S$.

\subsubsection{The pNGB Higgs}

The full global symmetry of the TC sector is $\SU{4}_\F\times \Sp{24}_\S$.
However, the strong TC interactions break the $\SU{4}_\F$ symmetry to $\Sp{4}_\F$ by forming the fermion bilinear condensate\footnote{%
In the absence of the techniscalars, this has been shown by lattice simulations for $N_\TC=2$~\cite{Lewis:2011zb}.
}
\begin{equation}\label{eq:MFPC:fermion_condensate}
\left<\tcf^{a} \epsilon_\TC\, \tcf^{b} \right>
=
\Lambda_\TC\, f_\TC^2\, \Sigma_{\theta}^{a b},
\end{equation}
where $a,b$ are $\SU{4}_\F$ indices, $\Lambda_\TC$ is the composite scale of the TC interaction, ${f_\TC\approx 4\,\pi\,\Lambda_\TC}$ is the NGB decay constant associated with the spontaneous symmetry breaking, and $\Sigma_{\theta}^{a b}$ is an antisymmetric matrix that specifies the vacuum alignment of the unbroken $\Sp{4}_\F$ group.
This vacuum alignment can be parameterized by the angle $\theta$ and is chosen such that $\theta=0$ leaves the EW symmetry unbroken\footnote{\label{fn:FPC:eta}%
Since the $\SU{4}/\Sp{4}$ coset contains two NGBs that cannot be gauged away by an EW gauge transformations, there are in principle two angles that determine the vacuum alignment.
However, any non-zero value for the second angle would break $C\!P$ spontaneously.
In this section, this second angle is assumed to be zero.
For an analysis of a model where it is explicitly allowed to be non-zero, see chapter~\ref{chap:direct_constraints}.
}.
The NGBs arising from the $\SU{4}_\F\to\Sp{4}_\F$ breaking
are parametrized by the NGB matrix
\begin{equation}\label{eq:MFPC:NGB_matrix}
\Sigma(x) = \exp\left[i \frac{2\sqrt{2}}{f_\TC}\,\pi^a(x)\,X^a_\theta \right] \Sigma_\theta.
\end{equation}
The NGB fields $\pi^a(x)$, $a\in\{1,2,3,4,5\}$ correspond to the fluctuations around the true vacuum $\Sigma_{\theta}$.
Hence, the misalignment is not parameterized by a pNGB VEV and the generators $X^a_\theta$ depend on the misalignment angle $\theta$
(cf.~\cite{Cacciapaglia:2014uja} and the discussion on different generator bases in section~\ref{sec:EWSB:vac_allign}).
Since $\SU{4}/\Sp{4}$ is a symmetric space, the leading order EFT Lagrangian for the NGBs is given by the simple form (cf. section~\ref{sec:symmetric_spaces})\footnote{%
While the NGB matrix used here is analogous to the matrix $\Sigma(x)$ introduced in section~\ref{sec:symmetric_spaces}, the normalization  of the generators is different.
While the $X^a$ in section~\ref{sec:symmetric_spaces} are normalized by $\tr\left[X^a\,X^b\right] = \delta^{ab}$, the $X^a_\theta$ used here satisfy $\tr\left[X^a_\theta\,X^b_\theta\right] = \frac{1}{2}\,\delta^{ab}$.
This is the reason for the prefactor $\frac{f^3_\TC}{8}$ in eq.~(\ref{eq:FPC:NGB_LO_Lagrangian}) being different from the prefactor~$\frac{f^2}{16}$ in eq.~(\ref{eq:CCWZ_L_2_Usquared}).
}
\begin{equation}\label{eq:FPC:NGB_LO_Lagrangian}
\mathcal{L}_\mathrm{EFT} \supset \dfrac{f_\TC^2}{8}\, \tr\left[ D_\mu \Sigma^\dagger \, D^\mu \Sigma \right].
\end{equation}
While the NGBs $\pi^1$, $\pi^2$ and $\pi^3$ eventually become would-be NGBs when $\theta\neq 0$ and the EW symmetry is broken, $\pi^4=h$ can be identified with the composite Higgs boson. $\pi^5=\eta$ is a scalar singlet that generically has a mass of order $m_\eta\approx m_h/\sin(\theta)$ and couples only weakly to the SM fields, at least in the case of a vacuum alignment that preserves $C\!P$ (cf.\ footnote~\ref{fn:FPC:eta} and chapter~\ref{chap:direct_constraints}).
Its phenomenologically implications are therefore negligible.
Like in the discussion in section~\ref{sec:EWSB:vac_allign}, the mass scale for the EW gauge bosons is set by the misalignment angle $\theta$, such that the SM Higgs VEV $v_{\rm SM}$ can be identified with
\begin{equation}\label{eq:FPC:v_SM}
 v_{\rm SM} = f_\TC\,s_\theta,
\end{equation}
where the short-hand notation $s_\theta = \sin(\theta)$ is introduced.

\subsubsection{Fermion partial compositeness}
While the possibility to get a pNGB Higgs from the technifermions present in the MFPC model has been considered many times in the literature (cf.~\cite{Kaplan:1983sm,Katz:2005au,Galloway:2010bp,Barnard:2013zea,Ferretti:2013kya,Cacciapaglia:2014uja,Sannino:2016sfx,Galloway:2016fuo,Agugliaro:2016clv}),
the central new ingredient in the MFPC model that allows for an implementation of partial compositeness are the techniscalars.
For partial compositeness to be realized, it is of course crucial that the TC sector is coupled to SM fermions.
This is actually the case as the chosen quantum numbers of technifermions and techniscalars allow for fundamental Yukawa couplings involving the SM field.
They are given by
\begin{equation}\label{eq:MFPC:L_yuk}
\begin{split}
\mathcal{L}_{\mathrm{yuk}} = \, &
y_Q\, Q_{\alpha}\, \tcs_q\epsilon_{\TC} \Fupdn^{\alpha}
-y_\ubar\, \ubar\, \tcs_q^{\ast} \Fdnbar
+ y_\dbar\, \dbar\, \tcs_q^{\ast} \Fupbar\\
&+ y_L\, L_\alpha\, \tcs_l \epsilon_{\TC} \Fupdn^\alpha
- y_\nubar\, \nubar\, \tcs_l^{\ast} \Fdnbar
+ y_\ebar\, \ebar\, \tcs_l^{\ast} \Fupbar
-y'_\nubar\, \nubar\, \tcs_l \Fupbar
\hc\,
\end{split}
\end{equation}
where $\alpha$ is an $\SUs{2}{L}$ index and the fundamental Yukawa couplings $y_f$ are $3\times3$ matrices in generation space.
When techniscalars and technifermions form fermionic bound states $\mathcal{O}_F\sim (\F\S)$, these bound states are coupled to the SM fields via the fundamental Yukawa couplings.
Using the notation employed in the effective models discussed above, this means that the composite-elementary mixings $\Delta_f$ are related to the fundamental Yukawa couplings $y_f$ by
\begin{equation}
 \Delta_f \propto y_f,
\end{equation}
i.e.\ the mixing terms of the composite operators $\mathcal{O}_F$ and the SM fields are nothing but the fundamental Yukawa couplings $y_f$.
This relation can be used to construct an effective theory containing the fermionic bound states $(\F\S)$ analogous to the discussion in section~\ref{sec:Fermions:partial_compositeness}.
Interestingly, exactly like in the purely effective models of partial compositeness, one finds that the SM fermions can couple to the $(\F\F)$ composite Higgs bound state only via the fundamental Yukawa terms, i.e.\ via the mixing with the composite fermions.

\subsection{The MFPC effective field theory}\label{sec:FPC:MFPC-EFT}
To derive the phenomenological consequences of the MFPC model, one needs a description at low energies accessible by experimental measurements.
It would certainly be interesting to construct an effective theory along the lines of the multi-site moose models, i.e.\ a theory containing an effective description of the composite bound states.
In particular, such an effective theory could be backed up by lattice calculations that fix some of the effective parameters\footnote{%
For 
lattice simulations of a model similar to MFPC but without techniscalars, see~\cite{Lewis:2011zb,Hietanen:2014xca,Arthur:2016dir,Bennett:2017kga,Lee:2017uvl,Drach:2017btk}.
For preliminary results on lattice simulations of a $G_\TC=\SU{2}\cong \Sp{2}$ gauge theory featuring technifermions and techniscalars, see~\cite{Hansen:2017mrt}.
}.
However, since there are no direct observations yet and masses of composite resonances could be out of reach for the LHC, a different approach might be useful for first explorations.
To this end, an effective field theory for MFPC (MFPC-EFT) containing only the SM fermions and vector bosons as well as the pNGBs as dynamical degrees of freedom has been constructed in~\cite{Cacciapaglia:2017cdi}.
All effects stemming from bound states heavier than the pNGBs are parameterized in terms of effective operators.
This approach is especially justified by indirect bounds on $\sin(\theta)$~\cite{Khachatryan:2016vau} indicating a quite large separation of the EW and the TC scale $\frac{v_{\rm SM}}{\Lambda_\TC}<\frac{1}{25}\ll 1$.

The Lagrangian of the effective theory can be written as
\begin{equation}
\mathcal{L}_{\mathrm{EFT}} = \mathcal{L}_{\SM-\mathrm{Higgs}+{\rm NGBs}} + \sum_{A} C_A \, \mathcal{O}_A + \left(\sum_{B}C_B\, \mathcal{O}_B \hc \right),
\end{equation}
where $\mathcal{L}_{\SM-\mathrm{Higgs}+{\rm NGBs}}$ contains the SM Lagrangian without the Higgs sector, plus the leading order NGB Lagrangian shown in eq.~(\ref{eq:FPC:NGB_LO_Lagrangian}).
The only parameters it depends on are the SM gauge couplings and the decay constant $f_\TC$.
All other effects are parameterized by the WCs of the hermitian operators $\mathcal{O}_A$ and the complex operators $\mathcal{O}_B$.

For constructing the MFPC-EFT, it is useful to employ the global symmetries of the strong sector.
They are only broken by the interactions with SM field, which can be treated as spurions formally transforming under $\SU{4}_\F\times \Sp{24}_\S$.
To this end, it is convenient to  promote all SM fermions including the fundamental Yukawa couplings to the spurion
\begin{equation} \label{eq:spurionpsi}
\spur{i}{a} \in \mathbf{\overline{4}}_{\tcf} \otimes \mathbf{24}_{\tcs},
\end{equation}
where $a$ is an $\SU{4}_\F$ index and $i$ is an $\Sp{24}_\S$ index.
It assumes the background value
\begin{equation}\label{eq:FPC:Psi_background_value}
\spur{i}{a}\to\begin{pmatrix}
0 & 0 & y_{\bar{d}}\, \bar{d} & -y_{\bar{u}}\, \bar{u} \\
0 & 0 & y_{\bar{e}}\, \bar{e} & -y_{\bar{\nu}}\, \bar{\nu} \\
y_Q\, d & - y_Q\, u & 0 & 0 \\
y_L\, e & - y_L\, \nu & y'_{\bar{\nu}}\, \bar{\nu} & 0
\end{pmatrix},
\end{equation}
where the components of $\SUs{2}{L}$ doublets are written explicitly and the color and generation indices are implicit.
Using that the technifermion multiplet~$\F$ and the techniscalar multiplet~$\Phi$ transform as (cf.\ eqs.~(\ref{eq:MFPC:F_fourplet}),(\ref{eq:MFPC:S_12plet}), and (\ref{eq:MFPC:Phi_24plet}))
\begin{equation}
\tcf^{a} \in \mathbf{4}_\tcf \otimes \mathbf{N}_\TC,
\quad
\Phi^{i} \in \mathbf{24}_\tcs \otimes \mathbf{N}_\TC,
\end{equation}
the fundamental Yukawa coupling terms in eq.~(\ref{eq:MFPC:L_yuk}) can be written compactly as
\begin{equation}
\mathcal{L}_{\mathrm{yuk}} = - \spur{i}{a}\, \epsilon_{ij}\, \Phi^{j}\, \epsilon_\TC\, \tcf^{a} \hc,
\end{equation}
where $\epsilon_{ij}$ is the antisymmetric invariant tensor of $\Sp{24}_\S$.
In the low-energy MFPC-EFT, these Yukawa coupling terms lead to operators containing the SM fermions in the form of the spurion~$\spur{i}{a}$.
The leading-order operator contains only two SM fermions and is given by
\begin{equation} \label{eq:MFPC_EFT:OYuk}
\mathcal{O}_{\mathrm{Yuk}} = -\dfrac{f_\TC}{8 \pi} \, (\spur{i_1}{a_1} \spur{i_2}{a_2})\, \Sigma^{a_1 a_2} \epsilon_{i_1 i_2 },
\end{equation}
where $\Sigma^{a_1 a_2}$ is the NGB matrix (cf.\ eqs.~(\ref{eq:MFPC:fermion_condensate}) and~(\ref{eq:MFPC:NGB_matrix})).
This operator actually contains terms that correspond to the SM Yukawa couplings.
As such, it yields the fermion-pNGB couplings as well as the mass terms of SM fermions.
Assuming unitary gauge and expanding the product of pNGB matrices in powers of external Higgs states up to the linear term, one finds\footnote{%
The neutrinos are treated as massless in the following, i.e.\ their fundamental Yukawa couplings are set to zero, $y_{\bar{\nu}}=y'_{\bar{\nu}}=0$.
}
\begin{equation}\label{eq:FPC:O_Yuk}
C_\mathrm{Yuk} \mathcal{O}_\mathrm{Yuk} =-\sum_{f\in\{u,d,e\}} \dfrac{C_\mathrm{Yuk}\, s_\theta\, f_\TC}{4\pi} \, (y_{f}\transpose\, y_{\bar{f}})_{ij} \left(f_{i} \bar{f}_{j} \right) \left(1 + \dfrac{c_\theta h} {v_{\mathrm{SM}}} + \dots \right),
\end{equation}
where $c_\theta = \cos(\theta)$ and a compact notation is used to simplify the sum:
the fundamental Yukawa couplings of the $\SUs{2}{L}$ doublets are labeled by the names of their doublet components, i.e.\ one can identify $y_{Q}=y_{u}=y_{d}$ and $y_{L}=y_{e}=y_{\nu}$.
The leading term in the expansion yields the masses of the SM fermions; their mass matrices are given by
\begin{equation}
m_{f,ij} = \dfrac{C_\mathrm{Yuk}\, s_\theta\, f_\TC}{4\pi}\,  \left( y_{f}\transpose\, y_{\bar{f}} \right)_{ij},
\end{equation}
where $f\in\{u,d,e\}$.

Among the operators of the MFPC-EFT, those with external fermions are especially interesting for the analysis of effects on flavor physics presented in chapter~\ref{chap:Flavor_MFPC}.
There are eight operators in the MFPC-EFT that contain four SM fermions.
These are the five hermitian operators
\begin{equation}\label{eq:FPC:O_4f_hermitian}
 \begin{aligned}
\mathcal{O}^{1}_{4f} &= \dfrac{1}{64\pi^2 \Lambda_\TC^2} (\spur{i_1}{a_1} \spur{i_2}{a_2} ) (\spurbar{i_3}{a_3} \spurbar{i_4}{a_4} ) \Sigma^{a_1 a_2} \Sigma^\dagger_{a_3 a_4} \epsilon_{i_1 i_2} \epsilon_{i_3 i_4}\ , \\
\mathcal{O}^{2}_{4f} &= \dfrac{1}{64 \pi^2 \Lambda_\TC^2} (\spur{i_1}{a_1} \spur{i_2}{a_2} ) (\spurbar{i_3}{a_3} \spurbar{i_4}{a_4} ) \left(\delta^{a_1}_{\enspace a_3} \delta^{a_2}_{\enspace a_4} - \delta^{a_1}_{\enspace a_4} \delta^{a_2}_{\enspace a_3} \right) \epsilon_{i_1 i_2} \epsilon_{i_3 i_4}\ , \\
\mathcal{O}^{3}_{4f} &= \dfrac{1}{64\pi^2 \Lambda_\TC^2} (\spur{i_1}{a_1} \spur{i_2}{a_2} ) (\spurbar{i_3}{a_3} \spurbar{i_4}{a_4} ) \Sigma^{a_1 a_2} \Sigma^\dagger_{a_3 a_4} \left(\epsilon_{ i_1 i_4} \epsilon_{ i_2 i_3} - \epsilon_{ i_1 i_3} \epsilon_{ i_2 i_4} \right)\ ,	\\
\mathcal{O}^{4}_{4f} &= \dfrac{1}{64 \pi^2 \Lambda_\TC^2} (\spur{i_1}{a_1} \spur{i_2}{a_2} ) (\spurbar{i_3}{a_3} \spurbar{i_4}{a_4} ) \left( \delta^{a_1}_{\enspace a_3} \delta^{a_2}_{\enspace a_4} \epsilon_{ i_1 i_3} \epsilon_{ i_2 i_4} + \delta^{a_1}_{\enspace a_4} \delta^{a_2}_{\enspace a_3} \epsilon_{ i_1 i_4} \epsilon_{ i_2 i_3}\right)\ , \\
\mathcal{O}^{5}_{4f} &= \dfrac{1}{64 \pi^2 \Lambda_\TC^2} (\spur{i_1}{a_1} \spur{i_2}{a_2} ) (\spurbar{i_3}{a_3} \spurbar{i_4}{a_4} ) \left( \delta^{a_1}_{\enspace a_3} \delta^{a_2}_{\enspace a_4} \epsilon_{ i_1 i_4} \epsilon_{ i_2 i_3} + \delta^{a_1}_{\enspace a_4} \delta^{a_2}_{\enspace a_3} \epsilon_{ i_1 i_3} \epsilon_{ i_2 i_4}\right)\ ,
 \end{aligned}
\end{equation}
and the three complex operators
\begin{equation}\label{eq:FPC:O_4f_complex}
 \begin{aligned}
\mathcal{O}^{6}_{4f} &= \dfrac{1}{128 \pi^2 \Lambda_\TC^2} (\spur{i_1}{a_1} \spur{i_2}{a_2} ) (\spur{i_3}{a_3} \spur{i_4}{a_4} )  \Sigma^{a_1 a_2} \Sigma^{a_3 a_4} \epsilon_{i_1 i_2} \epsilon_{i_3 i_4}\,,  \\
\mathcal{O}^{7}_{4f} &= \dfrac{1}{128 \pi^2 \Lambda_\TC^2} (\spur{i_1}{a_1} \spur{i_2}{a_2} ) (\spur{i_3}{a_3} \spur{i_4}{a_4} )  \left(\Sigma^{a_1 a_4} \Sigma^{a_2 a_3} - \Sigma^{a_1 a_3} \Sigma^{a_2 a_4}\right) \epsilon_{i_1 i_2} \epsilon_{i_3 i_4}\,,   \\
\mathcal{O}^{8}_{4f} &= \dfrac{1}{128 \pi^2 \Lambda_\TC^2} (\spur{i_1}{a_1} \spur{i_2}{a_2} ) (\spur{i_3}{a_3} \spur{i_4}{a_4} )  \Sigma^{a_1 a_2} \Sigma^{a_3 a_4} \left(\epsilon_{i_1 i_4} \epsilon_{i_2 i_3} - \epsilon_{i_1 i_3} \epsilon_{i_2 i_4}\right)\, .
 \end{aligned}
\end{equation}
At the EW scale, they can be matched to four-fermion operators of the WEH (cf.\ section~\ref{sec:FlavMFPC:flavor_obs}).
Another operator relevant for this matching is
\begin{equation} \label{eq:FPC:OPif}
\mathcal{O}_{\Pi f} = \dfrac{i}{32\pi^2 }(\spurbar{i_1}{a_1} \bar{\sigma}_\mu \spur{i_2}{a_2} )\  \Sigma _{a_1 a_3}^\dag  \overleftrightarrow{D}^\mu \Sigma ^{a_3 a_2}\  \epsilon _{i_1 i_2}\,.
\end{equation}
It modifies the couplings of SM fermions to the weak gauge bosons, which are contained in the covariant derivative.
When $W_\mu^\pm$ and $Z_\mu$ are integrated out at the EW scale, this operator induces contributions to four-fermion operators of the WEH.
In addition, the modified couplings to gauge bosons yield important constraints on the model.
In particular, they contribute to $Z$ boson observables measured with high precision at the Large Electron–Positron~(LEP) collider.

While other operators compatible with the global symmetries of the TC sector can be constructed, they are not of particular interest in the context of this thesis.
A complete list of the MFPC-EFT operators can be found in~\cite{Cacciapaglia:2017cdi}.

\chapter[Direct collider constraints in CHM\lowercase{s}]{Direct collider constraints\\ in composite Higgs models}\label{chap:direct_constraints} 

A generic feature of CHMs is the presence of composite bound states in the low-energy effective theory.
In particular, one expects vector and fermion resonances as discussed in sections~\ref{sec:Vectorres} and~\ref{sec:Fermions}.
If the $G/H$ coset of the spontaneous symmetry breaking in the strong sector contains NGBs in addition to the Higgs doublet, they are turned into pNGBs by the effective scalar potential and also join the list of heavy NP states.
The mass scale of these states is set by NGB decay constant $f$.
As discussed in section~\ref{sec:EWSB:vac_allign}, symmetry breaking by vacuum misalignment allows $f$ to be considerably larger than the EW scale $v$.
Consequently, also the masses of these heavy states could be much heavier than the SM particles.
However, to avoid tuning in the scalar potential and to generate the correct Higgs mass, many CHMs actually require relatively light resonances (cf.\ e.g.~\cite{Contino:2006qr,Matsedonskyi:2012ym,Pomarol:2012qf,Panico:2012uw}).
In this case, they could potentially be produced and observed at the LHC.
On the other hand, a non-observation might put stringent bounds on the parameters of these models, or would at least require a larger amount of tuning.

In view of this, interesting questions for analyses of CHMs are: What are the prospects for observing any of the heavy resonances at the LHC and what are the current experimental constraints?
And in particular, what are their masses, cross sections and branching ratios and how do they compare to the experimental bounds available so far?

Apart from direct collider searches, also indirect searches put important constraints on the properties of the composite resonances.
In addition, the masses and couplings of the resonances determine the structure of the effective scalar potential and thus play an important role in EWSB.
The above questions are therefore best answered in the context of a global analysis that takes into account both direct and indirect searches and considers radiative EWSB.
To this end, we have performed comprehensive numerical studies of CHMs featuring the minimal $\SO{5}/\SO{4}$ and the next-to-minimal $\SO{6}/\SO{5}$ coset in \cite{Niehoff:2015iaa} and \cite{Niehoff:2016zso}, respectively.
The treatment of direct collider searches in these analyses and the results we have obtained are discussed in detail in this chapter.
For a detailed discussion of the indirect constraints, see \cite{Niehoff:2015iaa,Niehoff:2016zso} and in particular also \cite{Niehoff:2017thesis}.
\nowidow[3]

\section{Global analyses of composite Higgs models}\label{sec:decays:global_analyses}

The aim of our analyses in~\cite{Niehoff:2015iaa,Niehoff:2016zso} was to perform a parameter scan of a CHM that features
\begin{itemize}
 \item a pNGB Higgs,
 \item a full quark flavor structure with partial compositeness, i.e.\ composite fermion partners for all three generations,
 \item a calculable scalar potential that makes it possible to relate the mass and the VEV of the pNGB Higgs to the model parameters.
\end{itemize}
While a parameter scan of such a model is already complicated due to the large number of parameters (between $30$ and $52$ for the models considered here), the specific structure of CHMs makes it even more challenging.
In contrast to many other NP models, the parameter space does not ``factorize'' into a SM part and a NP part.
In particular, due to fermion partial compositeness, all quark masses and the elements of the CKM matrix are complicated functions of many of the model parameters.
In addition, the radiatively generated scalar potential, which is responsible for EWSB, depends on the masses and couplings of all fermions and vector bosons in the theory.
In view of this, a naive brute-force scan of the parameter space is not applicable.
Instead, we have applied a new numerical method pioneered in~\cite{Straub:2013zca} and described in the following\footnote{%
Since the focus of this chapter lies on direct constraints, only the most important properties of the scanning procedure are discussed in the following.
For an in-depth description of the specific implementation used in~\cite{Niehoff:2015iaa,Niehoff:2016zso}, see~\cite{Niehoff:2017thesis}.
}.

\subsection{Numerical strategy}\label{sec:direct_constraints:numerical_strategy}
In order to avoid sampling the whole parameter space, only those regions are sampled that satisfy the experimental constraints applied in the analysis.
%
To this end, a $\chi^2$~function is constructed that depends on the experimental measurements of all considered observables, on their theory predictions at a given parameter point~$\vec{\Theta}$, and on the theoretical and experimental uncertainties.
Combining the measurements into a vector~$\vec{\rm O}_{\rm exp}$, the \mbox{$\vec{\Theta}$-dependent} theory predictions into a vector~$\vec{\rm O}_{\rm th}(\vec{\Theta})$, and all uncertainties\footnote{%
Correlations of uncertainties are also taken into account for theory predictions, as well as for those measurements for which they are publicly available.}
into a covariance matrix~$\hat{\rm C}$, the $\chi^2$ function can be written as
\begin{equation}\label{eq:decays:num_strategy:chi2}
 \chi^2(\vec{\Theta})
 =
 \left[\vec{\rm O}_{\rm exp}-\vec{\rm O}_{\rm th}(\vec{\Theta})\right]\transpose
 [\hat{\rm C}]^{-1}
 \left[\vec{\rm O}_{\rm exp}-\vec{\rm O}_{\rm th}(\vec{\Theta})\right].
\end{equation}
The value of this $\chi^2$ function is a measure for the agreement between theoretical predictions and experimental data; the smaller its value, the better the agreement.
Using the $\chi^2$ function, {\it viable parameter points} that satisfy all constraints are determined in a procedure consisting of four steps:
\begin{itemize}
 \item A point in the parameter space is randomly chosen that is only required to fulfill most basic consistency conditions like a non-zero misalignment angle.
 \item Using this point as a starting point, a numerical optimization algorithm (from the \texttt{NLopt} package~\cite{NLopt}) is used to find a region in the parameter space with a relatively low $\chi^2$ value.
 \item This region is then sampled by a Markov chain, for which the Markov-Chain-Monte-Carlo implementation from the \texttt{pypmc} package~\cite{pypmc} is used.
 \item Because a low total $\chi^2$ value does not automatically guarantee all constraints to be satisfied, points are discarded if they violate any individual constraint by more than~3$\sigma$.
\end{itemize}
To generate a large number of viable parameter points, the above steps are repeated many times (between $\mathcal{O}(10^3)$ and $\mathcal{O}(10^4)$ depending on the model).
This makes it possible to find points from very different local minima of the $\chi^2$~function.
However, it is of course not possible to sample all regions with low $\chi^2$.
The above described method is not intended to provide a sufficient coverage of the parameter space to make any statistical statements.
Rather, it is used to yield viable parameter points in a high-dimensional parameter space, where such points are tremendously difficult to find by simply choosing parameter values randomly.

\subsection{Constraints}\label{sec:direct_constraints:constraints}
The following observables are used in our global analyses in~\cite{Niehoff:2015iaa,Niehoff:2016zso} and enter the $\chi^2$~function in eq.~(\ref{eq:decays:num_strategy:chi2}).
\begin{itemize}
 \item {\bf SM masses and Higgs VEV}:
 The VEV and the mass of the Higgs are provided by the minimum of the scalar potential and the curvature at the minimum, respectively.
 The mass matrices of vector bosons and fermions are evaluated at the minimum of the scalar potential and then numerically diagonalized.
 The eigenvalues of the mass matrices that correspond to the SM fields are interpreted as tree-level $\overline{\text{MS}}$ running masses at the scale $m_t$.
 All masses are run to the scale where they can be compared to their PDG averages~\cite{Agashe:2014kda}.
 The Higgs VEV is compared to the tree-level value of the Fermi constant in muon decay.
 \item {\bf CKM elements}:
 Since the $3\times 3$ quark mixing matrix is not unitary in the presence of composite fermions that mix with the elementary ones, effective CKM elements are defined from the ratio of $W$ couplings of quarks and leptons.
 They are compared to experimental values obtained from measurements of tree-level semi-leptonic charged-current decays (for $|V_{ud}|$~\cite{Hardy:2014qxa}, $|V_{us}|$~\cite{Aoki:2013ldr}, $|V_{ub}|$~\cite{Lattice:2015tia,Lees:2011fv}, and $|V_{cb}|$~\cite{Bailey:2014tva,Alberti:2014yda}), t-channel single top production (for $|V_{tb}|$~\cite{Khachatryan:2014iya}), and $B\to D K$ decays (for the CKM angle $\gamma$~\cite{Charles:2015gya}).
 \item {\bf Electroweak \textit{S} and \textit{T} parameters}:
 The one-loop fermion contributions to the $T$~parameter and the tree-level contributions to the $S$~parameter are compared to values from a global fit to EW precision data~\cite{Baak:2014ora}.
 \item {\bf \textit{Z} decays}: Ratios of partial widths in $Z$-boson decay are calculated with tree-level NP contributions at zero momentum and higher-order SM contributions.
 They are compared to measurements at LEP~\cite{ALEPH:2005ab}.
 \item {\bf Higgs production and decays}:
 The partial widths of the Higgs boson are calculated at tree level for the decays to $WW$, $ZZ$, $b\bar b$, and $\tau^+\tau^-$, and at the one-loop level for decays to $gg$ and $\gamma\gamma$.
 They are compared to measurements by the ATLAS and CMS collaborations \cite{Khachatryan:2014jba,ATLAS:2015bea,Khachatryan:2016vau}.
 \item {\bf Meson-antimeson mixing}:
 Several observables in meson-antimeson mixing in the $K^0$, $B^0$, and $B_s$ systems are calculated and compared to their corresponding experimental values.
 In particular, these are the mass differences $\Delta M_K$~\cite{Agashe:2014kda}, $\Delta M_d$~\cite{Amhis:2014hma}, and $\Delta M_s$~\cite{Amhis:2014hma}, the observables $S_{\psi K_S}$~\cite{Amhis:2014hma} and $\phi_s$~\cite{Aaij:2014zsa} measuring mixing induced $C\!P$ asymmetry in the $B^0$ and $B_s$ system, respectively, and the observable $\epsilon_K$~\cite{Agashe:2014kda} measuring indirect $C\!P$ violation in kaon mixing.
 \item {\bf Rare \textit{B} decays}:
 In light of tensions between experimental measurements and SM predictions in semi-leptonic rare $B$ decays (cf.\ chapter~\ref{chap:anomalies}), they are not used as constraints.
 However, experimental measurements of $\text{BR}(B\to X_s\gamma)$~\cite{Amhis:2014hma} and ${\text{BR}(B_s\to\mu^+\mu^-)}$~\cite{CMS:2014xfa} are included as constraints.
 \item {\bf Contact interactions}:
 Significant degrees of compositeness of first-generation quarks can be constrained by four-quark contact interactions that contribute to the dijet angular distributions.
 Calculations of corresponding WCs are compared to LHC measurements~\cite{Aad:2015eha,ATLAS:2015nsi}.
 \item {\bf Neutron electric dipole moment}:
 The next-to-minimal CHM allows for spontaneous $C\!P$ violation in the scalar sector.
 To constrain this effect, the neutron electric dipole moment (EDM) in terms of the quarks' EDM and chromo-EDM  is calculated and compared to the experimental limit~\cite{Baker:2006ts}.
 \item {\bf Direct constraints}:
 All cross sections and branching ratios of heavy resonances are calculated and compared to experimental limits (see tables in appendix~\ref{app:seaches}).
 These constraints are discussed in detail in section~\ref{sec:direct_constraints:direct_constraints}.
\end{itemize}

\subsection{The models}\label{sec:direct_constraints:models}

To select the models to be investigated by our global analyses, we have first considered the requirements at the beginning of section~\ref{sec:decays:global_analyses}.
In particular, the requirements of a pNGB Higgs and a calculable potential can be satisfied in multi-site moose models.
The contributions to the scalar potential stemming from vector bosons are finite in a model with three or more sites if it contains one level of spin-1 resonances in the adjoint representation of the unbroken group $H$, or in a model with two or more sites if it contains only spin-1 resonances in the adjoint representation of the full group $G$ (cf.\ section~\ref{sec:EWSB:eff_pot_coll_breaking}).
Requiring a particle content as minimal as possible singles out the two-site model of the latter type.
This is the construction of the two-site 4DCHM~\cite{DeCurtis:2011yx}, for which the fermion sector also yields
a finite contribution to the scalar potential (cf.\ section~\ref{sec:EWSB:eff_pot_coll_breaking}).

In specifying a CHM, a central aspect is the choice of the NGB coset.
%
%
As discussed at the beginning of chapter~\ref{chap:CHMs}, the minimal choice that yields NGBs in a complex $\SUs{2}{L}$ doublet and a custodial $\SUs{2}{L+R}$ symmetry is the
NGB coset $\SO{5}/\SO{4}$.
We have chosen to analyze such a minimal CHM (MCHM) in~\cite{Niehoff:2015iaa}.
As detailed in section~\ref{sec:FPC}, the MCHM breaking pattern $\SO{5}\to\SO{4}$ cannot be realized by chiral symmetry breaking in a UV completion of an effective CHM.
However, this is actually possible for the only slightly less minimal
breaking pattern $\SO{6}\to\SO{5}$
, which yields a scalar singlet NGB in addition to the complex $\SUs{2}{L}$ doublet.
We have chosen to analyze such a next-to-minimal CHM (NMCHM) in~\cite{Niehoff:2016zso}.

In a CHM that contains an unbroken global symmetry $\SO{4} \cong \SUs{2}{L}\times \SUs{2}{R}$, the $\T^3_R$ generator\footnote{%
For the definition of the generators of $\SO{6}$, $\SO{5}$, $\SUs{2}{L}$, and $\SUs{2}{R}$ used in this chapter, see appendix~\ref{sec:generator_definitions}.
} of $\SUs{2}{R}$ plays the role of the hypercharge generator as long as only the Higgs sector and the lepton sector are considered.
However, it is well known that this assignment of hypercharge does not work for quarks that are embedded into a multiplet transforming under $\SO{4}$ (cf.\ e.g.~\cite{Agashe:2004rs}).
This problem can be solved if the quark $\SO{4}$ multiplets are charged under an additional $\Us{1}{X}$ symmetry and if the hypercharge generator $Y$ is defined in terms of the $\T^3_R$ generator and the $\Us{1}{X}$ generator $X$ as
\begin{equation}\label{eq:decays:global_analyses:Y}
 Y = \T^3_R + X.
\end{equation}
Consequently, the groups $G$ and $H$ have to contain the $\Us{1}{X}$ group as a subgroup.
The minimal choice is to simply consider a direct product group such that the breaking patterns in the MCHM and in the NMCHM become
\begin{equation}
 \SO{5}\times\Us{1}{X}\to\SO{4}\times\Us{1}{X}
 \quad
 \text{and}
 \quad
 \SO{6}\times\Us{1}{X}\to\SO{5}\times\Us{1}{X},
\end{equation}
respectively.
%
While this does not change the number of NGBs, it introduces an additional spin-1 resonance associated with the $\Us{1}{X}$ symmetry.

For a complete CHM, not only the EW gauge group should be embeddable into the global symmetry $H$, but also the QCD gauge group $\SUs{3}{C}$.
Again, the simplest way to achieve this is by considering a direct product group.
Consequently, the breaking patterns in the MCHM and in the NMCHM become
\begin{equation}
\begin{aligned}
 \SO{5}\times\Us{1}{X}\times\SUs{3}{C}&\to\SO{4}\times\Us{1}{X}\times\SUs{3}{C}
 \\
 &\text{and}
 \\
 \SO{6}\times\Us{1}{X}\times\SUs{3}{C}&\to\SO{5}\times\Us{1}{X}\times\SUs{3}{C},
\end{aligned}
\end{equation}
respectively.
Similar to the introduction of the $\Us{1}{X}$ group, this does not modify the NGB content of the models but introduces additional spin-1 resonances in the adjoint representation of $\SUs{3}{C}$.

Having specified the groups $G$ and $H$ for the MCHM and the NMCHM, the only thing that still remains to be fixed are the representations of $G$ under which the fermions transform.
While the representations of the $\Us{1}{X}\times\SUs{3}{C}$ group are fixed by the SM quantum numbers and eq.~(\ref{eq:decays:global_analyses:Y}), one is in principle free to choose any representations of $\SO{5}$ for the MCHM and of $\SO{6}$ for the NMCHM that allow for an embedding of the SM fermions.
%
However, the choice of representations has important phenomenological consequences.
%
In particular, if they are chosen such that they satisfy a discrete $P_{LR}$ symmetry, tree-level contributions to the $Z b_L b_L$ coupling can be avoided~\cite{Agashe:2006at}.
%
Without this so called {\it custodial protection}, the sizable composite-elementary mixing of the third generation's quark doublet, which is required by the large top quark mass, generically yields a significant tree-level contribution to the $Z b_L b_L$ coupling.
This in turn can lead to severe tensions with LEP measurements of the $Z$~boson's partial widths (cf.~\cite{Agashe:2005dk}).
In the MCHM, the simplest possibilities to achieve the custodial protection is to embed the SM fermions into incomplete multiplets transforming either under the fundamental $\mathbf{5}$ or the anti-symmetric $\mathbf{10}$ representation of $\SO{5}$, which are known as MCHM$_5$ and MCHM$_{10}$, respectively~\cite{Agashe:2006at,Contino:2006qr}.
Requiring minimal particle content, we have chosen the fundamental $\mathbf{5}$ representation of $\SO{5}$ for our MCHM analysis.
Similarly, for the NMCHM analysis, we have chosen the fundamental $\mathbf{6}$ representation of $\SO{6}$, which also implements the custodial protection mechanism with minimal particle content.
A notable consequence of quark partners in $\mathbf{5}$ or $\mathbf{6}$ multiplets is that the models feature heavy quark resonances with exotic electric charges -4/3 and +5/3. Due to their different charge compared to the up-type or down-type quark partners, they decay into distinct channels and dedicated experimental searches for these exotically charged resonances are available (cf.\ section~\ref{sec:direct_constraints:results:quarks}).

An important goal of the analyses is to study the quark flavor structure.
Consequently, three generations of composite quarks are considered.
On the contrary, the lepton sector is not studied in detail and only elementary leptons are included.
While it is beyond the scope of our analyses to consider partially composite leptons, they can have interesting phenomenological effects on the scalar potential~\cite{Carmona:2014iwa} or in the context of the $b\to s\,\ell^+\ell^-$ anomalies (cf.~\cite{Niehoff:2015bfa} and chapter~\ref{chap:LUFV_in_CHMs}).

Both models considered in our analyses can be expressed in terms of the following moose diagram (cf.\ eq.~(\ref{eq:fermions_2site_moose_ud}))
\begin{equation}
\begin{tabular}{c}
\xy
\xymatrix@R=.4pc@C=1.4pc{
\mathrm{Global:}
& G
&& G_1
&&
\\
& *=<20pt>[o][F]{} \doublerightxyarrow^{\mbox{\raisebox{1.5ex}{$\Omega_1$}}}
&& *=<20pt>[o][F]{} \doublerightxyarrow^{\mbox{\raisebox{1.5ex}{$\Omega_2$}}}
&& *=<0pt,20pt>[l][F]{} & *-<0pt,20pt>[l]{H}
\\
\mathrm{Gauged:}
& E
&& G_1
&&
\\
\mathrm{Quarks:}
&
{
\begin{matrix}
\begin{matrix}[1.5]
 \xi_{uR}\, [u_R^{(0)}]
\end{matrix}
\\
\begin{matrix}[1.5]
 \xi_{uL}\, [q_L^{(0)}]
 \\
 \xi_{dL}\, [q_L^{(0)}]
\end{matrix}
\\
\begin{matrix}[1.5]
 \xi_{dR}\, [d_R^{(0)}]
\end{matrix}
\end{matrix}
}
&&
{
\begin{matrix}
\left.\begin{matrix}[1.5]
  \widetilde{\Psi}_{uL}
& \widetilde{\Psi}_{uR}
 \\
  \Psi_{uR}
& \Psi_{uL}
\end{matrix}\ \ \hspace{20pt}\right>\,Y_u
\\
\left.\begin{matrix}[1.5]
  \Psi_{dR}
& \Psi_{dL}
 \\
  \widetilde{\Psi}_{dL}
& \widetilde{\Psi}_{dR}
\end{matrix}\ \ \hspace{20pt}\right>\,Y_d
\end{matrix}
}
\hspace{-20pt}
\hspace{-40pt}
}
\endxy
\end{tabular}\quad.
\end{equation}
%
The symmetry groups can be written as
\begin{equation}
 \begin{aligned}
 G &= \mathcal{G}\times\Us{1}{X}^{(0)}\times\SUs{3}{C}^{(0)},
 \quad\quad&
 H &= \mathcal{H}\times\Us{1}{X}^{(2)}\times\SUs{3}{C}^{(2)},
 \\
 G_1 &= \mathcal{G}_1\times\Us{1}{X}^{(1)}\times\SUs{3}{C}^{(1)},
 \quad\quad&
 E &= \SUs{2}{L}^{(0)}\times\Us{1}{Y}^{(0)}\times\SUs{3}{C}^{(0)},
 \end{aligned}
\end{equation}
where superscripts are used to distinguish fields and symmetry groups at different sites.
The groups $\mathcal{G}$, $\mathcal{G}_1$, and $\mathcal{H}$ depend on the considered model.
%
In the MCHM, they are defined by
\begin{equation}
\begin{aligned}
 \mathcal{G} &= \SO{5}^{(0)},&
 \quad\quad
 \mathcal{G}_1 &= \SO{5}^{(1)},&
 \quad\quad
 \mathcal{H} &= \SO{4}^{(2)},
\end{aligned}
\end{equation}
while in the NMCHM they are given by
\begin{equation}
\begin{aligned}
 \mathcal{G} &= \SO{6}^{(0)},&
 \quad\quad
 \mathcal{G}_1 &= \SO{6}^{(1)},&
 \quad\quad
 \mathcal{H} &= \SO{5}^{(2)}.
\end{aligned}
\end{equation}
The hidden local symmetry $H$ can be used to remove unphysical would-be NGBs in $\Omega_2$ such that only NGBs in the $\mathcal{G}_1/\mathcal{H}$ coset remain. 
$\Omega_1$
contains NGBs in the coset $(G\times G_1)/G_{D}$, where $G_D$ is the diagonal subgroup of $G\times G_1$.
It is possible to choose a gauge where the direct factors $\Us{1}{X}^{(1)}$ and $\SUs{3}{C}^{(1)}$ of the $G_1$ gauge group are used to remove the would-be NGBs in the
$\left(\Us{1}{X}^{(0)}\times\Us{1}{X}^{(1)}\right)\Big/\Us{1}{X}^{(0+1)}$
and the
$\left(\SUs{3}{C}^{(0)}\times\SUs{3}{C}^{(1)}\right)\Big/\SUs{3}{C}^{(0+1)}$
parts of $(G\times G_1)/G_{D}$, respectively\footnote{%
Here, the $(0+1)$ superscript denotes a diagonal subgroup of a direct product of groups with superscripts $(0)$ and $(1)$.
}.
Consequently, in this gauge, $\Omega_1$ only contains the NGBs in the $(\mathcal{G}\times\mathcal{G}_1)/\mathcal{G}_D$ coset, while mass and mixing terms for the $\SUs{3}{C}^{(0)}$ and $\SUs{3}{C}^{(1)}$ gauge bosons $G_\mu^{(0)}$ and $G_\mu^{(1)}$ as well as the $\Us{1}{Y}^{(0)}$ and $\Us{1}{X}^{(1)}$ gauge bosons $B_\mu^0$ and $X_\mu$ are introduced.
%
%
%
%
%
Denoting the $\mathcal{G}_1$ gauge bosons by $\rho_\mu$ and those of $\SUs{2}{L}^{(0)}$ by $W_\mu^{(0)}$,
the corresponding leading-order Lagrangian for the NGBs and vector bosons then reads (cf.\ eq.~(\ref{eq:Vectorres_N_moose_lagrangian}))
\begin{equation}\label{eq:decays:global_analyses:L_boson}
\begin{aligned}
 \mathcal{L}_{\rm boson} &=
 \frac{f_1^2}{4}\,\tr\left[D_\mu \Omega_1^{-1}(x)\,D^\mu \Omega_1(x)\right]
 +
 \frac{f_2^2}{4}\,\tr\left[D_\mu \Omega_2^{-1}(x)\,D^\mu \Omega_2(x)\right]
 \\
 &
 - \frac{1}{4}\,\tr\left[{W}^{(0)}_{\mu\nu}(x)\,W^{(0)\mu\nu}(x)\right]
 - \frac{1}{4}\,{B}^{(0)}_{\mu\nu}(x)\,B^{(0)\mu\nu}(x)
 - \frac{1}{4}\,\tr\left[{G}^{(0)}_{\mu\nu}(x)\,G^{(0)\mu\nu}(x)\right]
 \\
 &
 - \frac{1}{4}\,\tr\left[{\rho}_{\mu\nu}(x)\,\rho^{\mu\nu}(x)\right]
 - \frac{1}{4}\,{X}_{\mu\nu}(x)\,X^{\mu\nu}(x)
 - \frac{1}{4}\,\tr\left[{\rho_G}_{\mu\nu}(x)\,{\rho_G}^{\mu\nu}(x)\right]
 \\
 &
 + \frac{f_G^2}{4} \, \left(
g_{3(0)}
\, G_{\mu}^{(0)}(x) - g_{3(1)} \, G_\mu^{(1)}(x) \right)^2
+\frac{f_X^2}{4} \left( g'_{(0)} \, B^{(0)}_\mu(x) - g_X\, X_{\mu}(x) \right)^2
 ,
\end{aligned}
\end{equation}
where the decay constants $f_G$ and $f_X$ of the NGBs associated with the heavy ${\rho_G}_\mu$ and $X_\mu$ bosons have been allowed to be independent of each other and of $f_1$ and  $f_2$.
The gauge covariant derivatives are defined by
\begin{equation}
\begin{aligned}
 i\,D_\mu\,\Omega_1(x)
 &= i\,\partial_\mu\,\Omega_1(x)
 +\left(g_{(0)}\, W_\mu^{(0)a}(x)\,\T^a_L+g'_{(0)}\,B^{(0)}_\mu(x)\,\T^3_R\right)\,\Omega_1(x)
 -g_\rho\,\Omega_1(x)\,\rho_\mu(x),
 \\
 i\,D_\mu\,\Omega_2(x)
 &= i\,\partial_\mu\,\Omega_2(x)
 +g_\rho\,\rho_\mu(x)\,\Omega_2(x).
\end{aligned}
\end{equation}
In the above expressions, $g_{(0)}$, $g'_{(0)}$, $g_{3(0)}$, $g_{3(1)}$, $g_X$, and $g_\rho$ denote the gauge couplings of $\SUs{2}{L}^{(0)}$, $\Us{1}{Y}^{(0)}$, $\SUs{3}{C}^{(0)}$, $\SUs{3}{C}^{(1)}$, $\Us{1}{X}^{(1)}$, and $\mathcal{G}_1$, respectively.

The Lagrangian of elementary and composite quarks  is given by (cf.\ eq.~(\ref{eq:fermions_general_lagrangian}))
\begin{equation}\label{eq:decays:global_analyses:L_quark}
\begin{aligned}
 \mathcal{L_{\rm quark}}
 &=
 \bar{q}_{L}^{(0)}(x)\,i\slashed{D}\,q_{L}^{(0)}(x)
 +
 \bar{u}_{R}^{(0)}(x)\,i\slashed{D}\,u_{R}^{(0)}(x)
 \\
 &
 +\bar{\Psi}_u(x)\left(i\slashed{D}-m_{U}\right)\Psi_u(x)
 +
 \bar{\widetilde{\Psi}_u\hspace{-.5em}}\hspace{.5em}(x)\left(i\slashed{D}-m_{\widetilde{U}}\right)\widetilde{\Psi}_u(x)
 \\
 &+
 \Big\{
 \Delta_{u_L}\,\bar{\xi}_{uL}(x)\,\Omega_1(x)\,\Psi_{uR}(x)
 +
 \Delta_{u_R}\,\bar{\xi}_{uR}(x)\,\Omega_1(x)\,\widetilde{\Psi}_{uL}(x)+{\rm h.c.}\Big\}
 \\
 &-\Big\{
 Y_{u}\,\bar{\Psi}_{uL}(x)\,\Omega_2(x)\,
 \phi_0\,\phi_0^\dagger\,
 \Omega_2^\dagger(x)\,\widetilde{\Psi}_{uR}(x)
 +
 m_{Y_{u}}\,\bar{\Psi}_{uL}(x)\,\widetilde{\Psi}_{uR}(x)+{\rm h.c.}\Big\}
 \\
 &+ \left(\{u,U\}\rightarrow \{d,D\}\right)\ ,
\end{aligned}
\end{equation}
where the covariant derivatives are defined by
\begin{equation}
\begin{aligned}
 i D_\mu\,q_L^{(0)}(x) &=
    \left(i\,\partial_\mu
    +\frac{1}{6}\,g'_{(0)}\,B^{(0)}_\mu(x)
    +g_{3(0)}\, G_\mu^{(0)a}(x)\,\T^a_C
    +g_{(0)}\, W_\mu^{(0)a}(x)\,\T^a_L
    \right)
    q_L^{(0)}(x),
 \\
 i D_\mu\,u_R^{(0)}(x) &=
    \left(i\,\partial_\mu
    +\frac{2}{3}\,g'_{(0)}\,B^{(0)}_\mu(x)
    +g_{3(0)}\, G_\mu^{(0)a}(x)\,\T^a_C
    \right)
    u_R^{(0)}(x),
 \\
 i D_\mu\,d_R^{(0)}(x) &=
    \left(i\,\partial_\mu
    -\frac{1}{3}\,g'_{(0)}\,B^{(0)}_\mu(x)
    +g_{3(0)}\, G_\mu^{(0)a}(x)\,\T^a_C
    \right)
    d_R^{(0)}(x),
 \\
 i D_\mu\,\Psi_u(x) &=
    \left(i\,\partial_\mu
    +\frac{2}{3}\,g_X\,X_\mu(x)
    +g_{3(1)}\, G_\mu^{(1)a}(x)\,\T^a_C
    +g_\rho\, \rho_\mu^{a}(x)\,\T^a_\mathcal{G}
    \right)
    \Psi_u(x),
 \\
 i D_\mu\,\Psi_d(x) &=
    \left(i\,\partial_\mu
    -\frac{1}{3}\,g_X\,X_\mu(x)
    +g_{3(1)}\, G_\mu^{(1)a}(x)\,\T^a_C
    +g_\rho\, \rho_\mu^{a}(x)\,\T^a_\mathcal{G}
    \right)
    \Psi_d(x),
\end{aligned}
\end{equation}
and those for $\widetilde{\Psi}_u(x)$ and $\widetilde{\Psi}_d(x)$ are the same as for $\Psi_u(x)$ and $\Psi_d(x)$, respectively.
Here, $\T^a_C=\lambda^a/2$ are the generators of $\SUs{3}{C}$ and $\lambda^a$ are the Gell-Mann matrices, while $\T^a_\mathcal{G}$ are the generators of $\mathcal{G}\cong\mathcal{G}_1$.

As one of the aims of our numerical analysis was to study a model with full quark flavor structure, it is understood that the above quark fields have an implicit generation index and the composite-elementary mixings are $3\times 3$ matrices.
In our numerical analysis, we have considered different implementations of the flavor symmetries discussed in section~\ref{sec:Fermions:flavor} to suppress large contributions to meson-antimeson mixing.

\subsubsection{The minimal composite Higgs model}

In the MCHM, where $\mathcal{G}\cong\mathcal{G}_1\cong \SO{5}$, the composite vector bosons $\rho_\mu = \rho^a_\mu \, \T^a_\mathcal{G}$ can be decomposed
as
 \begin{equation}\label{eq:decays:global_analyses:MCHM_gauge_decomposition}
  \rho_\mu = {\rho_L}_\mu+ {\rho_R}_\mu+ \axial_\mu,
 \end{equation}
where ${\rho_L}_\mu = {\rho_L}^{a}_\mu \, \T^a_L$, ${\rho_R}_\mu = {\rho_R}^{a}_\mu \, \T^a_R$, and $\axial_\mu=\axial_\mu^{a}\, \T^a_{\hat{1}}$.
They transform under $\SUs{2}{L}\times\SUs{2}{R}$ as $(\mathbf{3},\mathbf{1})$, $(\mathbf{1},\mathbf{3})$, and $(\mathbf{2},\mathbf{2})$, respectively (cf.\ appendix~\ref{sec:generator_definitions}).
Thus, the vector bosons ${\rho_L}_\mu$ and ${\rho_R}_\mu$ can be identified with the gauge bosons of the $\SUs{2}{L}^{(1)}$ and $\SUs{2}{R}^{(1)}$ subgroups of $\SO{5}^{(1)}$, while $\axial_\mu$ is associated with the generators of the $\SO{5}/\SO{4}$ coset.

When discussing a concrete model, it is convenient to choose a gauge that reduces the number of NGBs by removing all would-be NGBs.
To this end, one can use the holographic gauge (cf.\ eq.~(\ref{eq:Vectorres_gauge_holographic_f1}))
\begin{equation}\label{eq:decays:global_analyses:MCHM_holo_gauge}
 \Omega_1(x) = U_1(x)=\exp\left[i\frac{\sqrt{2}}{f_1}\,\pi_a(x)\,\T^a_{\hat{1}}\right],
 \quad
 \Omega_2(x) = \mathds{1}.
\end{equation}
Employing in addition the SM unitary gauge, only $\pi_4(x)=h(x)$ is non-zero.
As noted in section~\ref{sec:vectorres_higher_levels}, the holographic gauge leads to a mixing between the NGBs and the $\axial_\mu$ vector bosons.
This mixing can be removed by the field shift\footnote{
$f^{-2}=f_1^{-2}+f_2^{-2}$ (cf.\ eq.~(\ref{eq:Vectorres_f_fk_relation})).
}
\begin{equation}\label{eq:decays:global_analyses:MCHM_field_shift}
 \axial_4^\mu(x) \rightarrow  \axial_4^\mu(x) - \frac{\sqrt{2}}{g_\rho}
\frac{f}{f_2^2} \partial^\mu h(x)\,.
\end{equation}
In addition, the holographic gauge as defined in eq.~(\ref{eq:decays:global_analyses:MCHM_holo_gauge}) requires the field redefinition
\begin{equation}
 h(x)  \rightarrow  \frac{f_1}{f} h(x)
\end{equation}
to get a canonically normalized kinetic term for $h(x)$.
The NGB matrix $\Omega(x)=\Omega_1(x)\,\Omega_2(x)$ then takes the explicit form
\begin{equation}\label{eq:decays:global_analyses:MCHM_Omega}
\Omega(x) = \left( \begin{array}{ccccc}
1& & &	      &\\
 &1& &	      &\\
 & &1&	      &\\
 & & &c_h(x)  & s_h(x)\\
 & & &-s_h(x) & c_h(x)\\
\end{array}
 \right),
\end{equation}
where the short-hand notation
\begin{equation}
 s_h(x) = \sin\left( \frac{h(x)}{f} \right),
 \quad
 c_h(x) = \cos\left( \frac{h(x)}{f} \right)
\end{equation}
is used.

Similarly to the vector bosons, also the fermions can be decomposed into $\SUs{2}{L}\times\SUs{2}{R}$ multiplets.
In particular, a field $\Psi(x)$ that transforms under the fundamental $\mathbf{5}$ representation of $\SO{5}$ decomposes into a bidoublet $Q(x)$ and a singlet $S(x)$.
Specifically, this can be expressed as
\begin{equation}\label{eq:decays:global_analyses:MCHM_Psi_decomposition}
 \Psi(x) =
 \frac{1}{\sqrt{2}}
 \begin{pmatrix}
     Q^{++}(x) +    Q^{--}(x)\\
  i\,Q^{++}(x) - i\,Q^{--}(x)\\
     Q^{+-}(x) -    Q^{-+}(x)\\
  i\,Q^{+-}(x) + i\,Q^{-+}(x)\\
  \sqrt{2}\,S(x)
 \end{pmatrix},
\end{equation}
where the superscripts on the four components of $Q(x)$ indicate their $\T_L^3$ and $\T_R^3$ charges.
While this decomposition can be used for each of the fields $\Psi_u(x)$, $\Psi_d(x)$, $\widetilde{\Psi}_u(x)$, and $\widetilde{\Psi}_d(x)$, it also determines the embedding of the elementary fields into incomplete $\SO{5}$ multiplets.
This embedding is given by
\begin{equation}\label{eq:decays:global_analyses:MCHM_xi_decomposition}
\begin{aligned}
 \xi_{u L}(x) &= \frac{1}{\sqrt{2}} \left( \begin{array}{c}
                                     d_L^{(0)}(x) \\
                                     -i d_L^{(0)}(x) \\
                                     u_L^{(0)}(x) \\
                                     i u_L^{(0)}(x) \\ 0
                                    \end{array} \right),
\quad&
 \xi_{u R}(x) &= \left( \begin{array}{c}
                                       0 \\ 0 \\ 0 \\ 0 \\ u_R^{(0)}(x)
                                      \end{array}
 \right),
\\
  \xi_{d L}(x) &= \frac{1}{\sqrt{2}} \left( \begin{array}{c}
                                     u_L^{(0)}(x) \\
                                     i u_L^{(0)}(x) \\
                                     -d_L^{(0)}(x) \\
                                     i d_L^{(0)}(x) \\ 0
                                    \end{array} \right),
\quad&
 \xi_{d R}(x) &= \left( \begin{array}{c}
                                       0 \\ 0 \\ 0 \\ 0 \\ d_R^{(0)}(x)
                                      \end{array}
 \right).
\end{aligned}
\end{equation}

For our analysis of the MCHM, we have considered four scenarios with different flavor symmetries in the quark sector:
$\U{3}^3$ left-compositeness, $\U{3}^3$ right-compositeness, $\U{2}^3$ left-compositeness, and $\U{2}^3$ right-compositeness, which in the following are denoted by $\Us{3}{LC}^3$,  $\Us{3}{RC}^3$,  $\Us{2}{LC}^3$, and $\Us{2}{RC}^3$, respectively (cf.\ section~\ref{sec:Fermions:flavor}).
The explicit expressions of the mixing matrices $\Delta_{u_L}$, $\Delta_{u_R}$, $\Delta_{d_L}$, and $\Delta_{d_R}$ are given in appendix~\ref{app:comp_elem:NMCHM} for all four cases.

Plugging the decompositions of the vector bosons $\rho_\mu$, eq.~(\ref{eq:decays:global_analyses:MCHM_gauge_decomposition}), and fermions, eqs.~(\ref{eq:decays:global_analyses:MCHM_Psi_decomposition}) and~(\ref{eq:decays:global_analyses:MCHM_xi_decomposition}), as well as the NGB matrices $\Omega_2(x)=\mathds{1}$ and $\Omega_2(x)=\Omega(x)$, eq.~(\ref{eq:decays:global_analyses:MCHM_Omega}), into the Lagrangians in eqs.~(\ref{eq:decays:global_analyses:L_boson}) and~(\ref{eq:decays:global_analyses:L_quark}), one gets $h(x)$-dependent mass matrices for the fermions and vector bosons.
They are explicitly given in appendix~\ref{app:Mass_matrices:MCHM}.
These matrices are the basis for the phenomenological study.
For a given parameter point, they can be used to calculate the scalar one-loop potential via eq.~(\ref{eq:EWSB:potential:finite}).
The minimum of the potential then determines the vacuum alignment in terms of the VEV of $h(x)$.
This VEV also enters the fermion and vector boson mass matrices via their $h(x)$-dependence.
Diagonalizing the mass matrices with $h(x)$ set to its VEV then yields all physical mass eigenstates after EWSB.
The masses and couplings of these mass eigenstates can finally be used for studying the phenomenology of the given parameter point.

%
%
%
%

\subsubsection{The next-to-minimal composite Higgs model}
As in the MCHM, it is convenient to decompose the vector and fermion fields in the NMCHM into $\SUs{2}{L}\times\SUs{2}{R}$ multiplets.
For $\mathcal{G}\cong\mathcal{G}_1\cong \SO{6}$, one finds the following decomposition of the $\rho_\mu$ vector bosons:
\begin{equation}
  \rho_\mu = \rho_L^\mu + \rho_R^\mu + \axial_1^\mu + \axial_2^\mu + \rho_S^\mu.
 \end{equation}
While $\rho_L^\mu$ and $\rho_R^\mu$ are exactly the same $\SUs{2}{L}$ and $\SUs{2}{R}$ triplets as in the MCHM, there are now two bidoublets $\axial_{1}^\mu = \axial_{1}^{a\,\mu}\, \T^a_{\hat{1}}$ and $\axial_{2}^\mu = \axial_{2}^{a\,\mu}\, \T^a_{\hat{2}}$.
While $\axial_{1}^\mu$ is actually associated with the same $\SO{5}$ generators as $\axial^\mu$ in the MCHM, these generators are the coset generators in the MCHM but are unbroken generators in the NMCHM.
Therefore, $\axial_{1\mu}^\mu$ and $\axial^\mu$ are quite different from each other.
On the other hand, since $\axial_{2}^\mu$ is a bidoublet associated with coset generators in the NMCHM, its mass, couplings, and mixing terms resemble those of $\axial^\mu$ in the MCHM.
In this sense, the presence of $\axial_1^\mu$ rather than $\axial_{2}^\mu$ should be considered as a main difference to the MCHM.
Another main difference is the presence of $\rho_S^\mu$, which is an $\SUs{2}{L}\times\SUs{2}{R}$ singlet and associated with the $\T_S$ generator of $\SO{6}$ (cf.\ appendix~\ref{sec:generator_definitions}).

For considering the NGBs, it is again convenient to use the holographic gauge, which yields
\begin{equation}\label{eq:decays:global_analyses:NMCHM_holo_gauge}
 \Omega_1(x) = U_1(x)=\exp\left[i\frac{\sqrt{2}}{f_1}\,\left(\pi_a(x)\,\T^a_{\hat{2}}+\pi_5(x)\,\T_S\right)\right],
 \quad
 \Omega_2(x) = \mathds{1}.
\end{equation}
Employing the SM unitary gauge removes the would-be NGBs $\pi_1(x)$, $\pi_2(x)$, and $\pi_3(x)$ and leaves the two physical NGBs $\pi_4(x)$ and $\pi_5(x)$.
It is convenient to parametrize them as (cf.~\cite{Redi:2012ha})
\begin{equation}
 \pi_4(x) = \gb{h}(x) \cos\left( \frac{\gb{\eta}(x)}{f_1} \right), \qquad \pi_5(x) = \gb{h}(x) \sin\left( \frac{\gb{\eta}(x)}{f_1} \right).
\end{equation}
It is interesting to note that $\gb{\eta}(x)$ and $\pi_5(x)$ are pseudoscalars and odd under $C\!P$, while $\gb{h}(x)$ and $\pi_4(x)$ are even under $C\!P$.
However, in the presence of $C\!P$ violating contributions to the scalar potential, $\gb{h}(x)$ and $\gb{\eta}(x)$ can mix with each other.
Consequently, the mass eigenstates, which will be denoted by $h(x)$ and $\eta(x)$, are not necessarily $C\!P$ eigenstates.
In any case, the mass eigenstate $h(x)$ will be identified with the Higgs boson.
In the absence of $C\!P$ violation, $h(x)=\gb{h}(x)$ and $\eta(x)=\gb{\eta}(x)$.

Like in the MCHM, the holographic gauge introduces mixing terms between NGBs and gauge fields.
These mixing terms are removed by field shifts\footnote{%
For a detailed discussion of the mixing terms and field shifts, see~\cite{Niehoff:2017thesis}.
} of $\axial_1^{4 \, \mu}$, $\axial_2^{4 \, \mu}$, and $\rho_S^\mu$.
To get canonically normalized kinetic terms for $\gb{h}(x)$ and $\gb{\eta}(x)$, the field shifts are succeeded by the field redefinitions
\begin{equation} \label{eq:ScalarRot}
 \gb{h}(x) \rightarrow \frac{f_1}{f} \gb{h}(x), \qquad \gb{\eta}(x) \rightarrow \frac{f_1}{f \sin\left( \frac{v_\gb{h}}{f} \right)} \gb{\eta}(x),
\end{equation}
where $v_\gb{h}$ denotes the VEV of $\gb{h}$.
Using the short-hand notation
\begin{equation} \label{eq:sh_setatilde_def}
\begin{aligned}
 \sh(x) &= \sin \left( \frac{\gb{h}(x)}{f} \right),
 \quad&
 \setatilde(x) &= \sin \left( \frac{\gb{\eta}(x)}{f \sin \left( \frac{v_\gb{h}}{f} \right)} \right),
 \\
 \ch(x) &= \sqrt{1-\sh^2(x)},
 \quad&
 \cetatilde(x) &= \sqrt{1-\setatilde^2(x)},
\end{aligned}
\end{equation}
the NGB matrix $\Omega(x)=\Omega_1(x)\,\Omega_2(x)$ can then be expressed as
\begin{equation}
 \Omega(x) =
 \left( \begin{array}{cccccc}
 1 &  &  &                     &                      &          \\
  & 1 &  &                     &                      &          \\
  &  & 1 &                     &                      &          \\
  &  &  &  \ch(x)\, \cetatilde^2(x)+\setatilde^2(x) &  -(1-\ch(x))\, \setatilde(x)\, \cetatilde(x) & \sh(x)\, \cetatilde(x) \\
  &  &  & -(1-\ch(x))\, \setatilde(x)\, \cetatilde(x) & \ch(x)\, \setatilde^2(x) + \cetatilde^2(x) & \sh(x)\, \setatilde(x) \\
  &  &  &           -\sh(x)\, \cetatilde(x) &            -\sh(x)\, \setatilde(x) &       \ch(x)
\end{array} \right).
\end{equation}

Turning to the fermion sector, a decomposition into $\SUs{2}{L}\times \SUs{2}{R}$ multiplets similar to the one in the MCHM can be performed.
%
In particular, a field $\Psi(x)$ transforming under the fundamental $\mathbf{6}$ representation of $\SO{6}$ can be expressed in terms of $\SUs{2}{L}\times \SUs{2}{R}$ multiplets as
\begin{equation}
 \Psi(x) =
 \frac{1}{\sqrt{2}}
 \begin{pmatrix}
     Q^{++}(x) +    Q^{--}(x)\\
  i\,Q^{++}(x) - i\,Q^{--}(x)\\
     Q^{+-}(x) -    Q^{-+}(x)\\
  i\,Q^{+-}(x) + i\,Q^{-+}(x)\\
  \sqrt{2}\,S_1(x)\\
  \sqrt{2}\,S_2(x)
 \end{pmatrix},
\end{equation}
where $Q(x)$ is the same bidoublet as in the MCHM, while in contrast to the MCHM there are two $\SUs{2}{L}\times \SUs{2}{R}$ singlets $S_1(x)$ and $S_2(x)$ present in the NMCHM.
This has the consequence that there is more freedom than in the MCHM for embedding the elementary fields into incomplete multiplets of the global symmetries.
The right-handed elementary fields can actually be embedded both in the fifth and in the sixth component of the $\mathbf{6}$ multiplet, which means they can have a mixing term with both $S_1(x)$ and $S_2(x)$.
To account for this, the right-handed composite-elementary mixings in the quark Lagrangian are replaced as
\begin{equation}
\begin{aligned}
\Delta_{u_R}\,\bar{\xi}_{uR}(x)
&\to
\left(
\Delta_{u_R}^5\,\bar{\xi}^5_{uR}(x)+\Delta_{u_R}^6\,\bar{\xi}^6_{uR}(x)
\right),
\\
\Delta_{d_R}\,\bar{\xi}_{dR}(x)
&\to
\left(
\Delta_{d_R}^5\,\bar{\xi}^5_{dR}(x)+\Delta_{d_R}^6\,\bar{\xi}^6_{dR}(x)
\right),
\end{aligned}
\end{equation}
where the matrices $\Delta_{u_R}^5$ and $\Delta_{d_R}^5$ are in general different from $\Delta_{u_R}^6$ and $\Delta_{d_R}^6$.
The embeddings of elementary fields into incomplete multiplets are then defined by
\begin{equation}
 \xi_{u L}(x) = \frac{1}{\sqrt{2}} \left( \begin{array}{c}
                                     d_L^{(0)}(x) \\
                                     -i d_L^{(0)}(x) \\
                                     u_L^{(0)}(x) \\
                                     i u_L^{(0)}(x) \\ 0\\0
                                    \end{array} \right),
\quad
 \xi_{u R}^5(x) = \left( \begin{array}{c}
                                       0 \\ 0 \\ 0 \\ 0 \\ u_R^{(0)}(x)\\ 0
                                      \end{array}
 \right),
\quad
 \xi_{u R}^6(x) = \left( \begin{array}{c}
                                       0 \\ 0 \\ 0 \\ 0 \\ 0 \\ u_R^{(0)}(x)
                                      \end{array}
 \right),
\end{equation}
\begin{equation}
  \xi_{d L}(x) = \frac{1}{\sqrt{2}} \left( \begin{array}{c}
                                     u_L^{(0)}(x) \\
                                     i u_L^{(0)}(x) \\
                                     -d_L^{(0)}(x) \\
                                     i d_L^{(0)}(x) \\ 0\\0
                                    \end{array} \right),
\quad
 \xi_{d R}^5(x) = \left( \begin{array}{c}
                                       0 \\ 0 \\ 0 \\ 0 \\ d_R^{(0)}(x) \\ 0
                                      \end{array}
 \right),
\quad
 \xi_{d R}^6(x) = \left( \begin{array}{c}
                                       0 \\ 0 \\ 0 \\ 0 \\ 0 \\ d_R^{(0)}(x)
                                      \end{array}
 \right).
\end{equation}
While the presence of two different mixing terms for right-handed quarks would introduce an additional source of flavor violation in models with a left-compositeness flavor symmetry, we have avoided this by assuming a $\Us{2}{RC}^3$ flavor symmetry in our analysis of the NMCHM.
Such a scenario has proved to be viable in our analysis of the MCHM, where different flavor symmetries have been compared.
The explicit mixing matrices used in the analysis of the NMCHM are given in appendix~\ref{app:comp_elem:NMCHM}.

Analogous to the MCHM, one can construct mass matrices for fermions and vector bosons using the above relations.
These matrices are given in appendix~\ref{app:Mass_matrices:NMCHM}.
An important difference to the MCHM is that they depend on both $\gb{h}$ and $\gb{\eta}$.
Consequently, also the effective potential is a function of $\gb{h}$ and $\gb{\eta}$.
This can lead to mass mixing between these two scalars.
In particular, the scalar mass matrix is given by the second derivatives at the minimum of the effective potential,
\begin{equation}
M^2_\text{scalar} = \left. \left( \begin{array}{cc}
                            \partial_\gb{h}^2 & \partial_\gb{h} \partial_\gb{\eta} \\ \partial_\gb{h} \partial_\gb{\eta} & \partial_\gb{\eta}^2
                           \end{array} \right) \,\, V_\text{eff}(\gb{h}, \gb{\eta}) \right|_{\gb{h}=v_\gb{h}, \gb{\eta}=v_\gb{\eta}},
\end{equation}
where $v_\gb{\eta}$ denotes the VEV of $\gb{\eta}$.
In the presence of non-zero off-diagonal terms, the mass eigenstates $h$ and $\eta$ have to be obtained by diagonalizing this matrix and are given by linear combinations of $\gb{h}$ and $\gb{\eta}$.

It is interesting to note that due to the structure of the potential, the off-diagonal terms vanish for $v_\gb{\eta}=0$.
Note that this does not imply a vanishing mass of $\eta$.
%
%
%
Interestingly, $v_\gb{\eta}$ strongly depends on the composite-elementary mixing matrices $\Delta_{u_R}^5$, $\Delta_{d_R}^5$, $\Delta_{u_R}^6$ and $\Delta_{d_R}^6$.
In particular, for vanishing $\Delta_{u_R}^5$  and $\Delta_{d_R}^5$, also $v_\gb{\eta}$ vanishes and there is no mixing between $\gb{\eta}$ and $\gb{h}$.
Furthermore, in the case $v_\gb{\eta}=0$, the mass matrices shown in appendix~\ref{app:Mass_matrices:NMCHM} resemble those of the MCHM and all particles in the NMCHM that are not present in the MCHM decouple.
In this sense, the MCHM is contained in the NMCHM as a limiting case in the limit $v_\gb{\eta}\to 0$.
The structure of the scalar potential in the NMCHM and this limiting case are discussed in further detail in~\cite{Niehoff:2017thesis}.
In general, it is assumed that $\gb{\eta}$ and $\gb{h}$ mix with each other.
In particular, the discussion of the collider constraints on $\eta$ presented in section~\ref{sec:direct_constraints:results:eta} is actually only meaningful if $v_\gb{\eta}\neq 0$ such that $\eta$ does not decouple and can be produced at particle colliders.

It should be noted that even in the absence of mixing, Wess-Zumino-Witten terms~\cite{Wess:1971yu,Witten:1983tw} could induce couplings of $\eta$ to gauge bosons.
In particular, they can contribute to the couplings $\eta Z Z$, $\eta W^+ W^-$, and $\eta G G$~\cite{Gripaios:2009pe,Bellazzini:2015nxw,Low:2015qep}.
However, these contributions strongly depend on the UV structure of the model, which is not specified in the effective approach used here.
Treating the contributions as free parameters in the numerical scan is pointless as far as they are not correlated with other parameters.
The minimization used in the scanning procedure (cf.\ section~\ref{sec:direct_constraints:numerical_strategy}) could simply tune them to zero to avoid experimental bounds.
Consequently, these contributions are neglected in our numerical analysis and our bounds should be considered as conservative.

\section{Direct collider constraints}\label{sec:direct_constraints:direct_constraints}


For being able to use direct constraints in the context of the numerical method described in section~\ref{sec:direct_constraints:numerical_strategy}, a central requirement on the numerical implementation of the direct constraints is that it is reasonably fast.
In particular, the time it takes to calculate the $\chi^2$ function for all direct constraints should be $\mathcal{O}(100\,{\rm ms})$.
This can actually be achieved by relying on experimental searches for narrow resonances that give bounds on the production cross section times the branching ratio as a function of the resonance mass for specific decay channels.
Constraining a given parameter point then requires the calculation of cross sections and branching ratios of all particles for which experimental searches should be considered.
In the above described models, this amounts to $\mathcal{O}(100)$ particles\footnote{%
This large number is mainly due to the full flavor structure in the quark sector that implies several composite partners for each SM quark.
}
(cf.\ the mass matrices in appendix~\ref{app:Mass_matrices}) and requires some simplifying assumptions.

\subsection{Simplifying assumptions}\label{sec:direct_constraints:direct_constraints:simplifying_assumption}
To make it possible to calculate all cross sections and branching ratios for $\mathcal{O}(100)$ particles in less than a second, the following simplifications are made:
\begin{itemize}
 \item For the production cross section of quark partners, only the model-independent NNLO QCD pair-production is considered.
 To this end, the cross section is computed over a wide range of quark partner masses with the \texttt{HATHOR} code~\cite{Aliev:2010zk}.
 The results are used to construct an interpolating function that allows for a very fast calculation of the pair-production cross section for arbitrary quark partner masses.
 However, this means that single production and pair production via heavy gluon resonances are neglected.

 Single production is relevant for heavy partners of SM quarks with a large degree of compositeness, i.e.\ usually for the partners of top and bottom, and can yield considerably larger cross sections than pair production (cf.~\cite{AguilarSaavedra:2009es,DeSimone:2012fs,Li:2013xba,Redi:2013eaa,Azatov:2013hya,
Delaunay:2013pwa,Backovic:2014uma,Aguilar-Saavedra:2013qpa,Matsedonskyi:2014mna,Buchkremer:2013bha}).
 In addition, since only one heavy resonance is produced, this requires less energy than pair production.
 Consequently, the experimental searches for singly produced quark partners are sensitive to higher resonance masses.
 While neglecting single production does not affect the results obtained from considering experimental searches for pair produced quark partners, it reduces the number of experimental analyses than can be used as constraints.

 While pair-production via heavy gluon resonances does in principle affect the results obtained from the considered experimental searches, these effects are assumed to be very small~\cite{Araque:2015cna}.
 Taking them into account would yield a slightly larger cross section and thus stronger bounds.
 In addition it would also allow to set additional indirect bounds on the heavy gluon partners (cf.~\cite{Carena:2007tn,Azatov:2015xqa,Araque:2015cna}).
 However, this is beyond the scope of the analyses presented here.

 \item In the calculations of branching ratios and boson cross sections, the narrow-width approximation (NWA) is used.
 While a narrow resonance is usually also assumed in the considered experimental analyses, heavy resonances that are kinematically allowed to decay to other heavy resonances can be very broad.
 Applying the same bounds to broad resonances as to narrow ones is problematic because the experimental searches are considerably less sensitive to broad resonances.
 Consequently, for such a broad resonance, the experimental bound obtained with the NWA would be too strong.
 This is taken into account by multiplying the $\chi^2$ value that corresponds to the bound on a given resonance with mass $m_R$ and width $\Gamma_R$ with a smooth function\footnote{%
 In the analysis in~\cite{Niehoff:2015iaa}, actually a hard cut at $\Gamma_R/m_R=5\%$ was used.
 This was changed in the analysis in~\cite{Niehoff:2016zso} because we found that it has the effect that for resonances with a width close but below 5\%, the scan tries to increase it above 5\% to avoid the experimental constraints.
 } that is close to one for $\Gamma_R/m_R<5\%$ and vanishes for $\Gamma_R/m_R\gg5\%$.

 \item All processes are only calculated to leading order.
 In particular, tree-level expressions are used for the branching ratios of vector bosons and fermions and for the vector boson production cross sections.
 For scalar branching ratios and production cross sections, the loop induced couplings to gluons, photons and $Z$ bosons are considered in addition to the tree-level couplings to other particles.
 Especially the loop induced coupling to gluons is essential for calculating the production cross section of neutral scalars that can couple to quarks.
 In particular, the scalar $\eta$ in the NMCHM is dominantly produced via gluon fusion.
 In this case, also higher order QCD corrections are approximately included by multiplying the $gg\to\eta$ production cross section by a K-Factor of 2.

 \item Only two-body decays are considered.
 While a coupling of a fermion to more than two particles correspond to an operator of dimension larger than four and is therefore suppressed, unsuppressed quartic couplings are in principle possible for vector bosons and scalars.
 However, the only scalar resonance considered is $\eta$ in the NMCHM, which would have to decay to three Higgses.
 Because the mass of $\eta$ is usually below 800~GeV (cf.\ section~\ref{sec:direct_constraints:results:eta}), such a decay would be phase-space suppressed.
 Decays of heavy vector bosons to three SM vector bosons on the other hand would require insertions of three composite-elementary mixings and are thus also suppressed.
 Taking these effects into account is beyond the scope of the analyses presented here.

 \item Only decays directly to SM particles are considered when setting bounds.
 In particular, decay chains involving several intermediate decays between heavy resonances are not considered.
 However, this is not a strong restriction because the lightest heavy resonances can only decay to SM particles for kinematical reasons.
 Since the experimental bounds are stronger for smaller masses and the production cross section is usually larger, these lightest heavy resonances usually yield the strongest bounds anyway.
 It should be noted that decays of heavy resonances to other heavy resonances are taken into account in the calculation of the total widths.
 This is important to derive reasonable branching ratios.


\end{itemize}

\subsection{Calculation of decay widths and branching ratios}\label{sec:direct_constraints:direct_constraints:widths_and_BRs}

A central aspect of deriving direct bounds on a given parameter point is to calculate decay widths and branching ratios.
The partial decay rate, or partial width, of a particle $R$ decaying to two particles $i$ and $j$ is given by~\cite{Agashe:2014kda,Eilam:1990zc}
\begin{equation}
 \Gamma_{R\to ij}=\frac{\sqrt{\lambda(m_R^2,m_i^2,m_j^2)}}{16\,\pi\,m_R^3}\,\overline{|\mathcal{M}_{R\to ij}|}^2,
\end{equation}
where $m_R$, $m_i$, and $m_j$ are the masses of $R$, $i$, and $j$, respectively, and the kinematic function $\lambda$ is defined as~\cite{Eilam:1990zc}
\begin{equation}\label{eq:triangle_lambda}
\lambda(a,b,c)=a^2+b^2+c^2-2(ab+ac+bc).
\end{equation}
In the above expression, $\overline{|\mathcal{M}_{R\to ij}|}^2$ denotes the squared amplitude of the process that has been averaged over the initial states and summed over the final states.
In the following, this will be just called the {\it amplitude squared} for convenience, but it should be understood that the averaging over the initial states and the summation over the final states is implied.
The total decay rate, or total width, of the particle $R$ is given by the sum over all of its partial widths, i.e.
\begin{equation}
 \Gamma_{R} = \sum_{ij}\Gamma_{R\to ij}.
\end{equation}
This expression relies on the assumption that the total width can be sufficiently good approximated by summing only over all partial widths of two-body decays (cf.\ section~\ref{sec:direct_constraints:direct_constraints:simplifying_assumption}).
The branching ratio for the decay of $R$ into the particles $i$ and $j$ is then simply the ratio of the corresponding partial width and the total width
\begin{equation}\label{eq:direct_constraints:BR}
 BR(R\to ij) = \Gamma_{R\to ij}/\Gamma_{R}.
\end{equation}
Consequently, even for calculating only the branching ratio of a single decay channel, it is necessary to calculate all partial widths to get the total width.
To this end, the following decay channels are considered:
\begin{itemize}
 \item A fermion resonance $R_F$ decaying to
 \begin{itemize}
  \item a fermion $i_F$ and a vector boson $j_V$,
  \item a fermion $i_F$ and a scalar $j_S$.
 \end{itemize}
 \item A vector resonance $R_V$ decaying to
 \begin{itemize}
  \item two fermions $i_F$ and $j_F$,
  \item two vector bosons $i_V$ and $j_V$,
  \item a vector bosons $i_V$ and a scalar $j_S$.
 \end{itemize}
 \item A scalar resonance $R_S$ decaying to
 \begin{itemize}
  \item two fermions $i_F$ and $j_F$,
  \item two scalars $i_S$ and $j_S$,
  \item two vector bosons $i_V$ and $j_V$.
 \end{itemize}
\end{itemize}
To calculate the decay widths for all these processes, the corresponding amplitudes are calculated at tree level, except for the decay $R_S\to i_V j_V$ with at least one massless vector boson in the final state, which is calculated at one loop.

\subsubsection{Amplitudes for decays of fermion resonances}
\renewcommand{\Re}{\operatorname{Re}}
\renewcommand{\Im}{\operatorname{Im}}

The generic tree-level matrix element for the decay $R_F\to i_F\,j_V$, can be written as
\begin{equation}
\mathcal{M}_{R_F \to i_F\,j_V} = \epsilon_{\mu}(q_j)\, \overline{i_F}(q_i)\, \gamma^{\mu} \left(g_{L}^{R_F\, i_F\, j_V}\, P_L + g_{R}^{R_F\, i_F\, j_V}\, P_R\right) R_F(q_R),
\end{equation}
where $g_L^{R_F\, i_F\, j_V}$ and $g_R^{R_F\, i_F\, j_V}$ are in general complex coupling constants and $q_R$, $q_i$, and $q_j$ are the momenta of $R_F$, $i_F$, and $j_V$, respectively.
Squaring, averaging over initial spins and summing final spins and polarizations yields
\begin{equation}
 \begin{aligned}
\overline{|\mathcal{M}_{R_F \to i_F\,j_V}|}^2=
 \frac{1}{2} \Bigg\{
&
\left(\left| g_L^{R_F\, i_F\, j_V}\right| ^2+\left| g_R^{R_F\, i_F\, j_V}\right| ^2\right) \left(\frac{\left(m_i^2-m_R^2\right)^2}{m_j^2}+m_i^2+m_R^2-2 \,m_j^2\right)
\\&
-12\,m_i\,m_R \left(\Re g_L^{R_F\, i_F\, j_V}\, \Re g_R^{R_F\, i_F\, j_V} + \Im g_L^{R_F\, i_F\, j_V}\, \Im g_R^{R_F\, i_F\, j_V}\right)
\Bigg\}.
 \end{aligned}
\end{equation}
The matrix element for the $R_F\to i_F\, j_S$ transition, where now the final state boson is not a vector but a scalar $j_S$, is given by
\begin{equation}
\mathcal{M}_{R_F \to i_F\,j_S} = \overline{i_F}(q_i) \left(g_{L}^{R_F\, i_F\, j_S}\, P_L + g_{R}^{R_F\, i_F\, j_S}\, P_R\right) R_F(q_R).
\end{equation}
After squaring, averaging over initial and summing over final spins, one finds
\begin{equation}
 \begin{aligned}
\overline{|\mathcal{M}_{R_F \to i_F\,j_S}|}^2=
 \frac{1}{2} \Bigg\{
&
\left(\left| g_L^{R_F\, i_F\, j_S}\right| ^2+\left| g_R^{R_F\, i_F\, j_S}\right| ^2\right) \left(m_i^2+m_R^2-m_j^2\right)
\\&
+4\,m_i\,m_R \left(\Re g_L^{R_F\, i_F\, j_S}\, \Re g_R^{R_F\, i_F\, j_S} + \Im g_L^{R_F\, i_F\, j_S}\, \Im g_R^{R_F\, i_F\, j_S}\right)
\Bigg\}.
 \end{aligned}
\end{equation}

\subsubsection{Amplitudes for decays of vector resonances}

The matrix element for a heavy vector boson $R_V$ decaying to a fermion $i_F$ and an anti-fermion $j_F$ is given by
\begin{equation}
\mathcal{M}_{R_V \to i_F\,j_F} = \epsilon_{\mu}^{*}(q_R)\, \overline{i_F}(q_{1})\, \gamma^{\mu} \left(g_{L}^{i_F\, j_F\, R_V} P_L + g_{R}^{i_F\, j_F\, R_V} P_R\right) j_F(q_{2}).
\end{equation}
Squaring the matrix element, averaging over the initial polarizations and summing over the final spins yields
\begin{equation}
 \begin{aligned}
\overline{|\mathcal{M}_{R_V \to i_F\,j_F}|}^2=
 \frac{1}{3} \Bigg\{
&
\left(\left| g_L^{i_F\, j_F\, R_V}\right| ^2+\left| g_R^{i_F\, j_F\, R_V}\right| ^2\right) \left(2 \,m_R^2-\frac{\left(m_{i}^2-m_{j}^2\right)^2}{m_R^2}-m_{i}^2-m_{j}^2\right)
\\&
+12\,m_{i}\,m_{j} \left(\Re g_L^{i_F\, j_F\, R_V}\, \Re g_R^{i_F\, j_F\, R_V} + \Im g_L^{i_F\, j_F\, R_V}\, \Im g_R^{i_F\, j_F\, R_V}\right)
\Bigg\}.
 \end{aligned}
\end{equation}
For the decay of a heavy vector boson $R_V$ to two light (but also massive) vector bosons $i_V$ and $j_V$, the matrix element is given by
\begin{equation}
\mathcal{M}_{R_V \to i_V\,j_V}
=
 g^{R_V\, i_V\, j_V}\,\epsilon _{\mu }^*(q_R)\, \epsilon _{\nu }(q_i)\, \epsilon _{\rho }(q_j)  \Big\{
\eta^{\nu  \rho } \left(q_i^{\mu }-q_j^{\mu }\right)
+
\eta^{\mu  \rho } \left(q_R^{\nu }+q_j^{\nu }\right)
+
\eta^{\mu  \nu } \left(-q_R^{\rho }-q_i^{\rho }\right)
\Big\},
\end{equation}
where $g$ is the coupling constant.
When squaring the matrix element, summing over the final polarizations and averaging over the initial ones, one finds
\begin{equation}
\begin{aligned}
\overline{|\mathcal{M}_{R_V \to i_V\,j_V}|}^2=
 \frac{\left(g^{R_V\, i_V\, j_V}\right)^2 \lambda(m_R^2,m_i^2,m_j^2)}{12\, m_R^2\, m_i^2\, m_j^2}
\Big\{&
m_R^4+m_i^4+m_j^4
\\&
+10  \left(m_R^2\, m_i^2+m_R^2\, m_j^2+m_i^2\, m_j^2\right)\Big\},
\end{aligned}
\end{equation}
where $\lambda(a,b,c)$ is defined in eq.~\eqref{eq:triangle_lambda}.
The matrix element for a heavy vector boson $R_V$ decaying into a light vector boson $i_V$ and a scalar $j_S$ can be written as
\begin{equation}
 \mathcal{M}_{R_V \to i_V\,j_S}
=
 g^{R_V\, i_V\, j_S}\,\epsilon _{\mu }^*(q_R)\, \epsilon _{\nu }(k_i)\,\eta^{\mu\nu},
\end{equation}
where in contrast to the dimensionless coupling constants used above, $g^{R_V\, i_V\, j_S}$ has mass dimension one.
Squaring this matrix element, summing over the final polarizations and averaging over the initial ones yields
\begin{equation}
\overline{|\mathcal{M}_{R_V \to i_V\,j_S}|}^2=
\frac{\left(g^{R_V\, i_V\, j_S}\right)^2}{12\,m_R^2\,m_i^2}
\Big\{
m_R^4+m_i^4+m_j^4+10\,m_R^2\, m_i^2-2\,m_j^2\left(m_R^2+m_i^2\right)
\Big\}.
\end{equation}

\subsubsection{Tree-level amplitudes for decays of scalar resonances}
The matrix element for a heavy scalar $R_S$ decaying to a fermion $i_F$ and an anti-fermion $j_F$ is given by
\begin{equation}
\mathcal{M}_{R_S \to i_F\,j_F} = \overline{i_F}(q_{1})\, \left(g_{L}^{i_F\,j_F\,R_S}\, P_L + g_{R}^{i_F\,j_F\,R_S}\, P_R\right) j_F(q_{2}).
\end{equation}
Squaring the matrix element, summing over final and averaging over initial states yields
\begin{equation}
\begin{aligned}
\overline{|\mathcal{M}_{R_V \to i_F\,j_F}|}^2=
&
\left(\left| g_L^{i_F\,j_F\,R_S}\right| ^2+\left| g_R^{i_F\,j_F\,R_S}\right| ^2\right) \left(m_R^2-m_i^2-m_j^2\right)
\\&
-4\,m_i\,m_R \left(\Re g_L^{i_F\,j_F\,R_S}\, \Re g_R^{i_F\,j_F\,R_S} + \Im g_L^{i_F\,j_F\,R_S}\, \Im g_R^{i_F\,j_F\,R_S}\right).
\end{aligned}
\end{equation}
For the matrix element of a heavy scalar $R_S$ decaying to two scalars $i_S$ and $j_S$, summing over final states and averaging over initial states is trivial.
The matrix element and the matrix element squared are simply given by
\begin{equation}
\mathcal{M}_{R_S \to i_S\,j_S}
=
g^{R_S\,i_S\,j_S},
\quad\quad\quad
|\mathcal{M}_{R_S \to i_S\,j_S}|^2= \left(g^{R_S\,i_S\,j_S}\right)^2,
\end{equation}
where $g^{R_S\,i_S\,j_S}$ has mass dimension one.
The matrix element for a heavy scalar $R_S$ decaying to two massive vector bosons $i_V$ and $j_V$ is given by
\begin{equation}
\mathcal{M}_{R_S \to i_V\,j_V}
=
g^{i_V\,j_V\,R_S}\,  \epsilon_{\mu }(q_i)\, \epsilon_{\nu}(q_j)\,\eta^{\mu  \nu },
\end{equation}
where $g^{i_V\,j_V\,R_S}$ has mass dimension one.
After squaring the matrix element, summing over final states and averaging over the initial ones, one finds
\begin{equation}
\overline{|\mathcal{M}_{R_S \to i_V\,j_V}|}^2=
\frac{\left(g^{i_V\,j_V\,R_S}\right)^2 }{4\, m_i^2\, m_j^2}\,\Big\{m_R^4+m_i^4+m_j^4+10\, m_i^2\, m_j^2-2\, m_R^2 \left(m_i^2+m_j^2\right)\Big\}.
\end{equation}

\subsubsection{One-loop amplitudes for decays of scalar resonances}
The dimension five operator that couples a scalar $R_S$ to vector bosons,
\begin{equation}
 R_S\,\left(
 g^{i_V\,j_V\,R_S}_{\rm eff}\,V^{\mu\nu}\,V_{\mu\nu}
 +
 \tilde{g}^{i_V\,j_V\,R_S}_{\rm eff}\,\tilde{V}^{\mu\nu}\,V_{\mu\nu}
 \right),
\end{equation}
yields the following matrix element for a decay of the scalar $R_S$ to vector bosons $i_V$ and $j_V$:
\begin{equation}
\mathcal{M}_{R_S \to i_V\,j_V}
=
4\,  \epsilon_{\mu }(q_i)\, \epsilon_{\nu}(q_j)\,\eta^{\mu  \nu }
\left(
 g^{i_V\,j_V\,R_S}_{\rm eff}\,\left(q_i^\nu\,q_j^\mu-\eta^{\mu\nu}\, q_i\cdot q_j\right)
-\tilde{g}^{i_V\,j_V\,R_S}_{\rm eff}\,\epsilon^{\alpha\beta\mu\nu}\,q_{i\,\beta}\,q_{j\,\alpha}
\right).
\end{equation}
Squaring this matrix element, averaging over the initial states and summing over the final ones yields
\begin{equation}
\overline{|\mathcal{M}_{R_S \to i_V\,j_V}|}^2
=
8\,m_R^4\,\left(\left(g^{i_V\,j_V\,R_S}_{\rm eff}\right)^2+\left(\tilde{g}^{i_V\,j_V\,R_S}_{\rm eff}\right)^2\right).
\end{equation}
The effective couplings $g^{i_V\,j_V\,R_S}_{\rm eff}$ and $\tilde{g}^{i_V\,j_V\,R_S}_{\rm eff}$ have mass dimension $-1$ and are generated at one loop.
The scalar coupling $g^{i_V\,j_V\,R_S}_{\rm eff}$ can receive contributions from both fermion and vector boson loops, while the pseudoscalar coupling $\tilde{g}^{i_V\,j_V\,R_S}_{\rm eff}$ only receives fermion contributions.
The different contributions will be indicated by an additional subscript such that
\begin{equation}
 g^{i_V\,j_V\,R_S}_{\rm eff} = g^{i_V\,j_V\,R_S}_{\rm eff,F} +g^{i_V\,j_V\,R_S}_{\rm eff,V},
\quad\quad\quad
\tilde{g}^{i_V\,j_V\,R_S}_{\rm eff} = \tilde{g}^{i_V\,j_V\,R_S}_{\rm eff,F}.
\end{equation}
The explicit expressions of the effective couplings depend on whether both final state vector bosons $i_V$ and $j_V$ are massless or one of them is massive.
To simplify the fermion contributions, it is convenient to define vector and axial vector couplings of the fermion~$k$ in the loop by
\begin{equation}
\begin{aligned}
g_V^{k_F\,k_F\,X}&=\frac{1}{2}\,\left(g_R^{k_F\,k_F\,X}+g_L^{k_F\,k_F\,X}\right),
\quad&
g_A^{k_F\,k_F\,X}&=\frac{1}{2}\,\left(g_R^{k_F\,k_F\,X}-g_L^{k_F\,k_F\,X}\right),
\end{aligned}
\end{equation}
where $X\in\{R_S,i_V,j_V\}$.
In addition, it is useful to define the color factor $N_C^k$ of the fermion~$k$, where $N_C^k=3$ for $k$ a quark and $N_C^k=1$ for $k$ a lepton.
The individual loop contributions depend on the kinematic variable
\begin{equation}
 x_k = 4\,\frac{m_k^2}{m_R^2}.
\end{equation}

For the case where both $i_V$ and $j_V$ are massless, the contributions from fermion loops are given by
\begin{equation}
\begin{aligned}
 g^{i_V\,j_V\,R_S}_{{\rm eff},F}\big|^{m_i=0}_{m_j=0}
 &=
 \frac{-1}{16\,\pi^2\,m_R}
 \sum_{k}\,
 N_C^k\ g_V^{k_F\,k_F\,R_S}\,g_V^{k_F\,k_F\,i_V}\,g_V^{k_F\,k_F\,j_V}\,
 \frac{A_F(x_k)}{\sqrt{x_k}},
 \\
 \tilde{g}^{i_V\,j_V\,R_S}_{{\rm eff},F}\big|^{m_i=0}_{m_j=0}
 &=
 \frac{-1}{16\,\pi^2\,m_R} \sum_{k}\,
 N_C^k\ g_A^{k_F\,k_F\,R_S}\,g_V^{k_F\,k_F\,i_V}\,g_V^{k_F\,k_F\,j_V}\,
 \frac{\tilde{A}_F(x_k)}{\sqrt{x_k}},
\end{aligned}
\end{equation}
while those from vector boson loops are
\begingroup
\postdisplaypenalty=100
\begin{equation}
\begin{aligned}
 g^{i_V\,j_V\,R_S}_{{\rm eff},V}\big|^{m_i=0}_{m_j=0}
 &=
 \frac{1}{16\,\pi^2\,m_R^2}
 \sum_{k}\,
 g^{k_V\,k_V\,R_S}\,g^{k_V\,k_V\,i_V}\,g^{k_V\,k_V\,j_V}\,
 \frac{A_V(x_k)}{x_k}.
\end{aligned}
\end{equation}
The loop functions $A_F(x_k)$, $\tilde{A}_F(x_k)$, and $A_V(x_k)$ are given in appendix~\ref{app:loop-functions}.
\endgroup

For the case where one of the vector bosons in the final state is massive, which is without loss of generality chosen to be $j_V$, the additional scale $m_j$ enters via the kinematic variable
\begin{equation}
 y_k = 4\,\frac{m_k^2}{m_j^2}.
\end{equation}
In this case, the fermion loop contributions are given by
\begin{equation}
\begin{aligned}
 g^{i_V\,j_V\,R_S}_{{\rm eff},F}\big|^{m_i=0}_{m_j\neq0}
 &=
 \frac{-1}{16\,\pi^2\,m_R}
 \left(1-\frac{m_j^2}{m_R^2}\right)
 \sum_{k}\,
 N_C^k\ g_V^{k_F\,k_F\,R_S}\,g_V^{k_F\,k_F\,i_V}\,g_V^{k_F\,k_F\,j_V}\,
 \frac{B_F(x_k,y_k)}{\sqrt{x_k}},
 \\
 \tilde{g}^{i_V\,j_V\,R_S}_{{\rm eff},F}\big|^{m_i=0}_{m_j\neq0}
 &=
 \frac{-1}{16\,\pi^2\,m_R}
 \left(1-\frac{m_j^2}{m_R^2}\right)
 \sum_{k}
 N_C^k\ g_A^{k_F\,k_F\,R_S}\,g_V^{k_F\,k_F\,i_V}\,g_V^{k_F\,k_F\,j_V}\,
 \frac{\tilde{B}_F(x_k,y_k)}{\sqrt{x_k}},
\end{aligned}
\end{equation}
while those from vector boson loops are
\begin{equation}
\begin{aligned}
 g^{i_V\,j_V\,R_S}_{{\rm eff},V}\big|^{m_i=0}_{m_j\neq0}
 &=
 \frac{1}{16\,\pi^2\,m_R^2}
 \left(1-\frac{m_j^2}{m_R^2}\right)
 \sum_{k}\,
 g^{k_V\,k_V\,R_S}\,g^{k_V\,k_V\,i_V}\,g^{k_V\,k_V\,j_V}\,
 \frac{B_V(x_k,y_k)}{x_k}.
\end{aligned}
\end{equation}
The loop functions $B_F(x_k)$, $\tilde{B}_F(x_k)$ and $B_V(x_k)$ are given in appendix~\ref{app:loop-functions}.

\subsection{Calculation of boson production cross sections}\label{sec:direct_constraints:direct_constraints:boson_xsec}

In contrast to the production cross section of quark partners, for which the model-independent QCD result is used, the cross section of vector and scalar bosons has to be calculated for each of them individually at each parameter point.
However, to simplify this task, it is possible to make use of the Breit-Wigner formula for the cross section of the $2\to 2$ process $ij\to kl$ mediated by a resonance $R$ with partial widths $\Gamma_{R\to ij}$ and $\Gamma_{R\to kl}$ and a total width~$\Gamma_R$ (cf.~\cite{Agashe:2014kda,Chivukula:2016hvp,Harris:2011bh}),
\begin{equation}
\sigma_{ij\to R\to kl}(\hat s)= 16\,\pi\,\frac{S_R\,c_R\,(1+\delta_{ij})}{S_i\,S_j\,c_i\,c_j}\,
\frac{\Gamma_{R\to ij}\,\Gamma_{R\to kl}}{(\hat s - m_R^2)^2+m_R^2\,\Gamma_R^2}
,
\end{equation}
where $\sqrt{\hat s}$ is the center of mass energy of $i$ and $j$.
The factors $S_R$, $S_i$, and $S_j$ denote the number of polarizations or spins of the particles $R$, $i$ and $j$, respectively, while $c_R$, $c_i$, and $c_j$ denote their color multiplicity factors.
In the calculation of $\Gamma_{R\to ij}$, the polarizations and colors of $R$ are averaged over while those of $i$ and $j$ are summed over.
In addition, a symmetry factor of $2$ is introduced if $i$ and $j$ are identical final states.
On the other hand, in the above cross section, $i$ and $j$ denote the initial states over which one wants to average and $R$ denotes intermediate states one wants to sum over.
This is taken into account by introducing the factors $S_R$, $S_i$, $S_j$, $c_R$, $c_i$, $c_j$, and $(1+\delta_{ij})$ in the above expression.
In the NWA employed here (cf.\ section~\ref{sec:direct_constraints:direct_constraints:simplifying_assumption}), one can use $\Gamma_R^2\ll m_R^2$ to approximate (cf.\ e.g.~\cite{Chivukula:2016hvp,Harris:2011bh})
\begin{equation}
\frac{1}{(\hat s - m_R^2)^2+m_R^2\,\Gamma_R^2}\approx
\frac{\pi}{\Gamma_R\,m_R}
\delta(\hat s - m_R^2).
\end{equation}
With this approximation, the above cross section simplifies to
\begin{equation}
\sigma_{ij\to R\to kl}(\hat s)=
\frac{16\,\pi}{m_R}
\,\frac{S_R\,c_R\,(1+\delta_{ij})}{S_i\,S_j\,c_i\,c_j}\,\Gamma_{R\to ij}\,
\delta(\hat s - m_R^2)\,
BR(R\to kl)
,
\end{equation}
where the branching ratio $BR(R\to kl)=\Gamma_{R\to kl}/\Gamma_R$ is introduced (cf.~\eqref{eq:direct_constraints:BR}).
This suggests to define the production cross section for the $2\to 1$ process $ij\to R$ by
\begin{equation}\label{eq:direct_constraints:partonic_xsec}
\sigma_{ij\to R}(\hat s)= \frac{16\,\pi^2}{m_R}\,\frac{S_R\,c_R\,(1+\delta_{ij})}{S_i\,S_j\,c_i\,c_j}\,\Gamma_{R\to ij}\,\delta(\hat s - m_R^2).
\end{equation}
This expression is very convenient since all partial widths of the resonance $R$ are calculated anyway in the derivation of the branching ratios (cf.\ section~\ref{sec:direct_constraints:direct_constraints:widths_and_BRs}).
Consequently, the additional computing time for evaluating the cross section $\sigma_{ij\to R}(\hat s)$ is negligible.
However, this is not the final result for the production of resonances at a hadron collider.
In this case, the elementary initial state particles $i$ and $j$ have to be partons of the colliding hadrons.
Using the parton luminosity $\mathcal L_{ij}^{p_1\,p_2}(s,\hat s)$ of partons $i$ and $j$ in a collision of the two hadrons $p_1$ and $p_2$ with center of mass energy $s$, one can express the hadronic cross section as
\begin{equation}\label{eq:direct_constraints:hadronic_xsec}
 \sigma_{p_1\,p_2\to R}(s)=\sum_{i,j}\,\int\!\frac{d{\hat s}}{s}\ \sigma_{ij\to R}(\hat s)\,\mathcal{L}_{ij}^{p_1\,p_2}(s,\hat s).
\end{equation}
The hadrons $p_1$ and $p_2$ are two protons for collisions at the LHC, i.e.\ $p_1\big|_{\rm LHC}=p_2\big|_{\rm LHC}=p$, while they are a proton and an anti-proton for collisions at the Tevatron, i.e.\ $p_1\big|_{\rm Tevatron}=p$ and $p_2\big|_{\rm Tevatron}=\bar{p}$.
In terms of the parton distribution function $f_i(x,\mu^2)$ of a parton $i$ with momentum fraction $x$ at the renormalization scale $\mu$, the parton luminosity can be expressed as
\begin{equation}
\mathcal L_{ij}^{p_1\,p_2}(s,\hat s)=
\int_{\hat s/s}^1 \frac{dx}{x}
f_{i}(x,\hat s)
f_{j}\!\left(\frac{\hat s}{x\,s},\hat s\right)
\left(1+\delta_{p_1\,p_2}\right)
\,,
\end{equation}
where, importantly, a factor 2 is introduced for the LHC, where $\delta_{p\,p}=1$, while this factor is not present for the Tevatron, where $\delta_{p\,\bar{p}}=0$.
The parton luminosity has to be calculated for each parton, each collider and each collider's center of mass energy $\sqrt{s}$ for the whole range of possible parton center of mass energies $\sqrt{\hat s}$.
This is done by employing the \texttt{LHAPDF} software~\cite{Buckley:2014ana}.
From the resulting parton luminosities\footnote{%
To be able to calculate the vector boson fusion (VBF) production cross section with the simplified approach presented here, additional effective parton luminosities for the electroweak gauge bosons obtained by means of the effective W approximation (EWA)~\cite{Dawson:1984gx,Chanowitz:1984ne,Altarelli:1987ue,Pappadopulo:2014qza} are included into our numerical code.
\label{fn:EWA}
}, an interpolating function is constructed
that can be used in the parameter scan to efficiently calculate the hadronic vector and scalar boson production cross sections.
Combining eqs.~\eqref{eq:direct_constraints:partonic_xsec} and~\eqref{eq:direct_constraints:hadronic_xsec}, the final result for the hadronic boson cross section used in our numerical code can be expressed as
\begin{equation}
\sigma_{p_1\,p_2\to R}(s)= \frac{16\,\pi^2\,S_R\,c_R}{m_R}\,\sum_{i,j} \frac{1+\delta_{ij}}{S_i\,S_j\,c_i\,c_j}\,\Gamma_{R\to ij}\,\frac{\mathcal L_{ij}^{p_1\,p_2}(s,m_R)}{s}.
\end{equation}

\subsection{Applying the experimental bounds}\label{sec:direct_constraints:direct_constraints:exp_bounds}
As soon as the production cross sections and branching ratios of the particles in the considered model are calculated for a given parameter point, one can use them to compare the predictions to the experimental data.
The experimental searches usually give bounds in terms of 95\% CL upper limits on the cross section $\sigma_{p_1\,p_2\to R}(s)$ times the branching ratio $BR(R\to ij)$ for a given decay $R\to ij$ as a function of the resonance mass $m_R$.
This mass dependent observed 95\% CL upper limit will be denoted by
\begin{equation}
 \left\{\sigma_{p_1\,p_2\to R}\times BR(R\to ij)\right\}^\text{95\% CL}_\text{observed}(m_R).
\end{equation}
To use this bound in the numerical method described in section~\ref{sec:direct_constraints:numerical_strategy}, it has to be converted into a $\chi^2$ value.
To this end, it is assumed that the $\chi^2$ value scales linearly with the ratio of the calculated $\sigma\times BR$ and the observed 95\% CL upper limit.
In the case where both are equal, i.e.\ where the calculated value is excluded at the 95\% CL, this is interpreted as $\chi^2=4$.
With these assumptions, the $\chi^2$ contribution of a single experimental search can be expressed as
\begin{equation}\label{eq:direct_constraints:chi2_direct}
 \chi^2 = 4\,\frac{\sigma_{p_1\,p_2\to R}(s)\, BR(R\to ij)}{
 \left\{\sigma_{p_1\,p_2\to R}\times BR(R\to ij)\right\}^\text{95\% CL}_\text{observed}(m_R)
 },
\end{equation}
where $\sigma_{p_1\,p_2\to R}(s)$ and $BR(R\to ij)$ are the calculated cross section and branching ratio for a given parameter point according to sections~\ref{sec:direct_constraints:direct_constraints:widths_and_BRs} and~\ref{sec:direct_constraints:direct_constraints:boson_xsec}.
It should be noted that in our numerical analyses, parameter points are only discarded if they violate any individual constraint by more than 3$\sigma$, which corresponds to $\chi^2=9$~(cf.\ section~\ref{sec:direct_constraints:numerical_strategy}).
Consequently, they are actually allowed to violate the experimentally observed 95\% CL upper limit by a small amount.
This can be seen in the plots shown in section~\ref{sec:direct_constraints:results}.

There is a peculiarity in the case of searches for pair produced quark partners.
Depending on the analysis, it is either assumed that exactly one, both, or at least one of the two quark partners decays in the analyzed decay channel.
In the first case, the expression in eq.~\eqref{eq:direct_constraints:chi2_direct} can readily be applied.
In the other two cases, this expression is corrected by the following replacements:
\begin{itemize}
 \item $BR(R\to ij)\to BR(R\to ij)^2$ \newline in the case of both quark partners assumed to decay to~$ij$,
 \item $BR(R\to ij)\to 1-\left(1-BR(R\to ij)\right)^2$\newline in the case where one or both of the quark partners are assumed to decay to $ij$.
\end{itemize}

To be able to apply direct constraints in as many decay channels as possible, a large number of experimental searches is included into our numerical code.
These searches are listed in appendix~\ref{app:seaches:MCHM} for the analysis of the MCHM and in appendix~\ref{app:seaches:NMCHM} for the one performed in the NMCHM.
The experimental data for all these searches is, up to few exceptions, only available in the form of so called ``Brazil band'' plots.
Digitizing the large number of plots by hand is tremendously tedious.
Hence, the open source code \texttt{svg2data}~\cite{peter_stangl_2017_292635} has been developed to automatize this task.

\section{Results}\label{sec:direct_constraints:results}

\subsection{Quark resonances}\label{sec:direct_constraints:results:quarks}
\begin{figure}[p!]
 \centering
\begin{subfigure}[b]{0.42\textwidth}
 \includegraphics[width=\textwidth]{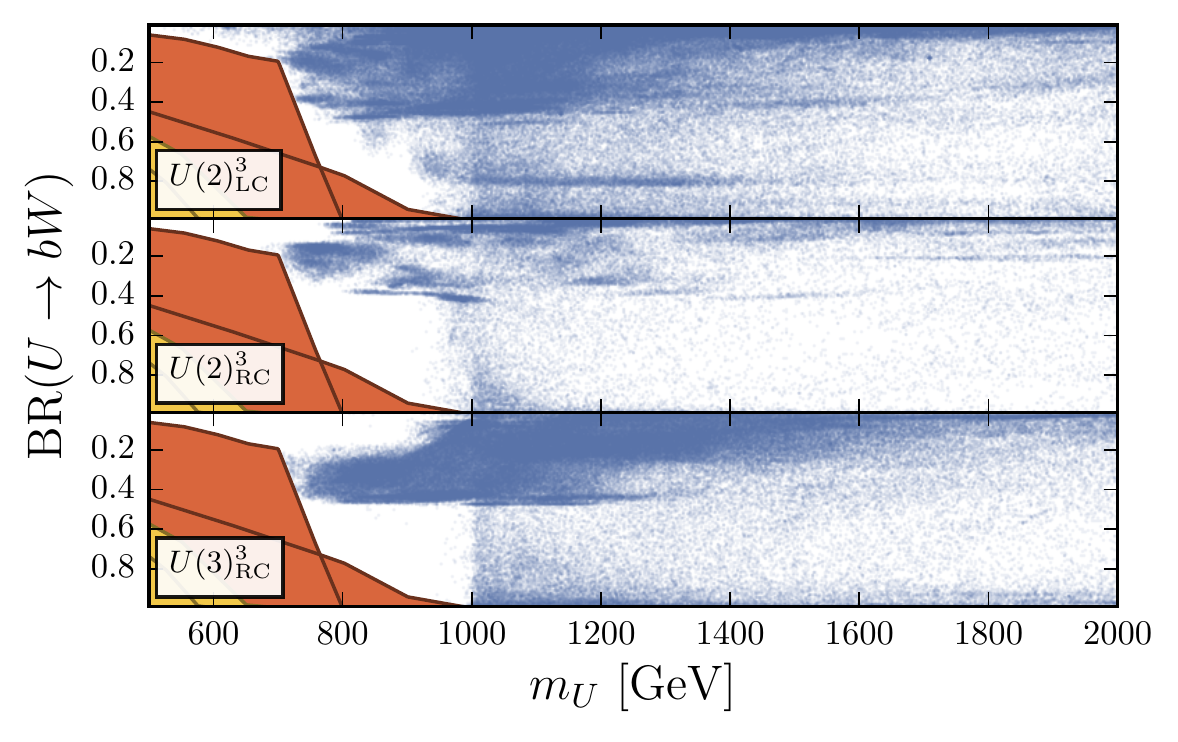}
 \includegraphics[width=\textwidth]{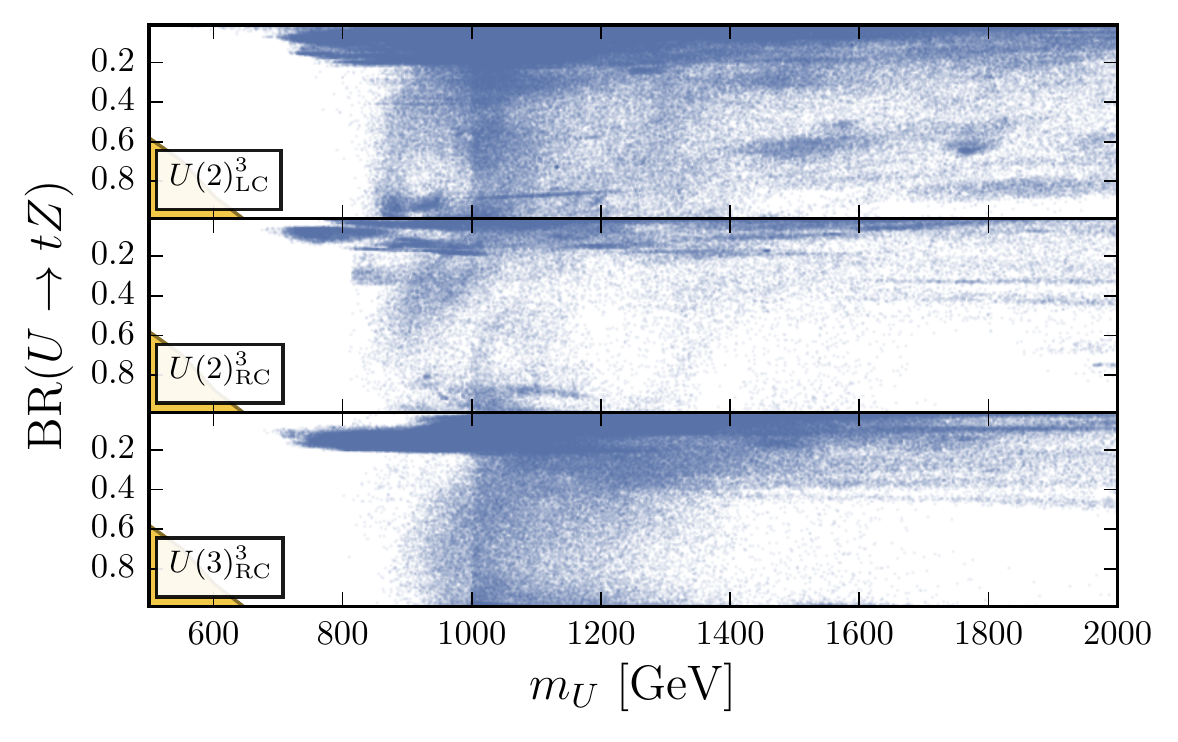}
 \includegraphics[width=\textwidth]{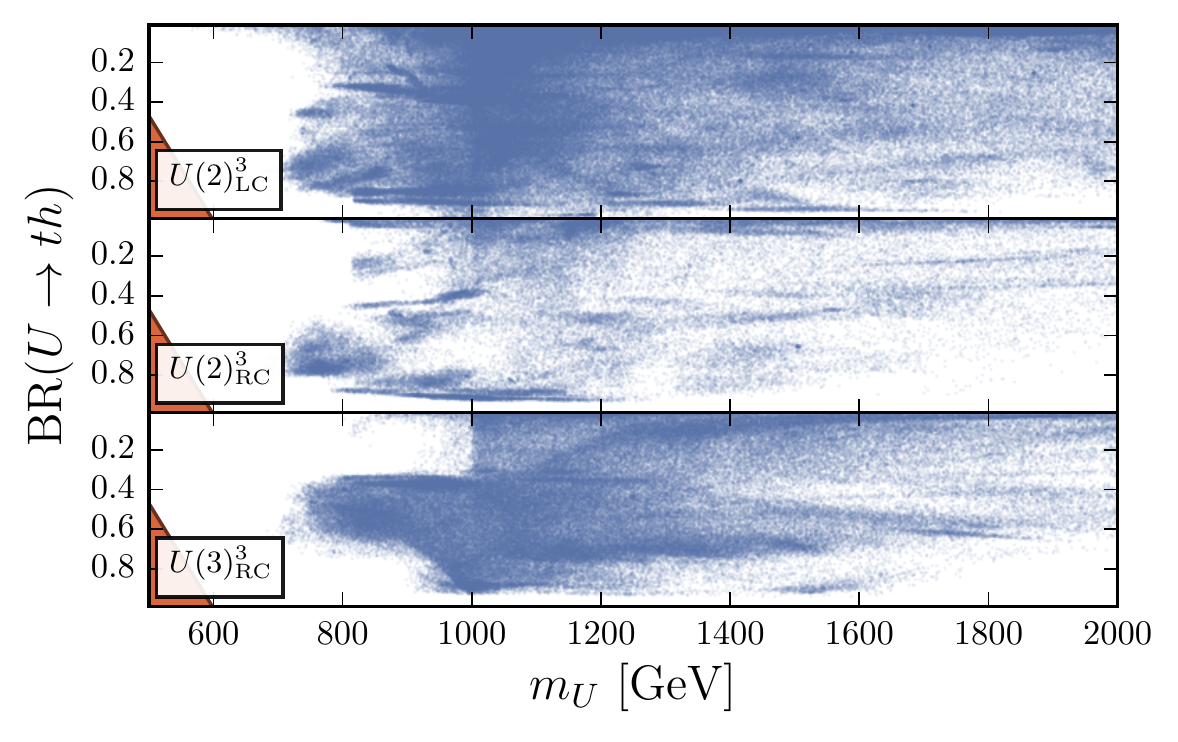}
 \includegraphics[width=\textwidth]{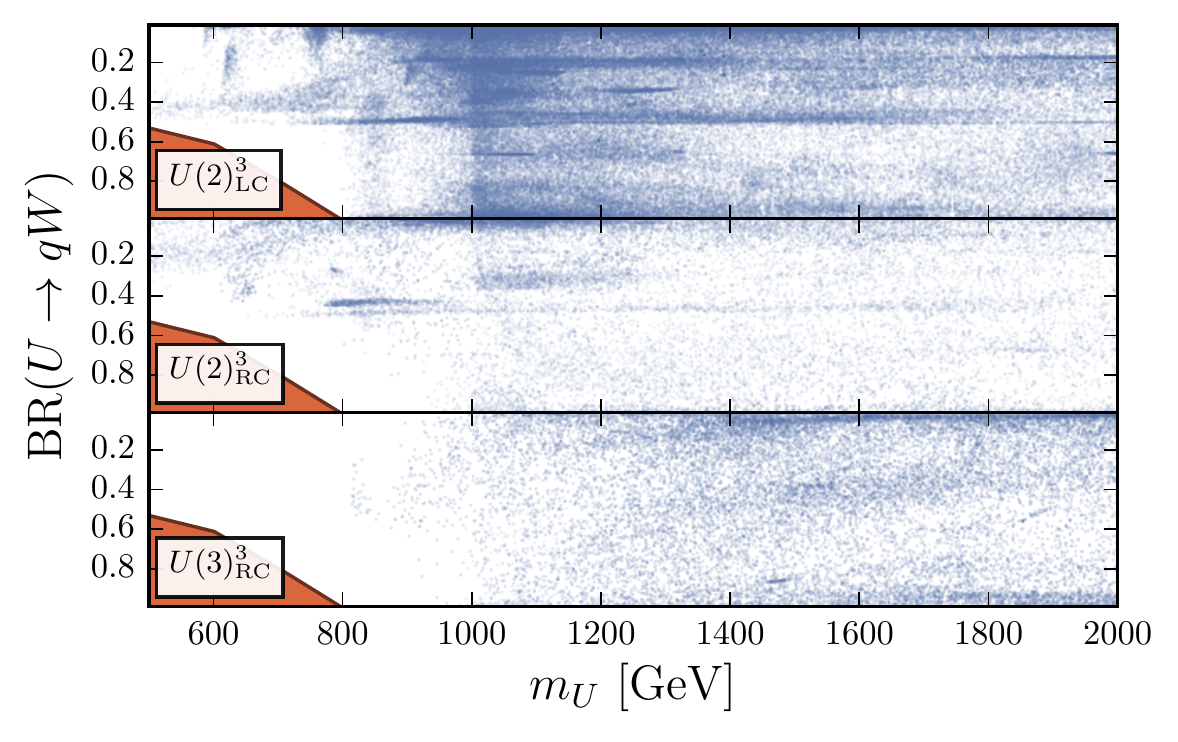}
 \caption{}\label{fig:direct_constraints:results:U_bounds_MCHM}
\end{subfigure}
\begin{subfigure}[b]{0.57\textwidth}
 \includegraphics[width=\textwidth]{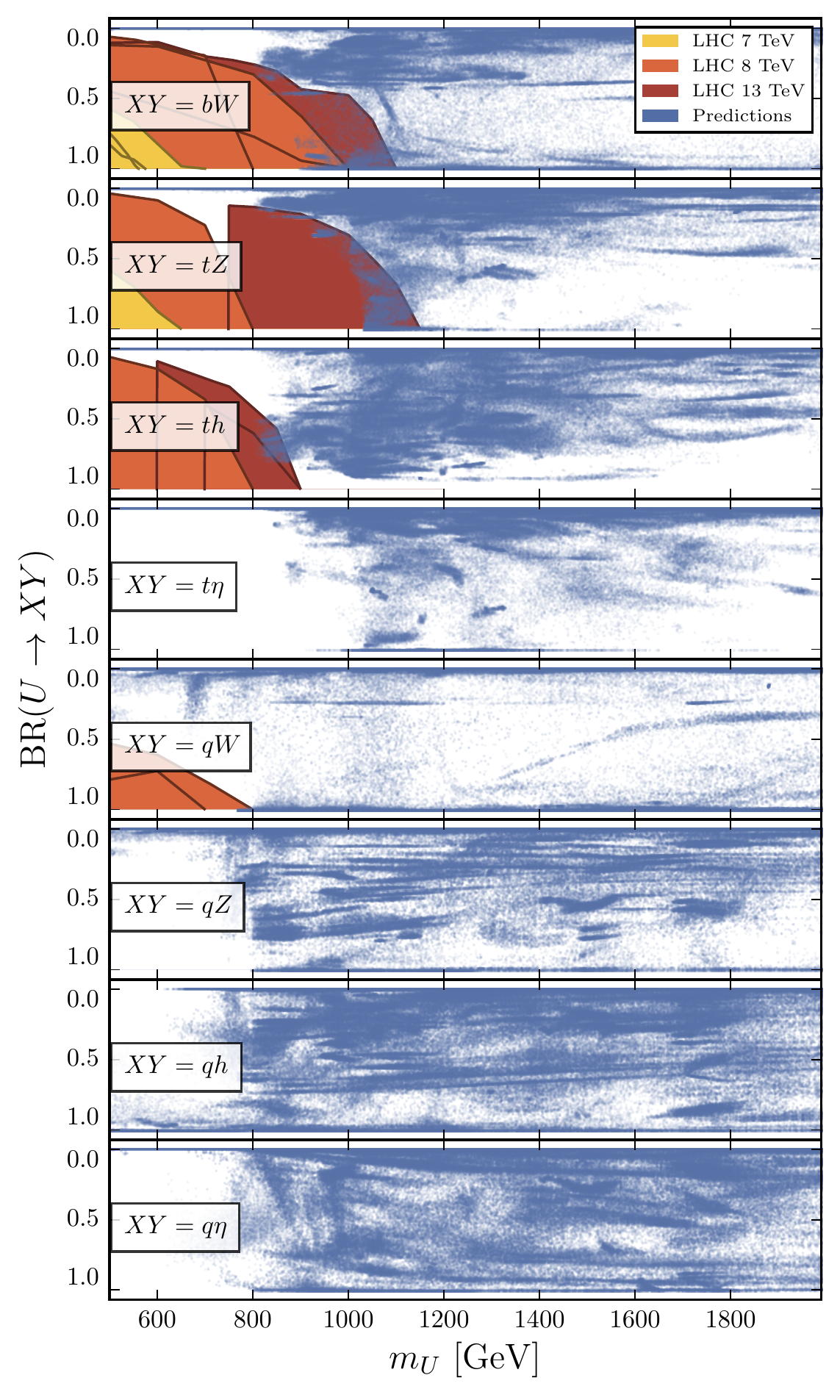}
 \vspace{7pt}
 \caption{}\label{fig:direct_constraints:results:U_bounds_NMCHM}
\end{subfigure}
 \caption{Predictions for masses and branching ratios of up-type quark resonances.
 Experimental bounds from the LHC running at different center of mass energies are shown as colored areas.
 Predictions and experimental bounds from the MCHM analysis are shown in (a), while those from the NMCHM analysis are shown in (b).
 The experimental searches included in (a) are listed in table~\ref{tab:app:MCHM:exp_quark_res}, while those included in (b) are shown in table~\ref{tab:app:NMCHM:exp_quark_res}.}
 \label{fig:direct_constraints:results:U_bounds_both_models}
\end{figure}
\begin{figure}[p!]
 \centering
\begin{subfigure}[b]{0.42\textwidth}
 \includegraphics[width=\textwidth]{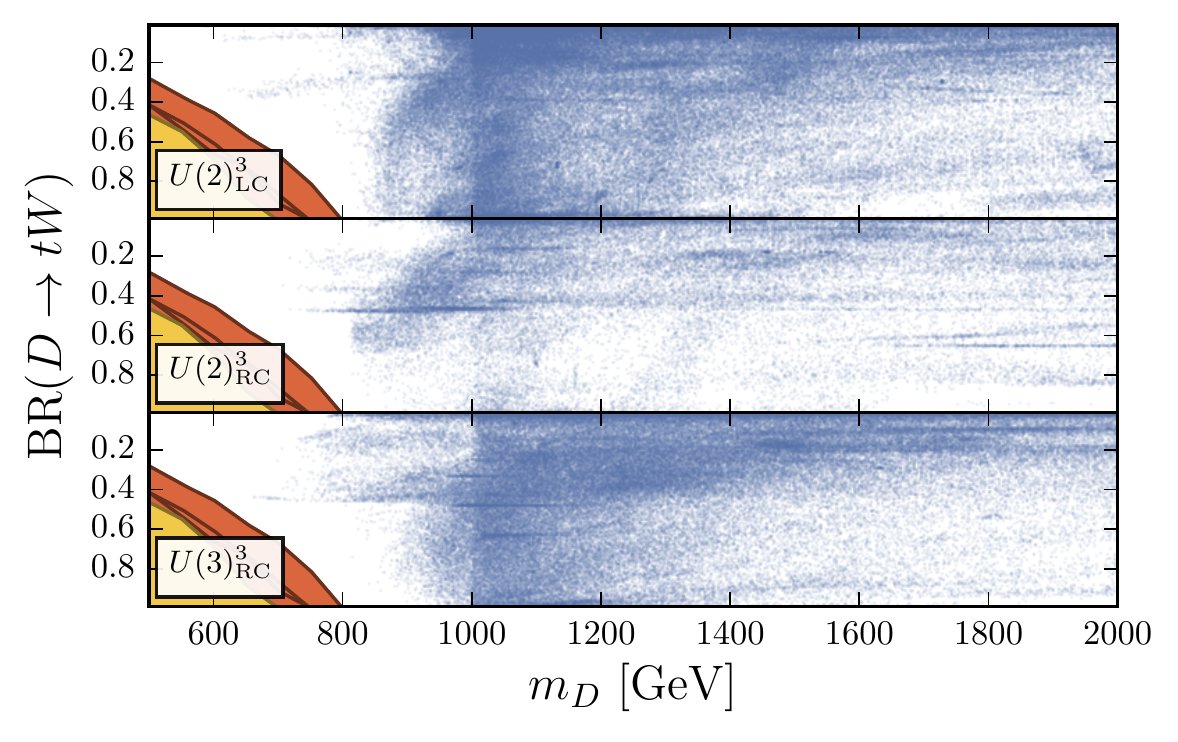}
 \includegraphics[width=\textwidth]{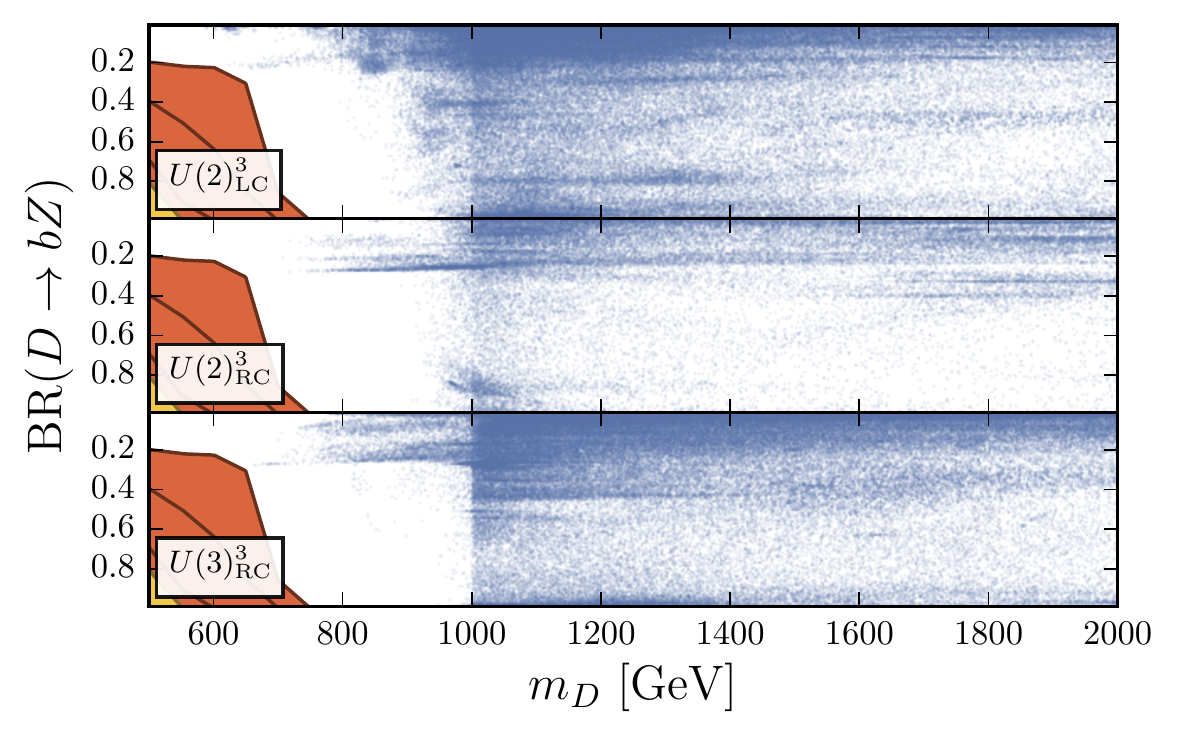}
 \includegraphics[width=\textwidth]{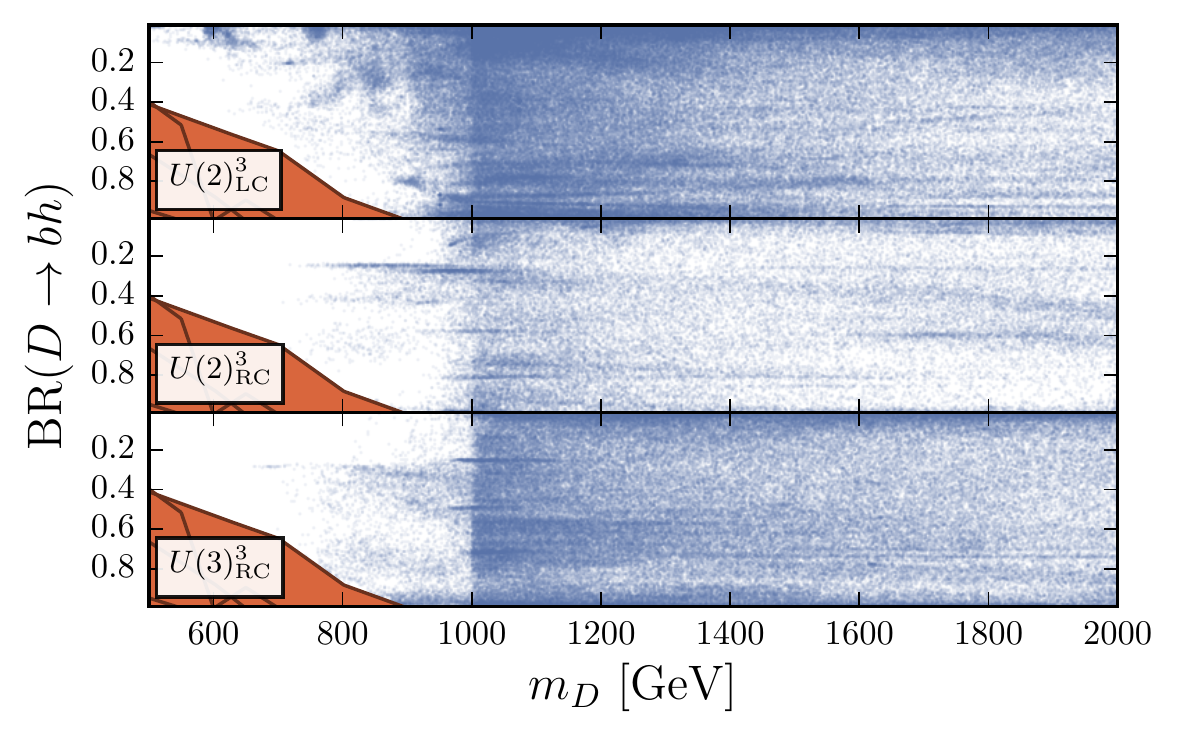}
 \includegraphics[width=\textwidth]{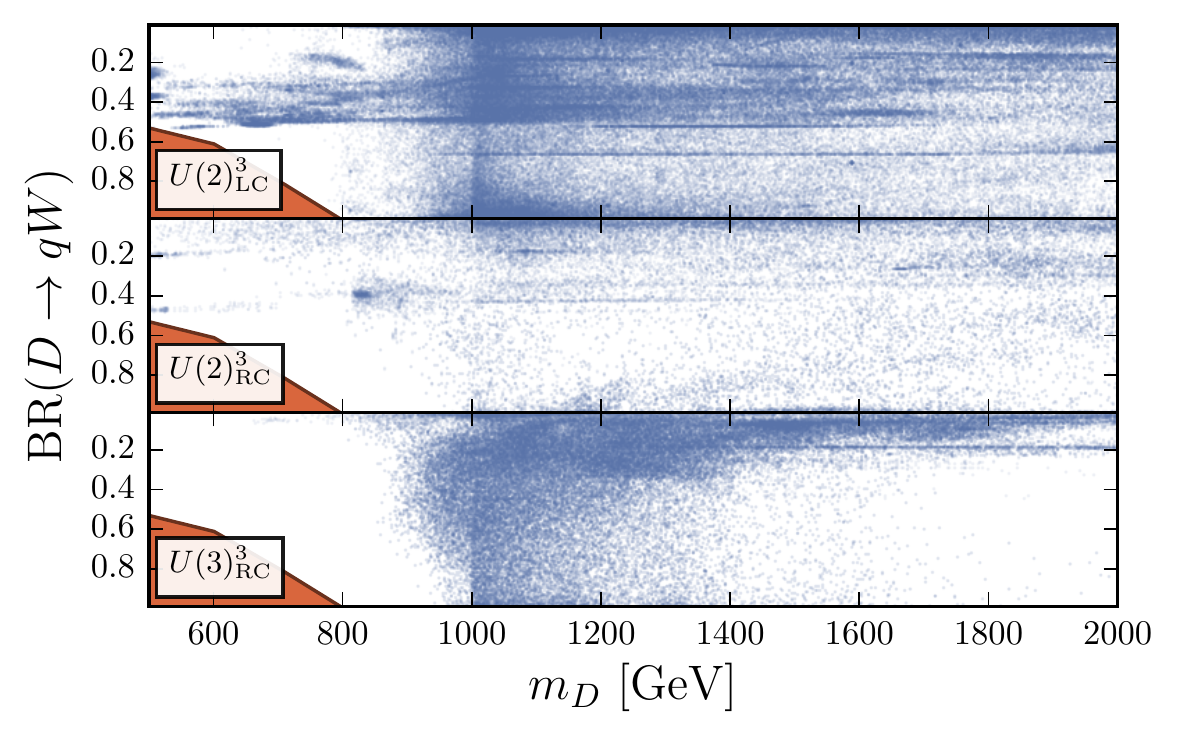}
 \caption{}\label{fig:direct_constraints:results:D_bounds_MCHM}
\end{subfigure}
\begin{subfigure}[b]{0.57\textwidth}
 \includegraphics[width=\textwidth]{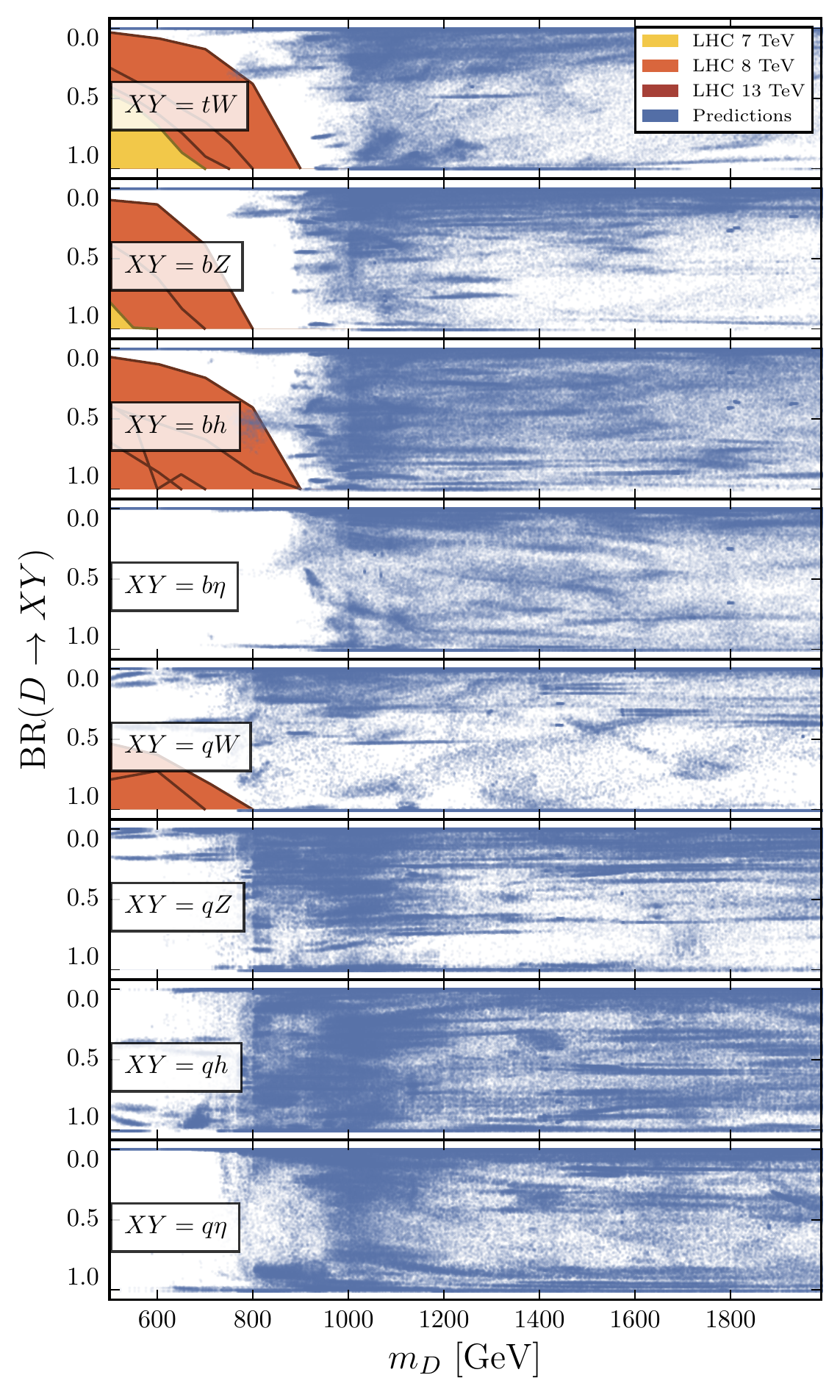}
 \vspace{7pt}
 \caption{}\label{fig:direct_constraints:results:D_bounds_NMCHM}
\end{subfigure}
 \caption{Predictions for masses and branching ratios of down-type quark resonances.
 Experimental bounds from the LHC running at different center of mass energies are shown as colored areas.
 Predictions and experimental bounds from the MCHM analysis are shown in (a), while those from the NMCHM analysis are shown in (b).
 The experimental searches included in (a) are listed in table~\ref{tab:app:MCHM:exp_quark_res}, while those included in (b) are shown in table~\ref{tab:app:NMCHM:exp_quark_res}.}
 \label{fig:direct_constraints:results:D_bounds_both_models}
\end{figure}
\begin{figure}[t]
 \centering
\begin{subfigure}[b]{0.44\textwidth}
 \includegraphics[width=\textwidth]{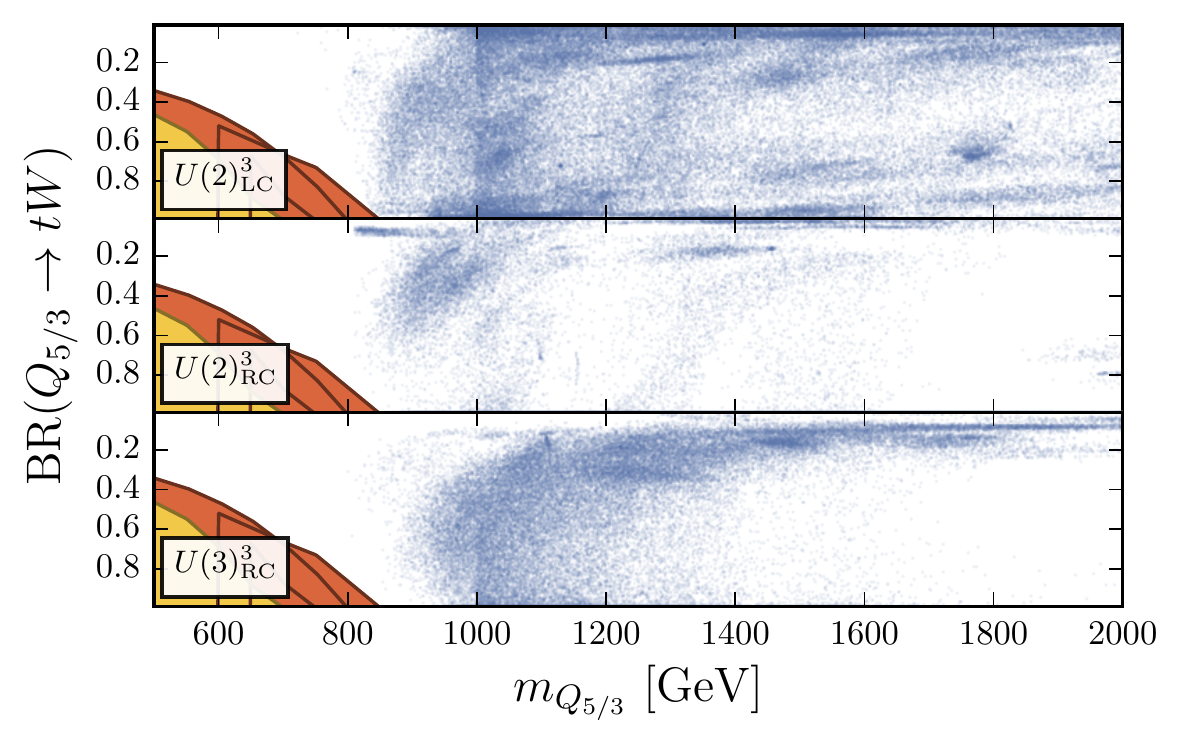}
 \includegraphics[width=\textwidth]{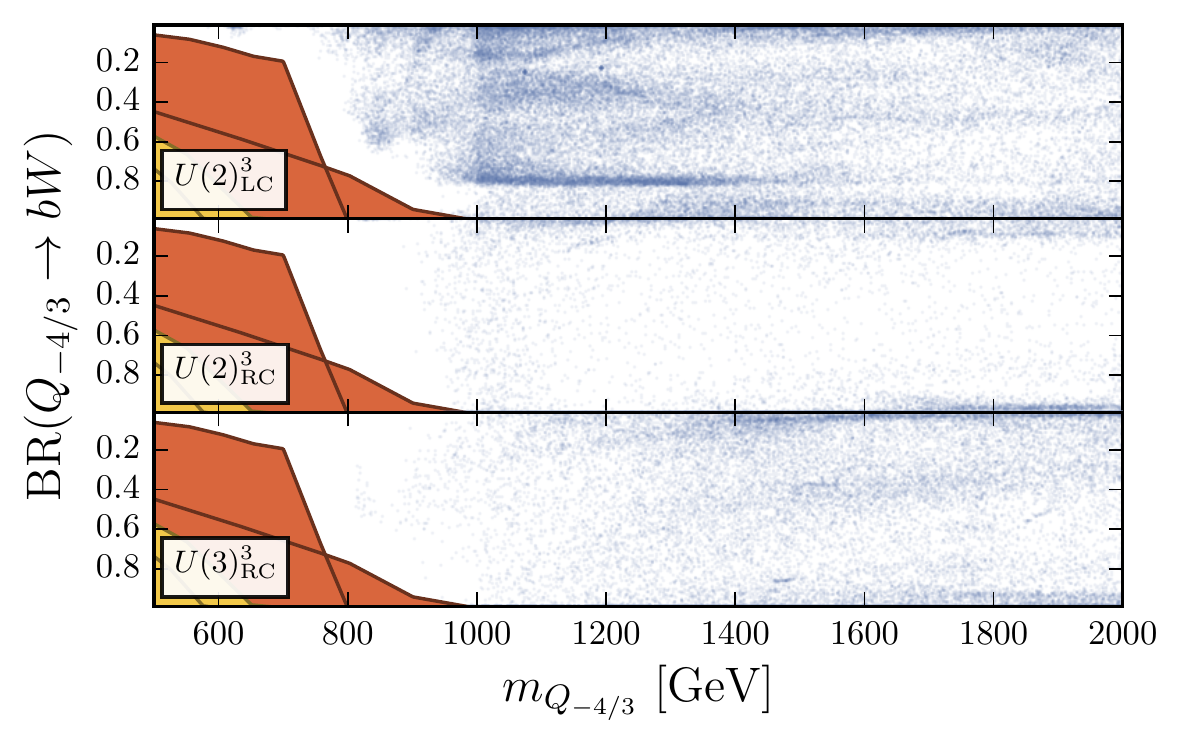}
 \vspace{-10pt}
 \caption{}\label{fig:direct_constraints:results:Q45_bounds_MCHM}
\end{subfigure}
\begin{subfigure}[b]{0.55\textwidth}
 \includegraphics[width=\textwidth]{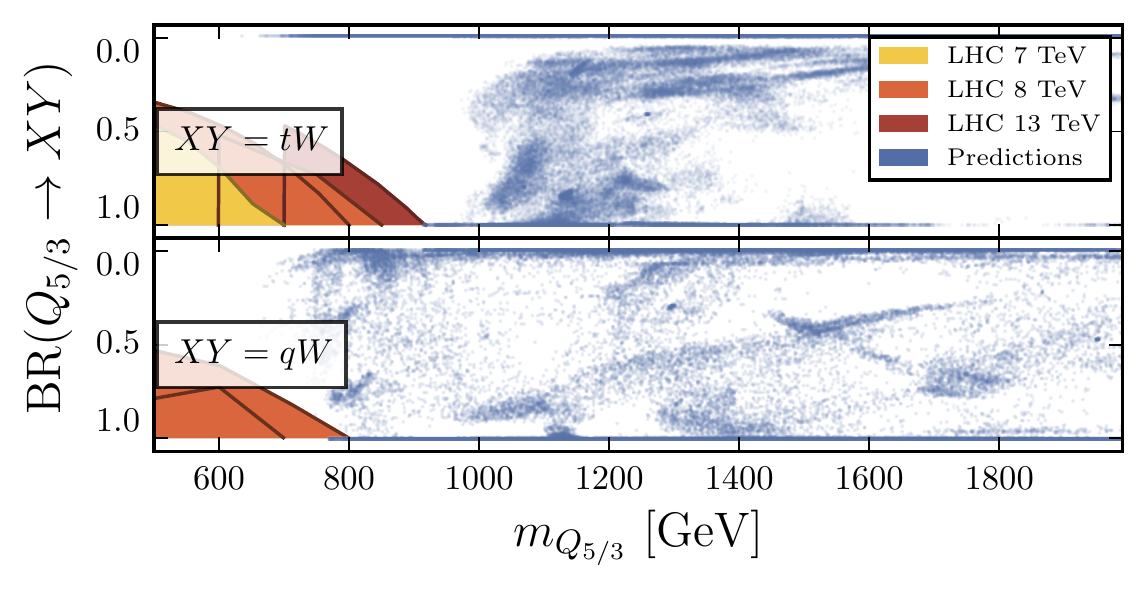}
 \includegraphics[width=\textwidth]{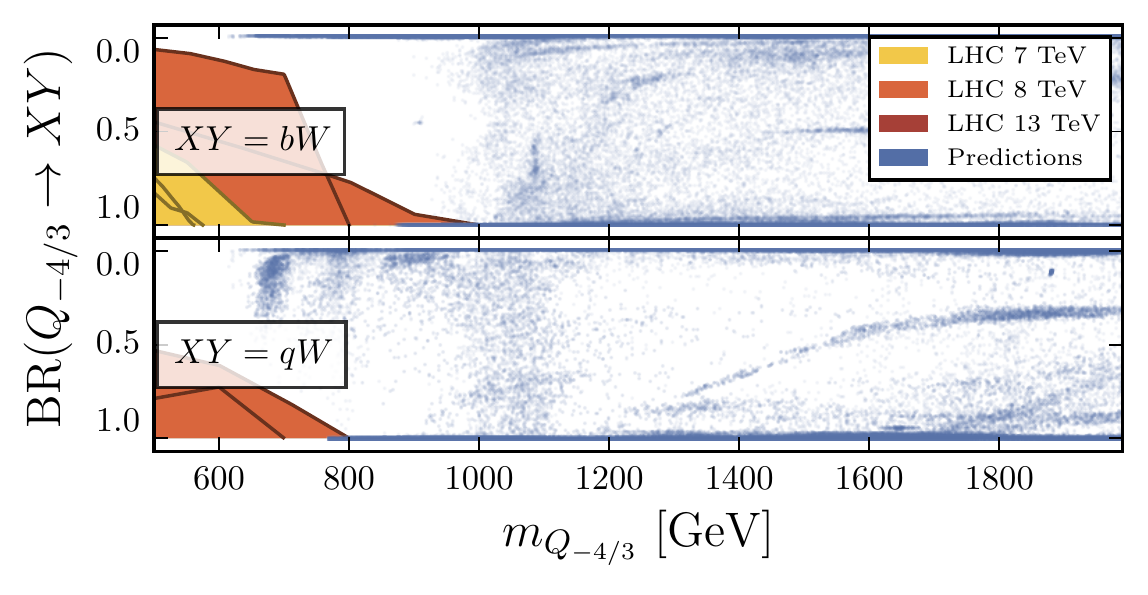}
 \caption{}\label{fig:direct_constraints:results:Q45_bounds_NMCHM}
\end{subfigure}
 \caption{
 Predictions for masses and branching ratios of charge $5/3$ (first row) and charge $-4/3$ (second row) quark resonances.
 Experimental bounds from the LHC running at different center of mass energies are shown as colored areas.
 Predictions and experimental bounds from the MCHM analysis are shown in (a), while those from the NMCHM analysis are shown in (b).
 The experimental searches included in (a) are listed in table~\ref{tab:app:MCHM:exp_quark_res}, while those included in (b) are shown in table~\ref{tab:app:NMCHM:exp_quark_res}.}
 \label{fig:direct_constraints:results:Q45_bounds_both_models}
\end{figure}

Experimental searches for quark partners provide important bounds on CHMs.
This is in particular the case in models
that require light quark partners in order
to obtain the correct Higgs and top masses (cf.\ e.g.~\cite{Contino:2006qr,Matsedonskyi:2012ym,Pomarol:2012qf,Panico:2012uw}).
The fact that we have only considered the model independent QCD pair production in our numerical analyses is very convenient for discussing the experimental bounds.
It allows for readily recasting the experimental limits on cross section times branching ratio into limits on the branching ratio alone.
In figures~\ref{fig:direct_constraints:results:U_bounds_both_models}, \ref{fig:direct_constraints:results:D_bounds_both_models}, and~\ref{fig:direct_constraints:results:Q45_bounds_both_models}, predictions for the branching ratios and masses of quark resonances lighter than 2~TeV are shown.
Only resonances corresponding to viable parameter points are included in these plots.
Various decay channels for up-type (figure~\ref{fig:direct_constraints:results:U_bounds_both_models}), down-type (figure~\ref{fig:direct_constraints:results:D_bounds_both_models}), and exotically charged (figure~\ref{fig:direct_constraints:results:Q45_bounds_both_models}) quark resonances are shown.
The experimentally observed 95\% CL upper limits are included as colored regions in the plots.
Note that due to the fact that we have only excluded parameter points that violate an individual constraint by more than 3$\sigma$, some of the predicted points
actually already lie above the observed 95\% CL upper limits.
Figures~\ref{fig:direct_constraints:results:U_bounds_MCHM}, \ref{fig:direct_constraints:results:D_bounds_MCHM}, and~\ref{fig:direct_constraints:results:Q45_bounds_MCHM} show the predictions and included experimental searches for our scan of the MCHM, where three\footnote{%
As already suggested by the discussion in section~\ref{sec:Fermions:flavor} and analytical analyses of similar models (cf.~\cite{Redi:2011zi,Barbieri:2012tu}), the $\Us{3}{LC}^3$ scenario suffers from very strong constraints imposed by electroweak precision tests and CKM unitarity.
No viable parameter point that is actually able to satisfy these constraints has been found for this scenario in our numerical scan.
} different flavor symmetry scenarios are displayed in each of the plots, while figures~\ref{fig:direct_constraints:results:U_bounds_NMCHM}, \ref{fig:direct_constraints:results:D_bounds_NMCHM}, and~\ref{fig:direct_constraints:results:Q45_bounds_NMCHM} show the results for our analysis of the NMCHM.

A main difference between the MCHM and the NMCHM is the presence of the scalar resonance $\eta$ in the latter.
In the NMCHM, this implies additional decay channels involving SM quarks.
In particular, up-type quark resonances can decay to SM up-type quarks and an $\eta$,
and analogously for down-type quarks.
With a new channel available, the branching ratios of all other channels are slightly reduced.
However, since there are many decay channels in total, the overall picture is not considerably changed compared to the MCHM analysis.

While there can be up to 30 different resonances of a given quark type in the models considered (cf.\ the mass matrices in appendix~\ref{app:Mass_matrices}), the lightest resonances are the most interesting ones from the phenomenological point of view.
They are required to decay to SM quarks for kinematical reasons and therefore have large branching ratios in the experimentally analyzed channels.
In addition, many of them are light enough to be already in reach of LHC run 2.
In particular, in the NMCHM analysis, we have observed that for 97\% of the viable parameter points, at least one quark resonance has a mass below 1.2~TeV.
For masses considerably above 1~TeV, most quark partners are therefore kinematically allowed to decay to the lightest quark resonances.
Due to these new decay channels opening up with higher resonance masses, the branching ratios to SM particles decrease.
This is a general feature that can be observed in all of the plots in figures~\ref{fig:direct_constraints:results:U_bounds_both_models}, \ref{fig:direct_constraints:results:D_bounds_both_models}, and~\ref{fig:direct_constraints:results:Q45_bounds_both_models}.

The strongest experimental bounds we have found are on up-type quark partners $U$ that decay to third generation quarks and a SM boson.
In particular, the relevant decay channels are $U\to b W$, $U\to t Z$ and $U\to t h$.
These three decay channels are the ones best covered by experimental searches.
It is interesting to compare the bounds used in the MCHM analysis to those used in the NMCHM analysis.
While only searches from LHC run 1 have been included in the former, the latter also considered searches from LHC run 2.
This is seen by comparing the plots in figure~\ref{fig:direct_constraints:results:U_bounds_MCHM} with those in figure~\ref{fig:direct_constraints:results:U_bounds_NMCHM}, where the dark red regions correspond to run~2 searches.
While the overall picture does not change considerably between the MCHM and the NMCHM analyses, the early run 2 searches included in the latter
actually already probe the viable parameter points.
One finds many points to be adjacent or even above the 95\% CL upper limits.
This is in contrast to the analysis of the MCHM, where LHC run 2 data was not yet available, and the parameter space is barely probed by the direct quark partner searches.
While also more experimental searches for down-type quark partners $D$ have been included into the NMCHM analysis compared to the one of the MCHM, corresponding data from LHC run 2 was not yet available.
However, one observes that the decay channels to third generation quarks, i.e.\ $D\to t W$, $D\to b Z$ and $D\to b h$, also offer good prospects to probe large parts of the viable parameter space with LHC run 2 searches (cf.\ figure~\ref{fig:direct_constraints:results:D_bounds_both_models}).

While many experimental searches are available for decay channels involving third generation SM quarks, only few consider decays to the light first and second generation quarks.
Actually, for the parameter scans of the MCHM and the NMCHM, only searches with a light quark and $W$ boson in the final state were available (cf.\ figures~\ref{fig:direct_constraints:results:U_bounds_both_models}, \ref{fig:direct_constraints:results:D_bounds_both_models}, and~\ref{fig:direct_constraints:results:Q45_bounds_both_models}).
However, any model with a full flavor structure, which features quark partners for all three generations, generically predicts also heavy quark resonances decaying to light SM quarks.
In particular, we find very light resonances with masses below 750~GeV that dominantly decay to quarks of the first two generations and are virtually unconstrained by direct experimentally searches.
This is interesting because they can presumably be probed by analyzing already available experimental data.
Interestingly, in our NMCHM analysis, essentially all quark resonances with very light masses below 750~GeV are found to be mainly composed of the singlets $S_2$ and $\widetilde{S}_2$ and to decay dominantly to a light SM quark and the Higgs\footnote{
The fact that the singlet resonances coupled to light quarks can themselves be very light and dominantly decay in virtually unconstrained channels involving the Higgs boson was already noted in~\cite{Delaunay:2013pwa}.
} (cf.\ the $XY$~$=$~$qh$-channels in figures~\ref{fig:direct_constraints:results:U_bounds_NMCHM} and~\ref{fig:direct_constraints:results:D_bounds_NMCHM}).
The $qh$ decay channels are thus by far the most promising ones to search for the very light unconstrained quark resonances that could have masses as low as 500~GeV.

In figure~\ref{fig:direct_constraints:results:Q45_bounds_both_models}, predictions and experimental bounds are shown for the exotically charged quark resonances.
Due to their charges, their decays to SM particles always has to involve a $W$ boson.
This considerably reduces the number of possible decay channels compared to the up-type and down-type quark resonances.
One might assume that this leads to very high branching ratios for decays to SM quarks, at last for the lightest exotically charged quark resonances.
However, since their mass cannot be lowered by mixing with SM quarks, the exotically charged quark resonances are generically heavier than the up-type and down-type quark partners.
Consequently, already the lightest exotically charged quark resonances are usually kinematically allowed to decay to up-type and down-type quark resonances.
Still, as can be seen in figure~\ref{fig:direct_constraints:results:Q45_bounds_both_models}, many of them have a considerably large branching ratio and a mass around 1~TeV, which allows them to be probed by LHC run 2 searches.

It is a very interesting general result of our analyses that most of the viable parameter points we have found can presumably be probed by direct searches for quark partners with LHC run 2 data.
Among all constraints considered in our analyses, these searches therefore have the arguably highest potential to observe or exclude our viable parameter points in the near future.

\subsection{Vector resonances}\label{sec:direct_constraints:results:vectors}
The models considered in our analyses contain neutral and charged electroweak resonances as well as a gluon resonance.
The mass of the latter depends on the NGB decay constant $f_G$ and is therefore independent of all other resonance masses (cf.\ section~\ref{sec:direct_constraints:models}).
Hence, the numerical scan can choose it to be relatively heavy.
In addition, it couples strongly to quark resonances, which usually makes it very broad and substantially decreases the branching ratios of decays to SM quarks.
Consequently, to search for pairs of quark partners produces via a gluon resonance~\cite{Azatov:2015xqa,Araque:2015cna} is arguably the best way to search for it.
However, as already mentioned in section~\ref{sec:direct_constraints:direct_constraints:simplifying_assumption}, this is beyond the scope of the analyses presented here.
The following discussion therefore focuses on the neutral and charged electroweak resonances.

In contrast to the quark sector, where one usually encounters strong mixing, the vector bosons only moderately mix with each other.
Hence, it is reasonable to associate each mass eigenstate with a corresponding gauge eigenstate it is mainly composed of.
In the following, the mass eigenstates will therefore simply be denoted\footnote{%
It might be useful to stress that while the same names are used for mass eigenstates and for the gauge eigenstates these mass eigenstates are mainly composed of, the mass eigenstates also contain other gauge eigenstates to a smaller degree.
In our numerical code, of course the full mixing matrix is diagonalized to yield the mass eigenstates.
} by the name of their corresponding gauge eigenstate.

\subsubsection{Charged electroweak resonances}
\begin{figure}[p!]
 \centering
 \includegraphics[keepaspectratio=true,width=0.48\textwidth]{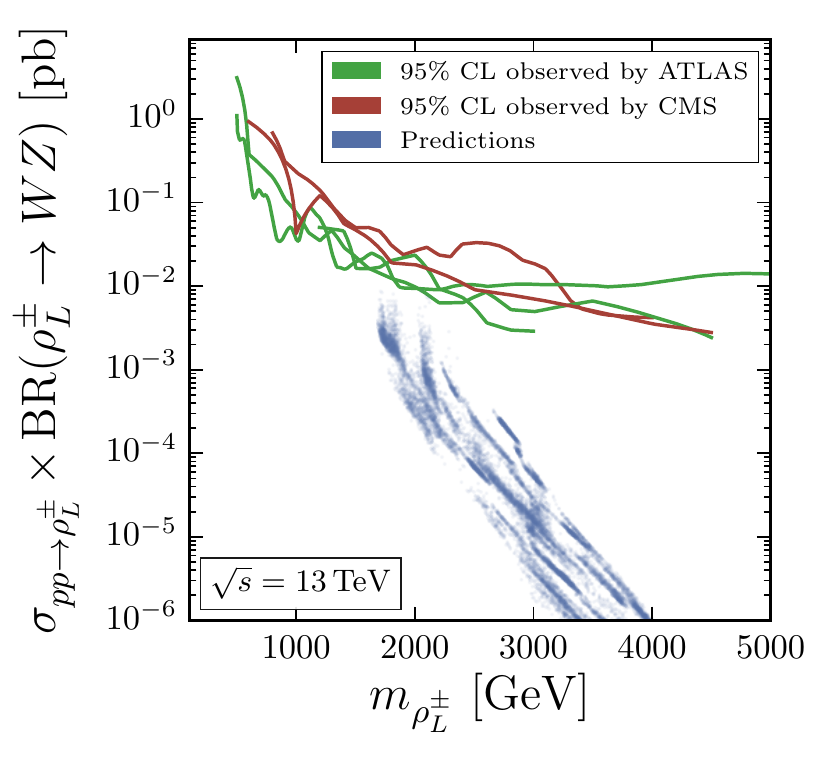}
 \quad
 \includegraphics[keepaspectratio=true,width=0.48\textwidth]{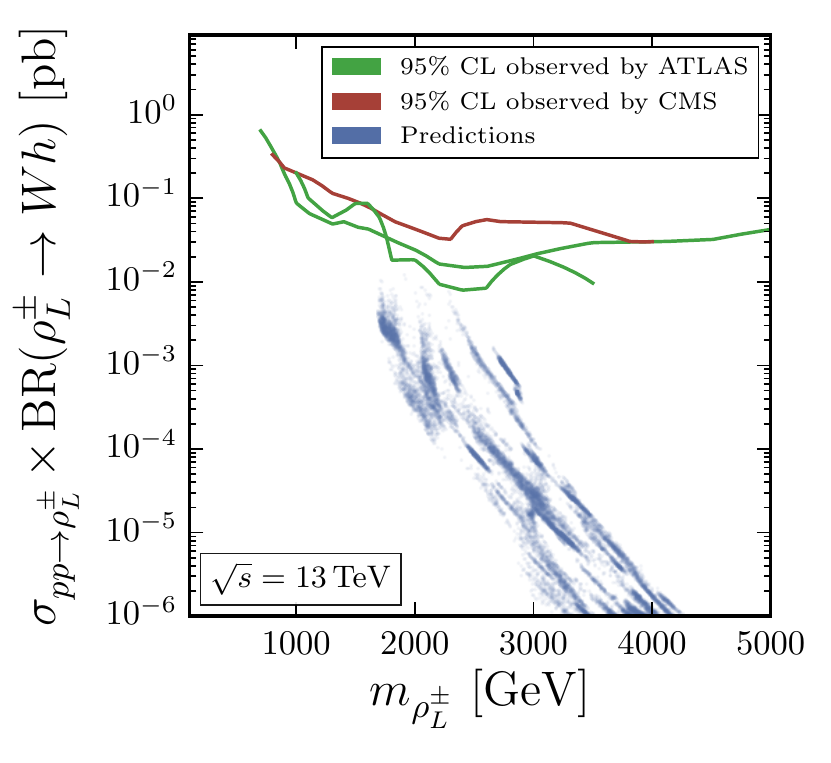}
 \\
 \includegraphics[keepaspectratio=true,width=0.48\textwidth]{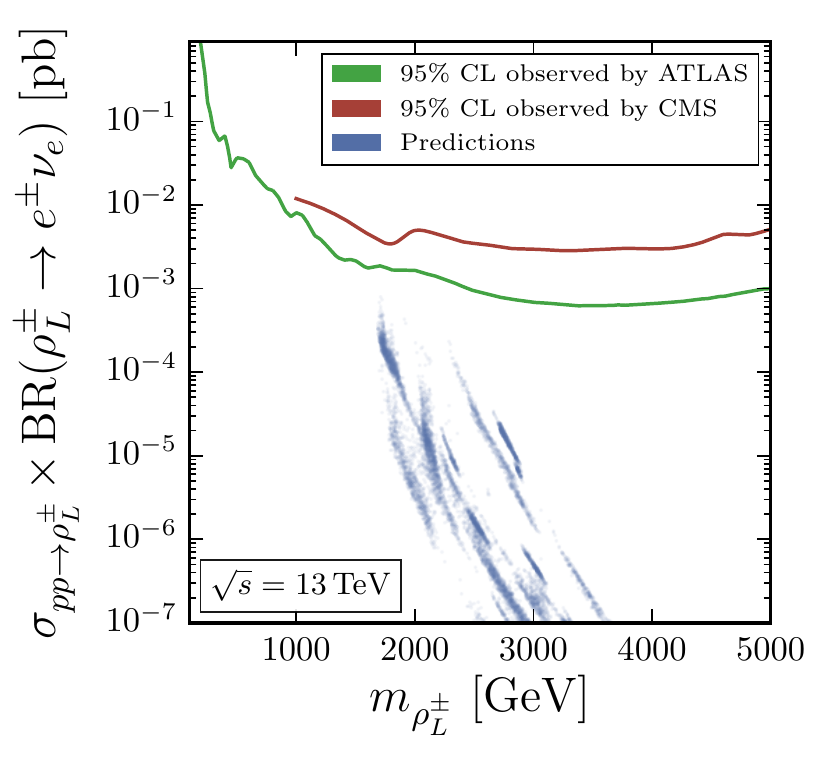}
 \quad
 \includegraphics[keepaspectratio=true,width=0.48\textwidth]{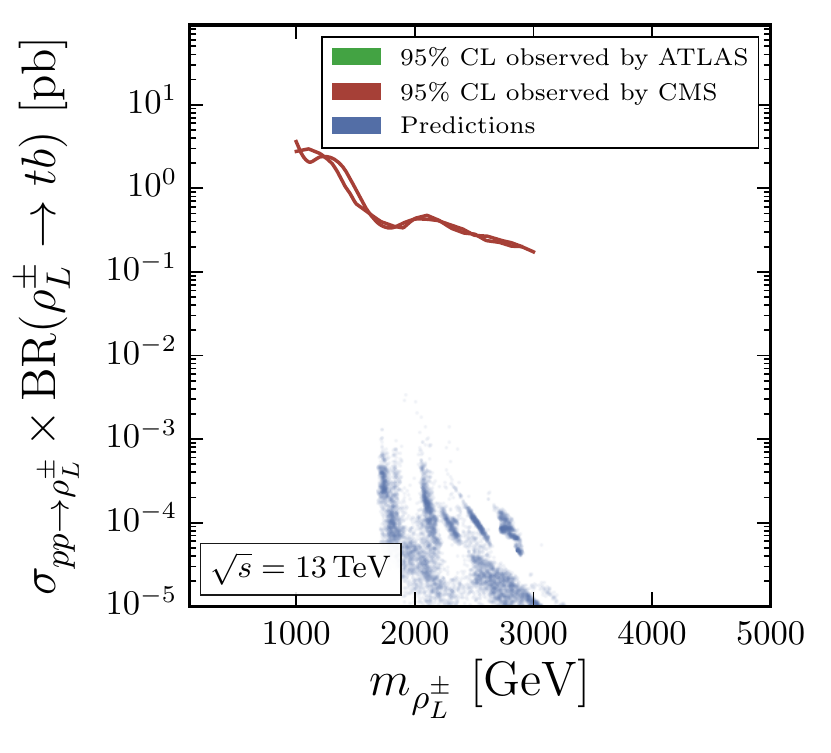}
 \caption{Experimental bounds from ATLAS and CMS and predictions from viable parameter points of the $\rho_{L\,\mu}^\pm$ production cross section times the branching ratio into $WZ$ (upper-left), $Wh$ (upper-right), $e^\pm\nu_e$ (lower-left) and $tb$ (lower-right). The experimental analyses shown in the plots are listed in table \ref{tab:app:NMCHM:exp_vector_res}.}
 \label{fig:direct_constraints:results:rhopm_bounds}
\end{figure}
The MCHM contains the three charged resonances $\rho_{L\,\mu}^\pm$, $\rho_{R\,\mu}^\pm$, and $a_{\mu}^\pm$.
While the first two have very similar masses, the latter is always heavier.
This can be understood by taking a look at the mass matrix in eq.~\eqref{eq:app:Mass_matrix:MCHM_EW_charged}, where one finds that the mass term of $a_{\mu}^\pm$ is enhanced by a factor $f_2^2/f^2$ compared to the other resonances\footnote{%
Note that $f^{-2}=f_1^{-2}+f_2^{-2}$ and thus $f_2>f$.
}.
In the NMCHM, one finds the same resonances $\rho_{L\,\mu}^\pm$ and $\rho_{R\,\mu}^\pm$ as in the MCHM but two resonances associated with a bidoublet (cf.\ section~\ref{sec:direct_constraints:models}): $a_{1\,\mu}^\pm$ and $a_{2\,\mu}^\pm$.
The resonance $a_{2\,\mu}^\pm$ is very similar to $a_{\mu}^\pm$ in the MCHM and its mass term is also enhanced by a factor $f_2^2/f^2$.
The resonance $a_{1\,\mu}^\pm$ on the other hand has a mass similar to those of $\rho_{L\,\mu}^\pm$ and $\rho_{R\,\mu}^\pm$ (cf.\ eq.~\eqref{eq:app:Mass_matrix:NMCHM_EW_charged}).

When it comes to the collider phenomenology, the only relevant resonance in both models is $\rho_{L\,\mu}^\pm$ as it is the only one that has a considerable Drell-Yan production cross section.
The cross sections of the other resonances are significantly smaller.
In a large part of the parameter space, the $\rho_{L\,\mu}^\pm$ is heavy enough to be kinematically allowed to decay to quark resonances; this then reduces its branching ratio to SM particles.
The largest branching ratios to the latter are found for $\rho_{L\,\mu}^\pm\to WZ$ and $\rho_{L\,\mu}^\pm\to Wh$.
Predicted values\footnote{%
Only plots containing predictions in the NMCHM are shown as they feature bounds from the LHC running at 13~TeV that were not available for the analysis of the MCHM.
However, it should be noted that the phenomenology of $\rho_{L\,\mu}$ is very similar in both models and the same conclusions can be drawn.
}
of cross section times branching ratio for these channels are shown in the two upper plots in figure~\ref{fig:direct_constraints:results:rhopm_bounds}.
One observers that while approximately equal, the branching ratios in the $Wh$ channel are slightly larger than those for $WZ$.
However, the experimental searches are slightly more sensitive in the $WZ$ channel, such that the bounds are very similar.
For both channels, the prospects are good to probe viable parameter points by future LHC searches.
While the branching ratios in the $\rho_{L\,\mu}^\pm\to e^\pm\nu_e$ channel are at least one order of magnitude smaller than those for decays to $WZ$ and $Wh$, the experimental searches are more sensitive by a similar factor (cf.\ lower-left plot in figure~\ref{fig:direct_constraints:results:rhopm_bounds}).
The situation in the $\rho_{L\,\mu}^\pm\to \mu^\pm\nu_\mu$ channel is essentially the same.
Consequently, in addition to the decays to $WZ$ and $Wh$, also decays to $e^\pm\nu_e$ and $\mu^\pm\nu_\mu$ are promising for probing viable parameter points.
While the branching ratios for the decay to $tb$ are similar to those of the leptonic decays, the experiments are far less sensitive, such that this decay channel is not very promising for probing the parameter space (cf.\ lower-right plot in figure~\ref{fig:direct_constraints:results:rhopm_bounds}).

\subsubsection{Neutral electroweak resonances}
The MCHM contains the five neutral uncolored resonances $\rho_{L\,\mu}^3$, $\rho_{R\,\mu}^3$, $a_{\mu}^3$, $a_{\mu}^4$, and $X_\mu$.
The resonances
$\rho_{L\,\mu}^3$ and $\rho_{R\,\mu}^3$ have masses very similar to each other.
Like the charged resonance $a_{\mu}^\pm$, also the neutral resonances $a_{\mu}^3$, and $a_{\mu}^4$ have a mass enhanced roughly by a factor of $f_2^2/f^2$ compared to $\rho_{L\,\mu}^3$ and $\rho_{R\,\mu}^3$ (cf.\ eq.~\eqref{eq:app:Mass_matrix:MCHM_EW_neutral}).
The fifth resonance $X_\mu$ is special in the sense that its mass depends on the NGB decay constant $f_X$ and on the coupling $g_X$, which are both independent of the decay constants and couplings entering the masses of the other resonances.
Therefore, it can be considerably lighter than the already relatively light $\rho_{L\,\mu}^3$ and $\rho_{R\,\mu}^3$, it can have a mass between those of the light $\rho_{L\,\mu}^3$ and $\rho_{R\,\mu}^3$ and the heavy $a_{\mu}^3$ and $a_{\mu}^4$, and it can also be the heaviest resonance.
In the NMCHM, the list of light resonances is extended by the resonances $a_{1\,\mu}^3$ and $a_{1\,\mu}^4$, while the heavy resonances $a_{2\,\mu}^3$ and $a_{2\,\mu}^4$ correspond to $a_{\mu}^3$ and $a_{\mu}^4$ in the MCHM.
The additional resonance $\rho_{S\,\mu}$ has a mass similar to the heavy resonances $a_{2\,\mu}^3$ and $a_{2\,\mu}^4$.

Not all of the above states are relevant for the collider phenomenology.
The resonance $a_{\mu}^4$ in the MCHM and $a_{1\,\mu}^4$, $a_{2\,\mu}^4$, and $\rho_{S\,\mu}$ in the NMCHM do not mix with any of the other states and are usually heavy (cf.\ eqs.~\eqref{eq:app:Mass_matrix:MCHM_EW_neutral} and~\eqref{eq:app:Mass_matrix:NMCHM_EW_neutral}).
This makes them irrelevant for the collider phenomenology.
While the resonances $a_{\mu}^3$ in the MCHM as well as $a_{1\,\mu}^3$ and $a_{2\,\mu}^3$ in the NMCHM do mix with $W_\mu^{(0)}$ and $B_\mu^{(0)}$, the mixings are suppressed by at least one factor of $s_h$ and their production cross section is very small.
Consequently, the only resonances relevant for the collider phenomenology are in both models $\rho_{L\,\mu}^3$, $\rho_{R\,\mu}^3$, and $X_\mu$.

\begin{figure}[p!]
 \centering
 \includegraphics[keepaspectratio=true,width=0.48\textwidth]{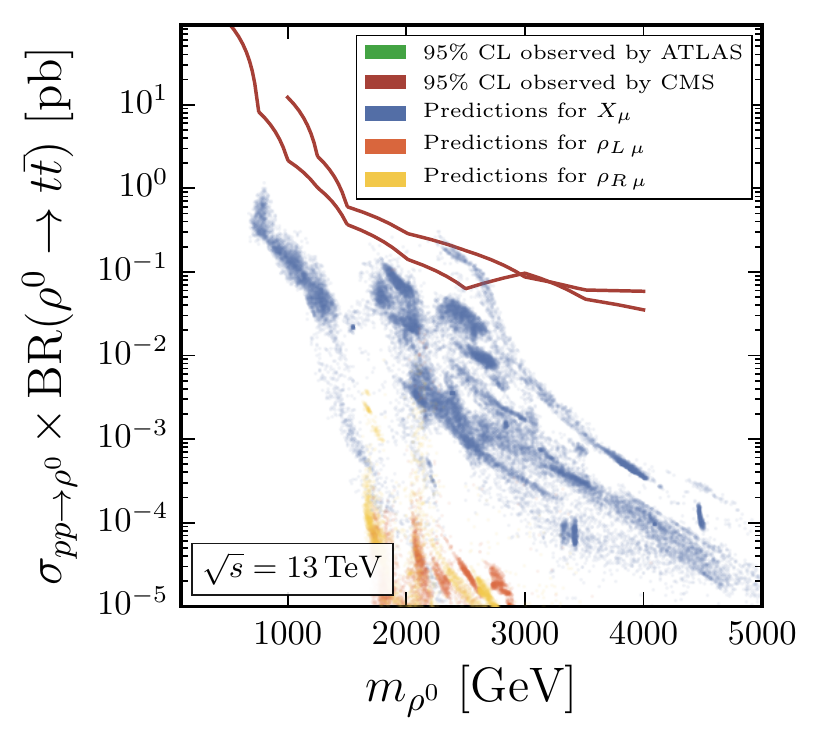}
 \quad
 \includegraphics[keepaspectratio=true,width=0.48\textwidth]{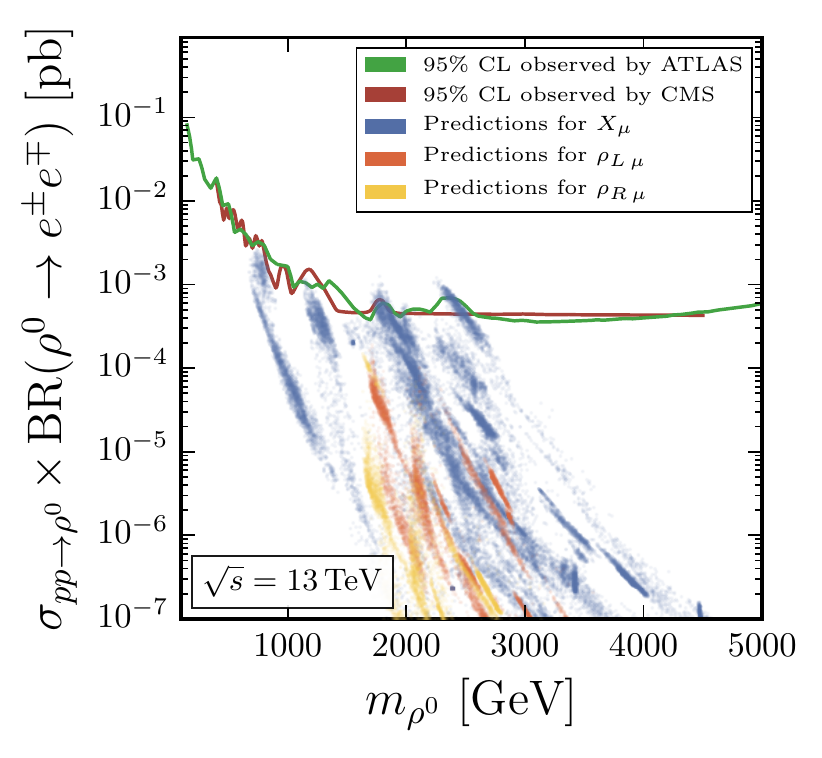}
 \\
 \includegraphics[keepaspectratio=true,width=0.48\textwidth]{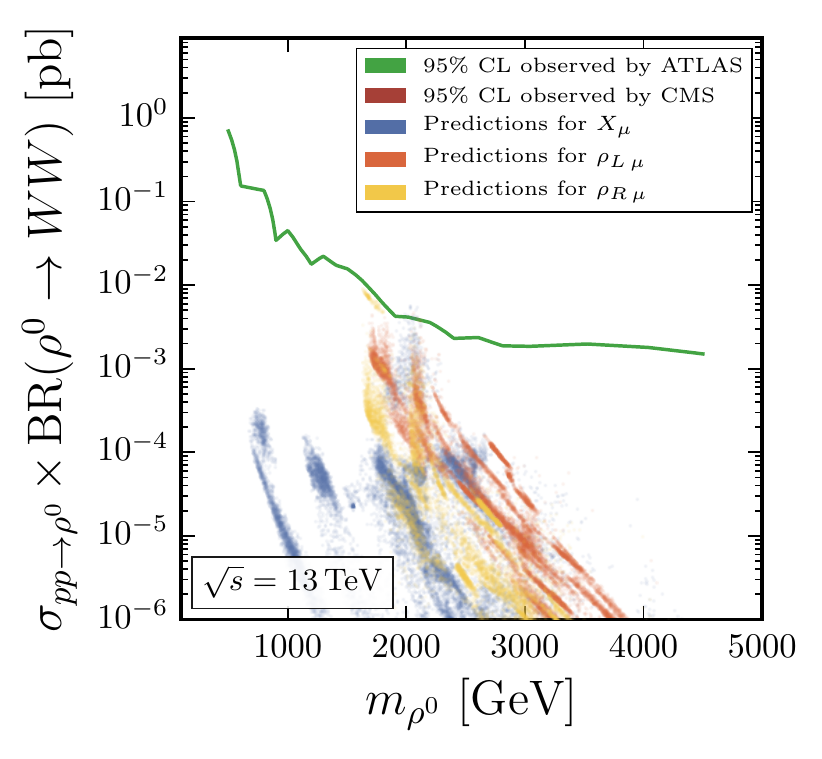}
 \quad
 \includegraphics[keepaspectratio=true,width=0.48\textwidth]{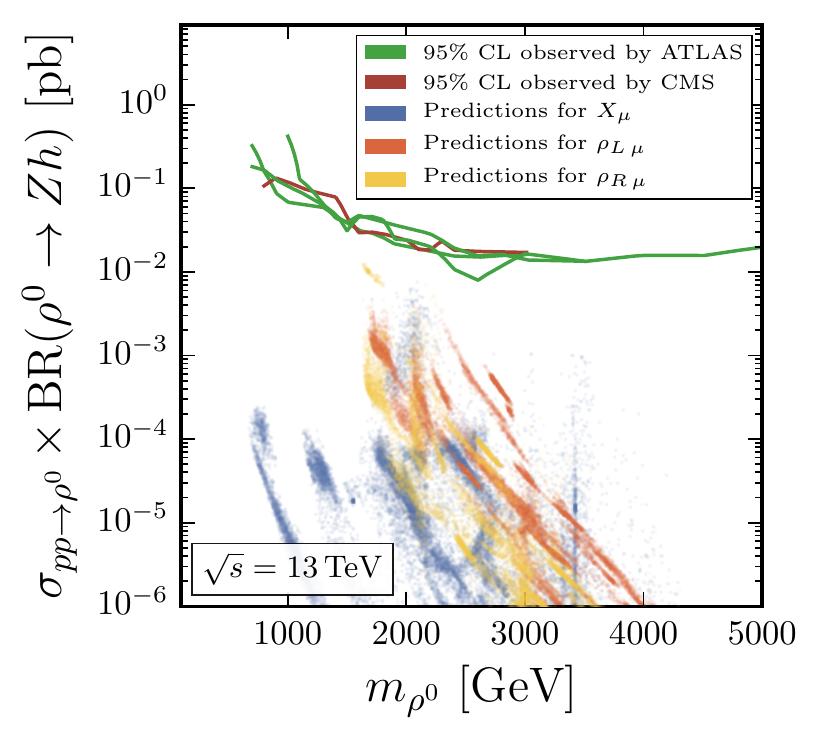}
 \caption{Experimental bounds from ATLAS and CMS and predictions for the neutral vector resonance production cross section times the branching ratio into $t\bar{t}$ (top-left), $e^+ e^-$ (top-right), $WW$ (bottom-left) and $ZH$ (bottom-right).
 We show values for the resonances $X_\mu$, $\rho_{L\,\mu}^3$ and $\rho_{R\,\mu}^3$ for the viable parameter points.
 The experimental analyses included in the plots are listed in table \ref{tab:app:NMCHM:exp_vector_res}.}
 \label{fig:direct_constraints:results:rho0_bounds}
\end{figure}
The case where $X_\mu$ is the lightest resonance is kind of special.
While the mass of heavy electroweak resonances is bounded from below due to their contributions to the electroweak $S$~parameter, this bound does not apply to a linear combination of resonances that couples to the same quantum numbers as the photon.
Interestingly, in the case $g_X\ll g_\rho$, such a linear combination is mainly composed of $X_\mu$ and at the same time $X_\mu$ is light.
This explains why $X_\mu$ can appear as a very light resonance even though the $S$~parameter is included as constraint in our global analyses (cf.\ section~\ref{sec:direct_constraints:constraints}).
In the case where $X_\mu$ is the lightest resonance, its production cross section is much larger than those of the other resonances.
The blue dots in the plots in figure~\ref{fig:direct_constraints:results:rho0_bounds} show predictions for the production cross section of $X_\mu$ times its branching ratio for the most important decay channel.
Since $X_\mu$ only mixes with $B_\mu^{(0)}$ and not with $W_\mu^{(0)}$, the branching ratio of its $WW$ decay channel is strongly suppressed.
A similar suppression is found for the $Zh$ channel since $X_\mu$ can couple to $h$ only via its mixing with $B_\mu^{(0)}$.
These suppressions can be seen in the two lower plots in figure~\ref{fig:direct_constraints:results:rho0_bounds}.
Because $X_\mu$ couples to composite quarks, and top quarks usually have a sizable degree of compositeness, the by far largest branching ratio is found for $X_\mu\to t\bar{t}$.
The upper-left plot in figure~\ref{fig:direct_constraints:results:rho0_bounds} shows that the predictions for many viable parameter points are close or even above the 95\% CL upper limits.
Thus, the $t\bar{t}$ channel is very promising for probing the region of parameter space where $X_\mu$ is the lightest resonance.
Even though $X_\mu$ can couple to leptons only via its mixing with $B_\mu^{(0)}$, the resulting suppression of the branching ratio can be compensated by the high sensitivity of experimental searches in the dilepton channel.
As a consequence, the bounds in this channel are even stronger than those in the $t\bar{t}$ case.
As shown in the upper-right plot in figure~\ref{fig:direct_constraints:results:rho0_bounds},
the experimental searches are already probing the viable parameter space with searches in the $e^\pm e^\mp$ channel.
The situation is very similar in the  $\mu^\pm \mu^\mp$ channel, for which no plot is shown here.

While the $t\bar{t}$ and dilepton channels are the most promising ones in the case of $X_\mu$ being the lightest resonance, this changes if $X_\mu$ is heavier than $\rho_{L\,\mu}^3$ and $\rho_{R\,\mu}^3$.
In this case, the latter two resonances can play the most important role in probing the parameter space.
Both mix with $W^{(0)}$ and can directly couple to the Higgs.
Consequently, the branching ratios in their $WW$ and $Zh$ channels are not strongly suppressed like they are for $X_\mu$.
The two lower plots in figure~\ref{fig:direct_constraints:results:rho0_bounds} show
that for the $WW$ and $Zh$ channels, the
predictions
of
cross section times branching ratio
are not far away from the experimental bounds, especially for $\rho_{L\,\mu}^3$.
While the predictions are approximately similar for both channels, the experimental analysis in the $WW$ channel has a higher sensitivity compared to those in the $Zh$ channel, such that the former channel might be more promising.
In general, one observes that cross section times branching ratio is slightly larger for $\rho_{L\,\mu}^3$ than for $\rho_{R\,\mu}^3$.
While this effect can be seen in the diboson channels, it es even more pronounced in the dilepton case.
As can be seen in the upper-right plot in figure~\ref{fig:direct_constraints:results:rho0_bounds}, the predictions for the $\rho_{L\,\mu}^3\to e^\pm e^\mp$ channel are already close to the experimental bounds.

To summarize the results found for the above two cases, the prospects for observing or excluding our viable parameter points are always high in the dilepton channel, while they are also high in the $t\bar{t}$ channel if $X_\mu$ is the lightest resonance and in the diboson channels if $\rho_{L\,\mu}^3$ is very light.

\subsection{The scalar resonance $\eta$ in the NMCHM}\label{sec:direct_constraints:results:eta}
\begin{figure}[t]
 \centering
\begin{subfigure}{0.49\textwidth}
 \includegraphics[width=\textwidth]{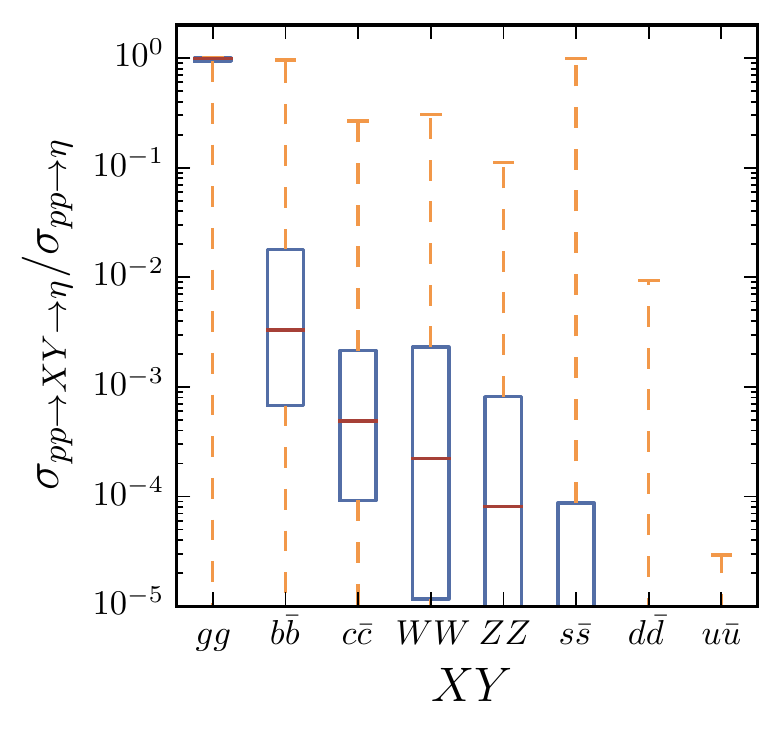}%
 \caption{}\label{fig:direct_constraints:results:eta_production}
\end{subfigure}
\begin{subfigure}{0.49\textwidth}
 \includegraphics[width=\textwidth]{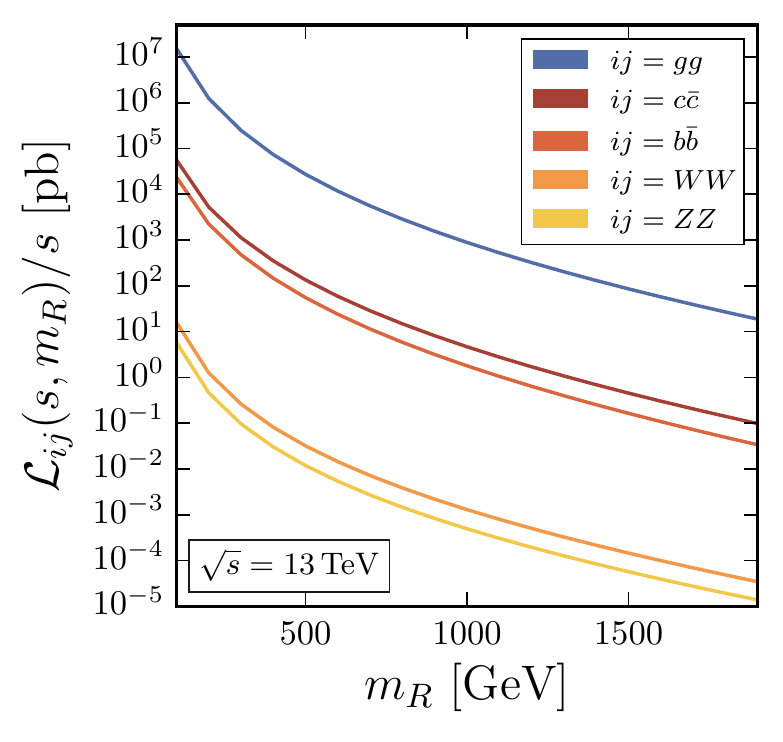}%
 \caption{}\label{fig:direct_constraints:results:partonlumis}
\end{subfigure}
 \caption{(a): Box plot of the $\eta$ production cross section in different channels relative to the total production cross section. For each channel, the plot shows the total range (as dashed orange line), the interquartile range (as blue box), and the median (as red line inside the box) of values from viable parameter points.
 For an explanation of {\it interquartile range} and {\it median}, see footnote~\ref{fn:box_plot}.
 (b): Parton luminosities for $ij=gg,c\bar{c},b\bar{b}$ and effective parton luminosities for $ij=WW,ZZ$ for the LHC with $\sqrt{s}=13\,\text{TeV}$.}
 \label{fig:direct_constraints:results:eta_production_partonlumis}
\end{figure}

While the phenomenology of the vector and quark resonances is very similar in the MCHM and the NMCHM, there is a clear distinction between both models.
This is the presence of the scalar resonance $\eta$ in the latter.
Since $\eta$ is a pNGB, it is usually much lighter than the other resonances.
Only in some small regions of the parameter space, very light quark partners can be slightly lighter than $\eta$.
Its mass $m_\eta$ is predicted to be below 790~GeV for 75\% of the viable parameter points, while no point with $m_\eta$ above 1300~GeV has been found by our numerical scan.
The couplings of $\eta$ to SM particles stem from the mixing with the Higgs.
As noted in section~\ref{sec:direct_constraints:models}, these couplings completely vanish in the absence of mixing in the scalar sector.
In the presence of mixing, however, they are always similar to those of the Higgs.
%
%
%
Hence, it is reasonable to expect that the dominant production mechanism of $\eta$ at a hadron collider is the same as for the Higgs, which is gluon fusion.
%
The box plot\footnote{%
In the box plot, the dashed orange lines show the total range of values that are predicted by the viable parameter points, the blue boxes show the range of values omitting the 25\% smallest and the 25\% largest values, which is also called the {\it interquartile range}.
50\% of the parameter points predict a smaller value than the one shown by the red line, which is called the {\it median}.
While statistics vocabulary is used here to describe the viable parameter points, it should be stressed that no statements are made about the probability of finding specific values.
\label{fn:box_plot}
} in figure~\ref{fig:direct_constraints:results:eta_production} shows that this assumption is correct.
%
%
This plot
shows ranges of values of the $\eta$ hadronic cross section in different production channels normalized to the total hadronic cross section of $\eta$.
In particular, $r_{\sigma}(gg)=\sigma(pp\to gg\to \eta)/\sigma(pp\to\eta)>0.93$ is found for 75\% of the viable parameter points, while for 50\% of the viable points, one even finds $r_{\sigma}(gg)>0.99$.
This clear dominance of the gluon fusion production mechanism is not due to a coupling of the $\eta$ to gluons that is orders of magnitude larger than those to other particles.
The reason is rather that gluons are abundantly available in a hadronic collision.
As detailed in section~\ref{sec:direct_constraints:direct_constraints:boson_xsec}, the hadronic cross section is calculated from the partonic cross sections and the parton luminosities (cf.\ eq.~\eqref{eq:direct_constraints:hadronic_xsec}).
The latter strongly depend on the considered parton.
To visualize this, figure~\ref{fig:direct_constraints:results:partonlumis} shows the parton luminosities for a partonic center of mass energy $\sqrt{\hat s}=m_R$ for different combinations of partons as a function of the mass of the produced resonance $m_R$.
%
One observes that the parton luminosity for two gluons is two to three orders of magnitudes above those of the $b\bar{b}$ and $c\bar{c}$ pairs, which usually give the second and third largest contribution to the hadronic cross section.
The effective parton luminosities for $WW$ and $ZZ$ are again around three orders of magnitudes below those of $b\bar{b}$ and $c\bar{c}$, such that their contributions to the production cross section are even more suppressed.
The contributions from light quarks on the other hand are negligible due to their tiny Yukawa couplings.
(cf.\ figure~\ref{fig:direct_constraints:results:eta_production}).

Another important property of the parton luminosities is that they substantially decrease for larger resonance masses $m_R$.
This dependence of the parton luminosities on the resonance mass is
one of the main reasons for
the difference between the phenomenology of $\eta$ and the Higgs.
While $\eta$ has couplings very similar to those of the Higgs, it is usually substantially heavier.
Consequently, its production cross section is suppressed due to comparatively small parton luminosities.
\begin{figure}[t]
 \centering
\begin{subfigure}{0.459\textwidth}
 \includegraphics[width=\textwidth]{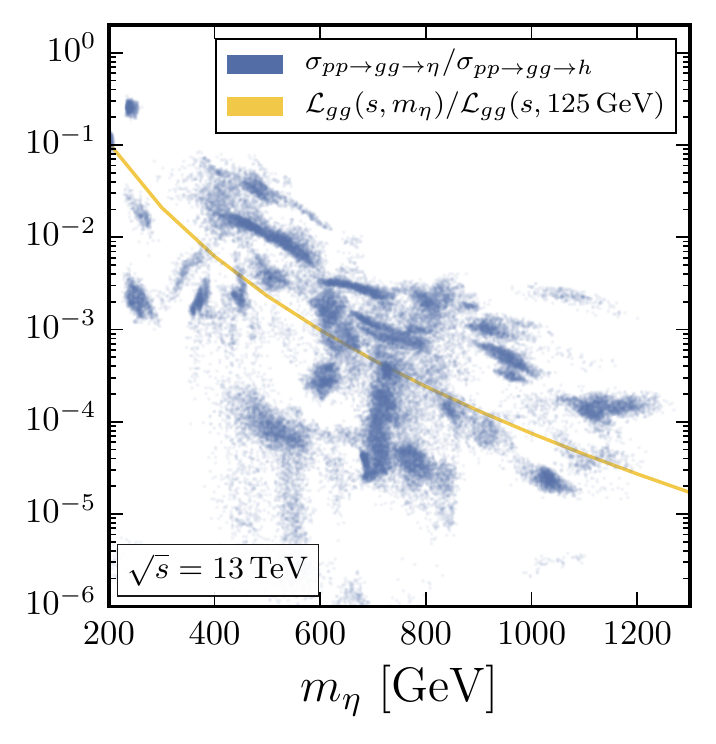}%
 \caption{}\label{fig:direct_constraints:results:eta_h_prod}
\end{subfigure}
\begin{subfigure}{0.506\textwidth}
 \includegraphics[width=\textwidth]{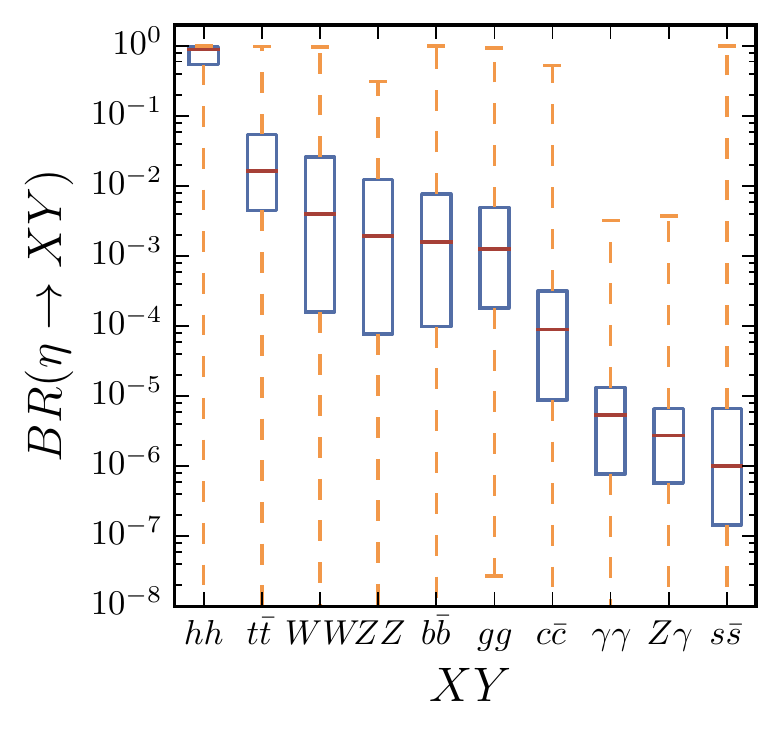}%
 \caption{}\label{fig:direct_constraints:results:eta_BR}
\end{subfigure}
 \caption{(a):
 Ratio of the hadronic production cross section via gluon fusion of $\eta$ to the one of the Higgs (blue dots) and ratio of gluon-gluon parton luminosities with $\sqrt{\hat{s}}=m_\eta$ to the one with $\sqrt{\hat{s}}=125$~GeV.
 (b):
 Box plot of the $\eta$ branching ratios of different decay channels. The usually very small up and down quark branching ratios are not included.
 For an explanation of the features of a box plot, see footnote~\ref{fn:box_plot}.}
 \label{fig:direct_constraints:results:eta_prod_decay}
\end{figure}
\begin{figure}[p]
 \centering
 \includegraphics[keepaspectratio=true,width=0.48\textwidth]{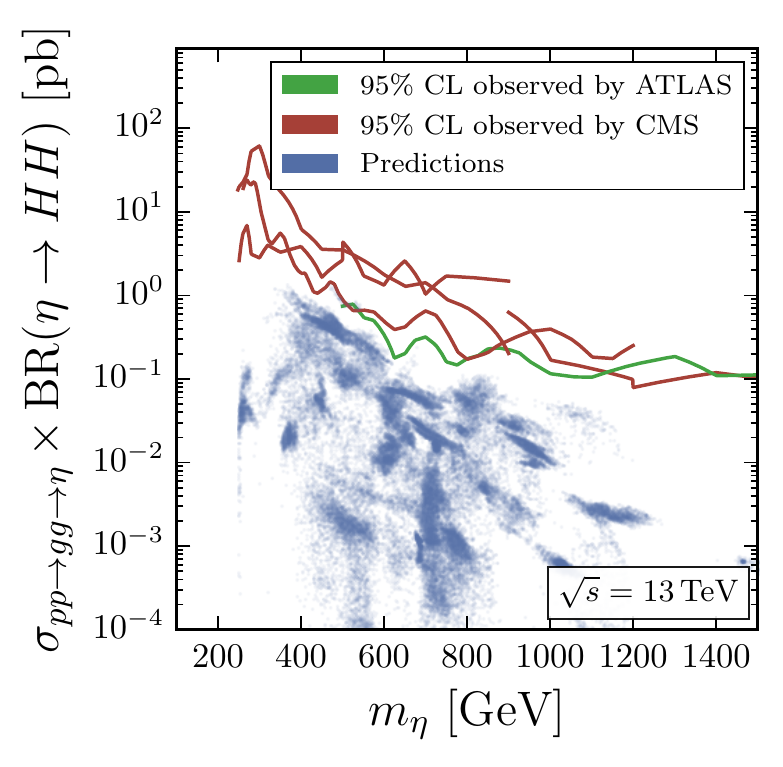}
 \quad
 \includegraphics[keepaspectratio=true,width=0.48\textwidth]{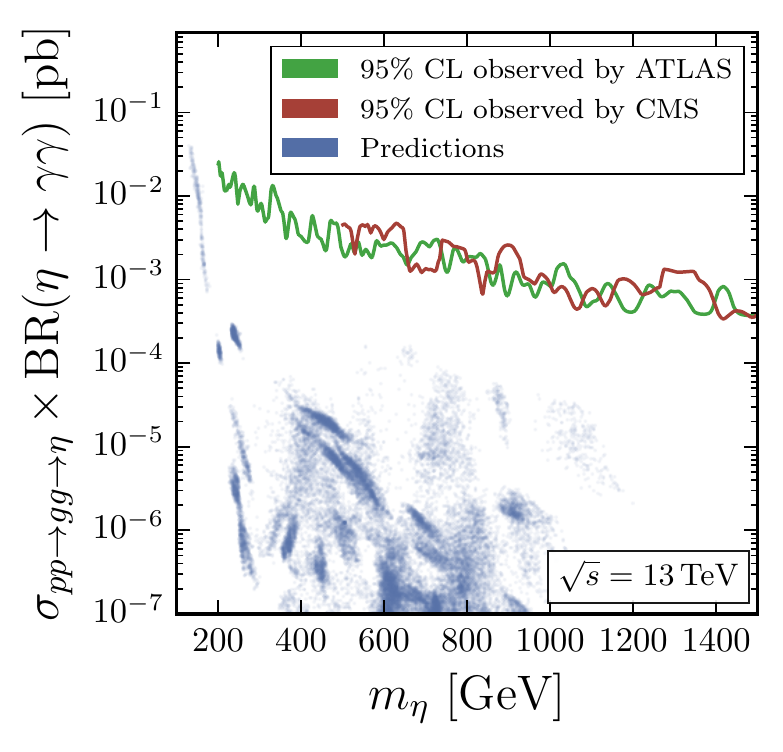}
 \\
 \includegraphics[keepaspectratio=true,width=0.48\textwidth]{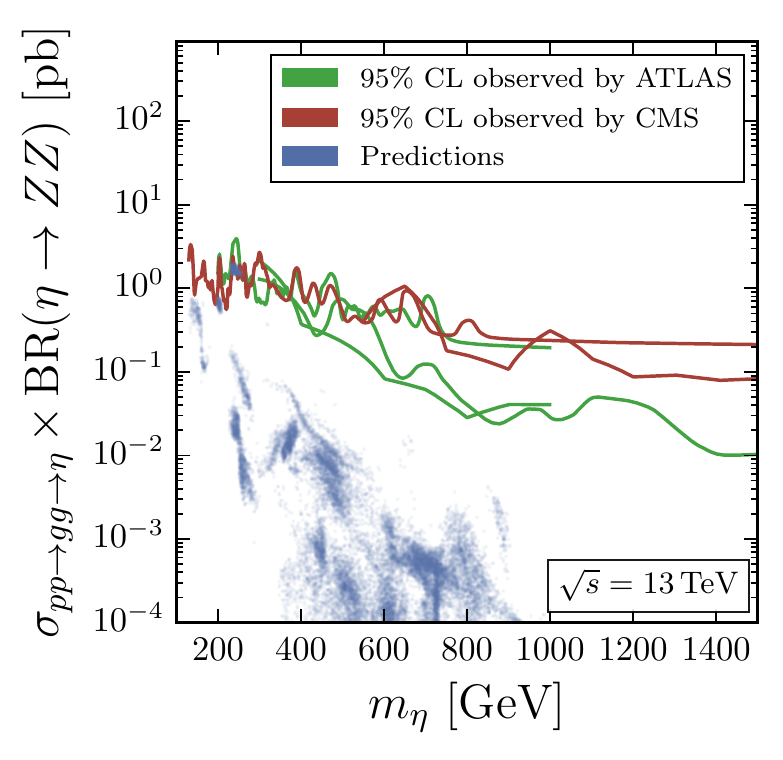}
 \quad
 \includegraphics[keepaspectratio=true,width=0.48\textwidth]{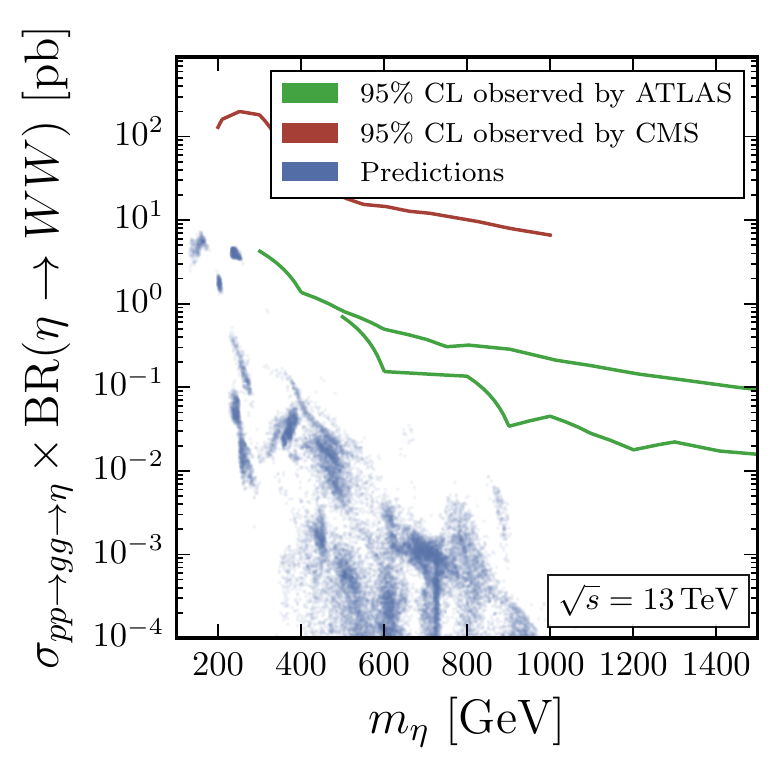}
 \caption{Experimental bounds from ATLAS and CMS and predictions from viable parameters points of the $\eta$ production cross section via gluon fusion times the branching ratio into $hh$ (top-left), $\gamma\gamma$ (top-right), $ZZ$ (bottom-left) and $WW$ (bottom-right). The experimental analyses shown in the plots are listed in table \ref{tab:app:NMCHM:exp_scalar_res}.}
 \label{fig:direct_constraints:results:eta_bounds}
\end{figure}
This can be further illustrated by
directly comparing cross sections and parton luminosities in $\eta$ production to those in Higgs production.
To this end, the blue dots in figure~\ref{fig:direct_constraints:results:eta_h_prod} show the ratio of the hadronic gluon fusion cross section of $\eta$ to the one of the Higgs as a function of $m_\eta$ for viable parameter points.
In addition, the yellow line shows the ratio of the $gg$ parton luminosity at $m_\eta$ to the one at the Higgs mass.
While the ratio of the hadronic cross sections can clearly be above or below the yellow line, which just means that the partonic gluon fusion cross section of $\eta$ is larger or smaller than the one of the Higgs, its maximal value at a given mass $m_\eta$ decreases with a similar slope as the parton luminosity ratio.
For the bulk of the viable parameter points, one finds
the gluon fusion cross section
of $\eta$ to be suppressed by at least one to two orders of magnitude compared to the gluon fusion cross section of the Higgs.

Since the couplings of $\eta$ are similar to those of the Higgs, also the decay channels are similar.
However, the fact that $m_\eta$ is generically larger than the Higgs mass has important implications.
In particular, an $\eta$ with a mass $m_\eta\gtrsim 200$~GeV can decay to on-shell $WW$ and $ZZ$ pairs.
With a mass $m_\eta\gtrsim 250$~GeV, it can decay to a pair of Higgses.
And for a mass $m_\eta\gtrsim 350$~GeV, the decay channel to a $t\bar{t}$ pair opens up.
For most viable parameter points, actually all of these decay channels are kinematically available.
Predictions of branching ratios in all relevant decay channels are shown as a box plot in figure~\ref{fig:direct_constraints:results:eta_BR}.
The by far largest branching ratio is usually the one of the $\eta\to hh$ channel.
The $\eta h h$ coupling is obtained from a third derivative of the effective potential and is usually large if there is considerable $\hat \eta - \hat h$ mixing.
The large branching ratio of the $\eta\to hh$ channel significantly reduces the branching ratios of the other channels.
The $h\to t\bar{t}$ channel has usually the second largest branching ratio, which for most viable parameter points is smaller than the one of the $\eta\to hh$ channel by one to two orders of magnitude.
The branching ratios of decays to weak gauge bosons\footnote{%
Contrary to what one might expect from the Goldstone boson equivalence theorem, the $\eta W W$ and $\eta ZZ$ couplings are substantially smaller than the $\eta hh$ coupling.
This is due to the facts that the latter stems from the $\hat \eta - \hat h$ mixing and the longitudinal polarizations of $W$ and $Z$ do not mix with $\hat \eta$.
}, $b\bar{b}$, and gluons are usually roughly one order of magnitude smaller than the one of the $t\bar{t}$ channel.
%
%
%
The decays to first and second generation quarks and the loop-induced decays to uncolored neutral vector bosons are even more strongly suppressed.

The large branching ratio found for $\eta\to hh$ suggest that this channel should be promising for probing the viable parameter points.
As shown in the upper-left plot in figure~\ref{fig:direct_constraints:results:eta_bounds}, this is indeed the case.
The blue dots in this plot show the predicted gluon fusion cross section of $\eta$ times the branching ratio in the $\eta\to hh$ channel as a function of $m_\eta$.
The green and red lines show the 95\% CL upper limits observed by ATLAS and CMS, respectively.
Many viable parameter points are actually found predict values close to the experimental bounds.
This decay channel is thus by far the most promising one to probe the parameter space already at LHC run 2.
However, even though the branching ratios for the other decay channels are significantly smaller, this can be compensated to some degree by a higher sensitivity of the experimental searches.
In particular, the channels $\eta\to ZZ$ and $\eta\to WW$ shown in the two lower plots in figure~\ref{fig:direct_constraints:results:eta_bounds} are especially promising for probing parameter points predicting a very light $\eta$ that is kinematically not allowed to decay to two Higgses.
Other decay channels are less promising.
In the upper-right plot in figure~\ref{fig:direct_constraints:results:eta_bounds}, predictions and experimental bounds for the $\eta\to\gamma\gamma$ channel are shown.
Even though the experimental searches are sensitive to much smaller values of cross section times branching ratio than in the afore mentioned promising channels, the predicted values for most viable parameter points are at least two orders of magnitude below the bounds.
However, given the tiny branching ratio of the $\eta\to\gamma\gamma$ channel, this is not surprising.
The situation is very similar in the $\eta\to Z\gamma$ channel for which no plot is shown.
While the decays to third generation quarks actually do have a much larger branching ratio, the experiments are less sensitive in these channels, such that the bounds are actually even further away from the predictions than in the $\gamma\gamma$ channel.

To summarize, the prospects for excluding or observing viable parameter points are by far the best in the $\eta\to h h$ channel, while the diboson channels are also interesting especially for a very light $\eta$ that cannot decay to two Higgses.

\chapter{Anomalies in rare \textit{B} decays}\label{chap:anomalies}
In the absence of direct evidence for new particles at the LHC, the arguably best way to search for NP is to look for indirect effects.
Among the most promising indirect probes of NP are rare meson decays involving FCNCs.
With their SM contribution being suppressed both by a loop factor and small CKM elements, these processes are very sensitive to NP contributions.
The rare $B$-meson decay $B\to K^*\mu^+\mu^-$ is an important example, whose key role in the search for NP at the LHC has been emphasized by several authors over the last two decades (cf.~\cite{Kruger:1999xa,Kruger:2005ep,Bobeth:2008ij,Egede:2008uy,Altmannshofer:2008dz}).
The angular distribution of its decay products yields several angular observables, among which the one called $S_5$ seems to be especially promising~\cite{Altmannshofer:2008dz,Bharucha:2010bb}.
%
%
%
To reduce its form factor uncertainties, a differently normalized version of $S_5$ has been suggested in~\cite{DescotesGenon:2012zf} and called $P_5'$.
Its first measurement by the LHCb collaboration in 2013 already showed a tension with the SM prediction at the level of about 3$\sigma$, only to be confirmed by an analysis of the full LHCb Run~1 data in 2015~\cite{Aaij:2015oid}.
In addition to $P_5'$, tensions with the SM predictions have also been found in branching ratio measurements of the decays $B\to K^{(*)}\mu^+\mu^-$ and $B_s\to\phi \mu^+\mu^-$~\cite{Aaij:2014pli,Aaij:2015esa}.
Analyses of these tensions by several groups~\cite{Descotes-Genon:2013wba,Altmannshofer:2013foa,Beaujean:2013soa,Hurth:2013ssa,Altmannshofer:2014rta,Descotes-Genon:2015uva,Hurth:2016fbr} have since shown that they are all compatible with a NP explanation in terms of a contribution to a single WC in the WEH.
%
%
%
Recently, also ATLAS~\cite{ATLAS:2017dlm} and CMS~\cite{CMS:2017ivg} presented preliminary results for their measurements of $B\to K^*\mu^+\mu^-$ angular observables, including the full Run 1 data set.
In~\cite{Altmannshofer:2017fio}, we have performed a numerical analysis of rare $B$ meson decays that are based on the $b\to s\mu\mu$ transition, where we have taken into account also the new results by ATLAS and CMS.
This analysis is presented in section~\ref{sec:anomalies:bsmumu}.

Unfortunately, the NP explanation of the $B\to K^*\mu^+\mu^-$ anomaly has some trouble.
Despite progress on improving the $B\to K^*$ form factors~\cite{Horgan:2013hoa,Horgan:2015vla,Straub:2015ica}, hadronic effects still cannot be ruled out as the origin of this anomaly.
%
In addition, the significance of the deviations depends on the uncertainties in both form factors and non-factorizable hadronic effects (cf.~\cite{Jager:2014rwa,Descotes-Genon:2014uoa,Capdevila:2017ert,Lyon:2014hpa,Ciuchini:2015qxb,Chobanova:2017ghn,Bobeth:2017vxj}).
However, in ratios of branching ratios~\cite{Hiller:2003js,Bobeth:2007dw,Hiller:2014ula} like
\begin{equation}
 R_K = \frac{\mathcal B(B \to K \mu^+\mu^-)}{\mathcal B(B \to K e^+e^-)}
\,,~
R_{K^*} = \frac{\mathcal B(B \to K^* \mu^+\mu^-)}{\mathcal B(B \to K^* e^+e^-)} \,,
\end{equation}
and differences of angular observables~\cite{Altmannshofer:2015mqa,Capdevila:2016ivx,Serra:2016ivr} like
\begin{align}
D_{P_4^\prime} &= P_4^\prime(B \to K^* \mu^+\mu^-) - P_4^\prime(B \to K^* e^+e^-) \,, \\
D_{P_5^\prime} &= P_5^\prime(B \to K^* \mu^+\mu^-) - P_5^\prime(B \to K^* e^+e^-) \,,
\end{align}
the dependence on hadronic effects and uncertainties cancel to a large degree, such that these observables are theoretically much cleaner.
A NP effect that affects all lepton generations equally, i.e. one that is lepton flavor universal, cannot be found by measurements of
%
these observables.
However, the SM itself satisfies LFU to an excellent degree over a large range of\footnote{%
$q^2$ is the dilepton invariant mass squared.
} $q^2$, where one finds $R_K=R_{K^*}=1$ and $D_{P_4^\prime}=D_{P_5^\prime}=0$ with only very small uncertainties (cf.~\cite{Bordone:2016gaq,Altmannshofer:2017fio} and section~\ref{sec:anomalies:LFUV:predictions}).
An observed deviation from these values would thus be a clear sign of NP.

While there is no 5$\sigma$ observation yet, several measurements actually show tensions with the SM prediction.
The LHCb collaboration has measured both $R_K$ and $R_{K^*}$.
Their $R_K$ measurement found~\cite{Aaij:2014ora},
\begin{equation}
R_{K}^{[1,6]} = 0.745 ^{+0.090}_{-0.074} \pm 0.036 \,,
\end{equation}
where the superscript specifies the range of the $q^2$ region, or bin, in which the measurement was performed.
This corresponds to a $2.6\sigma$ deviation from the SM prediction.
Their measurement of $R_{K^*}$ in two different $q^2$ bins found~\cite{Aaij:2017vbb},
\begin{align}
 R_{K^*}^{[0.045, 1.1]} &= 0.66 ^{+0.11}_{-0.07} \pm 0.03\,,\\
 R_{K^*}^{[1.1, 6]} &= 0.69 ^{+0.11}_{-0.07} \pm 0.05\,,
\end{align}
which corresponds to deviations from the SM prediction by $2.4$ and $2.5\sigma$, respectively.
The observables $D_{P_4^\prime}$ and $D_{P_5^\prime}$ have been measured by the Belle collaboration, finding~\cite{Wehle:2016yoi}
\begin{align}
 D_{P_4^\prime}^{[1, 6]} &= 0.498 \pm 0.553\,,\\
 D_{P_5^\prime}^{[1, 6]} &= 0.656 \pm 0.496\,,
\end{align}
which has still large uncertainties.
However, given that the three LHCb measurements all show tensions with the SM prediction, an interesting question is if they are compatible with one other and what implications a confirmation of these results would have.
To this end, we have performed a model-independent analysis\footnote{%
Several other groups have also performed similar analyses~\cite{Ciuchini:2017mik,Capdevila:2017bsm,Geng:2017svp,DAmico:2017mtc}.
} in~\cite{Altmannshofer:2017yso} to interpret these tantalizing hints for NP.
This analysis is presented in section~\ref{sec:anomalies:LFUV}.

\section{Weak effective Hamiltonian and numerical method}\label{sec:anomalies:WEH}

On the quark level, all of the above processes are due to $b\to s\ell\ell$ (with $\ell\in\{e,\mu\}$) transitions.
Assuming that any NP that could contribute to these transitions is sufficiently heavy, it can be described model-independently by the WEH $\mathcal{H}_\text{weak} = \mathcal{H}_\text{weak,SM}+\mathcal{H}_\text{weak,NP}$.
The part of the WEH that contains NP contributions to WCs of semi-leptonic operators relevant for $b\to s \ell\ell$ processes is
\begin{equation}
\label{eq:anomalies:Heff}
\mathcal{H}_\text{weak,NP}^{b\to s\ell\ell} = - \frac{4\,G_F}{\sqrt{2}} V_{tb}V_{ts}^* \frac{e^2}{16\pi^2}
\sum_k
(C_k^\ell O_k^\ell + C^{\prime\ell}_k O^{\prime\ell}_k) + \text{h.c.}\,.
\end{equation}
Here $C_k^\ell$ and $C^{\prime\ell}_k$ are defined such that they vanish in the SM.
The operators considered in the following analyses are
\begin{equation}
 \begin{aligned}
O_9^\ell &=
(\bar{s} \gamma_{\mu} P_{L} b)(\bar{\ell} \gamma^\mu \ell)\,,
&
O_9^{\prime\ell} &=
(\bar{s} \gamma_{\mu} P_{R} b)(\bar{\ell} \gamma^\mu \ell)\,,
\\
O_{10}^\ell &=
(\bar{s} \gamma_{\mu} P_{L} b)( \bar{\ell} \gamma^\mu \gamma_5 \ell)\,,
&
O_{10}^{\prime\ell} &=
(\bar{s} \gamma_{\mu} P_{R} b)( \bar{\ell} \gamma^\mu \gamma_5 \ell)\,.
 \end{aligned}
\end{equation}
While four-quark operators containing $b$ and $s$ can also contribute to $b\to s\ell\ell$, and especially certain $b\to c\bar{c}s$ operators might be interesting in light of the $P_5'$ tensions~\cite{Jager:2017gal}, they are not considered here.
Also not considered are scalar semi-leptonic operators and dipole operators.
The former are strongly constrained by measurements of the $B_s \to \mu\mu$ and $B_s \to ee$
branching ratios~\cite{Aaij:2017vad,Aaltonen:2009vr} (cf.\ also e.g.~\cite{Altmannshofer:2017wqy}), and the latter by inclusive radiative $B$~decays (cf.\ e.g.~\cite{Paul:2016urs}).
When considering LFU observables, the latter are irrelevant because they cannot lead to LFU violation (LFUV).

To find possible explanations of the $b\to s\,\mu^+\mu^-$ anomaly and the hints for LFUV in terms of NP contributions to the WCs $C_9^{(\prime)\ell}$ and $C_{10}^{(\prime)\ell}$, one can employ the open source code \texttt{flavio}~\cite{flavio}\footnote{%
For another open source code for flavor physics, see \texttt{EOS}~\cite{EOS}.
}.
This code is able to perform a $\chi^2$-fit that uses certain simplifying assumptions, implemented by the \texttt{FastFit} class and first proposed in~\cite{Altmannshofer:2014rta}.
This fit method can be described as follows.
First, a $\chi^2$ function is constructed that quantifies the difference between measured observables combined into a vector $\vec{\rm O}_{\rm exp}$ and theoretical predictions of these observables combined into a vector $\vec{\rm O}_{\rm th}$.
The latter in general depends on the NP contributions to the WCs one wants to include in the fit, which can be combined into the vector $\vec{C}^{\rm NP}$.
The $\chi^2$ function thus also depends on $\vec{C}^{\rm NP}$ and reads
%
\begin{equation}
 \chi^2(\vec{C}^{\rm NP})
 =
 \left[\vec{\rm O}_{\rm exp}-\vec{\rm O}_{\rm th}(\vec{C}^{\rm NP})\right]\transpose
 [\hat{\rm C}_{\rm exp}+\hat{\rm C}_{\rm th}]^{-1}
 \left[\vec{\rm O}_{\rm exp}-\vec{\rm O}_{\rm th}(\vec{C}^{\rm NP})\right],
\end{equation}
where $\vec{\rm O}_{\rm th}(\vec{C}^{\rm NP})$ and $\vec{\rm O}_{\rm exp}$ contain only the central values of the theory predictions and experimental measurements.
The experimental and theoretical uncertainties are taken into account in terms of the covariance matrices $\hat{\rm C}_{\rm exp}$ and $\hat{\rm C}_{\rm th}$.
These matrices contain all uncertainties and publicly known correlations of experimental measurements as well as all theory uncertainties and correlations.
The minimum of the $\chi^2$ function yields the best-fit point, i.e.\ the value of $\vec{C}^{\rm NP}$ for which the theory predictions have the best agreement with the experimental measurements.
In constructing the above $\chi^2$ function, the \texttt{FastFit} class in \texttt{flavio} makes simplifying assumptions concerning the covariance matrices:
\begin{enumerate}
 \item All uncertainties are assumed to be Gaussian when combining them in terms of the sum of $\hat{\rm C}_{\rm exp}$ and $\hat{\rm C}_{\rm th}$.
 \item\label{item:anomalies:fastfit} The non-zero NP contributions $\vec{C}^{\rm NP}$ are assumed to have a negligible impact on the theory uncertainties and correlations in $\hat{\rm C}_{\rm th}$.
\end{enumerate}
The latter assumption is the main reason for \texttt{FastFit} being fast.
It only requires to perform the time consuming numerical calculation of the covariance matrix $\hat{\rm C}_{\rm th}$ once for vanishing $\vec{C}^{\rm NP}$, i.e.\ at the SM point of the parameter space.
After this is done, $\hat{\rm C}_{\rm th}$ can be used for evaluating the $\chi^2$ function at arbitrary values of $\vec{C}^{\rm NP}$.
Without assumption \#\ref{item:anomalies:fastfit}, the covariance matrix would have to be calculated again for each value of $\vec{C}^{\rm NP}$, and thus the computing time would increase substantially.
While it is in general not guaranteed that the NP contributions to the covariance matrix $\hat{\rm C}_{\rm th}$ are negligible, this has been found to be a very good approximation in most cases (cf.~\cite{Altmannshofer:2014rta}).
However, especially for a best-fit point significantly differing from the SM point, one should check the viability of the method by at least recalculating $\hat{\rm C}_{\rm th}$ at the best-fit point.

With the minimum of the $\chi^2$ function denoted by $\chi^2_\text{best-fit}$, it is convenient to define
\begin{equation}
 \Delta \chi^2(\vec{C}^{\rm NP}) = \chi^2(\vec{C}^{\rm NP})-\chi^2_\text{best-fit},
\end{equation}
i.e.\ the difference between the value of the $\chi^2$ function at a given point $\vec{C}^{\rm NP}$ and its minimum at the best-fit point.
This difference $\Delta \chi^2$ can be converted into a {\it pull} in $\sigma$, which for the case of a one-dimensional $\vec{C}^{\rm NP}$ is simply given by $\sqrt{\Delta \chi^2}$.
For the $n$-dimensional case, the pull can be evaluated using the inverse cumulative distribution function of the $\chi^2$ distribution with $n$ degrees of freedom. Taking e.g.\ $n=2$, a pull of 1$\sigma$, 2$\sigma$, and 3$\sigma$ corresponds to $\Delta\chi^2 \approx 2.3$, $6.2$, and $11.8$.
Usually, one is mainly interested in the pull of the SM point, i.e.\ the pull of $\Delta \chi^2_{\rm SM}=\Delta \chi^2(\vec{0})$.

\section{The \lowercase{$b\to s\,\mu^+\mu^-$} anomaly}\label{sec:anomalies:bsmumu}
In~\cite{Altmannshofer:2017fio}, we have performed a numerical analysis of $b\to s\,\mu^+\mu^-$ processes using the method described above, where we have considered NP contributions to the WCs $C_9^{(\prime)\mu}$ and $C_{10}^{(\prime)\mu}$.
The observables that we have included are
\begin{itemize}
\item Angular observables in $B^0\to K^{*0}\mu^+\mu^-$ measured by LHCb~\cite{Aaij:2015oid}, ATLAS*~\cite{ATLAS:2017dlm}, CMS*~\cite{Khachatryan:2015isa,1385600,CMS:2017ivg}, and CDF~\cite{CDFupdate},
\item $B^{0,\pm}\to K^{*0,\pm}\mu^+\mu^-$ branching ratios by LHCb*~\cite{Aaij:2014pli,Aaij:2016flj},
CMS ~\cite{Khachatryan:2015isa,1385600},
and CDF~\cite{CDFupdate},
\item $B^{0,\pm}\to K^{0,\pm}\mu^+\mu^-$ branching ratios by LHCb~\cite{Aaij:2014pli} and CDF~\cite{CDFupdate},
\item $B_s\to\phi\mu^+\mu^-$ branching ratio by LHCb*~\cite{Aaij:2015esa} and CDF~\cite{CDFupdate},
\item $B_s\to\phi\mu^+\mu^-$ angular observables by LHCb*~\cite{Aaij:2015esa},
\item the branching ratio of the inclusive decay $B\to X_s\mu^+\mu^-$ measured
by BaBar~\cite{Lees:2013nxa},
\end{itemize}
where the collaborations marked by an asterisk have released new results since the global fit performed in~\cite{Altmannshofer:2014rta}.
A comment is in order concerning the angular observables in $B^0\to K^{*0}\mu^+\mu^-$ measured by LHCb and ATLAS.
They both have measured the $S_i$ observables and the $P_i'$ observables.
We have used the measurements of the $P_i'$ observables for our fit but have explicitly checked that the results are not significantly affected by this choice.

There are observables of $b\to s\,\mu^+\mu^-$ processes that we have explicitly not included into our fit.
These are
\begin{itemize}
 \item lepton-averaged observables, as we want to focus on NP in only $b\to s\,\mu^+\mu^-$,
 \item $B\to K\,\mu^+\mu^-$ angular observables, which are only relevant in the presence of scalar and tensor operators (cf.~\cite{Beaujean:2015gba}),
 \item the Belle measurement of $B\to K^*\mu^+\mu^-$ angular observables
~\cite{Wehle:2016yoi}, as it contains an unknown mixture of $B^0$ and $B^\pm$ decays,
 \item the LHCb measurement of $\Lambda_b\to\Lambda\mu^+\mu^-$~\cite{Aaij:2015xza}, as its central values are not compatible with a viable short-distance hypothesis, and its uncertainties are also still large~\cite{Meinel:2016grj},
 \item measurements of $B_s\to\mu^+\mu^-$, as it can be affected by scalar operators not taken into account in this analysis.
\end{itemize}

All the observables we have included are measured in bins of $q^2$.
We have only taken into account measurements in bins of $q^2$ where the theoretical predictions are reliable and where only the operators $O_9^{(\prime)\mu}$ and $O_{10}^{(\prime)\mu}$ dominate the effects.
In particular, we have excluded the following bins:
\begin{itemize}
\item Bins below the $J/\psi$ resonance that extend above 6~GeV$^2$, as calculations based on QCD factorization are not reliable in this region~\cite{Beneke:2001at}.
\item Bins above the $\psi(2S)$ resonance that are less than 4~GeV$^2$ wide. In this region,  theoretical predictions are only valid for observables integrated over a sufficiently large $q^2$ interval~\cite{Beylich:2011aq}.
\item Bins with upper boundary at or below 1~GeV$^2$, as this region is dominated by dipole operators.
\end{itemize}

The calculations that are used for the theoretical predictions are implemented in the \texttt{flavio} code.
They are discussed in detail in~\cite{Altmannshofer:2014rta,Straub:2015ica}.
Compared to the earlier analysis in~\cite{Altmannshofer:2014rta}, improved predictions for $B \to K^*$ and $B_s \to \phi$ form factors~\cite{Straub:2015ica} and $B \to K$ form factors~\cite{Bailey:2015dka} have been included into the code.
This significantly reduces the uncertainties in the $B \to K$ form factors.

\subsection{New physics in individual Wilson coefficients}

We have first performed one-dimensional fits in specific directions of the four-dimensional parameter space of NP contributions to $C_9^{(\prime)\mu}$ and $C_{10}^{(\prime)\mu}$.
These directions correspond to the four WCs $C_9^{(\prime)\mu}$ and $C_{10}^{(\prime)\mu}$ and four linear combinations of them.
All these scenarios with their best-fit points, 1 and 2$\sigma$ ranges and the pull at the SM point are shown in table~\ref{tab:anomalies:bsmumu:pulls_1D}.
The following observations can be made:
\begin{itemize}
 \item The scenario with a NP contribution only to $C_9^\mu$ has clearly the strongest pull, slightly above 5$\sigma$.
 The value of the best-fit point for this scenario is consistent with earlier fits that did not include the ATLAS and CMS measurements.
 While the significance has increased with respect to earlier analyses (in~\cite{Altmannshofer:2014rta}, a pull of 3.9$\sigma$ has been found for the same scenario), this is not mainly due to new experimental data included in the present analysis, but can be traced back to the updated form factors and their smaller uncertainties\footnote{%
 The fact that the updated form factors increase the tension was also pointed out in \cite{Du:2015tda}.
 }.
 \item The scenario with NP only in $C_{10}^\mu$ gives an improved fit compared to the SM. However, its significance is considerably smaller than in the $C_9^\mu$ scenario.
 It is interesting to note that the $B_s\to\mu^+\mu^-$ branching ratio that was not included in the fit due to its dependence on scalar operators is also affected by $C_{10}^\mu$. In particular, the best-fit value in the $C_{10}^\mu$ scenario would imply a suppression of the $B_s\to\mu^+\mu^-$ branching ratio by about 35\%.
 \item The scenario with NP in $C_{9}^\mu=-C_{10}^\mu$ has a significance slightly smaller, but similar to the $C_9^\mu$ case.
 This scenario corresponds to NP that only couples to left-handed muons, which can be realized in CHMs (cf.\ chapters~\ref{chap:LUFV_in_CHMs} and~\ref{chap:Flavor_MFPC}).
 \item The orthogonal direction $C_{9}^\mu=C_{10}^\mu$ provides only a marginally improved fit compared to the SM.
 \item All scenarios with NP in only one of the primed WC, i.e.\ right-handed quark currents, do not lead to a significantly better fit than the SM.
\end{itemize}
\renewcommand{\arraystretch}{1.2}
\begin{table}[t]
\begin{center}
\begin{tabularx}{\textwidth}{ccccX}
\hline\hline
 ~~~~~~ Coeff. ~~~~~~ & ~~~~~~~ best fit ~~~~~~~ & ~~~~~~~~~~ $1\sigma$ ~~~~~~~~~~ & ~~~~~~~~~~ $2\sigma$ ~~~~~~~~~~ & pull \\
\hline\hline
\rowcolor[gray]{.9} $C_9^\mu                  $ & $-1.21$ & [$-1.41$, $-1.00$] & [$-1.61$, $-0.77$] & $5.2\sigma$\\
                    $C_9^{\prime\mu}                     $ & $+0.19$ & [$-0.01$, $+0.40$] & [$-0.22$, $+0.60$] & $0.9\sigma$\\
\rowcolor[gray]{.9} $C_{10}^\mu               $ & $+0.79$ & [$+0.55$, $+1.05$] & [$+0.32$, $+1.31$] & $3.4\sigma$\\
                    $C_{10}^{\prime\mu}                  $ & $-0.10$ & [$-0.26$, $+0.07$] & [$-0.42$, $+0.24$] & $0.6\sigma$\\
\rowcolor[gray]{.9} $C_9^\mu=C_{10}^\mu$ & $-0.30$ & [$-0.50$, $-0.08$] & [$-0.69$, $+0.18$] & $1.3\sigma$\\
                    $C_9^\mu=-C_{10}^\mu$ & $-0.67$ & [$-0.83$, $-0.52$] & [$-0.99$, $-0.38$] & $4.8\sigma$\\
\rowcolor[gray]{.9} $C_9^{\prime\mu}=C_{10}^{\prime\mu}       $ & $+0.06$ & [$-0.18$, $+0.30$] & [$-0.42$, $+0.55$] & $0.3\sigma$\\
                    $C_9^{\prime\mu}=-C_{10}^{\prime\mu}      $ & $+0.08$ & [$-0.02$, $+0.18$] & [$-0.12$, $+0.28$] & $0.8\sigma$\\
\hline\hline
\end{tabularx}
\end{center}
\caption{Best-fit values with their 1$\sigma$ and $2\sigma$ ranges and pulls in sigma between the best-fit point and the SM point for scenarios with NP in one Wilson coefficient.}
\label{tab:anomalies:bsmumu:pulls_1D}
\end{table}
%
%
%
In light of the large tensions, it is interesting to investigate the contributions from different measurements.
To this end, we have repeated the fit for the scenario with NP only in $C_9^\mu$ for several subsets of the data:
\begin{itemize}
 \item Including only the measurements of the $B_s\to\phi\mu^+\mu^-$ branching ratios, one finds a pull of 3.5$\sigma$.
 \item Considering only the $B^0\to K^{*0}\mu^+\mu^-$ angular analysis by LHCb leads to a pull of 3.0$\sigma$.
 \item All branching ratio measurements together yield a pull of 4.6$\sigma$
 \item The new measurement of the $B^0\to K^{*0}\mu^+\mu^-$ angular observables by CMS reduces the pull, while the ATLAS measurement increases it.
\end{itemize}
Obviously, the branching ratio measurements play an important role in the large significance of the global fit.
They are, however, strongly dependent on the form factors.
Considerably underestimated uncertainties of these form factors might be a source of the discrepancies.
To estimate the impact of a possible underestimation, we have repeated the fit for the $C_9^\mu$ scenario with doubled uncertainties either of the form factors or of the non-factorizable hadronic corrections (see~\cite{Altmannshofer:2014rta} for details on these different uncertainties).
In the former case, we found a reduction of the significance from 5.2$\sigma$ to 4.0$\sigma$, and in the latter case from 5.2$\sigma$ to 4.4$\sigma$.
This indicates that underestimated uncertainties are probably not the only source of the discrepancies.

\subsection{New physics in a pair of Wilson coefficients}
\renewcommand{\arraystretch}{1.2}
\begin{table}[t]
\begin{center}
\begin{tabularx}{0.5\textwidth}{ccX}
\hline\hline
 ~~~~~ Coeff. ~~~~~ & ~~~~~~ best fit ~~~~~~ & pull \\
\hline\hline
\rowcolor[gray]{.9} $C_9^\mu,\ C_{10}^\mu$ & ($-1.15$, $+0.26$) & $5.0\sigma$\\
                    $C_9^\mu,\ C_{9}^{\prime\mu}    $ & ($-1.25$, $+0.59$) & $5.3\sigma$\\
\rowcolor[gray]{.9} $C_9^\mu,\ C_{10}^{\prime\mu}   $ & ($-1.34$, $-0.39$) & $5.4\sigma$\\
                    $C_9^{\prime\mu},\ C_{10}^\mu   $ & ($+0.25$, $+0.83$) & $3.2\sigma$\\
\rowcolor[gray]{.9} $C_9^{\prime\mu},\ C_{10}^{\prime\mu}      $ & ($+0.23$, $+0.04$) & $0.5\sigma$\\
                    $C_{10}^\mu,\ C_{10}^{\prime\mu}$ & ($+0.79$, $-0.05$) & $3.0\sigma$\\
\hline\hline
\end{tabularx}
\end{center}
\caption{Best-fit values and pulls in sigma between the best-fit point and the
SM point for scenarios with NP in two Wilson coefficients. For the first two cases, the best-fit regions are shown in figure~\ref{fig:anomalies:bsmumu:C9_C10_and_C9_C9p}.}
\label{tab:anomalies:bsmumu:pulls_2D}
\end{table}
%
Next, we have performed two-dimensional fits in planes of pairs of WCs.
The different scenarios together with their best-fit points and the pulls at the SM point are shown in table~\ref{tab:anomalies:bsmumu:pulls_2D}.
One observes that all scenarios that allow for a non-zero NP contribution to $C_9^\mu$ yield a pull of around 5$\sigma$, similar to the case with NP only in $C_9^\mu$.
So allowing for directions in addition to $C_9^\mu$ does not improve the fit considerably.
The scenarios allowing for NP in $C_{10}^\mu$ and one of the primed WCs improve the fit slightly compared to the SM, similar to the case with NP only in $C_{10}^\mu$.
A NP contribution only to the primed WCs cannot improve the fit significantly.

\begin{figure}[t]
\centering
\begin{subfigure}{0.49\textwidth}
 \includegraphics[width=\textwidth]{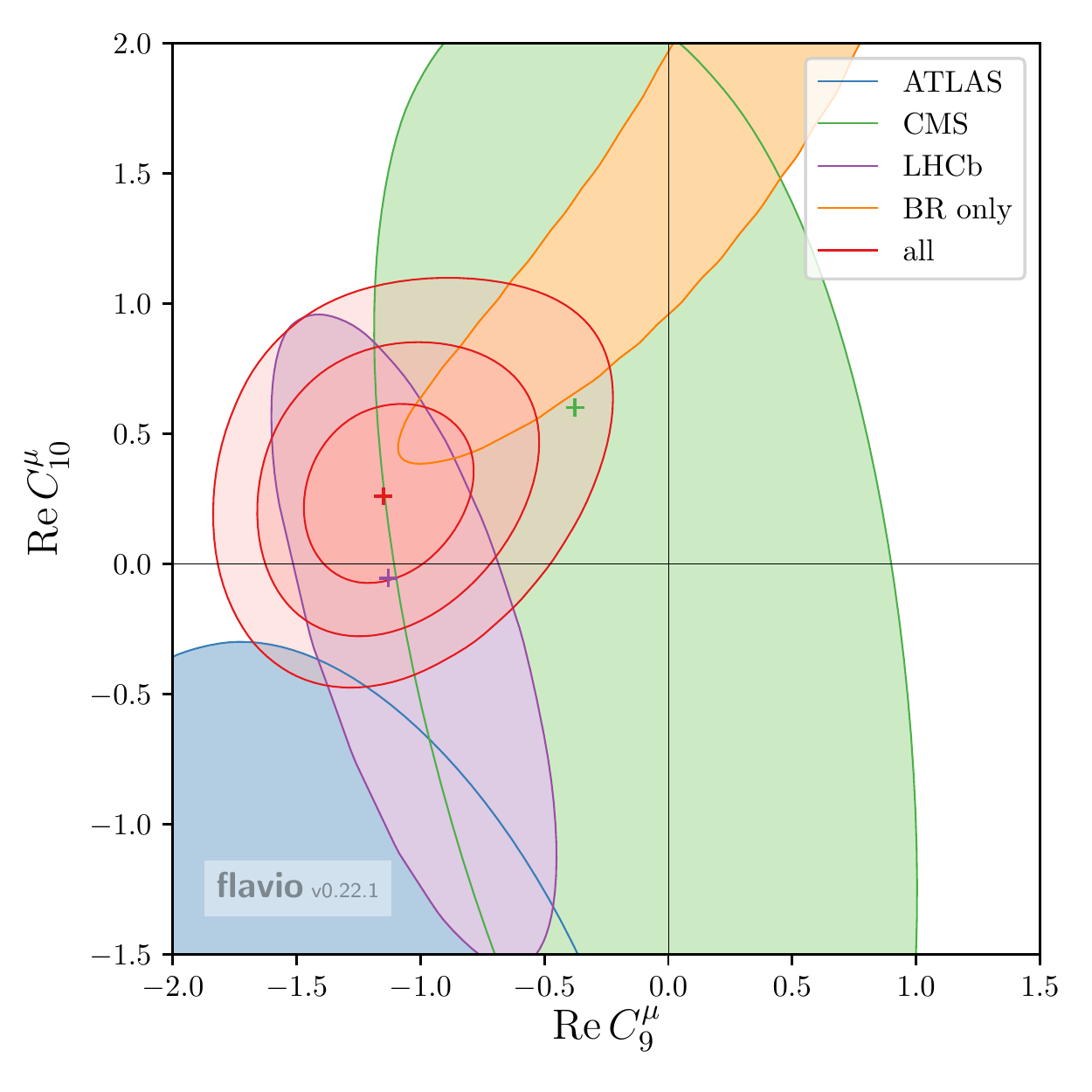}%
 \caption{}
\end{subfigure}
\begin{subfigure}{0.49\textwidth}
 \includegraphics[width=\textwidth]{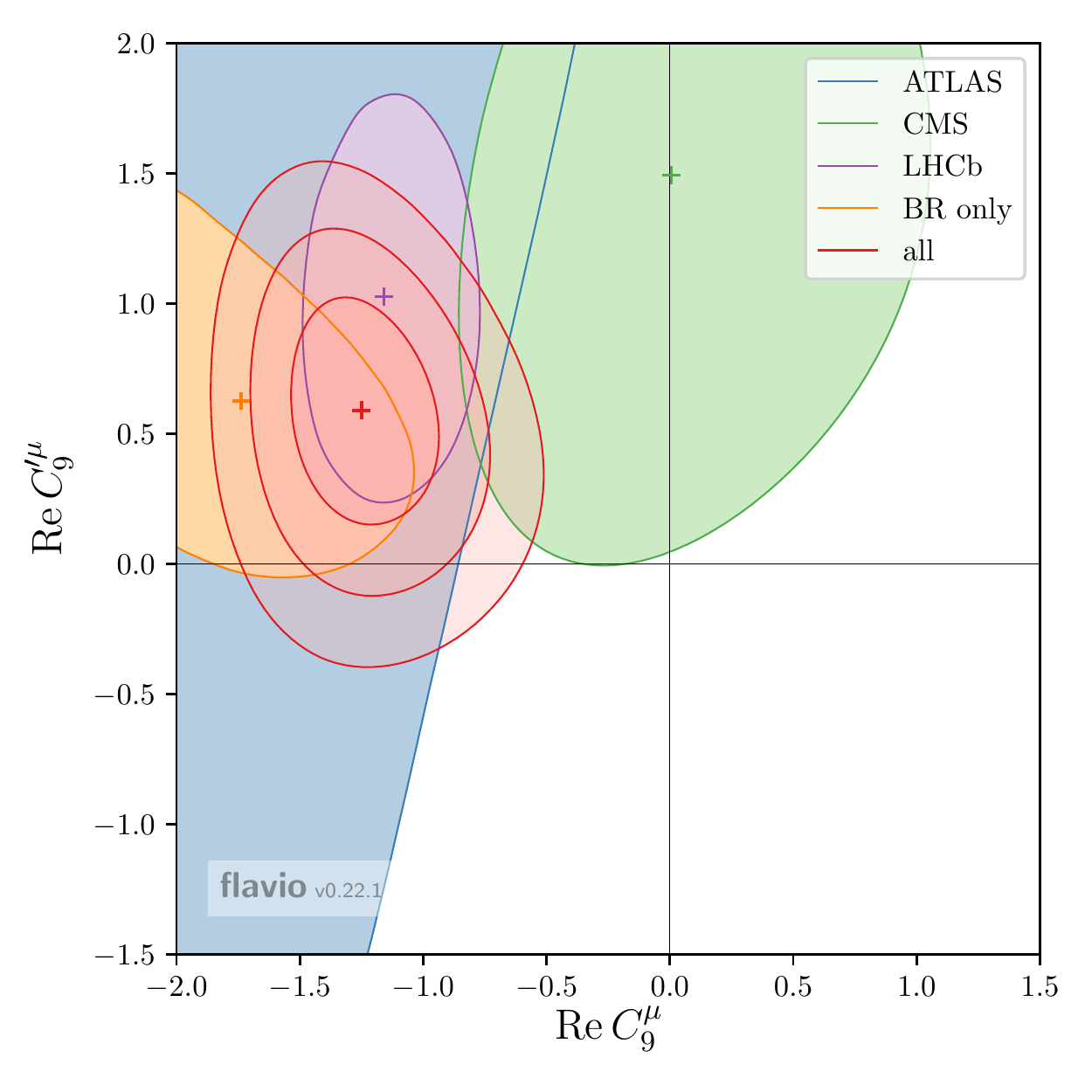}
 \caption{}
\end{subfigure}
\caption{Two-dimensional constraints in the plane of NP contributions to
the real parts of the Wilson coefficients $C_9^\mu$ and $C_{10}^\mu$~(a) or
$C_9^\mu$ and $C_9^{\prime\mu}$~(b), assuming all other Wilson coefficients to be SM-like.
For the constraints from the $B\to K^*\mu^+\mu^-$ and $B_s\to\phi\mu^+\mu^-$
angular observables from individual experiments as well as for the constraints
from branching ratio measurements of all experiments (``BR only''), the
$1\sigma$ ($\Delta\chi^2\approx 2.3$) contours are shown, while for the global fit (``all''), the 1, 2, and $3\sigma$ contours are shown.}
\label{fig:anomalies:bsmumu:C9_C10_and_C9_C9p}
\end{figure}
For the first two scenarios in table~\ref{tab:anomalies:bsmumu:pulls_2D}, i.e.\ NP in either $C_9^\mu$ and $C_{10}^\mu$ or in $C_9^\mu$ and $C_{9}^{\prime\mu}$, contours of constant $\Delta\chi^2$ are shown in figure~\ref{fig:anomalies:bsmumu:C9_C10_and_C9_C9p}.
The best-fit point and the 1$\sigma$, 2$\sigma$, and 3$\sigma$ contours are shown for the global fit.
In addition, the best-fit points and the 1$\sigma$ contours are shown for four fits with only a subset of the data.
These four fits only include respectively
\begin{itemize}
 \item the new measurement of $B^0\to K^{*0}\mu^+\mu^-$ angular observables by ATLAS,
 \item the new measurement of $B^0\to K^{*0}\mu^+\mu^-$ angular observables by CMS,
 \item the angular analysis of $B^0\to K^{*0}\mu^+\mu^-$ and $B_s\to\phi\mu^+\mu^-$ by LHCb,
 \item the branching ratio measurements by all experiments.
\end{itemize}
One observes that the cases including only a subset of the data are all compatible with the global fit at the 1$\sigma$ or 2$\sigma$ level.
While the angular analysis by CMS is compatible with the SM, all other measurements show deviations.
Due to their precision, the LHCb measurements of the angular observables and branching ratios dominate the global fit.
This leads to allowed regions similar to those in previous analyses irrespectively of the new measurements by ATLAS and CMS (cf.\ e.g.~\cite{Altmannshofer:2014rta}).
One finds no significant preference of the global fits for non-zero NP contributions to either $C_{10}^\mu$ or $C_{9}^{\prime\mu}$ in the two scenarios shown in figure~\ref{fig:anomalies:bsmumu:C9_C10_and_C9_C9p}.
This conclusion is similar to the one drawn above from comparing the pull of the two-dimensional cases including NP contributions to $C_9^\mu$ to the one-dimensional case with NP contributions only to $C_9^\mu$.

\begin{figure}[t]
\centering
\begin{subfigure}{0.49\textwidth}
\includegraphics[width=\textwidth]{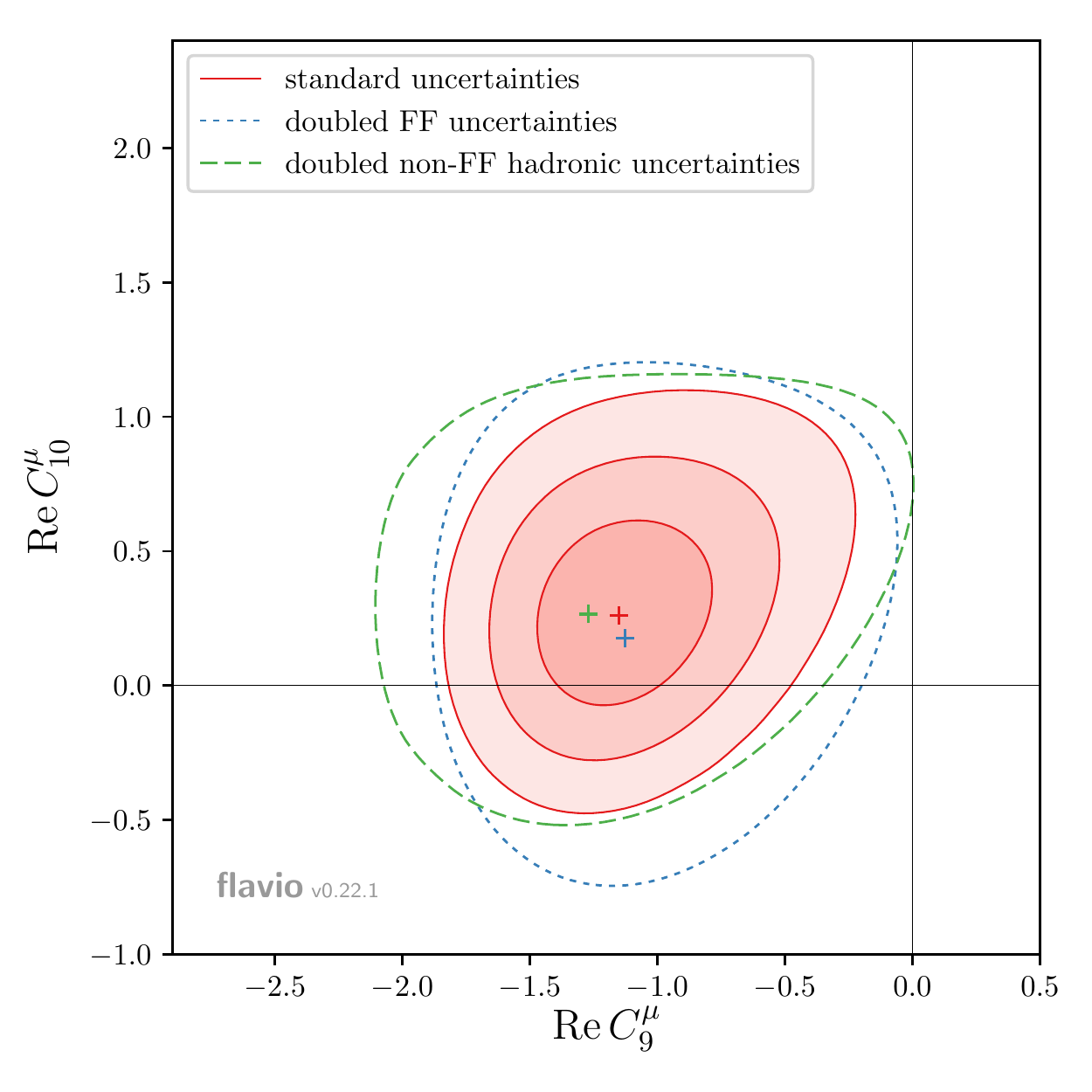}%
 \caption{}
\end{subfigure}
\begin{subfigure}{0.49\textwidth}
\includegraphics[width=\textwidth]{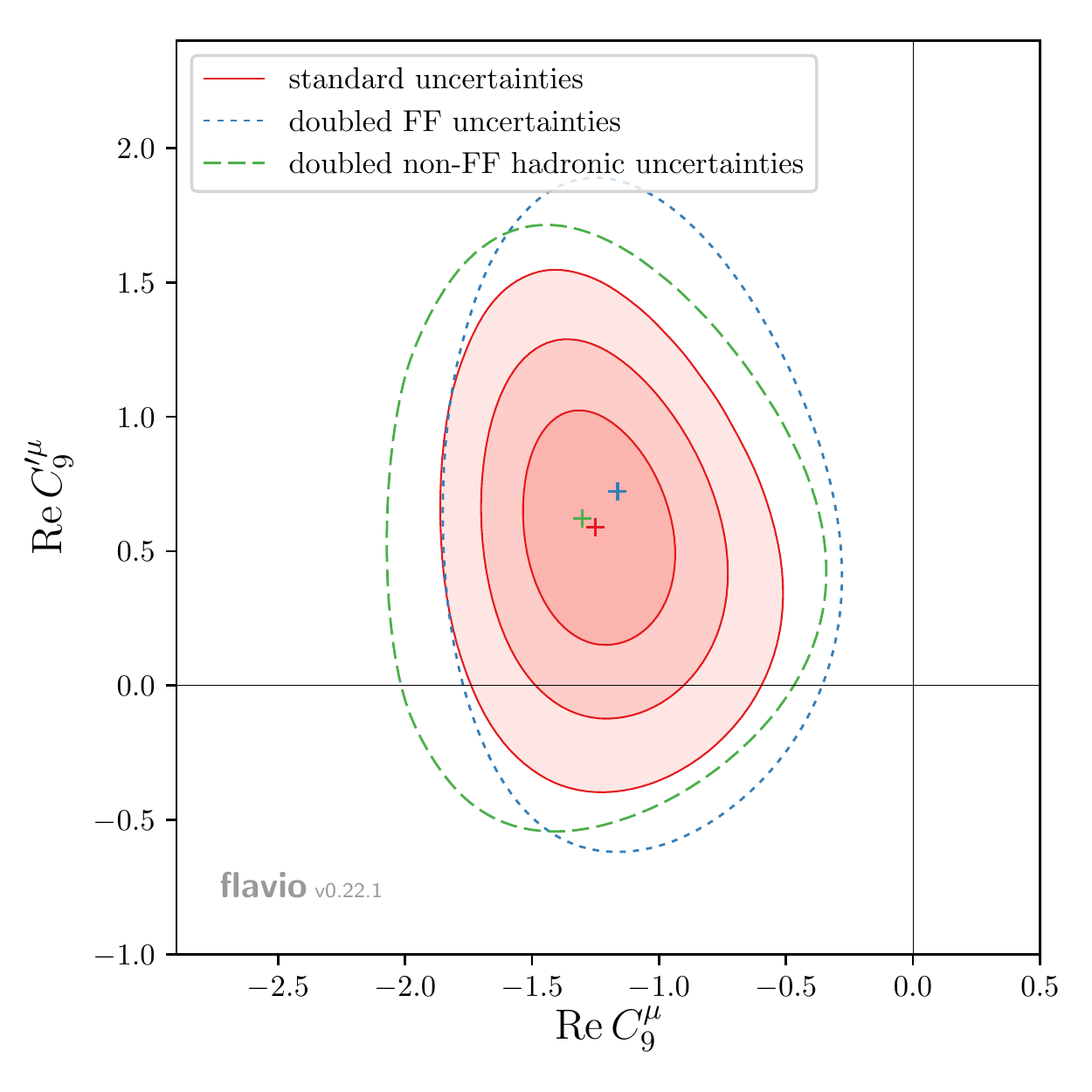}
 \caption{}
\end{subfigure}
\caption{Allowed regions in the Re$(C_9^\mu)$-Re$(C_{10}^\mu)$ plane~(a) and the Re$(C_9^\mu)$-Re$(C_9^{\prime\mu})$ plane~(b). In red the $1\sigma$, $2\sigma$, and $3\sigma$ best fit regions with nominal hadronic uncertainties. The green dashed and blue short-dashed contours correspond to the $3\sigma$ regions in scenarios with doubled uncertainties from non-factorizable corrections and doubled form factor uncertainties, respectively.}
\label{fig:anomalies:bsmumu:C9_C10_and_C9_C9p_uncertainties}
\end{figure}

To again estimate the impact of underestimated hadronic uncertainties, we have performed two-dimensional fits for the scenarios shown in figure~\ref{fig:anomalies:bsmumu:C9_C10_and_C9_C9p} but with doubled uncertainties either of the form factors or of the non-factorizable hadronic corrections.
In the scenario with NP allowed in $C_9^\mu$ and $C_{10}^\mu$, the pull is reduced from 5.0$\sigma$ to 3.7$\sigma$ and 4.1$\sigma$, respectively.
In the scenario with NP allowed in $C_9^\mu$ and $C_{9}^{\prime\mu}$, the pull is reduced from 5.3$\sigma$ to 4.1$\sigma$ and 4.4$\sigma$, respectively.
The best-fit points and 3$\sigma$ contours of the cases with doubled uncertainties are shown together with the best-fit points and the 1$\sigma$, 2$\sigma$, and 3$\sigma$ contours of the global fit in figure~\ref{fig:anomalies:bsmumu:C9_C10_and_C9_C9p_uncertainties}.
One observes that doubling the uncertainties is not sufficient for the SM point to lie inside the 3$\sigma$ contours.
Thus, like the one-dimensional fits, also the two-dimensional fits suggest that underestimated uncertainties are not the only source of the discrepancies.
\subsection{New physics or hadronic effects?}
\begin{figure}[t]
\centering
\begingroup
\sbox0{\includegraphics{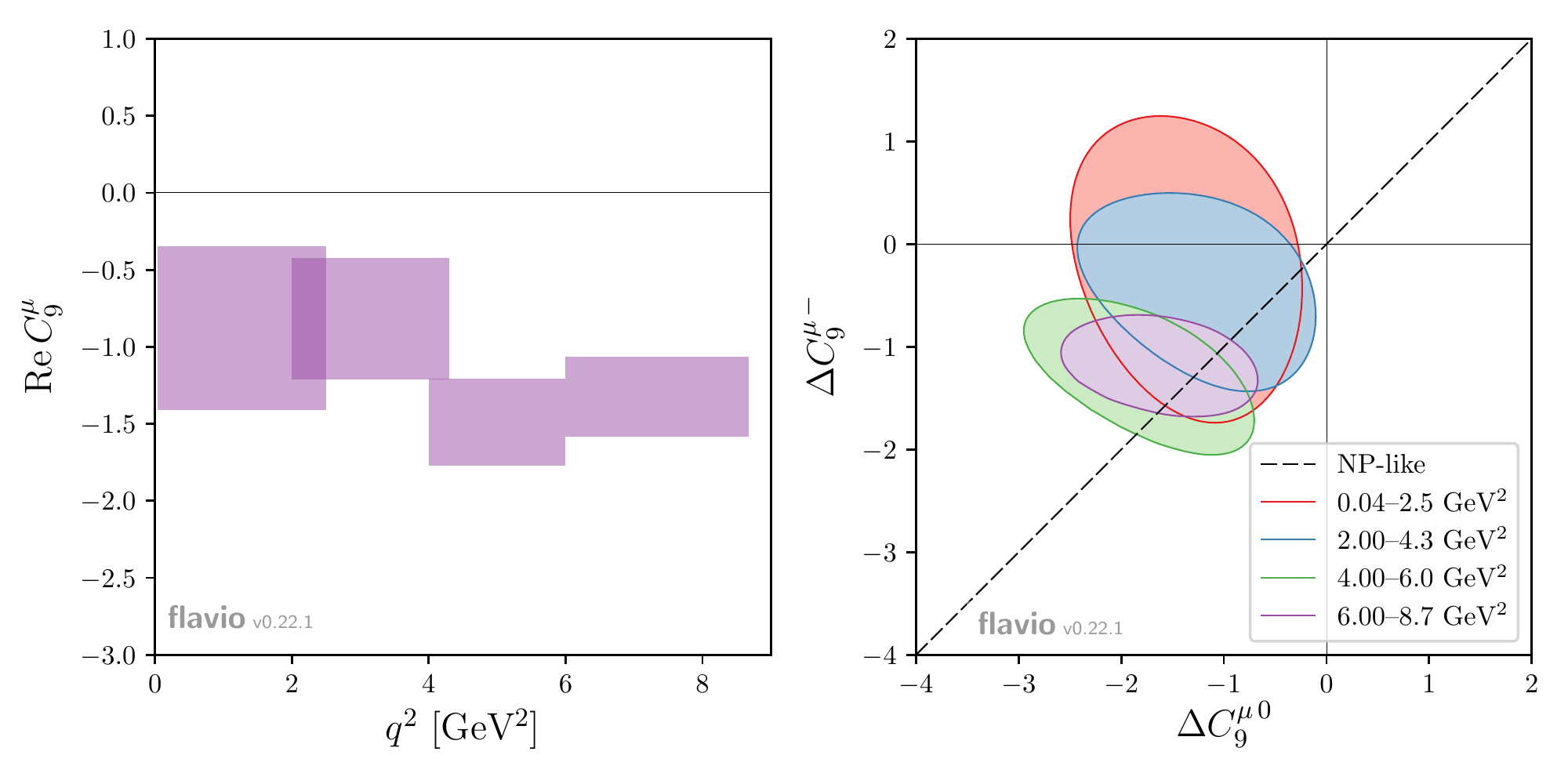}}%
\begin{subfigure}{0.49\textwidth}
\includegraphics[trim={0 0 {0.5\wd0} 0},clip,width=\textwidth]{figures/DeltaC9_thesis}
 \caption{}\label{fig:anomalies:bsmumu:C9_q2}
\end{subfigure}
\begin{subfigure}{0.49\textwidth}
\includegraphics[trim={{0.5\wd0} 0 0 0},clip,width=\textwidth]{figures/DeltaC9_thesis}
 \caption{}\label{fig:anomalies:bsmumu:deltaC9}
\end{subfigure}
\endgroup
\caption{(a):~Preferred $1\sigma$ ranges for a new physics contribution to $C_9^\mu$ from fits in different $q^2$ bins.
(b):~Preferred $1\sigma$ ranges for helicity dependent contributions to $C_9^\mu$ from fits in different $q^2$ bins. The dashed diagonal line corresponds to a helicity universal contribution, as predicted by new physics.}
\label{fig:anomalies:bsmumu:C9_q2_and_deltaC9}
\end{figure}
Any hadronic contribution to the $B^0\to K^{*0}\mu^+\mu^-$ helicity amplitudes that is photon mediated can in general be expressed by a $q^2$ and helicity dependent contribution to $C_9^\mu$: the photon couples to leptons via a vector current and the flavor-changing quark transition requires a left-handed current in the SM.
Underestimated hadronic effects could therefore mimic a NP contribution to $C_9^\mu$.
While a NP contribution is per definition $q^2$ and helicity independent, there is, however, no reason to expect that this is also the case for a hadronic contribution.
In fact, it is likely that hadronic effects in the $\lambda=+$ helicity amplitudes are suppressed~\cite{Jager:2012uw} and there is no reason why they should be of similar size in the $\lambda=0$ and $\lambda=-$ amplitudes.
So one would in general assume that an effect in $C_9^\mu$ due to hadronic effects is helicity dependent.
Furthermore, underestimated hadronic effects e.g.\ from charm loops are in general expected to show a non-trivial $q^2$ dependence.
Another interesting possibility that could mimic a NP effect in $C_9^\mu$ is NP contributions to $b\to c\bar{c}s$ operators (cf.~\cite{Lyon:2014hpa,Jager:2017gal}).
While the shift in $C_9^\mu$ would in this case be helicity independent up to correction of order $\alpha_s$ and $\Lambda_{QCD}/m_b$, it would have a non-trivial $q^2$ dependence.

To test whether the $B^0\to K^{*0}\mu^+\mu^-$ measurements actually show a preference for a $q^2$ or helicity dependent shift in $C_9^\mu$, we have performed fits in individual $q^2$ bins including only these measurements.
The bins of $q^2$ considered are\footnote{%
The overlaps in the bins are due to different experiments using different binning.
} $[0.04,2.5]$, $[2.0,4.3]$, $[4.0,6.0]$, $[6.0,8.7]$ in GeV$^2$, respectively.
While the latter bin is not included in the NP fits due to the unreliability of the estimation of hadronic effects in this region, it is used here to explicitly look for hadronic effects that mimic a shift in $C_9^\mu$.
In a first fit, equal contributions to the different helicity amplitudes have been assumed, while in a second fit also helicity dependent shifts have been allowed.

The results of the first fit are shown in figure~\ref{fig:anomalies:bsmumu:C9_q2}.
While the significance of the tension is more pronounced in the bins above 4 GeV$^2$, this is expected due to the higher sensitivity of the observables to $C_9^\mu$ in this region.
One observes that every individual bin shows a preference for a negative shift in $C_9^\mu$ that is compatible with a $q^2$ independent shift $C_9^\mu\approx-1.2$ at the 1$\sigma$ level.

In figure~\ref{fig:anomalies:bsmumu:deltaC9}, the 1$\sigma$ contours from the second fit are shown in the $\Delta C_9^{\mu\,-}$-$\Delta C_9^{\mu\,0}$ plane for each individual bin%
, where
$\Delta C_9^{\mu\,-}$ and $\Delta C_9^{\mu\,0}$ denote the contributions to the $\lambda=-$ and $\lambda=0$ helicity amplitudes, respectively.
The contours show perfect agreement with the assumption of a helicity universal shift, i.e.\ $\Delta C_9^{\mu\,-}=\Delta C_9^{\mu\,0}$.
Furthermore, the results for the individual contours corresponding to different $q^2$ bins are all consistent with each other.

Consequently, neither a preference for a dependence on $q^2$ nor on helicity is shown by the experimental data.
While this is an intriguing result, no robust prediction can be made at present about the precise properties of possible hadronic effects.
Therefore, they cannot be excluded as the actual source of the discrepancies in $b\to s\,\mu^+\mu^-$ transitions.


\section{Hints for violation of lepton flavor universality}\label{sec:anomalies:LFUV}
While hadronic effects could in principle be responsible for the $b\to s\,\mu^+\mu^-$ anomaly, deviations in LFU observables are clear evidence of NP.
Accordingly, in~\cite{Altmannshofer:2017yso} we have performed an analysis similar to the one presented in the previous section but taking into account all measurements of LFU observables available so far.
In particular, we have first performed ``LFU-only'' fits that only include the LFU observables
\begin{itemize}
 \item $R_K$ measured by LHCb~\cite{Aaij:2014ora},
 \item $R_{K^*}$ measured by LHCb~\cite{Aaij:2017vbb},
 \item $D_{P_4'}$ and $D_{P_5'}$ measured by Belle~\cite{Wehle:2016yoi}.
\end{itemize}
Subsequently, we have performed global fits, where we have considered all $b\to s\,\mu^+\mu^-$ observables included in the fits in section~\ref{sec:anomalies:bsmumu}, the LFU observables listed above, and in addition
\begin{itemize}
 \item the $B_s\to\mu^+\mu^-$ branching ratio measured by CMS~\cite{Chatrchyan:2013bka} and LHCb~\cite{Aaij:2017vad} (assuming no NP contribution to scalar operators),
 \item the $B\to X_s e^+ e^-$ branching ratio measured by BaBar~\cite{Lees:2013nxa}.
\end{itemize}

\subsection{New physics in one or two Wilson coefficients}\label{sec:anomalies:LFUV:NP_in_one_two_WCs}
\begin{figure}[t]
\centering
\begin{subfigure}{0.49\textwidth}
\includegraphics[width=\textwidth]{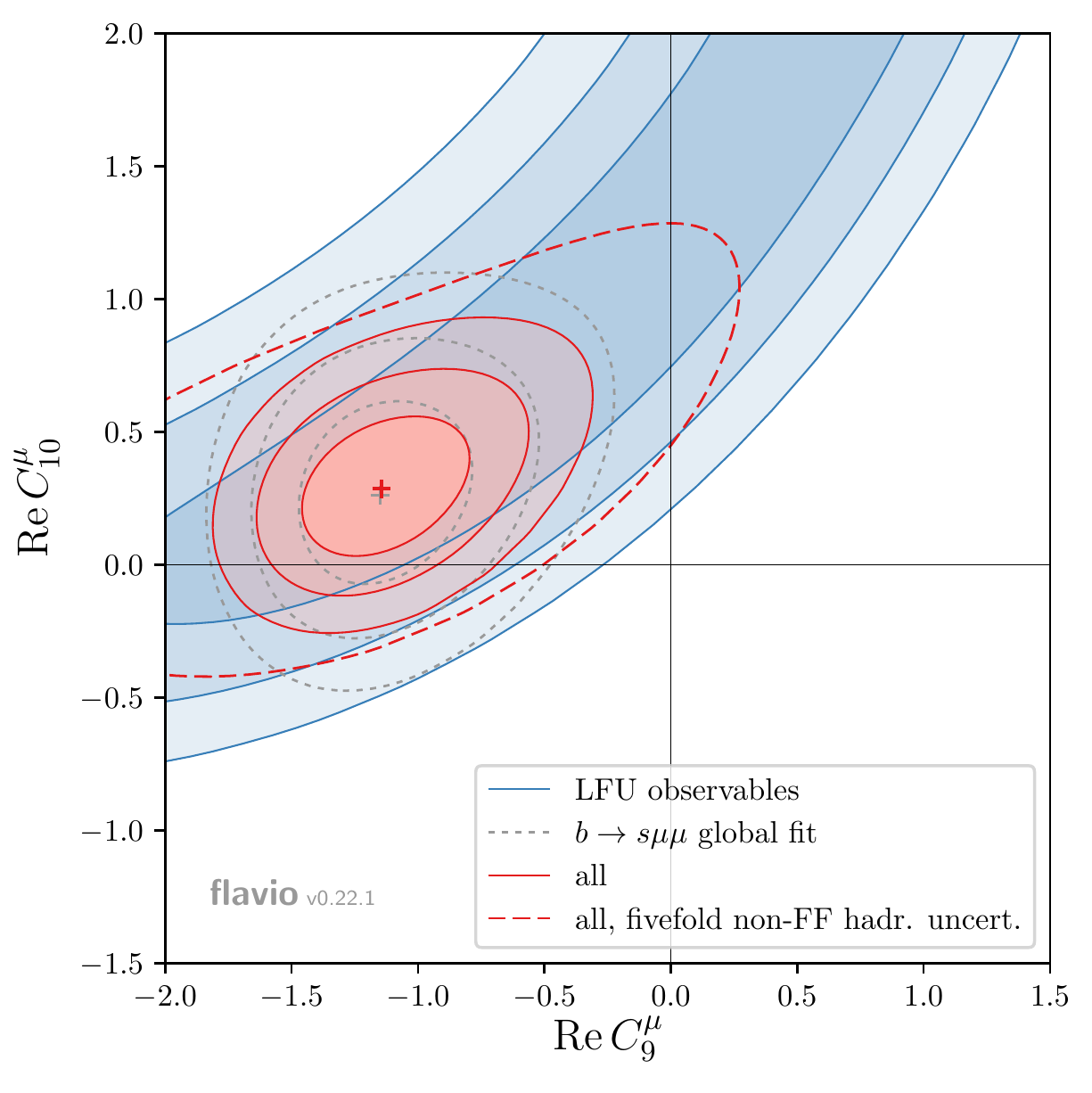}%
 \caption{}\label{fig:anomalies:LFUV:C9C10}
\end{subfigure}
\begin{subfigure}{0.49\textwidth}
\includegraphics[width=\textwidth]{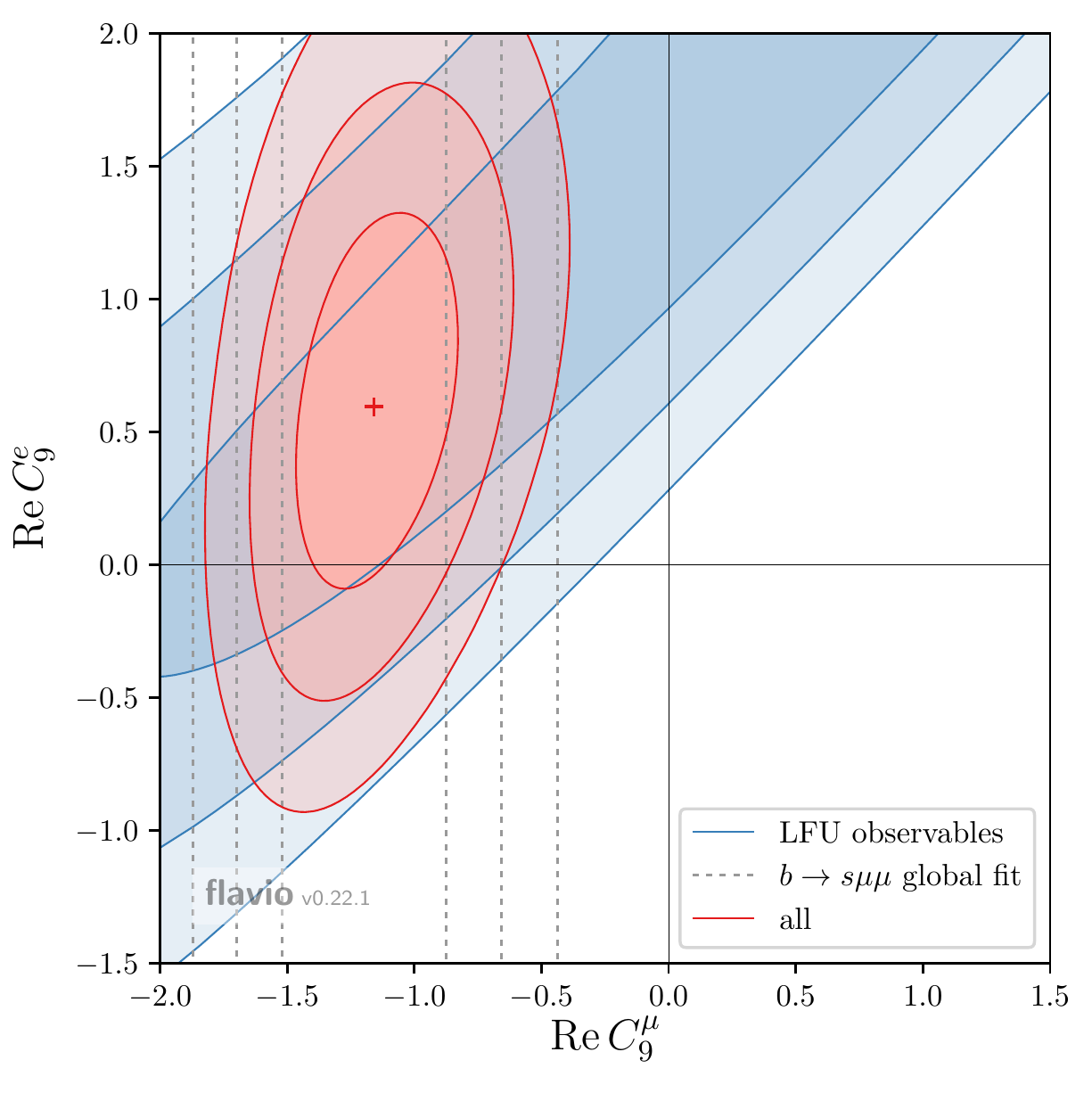}
 \caption{}\label{fig:anomalies:LFUV:C9C9e}
\end{subfigure}
\caption{Allowed regions in the Re$(C_9^\mu)$-Re$(C_{10}^\mu)$ plane~(a) and the Re$(C_9^\mu)$-Re$(C_9^e)$ plane~(b).
The 1$\sigma$, 2$\sigma$, and 3$\sigma$ contours are shown for the ``LFU-only'' (blue), the $b\to s \mu\mu$ (dotted gray), and the combined global fit (red).
The plot on the left also shows the 3$\sigma$ contour of a global fit with quintupled non-factorizable hadronic uncertainties (dashed red).
}
\label{fig:anomalies:LFUV:C9C10_C9C9e}
\end{figure}

\renewcommand{\arraystretch}{1.2}
\begin{table}[t]
\begin{center}
\begin{tabularx}{\textwidth}{ccccX}
\hline\hline
 ~~~~~~ Coeff. ~~~~~~ & ~~~~~~~ best fit ~~~~~~~ & ~~~~~~~~~~ $1\sigma$ ~~~~~~~~~~ & ~~~~~~~~~~ $2\sigma$ ~~~~~~~~~~ & pull \\
\hline\hline
\rowcolor[gray]{.9} $C_9^{\mu}                      $ & $-1.56$ & [$-2.12$, $-1.10$] & [$-2.87$, $-0.71$] & $4.1\sigma$\\
                    $C_{10}^{\mu}                   $ & $+1.20$ & [$+0.88$, $+1.57$] & [$+0.58$, $+2.00$] & $4.2\sigma$\\
\rowcolor[gray]{.9} $C_9^{e}                        $ & $+1.54$ & [$+1.13$, $+1.98$] & [$+0.76$, $+2.48$] & $4.3\sigma$\\
                    $C_{10}^{e}                     $ & $-1.27$ & [$-1.65$, $-0.92$] & [$-2.08$, $-0.61$] & $4.3\sigma$\\
\rowcolor[gray]{.9} $C_9^{\mu}=-C_{10}^{\mu}        $ & $-0.63$ & [$-0.80$, $-0.47$] & [$-0.98$, $-0.32$] & $4.2\sigma$\\
                    $C_9^{e}=-C_{10}^{e}            $ & $+0.76$ & [$+0.55$, $+1.00$] & [$+0.36$, $+1.27$] & $4.3\sigma$\\
\rowcolor[gray]{.9} $C_9^{e}=C_{10}^{e}             $ & $-1.91$ & [$-2.30$, $-1.51$] & [$-2.71$, $-1.10$] & $3.9\sigma$\\
\hline\hline
\end{tabularx}
\end{center}
\caption{Best-fit values with their 1$\sigma$ and 2$\sigma$ ranges and pulls in sigma between the best-fit point and the SM point for scenarios with NP in one Wilson coefficient when considering only LFU observables.
Scenarios with NP in only primed WCs are not shown; they cannot improve the fit compared to the SM (cf.\ discussion in main text and figure~\ref{fig:anomalies:LFUV:C9C9p}).
}
\label{tab:anomalies:LFUV:pulls_1D}
\end{table}
%
%
%
%
\begin{figure}[t]
\centering
\begin{subfigure}{0.49\textwidth}
\includegraphics[width=\textwidth]{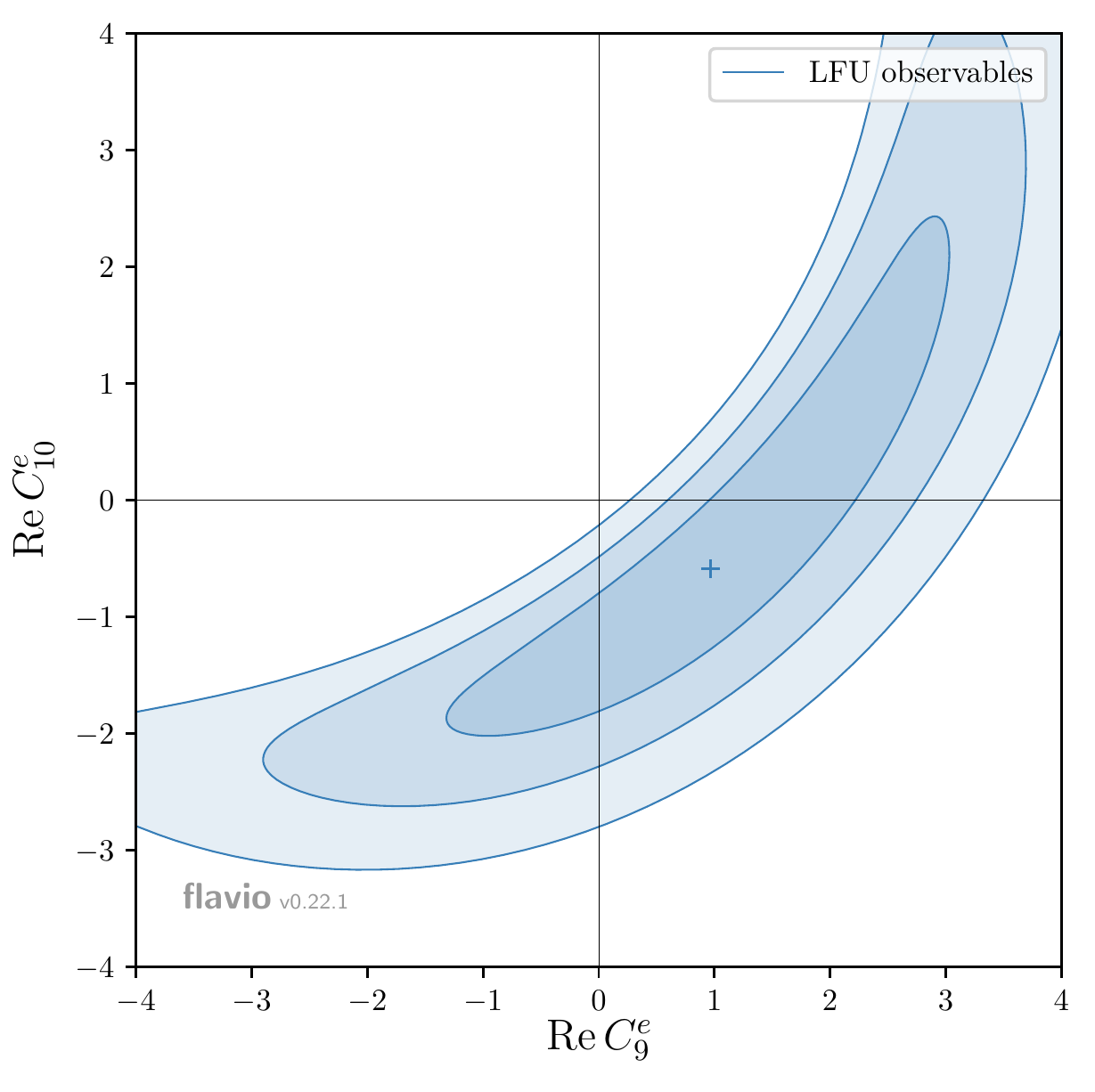}%
 \caption{}\label{fig:anomalies:LFUV:C9eC10e}
\end{subfigure}
\begin{subfigure}{0.49\textwidth}
\includegraphics[width=\textwidth]{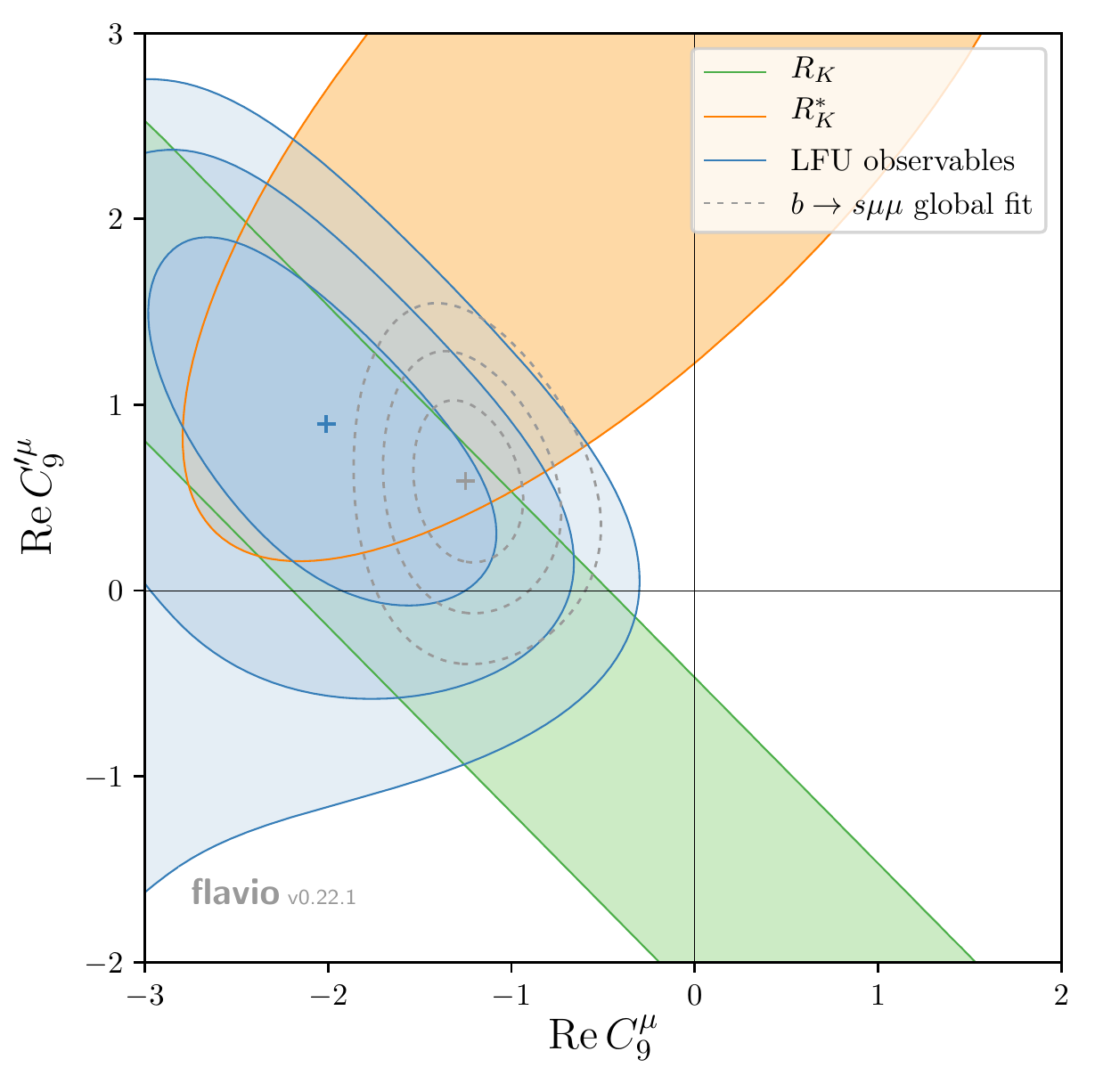}%
 \caption{}\label{fig:anomalies:LFUV:C9C9p}
\end{subfigure}
\caption{(a):~Re$(C_9^e)$-Re$(C_{10}^e)$ plane showing 1, 2, and 3$\sigma$ contours from the ``LFU-only'' fit.
(b):~Re$(C_9^\mu)$-Re$(C_{10}^{\prime\mu})$ plane showing the 1$\sigma$ contours of fits including only $R_K$ or $R_{K^*}$ in green and yellow, respectively, and the 1, 2, and 3$\sigma$ contours for the ``LFU-only'' fit in blue.
In addition, the contours of the $b\to s\mu\mu$ fit described in section~\ref{sec:anomalies:bsmumu} are shown in dotted gray.}
\label{fig:anomalies:LFUV:C9eC10e_C9C9p}
\end{figure}
In our fits, we have allowed for NP contributions to the eight WCs $C_9^{(\prime) \ell}$ and $C_{10}^{(\prime)\ell}$, with $\ell\in \{e,\mu\}$.
The results of one-dimensional ``LFU-only'' scenarios that can describe the data significantly better than the SM are collected in table~\ref{tab:anomalies:LFUV:pulls_1D}.
Contours of constant $\Delta\chi^2 \approx 2.3, 6.2, 11.8$ in the plane of two WCs are shown in figures~\ref{fig:anomalies:LFUV:C9eC10e}, \ref{fig:anomalies:LFUV:C9C9p}, \ref{fig:anomalies:LFUV:C9C10}, and~\ref{fig:anomalies:LFUV:C9C9e} for the scenarios with NP in $C_9^e$ and $C_{10}^e$, $C_9^\mu$ and $C_9^{\prime \mu}$, $C_9^\mu$ and $C_{10}^\mu$, as well as $C_9^\mu$ and $C_9^{e}$, respectively.
One observes that all scenarios in table~\ref{tab:anomalies:LFUV:pulls_1D} have a pull of around $4\sigma$ and involve NP contributions to WCs corresponding to left-handed quark currents.
In particular, a negative NP contribution to $C_9^{\mu}$ and/or a positive NP contribution to $C_{10}^{\mu}$ improves the agreement with the data significantly; this corresponds to a decrease in both $\mathcal B(B \to K \mu^+\mu^-)$ and $\mathcal B(B \to K^{*} \mu^+\mu^-)$.
A positive NP contribution to $C_9^{e}$ and/or a negative NP contribution to $C_{10}^{e}$ yields a similar result; this corresponds to an increase in both $\mathcal B(B \to K e^+e^-)$ and $\mathcal B(B \to K^{*} e^+e^-)$.
The above cases can be combined in terms of approximate flat directions that give an excellent description of the data,
\begin{equation}\label{eq:anomalies:LFUV:flat_directions}
 C_9^\mu - C_9^e - C_{10}^\mu + C_{10}^e \approx -1.4 ~,
\end{equation}
at least if the absolute value of a single WC is not much larger than 1.
These flat directions are also visible in the plot in figure~\ref{fig:anomalies:LFUV:C9eC10e} and both plots in figure~\ref{fig:anomalies:LFUV:C9C10_C9C9e}.
While right-handed muon currents ($C_9^\mu=C_{10}^\mu$) cannot describe the data (cf.\ figure~\ref{fig:anomalies:LFUV:C9C10}), a sizable contribution to right-handed electron currents ($C_9^e=C_{10}^e\approx-2$ or $+3$) yields a good fit (cf.\ table~\ref{tab:anomalies:LFUV:pulls_1D} and figure~\ref{fig:anomalies:LFUV:C9eC10e}).
NP contributions to only primed WCs, which correspond to right-handed quark currents, cannot improve the agreement with the data.
It is well known~\cite{Hiller:2014ula} that they shift $R_K$ and $R_{K^*}$ away from 1 into opposite directions.
However, the data prefers $R_K$ and $R_{K^*}$ both being smaller than 1.
For a NP contribution only to $C_9^{\prime \mu}$, the impossibility of accommodating the measurements of both $R_K$ and $R_{K^*}$ can be observed in figure~\ref{fig:anomalies:LFUV:C9C9p}:
a negative NP contribution to $C_9^{\prime \mu}$ is required for an agreement with the $R_K$ measurement, while a positive contribution is required to accommodate the $R_{K^*}$ measurement.
So while each individual measurement of either $R_K$ or $R_{K^*}$ could be explained by NP in right-handed quark currents, both measurements together exclude this possibility.
However, the plot in figure~\ref{fig:anomalies:LFUV:C9C9p} shows that in the presence of a sizable negative $C_9^\mu$, a non-zero $C_9^{\prime \mu}$ can improve the fit; similar improvements can be found for other combinations of primed and unprimed WCs.

In addition to the blue contours showing the results of ``LFU-only'' fits, the plots in figures~\ref{fig:anomalies:LFUV:C9eC10e_C9C9p} and~\ref{fig:anomalies:LFUV:C9C10_C9C9e} also show dotted gray contours of the $b\to s\mu\mu$ fits described in section~\ref{sec:anomalies:bsmumu} (except for the plot in figure~\ref{fig:anomalies:LFUV:C9eC10e} that does not involve muons).
Intriguingly, the ``LFU-only'' fits are fully compatible with the $b\to s\mu\mu$ fits.
Consequently, the combined global fits shown as red contours prefer a non-zero NP contribution with very high significance.
As detailed in section~\ref{sec:anomalies:bsmumu}, the global fit is, however, subject to possibly underestimated hadronic uncertainties.
To estimate their impact, the plot in figure~\ref{fig:anomalies:LFUV:C9C10} shows a red dashed line for the 3$\sigma$ contour of the global fit with non-factorizable hadronic uncertainties inflated by a factor 5 compared to the nominal uncertainties.
In this case, the global fit gets dominated by the LFU observables.
However, even with these huge uncertainties, there are still relevant constraints coming from to the $b\to s \mu\mu$ observables.
For example, the best-fit point of the ``LFU-only'' fit with NP only in $C_{10}^{\mu}$ implies a 50\% suppression of the $B_s\to\mu\mu$ branching ratio, which is already in tension with current measurements~\cite{Aaij:2017vad} (assuming no NP in scalar WCs interfering).
Furthermore, the inclusion of the $b\to s \mu\mu$ observables strongly favors NP in the muon WCs over NP in the electron WCs.
While a non-zero contribution to an electron WC can still improve the global fit in the presence of a sizable muon WC  (cf.\ figure~\ref{fig:anomalies:LFUV:C9C9e}), the $b\to s \mu\mu$ data of course cannot be explained by NP only in electron WCs.
In this sense, the ambiguity due to the flat directions, eq.~(\ref{eq:anomalies:LFUV:flat_directions}), is lifted by taking into account the $b\to s \mu\mu$ data, and a NP contribution to $C_9^{\mu}$ seems unavoidable for explaining the hints for LFUV and the $b\to s \mu\mu$ anomaly at once.

\subsection{Predictions for LFU observables}\label{sec:anomalies:LFUV:predictions}

\begin{figure}[t]
\centering
\begingroup
\sbox0{\includegraphics{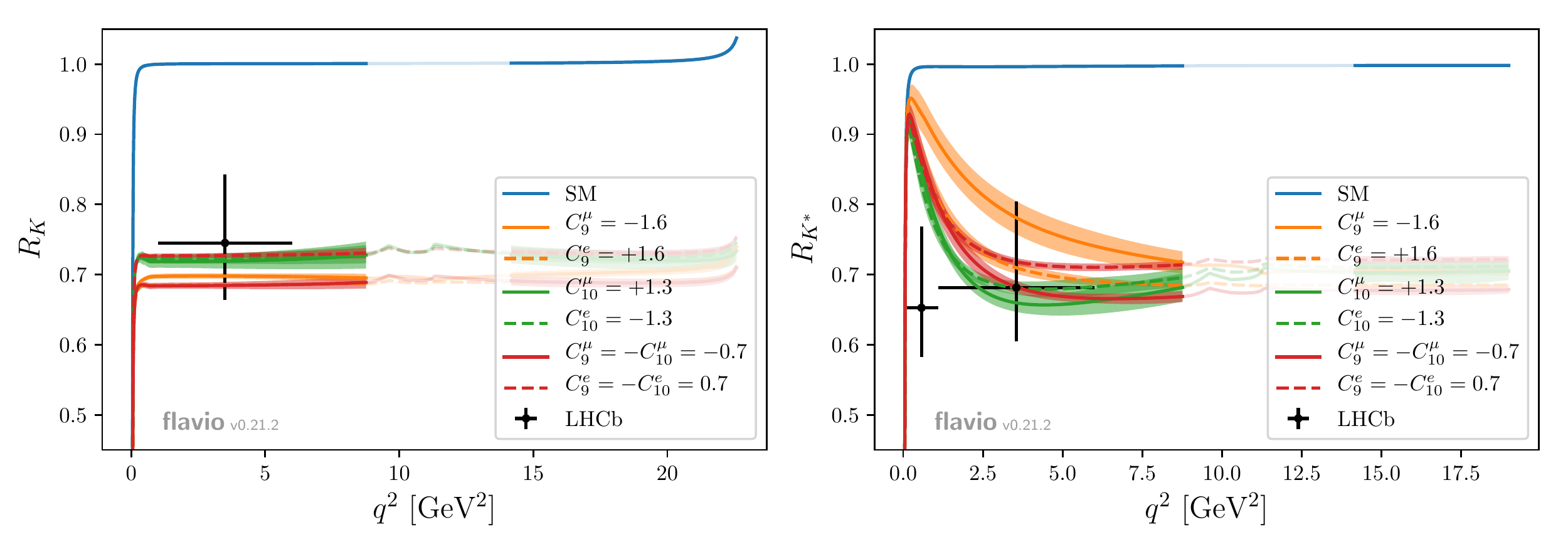}}%
\begin{subfigure}{0.49\textwidth}
\includegraphics[trim={0 0 {0.5\wd0} 0},clip,width=\textwidth]{figures/R-q2}
 \caption{}
\end{subfigure}
\begin{subfigure}{0.49\textwidth}
\includegraphics[trim={{0.5\wd0} 0 0 0},clip,width=\textwidth]{figures/R-q2}
 \caption{}
\end{subfigure}
\endgroup
\caption{The LFU ratios $R_{K^{(*)}}$ in the SM and various
NP benchmark models as function of $q^2$.
The error bands contain all theory uncertainties including
form factors and non-factorisable hadronic effects.
In the region of narrow charmonium resonances, only the
short-distance contribution without uncertainties is shown.}
\label{fig:anomalies:LFUV:RK_RKstar}
\end{figure}
\begin{figure}[t]
\centering
\begingroup
\sbox0{\includegraphics{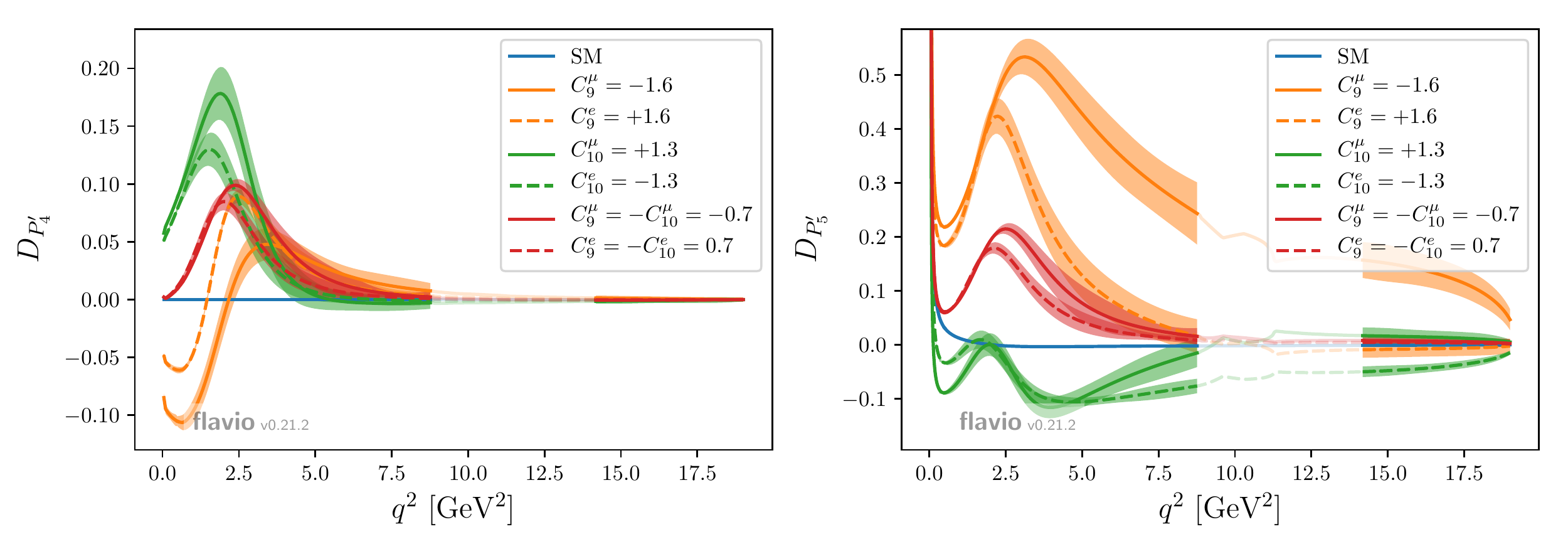}}%
\begin{subfigure}{0.49\textwidth}
\includegraphics[trim={0 0 {0.5\wd0} 0},clip,width=\textwidth]{figures/DP-q2}
 \caption{}
\end{subfigure}
\begin{subfigure}{0.49\textwidth}
\includegraphics[trim={{0.5\wd0} 0 0 0},clip,width=\textwidth]{figures/DP-q2}
 \caption{}
\end{subfigure}
\endgroup
\caption{The $B \to K^* \ell^+ \ell^-$ LFU differences
$D_{P_4^\prime}$ and $D_{P_5^\prime}$ in the SM and various
NP benchmark models as functions of $q^2$.
Concerning the error bands, the same comments as for Fig.~\ref{fig:anomalies:LFUV:RK_RKstar} apply.
}
\label{fig:anomalies:LFUV:DP4p_DP5p}
\end{figure}

While including $b\to s \mu\mu$ observables into the fit can single out certain scenarios, it is interesting to investigate if this can be done by LFU observables alone.
To this end, predictions of several NP scenarios for $R_K$ and $R_{K^*}$ as well as $D_{P_4'}$ and $D_{P_5'}$ are shown in figure~\ref{fig:anomalies:LFUV:RK_RKstar} and figure~\ref{fig:anomalies:LFUV:DP4p_DP5p}, respectively.

The plots in figure~\ref{fig:anomalies:LFUV:RK_RKstar} show that in the SM, a $q^2$ independent $R_K=R_{K^*}=1$ is a very good approximation over a large range of $q^2$.
For very low $q^2$, both $R_K$ and $R_{K^*}$ drop to zero due to phase space effects.
While NP contributions to any of the considered WCs yield a virtually constant, $q^2$ independent shift of $R_K$, the observable $R_{K^*}$ on the other hand shows a non-trivial $q^2$ dependence in the presence of NP.
The main reason for this difference between $R_K$ and $R_{K^*}$ is that in contrast to $B \to K \ell^+\ell^-$, the decays $B \to K^* \ell^+\ell^-$ are dominated by lepton flavor universal dipole operators at low $q^2$.
Hence, in this region, all NP scenarios yield a prediction for $R_{K^*}$ close to its SM value.
Comparing $R_K$ and $R_{K^*}$, it is possible to distinguish some of the different NP scenarios in the case of sufficiently precise experimental data.
For instance, while a NP contribution to $C_{10}$ predicts larger deviations from the SM in $R_{K^*}$ than in $R_K$, a NP contribution to $C_{9}$ has the opposite effect.
This has the consequence that the current measurements have a slight preference for the $C_{10}^\mu$ over the $C_{9}^\mu$ scenario: a positive $C_{10}^\mu$ predicts $1>R_K>R_{K^*}$, which is in accordance with the measurements, while a negative $C_{9}^\mu$ predicts $1>R_{K^*}>R_K$.
However, the differences between the $C_9^\mu$ and the $C_{10}^\mu$ scenarios are tiny and distinguishing them would require high experimental precision.

This is quite different for the observables $D_{P_4^\prime}$ and $D_{P_5^\prime}$, for which predictions are shown in figure~\ref{fig:anomalies:LFUV:DP4p_DP5p}.
While they are close to zero for a large range of $q^2$ in the SM, they both show a non-trivial $q^2$ dependence in the presence of NP.
Compared to the LFU ratios $R_K$ and $R_{K^*}$, the LFU differences $D_{P_4^\prime}$ and $D_{P_5^\prime}$ allow a much clearer distinction between different scenarios.
In particular, a positive $C_{10}^\mu$ predicts positive values for $D_{P_4^\prime}$ at low $q^2<2.5$~GeV$^2$, while a negative $C_{9}^\mu$ predicts negative values in the same region; a similar behavior can be seen for electron WCs of opposite sign.
Considering $D_{P_5^\prime}$, one observes that a positive $C_{10}^\mu$ yields small negative values, while a negative $C_{9}^\mu$ corresponds to sizable positive values.
Interestingly, $D_{P_5^\prime}$ even allows for distinguishing between $C_{9}^\mu$ and $C_{9}^e$:
for $q^2>5$, a negative $C_{9}^\mu$ can lead to a sizable increase in $P_5^{\prime \mu}$, while a positive $C_{9}^e$ can only slightly decrease $P_5^{\prime e}$. This is due to the SM value lying already close to the model-independent lower bound of $P_5^{\prime}=-1$, such that a NP contribution cannot decrease it much further.

Although the SM predictions for the LFU observables have only tiny uncertainties, form factor and other hadronic uncertainties actually do play a role in the presence of NP.
However, they are still small enough such that sufficiently precise measurements could allow for a clean distinction between different NP scenarios.

\chapter[Violation of lepton flavor universality in CHM\lowercase{s}]{Violation of lepton flavor universality\\ in composite Higgs models}\label{chap:LUFV_in_CHMs} 

In light of the significant tension between the SM and experimental measurements of LFU observables (cf.\ section~\ref{sec:anomalies:LFUV}), it is interesting to ask
which NP model is actually capable of explaining the discrepancies.
%
Many more or less ad hoc models are able to
do this
by generating the WCs $C_9^\ell$ and $C_{10}^\ell$ at tree level from the exchange of a heavy neutral gauge boson~\cite{Altmannshofer:2014cfa,Buras:2014fpa,Glashow:2014iga,Bhattacharya:2014wla,
Crivellin:2015mga,Altmannshofer:2014rta,Crivellin:2015lwa} or of spin-0 or spin-1 leptoquarks~\cite{Hiller:2014yaa,Biswas:2014gga,Buras:2014fpa,Sahoo:2015wya,Hiller:2014ula}.
However, it is more difficult to accommodate the experimental central values in more complete models that also solve the naturalness problem of the SM.
In particular, it has been shown that this is not possible in the Minimal Supersymmetric Standard Model (MSSM)~\cite{Altmannshofer:2014rta}.
While it has been known that CHMs featuring composite leptoquarks can explain the data~\cite{Gripaios:2014tna}, we presented an arguably more simple mechanism in~\cite{Niehoff:2015bfa} that is only based on partial compositeness of SM particles and allows for an explanation of both a violation of LFU and the $b\to s\,\mu^+\mu^-$ anomaly.
This mechanism was later found to be at work also in extra-dimensional constructions~\cite{Megias:2016bde,Megias:2017ove} and models with fundamental partial compositeness (cf.~\cite{DAmico:2017mtc,Sannino:2017utc} and chapter~\ref{chap:Flavor_MFPC}).
This mechanism is described in the present chapter.

\begin{figure}[t]
\centering
\begin{subfigure}{0.32\textwidth}
\centering
$
\begin{fmffile}{Zexchange}
\vcenter{\hbox{
\begin{fmfgraph*}(90,70)
\fmfset{arrow_len}{2mm}
\fmfset{thin}{.7pt}
\fmfpen{thin}
\fmfstraight

\fmfleft{x1,s,,,,,,,b,x2}
\fmfright{y1,mum,,,,,,,mup,y2}

\fmf{fermion,tension=2}{b,mb}
\fmf{double,tension=2}{mb,v1}
\fmf{fermion,tension=2}{ms,s}
\fmf{double,tension=2}{v1,ms}

\fmf{dbl_wiggly,tension=6/2}{v1,mZ}
\fmf{boson,tension=6/7}{mZ,v2}
\fmf{phantom,tension=0,label=$Z$}{v1,v2}

\fmf{fermion}{mup,v2}
\fmf{fermion}{v2,mum}

\fmfdot{v1,v2}
\fmfv{decor.shape=circle,decor.filled=empty,decor.size=5pt,label.dist=0,label=\bf\tiny{+}}{ms,mb,mZ}

\fmfv{lab=$b$,l.angle=-180}{b}
\fmfv{lab=$s$,l.angle=-180}{s}
\fmfv{lab=$\mu$,l.angle=0}{mup}
\fmfv{lab=$\mu$,l.angle=0}{mum}

\end{fmfgraph*}
}}
\end{fmffile}
$
\caption{}
\label{fig:LFUV:diags_bsll:Zexchange}
\end{subfigure}
\begin{subfigure}{0.32\textwidth}
\centering
$
\begin{fmffile}{RHOexchange1}
\vcenter{\hbox{
\begin{fmfgraph*}(90,70)
\fmfset{arrow_len}{2mm}
\fmfset{thin}{.7pt}
\fmfpen{thin}
\fmfstraight

\fmfleft{x1,s,,,,,,,b,x2}
\fmfright{y1,mum,,,,,,,mup,y2}

\fmf{fermion,tension=2}{b,mb}
\fmf{double,tension=2}{mb,v1}
\fmf{fermion,tension=2}{ms,s}
\fmf{double,tension=2}{v1,ms}

\fmf{dbl_wiggly,tension=6/7}{v1,mZ}
\fmf{boson,tension=6/2}{mZ,v2}
\fmf{phantom,tension=0,l.d=8,label=$\rho$}{v2,v1}

\fmf{fermion}{mup,v2}
\fmf{fermion}{v2,mum}

\fmfdot{v1,v2}
\fmfv{decor.shape=circle,decor.filled=empty,decor.size=5pt,label.dist=0,label=\bf\tiny{+}}{ms,mb,mZ}

\fmfv{lab=$b$,l.angle=-180}{b}
\fmfv{lab=$s$,l.angle=-180}{s}
\fmfv{lab=$\mu$,l.angle=0}{mup}
\fmfv{lab=$\mu$,l.angle=0}{mum}

\end{fmfgraph*}}}
\end{fmffile}
$
\caption{}
\label{fig:LFUV:diags_bsll:rhoexchange1}
\end{subfigure}
\begin{subfigure}{0.32\textwidth}
\centering
$
\begin{fmffile}{RHOexchange2}
\vcenter{\hbox{
\begin{fmfgraph*}(90,70)
\fmfset{arrow_len}{2mm}
\fmfset{thin}{.7pt}
\fmfpen{thin}
\fmfstraight

\fmfleft{x1,s,,,,,,,b,x2}
\fmfright{y1,mum,,,,,,,mup,y2}

\fmf{fermion,tension=2}{b,mb}
\fmf{double,tension=2}{mb,v1}
\fmf{fermion,tension=2}{ms,s}
\fmf{double,tension=2}{v1,ms}

\fmf{dbl_wiggly,tension=2/3,l.d=8,label=$\rho$}{v2,v1}

\fmf{fermion,tension=2}{mup,mmup}
\fmf{double,tension=2}{v2,mmup}
\fmf{fermion,tension=2}{mmum,mum}
\fmf{double,tension=2}{mmum,v2}

\fmfdot{v1,v2}
\fmfv{decor.shape=circle,decor.filled=empty,decor.size=5pt,label.dist=0,label=\bf\tiny{+}}{ms,mb,mmup,mmum}

\fmfv{lab=$b$,l.angle=-180}{b}
\fmfv{lab=$s$,l.angle=-180}{s}
\fmfv{lab=$\mu$,l.angle=0}{mup}
\fmfv{lab=$\mu$,l.angle=0}{mum}

\end{fmfgraph*}}}
\end{fmffile}
$
\caption{}
\label{fig:LFUV:diags_bsll:rhoexchange2}
\end{subfigure}
\caption{
(a): $Z$ exchange.
(b): $\rho$~exchange with $\rho$-muon coupling due to $Z$-$\rho$ mixing.
(c):~$\rho$~exchange with $\rho$-muon coupling due to muons mixing with their heavy partners.
}\label{fig:LFUV:diags_bsll}
\end{figure}
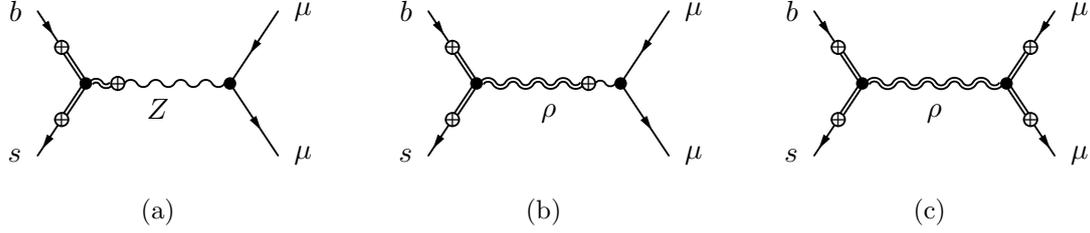

As is shown in chapter~\ref{chap:anomalies}, very good fits to the experimental data on $b\to s\,\mu^+\mu^-$ and LFU observables can be achieved by (cf.\ tables~\ref{tab:anomalies:bsmumu:pulls_1D} and~\ref{tab:anomalies:LFUV:pulls_1D})
\begin{itemize}
\item negative $C_{9}^{\mu}$ and all other WCs SM-like, corresponding to a vector-like muon current:
\begin{equation}
 C_{9}^{\mu}<0,
\end{equation}
\item negative $C_{9}^{\mu}$ and positive $C_{10}^{\mu}$ of the same absolute size, corresponding to a left-handed muon current:
\begin{equation}
 C_{9}^{\mu}=-C_{10}^{\mu}<0,
\end{equation}
\end{itemize}
where the same convention as in chapter~\ref{chap:anomalies} is used, i.e.\ $C_{9}^{\mu}$ and $C_{10}^{\mu}$ are defined such that they vanish in the SM.
The WCs of the WEH in eq.~(\ref{eq:anomalies:Heff}) that are due to FCNCs can arise only at one loop in the SM.
In models with partial compositeness, on the other hand, mixing of elementary and composite
fields allows them to be generated already at tree level.
For a $b\to s\,\ell^+\ell^-$ transition, there are the three possibilities shown in the diagrams in figure~\ref{fig:LFUV:diags_bsll}, where $\rho$ denotes a composite spin-one state that can mix with the $Z$ boson.
In the first case, figure~\ref{fig:LFUV:diags_bsll:Zexchange}, a flavor-changing $Z$ coupling for the left-handed quarks is induced by the mixing with composite states.
The lepton-$Z$ coupling is SM-like in this case and yields $C_{10}^\mu\gg C_{9}^\mu$.
This is obviously not the contribution that is able to explain the $b\to s\,\mu^+\mu^-$ anomaly.
Moreover, the coupling to leptons is flavor universal.
The second possibility, figure~\ref{fig:LFUV:diags_bsll:rhoexchange1}, is the exchange of a heavy $\rho$ resonance
with a coupling to quarks due to the same mixing terms as in the first case and a coupling to muons due to its mixing with the $Z$ boson.
Since this $\rho$-muon coupling has the same structure as the $Z$-muon coupling, one again gets a lepton-flavor universal contribution with $C_{10}^\mu\gg C_{9}^\mu$.
However, there is a third case, which is shown in figure~\ref{fig:LFUV:diags_bsll:rhoexchange2}.
While this again corresponds to the exchange of a heavy $\rho$ resonance like in the second case, the $\rho$-muon coupling in this case is not due to the $\rho$-$Z$ mixing but due to the muons mixing with composite lepton partners.
Interestingly, this coupling can actually violate lepton-flavor universality if the degrees of compositeness vary among the lepton generations.
In addition, different degrees of compositeness of left- and right-handed leptons allow for different chirality structures of the generated operators.
In particular, the case $C_9^\mu<0$ could be realized if the degrees of compositeness of left-handed muons $s_{\mu_L}$ and of right-handed muons $s_{\mu_R}$ are both sizable.
However, as detailed in section~\ref{sec:Fermions:partial_compositeness}, the SM-like effective Yukawa coupling of partially composite fermions is proportional to their left- and right-handed degrees of compositeness, i.e.
\begin{equation}
 Y_\mu^{\rm SM} \sim Y_\ell^{\rm comp}\,s_{\mu_L}\,s_{\mu_R},
\end{equation}
where $Y_\mu^{\rm SM}$ is the Yukawa coupling of muons in the SM and $Y_\ell^{\rm comp}$ is a composite sector lepton Yukawa coupling generically of $\mathcal{O}(1)$.
So to actually get a $Y_\mu^{\rm SM}$ of the correct size to reproduce the muon mass, $s_{\mu_L}$ and $s_{\mu_R}$ cannot both be sizable.
While this excludes the case with only $C_9^\mu<0$ and all other WCs SM-like, the second case $C_{9}^{\mu}=-C_{10}^{\mu}<0$ actually only requires a sizable degree of compositeness  $s_{\mu_L}$ of the left-handed muons.
This seems to be possible and is further investigated in the following.

\section{A simple model of partially composite muons}\label{sec:LFUV:model}
To explore the consequences of an explanation of the $b\to s\ell\ell$ flavor anomalies in terms of left-handed muons with a sizable degree of compositeness, it is useful to consider an explicit model.
However, it is not necessary to construct a complete multi-site CHM.
The following discussion merely requires partial compositeness of fermions and vector bosons and is independent of the actual structure of the Higgs sector.
Still, there are some basic properties that should be fulfilled by the model.
First, as discussed in the beginning of chapter~\ref{chap:CHMs}, any viable CHM should have a custodial symmetry that protects the ratio of the $W$ and $Z$ boson masses from large corrections.
It is therefore reasonable to assume a global $\SUs{2}{L}\times\SUs{2}{R}\cong \SO{4}$ symmetry only broken by hypercharge and fermion composite-elementary mixings.
As is well known, using the $\SUs{2}{R}$ generator $\T^3_R$ for generating $\Us{1}{Y}$ does not allow an embedding of the composite partners of quarks into $\SO{4}$ multiplets (cf.\ e.g.~\cite{Agashe:2004rs} and section~\ref{sec:direct_constraints:models}).
A commonly employed solution is to add an additional $\Us{1}{X}$ symmetry
with a generator $X$ and then to define the hypercharge generator $Y$ as in eq.~\eqref{eq:decays:global_analyses:Y}, i.e.\
$
Y = \T^3_R + X
$.
The global symmetry of the model is thus chosen to be $\SUs{2}{L}\times\SUs{2}{R}\times\Us{1}{X}$.
Like partial compositeness of quarks induces the flavor-changing $Z$ coupling in figure~\ref{fig:LFUV:diags_bsll:Zexchange}, a sizable degree of compositeness of muons generically modifies the $Z \mu \mu$ coupling.
But the $Z$-lepton couplings are strongly constrained by the $Z$~boson's partial widths measured at LEP~\cite{ALEPH:2005ab}.
This can result in a severe tension between experimental data and any model containing leptons with a sizable degree of compositeness.
A similar problem is encountered in section~\ref{sec:direct_constraints:models} in the quark sector: For the top quark to have a large Yukawa coupling, the composite-elementary mixing of the third generation's left-handed quark doublet has to be sizable.
This also affects the left-handed $b$ quark and generically leads to a tension between the predicted $Z b_L b_L$ coupling and LEP data.
As discussed in section~\ref{sec:direct_constraints:models}, the tree-level contributions to the $Z b_L b_L$ coupling can be avoided by a discrete $P_{LR}$ symmetry, also known as {\it custodial protection} of the $Z$ coupling~\cite{Agashe:2006at}.
Interestingly, the same kind of protection can be applied to the lepton sector~\cite{Agashe:2009tu} and allows for protecting the $Z \mu_L \mu_L$ coupling.
The custodial protection restricts the possible representations of $\SUs{2}{L}\times\SUs{2}{R}$ under which the composite leptons transform.
Following~\cite{Agashe:2006at}, the left-handed elementary muons are required to mix with composite leptons $L$ transforming as a $(\mathbf{2}, \mathbf{2})_0$ under $SU(2)_L \times SU(2)_R \times U(1)_X$, and the right-handed elementary muons mix with $(\mathbf{1}, \mathbf{3})_0$ composite leptons $E$.
The custodial $P_{LR}$ symmetry then requires the introduction of a second triplet $(\mathbf{3}, \mathbf{1})_0$, which is denoted by $E'$.
With this choice of representations, the Lagrangian of second generation leptons reads
\begin{equation}
 \begin{aligned}
\mathcal L_f =& \,\,
\bar l_L (i\slashed{ D}) l_L + \bar \mu_R (i \slashed{ D}) \mu_R
\\&
+
\bar L (i\slashed{ D}-m_L) L
+
\bar E (i\slashed{ D}-m_E) E
+
\bar E' (i\slashed{ D}-m_E) E',
 \end{aligned}
\end{equation}
where the covariant derivatives $D_\mu$ contain the couplings of elementary leptons to the elementary SM-like gauge fields and of composite leptons to the composite resonances associated with the $SU(2)_L \times SU(2)_R \times U(1)_X$ symmetry\footnote{%
The simplest realization of this model in terms of a multi-site CHM is a two-site model with one level of spin one resonances in an adjoint of $H=SU(2)_L \times SU(2)_R \times U(1)_X$, cf.\ section~\ref{sec:Vectorres_NLSM_HLS}.
}.
The corresponding mixing terms in the lepton sector are given by
\begin{equation}
 \begin{aligned}
\mathcal L_\text{mix} &=
    \Delta_L \, \text{tr}[ \bar\chi_L \, L_R  ]
  + \Delta_R \, \text{tr}[ \bar\chi_R \, E_L ] \\
& \quad + Y_{L}\, \text{tr}[\bar L_L \mathcal{H} E_R]
  + Y'_{L}\, \text{tr}[\mathcal{H} \bar L_L E'_R] \\
& \quad + Y_{R}\, \text{tr}[\bar L_R \mathcal{H} E_L]
  + Y'_{R}\, \text{tr}[\mathcal{H} \bar L_R E'_L] \\
& \quad + \text{h.c.} \,,
 \end{aligned}
\end{equation}
where $\chi_L$ and $\chi_R$ are incomplete $(\mathbf{2}, \mathbf{2})_0$ and $(\mathbf{1}, \mathbf{3})_0$ multiplets into which the elementary left- and right-handed muons are embedded.
For simplicity, the Higgs doublet is embedded into a $(\mathbf{2}, \mathbf{2})_0$ bidoublet $\mathcal{H}$ and not treated as a pNGB.
While the generalization to an actual pNGB CHM is straightforward, it is not necessary for the discussion of muon partial compositeness.
The composite-elementary mixings $\Delta_L$ and $\Delta_R$ yield, analogous to section~\ref{sec:Fermions:partial_compositeness}, the degrees of compositeness $s_{\mu_L}$ and $s_{\mu_R}$ (cf.\ eq.~\eqref{eq:fermions_degrees_of_compositeness}), and $Y_L$, $Y_L'$, $Y_R$, and $Y_R'$ are Yukawa couplings of the composite sector.
In the mass basis, the above Lagrangian induces a mass term for the muon,
\begin{equation}
m_\mu = \frac{Y_{L}}{2 \sqrt{2}} \, \left< h \right>  \,s_{\mu_L}\,s_{\mu_R} \,,
\end{equation}
where $\left<h\right>$ is the Higgs VEV.
Analogous mass terms for neutrinos as well as flavor mixing in the lepton sector are omitted here for simplicity.

\section{Constraints from electroweak precision tests}\label{sec:LFUV:EWPT}

\begin{figure}[t]
\centering
\begin{subfigure}[b]{0.49\textwidth}
\centering
\includegraphics[width=0.9\columnwidth]{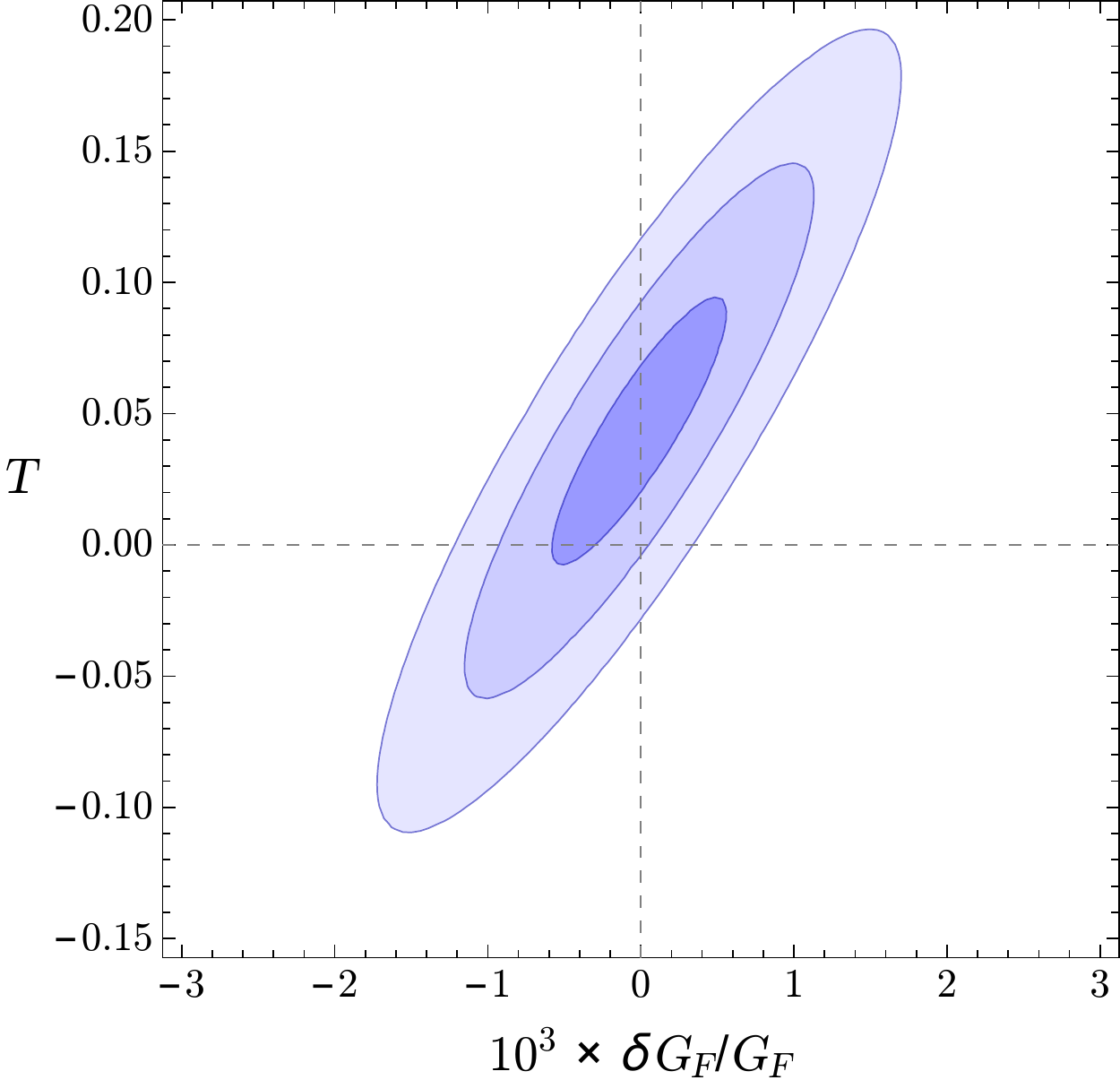}
\caption{}
\label{fig:LFUV:EWPT:S_T_correlation}
\end{subfigure}
\hfill
\begin{subfigure}[b]{0.49\textwidth}
\centering
$
\begin{fmffile}{Wexchange}
\vcenter{\hbox{
\begin{fmfgraph*}(135,105)
\fmfset{arrow_len}{2mm}
\fmfset{thin}{.7pt}
\fmfpen{thin}
\fmfstraight

\fmfleft{x1,s,,,,,,,b,x2}
\fmfright{y1,mum,,,,,,,mup,y2}

\fmf{fermion,tension=2}{b,mb}
\fmf{double,tension=2}{mb,v1}
\fmf{fermion,tension=2}{ms,s}
\fmf{double,tension=2}{v1,ms}

\fmf{dbl_wiggly,tension=6/2}{v1,mZ}
\fmf{boson,tension=6/7}{mZ,v2}
\fmf{phantom,tension=0,label=$W$}{v1,v2}

\fmf{fermion}{mup,v2}
\fmf{fermion}{v2,mum}

\fmfdot{v1,v2}
\fmfv{decor.shape=circle,decor.filled=empty,decor.size=5pt,label.dist=0,label=\bf\tiny{+}}{ms,mb,mZ}

\fmfv{lab=$\mu_L$,l.angle=-180}{b}
\fmfv{lab=$\nu_{\mu L}$,l.angle=-180}{s}
\fmfv{lab=$\nu_{e L}$,l.angle=0}{mup}
\fmfv{lab=$e_L$,l.angle=0}{mum}

\end{fmfgraph*}
}}
\end{fmffile}
$
\vspace{40pt}
\caption{}
\label{fig:LFUV:EWPT:Wexchange}
\end{subfigure}
\caption{
(a):~Constraints at 1$\sigma$, 2$\sigma$, and 3$\sigma$ on the modification of the Fermi
constant in muon decay relative to the SM versus a NP contribution to
the electroweak $T$ parameter.
(b):~Tree-level correction to the Fermi constant due to a shift
in the tree-level $W\mu_L\nu_{\mu L}$ coupling.
}\label{fig:LFUV:EWPT}
\end{figure}
While the $Z \mu_L \mu_L$ coupling can be protected from tree-level corrections\footnote{%
The custodial protection does not forbid loop-level corrections to the $Z \mu_L \mu_L$ coupling.
While they might be relevant in a complete analysis~\cite{Grojean:2013qca}, this is beyond the scope of the proof of concept presented here.
} by employing the discrete $P_{LR}$ symmetry, other couplings of muons and electroweak gauge bosons are also modified by partial compositeness and not protected by the $P_{LR}$ symmetry.
In particular, the custodial protection is not active for the $W\mu_L\nu_{\mu L}$ coupling.
This affects the muon lifetime and leads to a shift in the Fermi constant~$G_F$, which is extracted from muon decay.
The experimentally allowed shift in the Fermi constant depends on other possible deviations of electroweak precision observables~\cite{Wells:2014pga}.
In particular, the constraint on $G_F$ is strongly correlated with the constraint on the electroweak $T$ parameter.
Following~\cite{Wells:2014pga}, one finds the constraints on a shift in the Fermi constant and the $T$ parameter shown in figure~\ref{fig:LFUV:EWPT:S_T_correlation}.
In CHMs, the $T$ parameter receives loop-contributions that depend on details of the quark sector, which are not specified in the simple model presented here.
Anyway, a shift in the Fermi constant that is induced by a sizable $s_{\mu_L}$ can be translated into a required shift in the $T$ parameter.
For instance, allowing the Fermi constant to deviate by 3$\sigma$, the maximally allowed negative shift $\delta G_F/G_F\approx-1.6\cdot 10^{-3}$ suggests $T\approx-0.1$.
At tree level, the shift in the Fermi constant due to the modified $W\mu_L\nu_{\mu L}$ coupling can be calculated from the diagram in figure~\ref{fig:LFUV:EWPT:Wexchange}.
To leading order in $\xi=v^2/f^2$ and $s_{\mu_L}$, one finds\footnote{%
In this simplified model, $f$ plays the same role as in a full CHM; it sets the mass scale of the composite resonances. In particular, the masses of the vector resonances $\rho$ are assumed to fulfill $m_\rho^2=\frac{f^2\,g_\rho^2}{2}$, where $g_\rho$ is their gauge coupling (cf.\ eq.~(\ref{eq:Vectorres_gauge_boson_masses})).
}
\begin{equation}\label{eq:LFUV:WEPT:Wmun}
\frac{\delta G_F}{G_F}
= \frac{\delta g^L_{W\mu\nu}}{g^L_{W\mu\nu}}
=
-\frac{1}{4}\xi s_{L\mu}^2\left(1+\frac{m_L^2}{m_E^2}\right) \,.
\end{equation}
This shift is actually always negative, and for fixed $s_{\mu_L}$ and $\xi$ it has its smallest absolute value in the case $m_E\gg m_L$.
Assuming this favorable case, a maximally allowed negative shift $\delta G_F/G_F\approx-1.6\cdot 10^{-3}$ translates into an upper bound on the left-handed muon's degree of compositeness
\begin{equation}
 s_{\mu_L}\lesssim 0.08\,\xi^{-1/2}.
\end{equation}

There is yet another coupling of weak gauge bosons to leptons that is not custodially protected: the $Z\nu_{\mu L}\nu_{\mu L}$ coupling also receives corrections already at tree level.
At leading order in $s_{\mu_L}$ and $\xi$, they are equal to those of the $W\mu_L\nu_{\mu L}$ coupling,
\begin{equation}
\frac{\delta g^L_{Z\nu\nu}}{g^L_{Z\nu\nu}}=\frac{\delta g^L_{W\mu\nu}}{g^L_{W\mu\nu}}
\,,
\label{eq:Znn}
\end{equation}
which is a generic property of models with custodial protection of the $Z \mu_L \mu_L$ coupling (cf.~\cite{Agashe:2006at}).
The modification of the $Z\nu_{\mu L}\nu_{\mu L}$ coupling leads to a shift in the invisible $Z$ width that can be expressed in terms of the effective number of light neutrino species $N_\nu$ and is given by
\begin{equation}
N_\nu = 3 + 2 \frac{\delta g^L_{Z\nu\nu}}{g^L_{Z\nu\nu}} \,.
\end{equation}
Interestingly, its measurement at LEP shows a 2$\sigma$ deviation~\cite{ALEPH:2005ab},
\begin{equation}
N_\nu = 2.9840 \pm 0.0082 \,.
\label{eq:Nnuexp}
\end{equation}
Since the contribution from the modified $Z\nu_{\mu L}\nu_{\mu L}$ coupling is always negative in the model considered here (cf.\ eq.~(\ref{eq:LFUV:WEPT:Wmun})), it actually improves the agreement with the data.

\section{Constraints from quark flavor physics}\label{sec:LFUV:quark_flavor_contraints}
\begin{figure}[t]
\centering
\begin{subfigure}[b]{0.49\textwidth}
\centering
$
\begin{fmffile}{bsbs}
\vcenter{\hbox{
\begin{fmfgraph*}(135,105)
\fmfset{arrow_len}{2mm}
\fmfset{thin}{.7pt}
\fmfpen{thin}
\fmfstraight

\fmfleft{x1,s,,,,,,,b,x2}
\fmfright{y1,mum,,,,,,,mup,y2}

\fmf{fermion,tension=2}{b,mb}
\fmf{double,tension=2}{mb,v1}
\fmf{fermion,tension=2}{ms,s}
\fmf{double,tension=2}{v1,ms}

\fmf{dbl_wiggly,tension=2/3,l.d=8,label=$\rho_i$}{v1,v2}

\fmf{fermion,tension=2}{mup,mmup}
\fmf{double,tension=2}{v2,mmup}
\fmf{fermion,tension=2}{mmum,mum}
\fmf{double,tension=2}{mmum,v2}

\fmfv{decor.shape=circle,decor.filled=full,decor.size=2thick,
label=$g_\rho \Delta_{bs} X_i^q$,l.d=10}{v1}
\fmfv{decor.shape=circle,decor.filled=full,decor.size=2thick,
label=$g_\rho \Delta_{bs} X_i^q$,l.d=10}{v2}
\fmfv{decor.shape=circle,decor.filled=empty,decor.size=5pt,label.dist=0,label=\bf\tiny{+}}{ms,mb,mmup,mmum}

\fmfv{lab=$b_L$,l.angle=-180}{b}
\fmfv{lab=$s_L$,l.angle=-180}{s}
\fmfv{lab=$b_L$,l.angle=0}{mup}
\fmfv{lab=$s_L$,l.angle=0}{mum}

\end{fmfgraph*}}}
\end{fmffile}
$
\caption{}
\label{fig:LFUV:main_diags:bsbs}
\end{subfigure}
\begin{subfigure}[b]{0.49\textwidth}
\centering
$
\begin{fmffile}{bsll}
\vcenter{\hbox{
\begin{fmfgraph*}(135,105)
\fmfset{arrow_len}{2mm}
\fmfset{thin}{.7pt}
\fmfpen{thin}
\fmfstraight

\fmfleft{x1,s,,,,,,,b,x2}
\fmfright{y1,mum,,,,,,,mup,y2}

\fmf{fermion,tension=2}{b,mb}
\fmf{double,tension=2}{mb,v1}
\fmf{fermion,tension=2}{ms,s}
\fmf{double,tension=2}{v1,ms}

\fmf{dbl_wiggly,tension=2/3,l.d=8,label=$\rho_i$}{v1,v2}

\fmf{fermion,tension=2}{mup,mmup}
\fmf{double,tension=2}{v2,mmup}
\fmf{fermion,tension=2}{mmum,mum}
\fmf{double,tension=2}{mmum,v2}

\fmfv{decor.shape=circle,decor.filled=full,decor.size=2thick,
label=$g_\rho \Delta_{bs} X_i^q$,l.d=10}{v1}
\fmfv{decor.shape=circle,decor.filled=full,decor.size=2thick,
label=$g_\rho s_{\mu_L}^2 X_i^\mu$,l.d=10}{v2}
\fmfv{decor.shape=circle,decor.filled=empty,decor.size=5pt,label.dist=0,label=\bf\tiny{+}}{ms,mb,mmup,mmum}

\fmfv{lab=$b_L$,l.angle=-180}{b}
\fmfv{lab=$s_L$,l.angle=-180}{s}
\fmfv{lab=$\mu_L$,l.angle=0}{mup}
\fmfv{lab=$\mu_L$,l.angle=0}{mum}

\end{fmfgraph*}}}
\end{fmffile}
$
\caption{}
\label{fig:LFUV:main_diags:bsll}
\end{subfigure}
\caption{Tree-level contribution to (a) $B_s$ mixing and (b)
$b\to s\,\mu^+\mu^-$ transitions. $g_\rho$ is the coupling between composite fermions and vector resonances, $s_{\mu_L}$ the left-handed muon's degree of compositeness, $X_i^f$ is the charge under the global symmetry associated with vector resonance $\rho_i$ of the composite fermion mixing with $f$, and $\Delta_{bs}$ is a parameter depending on the flavor structure and the degrees of compositeness of $b$ and $s$ quark.
}\label{fig:LFUV:main_diags}
\end{figure}
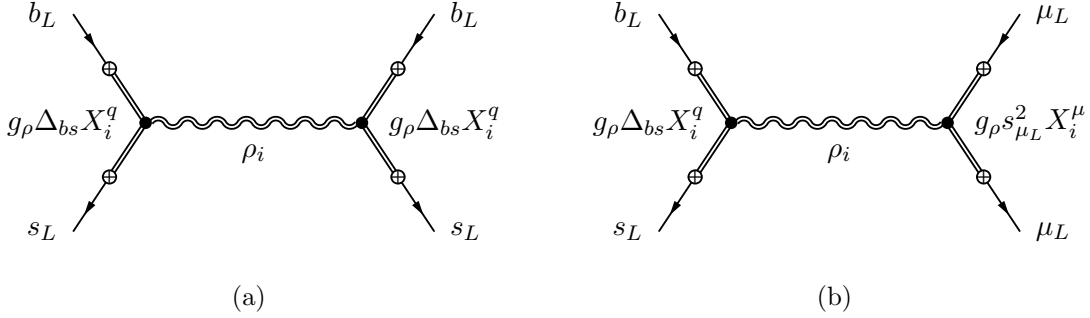

Any model that tries to explains the $b\to s\ell\ell$ anomalies in terms of a tree-level exchange of a heavy spin one resonance necessarily implies a flavor changing tree-level coupling of this resonance to left-handed quarks.
Such a coupling inevitably leads to the contribution to $B_s\text{-}\bar B_s$ mixing shown in the diagram in figure~\ref{fig:LFUV:main_diags:bsbs}.
This contribution can be parameterized in terms of the dimension-6
$\Delta B=2$ operator $O_V^{dLL}=(\bar s_{L}\gamma^\mu b_L)^2$.
Its WC $C_V^{dLL}$ can be inferred from the diagram in figure~\ref{fig:LFUV:main_diags:bsbs}, and one finds
\begin{equation}\label{eq:LFUV:bsbs:CVdll}
C_V^{dLL} = \frac{g_\rho^2}{m_\rho^2} \,\Delta_{bs}^2 \,c_V^{dLL} \,,
\end{equation}
where $\Delta_{bs}$ encodes the flavor structure and the degrees of compositeness of $b$ and $s$ quark.
$c_V^{dLL}$ is a numerical factor of $\mathcal{O}(1)$ that arises from the sum over the couplings of the quark partners to the heavy spin one resonances $\rho_i$.
Since the $\rho_i$ are gauge bosons associated with the unbroken symmetry $H$, the couplings can be given in terms of the $H$ charges $X_i^q$ of the heavy quark partners (cf.\ figure~\ref{fig:LFUV:main_diags:bsbs}).
Assuming a custodial protection of the $Z b_L b_L$ coupling, one finds $c_V^{dLL}=-23/36$ \cite{Barbieri:2012tu}.

Writing the $B_s\text{-}\bar B_s$ mixing amplitude $M_{12}$ in terms of the sum of a SM contribution $M_{12}^{\rm SM}$ and a NP contribution $M_{12}^{\rm NP}$, the mass difference $\Delta M_s$ in $B_s$ mixing is given by
\begin{equation}\label{eq:LFUV:bsbs:Delta_M_s}
 \Delta M_s
 = 2\, \left|M_{12}^{\rm SM}+M_{12}^{\rm NP}\right|
 = \Delta M_s^{\rm SM}\,\left|1+\frac{M_{12}^{\rm NP}}{M_{12}^{\rm SM}}\right|.
\end{equation}
The ratio $M_{12}^{\rm NP}/M_{12}^{\rm SM}$ can be expressed in terms of the WC $C_V^{dLL}$ and is given by~\cite{Altmannshofer:2014rta}
\begin{equation}
\frac{M_{12}^{\rm NP}}{M_{12}^{\rm SM}}
= -C_V^{dLL}\,v^2\,\left(\frac{g^2}{16\pi^2}\,(V_{tb}V_{ts}^*)^2\,S_0\right)^{-1},
\end{equation}
where $S_0$ is a loop function that evaluates to $S_0\approx2.3$.
By plugging in the expression for $C_V^{dLL}$ from eq.~(\ref{eq:LFUV:bsbs:CVdll}) and using $g_\rho^2/m_\rho^2=2/f^2=2\,\xi/v^2$, one finds
\begin{equation}\label{eq:LFUV:bsbs:M12_NP}
\frac{M_{12}^{\rm NP}}{M_{12}^{\rm SM}}
= - 2\,\xi\,\Delta_{bs}^2 \,c_V^{dLL}\,\left(\frac{g^2}{16\pi^2}\,(V_{tb}V_{ts}^*)^2\,S_0\right)^{-1}.
\end{equation}
Assuming $\Delta_{bs}$ to be real, the  negative value of $c_V^{dLL}$ implies\footnote{%
$V_{tb}$ is real and the imaginary part of $V_{ts}$ is negligible.
}
$M_{12}^{\rm NP}/M_{12}^{\rm SM}>0$.
This in turn leads to (cf.\ eq.~(\ref{eq:LFUV:bsbs:Delta_M_s}))
\begin{equation}\label{eq:LFUV:Delta_M_s_Delta_M_s_SM}
 \frac{M_{12}^{\rm NP}}{M_{12}^{\rm SM}}
 =\frac{\Delta M_s}{\Delta M_s^{\rm SM}}-1
 = \frac{\Delta M_s-\Delta M_s^{\rm SM}}{\Delta M_s^{\rm SM}}
 = \frac{|\Delta M_s-\Delta M_s^{\rm SM}|}{\Delta M_s^{\rm SM}},
\end{equation}
i.e.\ the model predicts a positive NP contribution to the $B_s$ meson mass difference $\Delta M_s$, where the magnitude of the relative deviation from the SM is equal to $M_{12}^{\rm NP}/M_{12}^{\rm SM}$.

Since the only free parameters on the right-hand side of eq.~(\ref{eq:LFUV:bsbs:M12_NP}) are $\xi=v^2/f^2$ and $\Delta_{bs}^2$, a bound on $\Delta M_s$ can be translated, for any given value of $f$, into a bound on $|\Delta_{bs}|$.
Since $\Delta_{bs}$ sets the strength of the $b$-$s$-$\rho$ coupling, the $\rho$ exchange contribution to $b\to s\,\mu^+\mu^-$ also depends on it (cf.\ figure~\ref{fig:LFUV:main_diags:bsll}), and the bound has an immediate consequence on a possible explanation of the anomalies.
%
%
%
%
%
%
%
The $\rho$ exchange contribution to $b\to s\,\mu^+\mu^-$ can be parameterized by the WC $C_{dl}$ of the $\Delta B=1$
operator $O_{dl}=(\bar s_{L}\gamma^\nu b_L)(\mu_L\gamma_\nu\mu_L)$.
From the diagram in figure~\ref{fig:LFUV:main_diags:bsll}, one gets
\begin{equation}\label{eq:LFUV:bsbs:Cdl}
C_{dl} = \frac{g_\rho^2}{m_\rho^2} \Delta_{bs} s_{L\mu}^2 \,c_{dl} \,,
\end{equation}
where $c_{dl}=-1/2$ for the choice of representations used here.
Using the notation of the WEH, eq.~(\ref{eq:anomalies:Heff}), $O_{dl}$ can be written as $O_{dl} = (O_9^\mu-O_{10}^\mu)/2$ and $C_{dl}$ is related to $C_9^\mu$ and $C_{10}^\mu$ by
\begin{equation}\label{eq:LFUV:bsbs:Cdl_C9_C10}
 C_{dl} = \mathcal{N}^{b\to s}\,(C_9^\mu-C_{10}^\mu),
 \quad\quad
 \text{where}
 \quad\quad
 \mathcal{N}^{b\to s} = \frac{4\,G_F}{\sqrt{2}} V_{tb}V_{ts}^* \frac{e^2}{16\pi^2}.
\end{equation}
Allowing for a 10\% deviation from the SM in $\Delta M_s$, the resulting bound on $|\Delta_{bs}|$ then implies a lower bound on $s_{\mu_L}$, which depends on the size of the desired effect in $C_9^\mu-C_{10}^\mu$.
For instance, a small but visible effect $C_9^\mu-C_{10}^\mu\approx 0.4$ requires
\begin{equation}
 s_{\mu_L}\gtrsim 0.15\,\xi^{-1/4}.
\end{equation}

\section{Explaining the \lowercase{$b\to s\,\ell^+\ell^-$} anomalies}\label{sec:LFUV:results}
\begin{figure}[t]
\centering
\includegraphics[width=\textwidth]{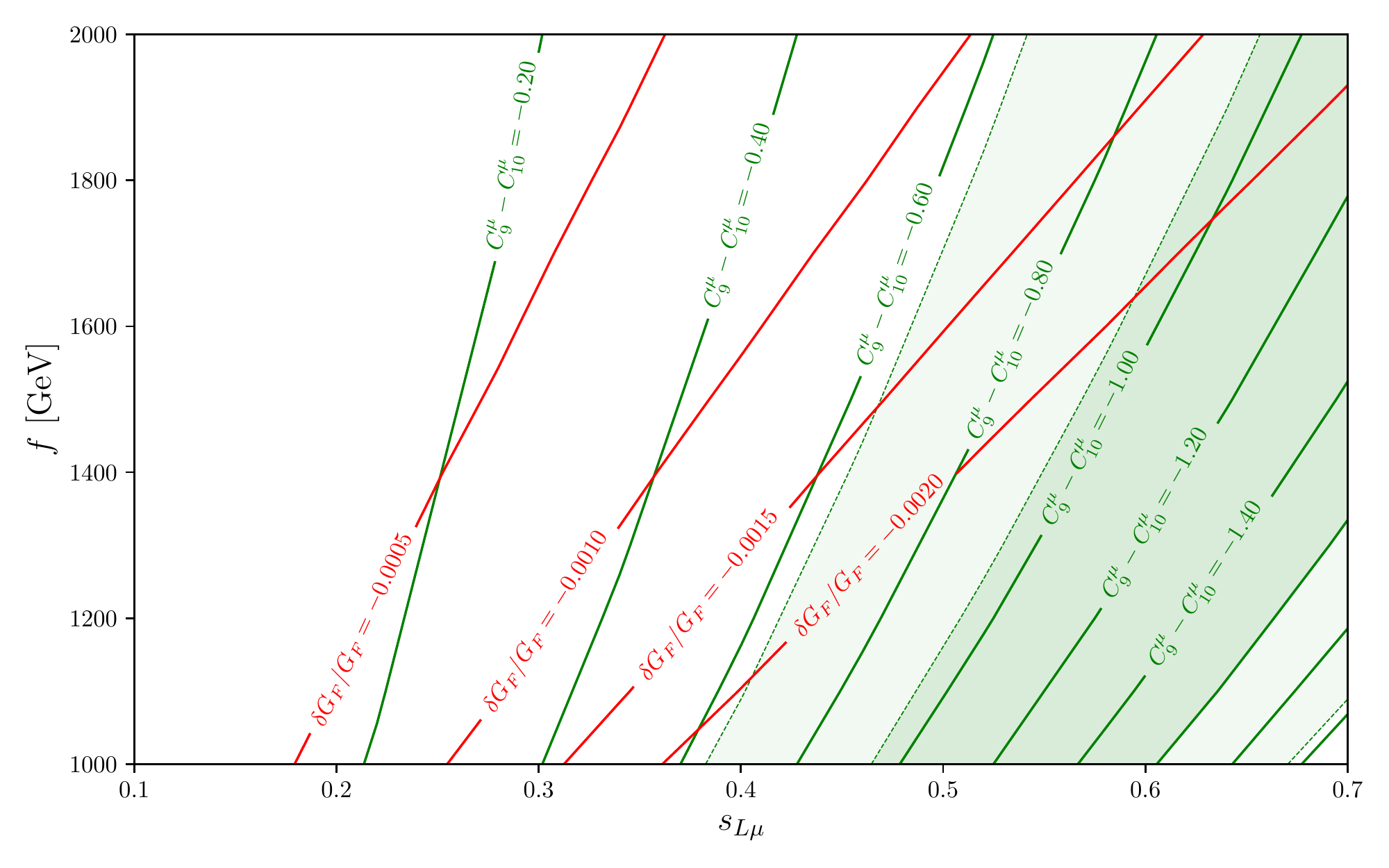}
\caption{Predictions for $C_9^\mu-C_{10}^\mu$ (green) and the relative shift in the Fermi constant (red) for a benchmark point with $m_L/m_E=0.3$. The flavor-changing coupling $\Delta_{bs}$ has been fixed to its maximum value allowing a 10\% shift in $\Delta M_s$. The green shaded regions correspond to the $1\sigma$ (dark green) and $2\sigma$ (light green) regions allowed by the $C_9^\mu=-C_{10}^\mu$ fit values in table~\ref{tab:anomalies:LFUV:pulls_1D}.
No contours are shown for $|{\delta G_F}/{G_F}|>0.002$, which is disfavored (cf.~fig.~\ref{fig:LFUV:EWPT:S_T_correlation}).}
\label{fig:LFUV:constraints}
\end{figure}
Given the lower bound on the muon's degree of compositeness due to the shift in the Fermi constant and the upper bound from $B_s\text{-}\bar B_s$ mixing, one might ask whether these bounds still allow for an explanation of the $b\to s\,\ell^+\ell^-$ anomalies.
To answer this question, it is useful to express the NP contribution to $C_9^\mu-C_{10}^\mu$ in terms of the deviation in $\Delta M_s$.
From eqs.~(\ref{eq:LFUV:bsbs:M12_NP}), (\ref{eq:LFUV:Delta_M_s_Delta_M_s_SM}), (\ref{eq:LFUV:bsbs:Cdl}), and (\ref{eq:LFUV:bsbs:Cdl_C9_C10}), one finds
\begin{equation}
C_9^\mu-C_{10}^\mu \approx
 \pm0.92
\left[\frac{1.7\,\text{TeV}}{f}\right]
\left[\frac{s_{\mu_L}}{0.6}\right]^2
\left[\frac{|\Delta M_s-\Delta M_s^\text{SM}|}{0.1\,\Delta
M_s^\text{SM}}\right]^{1/2}
\,,
\end{equation}
where the negative sign holds for positive $\Delta_{bs}$.
Consequently, assuming a 10\% deviation from the SM in $\Delta M_s$, a violation of LFU compatible with the measurements of $R_K$ and $R_{K^*}$ can be achieved with a sizable degree of compositeness of left-handed muons  $s_{\mu_L}\approx 0.6$ and a NP scale $f\approx 1.7$~TeV (cf.\ table~\ref{tab:anomalies:LFUV:pulls_1D}).
At the same time, this scenario can also explain the experimental data on $b\to s\,\mu^+\mu^-$ observables that is in tension with the SM prediction (cf.\ table~\ref{tab:anomalies:bsmumu:pulls_1D}).
These values for $s_{\mu_L}$ and $f$ lead to a relative shift in the Fermi constant that can still be inside the 3$\sigma$ contour shown in figure~\ref{fig:LFUV:EWPT:S_T_correlation}, depending on the mass ratio $m_L/m_E$ and the value of the $T$ parameter.
The possible values of $C_9^\mu-C_{10}^\mu$ under the assumption of a 10\% shift in $\Delta M_s$ are shown in the $s_{\mu_L}$-$f$ plane in figure~\ref{fig:LFUV:constraints}.
In addition, this figure also shows lines of constant $\delta G_F/G_F$ according to eq.~(\ref{eq:LFUV:WEPT:Wmun}), where $m_L/m_E=0.3$ has been assumed.

An explanation of the anomalies in $b\to s\,\ell^+\ell^-$ processes in terms of a sizable degree of compositeness of left-handed muons leads to several predictions that can be used to test the models presented here:
\begin{itemize}
 \item It predicts $C_9^\mu\approx-C_{10}^\mu<0$. While this is in perfect agreement with global fits to measurements of LFU and $b\to s\,\mu^+\mu^-$ observables (cf.\ tables~\ref{tab:anomalies:bsmumu:pulls_1D} and~\ref{tab:anomalies:LFUV:pulls_1D}), it can be tested by new measurements, e.g.\ of the LFU observables $D_{P_4'}$ and  $D_{P_5'}$ (cf.\ section~\ref{sec:anomalies:LFUV:predictions}).
 Moreover, the positive contribution to $C_{10}^\mu$ implies a suppression of $B_s\to\mu^+\mu^-$ (cf.~\cite{Altmannshofer:2017wqy,Glashow:2014iga}).
 \item It predicts a positive\footnote{%
 This might be problematic; a recent analysis indicates that $\Delta M_s^{\rm SM}$ is already 2$\sigma$ above the experimental value, i.e.\ experimental data seems to prefer a negative NP contribution~\cite{DiLuzio:2017fdq}.
 } NP contribution to the $B_s$ meson mass difference $\Delta M_s$.
 \item It requires a negative shift in the Fermi constant accompanied by a negative shift in the electroweak $T$ parameter.
 \item It implies a slightly smaller value of the effective number of light neutrino species $N_\nu$ compared to the SM.
 Interestingly, such a deficit is actually preferred by LEP data.
 \item It predicts spin one resonances with a sizable branching ratio into muons. However, they might be too heavy to be observable by direct searches at the LHC.
\end{itemize}
The above described model is incomplete in some ways: the precise structure of the quark sector is not specified and the lepton sector does not contain a mechanism for generating neutrino masses or to explain the absence of charged lepton flavor violation.
However, the presented mechanism for violation of LFU, which is primarily based on a sizable degree of compositeness of the muon, has proved to be compatible with current experimental bounds and may be implemented also in more complete models.

\chapter[Flavor physics and flavor anomalies in MFPC]{Flavor physics and flavor anomalies\\ in minimal fundamental partial compositeness}\label{chap:Flavor_MFPC} 

Models containing the partial compositeness mechanism for generating the masses of SM fermions have a rich flavor phenomenology.
They generically contain sources of flavor violation beyond those present in the SM, which can lead to strong constraints and may require flavor symmetries to make them phenomenologically viable (cf.\ section~\ref{sec:Fermions:flavor}).
On the other hand, the tree-level FCNCs present in these models not only lead to ``flavor problems'' but can also be used to explain experimental deviations from the SM that may be difficult to accommodate in other models.
A good example is the explanation of the $b\to s\,\ell^+\ell^-$ anomalies discussed in chapter~\ref{chap:LUFV_in_CHMs}.
There, another important property of models with partial compositeness is used: different degrees of compositeness for different lepton generations allow for couplings to heavy resonances that violate LFU.

Having at hand the UV complete model of partial compositeness described in section~\ref{sec:FPC}, interesting questions are how such an FPC model performs in a comprehensive analysis of low-energy flavor effects, if it is able to explain the $b\to s\,\ell^+\ell^-$ anomalies, and if it can even shed some light on other measured deviations from the SM that show up in processes involving the $b\to c\tau \nu$ transition.
To answer these questions, we have performed a comprehensive study of flavor constraints on the MFPC model in~\cite{Sannino:2017utc} and investigated its potential for explaining the flavor anomalies.
This study is presented in the following.

\section{Flavor and electroweak signals of the MFPC-EFT}\label{sec:FlavMFPC:observables}
For studying the effects of the MFPC model at and below the EW scale, it is convenient to employ the MFPC-EFT described in section~\ref{sec:FPC:MFPC-EFT}.
While observables at the EW scale can be studied directly in the MFPC-EFT, it is useful to consider the WEH for low-energy flavor observables.
This in turn requires matching the MFPC-EFT to the WEH.
In the following, the EW scale observables, the matching procedure for deriving the WEH, and the low-energy flavor observables are discussed in detail.


While the MFPC-EFT is defined with SM fermions in the gauge basis, the observables considered in the following are defined in the mass basis, where the fermion mass matrices have been diagonalized by biunitary transformations.
It is therefore useful to first fix the notation of the model parameters in the mass basis.
Recalling from section~\ref{sec:FPC:MFPC-EFT} that the SM fermion mass matrices in the MFPC-EFT are given by
\begin{equation}\label{eq:FlavMFPC:masses_gauge_basis}
m_{f,ij} = \dfrac{C_\mathrm{Yuk}\, s_\theta\, f_\TC}{4\pi}\,  \left( y_{f}\transpose\, y_{\bar{f}} \right)_{ij},
\end{equation}
where $f\in\{u,d,e\}$, the diagonalized matrices can be written as\footnote{%
The neutrinos are treated as massless. Hence, the charged lepton mass matrix can be chosen to be diagonal already in the gauge-basis, i.e.\ $U_{e}=U_{\nu}=U_{\bar{e}}=\mathds{1}_3$.}
\begin{equation}\label{eq:FlavMFPC:masses_mass_basis}
 m_{f}^{\text{diag}}=
 U_{f}^{\transpose}
 \,
 m_{f}
 \,
 U_{\bar{f}},
 \quad\quad
 f\in\{u,d,e\}.
\end{equation}
This relation between the gauge basis mass matrices $m_{f}$ and the diagonalized matrices $m_{f}^{\text{diag}}$ defines the unitary matrices $U_{f}$ and $U_{\bar{f}}$.
Like in the SM, the unitary matrices associated with the doublet components of up- and down-type quarks yield the CKM matrix, which is given by
 \begin{equation}\label{eq:FlavMFPC:CKM_matrix}
  V = U_{u}^\dagger\,U_{d}.
 \end{equation}
Inspecting eqs.~(\ref{eq:FlavMFPC:masses_gauge_basis}) and~(\ref{eq:FlavMFPC:masses_mass_basis}), one observes that the object that is actually transformed to the mass basis is the product of the matrix valued fundamental Yukawa couplings $\big( y_{f}\transpose\, y_{\bar{f}} \big)_{ij}$, while all other terms in eq.~(\ref{eq:FlavMFPC:masses_gauge_basis}) are flavor universal.
It might be useful to note that while the fundamental doublet Yukawa couplings are identical for the doublet components, i.e.\ $y_{Q}=y_{u}=y_{d}$ and $y_{L}=y_{e}=y_{\nu}$, the fundamental quark doublet Yukawa couplings $y_{u}$ and $y_{d}$ are rotated to the mass basis by different unitary matrices.
For later convenience, it is useful to introduce a notation for all possible products of fundamental Yukawa matrices in the mass basis.
%
There are two cases:
\begin{itemize}
 \item A product of two fundamental Yukawa matrices where one of them is complex conjugated and the other is not. This will be denoted by
 \begin{equation}\label{eq:FlavMFPC:X_ff}
  X_{f_1 f_2} = \dfrac{1}{4\pi} \, U_{f_1}^\dagger\,y_{f_1}^\dagger\,y_{f_2}\,U_{f_2},
  \quad\quad
  X_{f_1 f_2}^* = X_{f_2 f_1}\transpose.
 \end{equation}
 \item A product of two fundamental Yukawa matrices where both of them are either unconjugated or conjugated, which will be written as
 \begin{equation}\label{eq:FlavMFPC:Y_ff}
  Y_{f_1 f_2} =\dfrac{1}{4\pi} \, U_{f_1}\transpose\,y_{f_1}\transpose\,y_{f_2}\,U_{f_2},
  \quad\quad
  Y_{f_1 f_2}^* = \dfrac{1}{4\pi} \, U_{f_1}^\dagger\,y_{f_1}^\dagger\,y_{f_2}^*\,U_{f_2}^*.
 \end{equation}
\end{itemize}
Here, $f_1$ and $f_2$ denote a SM fermion, i.e. $f_1,f_2 \in \{u,d,e,\nu,\bar{u},\bar{d},\bar{e}\}$.
The second case is actually the one appearing in the fermion mass matrices, and using $Y_{f_1 f_2}$, one can write the mass basis mass matrices as (cf.\ eqs.~(\ref{eq:FlavMFPC:masses_gauge_basis}),~(\ref{eq:FlavMFPC:masses_mass_basis}), and~(\ref{eq:FlavMFPC:Y_ff}))
\begin{equation}\label{eq:FlavMFPC:m_F_diag}
 m_{f}^{\text{diag}}=C_\mathrm{Yuk}\, s_\theta\, f_\TC\, Y_{f \bar{f}}.
\end{equation}
Recalling from eq.~(\ref{eq:FPC:v_SM}) that the SM Higgs VEV $v_{\rm SM}$ can be identified with $v_{\rm SM}=f_\TC\,s_\theta$, the SM Yukawa coupling in the mass basis $Y_f^{\rm SM,diag}$ is given by
\begin{equation}
 Y_{f}^{\rm SM,diag} = \sqrt{2}\,C_\mathrm{Yuk}\,Y_{f \bar{f}}.
\end{equation}

\subsection{Constraints from observables at the electroweak scale}\label{sec:FlavMFPC:Constr_EW}
In the MFPC-EFT, the couplings of the Higgs and the electroweak gauge bosons are modified at tree level.
Therefore, experimental bounds on these couplings provide important constraints.

The modification of the Higgs couplings is due to its pNGB nature and the finiteness of the misalignment angle $\theta$.
The pNGB kinetic term, eq.~(\ref{eq:FPC:NGB_LO_Lagrangian}), leads to the modified couplings to weak gauge bosons
\begin{equation} \label{eq:Higgs_couplings}
\quad g_{ZZh} = c_\theta g_{ZZh}^{\mathrm{SM}}\,, \quad g_{WWh} = c_\theta g_{WWh}^{\mathrm{SM}}\,,
\end{equation}
while the fermion-Higgs couplings, which stem from the operator $\mathcal{O}_\mathrm{Yuk}$, eq.~(\ref{eq:FPC:O_Yuk}), are
\begin{equation}
 g_{ffh} = c_\theta g_{ffh}^{\mathrm{SM}}\,.
\end{equation}
Higgs coupling modifications of this kind are present in any model with a pNGB Higgs and have been discussed extensively in the literature (for a recent analysis, see e.g.~\cite{Sanz:2017tco}).
Experimental bounds on these couplings put a constraint on the size of the misalignment angle $\theta$.
A combination of ATLAS and CMS Run~1 data on the $hZZ$ coupling yields~\cite{Khachatryan:2016vau}
\begin{equation}
 s_\theta < 0.44 \quad @ \, 68\% \, \mathrm{CL},
\end{equation}
while bounds from other couplings are weaker.
In the analysis presented here, it is assumed that $f_{\rm TC}\geq1$.
Consequently, $ s_\theta < 0.25 $ and the bounds from modified Higgs couplings can always be satisfied.

The operator $\mathcal{O}_{\Pi f}$ modifies the couplings of fermions to the weak gauge bosons.
In particular, the $Z$ boson couplings receive a NP contribution
\begin{equation}
C_{\Pi f}\,\mathcal{O}_{\Pi f} \supset
\sum_{f\in\{u,d,e,\nu\}}
 \frac{g}{c_W} Z_\mu \left( {\delta g}_{f_L}^{ij}\, \bar{f}_{L}^i\, \gamma^\mu\, f_{L}^j
 + {\delta g}_{f_R}^{ij}\, \bar{f}_{R}^i\, \gamma^\mu\, f_{R}^j\, \right),
\end{equation}
where ${\delta g}_{f_L}^{ij}$ and ${\delta g}_{f_R}^{ij}$ are the deviations from the SM $Z$ couplings; they are given by
\begin{equation}\label{eq:FlavMFPC:Z_couplings}
 \begin{aligned}
{\delta g}_{u_L}^{ij} &=
 +\frac{C_{\Pi f}}{8\pi}\,s_\theta^2\, \big(X_{uu}\big)_{ij}\,,
 \quad\quad
{\delta g}_{u_R}^{ij} =
 -\frac{C_{\Pi f}}{8\pi}\,s_\theta^2\, \big(X_{\bar{u}\bar{u}}^*\big)_{ij}\,,
 \\
{\delta g}_{d_L}^{ij} &=
 -\frac{C_{\Pi f}}{8\pi}\,s_\theta^2\, \big(X_{dd}\big)_{ij}\,,
 \quad\quad
{\delta g}_{d_R}^{ij} =
 +\frac{C_{\Pi f}}{8\pi}\,s_\theta^2\, \big(X_{\bar{d}\bar{d}}^*\big)_{ij}\,,
 \\
{\delta g}_{e_L}^{ij} &=
 -\frac{C_{\Pi f}}{8\pi}\,s_\theta^2\, \big(X_{ee}\big)_{ij}\,,
 \quad\quad
{\delta g}_{e_R}^{ij} =
 +\frac{C_{\Pi f}}{8\pi}\,s_\theta^2\, \big(X_{\bar{e}\bar{e}}^*\big)_{ij}\,,
 \\
{\delta g}_{\nu_L}^{ij} &=
 +\frac{C_{\Pi f}}{8\pi}\,s_\theta^2\, \big(X_{\nu\nu}\big)_{ij}\,,
 \quad\quad
{\delta g}_{\nu_R}^{ij} = 0\,.
 \end{aligned}
\end{equation}
The terms diagonal in the flavor indices modify the partial widths of the $Z$ boson, which have been measured at LEP with high precision.
This can put strong constraints on the possible size of the fundamental Yukawa couplings that enter the $X_{ff}$ terms.
Very similar to the discussion in section~\ref{sec:LFUV:model}, the fundamental Yukawa couplings of the top quark need to be large to reproduce its mass.
Since the left-handed top and bottom quarks share the same fundamental Yukawa coupling, the $Z b_L b_L$ coupling can potentially receive large contributions.
In the model described in section~\ref{sec:LFUV:model}, they are avoided at tree level due to a discrete symmetry that serves as a custodial protection.
In MFPC, the $Z b_L b_L$ coupling is not protected\footnote{%
There are actually FPC models that feature a custodial protection of the $Z b_L b_L$ coupling, cf.~\cite{Sannino:2016sfx}.
}.
Hence, the LEP measurements of the $Z$ boson's partial widths are important constraints.
In our numerical analysis, we have therefore calculated the following observables for each parameter point,
\begin{equation}\label{eq:Z_observables_quarks}
 R_b=\frac{\Gamma(Z\to b\bar{b})}{\Gamma(Z\to q\bar{q})},
 \quad\quad
 R_c=\frac{\Gamma(Z\to c\bar{c})}{\Gamma(Z\to q\bar{q})},
\end{equation}
\begin{equation}\label{eq:Z_observables_leptons}
 R_e=\frac{\Gamma(Z\to q\bar{q})}{\Gamma(Z\to e\bar{e})},
 \quad\quad
 R_\mu=\frac{\Gamma(Z\to q\bar{q})}{\Gamma(Z\to \mu\bar{\mu})},
 \quad\quad
 R_\tau=\frac{\Gamma(Z\to q\bar{q})}{\Gamma(Z\to \tau\bar{\tau})},
\end{equation}
where $\Gamma(Z\to q\bar{q})$ implies a sum over all quarks except the top.
The calculations take into account higher order electroweak corrections~\cite{Freitas:2014hra} and the leading order QCD correction~\cite{Chetyrkin:1994js}, such that the correct SM predictions are reproduced in the limit $C_{\Pi f}=0$, where all terms in eq.~(\ref{eq:FlavMFPC:Z_couplings}) vanish.

The fundamental Yukawa couplings provide a source of breaking of the custodial $\SUs{2}{L+R}$ symmetry and thus contribute to the electroweak $S$ and $T$ parameters.
While $\SUs{2}{L+R}$ symmetry is also broken in the SM by the SM Yukawa couplings, the MFPC model modifies these SM contributions due to the modified fermion-Higgs couplings.
In addition, there are also contributions to $S$ and $T$ parameter stemming from the strong dynamics, which are encoded by WCs of effective operators in the MFPC-EFT.
The total contribution then strongly depends on these WCs, which are, however, independent of the WCs appearing in the flavor observables considered here.
Consequently, there is no strong correlation between the flavor observables and the $S$ and $T$ parameters.
Hence, the $S$ and $T$ parameters have not been considered in our numerical analysis.

\subsection{Low-energy probes of flavor and CP violation}\label{sec:FlavMFPC:flavor_obs}
Since models with partial compositeness can generate FCNCs already at tree level, precision measurements of processes like meson-antimeson mixing or rare decays of $K$ and $B$ mesons are important constraint that have to be taken into account.
Because flavor-changing charged currents are already generated at tree level in the SM, generic NP contributions to them are expected to be less pronounced than in FCNCs.
However, partial compositeness can also lead to violation of LFU or the unitarity of the CKM matrix.
It is therefore important to also consider charged-current observables as constraints.
%
%
%
%
\subsubsection{Matching the MFPC-EFT to the WEH}
Predictions for the flavor observables considered in this section are usually calculated in terms of WCs of operators in the WEH, which are evaluated at a hadronic scale of the order of a few GeV.
To derive these WCs, the MFPC-EFT is first matched to the WEH at the EW scale.
This scale is also called the {\it matching scale} and for our numerical analysis it has been chosen to be $160$~GeV.
The flavor observables are calculated by the \texttt{flavio} code, which is also employed in the analyses in chapter~\ref{chap:anomalies}.
This code implements the renormalization group~(RG) running necessary to evaluate the WEH WCs at the hadronic scale.
The matching of the MFPC-EFT to the WEH at the matching scale is done as follows.

The four-fermion operators in eqs.~(\ref{eq:FPC:O_4f_hermitian}) and~(\ref{eq:FPC:O_4f_complex}) are matched to four-fermion operators in the WEH in four steps:
\begin{enumerate}
 \item The background value of the spurion $\spur{i}{a}$, eq.~(\ref{eq:FPC:Psi_background_value}), is used to write the $\spur{i}{a}$ valued four-fermion operators in terms of SM fields in the two-component Weyl spinor notation (cf.\ table~\ref{tab:FPC:quantum_numbers}).
 \item By applying an assortment of Fierz transformations, these operators are then transformed to a {\it chiral basis} compatible with the WEH.
 This means that the non-chiral four-fermion operators in the WEH are simply given by a sum of operators in the chiral basis.
 In particular, no additional Fierz transformations are needed to get the operators in the WEH from those in the chiral basis.
 \item The Weyl spinor four-fermion operators in the chiral basis are translated to four-component Dirac spinors.
 If necessary, they are combined to constitute non-chiral four-fermion operators in the WEH.
 \item All four-fermion operators and fundamental Yukawa couplings are rotated to the mass basis by applying the unitary matrices defined in eq.~(\ref{eq:FlavMFPC:masses_mass_basis}).
 All products of fundamental Yukawa matrices can then be written in terms of the $X$ and $Y$ matrices defined in eqs.~(\ref{eq:FlavMFPC:X_ff}) and~(\ref{eq:FlavMFPC:Y_ff}).
\end{enumerate}

In addition to the four fermion operators of the MFPC-EFT, the operator $\mathcal{O}_{\Pi f}$ defined in eq.~(\ref{eq:FPC:OPif}) plays an important role.
Since it modifies the couplings between SM fermions and electroweak gauge bosons (cf.\ section~\ref{sec:FlavMFPC:Constr_EW}), it yields non-standard contributions to the operators in the WEH when the weak gauge bosons are integrated out.
To derive these contributions, it is convenient to integrate out the $W$ and the $Z$ already in the MFPC-EFT.
This yields new four-fermion operators in terms of the spurion $\spur{i}{a}$.
They can then be matched to the WEH by applying the four steps described above.

Since the operator $\mathcal{O}_{\rm Yuk}$ slightly modifies the Higgs couplings to fermions, the operators that are generated by integrating out the Higgs also slightly differ from those one gets by integrating out the Higgs in the SM.
However, the difference is always flavor diagonal and subleading in an expansion in $s_\theta$.
Modifications of four-fermion operators in the WEH due to $\mathcal{O}_{\rm Yuk}$ are therefore neglected here.

\subsubsection{Meson-antimeson mixing}
The part of the WEH containing the NP contributions to meson-antimeson mixing in the $K^0$, $B^0$, and $B_s$ systems is
\begin{equation}
\mathcal{H}_\mathrm{weak,NP}^{\Delta F=2} = - \sum_{k,ij} C_{k}^{ij} O_{k}^{ij} \,,
\end{equation}
where the sum runs over the following operators,
\begin{equation}
 \begin{aligned}
O_{VLL}^{ij} &= (\bar d^j_L \gamma^\mu d^i_L)(\bar d^j_L \gamma_\mu d^i_L)\,,
&
O_{VRR}^{ij} &= (\bar d^j_R \gamma^\mu d^i_R)(\bar d^j_R \gamma_\mu d^i_R)\,,
&
O_{VLR}^{ij} &= (\bar d^j_L \gamma^\mu d^i_L)(\bar d^j_R \gamma_\mu d^i_R) \,,
\\
O_{SLL}^{ij} &= (\bar d^j_R  d^i_L)(\bar d^j_R  d^i_L)\,,
&
O_{SRR}^{ij} &= (\bar d^j_L d^i_R)(\bar d^j_L  d^i_R)\,,
&
O_{SLR}^{ij} &= (\bar d^j_R  d^i_L)(\bar d^j_L  d^i_R)\,,
\\
O_{TLL}^{ij} &= (\bar d^j_R \sigma^{\mu\nu} d^i_L)(\bar d^j_R \sigma_{\mu\nu} d^i_L)\,,
&
O_{TRR}^{ij} &= (\bar d^j_L \sigma^{\mu\nu} d^i_R)(\bar d^j_L \sigma_{\mu\nu} d^i_R)\,,
 \end{aligned}
\end{equation}
and $ij=21,31,32$ for $K^0$, $B^0$, and $B_s$, respectively.
All of the above operators are generated from the MFPC-EFT, with contributions to their WCs coming from two sources.
%
The first source are simply the four-fermion operators in the MFPC-EFT, eqs.~(\ref{eq:FPC:O_4f_hermitian}) and~(\ref{eq:FPC:O_4f_complex}).
The second source are tree-level $Z$ exchange diagrams that involve two flavor-changing $Z$-couplings stemming from the operator $\mathcal{O}_{\Pi f}$.
These diagrams contribute to the WCs of the above operators when the $Z$ boson is integrated out.
However, since they require two insertions of $\mathcal{O}_{\Pi f}$, they are subleading in an expansion in $s_\theta$.
%
%
At leading order\footnote{%
In the numerical analysis discussed in section~\ref{sec:FlavMFPC:numerical_analysis}, also subleading contributions are included.
} in $s_\theta$, only four of the above operators are generated; their WCs read
\begin{equation}\label{eq:FlavMFPC:WCs_deltaF=2}
\begin{aligned}
C_{VLL}^{ij} &=
 \big(X_{dd}^*\big)_{ij}\,
 \big(X_{dd}^*\big)_{ij}\,
 \frac{ C^4_{4f}+C^5_{4f} }{\Lambda_\text{TC}^2},
\\
C_{VRR}^{ij} &=
 \big(X_{\bar{d}\bar{d}}\big)_{ij}\,
 \big(X_{\bar{d}\bar{d}}\big)_{ij}\,
 \frac{ C^4_{4f}+C^5_{4f} }{\Lambda_\text{TC}^2},
\\
C_{VLR}^{ij} &=
 \big(X_{dd}^*\big)_{ij}\,
 \big(X_{\bar{d}\bar{d}}\big)_{ij}\,
 \frac{ C^4_{4f} }{\Lambda_\text{TC}^2},
\\
C_{SLR}^{ij} &=
 \big(Y_{d\bar{d}}\big)_{ij}\,
 \big(Y_{\bar{d}d}^*\big)_{ij}\,
 \frac{ C^2_{4f} }{\Lambda_\text{TC}^2}.
\end{aligned}
\end{equation}
The left-right operators $O_{SLR}^{ij}$ and $O_{VLR}^{ij}$ are notorious for their role in the ``flavor problem'' of partial compositeness models with an anarchic flavor structure.
These operators are strongly suppressed in the SM but can be generated from heavy gluon resonance exchange in models with partial compositeness.
In the absence of flavor symmetries, the chiral enhancement of the hadronic matrix elements of the left-right operators in the kaon sector leads to a substantial contribution to $\epsilon_K$, which measures indirect $C\!P$ violation in kaon mixing and puts very strong constraints on these models (cf.~\cite{Csaki:2008zd,Blanke:2008zb,Bauer:2009cf}).
Interestingly, in the MFPC model, the NP contribution to the WC $C_{SLR}^{ij}$ always vanishes for $i\neq j$, i.e.\ it is flavor-diagonal and cannot contribute to meson-antimeson mixing.
This is also true for subleading terms in the $s_\theta$ expansion.
The reason for this is that the flavor structure of $C_{SLR}^{ij}$ depends only on $\big(Y_{d\bar{d}}\big)_{ij}$, which is proportional to the down-type quark mass matrix and by definition diagonal in the mass basis (cf.\ eq.~(\ref{eq:FlavMFPC:m_F_diag})).
This is in contrast to effective partial compositeness models and models with extra dimensions, where the heavy gluon resonance exchange generates also off-diagonal terms for $C_{SLR}^{ij}$  (cf.~\cite{Csaki:2008zd,Blanke:2008zb,Bauer:2009cf}).
The MFPC model is special in the sense that any heavy resonance in an adjoint of $\SUs{3}{C}$ necessarily has to be an $(\S_q\S_q^*)$ techniscalar bound state (cf.\ table~\ref{tab:FPC:quantum_numbers}).
The structure of the fundamental Yukawa couplings then guarantees that an exchange of such a bound state can only contribute to $C_{SLR}^{ij}$ with a term proportional to a product of quark mass matrices.
Consequently, these contributions are always flavor diagonal in the mass basis.
However, even for vanishing $C_{SLR}^{ij}$ at the matching scale, the QCD RG running leads to a sizable contribution to $C_{SLR}^{ij}$ proportional to $C_{VLR}^{ij}$ at the hadronic scale.
Therefore, the strongest bound from meson-antimeson mixing observables is still assumed to come from $\epsilon_K$.
In any case, in our numerical analysis, we have considered all of the following observables:
\begin{itemize}
\item The parameter $\epsilon_K$ measuring indirect $C\!P$ violation in $K^0$ mixing.
\item The mixing-induced CP asymmetry $S_{\psi K_S}$ in $B_d\to J/\psi \, K_S$.
\item The mixing-induced CP asymmetry $S_{\psi \phi}$ in $B_s\to J/\psi \, \phi$.
\item The mass differences $\Delta M_{d}$ and $\Delta M_{s}$ in the $B_d$ and $B_s$ systems.
\end{itemize}

\subsubsection{Neutral current semi-leptonic decays}
The rare neutral current $b\to s\,\ell\ell$ decays are of great interest in the light of the flavor anomalies discussed in chapter~\ref{chap:anomalies}.
In particular, a central aspect of the analysis presented here is to answer the question whether the MFPC model can account for violation of LFU in $B\to K^*\ell\ell$ and $B\to K\ell\ell$ decays.
The part of the WEH containing the operators that contribute dominantly to $b\to s\,\ell\ell$ processes is given in eq.~(\ref{eq:anomalies:Heff}) and repeated here for convenience,
%
\begin{equation}
\mathcal{H}_\text{weak,NP}^{b\to s\ell\ell} = - \mathcal{N}^{b\to s}
\sum_k
(C_k^\ell O_k^\ell + C^{\prime\ell}_k O^{\prime\ell}_k) + \text{h.c.}\,,
\end{equation}
where the normalization factor $\mathcal{N}^{b\to s}$ is given by
\begin{equation}
 \mathcal{N}^{b\to s} = \frac{4\,G_F}{\sqrt{2}} V_{tb}V_{ts}^* \frac{e^2}{16\pi^2}.
\end{equation}
As in section~\ref{sec:anomalies:WEH}, only NP contributions in form of the following operators are considered:
\begin{equation}
 \begin{aligned}
O_9^\ell &=
(\bar{s} \gamma_{\mu} P_{L} b)(\bar{\ell} \gamma^\mu \ell)\,,
&
O_9^{\prime\ell} &=
(\bar{s} \gamma_{\mu} P_{R} b)(\bar{\ell} \gamma^\mu \ell)\,,
\\
O_{10}^\ell &=
(\bar{s} \gamma_{\mu} P_{L} b)( \bar{\ell} \gamma^\mu \gamma_5 \ell)\,,
&
O_{10}^{\prime\ell} &=
(\bar{s} \gamma_{\mu} P_{R} b)( \bar{\ell} \gamma^\mu \gamma_5 \ell)\,.
 \end{aligned}
\end{equation}
In contrast to the $\Delta F=2$ meson-antimeson mixing observables, the semi-leptonic decays are $\Delta F=1$ processes, i.e.\ they only involve one flavor-changing coupling.
Consequently, contributions from $Z$ exchange diagrams to the above operators only require one insertion of $\mathcal{O}_{\Pi f}$.
Therefore, they enter at the same order of $s_\theta$ as the direct contributions stemming from four-fermion operators in the MFPC-EFT.
To leading order in $s_\theta$, the latter read
\begin{equation}\label{eq:FlavMFPC:WCS:bsll_4f}
\begin{aligned}
C_{9}^\ell \,\mathcal{N}^{b\to s}&\supset
-
\frac{1}{4}\,
\big(X_{dd}^*\big)_{bs}\,
\big(X_{\bar{e}\bar{e}}\big)_{\ell\ell}\,
\frac{ C^4_{4f}}{\Lambda_\text{TC}^2}
+
\frac{1}{4}\,
\big(X_{dd}^*\big)_{bs}\,
\big(X_{ee}\big)_{\ell\ell}\,
\frac{ C^4_{4f}+C^5_{4f} }{\Lambda_\text{TC}^2}
\,,\\
C_{9}^{\prime\ell} \,\mathcal{N}^{b\to s}&\supset
-
\frac{1}{4}\,
\big(X_{\dbar \dbar}\big)_{bs}\,
\big(X_{ee}\big)_{\ell\ell}\,
\frac{ C^4_{4f}}{\Lambda_\text{TC}^2}
+
\frac{1}{4}\,
\big(X_{\dbar \dbar}\big)_{bs}\,
\big(X_{\ebar \ebar}\big)_{\ell\ell}\,
\frac{ C^4_{4f}+C^5_{4f} }{\Lambda_\text{TC}^2}
\,,\\
C_{10}^\ell \,\mathcal{N}^{b\to s}&\supset
-
\frac{1}{4}\,
\big(X_{dd}^*\big)_{bs}\,
\big(X_{\bar{e}\bar{e}}\big)_{\ell\ell}\,
\frac{ C^4_{4f}}{\Lambda_\text{TC}^2}
-
\frac{1}{4}\,
\big(X_{dd}^*\big)_{bs}\,
\big(X_{ee}\big)_{\ell\ell}\,
\frac{ C^4_{4f}+C^5_{4f} }{\Lambda_\text{TC}^2}
\,,\\
C_{10}^{\prime\ell} \,\mathcal{N}^{b\to s}&\supset
+
\frac{1}{4}\,
\big(X_{\dbar \dbar}\big)_{bs}\,
\big(X_{ee}\big)_{\ell\ell}\,
\frac{ C^4_{4f}}{\Lambda_\text{TC}^2}
+
\frac{1}{4}\,
\big(X_{\dbar \dbar}\big)_{bs}\,
\big(X_{\ebar \ebar}\big)_{\ell\ell}\,
\frac{ C^4_{4f}+C^5_{4f} }{\Lambda_\text{TC}^2}
\,,
\end{aligned}
\end{equation}
and the contributions from integrating out the $Z$ boson are given by
\begin{equation}
 \begin{aligned}
C_{9}^\ell \,\mathcal{N}^{b\to s}&\supset
2\pi\,
\big(X_{dd}^*\big)_{bs}\,
(4\,s_w^2-1)\,
\frac{C_{\Pi f}}{\Lambda_\text{TC}^2}
\,,\\
C_{9}^{\prime\ell} \,\mathcal{N}^{b\to s}&\supset
-2\pi\,
\big(X_{\dbar \dbar}\big)_{bs}\,
(4\,s_w^2-1)\,
\frac{C_{\Pi f}}{\Lambda_\text{TC}^2}
\,,\\
C_{10}^\ell \,\mathcal{N}^{b\to s}&\supset
2\pi\,
\big(X_{dd}^*\big)_{bs}\,
\frac{C_{\Pi f}}{\Lambda_\text{TC}^2}
\,,\\
C_{10}^{\prime\ell} \,\mathcal{N}^{b\to s}&\supset
-2\pi\,
\big(X_{\dbar \dbar}\big)_{bs}\,
\frac{C_{\Pi f}}{\Lambda_\text{TC}^2}
\,.
\end{aligned}
\end{equation}
While the contributions from $Z$ boson exchange are LFU conserving at leading order in $s_\theta$, those stemming from four-fermion operators in the MFPC-EFT are actually expected to violate LFU.
To assess if this can explain the experimental hints for violation of LFU in neutral current decays, the following observables are predicted in our numerical analysis:
%
%
%
\begin{itemize}
 \item $R_K$ for $q^2\in[1,6]$ GeV$^2$,
 \item $R_{K^*}$ for $q^2\in[0.045,1.1]$ GeV$^2$ and for $q^2\in[1.1,6]$ GeV$^2$.
\end{itemize}

\subsubsection{Charged-current semi-leptonic decays}
In contrast to the rare FCNC decays, flavor-changing charged current decays are tree-level processes in the SM, mediated by the $W$ boson.
As such, they are far less sensitive to NP contributions from loop processes than FCNC decays.
Observables based on the $q\to q'\ell\nu$ transition are therefore used for determining the elements of the CKM matrix with the fewest possible pollution from NP effects.
In models with partial compositeness, however, NP contributions to the $q\to q'\ell\nu$ transition without loop-suppression are possible.
In the MFPC model, one source of contributions are again the four-fermion operators in the MFPC-EFT.
In addition, also diagrams with modified $W$ couplings due to the operator $\mathcal{O}_{\Pi f}$ can contribute at leading order in $s_\theta$.
In this case, the couplings of either quarks or leptons are modified, while a simultaneous modification of both couplings is subleading in $s_\theta$.
In our numerical analysis, we have focused on $d_i\to u_j\ell\nu$ processes.
The part of the WEH that describes NP contributions to these processes is
\begin{equation}
\mathcal{H}_\mathrm{weak,NP}^{d\to u\ell\nu} =\sum_{ij} \mathcal{N}^{d_i\to u_j} \sum_k C_k^{(\prime){d^iu^j\ell}} O_k^{(\prime){d^iu^j\ell}} + \text{h.c.},
\end{equation}
where the normalization factor is
\begin{equation}
 \mathcal{N}^{d_i\to u_j} = \frac{4\,G_F}{\sqrt{2}} V_{u_j d_i}\,,
\end{equation}
the sum runs over the following operators,
\begin{equation}
\begin{aligned}
O_V^{d^iu^j\ell} &= (\bar u^j_L \gamma^\mu d^i_L)(\bar \ell_L \gamma_\mu \nu_{\ell L})
\,, &
O_V^{\prime d^iu^j\ell} &= (\bar u^j_R \gamma^\mu d^i_R)(\bar \ell_L \gamma_\mu \nu_{\ell L})
\,, \\
O_S^{d^iu^j\ell} &= m_{d^i}(\bar u^j_L d^i_R)(\bar \ell_R \nu_{\ell L})
\,, &
O_S^{\prime d^iu^j\ell} &= m_{d^i}(\bar u^j_R d^i_L)(\bar \ell_R \nu_{\ell L})
\,, \\
O_T^{d^iu^j\ell} &= (\bar u^j_R \sigma^{\mu\nu} d^i_L)(\bar \ell_R \sigma_{\mu\nu}\nu_{\ell L}) \,
\end{aligned}
\end{equation}
and $ij=11,21,32$ for $d\to u \ell\nu$, $s\to u \ell\nu$, and $b\to c \ell\nu$, respectively.
The only operator that receives a contribution in the SM is $O_V^{d^iu^j\ell}$, and with the above normalization, its WC is simply $C_{V\,\text{SM}}^{d^iu^j\ell}=1$.
In MFPC, all of the above operators are generated.
To leading order in $s_\theta$, the direct contributions from four-fermion operators in the MFPC-EFT yield
\begin{equation}\label{eq:FlavMFPC:WCS:CC_4f}
\begin{aligned}
C_{V}^{d^iu^j\ell}\,\mathcal{N}^{d_i\to u_j} &\supset
\frac{1}{2}\,
\big(X_{du}^*\big)_{ij}\,
\big(X_{e\nu}\big)_{\ell\ell}\,
\frac{ C^5_{4f} - C^3_{4f} }{\Lambda_\text{TC}^2}
\,,\\
C_{V}^{\prime d^iu^j\ell}\,\mathcal{N}^{d_i\to u_j} &\supset 0
\,,\\
C_{S}^{d^iu^j\ell}\,\mathcal{N}^{d_i\to u_j} &\supset
\big(Y_{\dbar u}^*\big)_{ij}\,
\big(Y_{\ebar\nu}\big)_{\ell\ell}\,
\frac{ C^2_{4f} }{\Lambda_\text{TC}^2}
\,,\\
C_{S}^{\prime d^iu^j\ell}\,\mathcal{N}^{d_i\to u_j} &\supset
\frac{1}{2}
\big(Y_{d \ubar}\big)_{ij}\,
\big(Y_{\ebar\nu}\big)_{\ell\ell}\,
\frac{ C^{8*}_{4f}-2\,C^{7*}_{4f} }{\Lambda_\text{TC}^2}
\,,\\
C_{T}^{d^iu^j\ell}\,\mathcal{N}^{d_i\to u_j} &\supset
\frac{1}{8}
\big(Y_{d \ubar}\big)_{ij}\,
\big(Y_{\ebar\nu}\big)_{\ell\ell}\,
\frac{ C^{8*}_{4f} }{\Lambda_\text{TC}^2}
\,,
\end{aligned}
\end{equation}
while the contributions from integrating out the $W$ boson read
\begin{equation}\label{eq:FlavMFPC:WCS:CC_W}
\begin{aligned}
C_{V}^{d^iu^j\ell}\,\mathcal{N}^{d_i\to u_j} &\supset
-8\,\pi\,
\left(
\big(X_{du}^*\big)_{ij}\,
+
V_{u_j d_i}\,
\big(X_{e\nu}\big)_{\ell\ell}
\right)
\frac{ C_{\Pi f} }{\Lambda_\text{TC}^2}
\,,\\
C_{V}^{d^iu^j\ell\prime}\,\mathcal{N}^{d_i\to u_j} &\supset
8\,\pi\,
\big(X_{\dbar \ubar}\big)_{ij}\,
\frac{ C_{\Pi f} }{\Lambda_\text{TC}^2}
\,.
\end{aligned}
\end{equation}
Like in neutral current semi-leptonic decays, the former are actually expected to violate LFU.
As noted above, the contributions from $W$ exchange at leading order in $s_\theta$ involve either a modified lepton or quark coupling.
If only the quark coupling is modified, the resulting operators are lepton flavor universal.
On the other hand, if the lepton coupling is modified, they are expected to violate LFU.

A sizable fundamental Yukawa coupling of left-handed muons that might, similarly to chapter~\ref{chap:LUFV_in_CHMs}, explain the hints for violation of LFU in neutral current decays also enters the charged current WCs.
Thus, one has to ascertain that explaining violation of LFU in neutral current decays is not in conflict with experimental measurements of charged current decays.
Therefore, charged current decays with an electron or muon in the final state are taken into account in our numerical analysis as important constraints on the leptons' fundamental Yukawa couplings and the size of a possible violation of LFU.

But they are essential also for another reason:
%
%
%
%
%
they allow for consistently comparing the CKM measurements with the predictions for CKM elements obtained from diagonalizing the quark mass matrices.
In particular, in the presence of NP contributions to the charged current WCs, the experimental data might favor slightly different values of CKM elements than in the SM.
In our numerical analysis, the following processes are therefore considered as constraints:
\begin{itemize}
\item The branching ratio of $\pi^+\to e\nu$, which is based on the $d\to u\ell\nu$ transition.
\item The branching ratio of $K^+\to \mu\nu$ and the
 ratio of $K^+\to \ell\nu$ branching ratios with $\ell\in\{e,\mu\}$, which are based on the $s\to u\ell\nu$ transition.
\item
The branching ratios of $B\to D\ell\nu$ with $\ell\in\{e,\mu\}$, which are based on the $b\to c\ell\nu$ transition.
\end{itemize}
There are tensions between experimental data and the SM prediction of observables involving  the $b\to c\tau\nu$ transition.
In particular, measurements of the LFU ratios $R_{D^{(*)}}$, i.e.\ the ratios of the $B\to D^{(*)}\tau\nu$ and the $B\to D^{(*)}\ell\nu$ ($\ell=e,\mu$) branching ratios, show a deviation from the SM prediction at a combined level of around 4$\sigma$~\cite{Amhis:2016xyh}.
To assess whether these hints for violation of LFU in charged current decays can be explained by the MFPC model, the following observables are not considered as constraints but rather as predictions:
\begin{itemize}
\item The ratios $R_D$ and $R_{D^*}$, which are based on the $b\to c\tau\nu$ transition.
\end{itemize}

\section{Numerical analysis}\label{sec:FlavMFPC:numerical_analysis}
The effects of the MFPC model on the flavor and electroweak observables discussed in the previous section are investigated by calculating them from the parameters of the MFPC-EFT.
By varying these parameters, it is possible to find regions in the parameter space where all the applied constraints are satisfied.
The parameter points in those regions then yield predictions for the LFU observables $R_{K^{(*)}}$ and $R_{D^{(*)}}$, which are not considered as constraints.

In our numerical analysis, we have made some assumptions concerning the lepton sector that simplify the analysis or avoid additional constraints:
\begin{itemize}
 \item Strong constraints from charged lepton flavor violation (see e.g.\ \cite{KerenZur:2012fr}) are avoided by assuming that the fundamental Yukawa matrices $y_L$ and $y_{\bar{e}}$ can be both diagonalized in the same basis at the matching scale\footnote{%
 This assumption is not RG invariant if LFU is violated~\cite{Feruglio:2017rjo}.}.
 \item We have assumed right-handed neutrinos to be irrelevant for our analysis.
 Their effects are neglected by setting their fundamental Yukawa couplings $y_{\bar{\nu}}$ and $y'_{\bar{\nu}}$ to zero.
\end{itemize}

\subsection{Parameters}
Among all the parameters of the MFPC-EFT, only those entering the observables discussed above have to be considered.
In our numerical analysis, we have varied each of them over a specific range.
These parameters are
\begin{itemize}
 \item The NGB decay constant $f_{\TC}$, which is related to the strong coupling scale by $\Lambda_{\TC}=4\pi\,f_{\TC}$. $f_{\TC}$ is varied between 1~TeV and 3~TeV.
 \item The six real WCs $C^{1}_{4f}$, $C^{2}_{4f}$, $C^{3}_{4f}$, $C^{4}_{4f}$, $C^{5}_{4f}$ and $C_{\Pi f}$.
 Their absolute values are varied logarithmically between $0.1$ and $10$ and each of them is allowed to be positive or negative.
 \item The four complex WCs $C^{6}_{4f}$, $C^{7}_{4f}$, $C^{8}_{4f}$ and $C_{\rm Yuk}$.
 Their absolute values are varied logarithmically between $0.1$ and $10$ and their complex phases linearly between $0$ and $2\pi$.
 \item The four\footnote{%
 The assumption that $y_\ebar$ and $y_L$ are diagonal in the same basis at the matching scale allows for fixing the entries of $y_\ebar$ by requiring that the product of $y_L$ and $y_\ebar$ yields the correct masses of the charged leptons.}
 fundamental Yukawa matrices $y_Q$, $y_L$, $y_\ubar$, and $y_\dbar$.
 To parameterize them, it is convenient to define the effective Yukawa matrices
 \begin{equation}
  \tilde{y}_f = \sqrt{C_{\rm Yuk}}\,y_f.
 \end{equation}
 They have the advantage that they allow for expressing the fermion mass matrices independently of the WC $C_{\rm Yuk}$ (cf.\ eq.~(\ref{eq:FlavMFPC:masses_gauge_basis})).
 Each of the complex matrices $\tilde{y}_f$ can be decomposed by an SVD into one diagonal and two unitary matrices (cf.\ section~\ref{sec:Fermions:flavor}).
 This yields eight unitary and four diagonal matrices.
 The SM field $Q$, $L$, $\bar{u}$, and $\bar{d}$ and the techniscalar fields $\S_q$ and $\S_l$ can each absorb one unitary matrix.
 This leaves two physical unitary and four diagonal matrices.
 It is possible to choose the effective doublet Yukawa matrices $\tilde{y}_Q$ and $\tilde{y}_L$ to be diagonal,
 \begin{equation}
  \tilde{y}_Q = {\rm diag}(y_{Q1},y_{Q2},y_{Q3}),
  \quad\quad
  \tilde{y}_L = {\rm diag}(y_{L1},y_{L2},y_{L3}),
 \end{equation}
 while the effective singlet Yukawa matrices $\tilde{y}_{\bar{u}}$ and $\tilde{y}_{\bar{d}}$ then depend on one diagonal and one unitary matrix each.
 The two unitary matrices can be parameterized by in total six angles $t_{u}^{12}$ ,$t_{u}^{13}$, $t_{u}^{23}$, $t_{d}^{12}$ ,$t_{d}^{13}$, $t_{d}^{23}$ and four phases\footnote{
 While a general $3\times3$ unitary matrix has five independent phases, six of the ten phases of $\tilde{y}_\ubar$ and $\tilde{y}_\dbar$ can be absorbed by field redefinitions.} $\delta_d$, $\delta_u$, $a_d$, $b_d$.
 The effective Yukawa matrices $\tilde{y}_{\bar{u}}$ and $\tilde{y}_{\bar{d}}$ can then be expressed as
 \begin{equation}
 \begin{aligned}
  \tilde{y}_\ubar &= {\rm unitary}(t_{u}^{12},t_{u}^{13},t_{u}^{23},\delta_u)\cdot{\rm diag}(y_{u1},y_{u2},y_{u3}),
  \\
  \tilde{y}_\dbar &= {\rm unitary}(t_{d}^{12},t_{d}^{13},t_{d}^{23},\delta_d,a_d,b_d)\cdot{\rm diag}(y_{d1},y_{d2},y_{d3}).
 \end{aligned}
 \end{equation}
 The entries of the diagonal matrices are varied logarithmically between\footnote{%
 To ascertain that the diagonal entries of $y_{\bar{e}}$ stay below $4\pi$ when they are fixed by requiring the correct charged lepton masses, the lower boundaries of the diagonal entries of $\tilde{y}_L$ are adjusted accordingly.}
 $0.002$ and $4\pi$ and the phases and angles linearly between $0$ and $2\pi$.
\end{itemize}

\subsection{Strategy}
The large amount of parameters is a challenge for a parameter scan.
A naive random variation is problematic because only a very tiny fraction of the points in the high-dimensional parameter space is actually compatible with experimental measurements of quark masses and CKM elements.
However, the effective Yukawa matrices are defined such that the quark masses and CKM elements only depend on the 19 parameters of the matrices $\tilde{y}_Q$, $\tilde{y}_{\bar{u}}$, and $\tilde{y}_{\bar{d}}$.
This makes it possible to divide the parameter scan into two steps.
In the first step, only $\tilde{y}_Q$, $\tilde{y}_{\bar{u}}$, and $\tilde{y}_{\bar{d}}$ are varied to find regions in parameter space that yield predictions for the quark masses and CKM elements close to the experimental observations.
In the second step, $\tilde{y}_Q$, $\tilde{y}_{\bar{u}}$, and $\tilde{y}_{\bar{d}}$ are kept fixed while all other parameters are chosen randomly.

For the first step, the quark masses are predicted by constructing the quark mass matrices in eq.~(\ref{eq:FlavMFPC:masses_gauge_basis}) from $\tilde{y}_Q$, $\tilde{y}_{\bar{u}}$ and $\tilde{y}_{\bar{d}}$.
These mass matrices are then numerically diagonalized via eq.~(\ref{eq:FlavMFPC:masses_mass_basis}).
The entries of the resulting diagonal matrices are interpreted as $\overline{\rm MS}$ running masses at 160 GeV and are run down to the scale where they can be compared to their PDG average.
The numerical diagonalization also yields the unitary matrices $U_u$ and $U_d$, which define the CKM matrix via eq.~(\ref{eq:FlavMFPC:CKM_matrix}).
In contrast to the masses, the CKM elements cannot be directly compared to experimental measurements.
As described in section~\ref{sec:FlavMFPC:flavor_obs}, the CKM elements are measured in neutral current semi-leptonic decays.
These decays are subject to corrections from the WEH WCs in eqs.~(\ref{eq:FlavMFPC:WCS:CC_4f}) and~(\ref{eq:FlavMFPC:WCS:CC_W}), which depend on parameters that are not yet specified in the first step of the scanning procedure.
Consequently, the CKM elements can only be compared to experimental measurements in the second step.
However, in the first step, they are required to be close to certain input values that are chosen such that a high fraction of parameter points passes the constraints from CKM measurements applied in the second step.
To compare the predicted CKM elements to these input values and the predicted quark masses to their PDG values, the $\chi^2$ function $\chi^2_\text{mass, CKM}$ is constructed.
The scan is then carried out as follows:
\begin{itemize}
 \item After choosing a random starting point in the 19-dimensional parameter-subspace spanned by the parameters of $\tilde{y}_Q$, $\tilde{y}_{\bar{u}}$, and $\tilde{y}_{\bar{d}}$, the $\chi^2_\text{mass, CKM}$ function is numerically minimized.
 This yields a viable point that predicts correct quark masses and CKM elements close to the input values.
 \item This viable point is then used as starting point for a Markov chain that samples the region around this point and generates 10\,k viable points with a low value of $\chi^2_\text{mass, CKM}$.
 To this end, the Markov-Chain-Monte-Carlo implementation from the \texttt{pypmc} package~\cite{pypmc} is used.
 \item To reduce the auto-correlation of the 10\,k points generated in the previous step, only 1\,k points are selected.
\end{itemize}
In our scan, we have repeated the above steps 100\,k times to get 100\,M points from 100\,k local minima of the $\chi^2_\text{mass, CKM}$ function.
These points all predict correct quark masses and CKM elements close to the input values.

For these 100\,M viable points, the remaining 18 parameters are chosen randomly.
For each of the resulting points in the 37-dimensional parameter space, all observables discussed in section~\ref{sec:FlavMFPC:observables} are calculated.
For the flavor observables in section~\ref{sec:FlavMFPC:flavor_obs}, all calculations are performed by the \texttt{flavio} code, while a dedicated code is used for the $Z$~decay observables in section~\ref{sec:FlavMFPC:Constr_EW}.

To compare the calculated predictions to the experimental values shown in table~\ref{tab:FlavMFPC:observables}, the $\chi^2$ functions $\chi^2_\text{$Z$}$, $\chi^2_{\Delta F=2}$, and $\chi^2_\text{CC}$ are constructed from the observables in $Z$~decays, meson-antimeson mixing, and semi-leptonic charged-current decays, respectively.
The constraints are applied to the parameter points by requiring that the corresponding $\chi^2$ function stays below its 3$\sigma$ value.
This corresponds to $\chi^2_\text{$Z$}\leq 18.2$, $\chi^2_{\Delta F=2}\leq 18.2$, and $\chi^2_\text{CC}\leq 16.3$ (cf.\ table~\ref{tab:FlavMFPC:observables}).

\begin{table}
\renewcommand{\arraystretch}{1.3}
\begin{tabular}{|l|llll|}
\hline
$\chi^2$&Observable & measurement & & SM prediction \\
\hline
\multirow{5}{*}{$\chi^2_\text{$Z$}$} & $R_e$	& $20.804 \pm 0.050$ & \cite{ALEPH:2005ab} & $20.768 \pm 0.006$ \\
&$R_\mu$	& $20.785 \pm 0.033$ & \cite{ALEPH:2005ab} & $20.768 \pm 0.006$ \\
&$R_\tau$& $20.764 \pm 0.045$ & \cite{ALEPH:2005ab} & $20.813 \pm 0.006$ \\
&$R_b$	& $0.21629 \pm 0.00066$ & \cite{ALEPH:2005ab} & $0.21591 \pm 0.00004$ \\
&$R_c$	& $0.1721 \pm 0.0030$ & \cite{ALEPH:2005ab} & $0.17112 \pm 0.00002$ \\
\hline
\multirow{5}{*}{$\chi^2_{\Delta F=2}$} & $\Delta M_s$ & $(17.76 \pm 0.02) $ ps & \cite{Amhis:2014hma} & $(19.9\pm1.7)$ ps \\
&$\Delta M_d$ & $(0.505 \pm 0.002)$ ps & \cite{Amhis:2014hma} & $(0.64\pm0.09)$ ps \\
&$S_{\psi\phi}$ & $(3.3 \pm 3.3) \times 10^{-2}$ & \cite{Amhis:2014hma} & $(3.75\pm0.22) \times 10^{-2}$ \\
&$S_{\psi K_S}$ & $0.679 \pm 0.020$ & \cite{Amhis:2014hma} & $0.690\pm0.025$ \\
&$\vert\epsilon_K\vert$ & $(2.228 \pm 0.011) \times 10^{-3}$ & \cite{Agashe:2014kda} & $(1.76\pm0.22) \times 10^{-3}$ \\
\hline
\multirow{4}{*}{$\chi^2_\text{CC}$} & $\text{BR}(B^+\to D^0\ell^+\nu_\ell)$ & $(2.330 \pm 0.098) \times 10^{-2}$ & \cite{Amhis:2014hma} & $(2.92\pm0.21) \times 10^{-2}$ \\
&$\text{BR}(\pi^+\to e^+\nu)$ & $(1.234\pm0.002) \times 10^{-4}$ & \cite{Aguilar-Arevalo:2015cdf} & $(1.2341\pm0.0002) \times 10^{-4}$ \\
&$\text{BR}(K^+\to \mu^+\nu)$ & $0.6356 \pm 0.0011$ & \cite{Agashe:2014kda} & $0.6296\pm0.0066$ \\
&$R_{e\mu}(K^+\to \ell^+\nu)$ & $(2.488 \pm 0.009) \times 10^{-5}$ & \cite{Agashe:2014kda} & $(2.475\pm0.001) \times 10^{-5}$ \\
\hline
&$R_{D}$ & $0.397\pm0.049$ & \cite{Amhis:2016xyh} & $0.277\pm0.012$ \\
&$R_{D^*}$ & $0.316\pm0.019$ & \cite{Amhis:2016xyh} & $0.2512\pm0.0043$ \\
&$R_K^{[1, 6]}$ & $0.75 ^{+ 0.08}_{ - 0.10}$ & \cite{Aaij:2014ora} & $1.000\pm0.001$ \\
&$R_{K^*}^{[0.045, 1.1]}$ & $0.65 ^{+ 0.07}_{ - 0.12}$ & \cite{Aaij:2017vbb} & $0.926 \pm 0.004$ \\
&$R_{K^*}^{[1.1, 6.0]}$ & $0.68 ^{+ 0.08}_{ - 0.12}$ & \cite{Aaij:2017vbb} & $0.9965 \pm 0.0005$ \\
\hline
\end{tabular}
\caption{Measurements and SM predictions.
The first three blocks contain the $Z$~decay,
meson-antimeson mixing, and charged current observables used as constraints.
The last block contains the LFU observables considered as predictions.
The SM predictions for the flavor observables (last three blocks) are computed with \texttt{flavio} v0.23.
The SM predictions for the $Z$~decay observables are computed with a dedicated code%
.}
\label{tab:FlavMFPC:observables}
\end{table}

\section{Results}\label{sec:FlavMFPC:results}

\subsection{Meson-antimeson mixing}
\begin{figure}[t]
\centering
\includegraphics[width=0.6\textwidth]{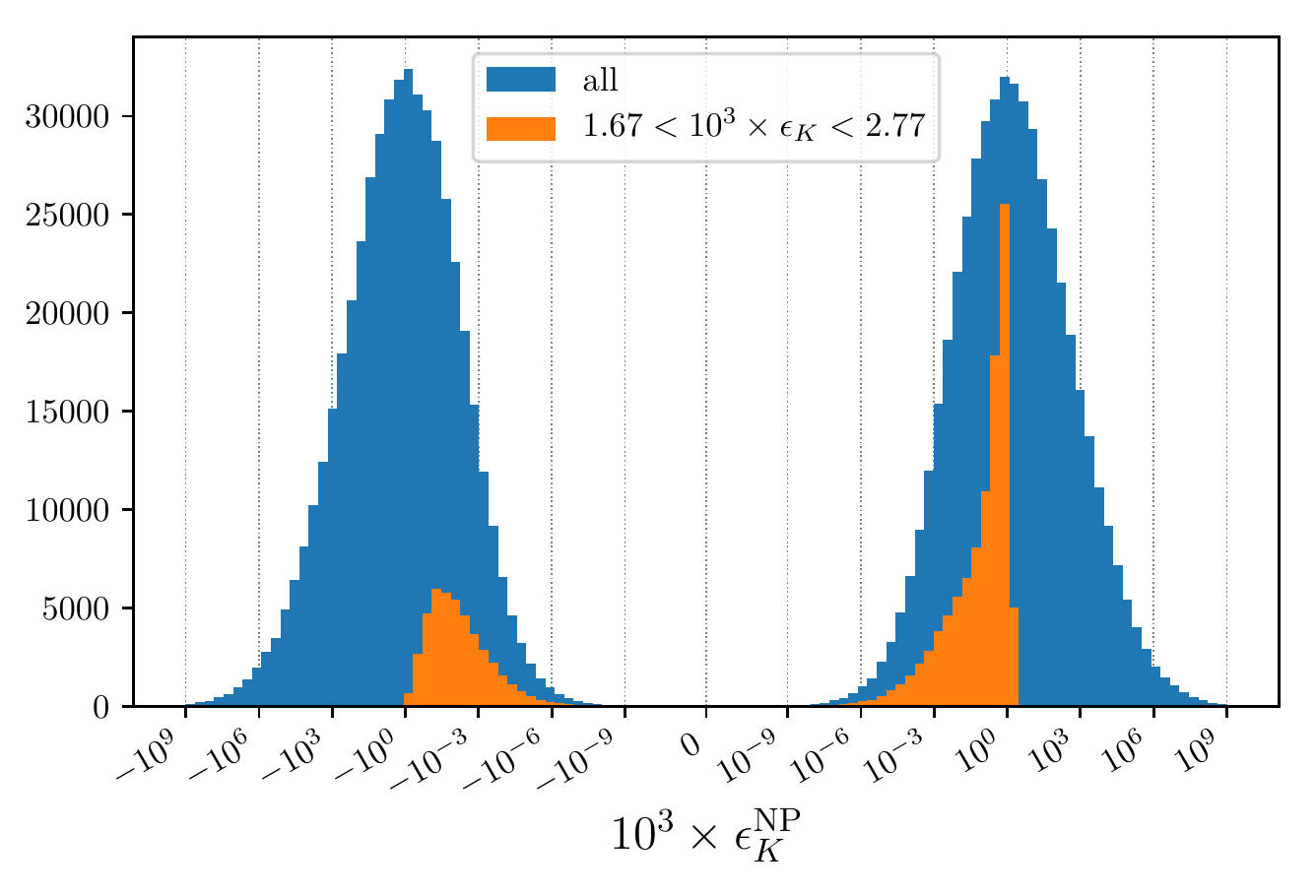}
\caption{Histogram showing the NP contribution to $\epsilon_K$ for a representative subset of all
points that feature the right masses and CKM elements,
compared to the points among those that pass the experimental constraint.
A positive NP contribution corresponds to constructive interference with the SM.}
\label{fig:epsKhist}
\end{figure}
As already discussed in section~\ref{sec:FlavMFPC:flavor_obs}, the meson-antimeson mixing observables, and in particular $\epsilon_K$, can put very strong constraints on models with partial compositeness and an anarchic flavor structure.
Because no flavor symmetries like those discussed in section~\ref{sec:Fermions:flavor} are considered here, these strong constraints are assumed to be present in the MFPC model.
In fact, our numerical analysis has found many of the points that predict correct quark masses and CKM elements to deviate from the measured value of $\epsilon_K$ by orders of magnitude.
However, the constraints strongly depend on the values of the fundamental Yukawa matrices.
We have actually found a significant number of points that lie inside the 3$\sigma$ region around the experimental measured value of $\epsilon_K$.
To get an impression of possible values $\epsilon_K$ can assume in MFPC, the histogram in figure~\ref{fig:epsKhist} shows the NP contributions due to the dimension six operators listed in section~\ref{sec:FlavMFPC:flavor_obs} for a representative subset of all points that predict correct quark masses and CKM elements.
It might be useful to recall that due to NP contributions to charged current semi-leptonic decays, the CKM elements for each point are in general different from those in the SM.
This has the effect that in addition to the NP contribution shown in figure~\ref{fig:epsKhist}, also the SM contribution $\epsilon_K^{\rm SM}$ varies due to varying CKM elements.
This is the main reason why there is actually a significant number of points with a NP contribution as small as $|\epsilon_K^{\rm NP}|=\mathcal{O}(10^{-6})$ that yields values for $\epsilon_K = \epsilon_K^{\rm SM} + \epsilon_K^{\rm NP}$ that are not compatible with the experimental measurement at the 3$\sigma$ level.
Another interesting effect that can be observed in figure~\ref{fig:epsKhist} is that the experimental data prefers positive values for $\epsilon_K^{\rm NP}$ over negative ones.
This can be traced back to the high sensitivity of $\epsilon_K^{\rm SM}$ to the value of the CKM element $V_{cb}$.
The experimental measurement of $\epsilon_K$ actually favors a value for $V_{cb}$ slightly larger than what is suggested by the exclusive charged current semi-leptonic decays $B\to D\ell\nu$ that are included as constraints.
This slight tension can be reduced by a positive NP contribution, i.e.\ $\epsilon_K^{\rm NP}>0$, which leads to the asymmetry visible in figure~\ref{fig:epsKhist}.
The histogram also shows that the NP contributions to $\epsilon_K$ can vary over several orders of magnitude.
This is mainly caused by the fundamental Yukawa matrices entering the WEH WCs in eq.~(\ref{eq:FlavMFPC:WCs_deltaF=2}), while the MFPC-EFT WCs, which are allowed to assume absolute values between 0.1 and 10, only have a minor effect.

\begin{figure}
\centering
\begingroup
\sbox0{\includegraphics{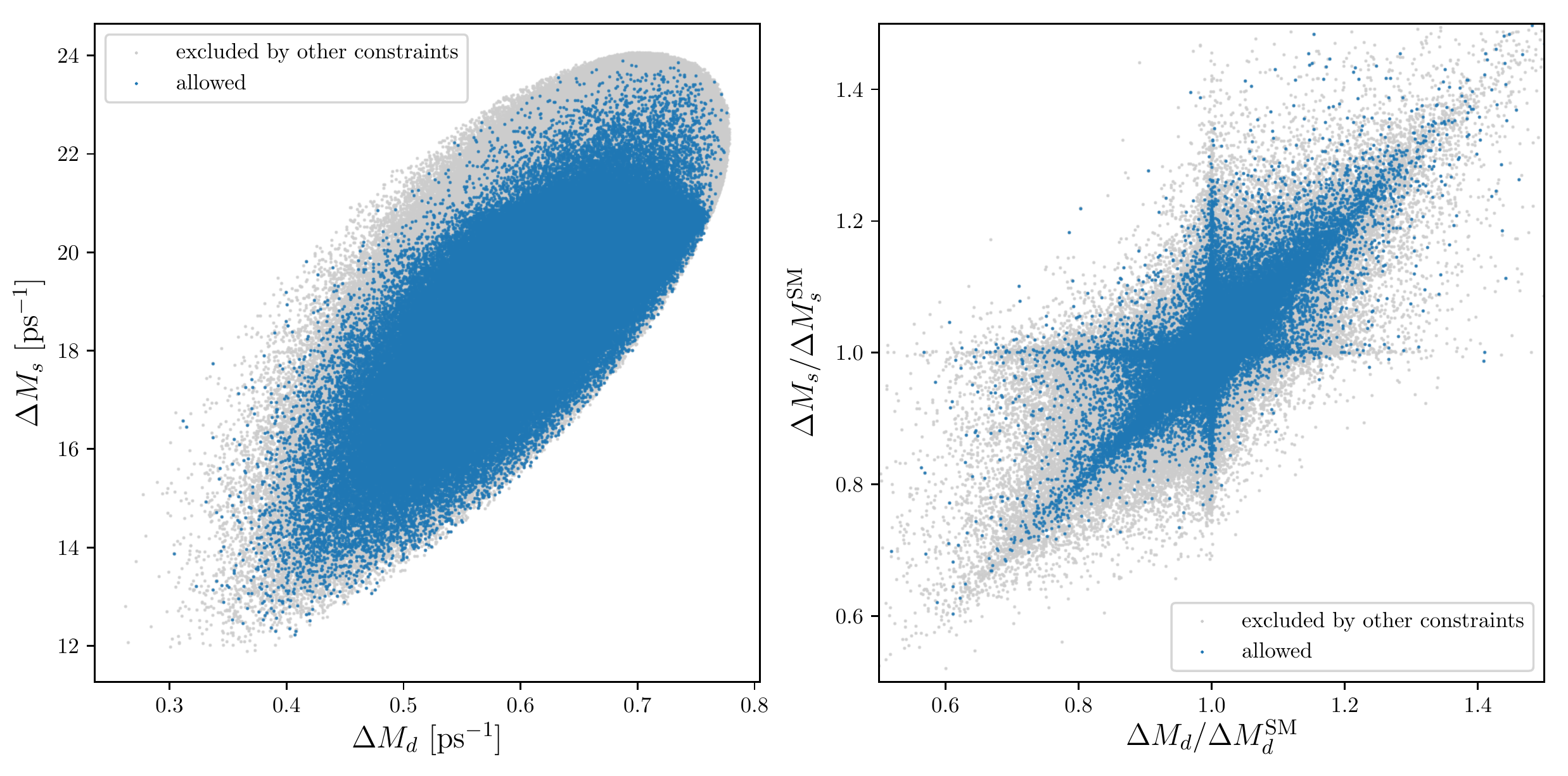}}%
\begin{subfigure}{0.49\textwidth}
\includegraphics[trim={0 0 {0.5\wd0} 0},clip,width=\textwidth]{figures/DeltaMq}
 \caption{}\label{fig:FlavMFPC:DeltaMq_abs}
\end{subfigure}
\begin{subfigure}{0.49\textwidth}
\includegraphics[trim={{0.5\wd0} 0 0 0},clip,width=\textwidth]{figures/DeltaMq}
 \caption{}\label{fig:FlavMFPC:DeltaMq_rel}
\end{subfigure}
\endgroup
\caption{Predictions for $\Delta M_d$ and $\Delta M_s$.
Gray points are excluded by constraints other than $\Delta F=2$.
Blue points are allowed by all constraints.}
\label{fig:DMq}
\end{figure}
In contrast to the kaon sector, a chiral enhancement of the hadronic matrix elements of the left-right operators $O_{SLR}^{ij}$ and $O_{VLR}^{ij}$ is not active in $B^0\text{-}\bar{B}^0$ and $B_s\text{-}\bar{B}_s$ mixing.
The constraints from meson-antimeson mixing observables involving $b$~quarks are therefore much weaker than those stemming from $\epsilon_K$.
However, visible NP effects are still generated.
The predictions of the mass differences $\Delta M_d$ and $\Delta M_s$ in $B^0\text{-}\bar{B}^0$ and $B_s\text{-}\bar{B}_s$ mixing are shown in figure~\ref{fig:FlavMFPC:DeltaMq_abs}.
All points lie in an ellipse corresponding to the 3$\sigma$ range around the experimentally measured values.
In addition to experimental uncertainties, the 3$\sigma$ range also takes into account theoretical uncertainties of the hadronic matrix elements from lattice QCD~\cite{Bazavov:2016nty}.
Only few points are excluded only by $\Delta M_d$ or $\Delta M_s$, which can be seen by the fact that the points allowed by all constraints (blue) do not fill out the whole ellipse.
Most points close to the edge of the ellipse are also excluded by other constraints (gray), i.e.\ $Z$~decays or charged current semi-leptonic decays.
In particular, points with relatively large values of $\Delta M_s$ favor high values of $V_{cb}$, which is however disfavored by the $B\to D\ell\nu$ branching ratios.
In general, one observes that both $\Delta M_d$ and $\Delta M_s$ can be suppressed or enhanced.

Deviations from the SM value in figure~\ref{fig:FlavMFPC:DeltaMq_abs} cannot be attributed solely to $\Delta F = 2$ operators because the CKM elements are not fixed but depend on the WCs of the charged-current operators.
%
To disentangle the different effects, figure~\ref{fig:FlavMFPC:DeltaMq_rel} shows the predictions of $\Delta M_d$ and $\Delta M_s$ relative to the SM values calculated from the predicted CKM elements at each parameter point.
Consequently, figure~\ref{fig:FlavMFPC:DeltaMq_rel} shows effects dominantly due to the $\Delta F = 2$ operators.
In contrast to what one might naively expect from figure~\ref{fig:FlavMFPC:DeltaMq_abs}, one finds relative deviations up to 40\%.
This is possible because large effects from $\Delta F = 2$ operators can be partially compensated by shifts in the CKM elements.
Figure~\ref{fig:FlavMFPC:DeltaMq_rel} also reveals that sizable effects due to $\Delta F = 2$ operators cluster in three regions, where either mostly $\Delta M_d$ is affected, mostly $\Delta M_s$ is affected, or both are affected similarly.

\begin{figure}
\centering
\begingroup
\sbox0{\includegraphics{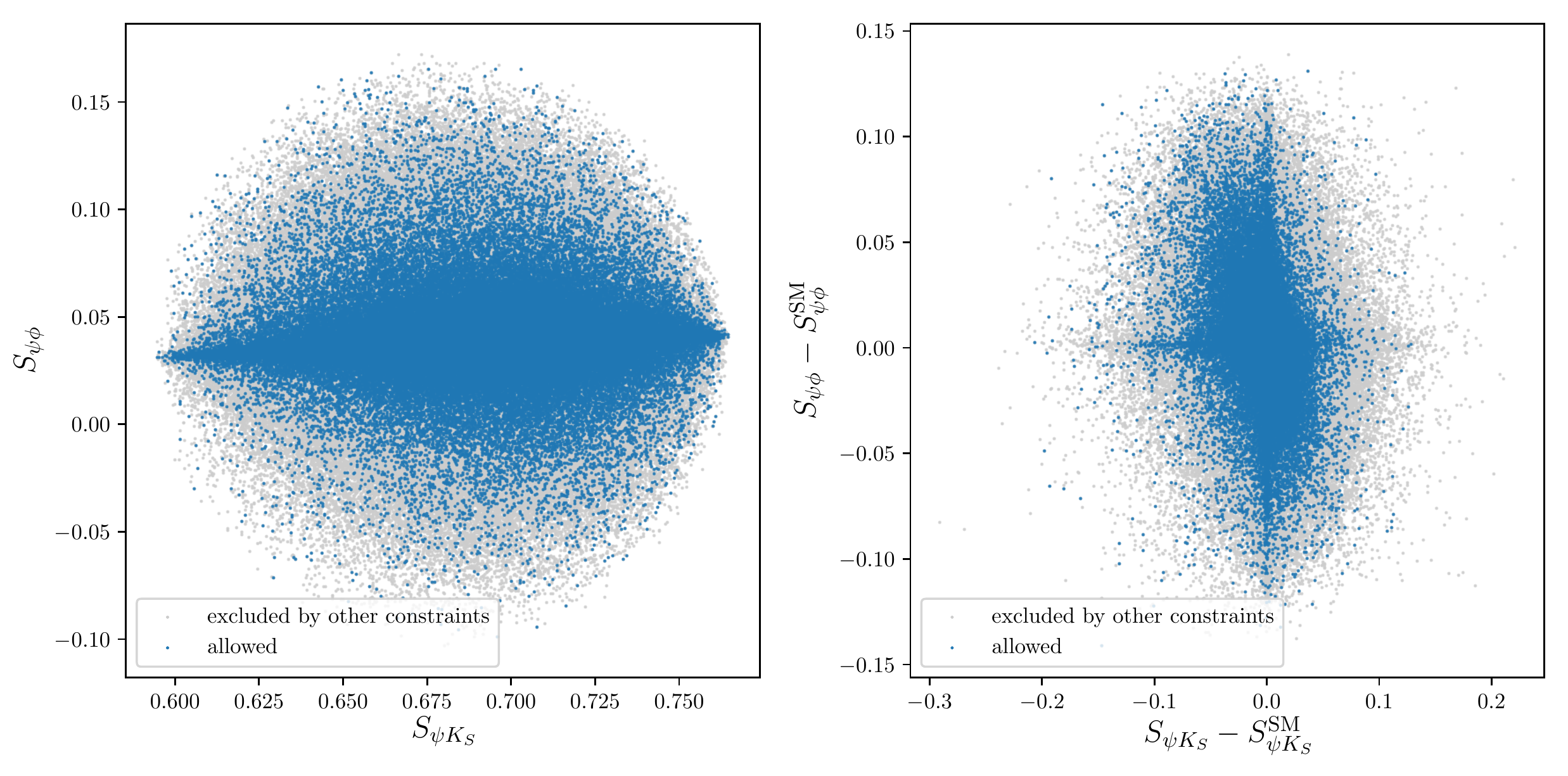}}%
\begin{subfigure}{0.49\textwidth}
\includegraphics[trim={0 0 {0.5\wd0} 0},clip,width=\textwidth]{figures/SpsiX}
 \caption{}\label{fig:FlavMFPC:SpsiX_abs}
\end{subfigure}
\begin{subfigure}{0.49\textwidth}
\includegraphics[trim={{0.5\wd0} 0 0 0},clip,width=\textwidth]{figures/SpsiX}
 \caption{}\label{fig:FlavMFPC:SpsiX_rel}
\end{subfigure}
\endgroup
\caption{Predictions for the mixing induced CP asymmetries
in $B^0\to J/\psi K_S$ and $B_s\to J/\psi\phi$,
sensitive to the $B^0$ and $B_s$ mixing phases.
Gray points are excluded by constraints other than $\Delta F=2$.
Blue points are allowed by all constraints.}
\label{fig:SpsiX}
\end{figure}
Because the WCs of $\Delta F = 2$ operators are in general complex valued, they introduce new $C\!P$-violating phases into the mixing amplitudes.
While these phases do not affect the mass differences $\Delta M_d$ and $\Delta M_s$, they can be probed by the observables $S_{\psi K_S}$ and $S_{\psi \phi}$, which correspond to the mixing induced CP asymmetries
in the decays $B^0\to J/\psi K_S$ and $B_s\to J/\psi\phi$, respectively.
Their predictions are shown in figure~\ref{fig:FlavMFPC:SpsiX_abs}.
Again, one can observe an ellipse corresponding to the 3$\sigma$ range around the experimentally measured values.
The effects in figure~\ref{fig:FlavMFPC:SpsiX_abs} are again due to both $\Delta F = 2$ operators and varying CKM elements.
Figure~\ref{fig:FlavMFPC:SpsiX_rel} shows the differences between the predicted values of the observables $S_{\psi K_S}$ and $S_{\psi \phi}$ and their SM values calculated from the predicted CKM elements at each point.
Thus, the effects in figure~\ref{fig:FlavMFPC:SpsiX_rel} are dominantly due to $\Delta F = 2$ operators.
One observes that deviations of around $0.1$ in either direction are possible for both observables.
Like in the plots of
the mass differences, clusters of points where mostly one of the observables is affected are visible.

\subsection{Charged current decays and lepton flavor universality}
\begin{figure}[t]
\centering
\begingroup
\sbox0{\includegraphics{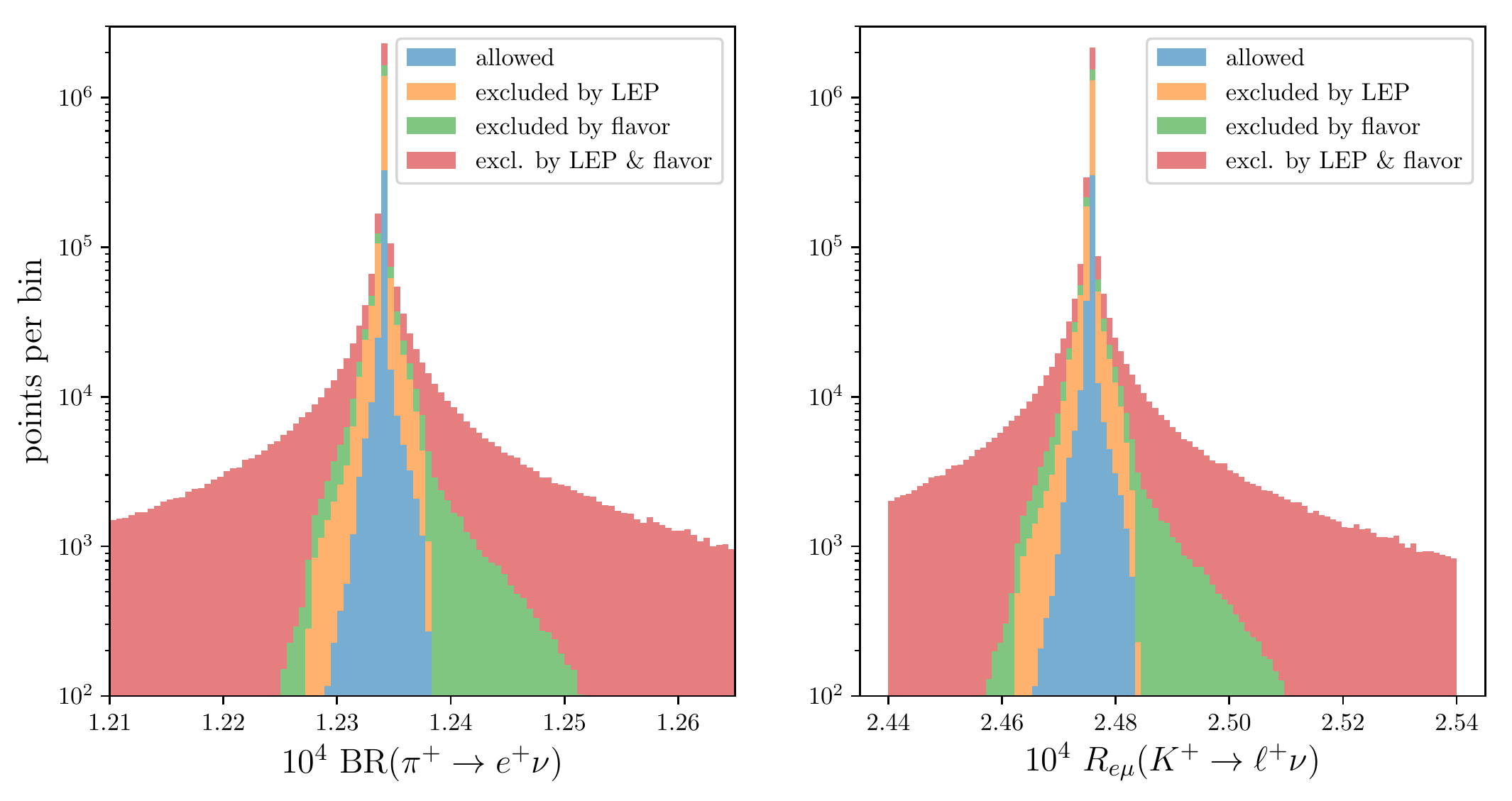}}%
\begin{subfigure}{0.49\textwidth}
\includegraphics[trim={0 0 {0.5\wd0} 0},clip,width=\textwidth]{figures/pienu_Klnu}
 \caption{}
\end{subfigure}
\begin{subfigure}{0.49\textwidth}
\includegraphics[trim={{0.5\wd0} 0 0 0},clip,width=\textwidth]{figures/pienu_Klnu}
 \caption{}
\end{subfigure}
\endgroup
\caption{Histogram showing the distribution of the predictions for
two observables probing $e\text{-}\mu$ universality violation in $Z$ couplings
for all points passing the meson-antimeson mixing constraints.
Points labeled ``excluded by LEP'' are excluded by the partial $Z$ width
measurements at LEP, while points labeled ``excluded by flavor''
are excluded by one of the charged-current decays imposed as constraints.}
\label{fig:FlavMFPC:pienu_Klnu}
\end{figure}
As already discussed in section~\ref{sec:FlavMFPC:flavor_obs}, an explanation of the hints for violation of LFU in neutral currents could also lead to LFU violation in semi-leptonic charged current decays.
Thus, the measurements of $\text{BR}(\pi\to e\nu)$ and
$R_{e\mu}(B\to K\ell\nu)=\text{BR}(K\to e\nu)/\text{BR}(K\to \mu\nu)$ have to be taken into account as important constraints on violation of $e\text{-}\mu$ universality.
Since most WCs of the charged current operators in eqs.~(\ref{eq:FlavMFPC:WCS:CC_4f}) and~(\ref{eq:FlavMFPC:WCS:CC_W}) are actually expected to violate LFU, it is not surprising that our numerical scan finds points that predict values of the two above observables that deviate by far more than 3$\sigma$ from the experimentally measured values.
These deviations are mainly due to the modified $W$-lepton coupling induced by the operator $\mathcal{O}_{\Pi f}$, which enters the WEH WC $C_{V}^{d^iu^j\ell}$ (cf.\ eq.~\eqref{eq:FlavMFPC:WCS:CC_W}).
This can be understood as follows.
Since the above observables are based on the $u\to d\ell\nu$ and $s\to u\ell\nu$ transitions, they involve light quarks.
The contributions to the WEH stemming from the MFPC-EFT four-fermion operators depend on the small fundamental Yukawa couplings of these light quarks and are thus strongly suppressed.
On the other hand, the contribution due to the modified $W$-lepton coupling involves the SM $W$-quark coupling and is not suppressed.
For parameter points that feature LFU violation from modified $W$ couplings, the $\SUs{2}{L}$ symmetry of the MFPC-EFT implies also LFU violation in $Z$ couplings, which is constrained by the LEP measurements of the $Z$ partial widths (cf.\ section~\ref{sec:FlavMFPC:Constr_EW}).
In figure~\ref{fig:FlavMFPC:pienu_Klnu}, the different constraints on $e\text{-}\mu$ universality are compared.
The histograms show all points that pass the meson-antimeson mixing constraints and divide them into four categories: points that are excluded by the charged current flavor observables used as constraints, points excluded by LEP, points excluded by LEP and flavor observables, and points that are allowed by all constraints.
While many points are excluded by both LEP and flavor observables, neither the LEP nor the flavor constraints are superior to the other.
Rather, there are points that are only excluded by either LEP or flavor constraints, such that they actually complement one another.
The resulting combined constraint is found to be at the per cent level.

\begin{figure}[t]
\centering
\includegraphics[width=0.6\textwidth]{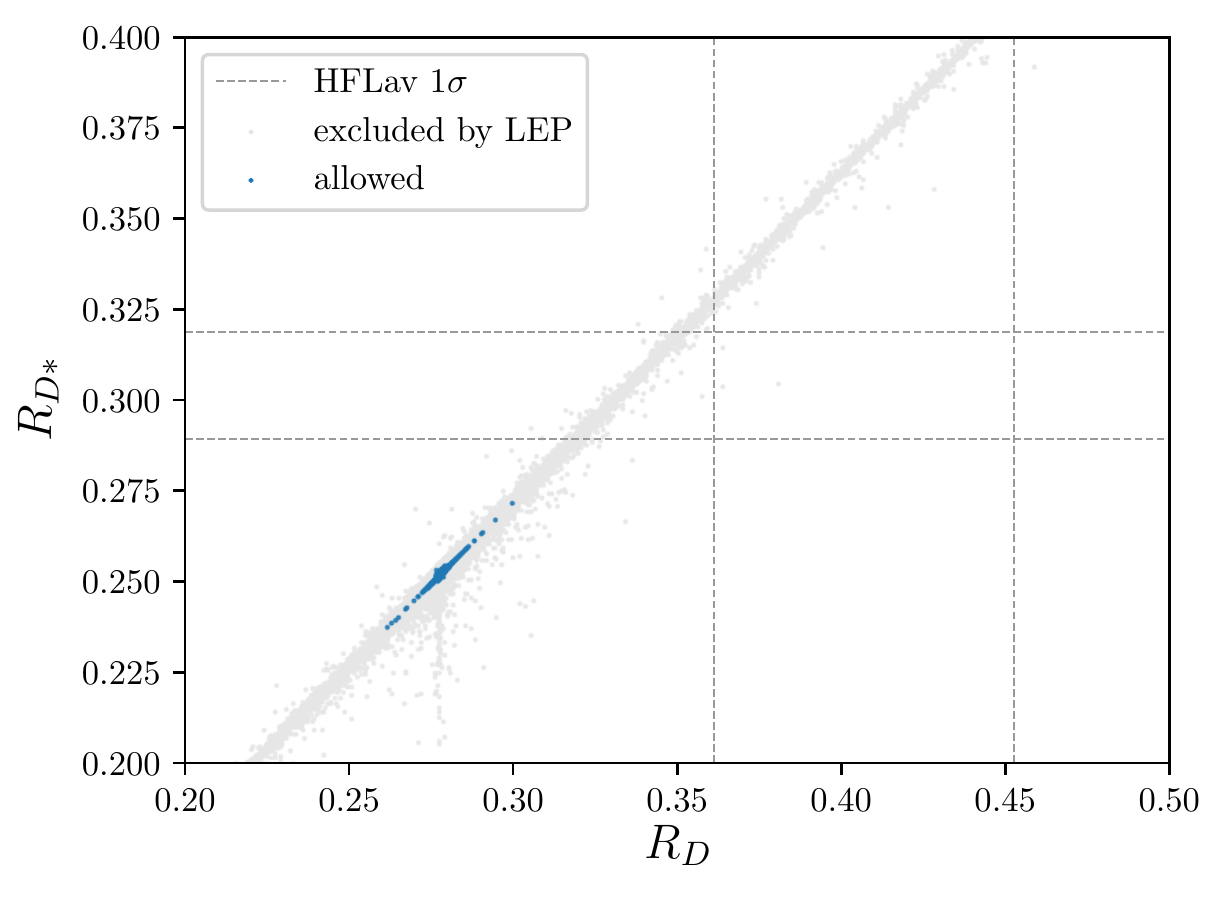}
\caption{Predictions for lepton flavor universality tests in $B\to D\tau\nu$ and $B\to D^*\tau\nu$ compared to the experimental world averages.
Points allowed by all constraints are shown in blue, while points excluded by LEP $Z$ pole constraints are shown in gray.}
\label{fig:FlavMFPC:RD}
\end{figure}
While the $B\to D^{(\ast)}\tau\nu$ decays based on the $b\to c\tau\nu$ transition are experimentally more challenging than
charged current decays with electrons or muons in the final state,
they allow for testing LFU in decays involving tau leptons.
Measurements by BaBar, Belle, and LHCb~\cite{Lees:2013uzd,Huschle:2015rga,Aaij:2015yra,Sato:2016svk,Hirose:2016wfn,Aaij:2017uff} of the ratios
\begin{equation}
R_{D^{(\ast)}} = \frac{\Gamma(B\to D^{(\ast)}\tau\nu)}{\Gamma(B\to D^{(\ast)}\ell\nu)}
\end{equation}
have actually shown deviations from the SM prediction at a combined level of 4$\sigma$~\cite{Amhis:2016xyh}.
Given the smallness of the theoretical uncertainties of the SM prediction as well as a possible connection to the hints for LFU violation in neutral currents, this is an intriguing result.
%
To assess if it is possible to explain the deviations in the MFPC model, the ratios $R_{D^{(\ast)}}$ are considered as predictions in our numerical analysis.
Figure~\ref{fig:FlavMFPC:RD} shows these predictions in the $R_{D}\text{-}R_{D^{\ast}}$ plane.
Although the MFPC model actually predicts a positive correlation between the deviations in $R_{D}$ and $R_{D^{\ast}}$ and such a pattern is also suggested by the experimental data, effects that would be large enough to be in agreement with the averaged measurements at the 1$\sigma$ level are excluded by LEP constraints.
This can be traced back to the fact that sizable contributions to $R_{D}$ and $R_{D^{\ast}}$ require a large fundamental Yukawa coupling of left-handed tau leptons.
This in turn modifies the $Z\tau_L \tau_L$ coupling (cf.\ eq.~\eqref{eq:FlavMFPC:Z_couplings}), which is constrained by LEP.
The parameter points that are excluded by LEP data are shown in gray in figure~\ref{fig:FlavMFPC:RD}.
While some of the blue points, which pass all constraints, can slightly reduce the tension with experiment compared to the SM, the corresponding effects are much too small to actually explain the measured values.
It is an open question if the MFPC model can be modified such that the $Z\tau_L \tau_L$ coupling is protected while the $R_{D}$ and $R_{D^{\ast}}$ measurements can be explained.

\subsection{Lepton flavor universality in neutral current decays}
\begin{figure}[t]
\centering
\begingroup
\sbox0{\includegraphics{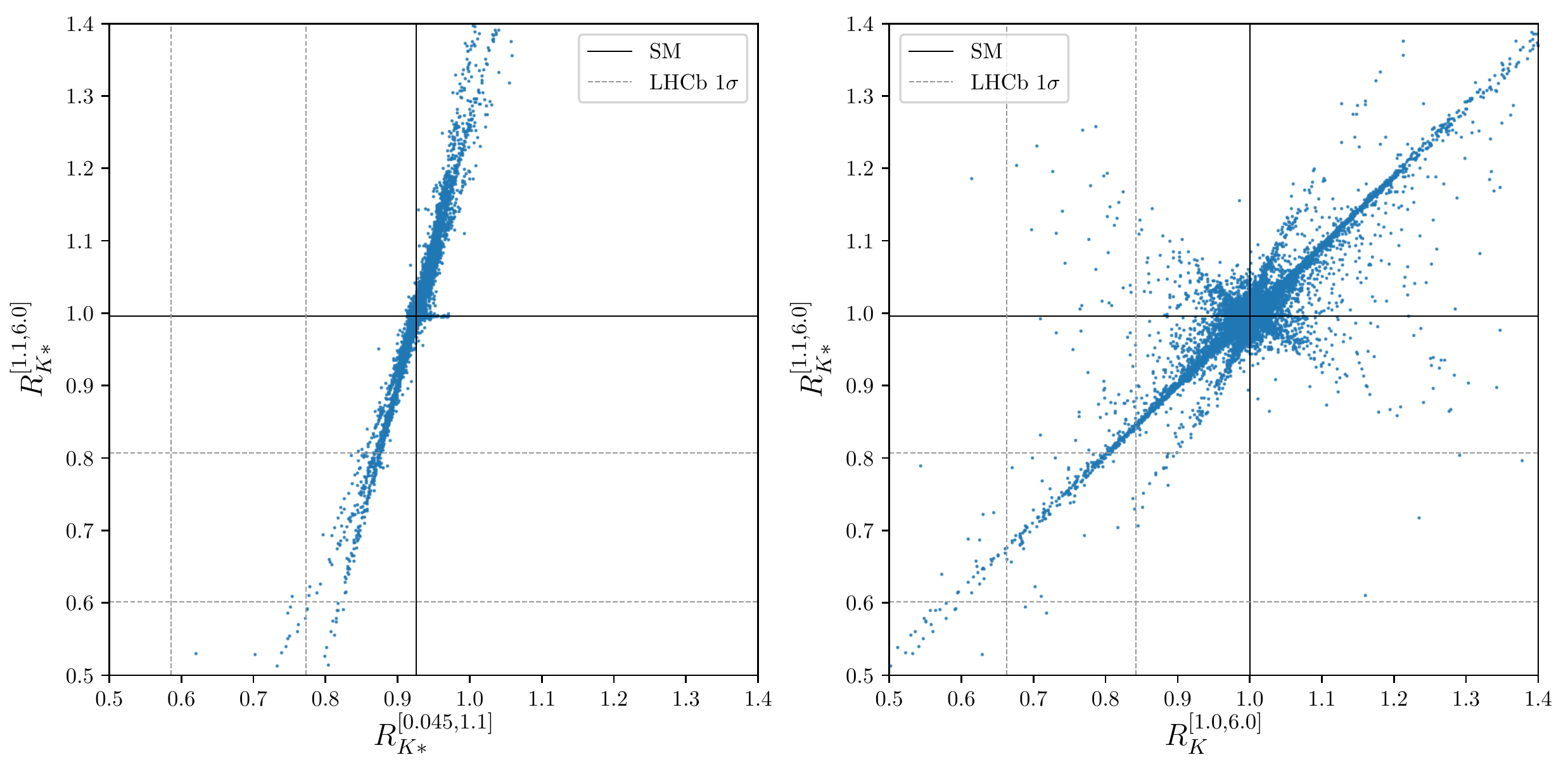}}%
\begin{subfigure}{0.48\textwidth}
\includegraphics[trim={{0.5\wd0} 0 0 0},clip,width=\textwidth]{figures/RK_RKstar}
 \caption{}\label{fig:FlavMFPC:RK_RKstar_scan}
\end{subfigure}
\begin{subfigure}{0.48\textwidth}
\includegraphics[width=1.04\textwidth]{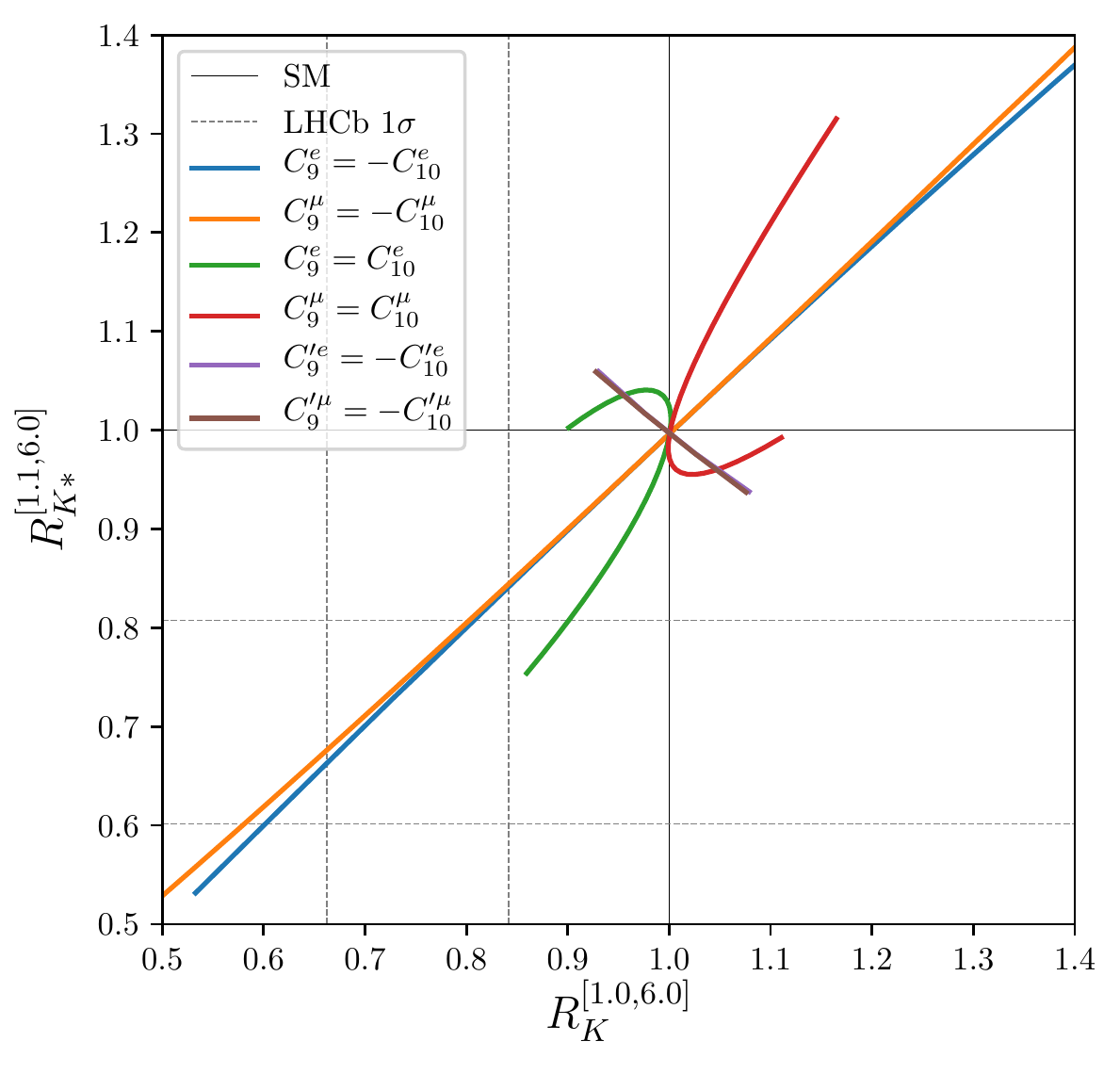}
 \caption{}\label{fig:FlavMFPC:RK_RKstar_explain}
\end{subfigure}
\endgroup
\caption{Predictions of $R_K$ for $q^2\in[1.0,6.0]$ and $R_{K^*}$ for $q^2\in[1.1,6.0]$
compared to the SM prediction and the LHCb measurements.
(a): The points found in our numerical analysis that are allowed by all of the applied constraints.
(b): Generic predictions for different scenarios of NP contributions to the WCs $C_{9}^{(\prime)\ell}$ and $C_{10}^{(\prime)\ell}$.
The unprimed WCs $C_{9}^{\ell}$ and $C_{10}^{\ell}$ corresponding to left-handed quark currents are varied between $-1.5$ and $1.5$, while the primed WCs $C_{9}^{\prime\ell}$ and $C_{10}^{\prime\ell}$ corresponding to right-handed quark currents are varied between $-0.15$ and $0.15$.
In the MFPC model, the latter are suppressed by relatively small fundamental Yukawa couplings (see text for details).
}
\label{fig:FlavMFPC:RK_RKstar}
\end{figure}

\begin{figure}[t]
\centering
\begingroup
\sbox0{\includegraphics{figures/RK_RKstar}}%
\begin{subfigure}{0.48\textwidth}
\includegraphics[trim={0 0 {0.5\wd0} 0},clip,width=\textwidth]{figures/RK_RKstar}
 \caption{}\label{fig:FlavMFPC:RKstar_scan}
\end{subfigure}
\begin{subfigure}{0.48\textwidth}
\includegraphics[width=1.04\textwidth]{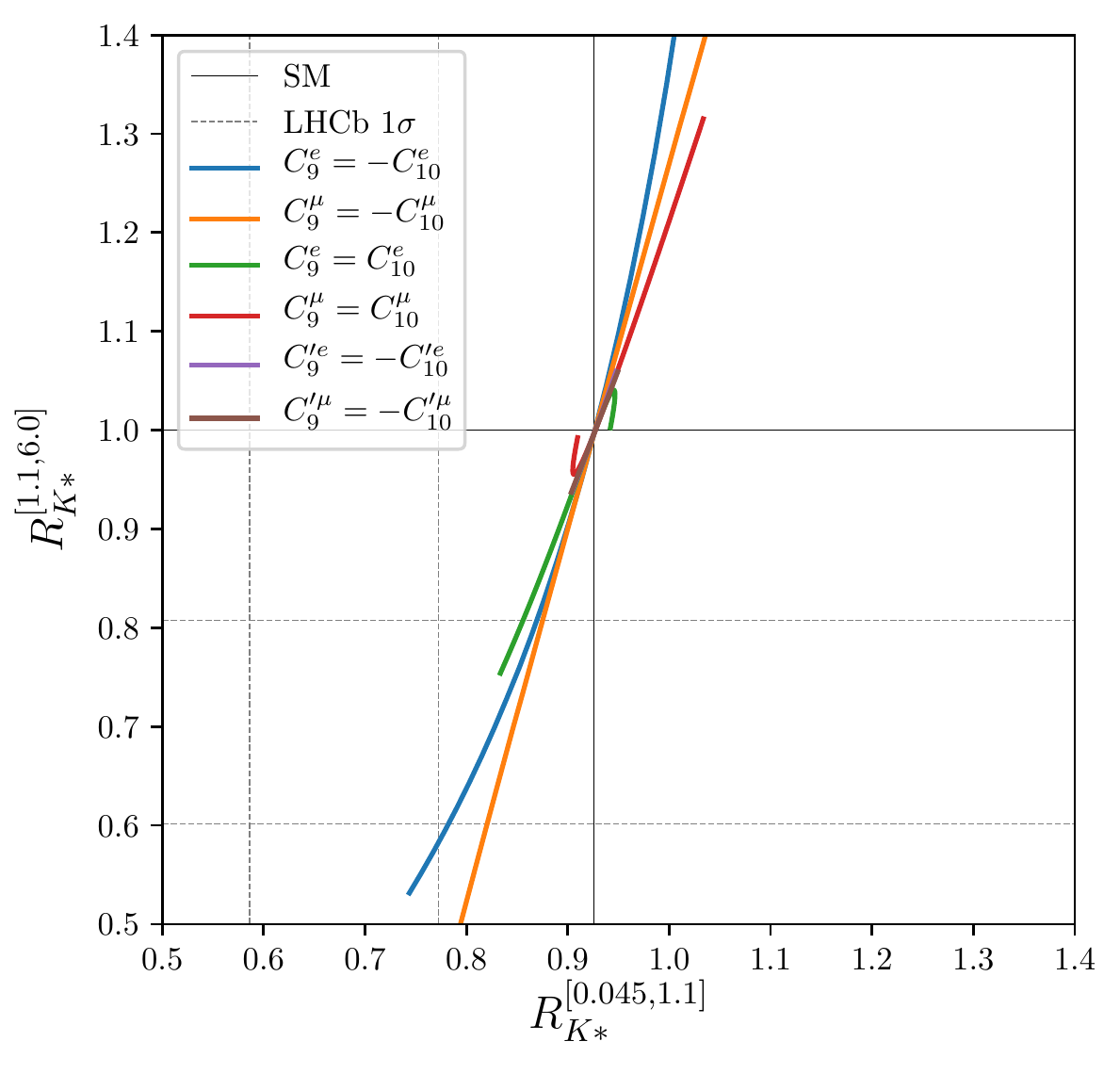}
 \caption{}\label{fig:FlavMFPC:RKstar_explain}
\end{subfigure}
\endgroup
\caption{Predictions of $R_{K^*}$ for $q^2\in[0.045,1.1]$ and $R_{K^*}$ for $q^2\in[1.1,6.0]$
compared to the SM prediction and the LHCb measurements.
(a): The points found in our numerical analysis that are allowed by all of the applied constraints.
(b): Generic predictions for different scenarios of NP contributions to the WCs $C_{9}^{(\prime)\ell}$ and $C_{10}^{(\prime)\ell}$.
Concerning the size of the WCs $C_{9}^{(\prime)\ell}$ and $C_{10}^{(\prime)\ell}$, the same comments as in figure~\ref{fig:FlavMFPC:RK_RKstar} apply.
}
\label{fig:FlavMFPC:RKstar}
\end{figure}

Since a possible explanation of LFU violation in $R_{D}$ and $R_{D^{\ast}}$ is spoiled by LEP constraints, one might expect something similar to happen to an explanation of LFU violation in the neutral current observables $R_K$ and $R_{k^*}$.
While it is demonstrated in chapter~\ref{chap:LUFV_in_CHMs}, that models with partial compositeness provide a mechanism that allows for explaining the $R_{k^{(*)}}$ anomaly, the simple model presented there features a custodial protection of the $Z\mu_L\mu_L$ coupling.
This allows for a sizable degree of compositeness of left-handed muons while satisfying LEP constraints, which in turn can explain the anomaly.
Translated to the MFPC model, an analogous mechanism would require a sizable fundamental Yukawa coupling of the left-handed muon.
In this case, however, the operator $\mathcal{O}_{\Pi f}$ induces a modification of the $Z\mu_L\mu_L$ coupling, which is not custodially protected (cf.\ eq.~\eqref{eq:FlavMFPC:Z_couplings}).
While at first sight, this is very similar to the problem with the $Z\tau_L \tau_L$ coupling in the explanation of the $R_{D}$ and $R_{D^{(\ast)}}$ measurements, it turns out that the LFU violating effects in neutral currents can actually be large enough to explain $R_K$ and $R_{k^*}$ and simultaneously pass the constraints from LEP measurements.
This is demonstrated in figures~\ref{fig:FlavMFPC:RK_RKstar} and~\ref{fig:FlavMFPC:RKstar}, where the predictions of $R_{K}$ and $R_{K^*}$ in the bins measured by LHCb are shown for points that pass all constraints imposed in our numerical analysis.
One observes that sizable effects are possible.
These effects actually predict the measured positive correlation between $R_{K}$ and $R_{K^*}$, while effects in the orthogonal direction in the $R_{K}\text{-}R_{K^*}$ plane are considerably smaller.

This can be understood as follows.
To yield the large top quark mass, the fundamental quark doublet Yukawa coupling has to be sizable for the third generation.
The hierarchy between the masses of top and bottom quark is then mainly generated by the fundamental quark singlet Yukawa couplings.
This implies that the fundamental doublet Yukawa coupling of the bottom quark is usually much larger than the one for the singlet.
A similar conclusion can be drawn for the second generation, where also the hierarchy between the charm and the strange quark is generated mainly by the fundamental quark singlet Yukawa couplings.
Since WCs of operators with a left-handed quark current depend on the fundamental doublet Yukawa couplings and WCs of operators with a right-handed quark current depend on the fundamental singlet Yukawa couplings, the unprimed WCs $C_{9}^\ell$ and $C_{10}^\ell$ usually receive considerably larger contributions than the primed WCs $C_{9}^{\prime\ell}$ and $C_{10}^{\prime\ell}$ (cf.\ eq.~(\ref{eq:FlavMFPC:WCS:bsll_4f}))\footnote{%
The suppression of right-handed currents involving only bottom and strange quarks is a general feature in partial compositeness models with an anarchic flavor structure.
This is analytically shown for $\Delta F=2$ operators in~\cite{Barbieri:2012tu}.
}.
This then yields the pattern in figure~\ref{fig:FlavMFPC:RK_RKstar}.

Next to the results from the numerical analysis in figures~\ref{fig:FlavMFPC:RK_RKstar_scan} and~\ref{fig:FlavMFPC:RKstar_scan}, predictions for several scenarios of NP contribution to the WCs $C_{9}^{(\prime)\ell}$ and $C_{10}^{(\prime)\ell}$ are shown.
%
In particular, figure~\ref{fig:FlavMFPC:RK_RKstar_explain} demonstrates that sizable contributions with a positive correlation of $R_{K}$ and $R_{K^*}$ can be achieved with NP in WCs of operators involving left-handed quark currents with $|C_{9}^{\ell}|=|C_{10}^{\ell}|\lesssim1.5$.
Assuming $|C_{9}^{\prime\ell}|=|C_{10}^{\prime\ell}|\lesssim0.15$ to take into account the suppression of right-handed quark currents, one finds only very small effects.
The points found in our numerical analysis that predict either sizable negative $C_{9}^{\mu}=-C_{10}^{\mu}$ or sizable positive $C_{9}^{e}=-C_{10}^{e}$ are compatible with the LHCb measurements at the 1-2$\sigma$ level.
While some of the points with negative $C_{9}^{e}=C_{10}^{e}$ are also in the region preferred by experimental data, sizable effects from right-handed electron currents require considerably larger WCs compared to left-handed lepton currents (cf.\ section~\ref{sec:anomalies:LFUV:NP_in_one_two_WCs}).
Points with a sizable negative $C_{9}^{\mu}=-C_{10}^{\mu}$, i.e. effects in left-handed muon currents, are also in good agreement with the global fits of $b\to s\mu\mu$ observables (cf.\ section~\ref{sec:anomalies:bsmumu}).
Consequently, our numerical analysis actually finds parameter points that are able to explain the hints for LFU violation in $R_K$ and $R_{k^*}$ as well as the $b\to s\mu\mu$ anomaly, while at the same time satisfying all imposed constraints.
\nowidow[4]

\chapter{Summary and Outlook}
\label{chap:concl}
The Higgs sector in the SM has a naturalness problem.
The observed Higgs mass requires an enormous fine-tuning without NP not too far above the EW scale that protects it from large quantum corrections.
An elegant solution to this problem is offered by CHMs:
if the Higgs is a composite bound state of a new strongly interacting sector, it only forms at energies below the new strong scale and cannot be plagued by quantum corrections at higher scales.
Models in which the Higgs is a pNGB even provide an explanation for a Higgs that is much lighter than other composite bound states that have to be present in such a model.
This is an important property since so far only the Higgs has been discovered at the LHC and nothing else.
However, there are some hints for NP from measurements of flavor observables that show a deviation from the SM predictions.
If these hints should turn into a discovery, then the NP that explains them has to be at a scale also not too far above the EW scale.
This suggests that this NP might be actually connected to the solution of the naturalness problem.
Or, stated differently, the solution of the naturalness problem might explain the hints for NP.
%
In view of this, it is an interesting question if composite Higgs models can actually do both, i.e.\ solve the naturalness problem of the SM and at the same time explain the hints for NP.
In this case, however, also other signs of the new strongly interacting sector are expected.
In particular, bound states in addition to the composite Higgs might be in reach of direct searches at the LHC.
So another interesting question is how are the prospects for observing or excluding composite Higgs models through direct searches.
The work done in this thesis provides some answers to both of these interesting questions.

The first part introduces composite Higgs models that feature a pNGB Higgs and the partial compositeness mechanism to generate masses for the SM fermions.
This introduction is presented in chapter~\ref{chap:CHMs} and considers the following concepts:
\begin{itemize}
 \item NGBs are discussed in detail. After starting with a concrete example, the formalism to describe them is introduced and important special cases are considered.
 These NGBs can describe light spin zero composite bound states formed after spontaneous symmetry breaking in a strongly interacting theory.
 They are eventually used to describe the composite Higgs.
 \item Hidden local symmetries are discussed as an alternative view of the formalism that describes NGBs.
 \item It is shown how these hidden local symmetries can be used to add vector resonances to a model that initially only contained NGBs.
 These vector resonances can be interpreted as spin one bound states of the strongly coupled theory.
 \item It is described how the hidden local symmetry construction can be further extended to add an arbitrary number of levels of vector resonances.
 This yields a so called multi-site moose model.
 \item It is shown how taking a continuum limit that corresponds to adding an infinite number of resonances leads to an extra dimensional theory that can be interpreted as a dual of the strongly coupled one.
 \item Fermion partial compositeness is introduced as a means to provide masses to the SM fermions in an effective description of a strongly coupled theory.
 \item It is shown how adding fermions in the extra dimensional theory leads to fermion resonances when the continuum limit is reversed by discretizing the extra dimension and that these fermion resonances automatically implement the partial compositeness mechanism in a multi-site moose model.
 \item The quark flavor structure in a model with partial compositeness is discussed and flavor symmetries are introduced to avoid stringent bounds from flavor observables.
 \item The Mechanism of electroweak symmetry breaking by vacuum misalignment is described in detail by way of concrete example.
 \item The effective radiatively generated potential responsible for the misalignment mechanism is described.
 It is shown that this potential can be finite in multi-site moose models due to a mechanism called collective breaking.
 \item The MFPC model is introduced as a UV completion of the effective models discussed so far. This model features both a pNGB Higgs and fermion partial compositeness.
 \item It is described how an effective low-energy description of the MFPC model, the MFPC-EFT, can be constructed.
\end{itemize}
The remainder of the thesis applies the concepts listed above in  several phenomenological studies.

Chapter~\ref{chap:direct_constraints} discusses direct collider constraints in CHMs in the context of comprehensive numerical global analyses.
After describing the numerical method and the considered constraints, the concrete models that have been analyzed are specified.
These are two multi-site moose models containing a pNGB Higgs as well as vector and fermion resonances. They implement fermion partial compositeness, flavor symmetries in the quark sector, and radiative EWSB by vacuum misalignment due to an effective potential that is finite by collective breaking.
One of the models, the MCHM, features NGBs in the minimal $\SO{5}/\SO{4}$ coset, while the other, the NMCHM, contains NGBs in the next-to-minimal $\SO{6}/\SO{5}$ coset.
The implementation of direct collider constraints in the numerical code used in the analyses is detailed.
In particular, all expressions used for the calculations of cross sections and branching ratios are given.
The results of the analyses are discussed.
In particular, the cross sections and branching ratios predicted by viable parameter points that satisfy all constraints are compared to experimental data.
Important conclusions are:
\begin{itemize}
 \item Experimental searches for quark resonances at LHC run 2 can probe nearly all of the viable parameter points we have found.
 \item Decays with light SM quarks in the final state are virtually unconstrained at the moment but can presumably be probed by analyzing existing data.
 The most promising decay channels to do this have a light SM quark and a Higgs boson in the final state.
 \item While vector resonances can be too heavy to be probed by LHC run 2, many of the viable parameter points we have found can be probed by near future analyses.
 There are two distinct cases:
 \begin{itemize}
  \item If the resonance $X_\mu$ is the lightest vector resonance, it can be lighter than the naive bound from the electroweak $S$ parameter suggests. It dominantly decays to $t\bar{t}$ but can also be probed in the dilepton channel.
  In this case, both of these decay channels have high prospects for observing or excluding viable parameter points.
  \item If $\rho_{L\,\mu}^3$ is lighter than $X_\mu$, the diboson as well as the dilepton channel have the highest prospects to probe the viable parameter points.
 \end{itemize}
 \item If mixing is allowed in the scalar sector of the NMCHM, the scalar resonance $\eta$ has couplings very similar to the Higgs and a mass usually below 1~TeV.
 The main features of its collider phenomenology are:
 \begin{itemize}
 \item $\eta$ is dominantly produced via gluon fusion; its hadronic cross section is suppressed compared to the Higgs cross section due to the larger mass and the consequently smaller gluon parton luminosity.
 \item If $\eta$ is heavier than roughly $250$~GeV, it dominantly decays to two Higgses.
 Direct searches for a neutral scalar decaying in this channel have by far the highest prospects for observing or excluding viable parameter points.
 \item The diboson channels are also promising, especially if $\eta$ is very light and kinematically forbidden to decay to two Higgses.
 \end{itemize}
\end{itemize}

The hints for NP found in measurements of rare $B$ decays are discussed in detail in chapter~\ref{chap:anomalies}.
In particular, two model independent studies of the tensions between experimental measurements and SM predictions are presented.
The first one analyzes the status of the $B\to K^*\mu^+\mu^-$ anomaly including new and updated measurements.
It is found that
\begin{itemize}
 \item A good fit is obtained with a negative NP contribution to the WC $C_9$, possibly accompanied by a positive contribution to $C_{10}$.
 \item NP in WCs of right-handed quark currents cannot explain the tensions.
 \item It is shown that increasing possibly underestimated hadronic uncertainties cannot fully account for the tensions.
 \item The measured data is compatible with a helicity and $q^2$ independent shift in $C_9$, suggesting a NP explanation.
\end{itemize}
The second study considers hints for violation of LFU from measurements of the theoretically very clean observables $R_{K^{(*)}}$.
It is found that
\begin{itemize}
 \item A NP contribution with $C_9^\mu - C_9^e - C_{10}^\mu + C_{10}^e \approx -1.4 ~$ provides a good fit to the data.
 \item NP in WCs of right-handed quark currents cannot explain the tensions.
 \item A NP explanation of the tensions found in LFU observables is fully compatible with an explanation of the $B\to K^*\mu^+\mu^-$ anomaly if NP yields a negative contribution to $C_9^{\mu}$, possibly accompanied by a positive contribution to $C_{10}^{\mu}$.
 \item Measurements of the LFU observables $D_{P_4^\prime}$ and $D_{P_5^\prime}$ could further distinguish between different NP scenarios.
\end{itemize}

A possible explanation of the tantalizing hints for NP in rare B decays is presented in chapter~\ref{chap:LUFV_in_CHMs}.
It is shown that partially composite left-handed muons can potentially explain both the $B\to K^*\mu^+\mu^-$ anomaly and the hints for LFU by generating negative contributions to $C_9^{\mu}=-C_{10}^{\mu}$.
A very simple model constructed to provide partially composite left-handed muons predicts:
\begin{itemize}
 \item A positive NP contribution to the $B_s$ meson mass difference $\Delta M_s$.
 \item A negative shift in the Fermi constant accompanied by a negative shift in the electroweak $T$ parameter.
 \item A slight reduction of the effective number of light neutrino species $N_\nu$, which is actually suggested by LEP data.
 \item Spin one resonances with sizable branching ratio into muons, but possibly too heavy to be directly observed at the LHC.
\end{itemize}

A much more ambitious model is analyzed in chapter~\ref{chap:Flavor_MFPC}.
A UV completion of CHMs in the form of the MFPC model is considered.
A numerical comprehensive study of the effects of this model on observables at the electroweak scale and on flavor observables at low energies is presented.
Using the MFPC-EFT, the possible contributions to observables at the electroweak scale is described.
For the low-energy flavor phenomenology, the MFPC-EFT is matched to the WEH and the possible contributions to observables in meson-antimeson mixing, semi-leptonic neutral current decays and semi-leptonic charged current decays are detailed.
The constraints applied in the numerical analysis are specified and the numerical strategy is described in detail.
The results found in this study are:
\begin{itemize}
 \item Indirect $C\!P$ violation in Kaon mixing provides a very strong constraint and the experimental measurement is in conflict with predictions in large parts of the parameter space.
 However, also a large number of viable parameter points is found that is in accordance with experimental data.
 \item The viable parameter points allow for sizable effects in $B^0$  and $B_s$ mixing observables close to the level probed by current experiments.
 \item Even though the absence of charged current flavor violation is imposed, violation of LFU is a generic prediction due to partial compositeness.
 Consequently, tests of $e$-$\mu$ universality violation in charged current decays are important constraints.
 \item Large LFU violation in $B\to D^{(\ast)}\tau\nu$, which is indicated by several experiments at the combined level of 4$\sigma$, cannot by explained by the model while satisfying LEP constraints on the $Z\tau\tau$ coupling.
 However, the tensions can be slightly ameliorated compared to the SM.
 \item Interestingly, the MFPC model can explain both anomalies in rare $B$ decays, i.e.\  the $B\to K^*\mu^+\mu^-$ anomaly and the hints for LFU in $R_{K^{(*)}}$.
 The explanation of both anomalies at once requires a sizable partial compositeness of left-handed muons, very similar to the mechanism discussed in chapter~\ref{chap:LUFV_in_CHMs}.
\end{itemize}

In view of the tensions in rare $B$ decays, but also in charged current LFU observables containing the $\tau$, it is a very exciting time for doing flavor physics.
The question if these hints for NP actually turn out to be first signs of a sector beyond the SM will be answered during the following years by measurements both performed by LHCb and at the upcoming Belle~2 experiment.
If they should actually confirm the long sought NP effects, it might also be possible to probe the currently still hypothetical NP sector by direct searches at the LHC.
While no direct effects have been observed so far, this might change with a substantially larger integrated luminosity over the forthcoming years.
If only indirect effects are seen and no direct detection is made, this might hint to relatively heavy NP particles that are strongly coupled, such that they still can produce sizable effects in non-renormalizable operators in the low-energy effective theory.
Composite Higgs models are among the prime candidates for describing these strongly coupled particles.

\begin{appendix}
\chapter{Appendix}

\section{Explicit vacuum states}\label{sec:vacuum_states}
While according to the CCWZ formalism, the NGB Lagrangian depends only on the $G/H$ coset parametrized by the NGB matrix $U(x)$ and not on the vacuum $\phi_0$, one may use the field
\begin{equation}
 \phi(x)=U(x)\,\phi_0
\end{equation}
to write down a Lagrangian that is equivalent to the CCWZ prescription.
It is assumed here that the vacuum that breaks $G\to H$ is parametrized by a fundamental%
\footnote{Constructions where the $G\to H$ breaking is parametrized by larger representations have been discussed e.g.\ in \cite{Bertuzzo:2012ya,Katz:2005au,Gripaios:2009pe,ArkaniHamed:2002qy,Chang:2003zn,Low:2002ws}.}
representation of $G$ (cf.\ e.g.\ \cite{Agashe:2004rs,Schmaltz:2004de,Chala:2012af,Bertuzzo:2012ya}).
For an $N\times N$ NGB matrix, the vacuum $\phi_0$ is thus considered to be an $N$-component vector.
Following \cite{Bertuzzo:2012ya}, two possible terms are found at leading order in derivatives,
\begin{equation}
 \mathcal{L}_2
 = c_1\, \Big(\partial_\mu \phi^\dagger(x)\Big)\Big(\partial^\mu \phi(x)\Big)
 + c_2\, \phi^\dagger(x)\,\Big(\partial_\mu \phi(x)\Big)\Big(\partial^\mu \phi^\dagger(x)\Big)\,\phi(x),
\end{equation}
where $c_1$ and $c_2$ are constants that depend on the $G/H$ coset and on the specific vacuum $\phi_0$.
Rewriting the above Lagrangian in terms of the NGB matrix $U(x)$ and $\phi_0$ yields
\begin{equation}
 \mathcal{L}_2
 = c_1\, \phi_0^\dagger\,\Big(\partial_\mu U^{-1}(x)\Big)\Big(\partial^\mu U(x)\Big)\,\phi_0
 + c_2\, \phi_0^\dagger\,U^{-1}(x)\,\Big(\partial_\mu U(x)\Big)\,\phi_0\,\phi_0^\dagger\,\Big(\partial^\mu U^{-1}(x)\Big)\,U(x)\,\phi_0.
\end{equation}
To see how this Lagrangian is equivalent to the CCWZ Lagrangian, eq.~(\ref{eq:CCWZ_L_2}), one can employ the relation from eq.~(\ref{eq:CCWZ_partial_UU}) to rearrange the $U(x)$ matrices such that the Lagrangian can be written in terms of the Maurer-Cartan-form $a_\mu[U]$ defined in eq.~(\ref{eq:CCWZ_Maurer_Cartan_form}).
It then reads
\begin{equation}
 \mathcal{L}_2
 = c_1\, \phi_0^\dagger\,a_\mu[U]\,a^\mu[U]\,\phi_0
 + c_2\, \phi_0^\dagger\,a_\mu[U]\,\phi_0\,\phi_0^\dagger\,a^\mu[U]\,\phi_0.
\end{equation}
Using the expansion of $a_\mu[U]$ in terms of the unbroken and broken generators, eq.~(\ref{eq:CCWZ_a_mu_expansion}), and noting that by definition the vacuum transforms trivially under the group elements associated with the unbroken generators, i.e.\ $T^a\,\phi_0=0$, one finds
\begin{equation}
 \mathcal{L}_2
 =
 d_\mu[U]^a\,d^\mu[U]^b
 \left(
 c_1\, \phi_0^\dagger\,X^a\,X^b\,\phi_0
 + c_2\, \phi_0^\dagger\,X^a\,\phi_0\,\phi_0^\dagger\,X^b\,\phi_0
 \right).
\end{equation}
Equating this with the NGB Lagrangian from the CCWZ prescription, eq.~(\ref{eq:CCWZ_L_2}), yields
\begin{equation}\label{eq:CCWZ_expl_vac_c1_c2}
 c_1\, \phi_0^\dagger\,X^a\,X^b\,\phi_0
 + c_2\, \phi_0^\dagger\,X^a\,\phi_0\,\phi_0^\dagger\,X^b\,\phi_0
 =\frac{f^2}{4}\,\delta^{ab},
\end{equation}
where for the derivation of the right-hand side, the normalization $\tr[X^a X^b]=\delta^{ab}$ is used.

As an explicit example, the spontaneous symmetry breaking ${\rm SO}(N)\to {\rm SO}(N-1)$ is now considered.
With an appropriate choice of generators, the vacuum $\phi_0$ can be written as $\phi_0=(0,\ldots, 0,1)^T$.
The above relation for $c_1$ and $c_2$, eq.~(\ref{eq:CCWZ_expl_vac_c1_c2}), thus simplifies to
\begin{equation}
 c_1\, [X^a\,X^b]_{NN}
 + c_2\, [X^a]_{NN}\,[X^b]_{NN}
 =\frac{f^2}{4}\,\delta^{ab}.
\end{equation}
Since the generators of ${\rm SO}(N)$ are antisymmetric, their diagonal elements vanish and one finds $[X^a]_{NN}=0$.
The second term in the Lagrangian, which is proportional to $c_2$, is thus absent in this case.
Considering an explicit basis for the generators $X^a$, e.g.\
\begin{equation}
 [X^a]_{IJ} = -\frac{i}{\sqrt{2}}\left(\delta^a_I\delta^N_J-\delta^N_I\delta^a_J\right),
 \quad
 a\in\{1,\dots,N-1\},
 \quad
 I,J\in\{1,\ldots,N\},
\end{equation}
one finds
\begin{equation}
 [X^a\,X^b]_{IJ} = \frac{1}{2}\left(
 \delta^a_I \delta^b_J
 +
 \delta^{ab}\delta^N_I \delta^N_J
 \right),
 \quad
 a,b\in\{1,\ldots,N-1\},
 \quad
 I,J\in\{1,\ldots,N\},
\end{equation}
and thus
\begin{equation}
 [X^a\,X^b]_{NN} = \frac{1}{2}\,\delta^{ab},
\end{equation}
such that $c_1=\frac{f^2}{2}$. Using the above choice of generators, the leading order NGB Lagrangian for a spontaneous breaking ${\rm SO}(N)\to {\rm SO}(N-1)$ is then given by
\begin{equation}
 \mathcal{L}_2 = \frac{f^2}{2}\,\Big[\partial_\mu U^{-1}(x)\,\partial^\mu U(x)\Big]_{NN}.
\end{equation}

\newpage
\section{Generators of SO(5) and SO(6)}\label{sec:generator_definitions}
The 10 generators of $\SO{5}$ can be grouped according to their transformation properties under the $\SUs{2}{L}\times \SUs{2}{R}\cong \SO{4}$ subgroup of $\SO{5}$:
\begin{itemize}
 \item The $\T^a_L$ transform as $(\mathbf{3},\mathbf{1})$ and generate the $\SUs{2}{L}$ subgroup.
 \item The $\T^a_R$ transform as $(\mathbf{1},\mathbf{3})$ and generate the $\SUs{2}{R}$ subgroup.
 \item The $\T^a_{\hat 1}$ transform as $(\mathbf{2},\mathbf{2})$ and correspond to the broken generators of the $\SO{5}\to\SO{4}$ symmetry breaking.
 These generators are associated with the $\SO{5}/\SO{4}$ Higgs doublet NGBs.
\end{itemize}
The generators can be defined as
\begin{equation}
\begin{aligned}
 \left[ \T^a_L \right]_{I J} &= -\frac{i}{2} \left[ \frac{1}{2} \epsilon^{abc} \left( \delta_{b I} \delta_{c J} - \delta_{b J} \delta_{c I} \right) + \left( \delta_{a I} \delta_{4 J} - \delta_{a J} \delta_{4 I} \right) \right],
 \quad a \in \{1,2,3\},
 \\
 \left[ \T^a_R \right]_{I J} &= -\frac{i}{2} \left[ \frac{1}{2} \epsilon^{abc} \left( \delta_{b I} \delta_{c J} - \delta_{b J} \delta_{c I} \right) - \left( \delta_{a I} \delta_{4 J} - \delta_{a J} \delta_{4 I} \right) \right],
 \quad a \in \{1,2,3\},
 \\
 \left[ \T^a_{\hat 1} \right]_{I J} &=  - \frac{i}{\sqrt{2}} \left( \delta_{a I} \delta_{5 J} - \delta_{a J} \delta_{5 I} \right),
 \quad a \in \{1,2,3,4\},
\end{aligned}
\end{equation}
where $I,J \in \{1,2,3,4,5\}$.

The 15 generators of $\SO{6}$ consist of the 10 generators of $\SO{5}$ and the five generators that are broken by the $\SO{6}\to\SO{5}$ symmetry breaking.
To define the former, the definition above can be simply extended by using $I,J \in \{1,2,3,4,5,6\}$.
The latter transform under $\SUs{2}{L}\times \SUs{2}{R}$ as follows:
\begin{itemize}
 \item The $\T^a_{\hat 2}$ transform as $(\mathbf{2},\mathbf{2})$.
 These generators are associated with the Higgs doublet NGBs in the $\SO{6}/\SO{5}$ coset.
 \item The $\T_S$ transforms as $(\mathbf{1},\mathbf{1})$.
 This generator is associated with the scalar singlet NGB in the $\SO{6}/\SO{5}$ coset.
\end{itemize}
They can be defined as
\begin{equation}
\begin{aligned}
 \left[ \T^a_{\hat 2} \right]_{I J} &=  - \frac{i}{\sqrt{2}} \left( \delta_{a I} \delta_{6 J} - \delta_{a J} \delta_{6 I} \right),
 \quad a \in \{1,2,3,4\},
 \\
 \left[ \T_S \right]_{I J} &=  - \frac{i}{\sqrt{2}} \left( \delta_{5 I} \delta_{6 J} - \delta_{5 J} \delta_{6 I} \right),
\end{aligned}
\end{equation}
where $I,J \in \{1,2,3,4,5,6\}$.

\newpage
\section{Loop functions}\label{app:loop-functions}
This appendix lists the loop functions introduced in section~\ref{sec:direct_constraints:direct_constraints:widths_and_BRs}.
The functions relevant for a decay of a heavy scalar to two massless vector bosons are (cf.\ e.g.~\cite{Weiler:1988xn,Djouadi:2005gi,Djouadi:2005gj})
\begin{equation}
\begin{aligned}
 A_F(x) &= 2\,x\,(1 + (1 - x)\,f(x)),
 \\
 \tilde{A}_F(x) &= 2\,x\,f(x),
 \\
 A_V(x) & = -2 - 3\,x - 3\,x\,(2 - x)\,f(x),
\end{aligned}
\end{equation}
where
\begin{equation}\label{eq:app:loop-functions:f_x}
 f(x) =
 \begin{cases}
 \arcsin^2\left(\frac{1}{\sqrt{x}}\right)
 & \text{if } x \geq 1
 \quad
 \\
 -\frac{1}{4}\,\left(\log\left(\frac{1 + \sqrt{1 - x}}{1 - \sqrt{1 - x}}\right) - i\,\pi\right)^2
 & \text{if } x < 1
 \end{cases}.
\end{equation}
In the case where one of the vector bosons in the final state is massive, the following functions apply (cf.\ e.g.~\cite{Weiler:1988xn,Bergstrom:1985hp,Djouadi:2005gi,Djouadi:2005gj}):
\begin{equation}
\begin{aligned}
 B_F(x,y) &= 4\,\left(I_1(x,y)-I_2(x,y)\right),
 \\
 \tilde{B}_F(x,y) &= 4\,I_2(x,y),
 \\
 B_V(x,y) & = \left(\frac{4}{y} + \frac{8}{x\,y} - 6 - \frac{4}{x} \right)\, I_1(x,y)
            +16\,\left(1-\frac{1}{y}\right)\,I_2(x,y).
\end{aligned}
\end{equation}
The functions $I_1(x,y)$ and $I_2(x,y)$ are defined by
\begin{equation}
\begin{aligned}
 I_1(x,y)&=
 \frac{x\,y}{2\,(x-y)}
+\frac{x^2\,y^2}{2\,(x-y)^2}\,
\big(f(x)-f(y)\big)
+\frac{x^2\,y}{(x-y)^2}\,
\big(g(x)-g(y)\big),
\\
 I_2(x,y)&=
 \frac{-x\,y}{2\,(x-y)}\,
 \big(f(x)-f(y)\big),
\end{aligned}
\end{equation}
where $f(x)$ is given in eq.~\eqref{eq:app:loop-functions:f_x} and $g(x)$ is
\begin{equation}
 g(x) =
 \begin{cases}
 \sqrt{x-1}\,\arcsin\left(\frac{1}{\sqrt{x}}\right)
 & \text{if } x \geq 1
 \quad
 \\
 \frac{\sqrt{1-x}}{2}\left(\log\left(\frac{1 + \sqrt{1 - x}}{1 - \sqrt{1 - x}}\right) - i\,\pi\right)
 & \text{if } x < 1
 \end{cases}.
\end{equation}

\newpage
\section{Mass matrices}\label{app:Mass_matrices}
\subsection{Minimal composite Higgs model}\label{app:Mass_matrices:MCHM}
\subsubsection{Vector bosons}
The mass matrix of the neutral vector bosons in the MCHM is
\begin{equation}\label{eq:app:Mass_matrix:MCHM_EW_neutral}
\begin{aligned}
 M^2_Z(h) =
 \left(
 \renewcommand{\arraystretch}{1.4}
 \begin{array}{c||c c|c c c c|c}
&  W_{\mu}^{(0)3} & B_\mu^{(0)} & \rho_{L \, \mu}^3 & \rho_{R \, \mu}^3 & \axial_{\mu}^3 & X_\mu  & \axial_{\mu}^4 \\
\hline \hline
W_{\mu}^{(0)3}
 & \frac{f_1^2 g_{(0)}^2}{2}
 &
 & \multicolumn{4}{c|}{\multirow{2}{*}{$v_Z^{\,\rm T}$}}
 & \multicolumn{1}{c}{\multirow{6}{*}{0}}
\\
B_\mu^{(0)}
 &
 & \frac{\left(f_1^2+f_X^2\right) g_{(0)}^{\prime \, 2}}{2}
 & \multicolumn{4}{c|}{}
 & \multicolumn{1}{c}{}
\\
\cline{1-7}
\rho_{L \, \mu}^3
 & \multicolumn{2}{c|}{\multirow{4}{*}{$v_Z$}}
 & \frac{f_1^2 g_\rho^2}{2}
 &
 &
 &
 & \multicolumn{1}{c}{}
\\
\rho_{R \, \mu}^3
 &
 &
 &
 & \frac{f_1^2 g_\rho^2}{2}
 &
 &
 & \multicolumn{1}{c}{}
\\
\axial_{\mu}^3
 &
 &
 &
 &
 & \frac{f_1^4 g_\rho^2}{2 \left(f_1^2-f^2\right)}
 &
 & \multicolumn{1}{c}{}
\\
X_\mu
 &
 &
 &
 &
 &
 & \frac{f_X^2 g_X^2}{2}
 & \multicolumn{1}{c}{}
\\
\hline
\axial_{2 \, \mu}^4
 & \multicolumn{6}{c|}{0}
 & \frac{f_1^4 g_\rho^2}{2 \left(f_1^2-f^2\right)}
\end{array} \right),
\end{aligned}
\end{equation}
where the $h$-dependent composite-elementary mixings are
\begin{equation}
v_Z = \left( \begin{array}{c||cc}
&  W_{\mu}^{(0)3} &  B_{\mu}^{(0)}
\\
\hline\hline
\rho_{L \, \mu}^3
 & -\frac{1}{4} f_1^2 \, g_{(0)} g_\rho \, \left(c_h +1\right)
 &  -\frac{1}{4} f_1^2 \, g_{(0)}^{\prime} g_\rho \, (1-c_h)
\\
\rho_{R \, \mu}^3
 & -\frac{1}{4} f_1^2 \, g_{(0)} g_\rho \, \left(1 - c_h\right)
 & -\frac{1}{4} f_1^2 \, g_{(0)}^{\prime} g_\rho \, \left(c_h +1\right)
\\
\axial_{\mu}^3
 & -\frac{f_1^2 \, g_{(0)} g_\rho \,  \sh }{2 \sqrt{2}}
 & \frac{f_1^2 \, g_{(0)}^{\prime} g_\rho \, \sh }{2 \sqrt{2}}
\\
X_\mu
 & 0
 & -\frac{1}{2} f_X^2 \, g_{(0)}^{\prime} g_X
\end{array} \right).
\end{equation}
\noindent
The mass matrix of the charged vector bosons $V^\pm_\mu = \frac{1}{\sqrt{2}} \left( V^1_\mu \mp i V^2_\mu \right)$ in the MCHM is
\begin{equation}\label{eq:app:Mass_matrix:MCHM_EW_charged}
  M^2_W(h) = \left(
\begin{array}{c||c|c c c c}
&  W^{(0)+}_\mu & \rho^+_{L \, \mu} & \rho^+_{R \, \mu} & \axial^+_{\mu} \\
\hline\hline
W^{(0)-}_\mu
 & \frac{f_1^2 g_{(0)}^2}{2}
 & \multicolumn{3}{c}{v_W^{\,\rm T}}
\\
\hline
\rho^-_{L \, \mu}
 & \multirow{4}{*}{$v_W$}
 & \frac{f_1^2 g_\rho^2}{2}
 &
 &
\\
\rho^-_{R \, \mu}
 &
 &
 & \frac{f_1^2 g_\rho^2}{2}
 &
\\
\axial^-_{\mu}
 &
 &
 &
 & \frac{f_1^4 g_\rho^2}{2 \left(f_1^2-f^2\right)}
\end{array} \right),
\end{equation}
where the $h$-dependent composite-elementary mixings are given by
\begin{align}
 v_W = \left( \begin{array}{c||c}
&  W^{(0)+}_\mu
\\
\hline \hline
\rho^-_{L \, \mu}
 & -\frac{1}{4} f_1^2 \, g_{(0)} g_\rho \, \left(c_h +1\right)
\\
\rho^-_{R \, \mu}
 & -\frac{1}{4} f_1^2 \, g_{(0)} g_\rho \, (1-c_h)
\\
\axial^-_{\mu}
 & -\frac{f_1^2 \, g_{(0)} g_\rho \, \sh}{2 \sqrt{2}}
\end{array} \right).
\end{align}
The mass matrix of the colored vector bosons in the MCHM and in the NMCHM is
\begin{align}
 M_G^2(h) =
 \left( \begin{array}{c||cc}
&  G^{(0)}_\mu & G^{(1)}_\mu
\\
\hline \hline
G^{(0)}_\mu
 & \frac{f_G^2 g_{3(0)}^2}{2}
 & -\frac{f_G^2 g_{3(0)} g_{3(1)}}{2}
\\
G^{(1)}_\mu
 & -\frac{f_G^2 g_{3(0)} g_{3(1)}}{2}
 & \frac{f_G^2 g_{3(1)}^2}{2}
\end{array} \right).
\end{align}

\subsubsection{Quarks}
The mass matrix of up-type quarks in the MCHM is
\begin{equation}
\begin{aligned}
 &M_u(h) =
 \\
 &\left(
\begin{array}{c||c|ccc|cc|ccc}
 & u_\text{R}^{(0)}
 & Q_{u \text{R}}^{+-}
 & Q_{u \text{R}}^{-+}
 & S_{u \text{R}}
 & Q_{d \text{R}}^{++}
 & \widetilde{Q}_{d \text{R}}^{++}
 & \widetilde{Q}_{u \text{R}}^{+-}
 & \widetilde{Q}_{u \text{R}}^{-+}
 & \widetilde{S}_{u \text{R}}
\\
\hline \hline
\overline{u}_\text{L}^{(0)}
 & 0
 & \multicolumn{3}{c|}{\underline{\Delta}_{u \text{L}}}
 & -\Delta_{dL}
 & 0
 & \multicolumn{3}{c }{0}
\\
\hline
\overline{Q}_{u \text{L}}^{+-}
 & \multirow{3}{*}{0}
 & m_U
 &&
 & \multicolumn{2}{c|}{\multirow{3}{*}{0}}
 & m_{Y_u}
\\
\overline{Q}_{u \text{L}}^{-+}
 &&
 & m_U
 &&&&
 & m_{Y_u}
\\
\overline{S}_{u \text{L}}
 &&&
 & m_U
 &&&&
 & m_{Y_u}+Y_u
\\
\hline
\overline{Q}_{d \text{L}}^{++}
 & \multirow{2}{*}{$0$}
 & \multicolumn{3}{c|}{\multirow{2}{*}{0}}
 & m_D
 & m_{Y_d}
 & \multicolumn{3}{c}{\multirow{2}{*}{0}}
\\
\overline{\widetilde{Q}}_{d \text{L}}^{++}
 &&&&&
 & m_{\widetilde{D}}
\\
\hline
\overline{\widetilde{Q}}_{u \text{L}}^{+-}
 & \multirow{3}{*}{$\underline{\Delta}_{u\text{R}}^{\dagger}$}
 & \multicolumn{3}{c|}{\multirow{3}{*}{0}}
 & \multicolumn{2}{c|}{\multirow{3}{*}{0}}
 & m_{\widetilde{U}}
\\
\overline{\widetilde{Q}}_{u \text{L}}^{-+}
 &&&&&&&
 & m_{\widetilde{U}}
\\
\overline{\widetilde{S}}_{u \text{L}}
 &&&&&&&&
 & m_{\widetilde{U}}
\\
\end{array}\right),
\end{aligned}
\end{equation}
where the $h$-dependent composite-elementary mixings are
\begin{equation}
\begin{aligned}
 &\underline{\Delta}_{u\text{R}}^{\dagger} =
 & \quad\quad
 &\underline{\Delta}_{u\text{L}}^{\dagger} =
 \\
 &\left( \begin{array}{c||c}
&  u_R^{(0)}
\\
\hline \hline
\overline{\widetilde{Q}}_{u \text{L}}^{+-}
 & -\frac{i}{\sqrt{2}} \Delta_{u \mathrm{R}}^{\dagger} \sh
\\
\overline{\widetilde{Q}}_{u \text{L}}^{-+}
 & -\frac{i}{\sqrt{2}} \Delta_{u \mathrm{R}}^{\dagger} \sh
\\
\overline{\widetilde{S}}_{u \text{L}}
 & - \Delta_{u \mathrm{R}}^{\dagger} c_h
\end{array} \right),
%
&
%
 &\left( \begin{array}{c||c}
&  u_L^{(0)}
\\
\hline \hline
\overline{Q}_{u \text{R}}^{+-}
 & -\frac{1}{2} \Delta_{u \mathrm{L}}^\dagger \left(c_h +1\right)
\\
\overline{Q}_{u \text{R}}^{-+}
 & \frac{1}{2} \Delta_{u \mathrm{L}}^\dagger  \left(1 - c_h \right)
\\
\overline{S}_{u \text{R}}
 & -\frac{i}{\sqrt{2}} \Delta_{u \mathrm{L}}^\dagger  \sh
\end{array} \right).
\end{aligned}
\end{equation}
An analogous matrix is found for down-type quarks.

The mass matrices of exotically charged quarks in the MCHM and in the NMCHM are
\begin{equation}
\begin{aligned}
 &M_{Q_{+5/3}}(h) =
 & \quad\quad
 &M_{Q_{-4/3}}(h) =
 \\
 &\left( \begin{array}{c||cc}
&  Q_{u \text{R}}^{++} &  \widetilde{Q}_{u \text{R}}^{++}
\\
\hline \hline
\overline{Q}_{u \text{L}}^{++}
 & m_U
 & m_{Y_u}
\\
\overline{\widetilde{Q}}_{u \text{L}}^{++}
 & 0
 & m_{\widetilde{U}}
\end{array} \right),
%
&
%
 &\left( \begin{array}{c||cc}
&  Q_{d \text{R}}^{--} &  \widetilde{Q}_{d \text{R}}^{--}
\\
\hline \hline
\overline{Q}_{d \text{L}}^{--}
 & m_D
 & m_{Y_d}
\\
\overline{\widetilde{Q}}_{d \text{L}}^{--}
 & 0
 & m_{\widetilde{D}}
\end{array} \right).
\end{aligned}
\end{equation}

\subsection{Next-to-minimal composite Higgs model}\label{app:Mass_matrices:NMCHM}
\subsubsection{Vector bosons}
The mass matrix of the neutral vector bosons in the NMCHM is
\begin{equation}\label{eq:app:Mass_matrix:NMCHM_EW_neutral}
\resizebox{\textwidth}{!}{$
\begin{aligned}
 &M^2_Z(\gb{h}, \gb{\eta}) =
 \\
 &\left(
 \renewcommand{\arraystretch}{1.4}
 \begin{array}{c||c c|c c c c c|c c c}
&  W_{\mu}^{(0)3} & B_\mu^{(0)} & \rho_{L \, \mu}^3 & \rho_{R \, \mu}^3 & \axial_{1 \, \mu}^3 & \axial_{2 \, \mu}^3 & X_\mu & \axial_{1 \, \mu}^4 & \axial_{2 \, \mu}^4 & \rho_{S \, \mu} \\
\hline \hline
W_{\mu}^{(0)3}
 & \frac{f_1^2 g_{(0)}^2}{2}
 &
 & \multicolumn{5}{c|}{\multirow{2}{*}{$v_Z^{\,\rm T}$}}
 & \multicolumn{3}{c}{\multirow{7}{*}{0}}
\\
B_\mu^{(0)}
 &
 & \frac{\left(f_1^2+f_X^2\right) g_{(0)}^{\prime \, 2}}{2}
 & \multicolumn{5}{c|}{}
 & \multicolumn{3}{c}{}
\\
\cline{1-8}
\rho_{L \, \mu}^3
 & \multicolumn{2}{c|}{\multirow{5}{*}{$v_Z$}}
 & \frac{f_1^2 g_\rho^2}{2}
 &
 &
 &
 &
 & \multicolumn{3}{c}{}
\\
\rho_{R \, \mu}^3
 &
 &
 &
 & \frac{f_1^2 g_\rho^2}{2}
 &
 &
 &
 & \multicolumn{3}{c}{}
\\
\axial_{1 \, \mu}^3
 &
 &
 &
 &
 & \frac{f_1^2 g_\rho^2}{2}
 &
 &
 & \multicolumn{3}{c}{}
\\
\axial_{2 \, \mu}^3
 &
 &
 &
 &
 &
 & \frac{f_1^4 g_\rho^2}{2 \left(f_1^2-f^2\right)}
 &
 & \multicolumn{3}{c}{}
\\
X_\mu
 &
 &
 &
 &
 &
 &
 & \frac{f_X^2 g_X^2}{2}
 & \multicolumn{3}{c}{}
\\
\hline
\axial_{1 \, \mu}^4
 & \multicolumn{7}{c|}{\multirow{3}{*}{0}}
 & \frac{f_1^2 g_\rho^2}{2}
 &
 &
\\
\axial_{2 \, \mu}^4
 & \multicolumn{7}{c|}{}
 &
 & \frac{f_1^4 g_\rho^2}{2 \left(f_1^2-f^2\right)}
 &
\\
\rho_{S \, \mu}
 & \multicolumn{7}{c|}{}
 &
 &
 & \frac{f_1^4 g_\rho^2}{2 \left(f_1^2-f^2\right)}
\end{array} \right),
\end{aligned}
$}
\end{equation}
where the $\gb{h}$- and $\gb{\eta}$-dependent composite-elementary mixings are
\begin{equation}
v_Z = \left( \begin{array}{c||cc}
&  W_{\mu}^{(0)3} &  B_{\mu}^{(0)}
\\
\hline\hline
\rho_{L \, \mu}^3
 & -\frac{1}{4} f_1^2 \, g_{(0)} g_\rho \, \left(c_h \cetatilde^2+\setatilde^2+1\right)
 &  -\frac{1}{4} f_1^2 \, g_{(0)}^{\prime} g_\rho \, (1-c_h) \cetatilde^2
\\
\rho_{R \, \mu}^3
 & -\frac{1}{4} f_1^2 \, g_{(0)} g_\rho \, \left(1 - c_h\right) \cetatilde^2
 & -\frac{1}{4} f_1^2 \, g_{(0)}^{\prime} g_\rho \, \left(c_h \cetatilde^2+\setatilde^2+1\right)
\\
\axial_{1 \, \mu}^3
 & \frac{f_1^2 \, g_{(0)} g_\rho \, (1-c_h) \setatilde \cetatilde}{2 \sqrt{2}}
 & -\frac{f_1^2 \, g_{(0)}^{\prime} g_\rho \, (1-c_h) \setatilde \cetatilde}{2 \sqrt{2}}
\\
\axial_{2 \, \mu}^3
 & -\frac{f_1^2 \, g_{(0)} g_\rho \,  \sh \cetatilde}{2 \sqrt{2}}
 & \frac{f_1^2 \, g_{(0)}^{\prime} g_\rho \, \sh \cetatilde}{2 \sqrt{2}}
\\
X_\mu
 & 0
 & -\frac{1}{2} f_X^2 \, g_{(0)}^{\prime} g_X
\end{array} \right).
\end{equation}
\noindent
The mass matrix of the charged vector bosons $V^\pm_\mu = \frac{1}{\sqrt{2}} \left( V^1_\mu \mp i V^2_\mu \right)$ in the NMCHM is
\begin{equation}\label{eq:app:Mass_matrix:NMCHM_EW_charged}
  M^2_W(\gb{h}, \gb{\eta}) = \left(
\begin{array}{c||c|c c c c}
&  W^{(0)+}_\mu & \rho^+_{L \, \mu} & \rho^+_{R \, \mu} & \axial^+_{1 \, \mu} & \axial^+_{2 \, \mu} \\
\hline\hline
W^{(0)-}_\mu
 & \frac{f_1^2 g_{(0)}^2}{2}
 & \multicolumn{4}{c}{v_W^{\,\rm T}}
\\
\hline
\rho^-_{L \, \mu}
 & \multirow{4}{*}{$v_W$}
 & \frac{f_1^2 g_\rho^2}{2}
 &
 &
 &
\\
\rho^-_{R \, \mu}
 &
 &
 & \frac{f_1^2 g_\rho^2}{2}
 &
 &
\\
\axial^-_{1 \, \mu}
 &
 &
 &
 & \frac{f_1^2 g_\rho^2}{2}
 &
\\
\axial^-_{2 \, \mu}
 &
 &
 &
 &
 & \frac{f_1^4 g_\rho^2}{2 \left(f_1^2-f^2\right)}
\end{array} \right),
\end{equation}
where the $\gb{h}$- and $\gb{\eta}$-dependent composite-elementary mixings are given by
\begin{align}
 v_W = \left( \begin{array}{c||c}
&  W^{(0)+}_\mu
\\
\hline \hline
\rho^-_{L \, \mu}
 & -\frac{1}{4} f_1^2 \, g_{(0)} g_\rho \, \left(c_h \cetatilde^2+\setatilde^2+1\right)
\\
\rho^-_{R \, \mu}
 & -\frac{1}{4} f_1^2 \, g_{(0)} g_\rho \, (1-c_h) \cetatilde^2
\\
\axial^-_{1 \, \mu}
 &  \frac{f_1^2 \, g_{(0)} g_\rho \, (1-c_h) \setatilde \cetatilde}{2 \sqrt{2}}
\\
\axial^-_{2 \, \mu}
 & -\frac{f_1^2 \, g_{(0)} g_\rho \, \sh \cetatilde}{2 \sqrt{2}}
\end{array} \right).
\end{align}

\newpage

\subsubsection{Quarks}

The mass matrix of up-type quarks in the NMCHM is
\begin{equation}
\begin{aligned}
 &M_u(\gb{h}, \gb{\eta}) =
 \\
 &\left(
\begin{array}{c||c|cccc|cc|cccc}
 & u_\text{R}^{(0)}
 & Q_{u \text{R}}^{+-}
 & Q_{u \text{R}}^{-+}
 & S^1_{u \text{R}}
 & S^2_{u \text{R}}
 & Q_{d \text{R}}^{++}
 & \widetilde{Q}_{d \text{R}}^{++}
 & \widetilde{Q}_{u \text{R}}^{+-}
 & \widetilde{Q}_{u \text{R}}^{-+}
 & \widetilde{S}^1_{u \text{R}}
 & \widetilde{S}^2_{u \text{R}}
\\
\hline \hline
\overline{u}_\text{L}^{(0)}
 & 0
 & \multicolumn{4}{c|}{\underline{\Delta}_{u \text{L}}}
 & -\Delta_{dL}
 & 0
 & \multicolumn{4}{c }{0}
\\
\hline
\overline{Q}_{u \text{L}}^{+-}
 & \multirow{4}{*}{0}
 & m_U
 &&&
 & \multicolumn{2}{c|}{\multirow{4}{*}{0}}
 & m_{Y_u}
\\
\overline{Q}_{u \text{L}}^{-+}
 &&
 & m_U
 &&&&&
 & m_{Y_u}
\\
\overline{S}^1_{u \text{L}}
 &&&
 & m_U
 &&&&&
 & m_{Y_u}
\\
\overline{S}^2_{u \text{L}}
 &&&&
 & m_U
 &&&&&
 & m_{Y_u}+Y_u
\\
\hline
\overline{Q}_{d \text{L}}^{++}
 & \multirow{2}{*}{$0$}
 & \multicolumn{4}{c|}{\multirow{2}{*}{0}}
 & m_D
 & m_{Y_d}
 & \multicolumn{4}{c}{\multirow{2}{*}{0}}
\\
\overline{\widetilde{Q}}_{d \text{L}}^{++}
 &&&&&&
 & m_{\widetilde{D}}
\\
\hline
\overline{\widetilde{Q}}_{u \text{L}}^{+-}
 & \multirow{4}{*}{$\underline{\Delta}_{u\text{R}}^{\dagger}$}
 & \multicolumn{4}{c|}{\multirow{4}{*}{0}}
 & \multicolumn{2}{c|}{\multirow{4}{*}{0}}
 & m_{\widetilde{U}}
\\
\overline{\widetilde{Q}}_{u \text{L}}^{-+}
 &&&&&&&&
 & m_{\widetilde{U}}
\\
\overline{\widetilde{S}}^1_{u \text{L}}
 &&&&&&&&&
 & m_{\widetilde{U}}
\\
\overline{\widetilde{S}}^2_{u \text{L}}
 &&&&&&&&&&
 & m_{\widetilde{U}}
\\
\end{array}\right),
\end{aligned}
\end{equation}
where the $\gb{h}$- and $\gb{\eta}$-dependent composite-elementary mixings are
\begin{equation}
\begin{aligned}
 &\underline{\Delta}_{u\text{R}}^{\dagger} =
 \ &
 &\underline{\Delta}_{u\text{L}}^{\dagger} =
 \\
 &\left( \begin{array}{c||c}
&  u_R^{(0)}
\\
\hline \hline
\overline{\widetilde{Q}}_{u \text{L}}^{+-}
 & -\frac{i}{\sqrt{2}} \left(\Delta_{u \mathrm{R}}^{5 \, \dagger} \left( (1-c_h) \setatilde \cetatilde \right)+\Delta_{u \mathrm{R}}^{6 \, \dagger} \sh \cetatilde\right)
\\
\overline{\widetilde{Q}}_{u \text{L}}^{-+}
 & -\frac{i}{\sqrt{2}} \left(\Delta_{u \mathrm{R}}^{5 \, \dagger} \left( (1-c_h) \setatilde \cetatilde \right)+\Delta_{u \mathrm{R}}^{6 \, \dagger} \sh \cetatilde\right)
\\
\overline{\widetilde{S}}^1_{u \text{L}}
 &  - \Delta_{u \mathrm{R}}^{5 \, \dagger} \left(\cetatilde^2+c_h \setatilde^2\right) + \Delta_{u \mathrm{R}}^{6 \, \dagger} \sh \setatilde
\\
\overline{\widetilde{S}}^2_{u \text{L}}
 & - \Delta_{u \mathrm{R}}^{5 \, \dagger} \sh \setatilde - \Delta_{u \mathrm{R}}^{6 \, \dagger} c_h
\end{array} \right),
%
&
%
 &\left( \begin{array}{c||c}
&  u_L^{(0)}
\\
\hline \hline
\overline{Q}_{u \text{R}}^{+-}
 & -\frac{1}{2} \Delta_{u \mathrm{L}}^\dagger \left(c_h \cetatilde^2+1\right)
\\
\overline{Q}_{u \text{R}}^{-+}
 & \frac{1}{2} \Delta_{u \mathrm{L}}^\dagger \cetatilde^2 \left(1 - c_h \right)
\\
\overline{S}^1_{u \text{R}}
 &  \frac{i}{\sqrt{2}} \Delta_{u \mathrm{L}}^\dagger \left( 1-c_h \right) \setatilde \cetatilde
\\
\overline{S}^2_{u \text{R}}
 & -\frac{i}{\sqrt{2}} \Delta_{u \mathrm{L}}^\dagger \cetatilde \sh
\end{array} \right).
\end{aligned}
\end{equation}
An analogous matrix is found for down-type quarks.

\newpage
\section{Composite-elementary mixings}
\subsection{Minimal composite Higgs model}\label{app:comp_elem:MCHM}

The explicit expressions of the $3\times3$ composite elementary mixings for the different flavor symmetries are
\begin{itemize}
\item In $\Us{3}{LC}^3$,
\begin{align}
\Delta_{u_L} &= \Delta_{Lt} ~\mathds{1} \,, &
\Delta_{u_R}^\dagger &= V^\dagger
\begin{pmatrix}
\Delta_{Ru} \\
& \Delta_{Rc} \\
&& \Delta_{Rt} \\
\end{pmatrix}
, \\
\Delta_{d_L} &= \Delta_{Lb} ~\mathds{1} \,, &
\Delta_{d_R}^\dagger &=
\begin{pmatrix}
\Delta_{Rd} \\
& \Delta_{Rs} \\
&& \Delta_{Rb} \\
\end{pmatrix}
. &
\end{align}
Here, $V$ is the CKM matrix with 3 angles and 1 phase.
\item In $\Us{3}{RC}^3$,
\begin{align}
\Delta_{u_L} &= V^\dagger
\begin{pmatrix}
\Delta_{Lu} \\
& \Delta_{Lc} \\
&& \Delta_{Lt} \\
\end{pmatrix}
, &
\Delta_{u_R}^\dagger &= \Delta_{Rt} ~\mathds{1} \,, \\
\Delta_{d_L} &=
\begin{pmatrix}
\Delta_{Ld} \\
& \Delta_{Ls} \\
&& \Delta_{Lb} \\
\end{pmatrix}
, &
\Delta_{d_R}^\dagger &= \Delta_{Rb} ~\mathds{1} \,.
\label{eq:u3rc-deltadL}
\end{align}
\item In $\Us{2}{LC}^3$,
\begin{align}
\Delta_{u_L} &= \begin{pmatrix}
\Delta_{Lu} \\
& \Delta_{Lu} \\
&& \Delta_{Lt} \\
\end{pmatrix} , &
\Delta_{u_R}^\dagger &=
\begin{pmatrix}
c_u \Delta_{Ru} & -s_u \Delta_{Rc} e^{i\alpha_u} \\
s_u \Delta_{Ru} e^{-i\alpha_u} & c_u \Delta_{Rc} & \epsilon_u\Delta_{Rt}
e^{i\phi_t}\\
&& \Delta_{Rt} \\
\end{pmatrix}
, \\
\Delta_{d_L} &= \begin{pmatrix}
\Delta_{Ld} \\
& \Delta_{Ld} \\
&& \Delta_{Lb} \\
\end{pmatrix} , &
\Delta_{d_R}^\dagger &=
\begin{pmatrix}
c_d \Delta_{Rd} & -s_d \Delta_{Rs} e^{i\alpha_d} \\
s_d \Delta_{Rd} e^{-i\alpha_d} & c_d \Delta_{Rs} &
\epsilon_d\Delta_{Rb}e^{i\phi_b}\\
&& \Delta_{Rb} \\
\end{pmatrix}
. &
\end{align}
\item In $\Us{2}{RC}^3$,
\begin{align}
\Delta_{u_L} &=
\begin{pmatrix}
c_u \Delta_{Lu} & -s_u \Delta_{Lc} e^{i\alpha_u} \\
s_u \Delta_{Lu} e^{-i\alpha_u} & c_u \Delta_{Lc} &
\epsilon_u\Delta_{Lt}e^{i\phi_t}\\
&& \Delta_{Lt} \\
\end{pmatrix}
, &
\Delta_{u_R}^\dagger &=
\begin{pmatrix}
\Delta_{Ru} \\
& \Delta_{Ru} \\
&& \Delta_{Rt} \\
\end{pmatrix}
, \\
\Delta_{d_L} &=
\begin{pmatrix}
c_d \Delta_{Ld} & -s_d \Delta_{Ls} e^{i\alpha_d} \\
s_d \Delta_{Ld} e^{-i\alpha_d} & c_d \Delta_{Ls} &
\epsilon_d\Delta_{Lb}e^{i\phi_b}\\
&& \Delta_{Lb} \\
\end{pmatrix}
, &
\Delta_{d_R}^\dagger &= \begin{pmatrix}
\Delta_{Rd} \\
& \Delta_{Rd} \\
&& \Delta_{Rb} \\
\end{pmatrix} .
\end{align}
\end{itemize}

\subsection{Next-to-minimal composite Higgs model}\label{app:comp_elem:NMCHM}

The explicit expressions of the $3\times3$ composite elementary mixings for the $\Us{2}{RC}^3$ flavor symmetry are
\begin{equation}
\Delta_{u \text{L}} =
\begin{pmatrix}
c_u\, \Delta_{u_1\text{L}} & -s_u\, \Delta_{u_2\text{L}}\, e^{ i \alpha_u} \\
s_u\, \Delta_{u_1\text{L}}\, e^{- i \alpha_u} & c_u\, \Delta_{u_2\text{L}} &
\epsilon_u\, \Delta_{u_3\text{L}}\, e^{ i \phi_u}\\
&& \Delta_{u_3\text{L}} \\
\end{pmatrix},
\end{equation}

\begin{equation}
\Delta_{d \text{L}} =
\begin{pmatrix}
c_d\, \Delta_{d_1\text{L}} & -s_d\, \Delta_{d_2\text{L}}\, e^{ i \alpha_d} \\
s_d\, \Delta_{d_1\text{L}}\, e^{- i \alpha_d} & c_d\, \Delta_{d_2\text{L}} &
\epsilon_d\, \Delta_{d_3\text{L}}\, e^{ i \phi_d}\\
&& \Delta_{d_3\text{L}} \\
\end{pmatrix},
\end{equation}

\begin{equation}
 \Delta_{u \text{R}}^{5 \, \dagger} =
\begin{pmatrix}
\Delta_{u_{12}\text{R}}^5 \\
& \Delta_{u_{12}\text{R}}^5 \\
&& \Delta_{u_3\text{R}}^5 \\
\end{pmatrix},
\quad
\Delta_{u \text{R}}^{6 \, \dagger} =
\begin{pmatrix}
\Delta_{u_{12}\text{R}}^6 \, e^{ i  \phi_{u_{12}\text{R}}^6} \\
& \Delta_{u_{12}\text{R}}^6 \, e^{ i  \phi_{u_{12}\text{R}}^6} \\
&& \Delta_{u_3\text{R}}^6 \, e^{ i  \phi_{u_3\text{R}}^6} \\
\end{pmatrix},
\end{equation}

\begin{equation}
\Delta_{d \text{R}}^{5 \, \dagger} = \begin{pmatrix}
\Delta_{d_{12}\text{R}}^5 \\
& \Delta_{d_{12}\text{R}}^5 \\
&& \Delta_{d_3\text{R}}^5 \\
\end{pmatrix},
\quad
\Delta_{d \text{R}}^{6 \, \dagger} =
\begin{pmatrix}
\Delta_{d_{12}\text{R}}^6 \, e^{ i  \phi_{d_{12}\text{R}}^6} \\
& \Delta_{d_{12}\text{R}}^6 \, e^{ i  \phi_{d_{12}\text{R}}^6} \\
&& \Delta_{d_3\text{R}}^6 \, e^{ i  \phi_{d_3\text{R}}^6} \\
\end{pmatrix}.
\end{equation}

\newpage
\section{Experimental searches included as direct constraints}\label{app:seaches}

\subsection{Analysis of the minimal composite Higgs model}\label{app:seaches:MCHM}
\begin{table}[H]
\centering
\renewcommand{\arraystretch}{1.15}
\begin{tabular}{llllll}
\hline
Decay		&Experiment	&$\sqrt{s}$ [TeV]	&Lum.
[fb$^{-1}$]&Analysis			&	\\
\hline
$\rho^{\pm} \to \ell^{\pm}\nu$	&	ATLAS	&	7		&
4.7
&	EXOT-2012-02		&	\cite{Aad:2012dm}	\\
\hline
\multirow{2}{*}{$\rho^{\pm} \to W^{\pm}h$}
			&	ATLAS	&	8		&	20.3

&	EXOT-2013-23		&	\cite{Aad:2015yza}	\\
			&	CMS		&	8		&
19.7		&	EXO-14-010		&	\cite{CMS:2015gla}
\\
\hline
\multirow{4}{*}{$\rho^{\pm} \to W^{\pm}Z$}
			&	ATLAS	&	8		&	20.3

&	EXOT-2013-01		&	\cite{Aad:2015ufa}	\\
			&	ATLAS	&	8		&	20.3

&	EXOT-2013-07		&	\cite{Aad:2014pha}	\\
			&	ATLAS	&	8		&	20.3

&	EXOT-2013-08		&	\cite{Aad:2015owa}	\\
			&	CMS		&	8		&
19.7		&	EXO-12-024		&
\cite{Khachatryan:2014hpa}	\\
\hline
$\rho^{\pm} \to tb$		&	CMS		&	8
&
19.5		&	B2G-12-010		&
\cite{Chatrchyan:2014koa}	\\
\hline
\multirow{2}{*}{$\rho^0 \to W^+W^-$}
			&	ATLAS	&	8		&	20.3

&	EXOT-2013-01		&	\cite{Aad:2015ufa}	\\
			&	CMS		&	8		&
19.7		&	EXO-13-009		&
\cite{Khachatryan:2014gha}	\\
\hline
\multirow{2}{*}{$\rho^0 \to Zh$}
			&	ATLAS	&	8		&	20.3

&	EXOT-2013-23		&	\cite{Aad:2015yza}	\\
			&	CMS		&	8		&
19.7		&	EXO-13-007		&
\cite{Khachatryan:2015ywa}	\\
\hline
\multirow{2}{*}{$\rho^0 \to \ell^+\ell^-$}
			&	ATLAS	&	8		&	20.3

&	EXOT-2012-23		&	\cite{Aad:2014cka}	\\
			&	CMS		&	8		&
20.6
&	EXO12061		&
\cite{Khachatryan:2014fba}	\\
\hline
\multirow{2}{*}{$\rho^0/\rho_G \to t\bar t$}
			&	ATLAS	&	8		&	20.3

&	CONF-2015-009	&	\cite{ATLAS:2015aka}	\\
			&	CMS		&	8		&
19.5		&	B2G-12-008		&	\cite{CMS:2013gqa}
\\

\hline
\end{tabular}
\caption{Experimental analyses included in our numerics for heavy vector
resonance decay.}
\label{tab:app:MCHM:exp_vector_res}
\end{table}
\newpage

\begin{table}[H]
\centering
\renewcommand{\arraystretch}{1.15}
\begin{tabular}{llllll}
\hline
Decay		&Experiment	&$\sqrt{s}$ [TeV]	&Luminosity
[fb$^{-1}$]&Analysis			&	\\
\hline
$Q \to tW$	&	CMS		&	7		&	5

	&	B2G-12-004		&	\cite{Chatrchyan:2012af}
\\
\hline
\multirow{2}{*}{$Q \to jW$}	&	ATLAS	&	7		&
1.04

&	EXOT-2011-28		&	\cite{Aad:2012bt}	\\
			&	CDF		&	1.96		&
4.6 & &\cite{CDF-PUB-TOP-PUBLIC-10110}	\\
\hline
$Q \to qW$	&	CMS		&	8		&	19.7

	&	B2G-12-017		&	\cite{CMS:2014dka}	\\
\hline
$Q \to jZ$ &	CDF		&	1.96		&
1.055 & &\cite{CDFnote8590}	\\
\hline
\multirow{2}{*}{$U \to tZ$}
			&	CMS		&	7		&
5			&	B2G-12-004		&
\cite{Chatrchyan:2012af}	\\
			&	CMS		&	7		&
1.1			&	EXO-11-005		&
\cite{Chatrchyan:2011ay}
\\\hline
\multirow{4}{*}{$D \to bH$}
			&	ATLAS	&	8		&	20.3

	&	CONF-2015-012	&	\cite{ATLAS:2015dka}	\\
			&	CMS		&	8		&
19.8			&	B2G-12-019		&
\cite{CMS:2012hfa}	\\
			&	CMS		&	8		&
19.5			&	B2G-13-003		&
\cite{CMS:2013una}	\\
			&	CMS		&	8		&
19.7			&	B2G-14-001		&
\cite{CMS:2014bfa}	\\
\hline
\multirow{4}{*}{$D \to bZ$}
			&	CMS		&	7		&
5			&	EXO-11-066		&
\cite{CMS:2012jwa}	\\
			&	CMS		&	8		&
19.8			&	B2G-12-019		&
\cite{CMS:2012hfa}	\\
			&	CMS		&	8		&
19.5			&	B2G-13-003		&
\cite{CMS:2013una}	\\
			&	CMS		&	8		&
19.6			&	B2G-12-021		&
\cite{CMS:2013zea}	\\
\hline
\multirow{4}{*}{$D \to tW$}
			&	ATLAS	&	8		&	20.3

	&	EXOT-2013-16		&	\cite{Aad:2015gdg}	\\
			&	CMS		&	8		&
19.8			&	B2G-12-019		&
\cite{CMS:2012hfa}	\\
			&	CMS		&	8		&
19.5			&	B2G-13-003		&
\cite{CMS:2013una}	\\
			&	CDF		&	1.96		&
2.7			&			&
\cite{Aaltonen:2009nr}	\\
\hline
\multirow{5}{*}{$Q \to bW$}
			&	CMS		&	7		&
5			&	EXO-11-050		&
\cite{CMS:2012ab}	\\
			&	CMS		&	7		&
5			&	EXO-11-099		&
\cite{Chatrchyan:2012vu}	\\
			&	ATLAS	&	7		&	4.7

	&	EXOT-2012-07		&	\cite{ATLAS:2012qe}	\\
			&	ATLAS	&	8		&	20.3

	&	CONF-2015-012	&	\cite{ATLAS:2015dka}	\\
			&	CMS		&	8		&
19.7			&	B2G-12-017		&
\cite{CMS:2014dka}	\\
\hline
\multirow{3}{*}{$Q_{5/3} \to tW$}
			&	ATLAS	&	8		&	20.3

	&	EXOT-2013-16		&	\cite{Aad:2015gdg}	\\
			&	ATLAS	&	8		&	20.3

	&	EXOT-2014-17		&	\cite{Aad:2015mba}	\\
			&	CMS		&	8		&
19.6			&	B2G-12-012		&	\cite{CMS:vwa}
\\
\hline
$U \to tH$		&	CMS		&	8		&
19.7			&	B2G-12-004		&
\cite{CMS:2014rda}	\\

\hline
\end{tabular}
\caption{Experimental analyses included in our numerics for heavy quark
partner decay. $Q$ stands for any quark partner where the decay in question is
allowed by electric charges, $j$ stands for a light quark or $b$ jet, and $q$
for a light quark jet.}
\label{tab:app:MCHM:exp_quark_res}
\end{table}

\subsection{Analysis of the next-to-minimal composite Higgs model}\label{app:seaches:NMCHM}

\begin{table}[H]
\centering
\resizebox{!}{238pt}{
\renewcommand{\arraystretch}{1.15}
\begin{tabular}{llllll}
\hline
Decay &Experiment &$\sqrt{s}$ [TeV] &Lum. [fb$^{-1}$] &Analysis &\\
\hline
$Q \to jZ$
&CDF	&1.96	&1.055 	&		&\cite{Aaltonen:2007je}\\%
\hline
\multirow{2}{*}{$Q \to jW$}
&ATLAS	&7	&1.04	&EXOT-2011-28	&\cite{Aad:2012bt}\\%
&CDF	&1.96	&4.6 	&		&\cite{CDF-PUB-TOP-PUBLIC-10110}\\%
\hline
\multirow{2}{*}{$Q \to qW$}
&CMS	&8	&19.7	&B2G-12-017	&\cite{CMS:2014dka}\\%
&ATLAS	&8	&20.3	&EXOT-2014-10	&\cite{Aad:2015tba}\\%
\hline
\multirow{5}{*}{$Q \to bW$}
&CMS	&7	&5	&EXO-11-050	&\cite{CMS:2012ab}\\%
&CMS	&7	&5	&EXO-11-099	&\cite{Chatrchyan:2012vu}\\%
&ATLAS	&7	&4.7	&EXOT-2012-07	&\cite{ATLAS:2012qe}\\%
&ATLAS	&8	&20.3	&CONF-2015-012	&\cite{ATLAS:2015dka}\\%
&CMS	&8	&19.7	&B2G-12-017	&\cite{CMS:2014dka}\\%
\hline
$Q \to tW$
&CMS	&7 	&5	&B2G-12-004	&\cite{Chatrchyan:2012af}\\
\hline
\multirow{3}{*}{$U \to tH$}
&CMS	&8	&19.7	&B2G-13-005	&\cite{Khachatryan:2015oba}\\%
&ATLAS	&13	&3.2	&CONF-2016-013	&\cite{ATLAS-CONF-2016-013}\\
&CMS	&13	&2.6	&PAS-B2G-16-011	&\cite{CMS:2016dmr}\\%
\hline
\multirow{4}{*}{$U \to tZ$}
&CMS	&7	&5	&B2G-12-004	&\cite{Chatrchyan:2012af}\\%
&CMS	&7	&1.1	&EXO-11-005	&\cite{Chatrchyan:2011ay}\\%
&CMS	&8	&19.7	&B2G-13-005	&\cite{Khachatryan:2015oba}\\%
&ATLAS	&13	&14.7	&CONF-2016-101	&\cite{ATLAS:2016qlg}\\%
\hline
\multirow{2}{*}{$U \to bW$}
&CMS	&8	&19.7	&B2G-13-005	&\cite{Khachatryan:2015oba}\\%
&ATLAS	&13	&14.7	&CONF-2016-102	&\cite{ATLAS:2016cuv}\\%
\hline
\multirow{4}{*}{$D \to bH$}
&ATLAS	&8	&20.3	&CONF-2015-012	&\cite{ATLAS:2015dka}\\%
&CMS	&8	&19.8	&B2G-12-019	&\cite{CMS:2012hfa}\\%
&CMS	&8	&19.5	&B2G-13-003	&\cite{CMS:2013una}\\%
&CMS	&8	&19.7	&B2G-14-001	&\cite{CMS:2014afa}\\%
\hline
\multirow{3}{*}{$D \to bZ$}
&CMS	&7	&5	&EXO-11-066	&\cite{CMS:2012jwa}\\%
&CMS	&8	&19.5	&B2G-13-003	&\cite{CMS:2013una}\\%
&CMS	&8	&19.7	&B2G-13-006	&\cite{Khachatryan:2015gza}\\%
\hline
\multirow{4}{*}{$D \to tW$}
&ATLAS	&8	&20.3	&EXOT-2013-16	&\cite{Aad:2015gdg}\\%
&CMS	&8	&19.5	&B2G-13-003	&\cite{CMS:2013una}\\%
&CMS	&8	&19.7	&B2G-13-006	&\cite{Khachatryan:2015gza}\\%
&CDF	&1.96	&2.7	&		&\cite{Aaltonen:2009nr}\\%
\hline
\multirow{3}{*}{$Q_{5/3} \to tW$}
&ATLAS	&8	&20.3	&EXOT-2014-17	&\cite{Aad:2015mba}\\%
&CMS	&8	&19.6	&B2G-12-012	&\cite{CMS:vwa}\\%
&CMS	&13	&2.2	&PAS-B2G-15-006	&\cite{CMS:2015alb}\\%
\hline
\end{tabular}
}
\caption{Experimental analyses included in our numerics for heavy quark
partner decay. $Q$ stands for any quark partner where the decay in question is
allowed by electric charges, $j$ stands for a light quark or $b$ jet, and $q$
for a light quark jet.}
\label{tab:app:NMCHM:exp_quark_res}
\end{table}

\newpage

\begin{table}[H]
\centering
\resizebox{!}{265pt}{
\renewcommand{\arraystretch}{1.15}
\begin{tabular}{llllll}
\hline
Decay &Experiment &$\sqrt{s}$ [TeV] &Lum. [fb$^{-1}$] &Analysis &\\
\hline
\multirow{6}{*}{$\eta \to hh$}
&CMS	&8	&19.7	&PAS-EXO-15-008		&\cite{CMS:2015zug}\\%
&ATLAS	&13	&3.2	&EXOT-2015-11		&\cite{Aaboud:2016xco}\\%
&CMS	&13	&2.3	&PAS-HIG-16-002		&\cite{CMS:2016tlj}\\%
&CMS	&13	&2.7	&PAS-B2G-16-008		&\cite{CMS:2016pwo}\\%
&CMS	&13	&12.9	&PAS-HIG-16-029		&\cite{CMS:2016knm}\\%
&CMS	&13	&2.7	&PAS-HIG-16-032		&\cite{CMS:2016vpz}\\%
\hline
\multirow{5}{*}{$\eta \to ZZ$}
&ATLAS	&13	&13.3	&CONF-2016-056		&\cite{ATLAS:2016bza}\\%
&ATLAS	&13	&14.8	&CONF-2016-079		&\cite{ATLAS:2016oum}\\%
&ATLAS	&13	&13.2	&CONF-2016-082		&\cite{ATLAS:2016npe}\\%
&CMS	&13	&12.9	&PAS-HIG-16-033		&\cite{CMS:2016ilx}\\%
&CMS	&13	&2.7	&PAS-B2G-16-010		&\cite{CMS:2016tio}\\%
\hline
\multirow{6}{*}{$\eta \to W^+W^-$}
&ATLAS	&8	&20.3	&EXOT-2013-01$^*$	&\cite{Aad:2015ufa}\\%
&CMS	&8	&19.7	&EXO-13-009$^*$		&\cite{Khachatryan:2014gha}\\%
&ATLAS	&13	&13.2	&CONF-2016-062$^*$	&\cite{ATLAS:2016cwq}\\%
&ATLAS	&13	&13.2	&CONF-2016-074		&\cite{ATLAS:2016kjy}\\%
&CMS	&13	&2.3	&PAS-HIG-16-023		&\cite{CMS:2016jpd}\\%
\hline
\multirow{2}{*}{$\eta \to \gamma\gamma$}
&ATLAS	&13	&15.4	&CONF-2016-059		&\cite{ATLAS:2016eeo}\\%
&CMS	&13	&16.2	&PAS-EXO-16-027		&\cite{CMS:2016crm}\\%
\hline
\multirow{5}{*}{$\eta \to Z\gamma$}
&ATLAS	&13	&13.3	&CONF-2016-044		&\cite{ATLAS:2016lri}\\%
&ATLAS	&13	&3.2	&EXOT-2016-02		&\cite{Aaboud:2016trl}\\%
&CMS	&13	&19.7	&PAS-EXO-16-025		&\cite{CMS:2016mvc}\\%
&CMS	&13	&12.9	&PAS-EXO-16-034		&\cite{CMS:2016pax}\\%
&CMS	&13	&12.9	&PAS-EXO-16-035		&\cite{CMS:2016cbb}\\%
\hline
\multirow{4}{*}{$\eta \to e^+e^-/\mu^+\mu^-$}
&ATLAS	&8	&20.3	&EXOT-2012-23$^*$	&\cite{Aad:2014cka}\\%
&CMS	&8	&20.6	&EXO-12-061$^*$		&\cite{Khachatryan:2014fba}\\%
&ATLAS	&13	&13.3	&CONF-2016-045$^*$	&\cite{ATLAS:2016cyf}\\%
&CMS	&13	&12.4	&PAS-EXO-16-031$^*$	&\cite{CMS:2016abv}\\%
\hline
\multirow{3}{*}{$\eta \to \tau^+\tau^-$}
&ATLAS	&8	&19.5	&EXOT-2014-05$^*$	&\cite{Aad:2015osa}\\%
&CMS	&8	&19.7	&EXO-12-046$^*$		&\cite{CMS:2015ufa}\\%
&CMS	&13	&2.2	&PAS-EXO-16-008$^*$	&\cite{CMS:2016zxk}\\%
&ATLAS	&13	&13.3	&CONF-2016-085		&\cite{ATLAS:2016fpj}\\%
&CMS	&13	&2.3	&PAS-HIG-16-006		&\cite{CMS:2016pkt}\\%
\hline
\multirow{4}{*}{$\eta \to t\bar t$}
&ATLAS	&8	&20.3	&CONF-2015-009$^*$	&\cite{ATLAS:2015aka}\\%
&CMS	&8	&19.7	&B2G-13-008$^*$		&\cite{Khachatryan:2015sma}\\%
&CMS	&13	&2.6	&PAS-B2G-15-002$^*$	&\cite{CMS:2016zte}\\%
&CMS	&13	&2.6	&PAS-B2G-15-003$^*$	&\cite{CMS:2016ehh}\\%
\hline
$\eta \to b\bar{b}$
&CMS	&13	&2.69	&PAS-HIG-16-025		&\cite{CMS:2016ncz}\\%
\hline
$\eta \to qq$
&CMS	&13	&12.9	&PAS-EXO-16-032$^*$	&\cite{CMS:2016wpz}\\%
\hline
$\eta \to gg$
&CMS	&13	&12.9	&PAS-EXO-16-032		&\cite{CMS:2016wpz}\\%
\hline
\multirow{2}{*}{$\eta \to jj$}
&ATLAS	&13	&3.6	&EXOT-2015-02$^*$	&\cite{ATLAS:2015nsi}\\%
&CMS	&13	&2.4	&EXO-15-001$^*$		&\cite{Khachatryan:2015dcf}\\%
\hline
\end{tabular}
}
\caption{Experimental analyses included in our numerics for $\eta$ decay.
The analyses marked with~$^*$ are actually searches for neutral vector resonances. Since for many channels there are no dedicated analyses searching for a neutral scalar resonance and the bounds should be similar, we include the spin-1 analyses in our numerics for $\eta$ decay.}
\label{tab:app:NMCHM:exp_scalar_res}
\end{table}

\begin{table}[H]
\centering
\resizebox{!}{287pt}{
\renewcommand{\arraystretch}{1.15}
\begin{tabular}{llllll}
\hline
Decay &Experiment &$\sqrt{s}$ [TeV] &Lum. [fb$^{-1}$] &Analysis &\\
\hline
\multirow{5}{*}{$\rho^{\pm} \to W^{\pm}h$}
&ATLAS	&8	&20.3	&EXOT-2013-23	&\cite{Aad:2015yza}\\%
&CMS	&8	&19.7	&EXO-14-010	&\cite{CMS:2015gla}\\%
&ATLAS	&13	&3.2	&EXOT-2015-18	&\cite{Aaboud:2016lwx}\\%
&ATLAS	&13	&13.3	&CONF-2016-083	&\cite{ATLAS:2016kxc}\\%
&CMS	&13	&2.17	&PAS-B2G-16-003	&\cite{CMS:2016dzw}\\%
\hline
\multirow{9}{*}{$\rho^{\pm} \to W^{\pm}Z$}
&ATLAS	&8	&20.3	&EXOT-2013-01	&\cite{Aad:2015ufa}\\%
&ATLAS	&8	&20.3	&EXOT-2013-07	&\cite{Aad:2014pha}\\%
&ATLAS	&8	&20.3	&EXOT-2013-08	&\cite{Aad:2015owa}\\%
&CMS	&8	&19.7	&EXO-12-024	&\cite{Khachatryan:2014hpa}\\%
&ATLAS	&13	&15.5	&CONF-2016-055	&\cite{ATLAS:2016yqq}\\%
&ATLAS	&13	&13.2	&CONF-2016-062	&\cite{ATLAS:2016cwq}\\%
&ATLAS	&13	&13.2	&CONF-2016-082	&\cite{ATLAS:2016npe}\\%
&CMS	&13	&2.2	&PAS-EXO-15-002	&\cite{CMS:2015nmz}\\%
&CMS	&13	&12.9	&PAS-B2G-16-020	&\cite{CMS:2016pfl}\\%
\hline
\multirow{4}{*}{$\rho^{\pm} \to tb$}
&CMS	&8	&19.5	&B2G-12-010	&\cite{Chatrchyan:2014koa}\\%
&CMS	&8	&19.7	&B2G-12-009	&\cite{Khachatryan:2015edz}\\%
&CMS	&13	&2.55	&PAS-B2G-16-009	&\cite{CMS:2016ude}\\%
&CMS	&13	&12.9	&PAS-B2G-16-017	&\cite{CMS:2016wqa}\\%
\hline
\multirow{2}{*}{$\rho^{\pm} \to \tau^{\pm}\nu$}
&CMS	&8	&19.7	&EXO-12-011	&\cite{Khachatryan:2015pua}\\%
&CMS	&13	&2.3	&PAS-EXO-16-006	&\cite{CMS:2016ppa}\\%
\hline
\multirow{3}{*}{$\rho^{\pm} \to e^{\pm}\nu/\mu^{\pm}\nu$}
&ATLAS	&7	&4.7	&EXOT-2012-02	&\cite{Aad:2012dm}\\%
&ATLAS	&13	&13.3	&CONF-2016-061	&\cite{ATLAS:2016ecs}\\%
&CMS	&13	&2.2	&PAS-EXO-15-006	&\cite{CMS:2015kjy}\\%
\hline
$\rho^{\pm} \to jj$
&ATLAS	&13	&3.6	&EXOT-2015-02	&\cite{ATLAS:2015nsi}\\%
\hline
\multirow{3}{*}{$\rho^0 \to W^+W^-$}
&ATLAS	&8	&20.3	&EXOT-2013-01	&\cite{Aad:2015ufa}\\%
&CMS	&8	&19.7	&EXO-13-009	&\cite{Khachatryan:2014gha}\\%
&ATLAS	&13	&13.2	&CONF-2016-062	&\cite{ATLAS:2016cwq}\\%
\hline
\multirow{6}{*}{$\rho^0 \to Zh$}
&ATLAS	&8	&20.3	&EXOT-2013-23	&\cite{Aad:2015yza}\\
&CMS	&8	&19.7	&EXO-13-007	&\cite{Khachatryan:2015ywa}\\
&ATLAS	&13	&3.2	&EXOT-2015-18	&\cite{Aaboud:2016lwx}\\
&ATLAS	&13	&3.2	&CONF-2015-074	&\cite{ATLAS-CONF-2015-074}\\
&ATLAS	&13	&13.3	&CONF-2016-083	&\cite{ATLAS:2016kxc}\\
&CMS	&13	&2.17	&PAS-B2G-16-003	&\cite{CMS:2016dzw}\\
\hline
$\rho^0 \to W^+W^-/Zh$
&CMS	&13	&2.2	&PAS-B2G-16-007	&\cite{CMS:2016wev}\\%
\hline
\multirow{4}{*}{$\rho^0 \to e^+e^-/\mu^+\mu^-$}
&ATLAS	&8	&20.3	&EXOT-2012-23	&\cite{Aad:2014cka}\\%
&CMS	&8	&20.6	&EXO-12-061	&\cite{Khachatryan:2014fba}\\%
&ATLAS	&13	&13.3	&CONF-2016-045	&\cite{ATLAS:2016cyf}\\%
&CMS	&13	&12.4	&PAS-EXO-16-031	&\cite{CMS:2016abv}\\%
\hline
\multirow{3}{*}{$\rho^0 \to \tau^+\tau^-$}
&ATLAS	&8	&19.5	&EXOT-2014-05	&\cite{Aad:2015osa}\\%
&CMS	&8	&19.7	&EXO-12-046	&\cite{CMS:2015ufa}\\%
&CMS	&13	&2.2	&PAS-EXO-16-008	&\cite{CMS:2016zxk}\\%
\hline
\multirow{4}{*}{$\rho^0/\rho_G \to t\bar t$}
&ATLAS	&8	&20.3	&CONF-2015-009	&\cite{ATLAS:2015aka}\\%
&CMS	&8	&19.7	&B2G-13-008	&\cite{Khachatryan:2015sma}\\%
&CMS	&13	&2.6	&PAS-B2G-15-002	&\cite{CMS:2016zte}\\%
&CMS	&13	&2.6	&PAS-B2G-15-003	&\cite{CMS:2016ehh}\\%
\hline
\multirow{2}{*}{$\rho^0/\rho_G \to jj$}
&ATLAS	&13	&3.6	&EXOT-2015-02	&\cite{ATLAS:2015nsi}\\%
&CMS	&13	&2.4	&EXO-15-001	&\cite{Khachatryan:2015dcf}\\%
\hline
$\rho^0/\rho_G \to qq$
&CMS	&13	&12.9	&PAS-EXO-16-032	&\cite{CMS:2016wpz}\\%
\hline
\end{tabular}
}
\caption{Experimental analyses included in our numerics for heavy vector
resonance decay.}
\label{tab:app:NMCHM:exp_vector_res}
\end{table}

\end{appendix}

\thispagestyle{empty}
\section*{Acknowledgements}

First and foremost, I thank David Straub so much for his great supervision of my PhD, for the numerous travel opportunities to schools, workshops, and conferences, for all the valuable advice on so many occasions, for proofreading and very useful comments on the manuscript, and for all I have learned about doing research over the past years.
I also thank him for the great open source software he is writing. Without \texttt{flavio}, large parts of this thesis would not have been possible, and \texttt{inspiretools} was a pleasant help in writing it.

Many thanks to Martin Beneke and Andreas Weiler for officially supervising my PhD.
I also want to thank Martin Beneke for his support and valuable advice when I applied for postdoc positions.

Special thanks to Christoph Niehoff for being such a nice officemate and collaborator, for the uncountable hours of discussion that helped in solving so many problems, and for proofreading of the manuscript and the many useful comments.
I also want to thank him again for telling me about the open PhD position in David's group four years ago.
Without this, I would never have started writing this thesis.

Many thanks to Wolfgang Altmannshofer, Francesco Sannino, and Anders Eller Thomsen for collaborations that led to three of the publications presented in this thesis.
I also want to thank Francesco Sannino for inviting me to Odense, for the pleasant time I spent there at CP$^3$-Origins, and for his support of my postdoc applications.

I thank Jason Aebischer for proofreading parts of the manuscript and for useful comments.

Special thanks to my flatmate Pascal for supporting me with his great cooking skills and for the numerous times dinner was ready when I came home late from working on this thesis.

Last but not least, I thank my family so much, in particular my parents, my sister, and my grandparents for their continuous support over all those years.

{\let\thefootnote\relax\footnotetext{{This work was supported by the DFG cluster of excellence ``Origin and Structure of the Universe``.}}}

\addcontentsline{toc}{chapter}{Bibliography}
\bibliographystyle{JHEP}
\bibliography{bibliography}

\end{document}